%% file: Thesis.tex
\documentclass[12pt,a4paper]{book}

\usepackage{amssymb, amsmath, amsfonts}
\usepackage{amsthm, amscd, mathrsfs, helvet, bm, type1cm}
\usepackage[usenames, dvipsnames]{color}
\usepackage{cite, graphicx, verbatim, psfrag, wrapfig, setspace}
\usepackage[all]{xy}
\usepackage[small,nohug,heads=littlevee]{diagrams}

\diagramstyle[labelstyle=\scriptstyle]

\usepackage{sectsty}
\partfont{\centering}

\theoremstyle{plain}
\newtheorem{theorem}{Theorem}[section]
\newtheorem{proposition}[theorem]{Proposition}
\newtheorem{definition}[theorem]{Definition}

\newtheorem{lemma}[theorem]{Lemma}
\newtheorem{corollary}[theorem]{Corollary}

\newcommand{\dub}[1]{{\bf #1}}

\DeclareMathOperator{\tr}{tr} \DeclareMathOperator{\rk}{rank}
\DeclareMathOperator{\im}{im} \DeclareMathOperator{\diag}{diag}
\DeclareMathOperator{\res}{res} \DeclareMathOperator{\ord}{ord}
\DeclareMathOperator{\Div}{Div} \DeclareMathOperator{\Pic}{Pic}

\definecolor{LightBlue}{rgb}{0.02,0.47,0.7}

\newcommand{\black}[0]{\color{Black}}
\newcommand{\red}[0]{\color{Red}}

\newcommand{\lightblue}[0]{\color{LightBlue}}
\newcommand{\green}[0]{\color{Green}}

\numberwithin{equation}{section}

\begin{document}

\input{epsf}

\include{Coverpage}

\pagenumbering{roman}

\include{Acknowledgements}

\include{Declaration}

\include{Abstract}

\tableofcontents

\setcounter{chapter}{-1}

\include{Introduction}


\part{Background} \label{part: Backgro}

    \include{RiemannSurf}
    \include{Semiclassical}


\part{Classical Integrability of String Theory on $\mathbb{R} \times S^3$} \label{part: Integra}

    \include{Strings_on_S3}
    \include{Hamiltonian}
    \include{Integrability}


\part{Finite-Gap Integration of String Theory on $\mathbb{R} \times S^3$} \label{part: Algebro}

    \include{Curves}
    \include{FiniteGap}
    \include{Symplectic}
    \include{Reality}


\part{Applications} \label{part: Semicla}

    \include{PerturbFG}


\part{Conclusions \& Outlook} \label{part: Conclus}

    \include{Conclusion}


\addcontentsline{toc}{chapter}
         {\protect\numberline{Bibliography\hspace{-96pt}}}
\bibliographystyle{utphys}
\bibliography{Thesis}

\end{document}

%% file: Coverpage.tex

\thispagestyle{empty}

\null\vfill
\begin{center}

  {\fontsize{40pt}{40pt}\selectfont \bf Finite-g Strings \par}

\vspace*{2cm}
  {\Large \sc Beno\^{\i}t Vicedo \par
\vspace*{4ex}}
  {\large {{Dissertation submitted for the Degree of} \par}
          {{Doctor of Philosophy} \par}
          {{at the University of Cambridge} \par}}

\vspace*{1.5cm}
  \includegraphics[height=15mm]{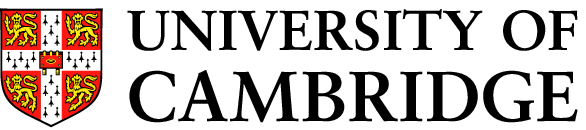} \par

\vspace*{0.5cm}
  {\large \it Department of Applied Mathematics and Theoretical Physics \\
              \& Trinity College\\
              University of Cambridge, UK \par}
\vspace{3ex}
  {\large July 2008}

\end{center}
\null\vfill


\newpage
\thispagestyle{empty}
\qquad
\newpage

%% file: Acknowledgements.tex
\newpage

\begin{center}
  \vspace*{1.5cm}
  {\Large {\bfseries Acknowledgements}}
\end{center}
\addcontentsline{toc}{chapter}{Acknowledgements}

\vspace{0.5cm}

\noindent First of all, I am deeply indebted to my supervisor
Nicholas Dorey, for all his invaluable advice and insight
throughout the whole of my Ph.D, for his guidance during our
collaborative work as well as his thoughtful input into my
independent work.

\noindent I am extremely grateful to Harry Braden, for taking
interest in my work, for reading my papers in great detail, for
raising many important points with regards to technical issues as
well as for the many useful and stimulating discussions on various
aspects of finite-gap integration. Without his ``rigour'' this
thesis would not be complete. I am also very grateful to Harry for
inviting me to give various talks in Edinburgh.

\noindent I would like to thank Marc Magro and Jean-Michel Maillet
from \'Ecole Normale Sup\'erieure de Lyon for giving me the
opportunity to present my work there and interact with the members of
the theoretical physics group. I am especially grateful to Marc for
his careful reading of various parts of my work as well as for
bringing up certain issues that needed elaboration.

\noindent I would also like to thank Keisuke Okamura and Ryo
Suzuki for the fruitful collaboration and the many interesting
email correspondences.

\noindent This work was supported by both a Trinity College
Internal Graduate Studentship and an Engineering and Physical
Sciences Research Council Grant.

\noindent Last but not least, I would like to thank my parents and
brother for all their moral support and constant encouragement
throughout my studies.


\newpage
\quad
\newpage

%% file: Declaration.tex
\newpage

\begin{center}
  \vspace*{1.5cm}
  {\Large {\bfseries Declaration}}
\end{center}
\addcontentsline{toc}{chapter}{Declaration}

\vspace{0.5cm}

\noindent This dissertation is the result of my own work and includes
nothing which is the outcome of work done in collaboration except
where specifically indicated in the text. The research described in
this dissertation was carried out in the Department of Applied
Mathematics and Theoretical Physics, Cambridge University, between
September 2004 and March 2008. Except where reference is made to the
work of others, all the results are original and based on the
following works of mine:

\begin{enumerate}
\item
{\sf \bfseries ``On the Dynamics of Finite-Gap Solutions in Classical String Theory''}
  \\[2mm]
  {}N.~Dorey and B.~Vicedo
  \\{}{\em JHEP} {\bf 0607}, 014 (2006)
  {{\tt hep-th/0601194}}
  \\[-3mm]
\item
{\sf \bfseries ``A Symplectic Structure for String Theory on Integrable Backgrounds''}
  \\[2mm]
  {}N.~Dorey and B.~Vicedo
  \\{}{\em JHEP} {\bf 0703}, 045 (2007)
  {{\tt hep-th/0606287}}
  \\[-3mm]
\item
{\sf \bfseries ``Semiclassical Quantisation of Finite-Gap Strings''}
  \\[2mm]
  {}B.~Vicedo
  \\{}{\em JHEP} {\bf 0806}, 086 (2008)
  {{\tt arXiv:0803.1605 [hep-th]}}
  \\[-3mm]
\end{enumerate}

\noindent These papers are referred to as \cite{Paper1},
\cite{Paper2} and \cite{Paper5} respectively in the bibliography.
The content of Part \ref{part: Integra} is taken mostly from
\cite{Paper2}. Part \ref{part: Algebro} is based on all three
papers \cite{Paper1, Paper2, Paper5} and Part \ref{part: Semicla}
is entirely based on \cite{Paper5}. None of the original works
contained in this dissertation has been submitted by me for any
other degree, diploma or similar qualification.

The following is a list of my other publications, referred to as
\cite{Paper3} and \cite{Paper4} in the bibliography. The main purpose
of these papers is not discussed in this thesis although certain minor
results from them are used:

\begin{enumerate}
\setcounter{enumi}{3}
\item
{\sf \bfseries ``Giant Magnons and Singular Curves''}
  \\[2mm]
  {}B.~Vicedo
  \\{}{\em JHEP} {\bf 0712}, 078 (2007)
  {{\tt hep-th/0703180}}
\\[-3mm]
\item
{\sf \bfseries ``Large winding sector of AdS/CFT''}
  \\[2mm]
  {}H.~Hayashi, K.~Okamura, R.~Suzuki and B.~Vicedo
  \\{}{\em JHEP} {\bf 0711}, 033 (2007)
  {{\tt arXiv:0709.4033 [hep-th]}}
\\[-3mm]
\end{enumerate}

\begin{flushright}
{Beno\^{\i}t Vicedo}\\
\textit{Cambridge, UK}\\
\textit{20th July 2008}
\end{flushright}


%% file: Abstract.tex
\newpage

\begin{center}
  \vspace*{1.5cm}
  {\Large {\bfseries Abstract}}
\end{center}
\addcontentsline{toc}{chapter}{Abstract}

\vspace{0.5cm}

\noindent In view of one day proving the AdS/CFT correspondence, a
deeper understanding of string theory on certain curved
backgrounds such as $AdS_5 \times S^5$ is required. In this
dissertation we make a step in this direction by focusing on
$\mathbb{R} \times S^3$.

\noindent It was discovered in recent years that string theory on
$AdS_5 \times S^5$ admits a Lax formulation. However, the complete
statement of integrability requires not only the existence of a Lax
formulation, but also that the resulting integrals of motion are in
pairwise involution. This idea is central to the first part of this
thesis.

\noindent Exploiting this integrability we apply algebro-geometric
methods to string theory on $\mathbb{R} \times S^3$ and obtain the
general finite-gap solution. The construction is based on an invariant
algebraic curve previously found in the $AdS_5 \times S^5$
case. However, encoding the dynamics of the solution requires
specification of additional marked points. By restricting the
symplectic structure of the string to this algebro-geometric data we
derive the action-angle variables of the system.

\noindent We then perform a first-principle semiclassical
quantisation of string theory on $\mathbb{R} \times S^3$ as a toy
model for strings on $AdS_5 \times S^5$. The result is exactly
what one expects from the dual gauge theory perspective, namely
the underlying algebraic curve discretises in a natural way. We
also derive a general formula for the fluctuation energies around
the generic finite-gap solution. The ideas used can be generalised
to $AdS_5 \times S^5$.


\newpage
\qquad
\newpage

%% file: Introduction.tex
\newpage

\pagenumbering{arabic}

\chapter{Introduction/Review} \label{chapter: intro}

\section{The AdS/CFT conjecture} \label{section: motivation}

Over the past thirty years there has been a fascinating rivalry
between string theory on the one hand and gauge theories on the
other in an attempt to describe the physics of the strong
interaction. Indeed, string theory was originally invented as a
way of describing some of the observed peculiarities of the strong
force between quarks, the quarks being thought of in this theory
as bound together by strings. But this theory of the strong force
never had much success and with the advent of gauge theories it
was soon discarded and replaced by the far more successful QCD
which describes the interaction between quarks in terms of gauge
fields. Later though string theory resurged as a possible
candidate for unifying all the forces of nature. In this modern
interpretation of string theory the strong force is now described
by encapsulating QCD as a low energy part of its dynamics. The
gauge fields however are now derived secondary objects of the
theory, the fundamental objects being the strings themselves.

There is however yet another use of string theory discovered by 't
Hooft \cite{tHooft:1973jz} who realised that perturbation expansions
of $SU(N)$ gauge field theory in the large $N$ limit resemble string
theory genus expansions (see \cite{Coleman:1980nk} for a
review). Loosely speaking, in the $N \rightarrow \infty$ limit (with
the 't Hooft coupling $\lambda \equiv g^2_{YM} N$ held fixed, $g_{YM}$
denoting the gauge theory coupling), each Feynman diagram of the
$SU(N)$ gauge theory can be attributed a topology and the Feynman
diagram expansion breaks up into a sum over topologies. Schematically
we have for example for the free energy
\begin{equation*}
\begin{split}
\mathcal{F} &= N^2 \begin{tabular}{c}
\includegraphics[height=1.1cm]{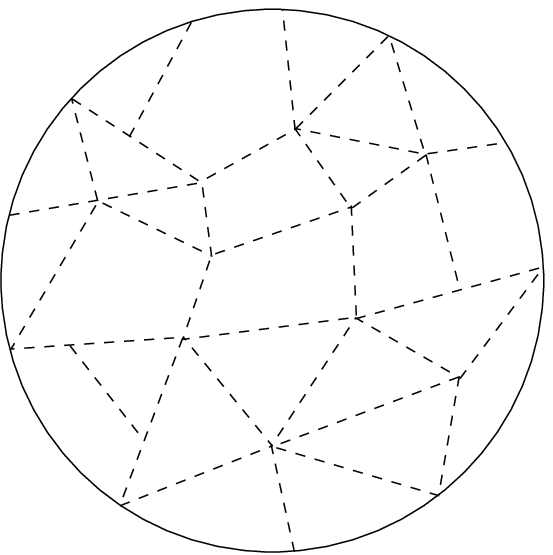}\end{tabular} + 1
\begin{tabular}{c} \includegraphics[height=1.1cm]{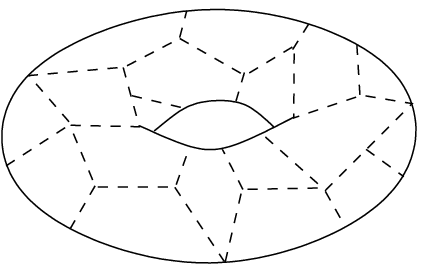}
\end{tabular} + \frac{1}{N^2} \begin{tabular}{c}
\includegraphics[height=1.1cm]{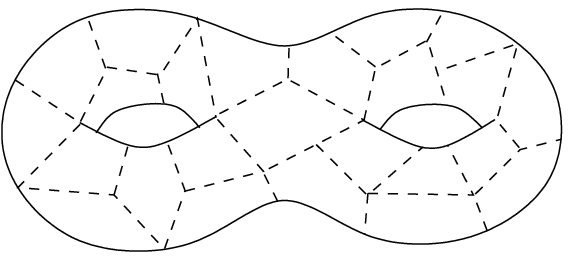} \end{tabular} +
\ldots\\
&= \sum_{g = 0}^{\infty} N^{2 - 2 g} \sum_{l = 0}^{\infty} c_{g,l}
\lambda^l,
\end{split}
\end{equation*}
where each picture in the equation represents the sum over Feynman
diagrams of the given topology. This reorganised sum of Feynman
diagrams resembles a string perturbation expansion over Riemann
surfaces with $1/N$ playing the role of the string coupling $g_S$
and the 't Hooft coupling $\lambda$ related to Planck's constant on the
world sheet. More generally the $N \rightarrow \infty$ limit of
correlation functions of $n$ (single-trace) gauge invariant operators
$\hat{\mathcal{O}}_j$ is schematically given by
\begin{equation*}
\begin{split}
\left\langle \prod_{j=1}^n \hat{\mathcal{O}}_j \right\rangle &= N^{2-n} \begin{tabular}{c}
\psfrag{d1}{\red $\vdots$}\includegraphics[height=1.1cm]{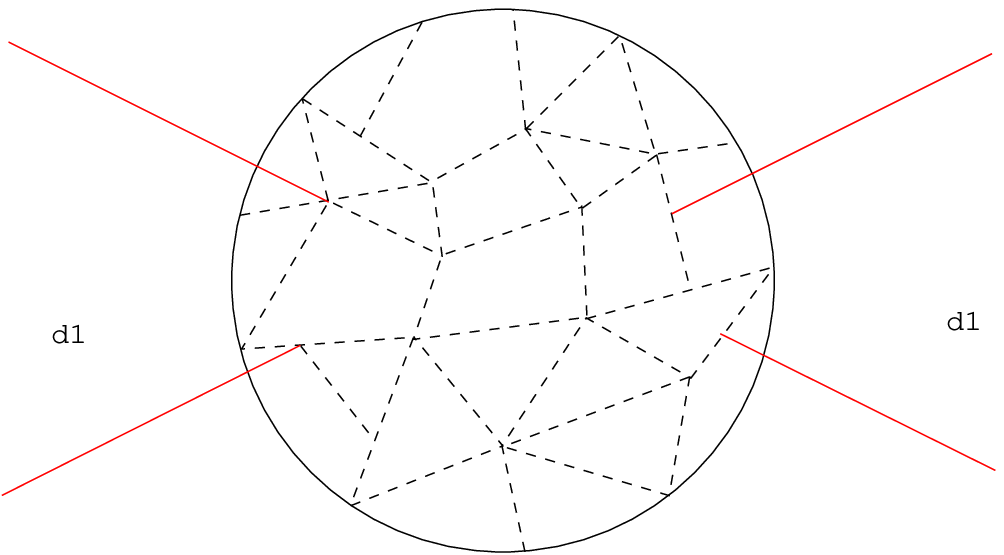}\end{tabular}
+ N^{-n} \begin{tabular}{c} \psfrag{d1}{\red $\vdots$}
\includegraphics[height=1.2cm]{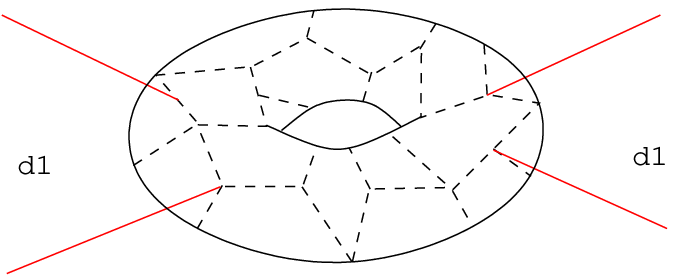}
\end{tabular}\\
&\qquad\qquad\qquad\qquad\qquad\qquad + \frac{1}{N^{2 + n}} \begin{tabular}{c}
\psfrag{d1}{\red $\vdots$}
\includegraphics[height=1.2cm]{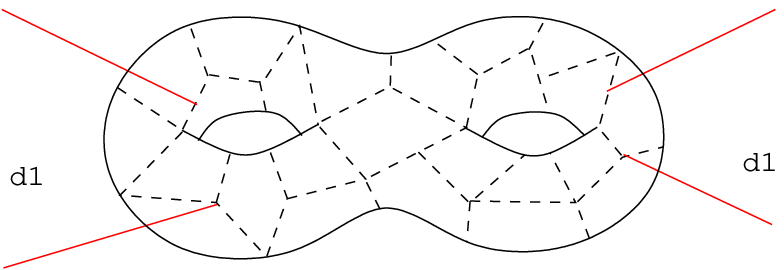} \end{tabular}
+ \ldots \\
&= \sum_{g = 0}^{\infty} N^{2 - 2 g - n} \sum_{l = 0}^{\infty}
c^{(n)}_{g,l} \lambda^l,
\end{split}
\end{equation*}
which in the string theory analogy resembles a correlation
function of $n$ vertex operator insertions on the world sheet. In
particular, any given gauge invariant operator
$\hat{\mathcal{O}}_j(x)$ should correspond to a certain string
theory state $| \mathcal{O}_j \rangle$. Of course the Feynman
diagrams in perturbative ($\lambda \ll 1$) gauge theory are not
literally smooth Riemann surfaces but the Feynman propagators
merely suggest simplicial decompositions of Riemann surfaces. One
can nevertheless imagine how in the $\lambda \gg 1$ regime, which
requires a nonperturbative formulation of the theory, the number
of vertices in a typical diagram would become huge and the Feynman
diagrams would more closely approximate smooth Riemann surfaces.
This beautiful observation about the large $N$ limit of gauge
theories is at the heart of the concept of string/gauge dualities.
Indeed, although the above analogy is far from rigourous it
strongly suggests that gauge theories are intimately related to
string theories on certain backgrounds, in that some gauge
theories may admit dual descriptions in terms of string theories.

The AdS/CFT correspondence due to Maldacena \cite{Maldacena:1997re} is
a conjectured realisation of such a duality for a supersymmetric
cousin of QCD, namely it relates four-dimensional $\mathcal{N} = 4$
supersymmetric Yang-Mills theory (SYM) with gauge group $SU(N)$ to
type IIB superstring theory on $AdS_5 \times S^5$ (see
\cite{Aharony:1999ti} for a review). Concretely, at
large 't Hooft coupling $\lambda \equiv g^2_{YM} N \gg 1$,
$\mathcal{N} = 4$ SYM theory is believed to have a dual description in
terms of type IIB superstring theory on $AdS_5 \times S^5$ with equal
radii of curvature $R$ such that $R / {\alpha'}^{\frac{1}{2}} =
\lambda^{\frac{1}{4}}$. The string coupling in the AdS/CFT
correspondence is not simply $1/N$ as above, but instead is given by
\begin{equation*}
g_S = 4 \pi g^2_{YM} = \frac{4 \pi \lambda}{N}.
\end{equation*}
The extra factor of $\lambda$ however does not affect the
interpretation of the gauge theory perturbation expansions as
genus expansions.

An important part of the AdS/CFT correspondence is establishing a
`dictionary' for translating the language of one theory into the
other. That is, given a gauge theory operator $\hat{\mathcal{O}}(x)$,
we need a way of determining its dual string theory state $|
\mathcal{O} \rangle$ and vice versa. For this it is helpful to
classify the states of both theories according to the global
symmetries present. Both theories share the global (bosonic)
symmetry group $SO(4,2) \times SO(6)$: in gauge theory $SO(4,2)$
corresponds to the conformal symmetry group (in $3+1$ dimensions) and
$SO(6)$ to the R-symmetry (acting for instance in the fundamental
representation on the scalar fields $\{ \phi_i \}_{i=1}^6$ of
$\mathcal{N} = 4$ SYM), whereas on the string theory side $SO(4,2)
\times SO(6)$ is the target space symmetry. States on either side thus
fall into representations of this global symmetry labelled by the
eigenvalues $(E = \Delta, S_1, S_2, J_1, J_2, J_3)$ of the six
Casimirs, the first three being for $SO(4,2)$ and the last three for
$SO(6)$. For instance the complex combinations $\mathcal{Z} = \phi_1 +
i \phi_2$, $\mathcal{W} = \phi_3 + i \phi_4$ and $\mathcal{Y} = \phi_5
+ i \phi_6$ of the $SO(6)$ scalars have R-charges $(J_1, J_2, J_3)$
equal to $(1,0,0)$, $(0,1,0)$ and $(0,0,1)$ respectively.

Note that one of the Casimirs of $SO(4,2)$ plays a distinguished
role. In string theory this is the energy eigenvalue $E$ of the
Hamiltonian $\mathcal{H}_{\text{string}}$ which generates time
translation in $AdS_5$. And according to the AdS/CFT conjecture,
it should be identified with the eigenvalue $\Delta$ of the
Dilation operator $\mathfrak{D}$ of $\mathcal{N} = 4$ SYM.
Therefore if $| \mathcal{O} \rangle$ is a string energy eigenstate
of energy $E$ and $\hat{\mathcal{O}}(x)$ its dual gauge invariant
conformal operator with anomalous dimension $\Delta$, namely
\begin{equation*}
\mathcal{H}_{\text{string}} |\mathcal{O} \rangle =
E\left(\frac{R^2}{\alpha'}, g_S\right) |\mathcal{O} \rangle,
\qquad \mathfrak{D} \hat{\mathcal{O}}(x) =
\Delta\left(\lambda,\frac{1}{N}\right) \hat{\mathcal{O}}(x)
\end{equation*}
then the AdS/CFT conjecture states that
\begin{equation} \label{AdS-CFT dictionary}
\Delta\left(\lambda,\frac{1}{N}\right) =
E\left(\frac{R^2}{\alpha'}, g_S\right).
\end{equation}

Checking \eqref{AdS-CFT dictionary} for arbitrary $N$ seems a
hopeless task since determining the energy spectrum of the string
to all orders in $g_S$ would be incredibly difficult. A more
modest goal, at least initially, would be to check the
correspondence in the 't Hooft limit $N \rightarrow \infty$ where
all diagrams on the gauge theory side become planar, and the
string theory becomes free, \textit{i.e.} the worldsheet is
topologically a sphere. Even with this simplification the duality
is still of strong/weak coupling type and is therefore very hard
to test since the weak coupling regions of both theories (in which
perturbative methods apply) are non-overlapping. Specifically, a
conformal operator in the strong coupling limit $\lambda \gg 1$
should admit an equivalent description in terms of a classical
string ($1/\sqrt{\lambda} \ll 1$), \textit{i.e.} a worldsheet
soliton. Conversely, a string moving on a highly curved background
$\sqrt{\lambda} = R^2/\alpha' \ll 1$ should have an equivalent
description as a weakly coupled ($\lambda \ll 1$) gauge field.
This makes the conjecture very hard to prove since we only have
access to perturbative methods on both sides of the
correspondence.

\section{The large Spin/R-charge limit} \label{section: spin vs r-charge}

Despite the strong/weak coupling obstruction, it was realised in the
work of Berenstein, Maldacena and Nastase \cite{Berenstein:2002jq}
that explicit tests of the correspondence could be made (beyond
sectors protected by supersymmetry) if one took the further limit $J
\rightarrow \infty$ where $J$ is a certain charge, say $J_1$. This
observation was later generalised in a series of papers by Frolov and
Tseytlin \cite{Frolov:2002av, Frolov:2003qc, Frolov:2003tu} to larger
sectors of the correspondence by taking multiple charges to infinity.

To first get an intuitive understanding of the significance of these
large charge limits we go back to the picture of the Feynman graphs
turning into Riemann surfaces. Focusing on the $SO(6)$ scalar sector
of $\mathcal{N} = 4$ SYM, consider single-trace conformal operators
\begin{equation*}
\raisebox{1mm}{$\hat{\mathcal{O}} = \tr(\hat{\Phi}_{i_1} \ldots
\hat{\Phi}_{i_n}) =$} \begin{tabular}{c}
\includegraphics[height=1.1cm]{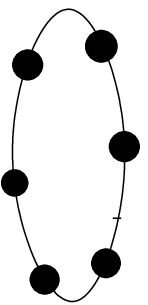}
\end{tabular} \quad \raisebox{1mm}{$(n = 6)$}
\end{equation*}
where in the pictorial representation the black dots each
represent a single operator $\hat{\Phi}_i \in \{ \hat{\mathcal{Z}},
\hat{\mathcal{W}}, \hat{\mathcal{Y}} \}$. They form a closed chain by
virtue of the trace in $\hat{\mathcal{O}}$. Now the 2-point
correlation function of $\hat{\mathcal{O}}$ can be written
symbolically as
\begin{equation*}
\raisebox{2.5mm}{$\langle \hat{\mathcal{O}}(x) \hat{\mathcal{O}}(y) \rangle =$}
\begin{tabular}{c} \psfrag{x}{\tiny $x$} \psfrag{y}{\tiny $y$}
\includegraphics[height=1.4cm]{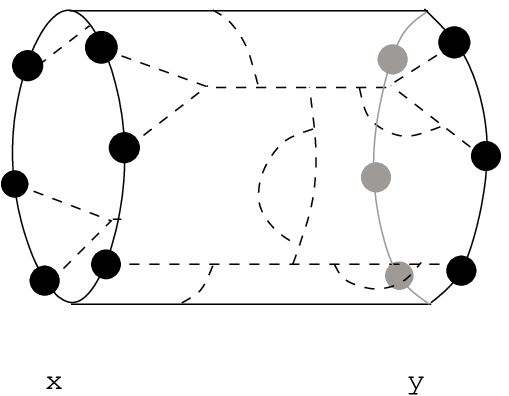}
\end{tabular} \raisebox{2.5mm}{,}
\end{equation*}
where the right hand side represents the sum of all possible Wick
contractions, \textit{i.e.} Feynman diagrams connecting the
operators $\hat{\mathcal{O}}$ at $x$ and $y$. As before the
Feynman diagrams suggest a simplicial decomposition of a Riemann
surface (with boundaries). This simplicial decomposition may be
refined in two ways: either one increases the coupling $\lambda$
as before to increase the number of vertices in these Feynman
diagrams, or one can also increase the number $n$ of constituent
operators $\hat{\Phi}_i$ in $\hat{\mathcal{O}}$.

For example a BMN operator $\tr (\hat{\mathcal{Z}}^{J-2}
\hat{\mathcal{W}} \hat{\mathcal{Y}} + \ldots)$ is made up of a
large number of reference fields $\hat{\mathcal{Z}}$ and a small
number of other ``impurity'' fields $\hat{\mathcal{W}}$ and
$\hat{\mathcal{Y}}$. Its string theory dual, the BMN string, is
almost point-like and has angular momentum $J_1 = J$ on $S^5$.
More generally an operator may contain a large amount of
impurities such as $\tr (\hat{\mathcal{Z}}^{J_1}
\hat{\mathcal{W}}^{J_2} \hat{\mathcal{Y}}^{J_3} + \ldots)$ with $J
= J_1 + J_2 + J_3$. Its string theory dual, the Frolov-Tseytlin
string, is spatially extended and spins with the three different
angular momenta $J_1$, $J_2$ and $J_3$ on $S^5$. As explained
above one expects such `long' ($J \rightarrow \infty$)
single-trace conformal operators to have a stringy behaviour even
at weak coupling $\lambda \ll 1$.

Concretely, suppose one can expand both sides of \eqref{AdS-CFT
dictionary} in terms of $\lambda/J^2$ and $1/J$. On the string
side this is achieved by doing a semiclassical expansion in $1/J
\propto 1/\sqrt{\lambda} \propto \alpha'$ with $\lambda/J^2$ held
fixed. On the gauge side one could first expand in $\lambda \ll 1$
and then further expand each coefficient in $1/J$. When such
expansions for the semiclassical energy $E$ and the perturbative
anomalous dimension $\Delta$ exist and take on the similar form
\begin{equation} \label{double expansion}
J \left[ 1 + \sum_{n = 1}^{\infty} \left( c_n +
\sum_{k=1}^{\infty} \frac{c_{nk}}{J^k}\right)
\left(\frac{\lambda}{J^2}\right)^n \right],
\end{equation}
then their respective coefficients, for $c_n$ say, could be
compared directly, even though they have been obtained differently
from both sides of the duality.

With this procedure for making quantitative tests of the
correspondence in place, the immediate goal from both sides of the
duality is clear. From the gauge theory perspective one
faces the problem of diagonalising the dilatation operator
$\mathfrak{D}$ on long single-trace conformal operators perturbatively
in $\lambda \ll 1$. Since it commutes with the Casimirs of $SO(4,2)
\times SO(6)$ it does not mix operators of different weights. For
instance, its action on the complete set of operators
$\hat{\mathcal{O}}^{J_1, J_2}_{\alpha}$ composed solely of the two
scalars $\hat{\mathcal{Z}}$ and $\hat{\mathcal{W}}$ is given by
\begin{subequations} \label{SU2 task}
\begin{equation} \label{SU2 gauge theory task}
\mathfrak{D} \hat{\mathcal{O}}^{J_1, J_2}_{\alpha}(x) = \sum_{\beta}
\mathfrak{D}_{\alpha \beta} \hat{\mathcal{O}}^{J_1, J_2}_{\beta}(x).
\end{equation}
The problem is therefore reduced to diagonalising the matrix
$\mathfrak{D}_{\alpha \beta}$. However, since we are interested in the
limit $J_1, J_2 \rightarrow \infty$ this simple diagonalisation task
quickly becomes intractable without recourse to numerical methods.

The task on the string theory side is to obtain the semiclassical
energy spectrum of strings on $AdS_5 \times S^5$ to leading order in
$1/\sqrt{\lambda} \ll 1$. This in turn requires complete knowledge of
the classical string motions on such a background. Restricting
attention to the $SU(2)$ sector corresponding to the operators
$\mathcal{O}^{J_1, J_2}_{\alpha}$ discussed above, the problem is
reduced to finding the general solution to the equations of motion for
a string moving on $\mathbb{R} \times S^3$. However, the equations of
motion for the fields $\{ X_i \}_{i=1}^4$ describing the embedding of
the string into $S^3$,
\begin{equation} \label{SU2 string theory task}
\partial_{\alpha} \partial^{\alpha} X_i + \left( \sum_j
\partial_{\alpha} X_j \partial^{\alpha} X_j \right) X_i = 0,
\end{equation}
\end{subequations}
are second order nonlinear partial differential equations subject to
the constraint $\sum_{i=1}^4 X_i^2 = 1$. Solving them exactly
therefore seems quite intractable as well.

\section{Classical/Quantum Integrability} \label{section: toolkit}

Fortunately, something of a miracle happens in both cases. By
computing the 1-loop planar dilatation operator on single-trace
operators of all six scalar fields of $\mathcal{N} = 4$ SYM, Minahan
and Zarembo \cite{Minahan:2002ve} discovered it was proportional to
the Hamiltonian of the $\mathfrak{so}(6)$ integrable spin chain with
nearest-neighbour interactions. Subsequently the complete one-loop
planar dilatation operator of $\mathcal{N} = 4$ SYM was computed by
Beisert \cite{Beisert:2003jj, Beisert:2004ry} and identified with an
$\mathfrak{su}(2,2 | 4)$ super spin chain by Beisert and Staudacher in
\cite{Beisert:2003yb}. Integrability also seems to persist at higher
loops \cite{Beisert:2003tq, Beisert:2003ys}. For the purpose of this
thesis we shall focus on the $SU(2)$ sector at one-loop where the
planar dilatation operator reduces to the famous Heisenberg
$\text{XXX}_{\frac{1}{2}}$ spin chain Hamiltonian which is
\textit{quantum integrable}. Specifically we have
\begin{equation} \label{planar dilatation operator}
\begin{split}
\mathfrak{D}_{SU(2)}^{\text{planar}} &= J + \frac{\lambda}{16 \pi^2}
\sum_{j = 1}^{J} (1 - \vec{\sigma}_j \cdot \vec{\sigma}_{j+1}) +
O(\lambda^2)\\ &= J + \frac{\lambda}{4 \pi^2}
\hat{\mathcal{H}}_{\text{XXX}_{\frac{1}{2}}} + O(\lambda^2),
\end{split}
\end{equation}
where $\vec{\sigma}_j = (\sigma^{\alpha}_j)_{\alpha = 1}^3$ is the
set of Pauli matrices acting on the $j^{\text{th}}$ site of the
spin chain. The tree-level term in \eqref{planar dilatation
operator} is just the common engineering dimension $J = J_1 + J_2$
of the operators $\hat{\mathcal{O}}^{J_1,J_2}$, which is also just
the length of the spin chain.

The fact that the one-loop planar dilatation operator
\eqref{planar dilatation operator} is integrable implies that it
can be diagonalised analytically for any length $J$. As usual, the
definition of quantum integrability requires the existence of a
maximal set of commuting operators which includes the Hamiltonian.
The construction of such operators in the Heisenberg
$\text{XXX}_{\frac{1}{2}}$ spin chain proceeds in the usual way
(see \cite{Chowdhury, Faddeev:1996iy, Nepomechie:1998jf} for a general
discussion on quantum integrable systems) by defining the Lax operator
$\hat{L}_{j,a}(u) = u {\bf 1}_j \otimes {\bf 1}_a + \frac{i}{2}
\sum_{\alpha} \sigma^{\alpha}_j \otimes \sigma^{\alpha}_a$ where $u
\in \mathbb{C}$ is called the \dub{spectral parameter}. Here the
subscript $j$ indicates that the matrix acts on the $j^{\text{th}}$
site of the spin chain and the subscript $a$ indicates that the matrix
acts on an extra `auxiliary' site. The main object of interest is the
\dub{monodromy matrix} $\hat{T}_a(u) = \hat{L}_{J,a}(u) \ldots
\hat{L}_{1,a}(u)$ (which acts on all $J$ sites as well as the
auxiliary site). Writing out the action on the auxiliary site in
matrix form it reads
\begin{equation*}
\hat{T}_a(u) = \left( \begin{array}{cc} \hat{A}(u) &
\hat{B}(u)\\ \hat{C}(u) & \hat{D}(u) \end{array}
\right).
\end{equation*}
Its trace over the auxiliary site $\hat{T}(u) = \tr_a
\hat{T}_a(u) = \hat{A}(u) + \hat{D}(u)$, the \dub{transfer matrix},
generates the desired family of commuting operators since one can show
\cite{Faddeev:1996iy}
\begin{equation*}
[ \hat{T}(u), \hat{T}(v) ] = 0, \qquad \forall u, v \in \mathbb{C}.
\end{equation*}
In particular the Hamiltonian can be extracted as
$\hat{\mathcal{H}}_{\text{XXX}_{\frac{1}{2}}} = \left. \frac{i}{2}
\frac{d}{d u} \log \hat{T}(u) \right|_{u = \frac{i}{2}} -
\frac{J}{2}$.

The diagonalisation of $\hat{\mathcal{H}}_{\text{XXX}_{\frac{1}{2}}}$
can therefore be achieved by simultaneously diagonalising the whole
family of operators $\hat{T}(u)$. For this one defines a reference
state $| \Omega \rangle$ on the spin chain by the condition
$\hat{C}(u) | \Omega \rangle = 0$ and looks for eigenvectors of the
form
\begin{equation} \label{Bethe state}
| u_1, \ldots, u_M \rangle = \hat{B}(u_1) \ldots \hat{B}(u_M) | \Omega
\rangle.
\end{equation}
This is akin to the Fock space construction where the operator
$\hat{B}(u)$ creates a magnon excitation on the spin chain with
rapidity $u$. One can show that \eqref{Bethe state} is an eigenstate
of the transfer matrix $\hat{T}(u)$ if and only if the parameters
$u_j$ satisfy the famous \dub{Bethe equations} which in this sector
read \cite{Faddeev:1996iy, Nepomechie:1998jf}
\begin{equation} \label{Bethe equations}
\left( \frac{u_j + \frac{i}{2}}{u_j - \frac{i}{2}} \right)^J =
\prod_{k = 1 \; (k \neq j)}^M \frac{u_j - u_k + i}{u_j - u_k + i}.
\end{equation}
The solutions $u_j \in \mathbb{C}$ of these equations are called
\dub{Bethe roots}.

\begin{figure}[ht]
\centering \psfrag{C1}{\footnotesize $\mathcal{C}_1$}
\psfrag{C2}{\footnotesize $\mathcal{C}_2$}
\psfrag{C3}{\footnotesize $\mathcal{C}_3$}
\includegraphics[height=25mm]{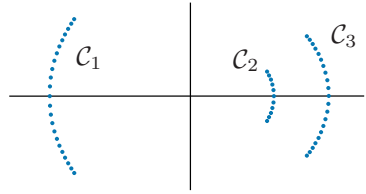}
\caption{Typical configuration of Bethe roots as $J \rightarrow
\infty$.} \label{fig: Bethe root cuts}
\end{figure}

To study the limit $J \rightarrow \infty$ of \eqref{Bethe equations}
one starts by taking its logarithm,
\begin{equation} \label{log Bethe equations}
J \log \frac{u_j + \frac{i}{2}}{u_j - \frac{i}{2}} = \prod_{k = 1 \;
(k \neq j)}^M \log \frac{u_j - u_k + i}{u_j - u_k + i} - 2 \pi i n_j,
\end{equation}
where the \dub{mode numbers} $n_j \in \mathbb{Z}$ specify the
branch of the logarithm. A careful study of these equations
determines the location of the Bethe roots in the limit $J
\rightarrow \infty$. Since all Bethe roots are of order $u_j \sim J$
it is convenient to introduce the scaled spectral parameter $x$ by $u
= J x$. If the number of mode numbers is finite, say $\{ n_I
\}_{I=1}^K$, and the number of Bethe roots with the same mode number
is of order $J$ then one finds that the Bethe roots of a given mode
number $n_I$ all agglomerate into a vertical `cut' $\mathcal{C}_I$ in
the complex plane, see Figure \ref{fig: Bethe root cuts}.
To characterise the density of the Bethe roots along the various
cuts one introduces a function $p(x)$ on the complex plane called the
\dub{quasi-momentum} which can then be shown to have a simple pole at
$x = 0$ and the property that its value jumps by $2 \pi n_I$ across
$\mathcal{C}_I$ (see \cite{KMMZ} for details). Moreover, its integral
around any cut $\mathcal{C}_I$ gives exactly the proportion of Bethe
root lying on $\mathcal{C}_I$ called the \dub{filling fraction},
\begin{equation} \label{filling fraction: gauge theory}
\mathcal{S}_I = \frac{1}{2 \pi i} \oint_{\mathcal{A}_I} p(x) dx,
\qquad I = 1, \ldots, K
\end{equation}
where $\mathcal{A}_I$ is a contour around the cut $\mathcal{C}_I$.
Now by construction, a distribution of Bethe roots like the one in
Figure \ref{fig: Bethe root cuts} characterises the $J \rightarrow
\infty$ limit of a single-trace eigen-operator of the
one-loop planar dilatation operator \eqref{planar dilatation
operator}. Therefore by the reasoning of section \ref{section: spin vs
r-charge} we expect it to match the description of a classical string
solution on $\mathbb{R} \times S^3$. To see this we now turn to the
string theory side.

Recall that the task there involves finding exact solutions to a set of
non-linear second order partial differential equations \eqref{SU2
string theory task} subject to a constraint, which in general is
impossible. Fortunately, it was discovered by Bena, Polchinski and
Roiban \cite{Bena:2003wd} that the equations of motion for a
superstring on $AdS_5 \times S^5$ can be formulated as a flatness
condition for a 1-parameter family of currents $J(x)$ depending on a
complex parameter $x \in \mathbb{C}$. This is a necessary condition
for the theory to be \textit{classically integrable}. In the $SU(2)$
sector the lightcone components of these currents $J(x)$ are
\begin{equation} \label{1-param flat connections}
J_{\pm}(x) = \frac{j_{\pm}}{1 \mp x}, \qquad \partial_+ J_-(x) -
\partial_- J_+(x) + [J_+(x), J_-(x)] = 0.
\end{equation}
This connection is built out of $j = - g^{-1} dg \in \mathfrak{su}(2)$
where $g \in SU(2)$ depends on the fields $\{ X_i \}_{i=1}^4$ and
specifies the embedding of the string into $SU(2) \simeq S^3$. The
flatness condition \eqref{1-param flat connections} is equivalent to
the equations of motion \eqref{SU2 string theory task}. As we will
show in this thesis, when written in this form \eqref{1-param flat
connections} the equations of motion can be solved exactly.

\begin{figure}[ht]
\centering \psfrag{C1}{\footnotesize $\mathcal{C}_1$}
\psfrag{C2}{\footnotesize $\mathcal{C}_2$}
\psfrag{C3}{\footnotesize $\mathcal{C}_3$}
\includegraphics[height=25mm]{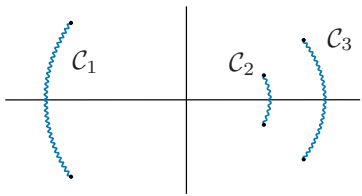}
\caption{Cut representation of a genus two Riemann surface.}
\label{fig: Cut plane genus 2}
\end{figure}

As we review in chapter \ref{chapter: curves}, the zero-curvature
representation \eqref{1-param flat connections} of the equations
of motion directly leads to the construction of an algebraic curve
$\hat{\Sigma}$ equipped with a meromorphic differential $dp$,
starting from a given solution $X^{\text{sol}}_i$ to \eqref{SU2
string theory task}. In other words \eqref{1-param flat
connections} provides an assignment
\begin{equation} \label{sol -> curve}
X_i^{\text{sol}} \quad \longrightarrow \quad ( \hat{\Sigma}, dp ).
\end{equation}
Moreover, the pair $( \hat{\Sigma}, dp )$ is independent of the
worldsheet $(\sigma, \tau)$-coordinates and therefore encodes the
integrals of motion of the solution $X^{\text{sol}}_i$. Thus all
solutions to \eqref{SU2 string theory task} on the string theory
side are classified by their respective algebraic curves. In the
$SU(2)$ sector these curves are all hyperelliptic and can be
represented in terms of cuts in the complex plane. In chapter
\ref{chapter: reality} we will give a proof of the usual assumption
that these cuts are all vertical in the complex plane, see Figure
\ref{fig: Cut plane genus 2} (note that the path taken by the cuts is
arbitrary as long as they join up all the branch points in pairs).
The remarkable similarity between Figures \ref{fig: Bethe root cuts}
and \ref{fig: Cut plane genus 2} was first discovered by Kazakov,
Marshakov, Minahan and Zarembo in their seminal paper \cite{KMMZ}
(see \cite{Zarembo:2004hp, Marshakov:2004py} for shorter reviews).
The quasi-momentum on the gauge theory side is identified here
with the Abelian integral $p(x) = \int^x dp$ since its value also
jumps across cuts $\mathcal{C}_I$ by $2 \pi n_I$, $n_I \in
\mathbb{Z}$. It also has simple poles but this time they are at $x =
\pm 1$ rather than $x = 0$. This is because to compare with the gauge
theory one needs to scale the spectral parameter on the string theory
side by setting $\tilde{x} = \frac{\sqrt{\lambda}}{4 \pi J} x$ so that
$p(\tilde{x})$ now has poles at $\tilde{x} = \pm \sqrt{T}$ where $T
\equiv \frac{\lambda}{16 \pi^2 J^2}$. In the limit
$\frac{\lambda}{J^2} \rightarrow 0$ the string theory then exactly
reproduces the one-loop gauge theory result \cite{KMMZ}.

As we discussed above, by virtue of quantum integrability
the one-loop planar dilatation operator
$\hat{\mathcal{H}}_{\text{XXX}_{\frac{1}{2}}}$ belongs to a whole
family of commuting operators encoded in the transfer matrix
$\hat{T}(u)$. Likewise, as we will see in chapter \ref{chapter:
integrability}, on the string theory side the energy is the first
member of a whole hierarchy of conserved Poisson commuting charges
encoded in a classical analogue $\tr \Omega(x)$ of the transfer
matrix. Now by construction, a distribution of Bethe roots
characterises an eigen-operator of $\hat{T}(u)$ and an algebraic curve
$( \hat{\Sigma}, dp )$ characterises a classical string
solution. Therefore the matching of the classical string theory
algebraic curve with the thermodynamic limit ($J \rightarrow \infty$)
of the one-loop Bethe root distribution provides a complete check in
the $SU(2)$ sector of the equality between the coefficients $c_1$ in
the expansion \eqref{double expansion} for the spectrum of the quantum
operator $\hat{T}(u)$ on the one hand and the range of the classical
phase-space function $\tr \Omega(x)$ on the other. The construction of
the algebraic curve was later generalised to the $SO(6)$ sector
\cite{Beisert:2004ag}, to the non-compact $SL(2, \mathbb{R})$ sector
\cite{Kazakov:2004nh} and eventually to the full supersymmetric case
\cite{Beisert:2005bm}. This curve was then successfully compared in
\cite{Beisert:2005di} against the full spectrum of $\mathcal{N} = 4$
SYM single-trace operators in the Frolov-Tseytlin limit.

To take the comparison to the next order in $\frac{\lambda}{J^2}
\propto T$ it was shown in \cite{KMMZ} that a further change of
spectral parameter was necessary on the gauge theory side. If one
first renames the spectral parameter $x$ as $\tilde{z}$, so that
equations such as \eqref{filling fraction: gauge theory} now read the
same with the relabelling $x \rightarrow \tilde{z}$,
\begin{equation} \label{filling fraction: gauge theory 2}
\mathcal{S}_I = \frac{1}{2 \pi i} \oint_{\mathcal{A}_I} p(\tilde{z})
d\tilde{z}, \qquad I = 1, \ldots, K
\end{equation}
then the change of spectral parameter $\tilde{z} \mapsto \tilde{x}$
required to match the string theory results (expressed in terms
$\tilde{x} = \sqrt{T} x$) is defined by the Zhukovsky map
\begin{equation} \label{Zhukovsky map}
\tilde{z} = \tilde{x} + \frac{T}{\tilde{x}}.
\end{equation}
This can also be written as $z = x + \frac{1}{x}$ in terms of the
unscaled variables $x = \frac{1}{\sqrt{T}} \tilde{x}$ and $z =
\frac{1}{\sqrt{T}} \tilde{z}$. As we will show in chapter
\ref{chapter: symplectic} the spectral parameter $z = x +
\frac{1}{x}$ is in fact the natural choice on the string theory side
since it brings the symplectic structure to the canonical Darboux
form. Furthermore, the filling fractions are also naturally expressed
in terms of it, as in \eqref{filling fraction: gauge theory 2}. With
this change of variables the two-loop gauge theory result was shown to
exactly match the next order in $T \propto \frac{\lambda}{J^2}$ of the
classical string theory algebraic curve (see \cite[p27]{KMMZ} for
details). This provides a test of the correspondence in the $SU(2)$
sector at the level of the coefficient $c_2$ in the expansion
\eqref{double expansion}. Despite this perfect agreement at two-loop,
the next coefficient $c_3$ in the expansion \eqref{double expansion}
on both sides of the correspondence were found to disagree, which has
become known as the `three-loop discrepancy'
\cite{Serban:2004jf}. This mismatch however is not in conflict with
the AdS/CFT correspondence and can be attributed to an order-of-limits
effect \cite{Beisert:2004hm, Beisert:2005cw}. Indeed, on the string
theory side one takes the classical limit $1/J \rightarrow 0$ before
expanding in $\lambda' \equiv \lambda / J^2$ whereas on the gauge
theory side the perturbation expansion in $\lambda$ precedes the
expansion in $1/J$. In other words, the procedures described in
section \ref{section: spin vs r-charge} for testing the AdS/CFT
correspondence rely on the assumption that the following diagram
\cite{Beisert:2004hm}
\begin{equation*}
\begin{CD}
\Delta(\lambda, J) = E(\lambda, J) @>J \rightarrow \infty>\lambda' =
\frac{\lambda}{J^2} \text{ fixed}> E(\lambda')\\
@V\lambda \text{ expansion}VV   @VV\lambda' \text{ expansion}V\\
\Delta_n(J) @>J \rightarrow \infty>> \Delta_n \overset{?}= E_n\\
\end{CD}
\end{equation*}
is commutative. Yet, assuming the AdS/CFT correspondence holds, the
mismatch $\Delta_3 \neq E_3$ at three-loop clearly shows otherwise and
with hindsight the agreement for the coefficients $c_1$ and $c_2$
seems quite fortuitous.

One way to circumvent this difficulty would be to directly quantise
string theory on $AdS_5 \times S^5$. The main objective of the work
presented in this thesis was to make a step towards obtaining the
leading semiclassical corrections to the string spectrum and possibly
gain some insight in view of one day performing an exact quantisation
of string theory on $AdS_5 \times S^5$. The more modest task of
obtaining the semiclassical string spectrum would provide the set of
coefficients $c_{n1}$ in the expansion \eqref{double expansion} from
the string theory side. These could then be perturbatively tested
against the corresponding coefficients obtained from the gauge theory
side. In this short introduction we have mostly been concerned with
the $SU(2)$ sector corresponding classically to bosonic strings moving
in an $\mathbb{R}\times S^3$ submanifold of $AdS_5 \times S^5$. This
restriction is legitimate because at the classical level it is a
consistent truncation of the full superstring theory on $AdS_5
\times S^5$. At the quantum level however, even if we
semiclassically quantise a solution in the subspace $\mathbb{R}
\times S^3$ we know that quantum fluctuations will leave this
subspace and so quantum mechanically one ought to consider the
full target-space $AdS_5 \times S^5$. Despite this, in this thesis
we will continue focusing on the subspace $\mathbb{R} \times S^3
\subset AdS_5 \times S^5$ as a toy model. The reason for doing
this is that the $SU(2)$ subsector is the only one for which the
complete set of solutions is explicitly know \cite{Paper1,
Paper2}, which is a necessary prerequisite for performing a
semiclassical study of any system.

\section{Outline of the thesis} \label{section: outline}

\noindent {\bf Part \ref{part: Backgro}} The first two chapters of
this thesis contain all the necessary background material on the
theory of Riemann surfaces \cite{Reyssat, Schlichenmaier, Farkas,
Jost, Miranda, Kirwan, Fay, Bobenko, Fedorov} and semiclassical
quantisation of finite-dimensional systems \cite{Bates:1997kc,
Sjostrand, Martinez, BerryTabor1, BerryTabor2, Voros, Voros1,
Voros2, VuNgoc1, VuNgoc2} required for Parts \ref{part: Algebro}
and \ref{part: Semicla} respectively. Since the theory of Riemann
surfaces plays such an important role in Part \ref{part: Algebro},
for completeness we cover the relevant aspects of it in some detail in
chapter \ref{chapter: Riemann surfaces}.

\noindent {\bf Part \ref{part: Integra}} In chapter \ref{chapter:
strings on RxS3} we give a review of bosonic strings theory on
$\mathbb{R} \times S^3$ from the Lagrangian point of view and
express it in terms of the $SU(2)$ principal chiral model subject
to the Virasoro constraints. In chapter \ref{chapter: hamiltonian}
we rephrase everything from the Hamiltonian perspective discussing
the implementation of the Virasoro and static gauge constraints in
the Dirac formalism. Finally, in chapter \ref{chapter:
integrability} we tackle the question of integrability of bosonic
strings on $\mathbb{R} \times S^3$. We start by reviewing the
construction of the Lax connection and monodromy matrix in section
\ref{section: conserved charges} and the extraction of the local
conserved charges in section \ref{section: local charges}. Section
\ref{section: involution} is based on \cite{Paper2} in which we
show that the integrals of motion previously obtained are also in
involution. This is the complete statement of integrability of
string theory on $\mathbb{R} \times S^3$. We then exploit this in
section \ref{section: hierarchy} to construct the integrable
hierarchy of the string as in \cite{Paper5}.

\noindent {\bf Part \ref{part: Algebro}} In this Part we put to
full use the integrability unveiled in Part \ref{part: Integra} to
construct the general solution to the equations of motion for a
string on $\mathbb{R} \times S^3$ following \cite{Paper1, Paper2,
Paper5} as well as \cite{Paper3} for the last section. Section
\ref{chapter: curves} is a review of the construction of the KMMZ
curve \cite{KMMZ} encoding the integrals of motion of a finite-gap
solution. We show in section \ref{chapter: finite-gap} that the
reconstruction of the solution requires additional data, namely a
finite set of points on the KMMZ curve. This completes the set of
so called algebro-geometric data. We express the general
finite-gap solution explicitly in terms of this data using Riemann
$\theta$-functions on the curve. In section \ref{chapter:
symplectic} we derive the restriction of the symplectic structure
of the string to the algebro-geometric data. The resulting
finite-dimensional symplectic structure is canonical if the
spectral parameter used is given by the Zhukovsky map. We then
perform a standard change of variables to action-angle variables,
obtaining explicit expressions for these in terms of the
algebro-geometric data. In section \ref{chapter: reality} we
discuss the necessary constraints on the data to obtain physical
finite-gap solutions. In particular we derive the reality
conditions on the KMMZ curve, showing that all the branch points
must lie off the real axis in the $SU(2)$ sector.

\noindent {\bf Part \ref{part: Semicla}} In chapter \ref{chapter:
semi} we use the knowledge of classical solutions acquired in Part
\ref{part: Algebro} to perform a semiclassical analysis of bosonic
string theory on $\mathbb{R} \times S^3$ from first principles. We
derive a general and simple formula for extracting the fluctuation
energies from the KMMZ curve in terms of a well defined
meromorphic differential on the curve, namely the quasi-energy. We
use these fluctuation energies to show formally (without
regularising) that their sum leads to the discretisation of the
KMMZ curve in the sense that all the fillings get half-integer
quantised, including those of the singular points which are
classically empty. The calculation therefore serves as a toy model
for understanding from the finite-gap perspective the origin of
the discretisation of the algebraic curve when leading order
semiclassical corrections are included.

\begin{figure}[ht]
\centering
\psfrag{1}{\footnotesize ${\bf 1}$}
\psfrag{2}{\footnotesize ${\bf 2}$}
\psfrag{3}{\footnotesize ${\bf 3}$}
\psfrag{4}{\footnotesize ${\bf 4}$}
\psfrag{6}{\footnotesize ${\bf 6}$}
\psfrag{7}{\footnotesize ${\bf 7}$}
\psfrag{8}{\footnotesize ${\bf 8}$}
\psfrag{9}{\footnotesize ${\bf 9}$}
\psfrag{10}{\footnotesize ${\bf 10}$}
\psfrag{51}{\footnotesize ${\bf 5.1}$}
\psfrag{52}{\footnotesize ${\bf 5.2}$}
\psfrag{53}{\footnotesize ${\bf 5.3}$}
\psfrag{54}{\footnotesize ${\bf 5.4}$}
\includegraphics[height=50mm]{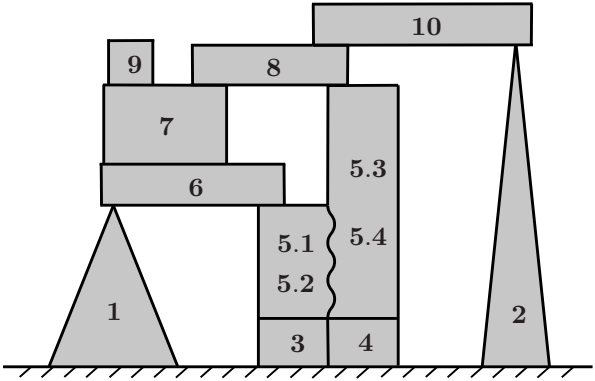}
\caption{Chapter dependence guideline.}
\end{figure}

%% file: RiemannSurf.tex
\newpage

\chapter{Riemann surfaces} \label{chapter: Riemann surfaces}

\begin{flushright}
{\small \textit{``Donuts. Is there anything they can't do?''}}\\
{\small Homer Simpson}
\end{flushright}
\vspace{1cm}

\noindent This chapter is intended as a self contained review, based
on \cite{Reyssat, Schlichenmaier, Farkas, Jost, Miranda, Kirwan, Fay,
Bobenko, Fedorov}, of those aspects from the theory of Riemann surfaces
relevant to Part \ref{part: Algebro} of this thesis. The most
important concepts and results required in the theory of finite-gap
integration are found in section \ref{section: analytic structure}.
Section \ref{section: algebraic curves} is a
discussion of singular algebraic curves which are fundamental to
chapters \ref{chapter: curves} and \ref{chapter: semi}. Finally,
section \ref{section: jacobian} discusses the relation of a curve to
its Jacobian, an object of great importance in Parts \ref{part:
Algebro} and \ref{part: Semicla}.


\section{Definition \& Examples} \label{section: definition}


Consider a real two-dimensional (connected) topological manifold
$M$, that is a second-countable Hausdorff space locally
homeomorphic to $\mathbb{R}^2$, and let $\{ U_{\alpha} \}_{\alpha
\in A}$ be an open cover of $M$, \textit{i.e.} $\cup_{\alpha \in
A} U_{\alpha} = M$. Then the fact that $M$ is locally homeomorphic
to $\mathbb{R}^2$ means we can find homeomorphisms $z_{\alpha} :
U_{\alpha} \rightarrow V_{\alpha} \subset \mathbb{R}^2$ called
\dub{local charts} from each $U_{\alpha}$ to open subsets
$V_{\alpha} \subset \mathbb{R}^2$. We are interested in doing
complex analysis on $M$ and so we use the homeomorphisms
$z_{\alpha}$ to locally equip $M$ with the analytic structure of
$V_{\alpha} \subset \mathbb{R}^2 \simeq \mathbb{C}$. For instance,
a function $f : U_{\alpha} \rightarrow \mathbb{C}$ will be called
\dub{holomorphic} if $f \circ z_{\alpha}^{-1} : \mathbb{C}
\rightarrow \mathbb{C}$ is a holomorphic map in the usual sense.
\begin{figure}[ht]
\centering \psfrag{z1}{\footnotesize $z_{\alpha}$}
\psfrag{z2}{\footnotesize $z_{\beta}$} \psfrag{U1}{\footnotesize
$U_{\alpha}$} \psfrag{U2}{\footnotesize $U_{\beta}$}
\psfrag{t21}{\footnotesize $t_{\beta \alpha}$}
\psfrag{M}{\footnotesize $M$}
\includegraphics[height=50mm]{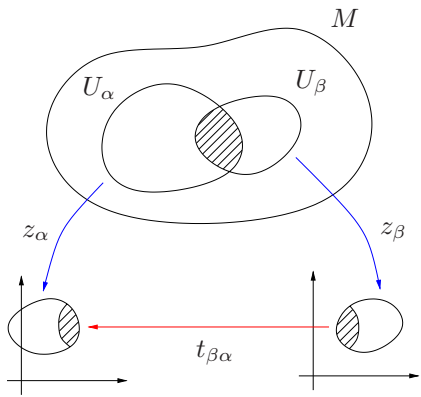}
\caption{Compatibility conditions.}
\end{figure}
But for this analytic structure to have any meaning globally on
$M$ we need a compatibility condition between charts on
overlapping sets $U_{\alpha} \cap U_{\beta} \neq \varnothing$
ensuring that $f \circ z_{\alpha}^{-1}$ is holomorphic iff $f
\circ z_{\beta}^{-1}$ is, for any $f : U_{\alpha} \cap U_{\beta}
\rightarrow \mathbb{C}$. Thus we say that two charts $(U_{\alpha},
z_{\alpha})$ and $(U_{\beta}, z_{\beta})$ are
\dub{(holomorphically) compatible} if
\begin{equation*}
t_{\alpha \beta} = z_{\beta} \circ z_{\alpha}^{-1} :
z_{\alpha}(U_{\alpha} \cap U_{\beta}) \rightarrow
z_{\beta}(U_{\alpha} \cap U_{\beta})
\end{equation*}
called the \dub{transition function}, is holomorphic as a function
from $\mathbb{C}$ to $\mathbb{C}$, \textit{c.f.} for a
differentiable manifold $t_{\alpha \beta}$ is required to be
differentiable. If the charts $\{ (U_{\alpha}, z_{\alpha})
\}_{\alpha \in A}$ are all compatible they are said to form a
\dub{complex atlas} $\mathcal{A}$ and two complex atlases
$\mathcal{A}, \tilde{\mathcal{A}}$ are compatible if $\mathcal{A}
\cup \tilde{\mathcal{A}}$ is a complex atlas. Any atlas
$\mathcal{A}$ can be extended to a \dub{maximal atlas}
$\overline{\mathcal{A}}$ consisting of all charts compatible with
$\mathcal{A}$. A maximal atlas is also called a \dub{complex
structure}.

\begin{definition}
A \dub{Riemann surface} is a real two-dimensional (complex
one-dimensional) connected manifold $M$ equipped with a complex
structure.
\end{definition}

\begin{remark}
One great advantage of working with a Riemann surface as opposed
to simply dealing with the underlying two-dimensional
differentiable manifold is that one can apply all the local
concepts and powerful theorems of complex analysis using the local
homeomorphisms with $\mathbb{C}$. However, just as with
differentiable manifolds, these local homeomorphisms are not
canonical because they depends on the choice of chart
$z_{\alpha}$, and so the only objects one can consider on a
Riemann surface are ones whose definitions are chart invariant.
\end{remark}
\newline

\noindent {\bf Examples}

\noindent The following are basic examples of Riemann surfaces
that will be important later:
\begin{itemize}
  \item Any connected open domain $U \subset \mathbb{C}$ equipped with
a single chart $(U, \textit{id})$.
  \item The Riemann sphere $\mathbb{C}P^1 = \mathbb{C} \cup \{ \infty
\}$ (the one-point compactification of $\mathbb{C}$) equipped with
two charts $(U_1, z_1), (U_2, z_2)$
\begin{equation*}
U_1 = \mathbb{C}, \; z_1 = z, \quad \text{and} \quad U_2 =
(\mathbb{C} \setminus \{ 0 \}) \cup \{ \infty \}, \; z_2 = 1/z,
\end{equation*}
with holomorphic transition functions $t_{12}, t_{21} : \mathbb{C}
\setminus \{ 0 \} \rightarrow \mathbb{C} \setminus \{ 0 \}, z
\mapsto 1/z$.
  \item Any \textit{non-singular} algebraic curve $C \subset \mathbb{C}^2$
defined by the zero-locus
\begin{equation*}
C = \{ (x,y) \in \mathbb{C}^2 | P(x,y) = 0 \}
\end{equation*}
of a polynomial $P$ in $x$ and $y$. The non-singular criteria
means that $\partial P/\partial x$ and $\partial P/\partial y$
never both vanish on $C$. By the implicit function theorem the
variable $y$ (resp. $x$) can be taken as a local chart near points
where $\partial P/\partial x \neq 0$ (resp. $\partial P/\partial y
\neq 0$) and $x(y)$ (resp. $y(x)$) is analytic so this defines a
complex structure on $C$.
\end{itemize}

\begin{remark}
In the neighbourhood of a singular point $(x,y) \in C$, the curve
$C$ looks like an intersection of several complex-lines and so
there is no neighbourhood of $(x,y)$ locally homeomorphic to
$\mathbb{C}$. When encountering singular algebraic curves we will
therefore have to desingularise them by a process to be explained
later.
\end{remark}


\section{Holomorphic maps} \label{section: Holomorphic maps}


\begin{definition}
A continuous mapping
\begin{equation*}
f : M \rightarrow N
\end{equation*}
between Riemann surfaces is called \dub{holomorphic} (or
\dub{analytic}) if for every local chart $(U,z)$ on $M$ and every
local chart $(V,w)$ on $N$ with $U \cap f^{-1}(V) \neq
\varnothing$, the mapping
\begin{equation*}
w \circ f \circ z^{-1} : z(U \cap f^{-1}(V)) \rightarrow w(V)
\end{equation*}
is holomorphic as a map from $\mathbb{C}$ to $\mathbb{C}$.
\end{definition}

\begin{remark}
This definition is independent of the choice of charts $z$ and $w$
by holomorphicity of the transition functions to another set of
charts $z'$ and $w'$. Moreover, because holomorphicity is a local
concept, all the usual local properties of holomorphic functions
on $\mathbb{C}$ will persist for holomorphic maps. For instance,
any holomorphic map $f$ is open, \textit{i.e.} $f$ sends open sets
$U \subset M$ to open sets $f(U) \subset N$.
\end{remark}

A holomorphic mapping into $\mathbb{C}$ is called a
\dub{holomorphic function}. A holomorphic mapping into $\mathbb{C}
\cup \{ \infty \}$ is called a \dub{meromorphic function}. The
ring of holomorphic functions on $M$ is denoted by
$\mathcal{H}(M)$ and the field of meromorphic functions on $M$ by
$\mathcal{K}(M)$.

\subsection*{Local behaviour}

A holomorphic function $f : M \rightarrow N$ is locally injective
around all but finitely many points of $M$. That is, there exists
a finite collection of points $P_1, \ldots, P_n \in M$ such that
for all other points $P \in M \setminus \{ P_1, \ldots, P_n \}$
the restriction $f|_U$ to a neighbourhood $U \subset M \setminus
\{ P_1, \ldots, P_n \}$ of $P$ is injective. The points $P_1,
\ldots, P_n$ around which $f$ fails to be locally injective are
called branch points. These statements are made precise by the
following Lemma:

\begin{lemma} \label{Theorem: local holomorphic maps}
Let $f : M \rightarrow N$ be a holomorphic map and $P \in M$. Then
there exists local charts $(U,z),(V,w)$ near $P \in U$, $f(P) \in
V$ such that $F \equiv w \circ f \circ z^{-1}$ is given by
\begin{equation*}
F(z) = z^k, \quad k \in \mathbb{N}.
\end{equation*}
\begin{proof}
Choose local charts $\tilde{z}$ on $M$ vanishing at $P$ and $w$ on
$N$ vanishing at $f(P)$. Now $F$ is holomorphic with $F(0) = 0$ so
we can write it as $F(\tilde{z}) = \tilde{z}^k g(\tilde{z})$ for
some $g$ holomorphic with $g(0) \neq 0$. Since $g$ is
non-vanishing on a disc around the origin it has a $k^{\text{th}}$
root and so $g(\tilde{z}) = h(\tilde{z})^k$. Defining a new coordinate
$z = \tilde{z} h(\tilde{z})$ the result follows.
\end{proof}
\end{lemma}

Thus a holomorphic map locally looks like the map $z \mapsto z^k$.
Hence in a small neighbourhood $U \ni P$ the number of solutions
to the equation $f(Q) = R$ when $R \in N$ approaches $f(P)$ is
$k$. We see that the number $k$ appearing in Lemma \ref{Theorem:
local holomorphic maps} has an invariant geometrical meaning for
the map $f$ and cannot depend of the choice of chart used to
represent $f$. It is called the \dub{valency} or the
\dub{ramification number} of $f$ at $P \in M$. The number $b_f(P)
= k - 1$ is called the \dub{branch number} of $f$ at $P \in M$.

\begin{definition}
A point $P \in M$ for which $b_f(P)  > 0$ is called a \dub{branch
point} of $f$.
\end{definition}

\begin{lemma}
The branch points of a holomorphic map $f : M \rightarrow N$ are
isolated.
\begin{proof}
Let $P \in M$ be a branch point of $f$. Then by Lemma
\ref{Theorem: local holomorphic maps}, there exists a
neighbourhood $U \ni P$ and coordinate $z$ with $z(P) = 0$ for
which $f$ takes the local form $F(z) = z^k, k > 1$. But the map $z
\mapsto z^k$ is locally injective for $z \neq 0$ so $b_f(Q) = 0$
for any $Q \in U \setminus \{ P \}$.
\end{proof}
\end{lemma}

\begin{corollary} \label{corollary: finite branch points}
If $M$ is compact, then $f : M \rightarrow N$ has finitely many
branch points.
\end{corollary}

\subsection*{Global behaviour}

The local property that a holomorphic map is open (which follows
from Lemma \ref{Theorem: local holomorphic maps}) implies a far
reaching global property of holomorphic maps on compact Riemann
surfaces:

\begin{theorem} \label{Theorem: holomorphic maps}
Let $M$ be compact and $f : M \rightarrow N$ a non-constant
holomorphic map. Then $f$ is surjective ($f(M) = N$) and N is
compact.
\begin{proof}
Since $f$ is not constant, $f(M)$ is open (a holomorphic mapping
is open). But $M$ is compact so $f(M)$ is compact (the continuous
image of a compact set is compact) and hence closed (a compact
subset of a Hausdorff space is closed). So $f(M)$ is a non-empty
open and closed subset of $N$, and since $N$ is connected we have
$f(M) = N$.
\end{proof}
\end{theorem}

In fact one can be a lot more precise. Not only is any $Q \in N$
attained by $f : M \rightarrow N$, but every $Q \in N$ is assumed
the same number of times, counting multiplicities.

\begin{theorem} \label{Theorem: branched cover}
Let $f : M \rightarrow N$ be a non-constant holomorphic function
with $M,N$ compact. Then there exists $m \in \mathbb{N}$ such that
for any $Q \in N$ the equation $f(P) = Q$ has precisely $m$
solutions (counting multiplicities), \textit{i.e.}
\begin{equation*}
\sum_{P \in f^{-1}(Q)} (b_f(P) + 1) = m, \qquad \forall Q \in N.
\end{equation*}
\begin{proof}
Let $Q \in N$. By Theorem \ref{Theorem: holomorphic maps} the
equation $f(P) = Q$ has at least one solution. The number of
solutions $m(Q)$ is finite because otherwise they would accumulate
in $N$ and hence $f$ would be the constant map $f : M \rightarrow
Q$ (since a non-zero holomorphic function has isolated zeroes).
Now by Lemma \ref{Theorem: local holomorphic maps} there exists
neighbourhoods $V_Q$ of $Q$ and $U_i$ of $P_i \in f^{-1}(Q)$ with
respect to which $f$ is of the local form $z \mapsto z^{b_f(P_i) +
1}$ in $U_i$. Since $z \mapsto z^k$ has $k$ zeroes near $z = 0$ it
follows that $m(Q) = \sum_i (b_f(P_i) + 1)$ is constant in $V_Q$.
By compactness one can cover $N$ by finitely many $V_Q$ and so
$m(Q) =: m$ remains constant over $N$.
\end{proof}
\end{theorem}

\begin{figure}[ht]
\centering \psfrag{M}{$M$} \psfrag{N}{$N$} \psfrag{f}{$f$}
\psfrag{b1}{\tiny $b = 1$} \psfrag{b2}{\tiny $b = 2$}
\includegraphics[height=25mm]{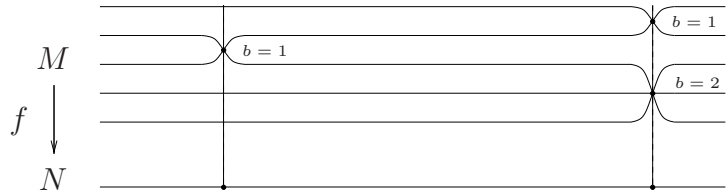}
\caption{Branched covering.} \label{Figure: branched covering}
\end{figure}

We say that $f : M \rightarrow N$ is an \dub{$\bm{m}$-sheeted
`branched' covering} of $N$, referring to the fact that branch
points are the multiple solutions of $f(P) = Q$, see Figure
\ref{Figure: branched covering}.

\begin{definition}
The number $m$ is called the \dub{degree} of $f$ and we write $m =
\text{deg}\, f$.
\end{definition}

Applying Theorem \ref{Theorem: branched cover} with $N =
\mathbb{C} \cup \{ \infty \}$ implies that a non-constant
meromorphic function $f : M \rightarrow \mathbb{C}P^1$ on a
compact Riemann surface $M$ assumes every value in $\mathbb{C}P^1$
the same number of times. In particular, $f$ has as many zeroes as
poles, provided they are counted correctly with multiplicities.

\begin{remark}
A single non-constant meromorphic function $f : M \rightarrow
\mathbb{C}P^1$ completely determines the complex structure of $M$.
Indeed, using Lemma \ref{Theorem: local holomorphic maps} and the
charts of $\mathbb{C}P^1$, a local chart vanishing at $P_0 \in M$
is constructed as follows (with $n = b_f(P_0) + 1$)
\begin{equation*}
z(P) = (f(P) - f(P_0))^{\frac{1}{n}} \quad \text{if} \; f(P_0)
\neq \infty, \qquad \text{or} \qquad z(P) = f(P)^{-\frac{1}{n}}
\quad \text{if} \; f(P_0) = \infty.
\end{equation*}
\end{remark}


\section{Topology} \label{section: topology}


In this section we temporarily forget about the complex structure
of Riemann surfaces and describe their topologies as real
two-dimensional topological manifolds. Accordingly, all the charts
on a surface $M$ in this section are homeomorphisms into
$\mathbb{R}^2$, that is $z_{\alpha} : U_{\alpha} \rightarrow
\mathbb{R}^2$. As before we still assume the surface is connected
and hence path connected.

\begin{definition}
A manifold $M$ is \dub{orientable} if there exists an atlas
$(U_{\alpha}, z_{\alpha})$ such that the transition functions
$t_{\alpha \beta} = z_{\beta} \circ z_{\alpha}^{-1}$ preserve
orientation.
\end{definition}

\begin{proposition} \label{prop: every RS orientable}
Every Riemann surface is orientable.
\begin{proof}
Holomorphic functions preserve orientation since by the Cauchy-Riemann
equations the Jacobian of such a transformation $(x,y) \mapsto
(x',y')$ is positive,
\begin{equation*}
dx \wedge dy = \left[ \frac{\partial
x}{\partial x'} \frac{\partial y}{\partial y'} - \frac{\partial
x}{\partial y'} \frac{\partial y}{\partial x'} \right] dx' \wedge
dy' = \left[ \left( \frac{\partial x}{\partial x'} \right)^2 +
\left( \frac{\partial x}{\partial y'} \right)^2 \right] dx' \wedge
dy'.
\end{equation*}
\end{proof}
\end{proposition}

The following theorem and corollary give a complete classification
of the possible topologies for a Riemann surface. The proof of
Theorem \ref{Theorem: normal form}, which we omit, usually relies
on the fact that every compact surface is triangulable
\cite{Reyssat} and proceeds by cutting and gluing the
triangulation to arrive at the final desired
polygon form:\\

\hspace{-32pt}
\begin{minipage}{0.70\linewidth}
\begin{theorem} \label{Theorem: normal form}
\cite{Reyssat} Every compact orientable surface $M$ is homeomorphic
either to the sphere $S^2$ or to a polygon with $4g$ edges
($\tilde{a}_i, \tilde{a}'_i, \tilde{b}_i, \tilde{b}'_i, i = 1, \ldots,
g$) identified pairwise in such a way that the orientations of these
edges with respect to $M$ are opposite ($\tilde{a}'_i =
\tilde{a}_i^{-1}, \tilde{b}'_i = \tilde{b}_i^{-1}$) and with all
vertices identified.
\end{theorem}
\end{minipage}
\begin{minipage}{0.30\linewidth}
\centering \psfrag{a1}{\tiny $\tilde{a}_g$} \psfrag{b1}{\tiny
$\tilde{b}_g$} \psfrag{a11}{\tiny $\tilde{a}_g^{-1}$}
\psfrag{b11}{\tiny $\tilde{b}_g^{-1}$} \psfrag{a2}{\tiny
$\tilde{a}_1$} \psfrag{b2}{\tiny $\tilde{b}_1$} \psfrag{a22}{\tiny
$\tilde{a}_1^{-1}$} \psfrag{b22}{\tiny $\tilde{b}_1^{-1}$}
\includegraphics[height=25mm]{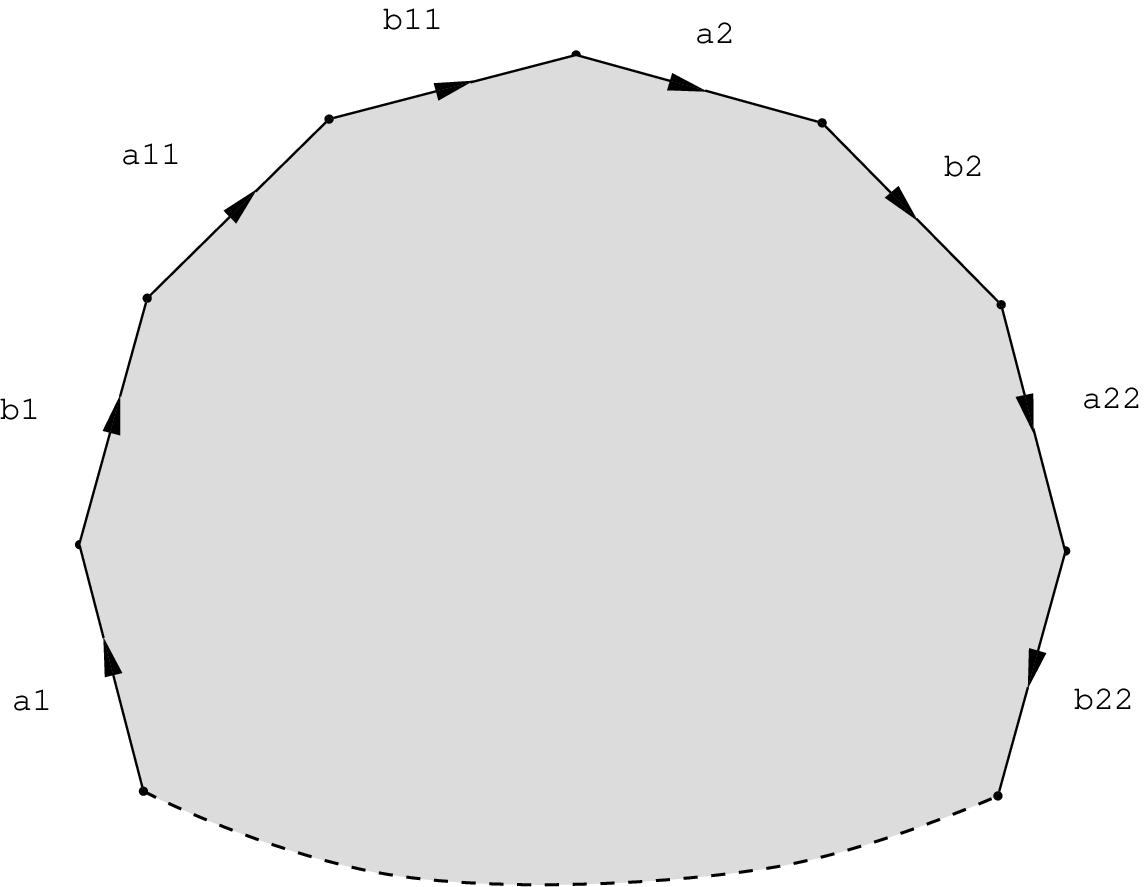}
\end{minipage}
\hspace{32pt}

\begin{remark}
The $4g$-gon described by Theorem \ref{Theorem: normal form} is a
lift of $M$ to its universal covering space $\tilde{M}$. We shall
denote it $M_{\text{cut}}$ since it can be obtained from $M$ by
cutting along certain cycles. The identification process described
in Theorem \ref{Theorem: normal form} corresponds to applying the
covering map $\tilde{\pi} : \tilde{M} \rightarrow M$, in other
words $\tilde{\pi}(M_{\text{cut}}) \approx M$. The simply
connected domain $M_{\text{cut}}$ will come in handy later for
defining branches of multi-valued functions on $M$ and so we give
it a name:
\end{remark}

\begin{definition} \label{def: normal form of RS}
The $4g$-gon $M_{\text{cut}}$ of Theorem \ref{Theorem: normal
form} is called the \dub{normal form} of $M$.
\end{definition}

In its normal form representation, the topology of $M$ is not very
transparent since the edges and vertices still need to be
identified following the prescription in Theorem \ref{Theorem:
normal form}. The next corollary describes the closed surface
resulting from these identifications.\\

\hspace{-32pt}
\begin{minipage}{0.70\linewidth}
\begin{corollary} \label{Theorem: sphere handles}
\cite{Reyssat} Every compact orientable surface $M$ is homeomorphic to a sphere
with $g$ handles, that is to $S^2$ when $g = 0$ or to the $g$-fold
connected sum of torii $\mathbb{T}^1 \# \mathbb{T}^1 \# \ldots \#
\mathbb{T}^1$ when $g \geq 1$.
\end{corollary}
\end{minipage}
\begin{minipage}{0.30\linewidth}
\centering \psfrag{ddd}{\tiny $\cdots$}
\includegraphics[height=9mm]{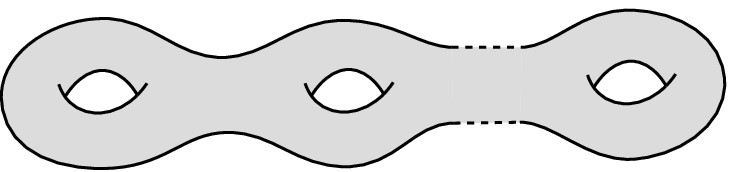}
\end{minipage}
\hspace{32pt}
\begin{proof}
Using Theorem \ref{Theorem: normal form} we just have to show that
the normal form is homeomorphic to a $g$-fold connected sum of
torii (a \dub{$\bm{g}$-fold torus}). We proceed by induction on
$g$. We start by cutting the $4g$-gon into two polygons. The first
has the 4 edges
$\tilde{a}_1,\tilde{b}_1,\tilde{a}_1^{-1},\tilde{b}_1^{-1}$ and a
new edge $\tilde{c}$. The second has the $4(g-1)$ remaining edges
and the edge $\tilde{c}^{-1}$.
\begin{center}
\begin{tabular}{ccc}
\psfrag{agg}{\tiny $\tilde{a}_g^{-1}$} \psfrag{bgg}{\tiny
$\tilde{b}_g^{-1}$} \psfrag{a11}{\tiny $\tilde{a}_1^{-1}$}
\psfrag{b11}{\tiny $\tilde{b}_1^{-1}$} \psfrag{a1}{\tiny
$\tilde{a}_1$} \psfrag{b1}{\tiny $\tilde{b}_1$} \psfrag{a2}{\tiny
$\tilde{a}_2$} \psfrag{b2}{\tiny $\tilde{b}_2$}
\includegraphics[height=30mm]{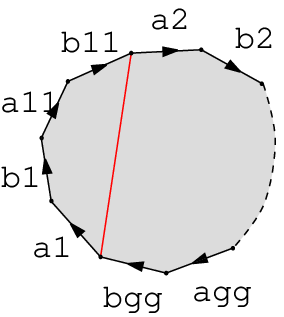} &
\raisebox{15mm}{$\quad \longrightarrow \quad$} &
\psfrag{agg}{\tiny $\tilde{a}_g^{-1}$} \psfrag{bgg}{\tiny
$\tilde{b}_g^{-1}$} \psfrag{a11}{\tiny $\tilde{a}_1^{-1}$}
\psfrag{b11}{\tiny $\tilde{b}_1^{-1}$} \psfrag{a1}{\tiny
$\tilde{a}_1$} \psfrag{b1}{\tiny $\tilde{b}_1$} \psfrag{c}{\tiny
$\tilde{c}$} \psfrag{cc}{\tiny $\tilde{c}^{-1}$} \psfrag{a2}{\tiny
$\tilde{a}_2$} \psfrag{b2}{\tiny $\tilde{b}_2$}
\includegraphics[height=30mm]{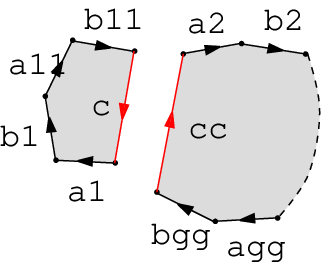}
\end{tabular}
\end{center}
Next we make the identification of edges and vertices in each of
these two polygons using the induction hypothesis. We end up on
the one hand with a torus with a disc cut out, whose boundary is
$\tilde{c}$, and on the other hand a $(g-1)$-fold torus with a
disc cut out, whose boundary is $\tilde{c}^{-1}$.
\begin{center}
\begin{tabular}{ccc}
\psfrag{agg}{\tiny $\tilde{a}_g^{-1}$} \psfrag{bgg}{\tiny
$\tilde{b}_g^{-1}$} \psfrag{a11}{\tiny $\tilde{a}_1^{-1}$}
\psfrag{b11}{\tiny $\tilde{b}_1^{-1}$} \psfrag{a1}{\tiny
$\tilde{a}_1$} \psfrag{b1}{\tiny $\tilde{b}_1$} \psfrag{a2}{\tiny
$\tilde{a}_2$} \psfrag{b2}{\tiny $\tilde{b}_2$} \psfrag{c}{\tiny
$\tilde{c}$} \psfrag{cc}{\tiny $\tilde{c}^{-1}$}
\includegraphics[height=25mm]{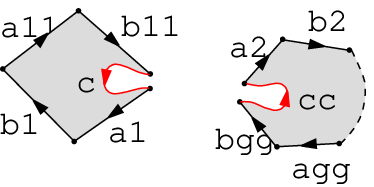} &
\raisebox{12mm}{$\quad \longrightarrow \quad$} &
\raisebox{7mm}{\psfrag{c}{\tiny $\tilde{c}$} \psfrag{cc}{\tiny
$\tilde{c}^{-1}$}
\includegraphics[height=15mm]{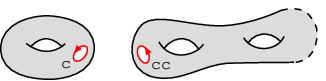}}
\end{tabular}
\end{center}
It is clear from the last figure that gluing the $\tilde{c}$
cycles back together corresponds to taking the connected sum of
the torus with the $(g-1)$-fold torus, which results in a $g$-fold
torus.
\end{proof}

\begin{definition}
The topological invariant $g \in \mathbb{N}$ is called the
\dub{genus} of M.
\end{definition}

\subsection*{Fundamental group}

A \dub{curve} $\gamma$ in $M$ is a continuous map $\gamma : [0,1]
\rightarrow M$. It \dub{starts} at $\gamma(0)$ and \dub{ends} at
$\gamma(1)$. If the start and end points coincide $\gamma(0) =
\gamma(1)$ then there is a natural multiplication between closed
curves starting and ending at $P \in M$, namely
\begin{equation} \label{product of cycles}
\begin{tabular}{ccc}
\raisebox{4mm}{$\gamma_1 \cdot \gamma_2(t) = \left\{
\begin{array}{ll}
\gamma_1(2t) & 0 \leq t \leq \frac{1}{2}\\
\gamma_2(2t - 1) & \frac{1}{2} \leq t \leq 1.
\end{array} \right.$}
& $\qquad$ & \psfrag{g1}{\footnotesize $\gamma_1$}
\psfrag{g2}{\footnotesize $\gamma_2$} \psfrag{P}{\footnotesize
$P$}
\includegraphics[height=12mm]{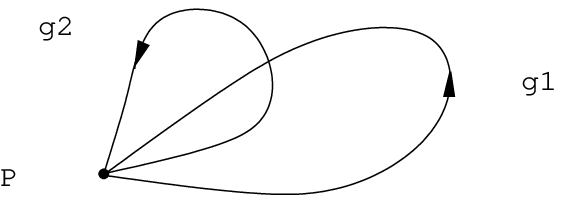}
\end{tabular}
\end{equation}
If we allow reparametrisations of curves ($\gamma \mapsto \gamma
\circ s$ where $s : [0,1] \rightarrow [0,1]$ with $\dot{s} > 0$),
the above product has an obvious identity, $\iota : [0,1]
\rightarrow P$ and every curve $t \mapsto \gamma(t)$ has as
inverse the same curve traversed in the opposite direction, $t
\mapsto \gamma(1-t)$. The resulting group however is far too big
and not very useful. One can reduce its size considerably by
taking a quotient:

\begin{definition} \label{Definition: homotopic}
Two curves $\gamma_1$ and $\gamma_2$ in $M$ both starting at $P$
and ending at $Q$ are called \dub{homotopic} if there exists a
continuous map $\gamma : [0,1] \times [0,1] \rightarrow M$ such
that
\begin{center}
\begin{tabular}{ccc}
\raisebox{4mm}{$\begin{array}{ll}
\gamma(t,0) = \gamma_1(t), &\gamma(t,1) = \gamma_2(t),\\
\gamma(0,\lambda) = P, &\gamma(1,\lambda) = Q.
\end{array}$}
& $\qquad$ & \psfrag{g1}{\footnotesize $\gamma_1$}
\psfrag{g2}{\footnotesize $\gamma_2$} \psfrag{P}{\footnotesize
$P$} \psfrag{Q}{\footnotesize $Q$}
\includegraphics[height=12mm]{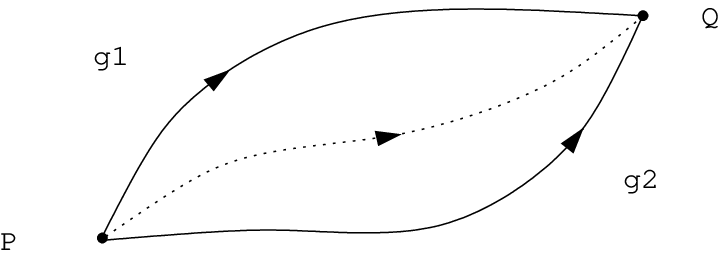}
\end{tabular}
\end{center}
The equivalence class of curves homotopic to a given curve
$\gamma$ is called its \dub{homotopy class} and denoted
$[\gamma]_{\pi}$.
\end{definition}

The definition \eqref{product of cycles} now descends to the
quotient and gives a well-defined product between homotopy classes
of curves based at $P$ by $[\gamma_1]_{\pi} \cdot [\gamma_2]_{\pi}
= [\gamma_1 \cdot \gamma_2]_{\pi}$. The identity corresponds to
the equivalence class $[\iota]_{\pi}$ of curves homotopic to the
point $P$. With this multiplication, the homotopy classes of
curves based at $P$ thus form a group denoted $\pi_1(M, P)$. Since
$M$ is always assumed to be connected, any two points $P,Q \in M$
are connected by a path and the groups $\pi_1(M, P)$ and $\pi_1(M,
Q)$ are isomorphic, although the isomorphism is not canonical
because it depends on the homotopy class of the path joining $P$
and $Q$.

\begin{definition}
The \dub{fundamental group} $\pi_1(M)$ of $M$ is the group
$\pi_1(M,P)$ with any choice of base point $P \in M$.
\end{definition}

Let us now identify the fundamental group of a compact orientable
surface $M$ by making use of the normal form representation
$M_{\text{cut}}$ of Theorem \ref{Theorem: normal form} which lives
in the universal cover $\tilde{M}$. Let $P \in M$ be the common image
of the vertices of the $4g$-gon under the covering map
$\tilde{\pi} : \tilde{M} \rightarrow M$. Define the projections
$a_i := \tilde{\pi}(\tilde{a}_i)$ and $b_i :=
\tilde{\pi}(\tilde{b}_i)$ for $i = 1, \ldots, g$ under
$\tilde{\pi}$ of all the edges of the $4g$-gon. These are all
closed loops in $M$ through $P$ so it is convenient to use the
base point $P$ to determine the fundamental group.

\begin{theorem}
$\pi_1(M)$ is generated by $[a_i]_{\pi}, [b_i]_{\pi}, i = 1,
\ldots, g$ subject to the single relation
\begin{equation} \label{generators of pi_1}
\prod_{i=1}^g [a_i]_{\pi} \cdot [b_i]_{\pi} \cdot [a_i]_{\pi}^{-1}
\cdot [b_i]_{\pi}^{-1} = 1.
\end{equation}
\begin{proof}
The lift $\tilde{c}$ to $\tilde{M}$ of any closed loop $c$ through
$P$ is a sum of paths in $M_{\text{cut}}$ starting and ending on
vertices. Such paths can clearly be retracted to portions of the
boundary $\partial M_{\text{cut}}$ which is spanned by the edges
$\tilde{a}_i, \tilde{a}_i^{-1}, \tilde{b}_i, \tilde{b}_i^{-1}, i =
1, \ldots, g$. We deduce that the homotopy classes $[a_i]_{\pi},
[b_i]_{\pi}, i = 1 ,\ldots, g$ generate the fundamental group
$\pi_1(M,P) \approx \pi_1(M)$. The existence of a non-trivial lift
$\tilde{c}$ which is contractible to an arbitrary point in
$M_{\text{cut}}$ gives rise to a relation amongst these
generators. Since $M_{\text{cut}}$ is simply connected, the only
such cycle is $\partial M_{\text{cut}} = \prod_{i=1}^g \tilde{a}_i
\cdot \tilde{b}_i \cdot \tilde{a}_i^{-1} \cdot \tilde{b}_i^{-1}$
which leads to the relation $[\partial M_{\text{cut}}]_{\pi} = 1$,
namely \eqref{generators of pi_1}.
\end{proof}
\end{theorem}

\subsection*{First homology group}

A triangulation of $M$ consists of oriented vertices, edges and
faces called $0$-, $1$- and $2$-simplices respectively. A
\dub{$\bm{0}$-simplex} is a point $P$ with an orientation, so
either $(P)$ or $-(P)$. A \dub{$\bm{1}$-simplex} is a segment with
endpoints $P_1, P_2$ and one of two possible orientations, either
$(P_1, P_2)$ or $(P_2, P_1) = - (P_1, P_2)$. A
\dub{$\bm{2}$-simplex} is a triangle with vertices $P_1, P_2, P_3$
and one of two possible orientations, either $(P_1, P_2, P_3)$ or
$(P_1, P_3, P_2) = -(P_1, P_2, P_3)$. Formal sums $\sum m_i t_i$
($m_i \in \mathbb{Z}$) of $n$-simplices $t_i$ are called
\dub{$\bm{n}$-chains} and form a free abelian group $C_n(M)$ under
addition. The requirement that simplices be oriented ensures that
$C_n(M)$ is indeed a group, where the negative $-t_i$ is the
simplex $t_i$ taken with opposite orientation.

One can define a natural sequence of \dub{boundary} operations
$\partial_n$ (all denoted $\partial$ when there is no ambiguity)
\begin{equation} \label{homology sequence}
0 \overset{\partial_3}\longrightarrow C_2(M)
\overset{\partial_2}\longrightarrow C_1(M)
\overset{\partial_1}\longrightarrow C_0(M)
\overset{\partial_0}\longrightarrow 0,
\end{equation}
given explicitly on $0$-, $1$- and $2$-simplices by
\begin{equation} \label{boundary operator}
\partial (P) = 0, \quad \partial (P_1,P_2) = (P_1) - (P_2), \quad
\partial (P_1,P_2,P_3) = (P_1,P_2) + (P_2, P_3) + (P_3,P_1),
\end{equation}
and extended to $0$-, $1$- and $2$-chains by linearity. We define
the subgroups of \dub{boundaries} and \dub{cycles} as $B_n(M) =
\im (\partial_{n+1}) = \partial C_{n+1}(M)$, and $Z_n(M) = \ker
(\partial_n) = \{ c \in C_n(M) | \partial c = 0 \}$ respectively.
It follows that the homomorphism $\partial_{n+1} : C_{n+1}(M)
\rightarrow B_n(M)$ is surjective with kernel $Z_{n+1}(M)$ so
$B_n(M) \approx C_{n+1}(M)/Z_{n+1}(M)$. It is trivial to check
using \eqref{boundary operator} that $\partial^2 = \partial_n
\partial_{n+1} = 0$ so that $B_n(M) \subset Z_n(M)$. Since these
groups are abelian, $B_n(M)$ is normal in $Z_n(M)$ and their
quotient $H_n(M) = Z_n(M)/B_n(M)$ is a group, called the
\dub{$\bm{n^{\text{th}}}$ homology group}. It measures the
deviation from exactness at the $n^{\text{th}}$ site of the
sequence \eqref{homology sequence}.

Now given $P \in M$, by definition $\partial (P) = 0$ so $Z_0(M) =
C_0(M)$. But since any two points $P,Q \in M$ are related by a
boundary $(P) = (Q) + \partial (P,Q)$ this means that $H_0(M)$ is
generated by a single point $(P)$ and hence $H_0(M) = (P) \cdot
\mathbb{Z} \approx \mathbb{Z}$. Next suppose the $2$-chain $c =
\sum m_i t_i$ is without boundary, $\partial c = 0$. Then $m_i =
m_j$ when two triangles $t_i, t_j$ in the sum have adjacent edges.
Since $c$ must be connected it follows that all the $m_i$ are
equal so $Z_2(M)$ is generated by $M = \sum t_i$. Since also
$B_2(M) = \{ 0 \}$ it follows that $H_2(M) = Z_2(M) \approx
\mathbb{Z}$. From now on we focus on the remaining homology
group,

\begin{definition}
The \dub{first homology group} of $M$ is defined as $H_1(M) =
Z_1(M)/B_1(M)$.
\end{definition}

\begin{remark}
$H_1(M)$ can be shown not to depend on the triangulation used for
$M$. Therefore from now on the word `curve' will refer to both
continuous maps $\gamma : [0,1] \rightarrow M$ and to $1$-chains,
the word `closed curve' refers to continuous maps with $\gamma(0)
= \gamma(1)$ as well as $1$-cycles and we use the word `boundary'
to designate curves which are $1$-dimensional boundaries of
domains in $M$.
\end{remark}

\begin{definition} \label{Definition: homologous}
Two closed curves $\gamma_1$ and $\gamma_2$ in $M$ are said to be
\dub{homologous} if
\begin{center}
\begin{tabular}{ccc}
\raisebox{4mm}{$\gamma_1 - \gamma_2 \in B_1(M).$} & $\qquad$ &
\psfrag{g1}{\footnotesize $\gamma_1$} \psfrag{g2}{\footnotesize
$\gamma_2$}
\includegraphics[height=12mm]{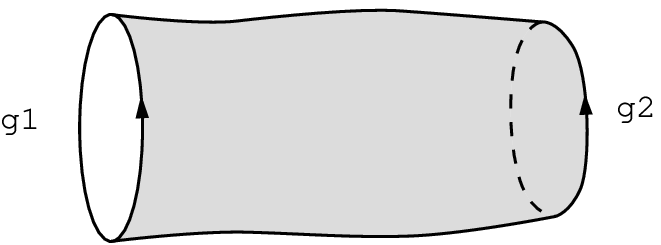}
\end{tabular}
\end{center}
The equivalence class of curves homologous to a given $\gamma$ is
called its \dub{homology class} which is an element of $H_1(M)$
denoted $[\gamma]$.
\end{definition}

We now have two different equivalence relations on closed curves
of $M$: homotopy equivalence (Definition \ref{Definition:
homotopic}) on the one hand and homology equivalence (Definition
\ref{Definition: homologous}) on the other. An obvious question to
ask is whether or not these are related. It is obvious that
homotopic curves $\gamma_1, \gamma_2$ are homologous since the
homotopy is a continuous map $\gamma : [0,1] \times [0,1]
\rightarrow M$ which defines a tubular cobordism on $M$ joining
$\gamma_1$ and $\gamma_2$. The converse is false however since the
cobordism from $\gamma_1$ to $\gamma_2$
\begin{figure}[ht]
\centering \psfrag{g1}{\footnotesize $\gamma_1$}
\psfrag{g2}{\footnotesize $\gamma_2$}
\includegraphics[height=15mm]{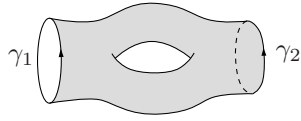}
\caption{Example of homologous cycles $\gamma_1, \gamma_2$ that
are not homotopic.} \label{homology neq homotopy}
\end{figure}
can be more general (Figure \ref{homology neq homotopy}). Recall
from their respective definitions that the fundamental group
$\pi_1(M)$ is non-abelian whereas the first homology group
$H_1(M)$ is abelian. As it turns out the first homology group
$H_1(M)$ is the \dub{abelianisation} of the fundamental group
$\pi_1(M)$. Specifically, defining the \dub{commutator subgroup}
$[\pi_1(M), \pi_1(M)] = \langle a \cdot b \cdot a^{-1} \cdot b^{-1} |
a,b \in \pi_1(M) \rangle$ we have,

\begin{theorem}
$H_1(M) \approx \pi_1(M)/[\pi_1(M),\pi_1(M)]$.
\begin{proof}
Since two homotopic curves are homologous, the map $\varphi :
\pi_1(M) \rightarrow H_1(M)$, $[\gamma]_{\pi} \mapsto [\gamma]$ is
well defined. It is clearly a homomorphism since
$\varphi([\gamma_1]_{\pi} \cdot [\gamma_2]_{\pi}) =
\varphi([\gamma_1 \cdot \gamma_2]_{\pi}) = [\gamma_1 \cdot
\gamma_2] = [\gamma_1] + [\gamma_2]$. Moreover $\varphi$ is
surjective. Its kernel consists of $[\gamma]_{\pi}$ such that
$[\gamma] = 0$. That is, $\gamma^{-1}$ can be chosen (up to
homology) as the single boundary of a surface of genus $n$
(arbitrary) with a small disc removed
\begin{center}
\begin{tabular}{ccc}
\raisebox{2mm}{\psfrag{cc}{\tiny $\gamma^{-1}$}
\includegraphics[height=15mm]{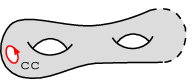}}
& \raisebox{8mm}{$\quad \cong \quad$} & \psfrag{a2}{\tiny $a_1$}
\psfrag{b2}{\tiny $b_1$} \psfrag{agg}{\tiny $a_n^{-1}$}
\psfrag{bgg}{\tiny $b_n^{-1}$} \psfrag{cc}{\tiny $\gamma^{-1}$}
\includegraphics[height=20mm]{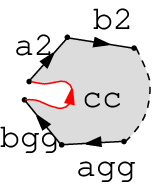}
\end{tabular}
\end{center}
The boundary $\gamma^{-1} \cdot \prod_{i=1}^n a_i \cdot b_i \cdot
a_i^{-1} b_i^{-1}$ of the normal form of this surface being
homotopic to a point implies $[\gamma]_{\pi} = \prod_{i=1}^n
[a_i]_{\pi} \cdot [b_i]_{\pi} \cdot [a_i]_{\pi}^{-1}
[b_i]_{\pi}^{-1} \in [\pi_1(M),\pi_1(M)]$.
\end{proof}
\end{theorem}

With the knowledge of the fundamental group it is now easy to
compute the first homology group by the process of `abelianising',
that is, forgetting about the order in which we multiply cycles.
In fact, since the relation \eqref{generators of pi_1} on the
generators of $\pi_1(M)$ becomes trivial in the abelian case, the
constraint disappears for $H_1(M)$ and we have

\begin{corollary} \label{Corollary: H_1 generators}
$H_1(M)$ is the free abelian group generated by $[a_i], [b_i], i =
1, \ldots, g$.
\end{corollary}

\begin{definition}
The rank of $H_n(M)$ is called the \dub{Betti number} and is
denoted $b_n$. The \dub{Euler characteristic} of $M$ is defined as
$\chi(M) = b_0 - b_1 + b_2$.
\end{definition}

\begin{remark}
The Euler characteristic is a topological invariant of $M$. Since
$H_0(M) \approx \mathbb{Z}$ and $H_2(M) \approx \mathbb{Z}$ we
have that $b_0 = b_2 = 1$. Furthermore, Corollary \ref{Corollary:
H_1 generators} implies $b_1 = 2g$. This leads to a simple
expression for $\chi(M)$ in terms of the other known topological
invariant, the genus $g$.
\end{remark}

\begin{corollary} \label{corollary: chi vs genus}
A compact Riemann surface $M$ of genus $g$ has $\chi(M) = 2 - 2g$.
\end{corollary}

\subsection*{Coverings}

In Theorem \ref{Theorem: branched cover} we saw that every
non-constant holomorphic map $f : M \rightarrow N$ between compact
Riemann surfaces $M,N$ was a branched covering of $N$. In the
present section we will give a topological property of branched
coverings relating the Euler characteristics (and hence the genus)
of the two surfaces $M,N$.

There is a simple way of computing the Euler characteristic of a
compact surface $M$ using a triangulation of $M$.

\begin{proposition} \label{proposition: calculate chi}
If a triangulation of $M$ has $F$ faces, $E$ edges and $V$
vertices then
\begin{equation*}
\chi(M) = F - E + V.
\end{equation*}
\begin{proof}
Let $c_n = \rk C_n(M)$, the number of $n$-simplices, and
$z_n = \rk Z_n(M)$. Then $\rk B_n(M) = c_{n+1} - z_{n+1}$ so that
$b_n = \rk H_n(M) = z_n - c_{n+1} + z_{n+1}$. Hence $\chi(M) = z_0
- c_1 + c_2$, but $Z_0(M) = C_0(M)$ implies $z_0 = c_0$.
\end{proof}
\end{proposition}

Recall from section \ref{section: Holomorphic maps} that a
covering $f : M \rightarrow N$ can have only finitely many branch
points, which are points $P \in M$ with positive branching number
$b_f(P) > 0$. We define the \dub{total branching number} as
\begin{equation*}
b = \sum_{P \in M} b_f(P).
\end{equation*}

\begin{theorem}[Riemann-Hurwitz]
Let $f : M \rightarrow N$ be a branched covering of degree $m$
between compact surfaces $M$ and $N$, then
\begin{equation*}
\chi(M) = m \chi(N) - b.
\end{equation*}
\begin{proof}
Let $B = \{ P \in M | b_f(P) > 0 \}$ be the set of branch points
of $f$. Since $B$ is finite (Corollary \ref{corollary: finite
branch points}) we can choose a triangulation $\mathcal{T}$ of $N$
which includes all the points of $f(B)$ as vertices. Assume
$\mathcal{T}$ has $F$ faces, $E$ edges and $V$ vertices. Then the
lift $f^{-1}(\mathcal{T})$ of $\mathcal{T}$ to $M$ is a
triangulation of $M$ with $m F$ faces, $m E$ edges and $m V - b$
vertices. Proposition \ref{proposition: calculate chi} implies
$\chi(N) = F - E + V$, $\chi(M) = m F - m E + m V - b$ and the
result follows.
\end{proof}
\end{theorem}

\begin{remark}
If we call $g$ the genus of $M$ and $\gamma$ the genus of $N$,
then using Corollary \ref{corollary: chi vs genus} the
Riemann-Hurwitz formula can be rewritten as
\begin{equation} \label{Riemann-Hurwitz formula}
g = m (\gamma - 1) + 1 + \frac{b}{2}.
\end{equation}
\end{remark}


\section{Differential structure} \label{section: differential calculus}


In this section we will exploit the real-differentiability of the
local charts $z_{\alpha} : U_{\alpha} \rightarrow \mathbb{C}$ and
use them to introduce differential calculus on $M$.

When working over the reals it is best to specify a local chart
using real coordinates. So given a local complex coordinate $z : U
\rightarrow \mathbb{C}$, the real and imaginary parts $z = x + i
y$ define corresponding real coordinates $(x, y) : U \rightarrow
\mathbb{R}^2$. In this chart, a local basis for the tangent space
$T_P(M)$ at a point $P \in U$ is given by
\begin{equation} \label{basis of TM}
\frac{\partial}{\partial x}, \quad \frac{\partial}{\partial y}
\end{equation}
and the dual basis of the cotangent space $T^{\ast}_P(M)$ is $\{
dx, dy \}$. The local expression of a real-valued $1$-form
$\omega_{\mathbb{R}}$ is given in terms of two differentiable
functions $f,g : U \rightarrow \mathbb{R}$ as
\begin{subequations} \label{form in basis}
\begin{equation} \label{one-form in basis}
\omega_{\mathbb{R}} = f(x,y) dx + g(x,y) dy.
\end{equation}
Under a change of chart the components of a $1$-form transform in
such a way that the expression \eqref{one-form in basis} for the
$1$-form in terms of its components remains valid in the new
chart. On a two dimensional manifold one can also define
$2$-forms. A local basis for these is given by $dx \wedge dy$ and
a real-valued $2$-form $\lambda_{\mathbb{R}}$ is specified by a
single function $h : U \rightarrow \mathbb{R}$ as
\begin{equation} \label{two-form in basis}
\lambda_{\mathbb{R}} = h(x,y) dx \wedge dy.
\end{equation}
\end{subequations}
The component transforms under a change of chart in such a way
that \eqref{two-form in basis} remains true.

However, when we come to treat the Riemann surface $M$ as a
complex manifold, it will be natural to consider complex-valued
functions $f : M \rightarrow \mathbb{C}$ rather than just
functions into $\mathbb{R}$. It is therefore more appropriate to
consider the \dub{complexifications} $T(M)^{\mathbb{C}} \equiv
T(M) \otimes \mathbb{C}$ and $T^{\ast}(M)^{\mathbb{C}} \equiv
T^{\ast}(M) \otimes \mathbb{C}$ of the tangent and cotangent
bundles respectively. The vectors \eqref{basis of TM} still define
a basis of $T_P(M)^{\mathbb{C}}$ over the complex numbers and $\{
dx, dy\}$ still provides a basis for $T^{\ast}_P(M)^{\mathbb{C}}$
over $\mathbb{C}$. Complex-valued differentials can now be
specified using complex-valued components such as $f,g,h : U
\rightarrow \mathbb{C}$ in \eqref{form in basis}. An alternative
basis for $T(M)^{\mathbb{C}}$ is
\begin{equation} \label{partial_z}
\frac{\partial}{\partial z} = \frac{1}{2} \left(
\frac{\partial}{\partial x} - i \frac{\partial}{\partial y}
\right), \quad \frac{\partial}{\partial \bar{z}} = \frac{1}{2}
\left( \frac{\partial}{\partial x} + i \frac{\partial}{\partial y}
\right).
\end{equation}
Likewise we define the new dual basis of complex-valued $1$-forms by
\begin{equation*}
dz = dx + i dy, \quad d\bar{z} = dx - i dy.
\end{equation*}
These two differentials are independent since $dz \wedge d\bar{z}
= -2 i dx \wedge dy \neq 0$. In this new basis a complex-valued
$1$-form $\omega$ is locally expressed in terms of two
real-differentiable functions $u,v : U \rightarrow \mathbb{C}$ as
\begin{subequations} \label{form in basis C}
\begin{equation} \label{one-form in basis C}
\omega = u(z) dz + v(z) d \bar{z}.
\end{equation}
For instance, the components of the real-valued $1$-form
$\omega_{\mathbb{R}}$ in \eqref{one-form in basis} with respect to
this basis are $u = \frac{1}{2}(f - i g)$ and $v = \bar{u}$.
Likewise, in the new basis a complex-valued $2$-form $\lambda$ can
be locally expressed in terms of a single function $w : U
\rightarrow \mathbb{C}$ as
\begin{equation} \label{two-form in basis C}
\lambda = w(z) dz \wedge d \bar{z}.
\end{equation}
\end{subequations}
The component of the real-valued $2$-form $\lambda_{\mathbb{R}}$ in
\eqref{two-form in basis} with respect to this basis is $w =
\frac{i}{2} h$. Let us denote the spaces of complex-valued
functions, $1$-forms and $2$-forms by $\Omega^0(M)$, $\Omega^1(M)$
and $\Omega^2(M)$ respectively.

\begin{remark}
The notation in \eqref{form in basis C} is slightly misleading:
although the components $u,v$ are functions of the local complex
parameter $z$, one can still have $\frac{\partial u}{\partial
\bar{z}} \neq 0$ and $\frac{\partial v}{\partial \bar{z}} \neq 0$.
Indeed, the statement that $\frac{\partial w}{\partial \bar{z}} =
0$ for a complex valued function $w = f + i g$ is equivalent to
the Cauchy-Riemann equations $\frac{\partial f}{\partial x} =
\frac{\partial g}{\partial y}, \frac{\partial f}{\partial y} = -
\frac{\partial g}{\partial x}$.
\end{remark}

\subsection*{Differentials and integration}

Given a function $f \in \Omega^0(M)$, its \dub{exterior
derivative} is a $1$-form defined locally as
\begin{equation*}
df \equiv f_x dx + f_y dy = f_z dz + f_{\bar{z}} d\bar{z}.
\end{equation*}
This definition is chart independent and so indeed defines a
$1$-form. We can extend this notion of exterior derivative to
$1$-forms $\omega \in \Omega^1(M)$ given locally in
\eqref{one-form in basis C} by defining
\begin{equation} \label{exterior derivative}
d \omega \equiv du \wedge dz + dv \wedge d\bar{z} = (v_z -
u_{\bar{z}}) dz \wedge d\bar{z}.
\end{equation}
The second equality follows from the definition of exterior
differential on functions. Finally, since the top forms on $M$ are
$2$-forms, their exterior derivative must be zero. It is obvious
from these definitions that the exterior derivative satisfies the
usual cohomology property
\begin{equation} \label{cohomology}
d^2 = 0.
\end{equation}
A $1$-form $\omega$ is \dub{closed} if $d \omega = 0$ and it is
\dub{exact} if $\omega = df$ for some function $f$. Denoting the
set of closed $1$-forms as $Z^1(M) = \{ \omega \in \Omega^1(M) | d
\omega = 0 \}$ and the set of exact $1$-forms as $B^1(M)= d
\Omega^0(M)$, the above condition \eqref{cohomology} means that
$B^1(M) \subset Z^1(M)$, and since these are both vector spaces, the
vector space quotient $H^1_{\text{dR}}(M) = Z^1(M)/B^1(M)$ is also a
vector space, called the \dub{first de-Rham cohomology group} of
$M$. In fact we have a sequence
\begin{equation} \label{cohomology sequence}
0 \overset{d}\longrightarrow \Omega^0(M)
\overset{d}\longrightarrow \Omega^1(M)
\overset{d}\longrightarrow \Omega^2(M)
\overset{d}\longrightarrow 0,
\end{equation}
and $H^1_{\text{dR}}(M)$ is the obstruction to this sequence being
exact at the middle site.

As usual one can define integration of $n$-forms over
$n$-chains. Integration therefore provides a natural pairing between
$\Omega^n(M)$ and $C_n(M)$,
\begin{equation} \label{integration pairing}
\Omega^n(M) \times C_n(M) \rightarrow \mathbb{C}, \qquad
(\omega, c) \mapsto \int_c \omega.
\end{equation}
A $0$-form $f \in \Omega^0(M)$ is just a function and a $0$-chain
$c \in C_0(M)$ is a finite sum of points $c = \sum_{\alpha}
n_{\alpha} P_{\alpha}$, $n_{\alpha} \in \mathbb{Z}$, $P_{\alpha}
\in M$. In this case integration is defined as the evaluation map,
\begin{equation*}
\int_c f = \sum_{\alpha} n_{\alpha} f(P_{\alpha}).
\end{equation*}
The integral of a $1$-form $\omega \in \Omega^1(M)$ along a
$1$-chain $\gamma \in C_1(M)$ given by $\gamma : [0,1] \rightarrow
M$ is also defined in the obvious way using local coordinates. If
the path $\gamma$ lies entirely inside a single chart $z : U
\rightarrow \mathbb{C}$ with respect to which $\omega$ has the
local expression given in \eqref{one-form in basis C} then we
define
\begin{equation*}
\int_{\gamma} \omega = \int_0^1 \left[(u(x,y) + v(x,y))
\frac{dx}{ds} + i (u(x,y) - v(x,y)) \frac{dy}{ds} \right] ds,
\end{equation*}
which is independent of both the choice of local chart on $U$ and the
parameter along $\gamma$. If $\gamma$ cannot be covered by a single
chart we define the integral $\int_{\gamma} \omega$
piecewise. Finally, one defines the integration of a $2$-form $\lambda
\in \Omega^2(M)$ given locally as in \eqref{two-form in basis} over a
domain $D$ in the usual way by proceeding patchwise, where if $D
\subset U$ is contained in a single chart $z : U \rightarrow \mathbb{C}$
\begin{equation*}
\int_D \lambda = \int_{z(D)} -2 i w(x,y) dx \wedge dy.
\end{equation*}

One of the most interesting properties of integration is that the
boundary operator $\partial$ defined on chains $C_n(M)$ and the
exterior differential $d$ defined on forms $\Omega^n(M)$ are adjoint of
each other with respect to the pairing \eqref{integration pairing}.
\begin{theorem}[Stokes]
Let $\omega \in \Omega^n(M)$ and $\gamma \in C_{n+1}(M)$ then
\begin{equation*}
\int_{\gamma} d\omega = \int_{\partial \gamma} \omega.
\end{equation*}
\end{theorem}
\noindent As an immediate consequence the pairings \eqref{integration
pairing} descend to pairings between cohomology and homology
groups. The most important of these is
\begin{equation} \label{cohomology-homology}
H^1_{\text{dR}}(M) \times H_1(M) \rightarrow \mathbb{C}, \qquad
(\omega, c) \mapsto \int_c \omega.
\end{equation}
Given a closed $1$-form $\omega \in Z^1(M)$ we define,
\begin{definition}
The integral $\int_c \omega$ over a closed path $c$ is called a
\dub{period} of $\omega$.
\end{definition}

The following lemma asserts that a closed $1$-form is uniquely
specified, up to exact forms, by its periods.
\begin{lemma} \label{lemma: vanishing periods}
A closed $1$-form $\omega$ is exact if and only if all its periods
vanish.
\begin{proof}
The `only if' direction is obvious. To prove the `if' statement, assume $\omega$
is closed and $\int_c \omega = 0$ for all $c$ with $\partial c =
0$. Then $f(P) \equiv \int^P_{P_0} \omega$ is well defined since it is
independent of the path chosen, and by the fundamental theorem of
calculus $df = d \left( \int^P \omega \right) = \omega$.
\end{proof}
\end{lemma}

It follows that if we consider the homology group with complex coefficients
$H_1(M,\mathbb{C})$ as a vector space over $\mathbb{C}$ then the pairing
\eqref{cohomology-homology} is non-degenerate and we have the following
duality
\begin{equation*}
H^1_{\text{dR}}(M) = (H_1(M,\mathbb{C}))^{\ast},
\end{equation*}
between vector spaces over $\mathbb{C}$. Thus in particular
$H^1_{\text{dR}}(M)$ is $2g$-dimensional.

If $c$ is a closed path it follows from corollary \ref{Corollary: H_1
generators} that it can be written as
\begin{equation*}
c \sim \sum_{i=1}^g n_i a_i + \sum_{i=1}^g m_i b_i, \quad n_i, m_i \in
\mathbb{Z},
\end{equation*}
modulo boundaries, indicated by the symbol $\sim$ for homology
equivalence. But it follows that for any closed $1$-form $\omega$ we
have the equality
\begin{equation} \label{basis of periods}
\int_c \omega = \sum_{i=1}^g n_i \int_{a_i} \omega + \sum_{i=1}^g m_i
\int_{b_i} \omega.
\end{equation}
Therefore the set of $2g$ periods $\int_{a_i} \omega$ and
$\int_{b_i} \omega$ form a basis of periods for $\omega$. They are
called respectively \dub{$\bm{a}$- and $\bm{b}$-periods} of
$\omega$. Specifying these uniquely determines a cohomology class:
indeed if two closed $1$-forms $\omega_1, \omega_2$ have the same
$\bm{a}$- and $\bm{b}$-periods then $\int_c \omega_1 = \int_c
\omega_2$ for any closed curve by \eqref{basis of periods} and hence
$\omega_1 - \omega_2$ is exact by lemma \ref{lemma: vanishing periods}
so $\omega_1$ and $\omega_2$ define the same cohomology class.

\subsection*{Riemann bilinear identities}

There is a natural anti-symmetric inner-product between $1$-forms
on $M$ defined by,
\begin{equation*}
\Omega^1(M) \times \Omega^1(M) \rightarrow \mathbb{C}, \qquad
(\omega_1, \omega_2) \mapsto \int_M \omega_1 \wedge \omega_2.
\end{equation*}
If both forms $\omega_1, \omega_2$ are closed then this
inner-product depends only on their cohomology classes since for
example $\int_M df \wedge \omega_2 = \int_M d(f \omega_2) =
\int_{\partial M} f \omega_2 = 0$ using $\partial M =
\varnothing$. The following proposition expresses this
inner-product in terms of the $\bm{a}$- and $\bm{b}$-periods of
the two $1$-forms. The important relations \eqref{Riemann
bilinear} are know as the \dub{Riemann bilinear identities}.

\begin{proposition} \label{proposition: Riemann bilinear}
Let $\omega_1, \omega_2 \in \Omega^1(M)$ be two closed $1$-forms
on $M$, then
\begin{equation} \label{Riemann bilinear}
\int_M \omega_1 \wedge \omega_2 = \sum_{i=1}^g \left[ \int_{a_i}
\omega_1 \int_{b_i} \omega_2 - \int_{b_i} \omega_1 \int_{a_i} \omega_2
\right].
\end{equation}
\begin{proof}
Consider the normal form $M_{\text{cut}}$ of $M$. Since
$M_{\text{cut}}$ is star-shaped and $\omega_1$ is closed we can write
$\omega_1 = df$ in $M_{\text{cut}}$ where $f(P) = \int^P_{P_0}
\omega_1$ with $P_0 \in M_{\text{cut}}$. Now using also the fact that
$d \omega_2 = 0$ (in the second last equality) we have
\begin{equation*}
\int_M \omega_1 \wedge \omega_2 = \int_{M_{\text{cut}}} \omega_1
\wedge \omega_2 = \int_{M_{\text{cut}}} df \wedge \omega_2 =
\int_{M_{\text{cut}}} d(f \omega_2) = \int_{\partial {M_{\text{cut}}}}
f \omega_2.
\end{equation*}
But the boundary $\partial {M_{\text{cut}}}$ consists of all the edges
$\{ \tilde{a}_i, \tilde{b}_j, \tilde{a}_i^{-1}, \tilde{b}_i^{-1}
\}_{i=1}^g$ so the last term on the right hand side can be written
more explicitly as
\begin{equation*}
\sum_{i=1}^g \left[ \int_{\tilde{a}_i} f \omega_2 + \int_{\tilde{b}_i}
f \omega_2 + \int_{\tilde{a}_i^{-1}} f \omega_2 +
\int_{\tilde{b}_i^{-1}} f \omega_2 \right].
\end{equation*}
The contribution from the cycles $\tilde{a}_i$ and $\tilde{a}_i^{-1}$
can be written as
\begin{equation*}
\int_{\tilde{a}_i} \left( \int^{P_i}_{P_0} \omega_1 -
\int^{P'_i}_{P_0} \omega_1 \right) \omega_2,
\end{equation*}
where $P_i$ denotes the integration point along the cycle
$\tilde{a}_i$ and $P'_i$ the integration point on the cycle
$\tilde{a}_i^{-1}$ which is identified with $P_i$ on $M$ (see
Figure \ref{Figure: Riemann bilinear}).
\begin{figure}[ht]
\centering \psfrag{a}{\footnotesize $\tilde{a}_i$}
\psfrag{b}{\footnotesize $\tilde{b}_i$} \psfrag{a1}{\footnotesize
$\tilde{a}_i^{-1}$} \psfrag{b1}{\footnotesize $\tilde{b}_i^{-1}$}
\psfrag{P}{\footnotesize $P_i$} \psfrag{Pp}{\footnotesize $P'_i$}
\psfrag{ca}{\footnotesize \red{$c_a$}} \psfrag{cb}{\footnotesize
\green{$c_b$}} \psfrag{M}{\footnotesize $M_{\text{cut}}$}
\includegraphics[width=0.3\textwidth]{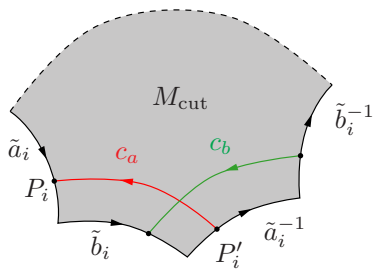}
\caption{The red contour $c_a$ joining $P'_i$ to $P_i$ is
homologous to $-b_i$ on $M$ whereas the green contour $c_b$ is
homologous to $a_i$.} \label{Figure: Riemann bilinear}
\end{figure}
But this is just equal to
\begin{equation*}
\int_{\tilde{a}_i} \left( \int^{P_i}_{P'_i} \omega_1 \right) \omega_2
= \int_{\tilde{a}_i} \left( - \int_{\tilde{b}_i} \omega_1 \right)
\omega_2,
\end{equation*}
or equivalently $- \int_{\tilde{b}_i} \omega_1 \int_{\tilde{a}_i}
\omega_2$. Likewise, the contribution from the cycles $\tilde{b}_i$
and $\tilde{b}^{-1}_i$ is $\int_{\tilde{a}_i} \omega_1
\int_{\tilde{b}_i} \omega_2$. Putting everything together, equation
\eqref{Riemann bilinear} now follows.
\end{proof}
\end{proposition}


\section{Analytic structure} \label{section: analytic structure}


At last we exploit the analyticity of the transition functions
between charts of $M$. Of course, everything up no now still holds
but as we will see, the simple analyticity requirement will lead
to a wealth of extra structure on $M$.

\subsection*{Abelian differentials}

\begin{definition}
A differential $\omega \in \Omega^1(M)$ is called \dub{holomorphic}
(or \dub{Abelian of the first kind}) if in any local chart $z : U
\rightarrow \mathbb{C}$ it is given by a holomorphic function
\begin{equation*}
\omega = f(z) dz, \qquad f \in \mathcal{H}(U).
\end{equation*}
The differential $\bar{\omega}$ is called \dub{anti-holomorphic}.
\end{definition}

\begin{remark}
This is well defined because in a different chart $z'$ we have $\omega
= f(z(z')) (\partial_{z'} z) dz'$ and $f(z(z')) (\partial_{z'}
z)$ is also holomorphic using the fact that $z' \mapsto z$ is.
\end{remark}

The general complex-valued differential $\omega = u dz + v
d\bar{z}$ in \eqref{one-form in basis C} is holomorphic if
\begin{equation} \label{holomorphic differential}
v = 0, \qquad u_{\bar{z}} = 0.
\end{equation}
As we saw in the remark following \eqref{one-form in basis C} the
condition on $u$ is equivalent to the Cauchy-Riemann equations for $u$
and hence is equivalent to $u$ being holomorphic. Equation
\eqref{holomorphic differential} together with \eqref{exterior
derivative} imply that every holomorphic differential is automatically
closed
\begin{equation*}
d \omega = 0.
\end{equation*}
We can therefore apply the Riemann bilinear identities \eqref{Riemann
bilinear} to holomorphic differentials. In particular,

\begin{lemma} \label{lemma: Rbilinear 2}
Let $\omega \not \equiv 0$ be a non-zero holomorphic differential
on $M$, then
\begin{equation*}
\textup{Im} \sum_{i=1}^g \int_{a_i} \omega \overline{\int_{b_i} \omega} < 0.
\end{equation*}
\begin{proof}
Let $\omega_1 = \omega$ and $\omega_2 = \bar{\omega}$ in
\eqref{Riemann bilinear} which in a local chart $U \subset M$ read
$\omega = f(z) dz$ and $\bar{\omega} = \overline{f(z)} d\bar{z}$.
Then $\omega \wedge \bar{\omega} = |f(z)|^2 dz \wedge d\bar{z} =
-2i |f|^2 dx \wedge dy$, so that $i \int_U \omega \wedge
\bar{\omega} > 0$ (this statement is coordinate independent
by proposition \ref{prop: every RS orientable}). The integral over
$M$ is defined patchwise so $i \int_M \omega \wedge \bar{\omega}
> 0$. The result follows after rewriting the right hand side of
\eqref{Riemann bilinear} as $2 i \sum_{i=1}^g \textup{Im}
\int_{a_i} \omega \int_{b_i} \bar{\omega}$.
\end{proof}
\end{lemma}

\begin{corollary} \label{corollary: holo vanishing a}
If $\omega \in \Omega^1(M)$ is holomorphic then
\begin{equation*}
\int_{a_i} \omega = 0, \quad i = 1, \ldots, g \qquad \Rightarrow
\qquad \omega \equiv 0.
\end{equation*}
\end{corollary}

The set of all holomorphic differentials obviously forms
a vector space over $\mathbb{C}$, which we denote
$\mathcal{H}^1(M)$. Denote also the set of anti-holomorphic
differentials as $\overline{\mathcal{H}}^1(M)$. Corollary
\ref{corollary: holo vanishing a} implies that $\dim
\mathcal{H}^1(M) \leq g$ since if $\omega_1, \ldots, \omega_{g+1}
\in \mathcal{H}^1(M)$ then some linear combination $\sum_{I =
1}^{g+1} \alpha_I \omega_I$ must have vanishing $\bm{a}$-periods
and hence must itself vanish, $\sum_{I = 1}^{g+1} \alpha_I
\omega_I = 0$. In fact, as we will see later $\dim
\mathcal{H}^1(M) = g$, and hence also $\dim
\overline{\mathcal{H}}^1(M) = g$. But corollary \ref{corollary:
holo vanishing a} also implies that there are no non-zero exact
holomorphic differentials on a compact Riemann surface $M$ without
boundary\footnote{The assumption that $M$ is compact and without
boundary is essential: if $D$ is the unit disc in $\mathbb{C}$ and $f$
is a function holomorphic in $D$ then $df$ is holomorphic and
exact. Equally, if $f$ is entire in $\mathbb{C}$ then $df$ is
holomorphic and exact in $\mathbb{C}$.}. So since $\mathcal{H}^1(M)
\cap \overline{\mathcal{H}}^1(M) = \varnothing$, it follows that
\begin{equation*}
H^1_{\text{dR}}(M) \cong \mathcal{H}^1(M) \oplus
\overline{\mathcal{H}}^1(M).
\end{equation*}
Differentials of the form $\alpha = \omega_1 + \bar{\omega}_2$
with $\omega_1, \omega_2$ holomorphic are called \dub{harmonic}.
They can be expressed locally as $\alpha = dh$ where $h$ is a
harmonic function.

\begin{lemma} \label{lemma: unique holo basis}
Let $\{a_i, b_i\}_{i=1}^g$ be a basis of $H_1(M)$. Then there exists a
unique dual basis $\{ \omega_i \}_{i=1}^g$ of $\mathcal{H}^1(M)$
which is \dub{normalised} by the condition
\begin{equation} \label{normalised holo basis}
\int_{a_i} \omega_j = \delta_{ij}.
\end{equation}
\begin{proof}
Let $\tilde{\omega}_1, \ldots, \tilde{\omega}_g$ be any basis of
$\mathcal{H}^1(M)$. By corollary \ref{corollary: holo vanishing a}
the $g \times g$ matrix $A_{ij} = \int_{a_i} \tilde{\omega}_j$ is
invertible (otherwise there exists $\alpha_j$ \textit{s.t.} $\sum_j
A_{ij} \alpha_j = 0$ and thus $\sum_j \alpha_j \tilde{\omega}_j =
0$). Then $\omega_j \equiv \tilde{\omega}_k A^{-1}_{kj}$ is another
basis of $\mathcal{H}^1(M)$ with the desired property
\eqref{normalised holo basis}.
\end{proof}
\end{lemma}

In order to get non-zero exact differentials we must therefore
allow for singularities.

\begin{definition}
A differential $\omega \in \Omega^1(M)$ is \dub{meromorphic} if
$\omega$ is holomorphic in $M \setminus \{ P_1, \ldots, P_m \}$ and
the behaviour around any $P_i \in U$ in a local chart $z : U
\rightarrow \mathbb{C}$ (with $z(P_i) = 0$) is given by a meromorphic
function
\begin{equation} \label{omega laurent series}
\omega = f(z) dz, \qquad f(z) = \sum_{j = -N_i}^{\infty} f_j z^j, \quad
N_i > 0, \; f_{N_i} \neq 0.
\end{equation}
\end{definition}

The set of all meromorphic differentials forms a vector space over
$\mathbb{C}$ which we denote $\mathcal{K}^1(M)$. Note that the set $S$
of poles of a meromorphic differential is discrete since meromorphic
functions on $\mathbb{C}$ have isolated poles. Moreover $S$ is finite
by compactness of $M$, \textit{i.e.} $S = \{ P_1, \ldots, P_m \}$.

\begin{remark}
One could have defined a meromorphic differential $\omega$ more
concisely as one that has a local representation of the form
\begin{equation} \label{mero diff 2}
\omega = f(z) dz, \qquad f \in \mathcal{K}(M).
\end{equation}
However, since we did not allow differentials to take the value
$\infty$ in the previous section one must be careful.
As a $\mathbb{C}$-valued differential, $\omega$ in \eqref{mero diff 2}
is only defined on $M' = M \setminus S$.
\end{remark}

\begin{definition} \label{definition: order, res, pole part}
With the notation of \eqref{omega laurent series} the \dub{order}
and \dub{residue} of $\omega$ at $P_i \in S$ are $\ord_{P_i}
\omega = - N_i$ and $\res_{P_i} \omega = f_{-1}$ respectively. The
\dub{singular part} of $\omega$ at $P_i$ is
\begin{equation*}
\sum_{j = -N_i}^{-1} f_j z^j, \qquad \text{where} \quad z(P_i) =
0.
\end{equation*}
\end{definition}

\begin{remark}
The order is well defined as it has an invariant geometrical
meaning (for much the same reason that the ramification number
of a branch point was well defined, see the discussion after lemma
\ref{Theorem: local holomorphic maps}), and the residue is chart
independent because $\res_{P_i} \omega = \frac{1}{2 \pi i} \int_{c_i}
\omega$, where $c_i$ is a counterclockwise cycle around $P_i$. In
general however the singular parts depend on the chart.
\end{remark}

\begin{proposition} \label{proposition: sum of residues}
Let $\omega$ be a meromorphic differential on a compact Riemann surface
$M$, then
\begin{equation*}
\sum_{P \in S} \res_P \omega = 0.
\end{equation*}
\begin{proof}
Consider the normal form $M_{\text{cut}}$ of $M$. Then
\begin{equation*}
\sum_{P \in S} \res_P \omega = \frac{1}{2 \pi i} \sum_{j = 1}^m
\int_{c_j} \omega = \frac{1}{2 \pi i} \int_{\partial M_{\text{cut}}}
\omega = 0,
\end{equation*}
using holomorphicity of $\omega$ on $M \setminus S$ in the second
equality. The last equality follows from the fact that $\omega$ is
single-valued on $M$ so for instance $\int_{\tilde{a}_i} \omega +
\int_{\tilde{a}_i^{-1}} \omega = 0$.
\end{proof}
\end{proposition}

\begin{definition}
An \dub{Abelian differential} is of the \dub{first kind} if it is
holomorphic, of the \dub{second kind} if it is meromorphic with
vanishing residues and of the \dub{third kind} otherwise.
\end{definition}

Since an Abelian differential $\omega$ is closed on $M \setminus S$,
its primitive is locally well defined
\begin{equation} \label{Abelian integral}
\Omega(P) = \int^P_{P_0} \omega.
\end{equation}
One can recover the Abelian differential from it by $\omega =
d\Omega$. It follows that $\Omega(P)$ defines a meromorphic
function on the whole of $M$ only if $\omega$ is exact. More
generally the \dub{Abelian integral} $\Omega(P)$ defined by
\eqref{Abelian integral} on $M$ will be multi-valued precisely
when the cohomology class of $\omega$ is non-trivial which
corresponds by lemma \ref{lemma: vanishing periods} to some of the
periods of $\omega$ being non-zero. So consider a closed cycle $c$
on $M' = M \setminus S$. Because $M'$ has extra `punctures' at the
set $S$, a closed path on $M'$ is of the form
\begin{equation*}
c \sim \sum_{i=1}^g n_i a_i + \sum_{i=1}^g m_i b_i + \sum_{j=1}^m k_j
c_j, \quad n_i, m_i, k_j \in \mathbb{Z},
\end{equation*}
modulo boundaries, where $c_j$ is a cycle around $P_j$. In other words
$\{ a_i, b_i \}_{i=1}^g$ together with $\{ c_j \}_{j=1}^m$ form a
basis of $H_1(M')$. It follows that for the closed Abelian
differential $d \Omega$ we have the equality
\begin{equation} \label{basis of periods third kind}
\int_c d \Omega = \sum_{i=1}^g n_i \int_{a_i} d \Omega + \sum_{i=1}^g
m_i \int_{b_i} d \Omega + 2 \pi i \sum_{j=1}^m k_j \res_{P_j} d
\Omega.
\end{equation}
This equation is to be contrasted with the analogous formula
\eqref{basis of periods} for the periods of regular differentials.
Note however that the new term involving residues is only present
when $d \Omega$ is of the third kind, and so in this case the
multi-valuedness of the Abelian integral $\Omega$ is specified by
the $\bm{a}$- and $\bm{b}$-periods of $d \Omega$ along with its
residues.

Due to lemma \ref{corollary: holo vanishing a}, not all
$\bm{a}$-periods of an Abelian integral of the first can be zero.
Now suppose $d \Omega$ is an Abelian differential of the second or
third kind. In general its $\bm{a}$-periods are non-trivial, say
\begin{equation*}
A_i = \int_{a_i} d \Omega.
\end{equation*}
Consider subtracting from $d\Omega$ a combination of holomorphic
differentials, by defining $d \hat{\Omega} = d \Omega - \sum_{j = 1}^g
\alpha_j \omega_j$. Clearly $d \hat{\Omega}$ has the same singular
behaviour as $d \Omega$. However, the $\bm{a}$-periods get shifted
\begin{equation*}
\int_{a_i} d \hat{\Omega} = A_i - \alpha_i.
\end{equation*}
Therefore by choosing $\alpha_i = A_i$ one can set all the
$\bm{a}$-periods of $d \hat{\Omega}$ to zero.
\begin{definition}
We will say that an Abelian differential $d\Omega$ of the second
or third kind is \dub{normalised} if all its $\bm{a}$-periods
vanish, \textit{i.e.} $\int_{a_i} d\Omega = 0, i = 1, \ldots, g$.
\end{definition}

\begin{remark}
By the discussion following equation \eqref{basis of periods third
kind}, an Abelian differential $d \Omega$ of the third kind must be
normalised with respect to a choice of $\bm{a}$-cycles in the homology
group $H_1(M')$ and \textit{not} $H_1(M)$. Indeed, two $\bm{a}$-cycles
$a_i$ and $a'_i$ which are homologous in $H_1(M)$ are not necessarily
homologous in $H_1(M')$ but $a'_i \sim a_i + \sum_{j = 1}^m k_j c_j$
so that $\int_{a_i} d \Omega \neq \int_{a'_i} d \Omega$.
\end{remark}

By the previous argument, any Abelian differential $d \Omega$ can be
normalised by adjusting its holomorphic part. Moreover, the normalised
differential is zero (\textit{i.e.} $d \hat{\Omega} = 0$) if and only
if $d \Omega$ was holomorphic. The following lemma shows that the
normalised part $d \hat{\Omega}$ uniquely characterises the singular
part of $d \Omega$.

\begin{lemma} \label{lemma: unique normalised}
A normalised meromorphic differential $d \Omega$ is uniquely
defined by the singular parts at each of its poles.
\begin{proof}
Suppose $d \Omega_1$ and $d \Omega_2$ are two normalised meromorphic
differentials with the same set of poles and the same singular parts
at these poles. Then $\omega = d \Omega_1 - d \Omega_2$ is holomorphic
since the poles parts cancel out. But $\int_{a_i} \omega = 0$ since $d
\Omega_1$ and $d \Omega_2$ are both normalised. It follows by lemma
\ref{corollary: holo vanishing a} that $\omega = 0$, namely $d
\Omega_1 = d \Omega_2$.
\end{proof}
\end{lemma}

\noindent {\bf Examples}

\noindent We give two important examples of Abelian differentials
denoted $\omega_P^{(n)}$ and $\omega_{PQ}$ of the second and third
kinds respectively.

\begin{itemize}
  \item Let $P \in M$ and $z$ a local coordinate around $P$ with
$z(P) = 0$. Define a normalised Abelian differential of the second
kind $\omega_P^{(n)}$ with singular parts at $P$ of the form
\begin{equation*}
\omega_P^{(n)} = \frac{dz}{z^n}, \quad n \geq 2.
\end{equation*}
Such a differential can be shown to exist and it is unique by lemma
\ref{lemma: unique normalised}. Note however that its definition
depends on the local coordinate $z$ at $P$.

  \item Let $P,Q \in M$. Introduce a normalised Abelian differential
of the third kind $\omega_{PQ}$ with singular parts at $P$ and $Q$
such that
\begin{gather*}
\ord_P \omega_{PQ} = \ord_Q \omega_{PQ} = - 1\\
\res_P \omega_{PQ} = 1, \quad \res_Q \omega_{PQ} = - 1.
\end{gather*}
Such a differential can also be shown to exist and once again it is
uniquely specified according to lemma \ref{lemma: unique
normalised}. This time however it does not depend on a choice of
coordinates since it was defined in terms of invariants.
\end{itemize}

These differentials together with the $g$ basis holomorphic
differentials $\omega_i$ form a complete basis of Abelian
differentials on $M$ in the sense that any Abelian differential
$d \Omega$ can be written as a finite linear combination of those
\begin{equation} \label{Abelian differential basis}
d \Omega = \sum_{i = 1}^g \alpha_i \omega_i + \sum_{P \in M} \sum_{n =
2}^{N_P} \beta_{P,n} \omega_P^{(n)} + \sum_{Q,R \in M} \gamma_{Q,R}
\omega_{QR},
\end{equation}
where all but finitely many of the constants $\alpha_i, \beta_{P,n},
\gamma_{Q,R} \in \mathbb{C}$ are zero. To arrive at \eqref{Abelian
differential basis} one first normalises $d\Omega$ to obtain $d
\hat{\Omega}$ by subtraction of a (unique) linear combination of
holomorphic differentials. One then reconstructs the finite singular
part of $d \hat{\Omega}$ from a linear combinations of the
$\omega_P^{(n)}, \omega_{QR}$ and invokes lemma \ref{lemma: unique
normalised}. The $\gamma$ coefficients are note quite unique since for
instance $\omega_{PQ} + \omega_{QR} = \omega_{PR}$.

\subsection*{More Riemann bilinear identities}

In section \ref{section: differential calculus} we derived the
Riemann bilinear identities \eqref{Riemann bilinear} for closed
differentials. Since holomorphic differentials are closed on $M$
one can readily apply \eqref{Riemann bilinear} to them. In fact,
for any $\omega_1, \omega_2 \in \mathcal{H}^1(M)$ we have
$\omega_1 \wedge \omega_2 = 0$ and so
\begin{equation} \label{Riemann bilinear holo}
\sum_{i=1}^g \left[ \int_{a_i} \omega_1 \int_{b_i} \omega_2 -
\int_{b_i} \omega_1 \int_{a_i} \omega_2 \right] = 0.
\end{equation}

But now we must also allow for $\omega_1$ and $\omega_2$ to have
singularities. In this case the Riemann bilinear identities
receive extra contributions from the singularities.
\begin{proposition} \label{proposition: Riemann bilinear 2}
Let $d \Omega_1, d \Omega_2$ be Abelian differentials on $M$ where $d
\Omega_1$ is not of the third kind, then
\begin{equation} \label{Riemann bilinear 3}
\sum_{i=1}^g \left[ \int_{a_i} d \Omega_1 \int_{b_i} d \Omega_2 -
\int_{b_i} d \Omega_1 \int_{a_i} d \Omega_2 \right] = 2 \pi i
\sum_{P \in M} \res_P \Omega_1 d \Omega_2.
\end{equation}
\begin{proof}
Consider once again the normal form $M_{\text{cut}}$ of $M$. Since $d
\Omega_1$ is not of the third kind its Abelian integral $\Omega_1$ is
single-valued in $M_{\text{cut}}$. Thus consider the meromorphic
differential $\Omega_1 d\Omega_2$ on $M_{\text{cut}}$. Its integral
around the boundary $\partial M_{\text{cut}}$ is
\begin{equation*}
\int_{\partial M_{\text{cut}}} \Omega_1 d\Omega_2 = \sum_{i=1}^g
\left[ \int_{\tilde{a}_i} \Omega_1 d\Omega_2 + \int_{\tilde{b}_i}
\Omega_1 d\Omega_2 + \int_{\tilde{a}_i^{-1}} \Omega_1 d\Omega_2 +
\int_{\tilde{b}_i^{-1}} \Omega_1 d\Omega_2 \right],
\end{equation*}
which by the exact same reasoning as in the proof of proposition
\ref{proposition: Riemann bilinear} gives the left hand side of
\eqref{Riemann bilinear 3}. On the other hand, $\Omega_1 d\Omega_2$ is
holomorphic on $M_{\text{cut}} \setminus S$ where $S = \{ P_1, \ldots,
P_m\}$ is the finite set of singular points of $\Omega_1
d\Omega_2$. Therefore
\begin{equation*}
\int_{\partial M_{\text{cut}}} \Omega_1 d\Omega_2 = \sum_{j = 1}^m
\int_{c_j} \Omega_1 d\Omega_2,
\end{equation*}
where $c_j$ is a small counterclockwise cycle around $P_j$. This last
sum of integrals produces the right hand side of \eqref{Riemann
bilinear 3}.
\end{proof}
\end{proposition}

\begin{corollary}
\begin{equation} \label{Riemann bilinear 4}
\int_{b_i} \omega_{PQ} = 2 \pi i \int_Q^P \omega_i.
\end{equation}
\begin{proof}
Apply proposition \ref{proposition: Riemann bilinear 2} to $d \Omega_1 =
\omega_i$ and $d \Omega_2 = \omega_{PQ}$ and use
\begin{equation*}
\sum_{P' \in M} \res_{P'} \Omega_1 \omega_{PQ}
= \Omega_1(P) - \Omega_1(Q) = \int_Q^P d\Omega_1. \vspace{-15mm}
\end{equation*}
\end{proof}
\end{corollary}

If the Abelian differentials $d \Omega_1, d \Omega_2$ are both of the
third kind we cannot make use of proposition \ref{proposition: Riemann
bilinear 2}. Yet there is also a Riemann bilinear identity relating
their periods. We will only need the case when $d \Omega_1 =
\omega_{PQ}$ and $d \Omega_2 = \omega_{RS}$.

\begin{proposition} \label{proposition: Riemann bilinear 3}
\begin{equation} \label{Riemann bilinear 5}
\int_S^R \omega_{PQ} = \int_Q^P \omega_{RS}.
\end{equation}
\begin{proof}
Because $\omega_{PQ}$ has residues at the points $P, Q$ (assumed
\textit{w.l.o.g.} to lie in the interior of $M_{\text{cut}}$) we
cannot write $\omega_{PQ} = df$ for some $f$ in $M_{\text{cut}}$. Yet
if we introduce an extra `cut' $[P,Q]$ between the points $P$ and $Q$
then $\omega_{PQ} = df$ is now exact on $M'_{\text{cut}} \equiv
M_{\text{cut}} \setminus [P, Q]$.
\begin{figure}[ht]
\centering \psfrag{a}{\footnotesize $\tilde{a}_i$}
\psfrag{b}{\footnotesize $\tilde{b}_i$} \psfrag{a1}{\footnotesize
$\tilde{a}_i^{-1}$} \psfrag{b1}{\footnotesize $\tilde{b}_i^{-1}$}
\psfrag{P}{\footnotesize $P$} \psfrag{Q}{\footnotesize $Q$}
\psfrag{R}{\footnotesize $R$} \psfrag{S}{\footnotesize $S$}
\psfrag{M}{\footnotesize $M'_{\text{cut}}$}
\includegraphics[width=0.3\textwidth]{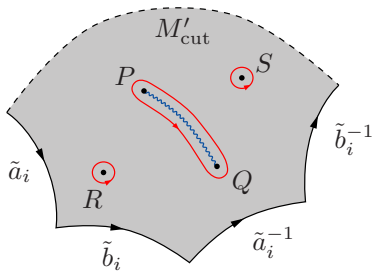}
\caption{The normal form $M_{\text{cut}}$ cut along $[P,Q]$.}
\label{Figure: Riemann bilinear 2}
\end{figure}
Consider the single-valued differential $f \omega_{RS}$ which is
holomorphic on $M'_{\text{cut}} \setminus \{ R, S\}$. As usual its
integral around the boundary $\partial M_{\text{cut}}$ is
\begin{equation} \label{RB 3 proof 1}
\int_{\partial M_{\text{cut}}} f \omega_{RS} = \sum_{i=1}^g
\left[ \int_{a_i} \omega_{PQ} \int_{b_i} \omega_{RS} - \int_{a_i}
\omega_{RS} \int_{b_i} \omega_{PQ} \right] = 0,
\end{equation}
where the last equality follows because $\omega_{PQ}$ and
$\omega_{RS}$ are both normalised. On the other hand, since $f
\omega_{RS}$ is holomorphic on $M'_{\text{cut}} \setminus \{ R, S\}$
we have
\begin{equation} \label{RB 3 proof 2}
\int_{\partial M_{\text{cut}}} f \omega_{RS} = 2 \pi i \res_R f
\omega_{RS} + 2 \pi i \res_S f \omega_{RS} + \int_c f \omega_{RS},
\end{equation}
where $c$ is the keyhole contour around the cut $[P,Q]$.
Since the function $f$ jumps by $2 \pi i \res_P \omega_{PQ} = 2
\pi i$ across this cut the right hand side of \eqref{RB 3 proof 2}
evaluates to
\begin{equation*}
2 \pi i \left(f(R) - f(S)\right) + 2 \pi i \int_P^Q \omega_{RS} = 2
\pi i \left( \int_S^R \omega_{PQ} + \int_P^Q \omega_{RS} \right).
\end{equation*}
Putting this result together with \eqref{RB 3 proof 1} yields
\eqref{Riemann bilinear 5}.
\end{proof}
\end{proposition}

\subsection*{Divisors}

\begin{definition} \label{definition: divisor}
A \dub{divisor} on $M$ is a formal finite sum of points
\begin{equation*}
D = \sum_{P \in M} m_P P, \qquad m_P \in \mathbb{Z}.
\end{equation*}
where $m_P = 0$ for all but finitely many points $P \in M$.
\end{definition}
\noindent We denote by $\Div(M)$ the \dub{group of divisors} on $M$,
\textit{i.e.} the free Abelian group generated by the points of
$M$. If $D' = \sum_{P \in M} n_P P$ is another divisor the group
operations are defined by,
\begin{equation*}
D + D' = \sum_{P \in M} (m_P + n_P) P, \qquad -D = \sum_{P \in M}
(-m_P) P,
\end{equation*}
and the identity divisor is denoted by $0$. This group is endowed with
a natural homomorphism, called the \dub{degree}
\begin{equation*}
\deg : \Div(M) \rightarrow \mathbb{Z}, \qquad \sum_{P \in M} m_P P
\mapsto \sum_{P \in M} m_P.
\end{equation*}
There is an obvious partial ordering on the set of divisors defined by
\begin{equation*}
D \geq D' \quad \Leftrightarrow \quad m_P \geq n_P \; \forall P \in M.
\end{equation*}
A divisor $D$ is said to be \dub{positive} (or \dub{integral} or
\dub{effective}) if $D \geq 0$.

A meromorphic function $f \in \mathcal{K}(M)$ on $M$ defines a divisor
$(f)$ called a \dub{principal divisor} as
\begin{equation} \label{principal divisor}
(f) = \sum_{P \in M} (\ord_P f) P,
\end{equation}
where $\ord_P f$ is the order of $P$ if $f$ has a pole at
$P$ or the multiplicity of $P$ if $f$ has a zero at $P$.
Since $M$ is compact, theorem \ref{Theorem: branched cover} implies
that principal divisors have degree zero,
\begin{equation} \label{deg principal divisor}
\deg (f) = 0.
\end{equation}

As it stands, the group $\Div(M)$ does not have much
structure and is rather huge. So consider the \dub{linear equivalence}
on the set of divisors defined as follows
\begin{equation*}
D \sim D' \quad \Leftrightarrow \quad \exists f \in
\mathcal{K}(M) \textit{ s.t. } (f) = D - D',
\end{equation*}
and define the \dub{divisor class group} $\Pic(M)$ as the quotient
$\Pic(M) \equiv \Div(M) / \!\! \sim$. In the same way that a
function on $M$ defined a natural divisor by equation
\eqref{principal divisor}, a $1$-form $\omega$ on $M$ also defines
a divisor $(\omega)$ as
\begin{equation*}
(\omega) = \sum_{P \in M} (\ord_P \omega) P.
\end{equation*}
Note that the ratio of two meromorphic $1$-forms $\omega_1, \omega_2
\in \mathcal{K}^1(M)$ is a meromorphic function $\omega_1/\omega_2 \in
\mathcal{K}(M)$ with divisor $(\omega_1) - (\omega_2)$ and thus
$(\omega_1) \sim (\omega_2)$. Therefore any meromorphic $1$-form
$\omega$ defines the same divisor class $K = (\omega) \in
\Pic(M)$ called the \dub{canonical divisor} or \dub{canonical
class}. Equation \eqref{deg principal divisor} also implies that the
degree of the canonical class is well defined since $\deg (\omega_1) =
\deg (\omega_2)$.

Given a meromorphic function $f$, by definition its divisor of poles
is equivalent to its divisor of zeroes. Conversely, given two
\textit{equivalent} divisors $D_0 = \sum_{i = 1}^n P_i$ and
$D_{\infty} = \sum_{i = 1}^n Q_i$ one can ask what meromorphic
function $f$ has the property that $(f) = D_0 - D_{\infty}$. This
question is answered by the following lemma. Equation \eqref{gen Abel
useful equality} will also be crucial later in discussions of section
\ref{section: jacobian} in relation to the generalised Abel map and
generalised Jacobians.

\begin{lemma} \label{lemma: f(P)/f(Q)=r}
Let $f$ be meromorphic with divisor $(f) = \sum_{i=1}^n (P_i - Q_i)$,
then
\begin{equation} \label{gen Abel useful equality}
\frac{f(P)}{f(Q)} = \exp \sum_{i = 1}^n \int_{Q_i}^{P_i} \omega_{PQ},
\end{equation}
for any two points $P,Q \in M$.
\begin{proof}
Using the Riemann bilinear identities \eqref{Riemann bilinear 5} the
quantity in the exponent can be rewritten as $\sum_{i = 1}^n \int_Q^P
\omega_{P_i Q_i}$. Since $(f) = \sum_{i=1}^n (P_i - Q_i)$ the
differential $\frac{df}{f}$ has poles only at $P_i$ with residue $+1$
and at $Q_i$ with residue $-1$. But then
\begin{equation} \label{gen Abel proof 1}
\frac{df}{f} - \sum_{i = 1}^n \omega_{P_i Q_i} = \sum_{j = 1}^g c_j
\omega_j,
\end{equation}
for some $c_j \in \mathbb{C}$. Taking the $\bm{a}$-periods
of this equation leads to $c_j = \int_{a_j} d \log f = 2 \pi i m_j$,
$m_j \in \mathbb{Z}$. On the other hand taking the integral from $Q$
to $P$ leads to
\begin{equation} \label{gen Abel proof 2}
\log \left( \frac{f(P)}{f(Q)} \right) = \sum_{i = 1}^n \int_Q^P
\omega_{P_i Q_i} + 2 \pi i \sum_{j = 1}^g m_j \int_Q^P \omega_j,
\end{equation}
which holds as an equality modulo $2 \pi i$. However in the limit $P_i
\rightarrow Q_i$ we have $f \rightarrow 1$ and so the left hand side
tends to zero modulo $2 \pi i$. Likewise the first sum on the right
hand side tends to zero in this limit because it can be written as
$\sum_{i = 1}^n \int_{Q_i}^{P_i} \omega_{PQ}$. Since the very last
term is discrete it must therefore always vanish modulo $2 \pi i$, so
we may set it to zero in \eqref{gen Abel proof 2}. Taking the
exponential proves the lemma.
\end{proof}
\end{lemma}

If we choose the function $f$ to be normalised at $Q$ say, so that
$f(Q) = 1$, then \eqref{gen Abel useful equality} gives a closed
formula for the function $f$ with $(f) = \sum_{i=1}^n (P_i -
Q_i)$, namely
\begin{equation} \label{function for Abel thm}
f(P) = \exp \sum_{i = 1}^n \int_{Q_i}^{P_i} \omega_{PQ}.
\end{equation}
Of course, if the divisors $D_0 = \sum_{i = 1}^n P_i$ and $D_{\infty}
= \sum_{i = 1}^n Q_i$ are not equivalent then \eqref{function for Abel
thm} should not define a single valued function on the Riemann surface
$M$.

\subsection*{The Riemann-Roch theorem}

Let $D$ be an arbitrary divisor. We introduce the following vector
space of meromorphic functions with prescribed zeroes and allowed
poles,
\begin{equation*}
L(D) = \{ f \in \mathcal{K}(M) \;|\; (f) \geq D \}.
\end{equation*}
The content of this vector space is determined by the divisor $D$ as
follows: if a point $P \in M$ figures in $D$ with coefficient $n > 0$
then every $f \in L(D)$ is \textit{forced} to have a zero of order $n$
at $P$. If however $Q \in M$ figures in $D$ with coefficient $m < 0$
then any $f \in L(D)$ is \textit{allowed} to have at most a pole of
order $-m$ at $Q$. In other words, if we split $D = D_0 - D_{\infty}$
into two positive divisors $D_0 = \sum_j n_j P_j \geq 0$ and
$D_{\infty}= \sum_k m_k Q_k  \geq 0$ then a meromorphic function $f$
is in $L(D)$ provided it has zeroes of order at \textit{least} $n_j$
at $P_j$ and poles of order at \textit{most} $m_k$ at $Q_k$.
We denote the dimension of this space as
\begin{equation*}
r(D) = \dim L(D).
\end{equation*}
Let us introduce a second vector space, containing meromorphic
differentials with prescribed zeroes and allowed poles,
\begin{equation*}
\Omega(D) = \{ \omega \in \mathcal{K}^1(M) \;|\; (\omega) \geq D \}.
\end{equation*}
The description of this space is identical to $L(D)$ but with
the word `function' replaced by the word `differential'. Its
dimension we denote by
\begin{equation*}
i(D) = \dim \Omega(D).
\end{equation*}
It is clear that $r(D)$ and $i(D)$ only depend on the divisor class of
$D$: if $D_1 \sim D_2$ then there exists $h \in \mathcal{K}(M)$ with
$(h) = D_1 - D_2$ and multiplication by $h$ defines vector space
isomorphisms $L(D_2) \rightarrow L(D_1)$ and $\Omega(D_2) \rightarrow
\Omega(D_1)$ and thus $r(D_1) = r(D_2)$ and $i(D_1) =
i(D_2)$. Furthermore, these dimensions are related as follows
\begin{equation} \label{relation i and r}
i(D) = r(D - K).
\end{equation}
Indeed, if $\omega_0$ is any meromorphic differential its divisor is
the canonical divisor $(\omega_0) = K$ so that $\omega \mapsto
\omega/\omega_0$ defines a vector space isomorphism $\Omega(D)
\rightarrow L(D-K)$.

We are now in a position to state one of the most important theorems
on compact Riemann surfaces,
\begin{theorem}[Riemann-Roch] \label{Theorem: Riemann-Roch}
Let $M$ be a compact Riemann surface of genus $g$ and $D$ a divisor on
$M$. Then
\begin{equation} \label{Riemann-Roch}
r(-D) = \deg D - g + 1 + i(D).
\end{equation}
\end{theorem}

\begin{corollary} \label{corollary: dim H^1}
$\dim \mathcal{H}^1(M) = g$.
\begin{proof}
Let $D = 0$ in \eqref{Riemann-Roch}. Since a meromorphic function $f :
M \rightarrow \mathbb{C}P^1$ on a compact Riemann surface $M$ is
either constant or surjective by theorem \ref{Theorem: holomorphic
maps} it follows that $L(0) = \mathbb{C}$, \textit{i.e.} $r(0) =
1$. But then $i(0) = g$, so the space $\Omega(0) = \mathcal{H}^1(M)$
of holomorphic differentials is $g$ dimensional.
\end{proof}
\end{corollary}

\begin{corollary} \label{corollary: i(-D)}
If $\deg D < 0$ then $i(D) = - \deg D - 1 + g$.
\begin{proof}
Again using theorem \ref{Theorem: holomorphic maps} we find that $r(-D)
= 0$ since a meromorphic function $f : M \rightarrow \mathbb{C}P^1$
cannot have strictly more zeroes than poles.
\end{proof}
\end{corollary}

\begin{corollary} \label{cor: canonical degree}
$\deg K = 2 g - 2$.
\begin{proof}
Let $D = K$ in \eqref{Riemann-Roch}. Using \eqref{relation i and r} we
have that $i(K) = r(0) = 1$ and $r(-K) = i(0) = g$ by corollary
\ref{corollary: dim H^1}.
\end{proof}
\end{corollary}

\begin{corollary} \label{cor: Riemann sphere}
Every compact Riemann surface $M$ of genus zero is conformally
equivalent to the Riemann sphere $\mathbb{C}P^1$.
\begin{proof}
Let $P \in M$ then clearly $r(-P) = 2$ (since $g=0$ and $i(P) = 0$ as
$\mathcal{H}^1(M) = \varnothing$) so there exists a non-constant
meromorphic function of degree one on $M$ which is is a bijection by
theorem \ref{Theorem: holomorphic maps}.
\end{proof}
\end{corollary}

It is obvious from theorem \ref{Theorem: holomorphic maps} that if $D
< 0$ then $-D$ is strictly positive and $r(-D) = 0$. Given a generic
divisor $D \geq 0$ we would like to use the Riemann-Roch theorem to
compute $r(-D)$. According to \eqref{Riemann-Roch} we need only
determine $i(D)$. Since we are assuming $D \geq 0$, this is the
dimension of the space $\Omega(D)$ of holomorphic differentials
vanishing at $D$. If $\deg D \geq 2g - 1$ then by corollary \ref{cor:
canonical degree} there is no such differential and so $i(D) =
0$. Thus we have
\begin{equation*}
r(-D) \left\{ \begin{array}{ll} = 0, & \quad \deg D < 0\\ \geq 1 - g +
\deg D, & \quad 0 \leq \deg D < 2g - 1\\ = 1 - g + \deg D, & \quad
\deg D \geq 2 g - 1 \end{array} \right.
\end{equation*}
It remains to discuss positive divisors of the form $D = P_1 +
\cdots + P_n$ of degree $\deg D = n$ in the range $0 \leq n < 2 g
- 1$. Since the space of holomorphic differentials is of dimension
$g$ by corollary \ref{corollary: dim H^1}, the space $\Omega(D)$
consists of the solutions $\bm{c} = (c_i)_{i = 1}^g$ to the linear
system
\begin{equation} \label{linear system for i(D)}
\sum_{i = 1}^g c_i \omega_i(P_j) = 0, \quad j = 1, \ldots, n.
\end{equation}
Now when $n \leq g$, the $n \times g$ matrix $M_{ji} = \omega_i(P_j)$
will typically be of rank $n$ except for very specific divisors
$D$. Therefore generically in this case we will have $i(D) = g - n$
and hence $r(-D) = 1$. If however $n > g$, then the system
\eqref{linear system for i(D)} is over determined and generically has
no solutions, except once again for very specific divisors $D$. So
generically in this case we have $i(D) = 0$ and hence $r(-D) = n - g +
1$.

\begin{definition} \label{def: non-special divisors}
A positive divisor $D \geq 0$ is \dub{special} if either $\deg D \leq
g$, $r(-D) > 1$ or $\deg D \geq g$, $i(D) > 0$. It is
\dub{non-special} (or \dub{generic} or \dub{in general position}) if
either $\deg D \leq g$, $r(-D) = 1$ or $\deg D \geq g$, $i(D) = 0$.
\end{definition}

Of particular interest will be the case $\deg D = g$. Note also that
if $D' = D + Q$ with $Q \in M$ then $D'$ is non-special whenever $D$
is non-special because $i(D) = 0 \Rightarrow i(D') = 0$ (from
observing \eqref{linear system for i(D)}).

\subsection*{Moduli space at genus g}

Topologically speaking, the only invariant of a compact Riemann
surface is its genus. That is, by theorem \ref{Theorem: sphere
handles} any two Riemann surfaces $M, M'$ are homeomorphic if and
only if they have the same genus. As it turns out, in two
dimensions any two compact orientable surfaces $M, M'$ that are
homeomorphic are also diffeomorphic. Thus at every genus $g \geq
0$ there is a unique differential structure up to diffeomorphisms.
When it comes to analytic structures however things are very
different. If we consider two Riemann surfaces $M,M'$ as
equivalent when there is a biholomorphic mapping
\begin{equation*}
f : M \rightarrow M',
\end{equation*}
\textit{i.e.} a bijection with $f$ and $f^{-1}$ holomorphic, then
it turns out that at every genus $g \geq 1$ there is a continuous
family of inequivalent Riemann surfaces. Denoting by
$\mathcal{M}_g$ the \dub{moduli space} of inequivalent Riemann
surfaces at genus $g$, the following important proposition is also
a direct consequence of the Riemann-Roch theorem.
\begin{proposition}
\begin{equation*}
\dim \mathcal{M}_g = \left\{ \begin{array}{ll} 0, & \quad g = 0\\
1, & \quad g = 1\\ 3g - 3, & \quad g \geq 2. \end{array} \right.
\end{equation*}
\end{proposition}

More generally it will be important to consider punctured Riemann
surfaces. A \dub{punctured} Riemann surface is simply a Riemann
surface $M$ marked at a finite set of ordered points $(P_{\alpha}
\in M)_{\alpha = 1}^N$. We denote by $\mathcal{M}_{g,N}$ the
moduli space of punctured Riemann surfaces of genus $g$ with $N$
punctures.
\begin{corollary} \label{corollary: moduli punctured RS}
For any genus $g$ and number $N$ of punctures we have
\begin{equation} \label{moduli punctured RS}
\dim \mathcal{M}_{g,N} = 3g - 3 + N.
\end{equation}
\begin{proof}
At genus zero there is only one Riemann sphere but its automorphism
group is the M\"obius group which has three complex parameters and
hence allows one to fix three of the punctures to say $0$,$1$ and
$\infty$. This leaves $N-3$ free parameters.

At genus one there is a one parameter family of conformally
inequivalent torii but one can fix a puncture to say $0$ so $\dim
\mathcal{M}_{1,N} = 1 + (N-1) = N$.

Finally when $g > 1$ the automorphism group is finite so the dimension
of the moduli space of Riemann surfaces of genus $g$ with $N$
punctures is simply $3g - 3 + N$.
In every case the formula $3g - 3 + N$ gives the correct count for
$\dim \mathcal{M}_{g,N}$.
\end{proof}
\end{corollary}


\section{Algebraic curves} \label{section: algebraic curves}


Most examples of Riemann surfaces we will need are non-singular
algebraic curves. These were already introduced in section
\ref{section: definition} as the zero-locus of a polynomial $P$ in two
complex variables $x,y$,
\begin{equation} \label{algebraic curve}
C = \{ (x,y) \in \mathbb{C}^2 | P(x,y) = 0 \}.
\end{equation}
The non-singular condition is the requirement that at any point
$(a,b) \in C$ the gradient of $P$ is non-vanishing, namely $d P(a,b)
\neq 0$. Therefore in the immediate neighbourhood of any point $(a,b)
\in C$ the curve \eqref{algebraic curve} looks locally like
\begin{equation} \label{tangent line}
(x - a) \frac{\partial P}{\partial x}(a,b) + (y - b) \frac{\partial
P}{\partial y}(a,b) = 0.
\end{equation}
This is the equation for a line in $\mathbb{C}^2$, namely a copy of
$\mathbb{C}$. In other words the non-singular condition means that $C$
is locally homeomorphic to $\mathbb{C}$ and an obvious local parameter
is $x$ if $\partial P/ \partial y \neq 0$ or $y$ if $\partial P/
\partial x \neq 0$. In a neighbourhood where either local parameter
works the transition functions $x(y)$ and $y(x)$ are holomorphic by
the implicit function theorem. Therefore non-singular algebraic curves
satisfy all the requirements of a Riemann surface.

\subsection*{Singularities}

Oftentimes however an algebraic curve defined by \eqref{algebraic
curve} will be singular.

\begin{definition}
A point $(a,b) \in C$ is \dub{singular} if $d P(a,b) = 0$.
\end{definition}

In the neighbourhood of such a point the curve $C$ no longer looks
like \eqref{tangent line} since one has to look at subleading
terms. The \dub{multiplicity} of a singular point is the smallest
integer $m$ such that
\begin{equation*}
\frac{\partial^m P}{\partial x^i \partial y^j}(a,b) \neq 0,
\end{equation*}
for some $0 \leq i, j \leq m$ such that $i + j = m$. The curve $C$ is
then locally described by a homogeneous polynomial of degree $m$ and
\eqref{tangent line} is replaced by
\begin{equation} \label{tangent lines at singular}
\sum_{i+j = m} \frac{\partial^m P}{\partial x^i \partial y^j}(a,b)
\frac{(x - a)^i (y - b)^j}{i! j!} = 0.
\end{equation}
Since the left hand side polynomial is homogeneous in $(x - a)$
and $(y - b)$ of degree $m$ it can be factored into a product of
$m$ linear polynomials and \eqref{tangent lines at singular} is
equivalent to a set of $m$ linear equations $\alpha_i (x - a) +
\beta_i (y - b) = 0$ where $i = 1, \ldots, m$ and $(\alpha_i,
\beta_i) \neq (0,0)$. Each of these linear equations defines a
complex line in $\mathbb{C}^2$ which means that locally near a
singular point the curve $C$ looks like the intersection of
several copies of $\mathbb{C}$. The singular point is
\dub{ordinary} if the polynomial in \eqref{tangent lines at
singular} has no repeated factor. In this case the curve $C$ looks
locally like the intersection of $m$ distinct lines.

\begin{definition}
A \dub{node} is an ordinary singular point of multiplicity two.
\end{definition}

\noindent By performing the birational change of variables $X =
\alpha_1 (x - a) + \beta_1 (y - b)$ and $Y = \alpha_2 (x - a) +
\beta_2 (y - b)$ a node can always be brought to the canonical form
\begin{equation} \label{node}
\raisebox{-14mm}{
\includegraphics[height=30mm]{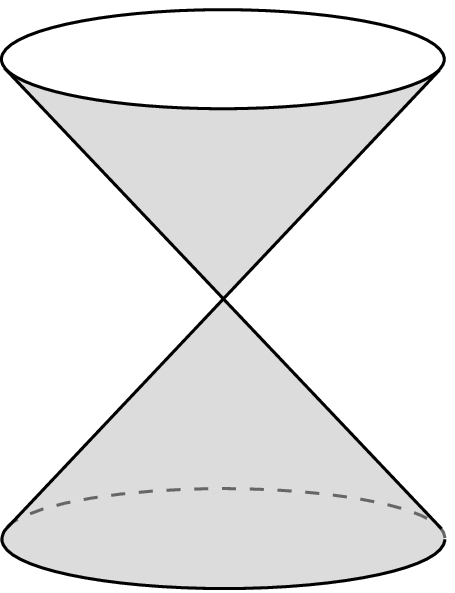}} \qquad \qquad
X Y = 0.
\end{equation}
As depicted in the picture this consists of two copies of the
complex line, namely $X = 0$ and $Y = 0$, intersecting at the
common node $(X,Y) = (0,0)$. Performing the birational change of
coordinates $X = z - w$ and $Y = z + w$ leads to an equivalent
representation of the node \eqref{node}, namely $z^2 = w^2$. A
singularity of the form $z^2 = w^3$ is called a \dub{cusp}. More
generally,
\begin{definition} \label{definition: cusps and nodes}
A singularity that can be brought to the local form
\begin{equation} \label{cusps and nodes}
z^2 = w^m, \quad m \geq 4
\end{equation}
will be called a \dub{higher cusp} if $m$ is odd and a \dub{higher
node} if $m$ is even.
\end{definition}

Given a singular algebraic curve $C$, there are two standard ways
of \dub{resolving} singularities so as to obtain a Riemann surface
which we now turn to. Afterwards we will describe the reverse
procedures whereby one obtains singular curves from non-singular
ones.

\subsection*{Normalisation}

The first procedure for resolving singularities, known as
\dub{normalisation} (or \dub{desingularisation}) consists of `blowing
up' each singular point into a finite set $S$ of points. The singular
curve in this case is recovered by identifying each set of points $S$
to single points. In the case of the node \eqref{node} the singular
point $(X,Y) = (0,0)$ is doubled
\begin{equation} \label{node normalised}
\raisebox{-14mm}{
\includegraphics[height=30mm]{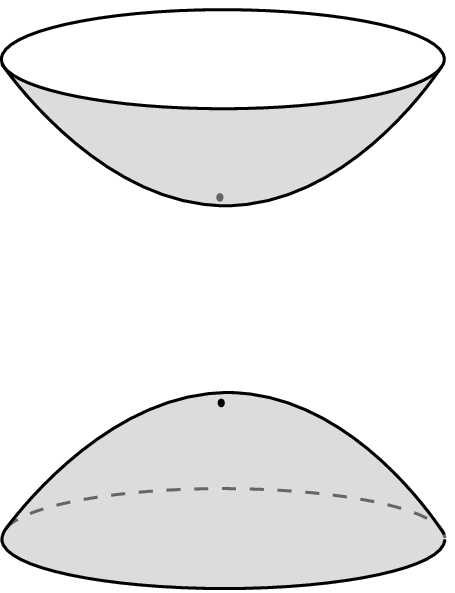}} \qquad \qquad
X = 0, \quad Y = 0.
\end{equation}
This results in two copies of the complex line which is obviously
a Riemann surface. The line $X = 0$ admits $Y$ as a local
parameter whereas $X$ is a local parameter on the line $Y = 0$.
Concretely, normalisation can be achieved using a birational
transformation\footnote{Another way to define the normalisation of
a singular curve $C$ is as the space of germs on $C$.} as follows.
Considering the node in the form $z^2 = w^2$, we perform a
birational transformation $(z,w) \mapsto (u,v)$ defined by $z = u
v$ and $w = v$. This transformation has the desired feature that
it is invertible except at the singular point itself $(z,w) =
(0,0)$. Indeed it transforms the node to $u^2 = 1$ and so the
singular point has been `blown up' to a pair of points $(u,v) =
(\pm 1, 0)$.

The normalisation of a singular point does not always result in
the addition of points. Consider for example the cusp singularity
$z^2 = w^3$. It may be desingularised by the same birational
transformation as we used for the node, resulting in the
non-singular curve $u^2 = v$. This time the singular point $(z,w)
= (0,0)$ gets mapped to the single point $(u,v) = (0,0)$ which is
a branch point of the map $(u,v) \mapsto v$. We conclude therefore
that a cusp resolves into a branch point.

More generally, a higher node $z^2 = w^{2r}$ may be desingularised
by using the birational transformation $z = u v^r$ and $w = v$
which transforms it to $u^2 = 1$. Thus as in the case of a node,
the singular point has been `blown up' to a pair of points $(u,v)
= (\pm 1, 0)$. The case of a higher cusp $z^2 = w^{2r + 1}$ can
also be desingularised by the same birational transformation
yielding the non-singular curve $u^2 = v$. So just as for the
cusp, the singular point doesn't get blown up but instead resolves
into a single branch point. Since the birational transformations used
to resolve singularities are always invertible away from the
singular points in question we may resolve each of the finitely many
singular points of an algebraic curve $C$ by proceeding one at a
time. This finite procedure results in a Riemann surface $\hat{C}$
known as the \dub{normalisation} of $C$. Moreover, there is a
continuous surjection
\begin{equation*}
\pi : \hat{C} \rightarrow C,
\end{equation*}
which restricts to a biholomorphic map $\pi : \hat{C} \setminus
\pi^{-1}(S) \rightarrow C \setminus S$, where $S$ is the finite set of
singular points of $C$. In the present case $\pi^{-1}(S)$ is also
finite and consists of at most twice as many points as $S$.

\subsection*{Smoothing}

The other procedure for resolving singularities, known as
\dub{smoothing} (or \dub{deformation}) consists of `perturbing' the
algebraic curve $C$ by a small parameter $t$. The original singular
curve is recovered in the limit $t \rightarrow 0$. An example of a
smoothing of the node
\eqref{node} is
\begin{equation} \label{node smoothed}
\raisebox{-14mm}{
\includegraphics[height=30mm]{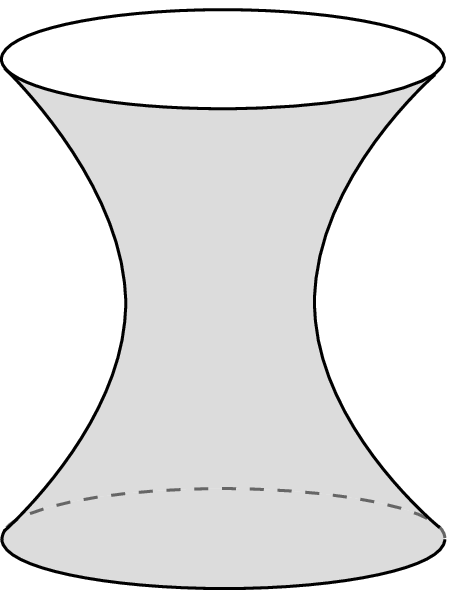}}
\qquad \qquad X Y = t, \quad t \in \mathbb{C}.
\end{equation}
For $t \neq 0$ this curve is no longer singular and either $X$ or
$Y$ maybe be used as local charts with the transition function $X
= t/Y$ being holomorphic. The smoothed out node \eqref{node
smoothed} is therefore a Riemann surface. To describe this surface
locally it is sufficient to restrict the coordinates to within the
unit disc $D = \{ t \in \mathbb{C} | \; |t| < 1 \}$. That is, given $t
\in D$ we define
\begin{equation} \label{node smoothed local}
N_t = \{ (X,Y) \in D^2 \; | \; XY = t \}.
\end{equation}
Because $|Y| < 1$ it follows that $|X| > |t|$, and likewise $|Y| >
|t|$ as a consequence of $|X| < 1$. Thus topologically the
deformed node $N_t$ is the annulus (or cylinder) given by $|t| <
|X| < 1$. To make use of both coordinates, it can also be modelled
topologically as the two annuli $\sqrt{|t|} < |X| < 1$ and
$\sqrt{|t|} < |Y| < 1$ glued together through the interior circle.
In this picture the limit $t \rightarrow 0$ is more apparent and
clearly gives two discs glued together at the origin, as in
\eqref{node}.

Defining the \dub{topological genus} of a singular algebraic curve as
the topological genus of its normalisation, smoothing a curve
will increase its topological genus. For singular algebraic curves
one can introduce an alternative genus to the topological genus,
called the \dub{algebraic genus}, defined as the genus of the
smoothed out curve. It follows that the algebraic genus is
invariant under the smoothing operation.

\subsection*{Indentifying points}

The reverse process to normalisation consists in \dub{identifying}
certain points of a smooth algebraic curve $C$. Following \cite{Serre}
we define a \dub{modulus}\footnote{The term `modulus' makes sense with
  regards to corollary \ref{corollary: moduli punctured RS} since
  marking a point on a Riemann surface generically increases the
  dimension of the moduli space by one.} to be an effective divisor
$\mathfrak{m} = \sum_{P \in C} n_P P, n_P \geq 0$. We refer to the
finite set of points $P \in C$ for which $n_P > 0$ as the
\dub{support} $S$ of $\mathfrak{m}$. Then in the simplest case, a
singular curve is obtained by collapsing the entire set $S$ to a
single point $Q$ (more generally $S$ collapses to a smaller set
$S'$). That is we define a singular curve as the set
$C_{\mathfrak{m}} = (C \setminus S) \cup \{ Q \}$. Notice that at
the level of the curve no use was made of the multiplicities $n_P$
of each point $P$ in the modulus. These multiplicities enter in
the definition of the allowed functions on the singular curve
$C_{\mathfrak{m}}$. For instance \cite[pp.61--62]{Serre}, if
$\mathfrak{m}$ consists of two distinct points, namely
$\mathfrak{m} = P_1 + P_2$ with $P_1 \neq P_2$ then it turns out
that the resulting singular point $Q \in C_{\mathfrak{m}}$ is a
node. In this case, a function on $C_{\mathfrak{m}}$ regular at
$Q$ should arise from a function $f$ on $C$ which is regular at
$P_1$ and $P_2$ but since these points are identified on
$C_{\mathfrak{m}}$ we must also request that $f(P_1) = f(P_2)$ for
$f$ to be single-valued on $C_{\mathfrak{m}}$. As another example,
if $\mathfrak{m} = 2 P$ then the curve $C_{\mathfrak{m}}$ is
identical to $C$ since $S = \{ P \}$ is a single point, however
functions on $C_{\mathfrak{m}}$ are taken to be functions on $C$
with a vanishing first derivative at $P$. The singular point $Q
\in C_{\mathfrak{m}}$ in this case turns out to be a cusp. In each
case the original curve $C$ is the normalisation of the resulting
singular curve $C_{\mathfrak{m}}$.

Recall that any two divisors $D,D' \in \Div(C)$ are said to be
equivalent $D \sim D'$ if there exists a meromorphic function $f$
on $C$ with divisor $(f) = D - D'$. On singular curves defined by
a modulus $\mathfrak{m}$ as above we can also define an
equivalence relation between divisors by defining a more stringent
equivalence on $\Div(C)$. First of all we say that a divisor $D
\in \Div(C)$ is \dub{prime} to $S$ if it has no points in common
with $S$. Two such divisors $D,D'$ are then said to be
\dub{$\bm{\mathfrak{m}}$-equivalent}, written $D
\sim_{\mathfrak{m}} D'$, if there exists a function $f$ on $C$
such that
\begin{equation} \label{m-equivalence}
(f) = D - D', \qquad \forall P \in S, \; \ord_P (f - 1) \geq n_P.
\end{equation}
The new second condition says that $f$ must take the value one at any
$P \in S$ with multiplicity $n_P$. In particular $f$ takes the same
value at all the points of $S$ which is required for $f$ to define a
single-valued function on $C_{\mathfrak{m}}$. This new equivalence
relation on $\Div(C \setminus S)$ allows us to define the
\dub{generalised divisor class group} relative to $\mathfrak{m}$,
denoted $\Pic_{\mathfrak{m}}(C) \equiv \Div(C \setminus S) /
\sim_{\mathfrak{m}}$, of divisors prime to $S$ modulo
$\mathfrak{m}$-equivalence. The main example we will need is that of a
nodal curve (with a single node) for which $\mathfrak{m} = P_1 + P_2$
with $P_1 \neq P_2$. In this case \eqref{m-equivalence} reads
\begin{equation} \label{m-equivalence 2}
(f) = D - D', \qquad f(P_1) = f(P_2) = 1.
\end{equation}

\subsection*{Degeneration}

Recall that the smoothing procedure resulted in a 1-parameter family
of Riemann surfaces $C_t$ for $t \neq 0$, with the original singular
curve $C_0$ sitting at the limiting point $t = 0$. The reverse process
of smoothing thus consists in \dub{pinching} the family $C_t$ of
Riemann surfaces by taking the limit $t \rightarrow 0$ to recover the
singular curve $C_0$. One therefore has to construct a family $C_t$ of
Riemann surfaces fibred over the unit disc $D = \{ t \in \mathbb{C} \;
| \; |t| < 1 \}$ which is locally modelled on the smoothed node
\eqref{node smoothed local}. There are two different ways of obtaining
a family $C_t$ of Riemann surface with a local neighbourhood modelled
on the smoothed node $N_t$ (see \cite[chapter III]{Fay}):
\begin{itemize}
  \item One can either take two distinct Riemann surfaces $M_1$ and
$M_2$ punctured at $P_1$ and $P_2$ respectively with local
coordinates $z_1$ and $z_2$ near these punctures and define $C_t =
M_1 \sqcup N_t \sqcup M_2 / \mathcal{R}$. The quotient serves to
specify the overlaps between the three surfaces $M_1$, $N_t$ and
$M_2$ in the disjoint union. Specifically the relation $\mathcal{R}$
is defined as follows. A point near $P_1$ with local coordinate $z_1$
on $M_1$ is to be identified with the point of local coordinate $X =
z_1$ on $N_t$. Similarly points of $M_2$ with local coordinate $z_2$
are identified with points of $N_t$ with local coordinate $Y =
z_2$. Thus in the overlap we have by construction $z_2 = t/z_1$. The
family $C_t$ then describes the \dub{pinching of a cycle homologous to
zero}.
\begin{figure}[ht]
\centering
\includegraphics[height=15mm]{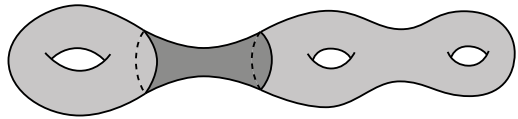}
\caption{Pinching a cycle homologous to zero.} \label{Figure:
pinching zero}
\end{figure}
  \item One can also take the two punctures $P_1$ and $P_2$ to be on
the \textit{same} Riemann surface $M$. In this case we define $C_t = M
\sqcup N_t / \mathcal{R}'$. Once again the quotient specifies the
overlap between the component surfaces $M$ and $N_t$ of the disjoint
union. Here the relation $\mathcal{R}'$ is defined as follows. A point
near $P_1$ with local coordinate $z_1$ on $M$ is to be identified with
the point with local coordinate $X = z_1$ on $N_t$. Similarly points
near $P_2$ with coordinate $z_2$ on $M$ are identified with points on
$N_t$ with coordinate $Y = z_2$. Once more in the overlap we have $z_2
= t/z_1$. Here the family $C_t$ describes the \dub{pinching of a
non-zero homology cycle}.
\begin{figure}[ht]
\centering
\includegraphics[height=30mm]{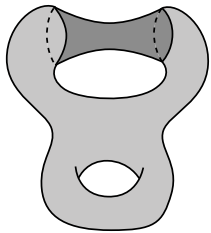}
\caption{Pinching a non-zero homology cycle.} \label{Figure:
pinching a}
\end{figure}
\end{itemize}
We will be mostly concerned with the second possibility of
pinching cycles homologous either to $\bm{a}$- or $\bm{b}$-cycles
on $M$.


\section{Jacobians} \label{section: jacobian}


Consider the dual space $\mathcal{H}^1(M)^{\ast}$ of linear
functionals $\mathcal{H}^1(M) \rightarrow \mathbb{C}$ on the space of
holomorphic forms. By the integration mapping, every closed cycle
$\gamma \in H_1(M,\mathbb{Z})$ defines an element of
$\mathcal{H}^1(M)^{\ast}$ through its periods, namely
\begin{equation*}
H_1(M,\mathbb{Z}) \rightarrow \mathcal{H}^1(M)^{\ast}, \qquad \gamma
\mapsto \left( \omega \mapsto \int_{\gamma} \omega \right).
\end{equation*}
An object of fundamental importance in the study of Riemann surfaces
is the quotient of $\mathcal{H}^1(M)^{\ast}$ by the subgroup of
periods $H_1(M,\mathbb{Z})$.

\begin{definition}
The \dub{Jacobian} of $M$ is the quotient $J(M) =
\mathcal{H}^1(M)^{\ast}/H_1(M, \mathbb{Z})$.
\end{definition}

We can describe the Jacobian more explicitly using bases. So let $\{
a_i, b_j \}$ be a canonical basis of $H_1(M)$ and $\omega_i$ the dual
`normalised' basis \eqref{normalised holo basis} for the space of
holomorphic differentials $\mathcal{H}^1(M)$. Since $\mathcal{H}^1(M)$
is a complex vector space of dimension $g$, its dual can be identified
using the basis $\{ \omega_j \}$ with the space $\mathbb{C}^g$ of
complex column vectors. As for the period subgroup, the
$\bm{a}$-periods of the basis differentials $\omega_j$ being
normalised by the condition $\int_{a_i} \omega_j = \delta_{ij}$ they
define $g$ independent vectors in $\mathbb{C}^g$. The $g$ remaining
$\bm{b}$-periods of the $g$ basis holomorphic differentials are
non-trivial and define an important $g \times g$ matrix.
\begin{definition}
The \dub{period matrix} $\Pi$ is $g \times g$ with components
$\Pi_{ij} = \int_{b_i} \omega_j$.
\end{definition}

The period matrix has the following important properties,

\begin{lemma} \label{lemma: imaginary Pi}
$\Pi$ is symmetric and has positive definite imaginary part.
\begin{proof}
To show symmetry, apply the Riemann bilinear identities
\eqref{Riemann bilinear holo} to the normalised holomorphic
differentials $\omega_1 = \omega_i$, $\omega_2 = \omega_j$. To show
positive definiteness of $\textup{Im} \, \Pi$, namely $\sum_{i,j} c_i
(\textup{Im} \, \Pi_{ij}) c_j > 0$, apply lemma \ref{lemma: Rbilinear
2} to $\omega = \sum_{i = 1}^g c_j \omega_j$, $c_j \in \mathbb{R}$.
\end{proof}
\end{lemma}

In particular, since $(\textup{Im} \, \Pi)$ is positive definite it is
invertible so that,

\begin{corollary} \label{cor: lattice of periods}
The $2g$ columns of the full $g \times 2g$ matrix of periods $({\bf 1}
, \Pi)$ are linearly independent over $\mathbb{R}$.
\end{corollary}

Hence the Jacobian is a \dub{complex $g$-dimensional torus},
namely it is the quotient of $\mathbb{C}^g$, viewed as a real
vector space, by a real $2g$-dimensional lattice\footnote{The
factors of $2 \pi$ are conventions we adopt to simplify some of
the notation later.}
\begin{equation} \label{Jacobian explicit}
J(M) = \mathbb{C}^g / \Lambda, \qquad \Lambda \equiv 2 \pi
\mathbb{Z}^g \oplus 2 \pi \Pi \mathbb{Z}^g.
\end{equation}
Note that the Jacobian has an obvious Abelian group structure.
Thus every Riemann surface $M$ has associated with it a natural
Abelian group $J(M)$. Recall that we have already assigned an
Abelian group to every Riemann surface $M$, namely the divisor
class group $\Pic(M)$, also called the \dub{Picard group}. The
Abel-Jacobi theorem states that the group $\Pic^0(M)$ of degree
zero divisors modulo principal divisors and the Jacobian $J(M)$
are isomorphic. The isomorphism is constructed using the Abel map
which we now turn to.

\subsection*{The Abel map}

\begin{definition}
The \dub{Abel map} $\bm{\mathcal{A}} : M \rightarrow J(M)$ is defined
relative to some base point $P_0 \in M$ by
\begin{equation} \label{Abel map}
P \mapsto \bm{\mathcal{A}}(P) = 2 \pi \int_{P_0}^P \bm{\omega} \; \mod
\Lambda,
\end{equation}
where $\bm{\omega} = (\omega_1, \ldots, \omega_g)^{\sf T}$ is the
vector of basis holomorphic forms.
\end{definition}
\begin{remark}
The integrals $\int_{P_0}^P \bm{\omega}$ themselves are not well
defined as they depend on the path $\gamma$ joining the base point
$P_0$ to $P$. But if $\gamma'$ is another such path then $\gamma -
\gamma'$ is closed so that the difference $2 \pi \int_{\gamma}
\bm{\omega} - 2 \pi \int_{\gamma'} \bm{\omega} = 2 \pi \int_{\gamma -
\gamma'} \bm{\omega} \in \Lambda$. For this reason equalities
involving the Abel map should always be understood to be $\text{mod}
\Lambda$ unless otherwise stated.
\end{remark}

\begin{remark}
The Abel map doesn't depend on the choice of basis holomorphic forms
since it can be written in a coordinate independent way as $\mathcal{A} :
P \mapsto \left( \omega \mapsto 2 \pi \int_{P_0}^P \omega \right)$.
\end{remark}

The Abel map can be extended to the group of divisors $\Div(M)$
by setting
\begin{equation*}
\bm{\mathcal{A}}\left( \sum_{P \in M} m_P P \right) = \sum_{P \in M}
m_P \bm{\mathcal{A}}(P),
\end{equation*}
which defines a group homomorphism $\bm{\mathcal{A}} : \Div(M)
\rightarrow J(M)$. In particular, when acting on divisors of
degree zero the Abel map $\bm{\mathcal{A}} : \Div^0(M) \rightarrow
J(M)$ is easily see not to depend on the base point $P_0$. Indeed,
for $D = \sum_{\alpha = 1}^n (P_{\alpha} - Q_{\alpha})$ we have
\begin{equation*}
\bm{\mathcal{A}}\left( \sum_{\alpha = 1}^n (P_{\alpha} - Q_{\alpha})
\right) = \sum_{\alpha = 1}^n 2 \pi \int_{P_0}^{P_{\alpha}} \bm{\omega} -
2 \pi \int_{P_0}^{Q_{\alpha}}\bm{\omega} = \sum_{\alpha = 1}^n
2 \pi \int_{Q_{\alpha}}^{P_{\alpha}} \bm{\omega}.
\end{equation*}
It is a consequence of Abel's theorem below that the Abel map on
$\Div^0(M)$ descends to a homomorphism
\begin{equation} \label{Pic iso J}
\bm{\mathcal{A}} : \Pic^0(M) \longrightarrow J(M)
\end{equation}
between the groups $\Pic^0(M)$ and $J(M)$. Moreover, this
homomorphism is also injective as a consequence of Abel's theorem
and surjective by Jacobi's theorem. Thus the Abel map \eqref{Pic
iso J} provides an isomorphism between the degree zero Picard
group $\Pic^0(M)$ on the one hand and the Jacobian $J(M)$ on the
other.
\begin{theorem}[Abel] \label{theorem: Abel}
A divisor $D \in \Div(M)$ is principal if and only if $\deg D = 0$
and $\bm{\mathcal{A}}(D) = 0$.
\begin{proof}
The condition $\deg D = 0$ is obvious from \eqref{deg principal
divisor}. Let $D = \sum_{i = 1}^n (P_i - Q_i)$ and consider the
function $f(P) = \exp \sum_{i = 1}^n \int_Q^P \omega_{P_i Q_i}$ in
\eqref{function for Abel thm} which by lemma \ref{lemma: f(P)/f(Q)=r}
has the right divisor $(f) = \sum_{i = 1}^n (P_i - Q_i)$. However this
divisor is principal if and only if $f$ is single-valued on $M$. Since
$\omega_{P_i Q_i}$ is normalised with unit residues at its poles, this
is the case if and only if $\int_{b_i} \sum_{i = 1}^n \omega_{P_i
Q_i} \in 2 \pi i \mathbb{Z}$. And by the Riemann bilinear identity
\eqref{Riemann bilinear 4} this is equivalent to $\sum_{i = 1}^n
\int_{Q_i}^{P_i} \omega_j \in \mathbb{Z}$.
\end{proof}
\end{theorem}
\begin{theorem}[Jacobi] \label{theorem: Jacobi}
Every point in $J(M)$ is the image of an integral divisor of
degree $g$.
\end{theorem}

\subsection*{Generalised Jacobians}

Consider the singular algebraic curve $C_{\mathfrak{m}}$ described by
a modulus $\mathfrak{m} = P_1 + P_2$, $P_1 \neq P_2$ on its
normalisation $C$. If the above construction of Jacobians for Riemann
surfaces is to carry over to singular algebraic curves then the Abel
map should be generalised. Indeed we would still like the Abel map to
characterise divisors up to equivalence on $C_{\mathfrak{m}}$. But we
saw that divisors on $C_{\mathfrak{m}}$ can be described as divisors
on $C \setminus S$ (where $S$ was the support of $\mathfrak{m}$)
subject to the stronger $\mathfrak{m}$-equivalence.

\begin{figure}[ht]
\centering
\begin{tabular}{ccccc}
\includegraphics[width=0.25\textwidth]{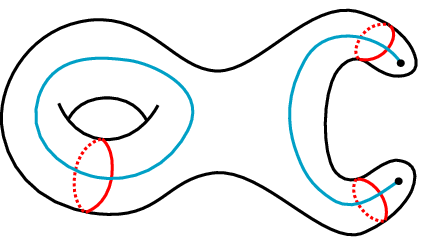} & $\quad$ &
\includegraphics[width=0.25\textwidth]{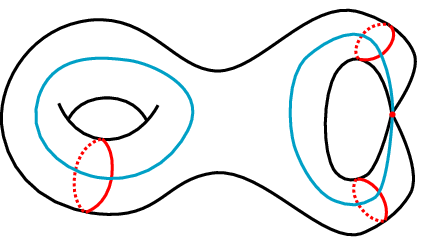} & $\quad$ &
\includegraphics[width=0.25\textwidth]{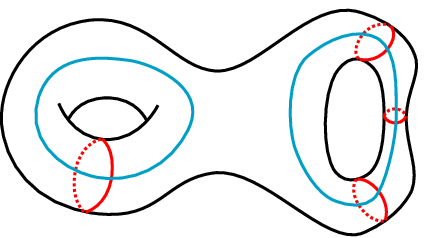}\\
$(a)$ Normalisation $C$ & & $(b)$ Singular $C_{\mathfrak{m}}$ & &
$(c)$ Smoothing $C_t$
\end{tabular}
\caption{Normalisation and smoothing of the singular curve
$C_{\mathfrak{m}}$.} \label{Figure: extra canonical cycles}
\end{figure}

As we have seen, the nodal curve $C_{\mathfrak{m}}$ can be
resolved into two different Riemann surfaces: it can be
desingularised to produce its normalisation $C$ or it can be
smoothed out to form a one-parameter family $C_t$. In the first
case the singular curve $C_{\mathfrak{m}}$ is recovered by
identifying $P_1$ with $P_2$ and in the second case by taking $t
\rightarrow 0$ to pinch off the extra handle. Both resolved curves
being Riemann surfaces the above analysis applies to these, see
Figure \ref{Figure: extra canonical cycles}.

In particular, each member of the family $C_t$, $t \neq 0$ can be
assigned a Jacobian $J(C_t)$. We shall define the \dub{generalised
Jacobian} $J_{\mathfrak{m}}(C)$ associated with the singular curve
$C_{\mathfrak{m}}$ as the limit of $J(C_t)$ as we take $t
\rightarrow 0$. We now aim to give a more explicit description of
$J_{\mathfrak{m}}(C)$ as a quotient much like equation
\eqref{Jacobian explicit} for the usual Jacobian. Recall that the
construction of $C_t$ using two punctures on the same Riemann
surface, as in Figure \ref{Figure: pinching a}, lead to a Riemann
surface with genus one higher since the smoothed out node gives it
one extra handle. Let us define the canonical homology basis $\{
a_I(t) \}_{I = 0}^g$ of $C_t$ so that the extra $a_0(t)$-cycle
goes around the smoothed out node with the extra $b_0(t)$-cycle
intersecting $a_0(t)$ once, as illustrated in Figure \ref{Figure:
extra canonical cycles} in the elliptic case $g = 1$.
\begin{figure}[ht]
\centering
\begin{tabular}{cccc}
$\;$ & \psfrag{a1}{\green $a_0(t)$} \psfrag{a2}{\red
$\tilde{a}_0(t)$} \psfrag{a3}{\green ${a'}_0(t)$}
\psfrag{M}{$C_t$}
\includegraphics[height=30mm]{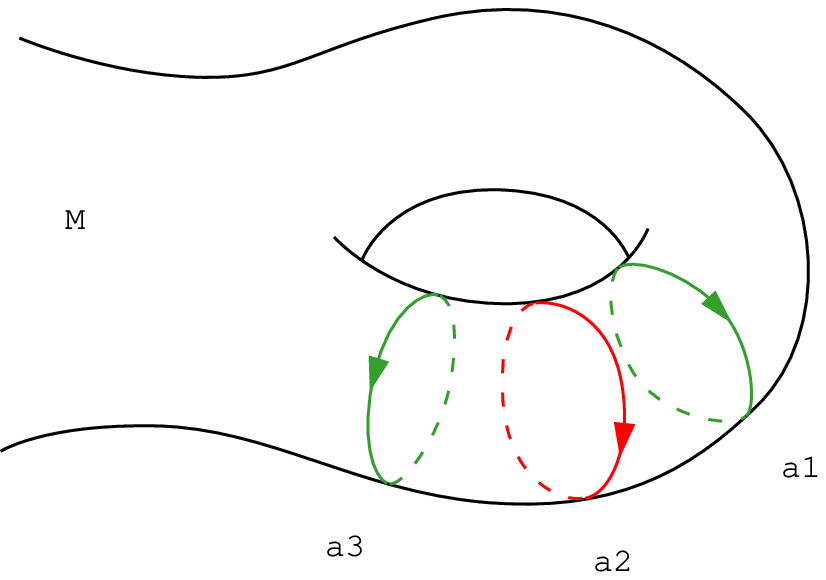} &
\raisebox{16mm}{$\qquad \longrightarrow \qquad$} &
\psfrag{a1}{\green $a_0$} \psfrag{a2}{\green ${a'}_0$}
\psfrag{ap}{\footnotesize \red $P_1$} \psfrag{am}{\footnotesize
\red $P_2$} \psfrag{M}{$C$}
\raisebox{7mm}{\includegraphics[height=20mm]{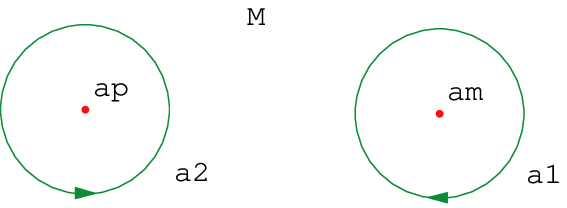}}
\end{tabular}
\caption{Pinching the cycle $a_0(t)$ on $C_t$.} \label{Figure:
pinching a cycles}
\end{figure}
The singular limit $t \rightarrow 0$
corresponds then to pinching a particular $a_0$-cycle
$\tilde{a}_0(t)$ to a point $Q$. We shall call $a_0(t)$ and $a'_0(t)$
the two cycles on either side of the shrinking cycle
$\tilde{a}_0(t)$, as depicted in Figure \ref{Figure: pinching a
cycles}.
Let $\{ \omega_I(t) \}_{I = 0}^g$ be the dual basis of holomorphic
1-forms canonically normalised as usual by the condition
\begin{equation} \label{normalised holo diff}
\int_{a_I(t)} \omega_J(t) = \delta_{IJ}, \quad I,J = 0, \ldots, g.
\end{equation}
It is clear from these relations that in the limit $t \rightarrow
0$ the 1-form $\omega_0(t)$ acquires poles at the points $P_1$ and
$P_2$ on $C$ corresponding to the desingularisation of $Q$ with the
following residues (For quantities taken at $t = 0$ we drop the
argument for clarity and write for instance $a_I \equiv a_I(0)$, $b_I
\equiv b_I(0)$, $\omega_I \equiv \omega_I(0)$, \textit{etc})
\begin{equation*}
\res_{P_1} \omega_0 = \frac{1}{2 \pi i} \int_{a_0} \omega_0 =
\frac{1}{2 \pi i}, \qquad \res_{P_2} \omega_0 = - \frac{1}{2 \pi i}
\int_{a'_0} \omega_0 = - \frac{1}{2 \pi i}.
\end{equation*}
Since $\omega_0$ has no further poles and $\int_{a_i}
\omega_0 = 0$ for $i = 1, \ldots g$ it uniquely determines the
normalised Abelian differential of the third kind $\omega_0 =
\frac{1}{2 \pi i} \omega_{P_1 P_2}$ on $C$. Moreover, the remaining
differentials $\{ \omega_i \}_{i = 1}^g$ form a basis of
holomorphic 1-forms on $C$ dual to the homology basis $\{ a_i,
b_i \}_{i = 1}^g$ for $C$ by \eqref{normalised holo diff}.

To identify the Jacobian $J(C_t)$ in the singular limit consider its
period matrix $\Pi_{IJ}(t) = \int_{b_I(t)} \omega_J(t)$. Since the
curve $b_0$ starts at $P_1$ and ends at $P_2$, the component
$\Pi_{00}(t) = \int_{b_0(t)} \omega_0(t)$ will diverge in the limit $t
\rightarrow 0$. All the other components $\Pi_{ij}(t)$ and
$\Pi_{0j}(t)$ of the period matrix stay finite in this limit. It is
clear now that the first column $\Pi_{I0}(t)$ tends to (an infinite
multiple of) the unit vector $(1, 0, \ldots, 0)^{\sf T}$. The lattice
of periods $({\bf 1}, \Pi(t) )$ from corollary \ref{cor: lattice of
periods} thus becomes degenerate in the singular limit $t \rightarrow
0$ and is only spanned by $2 g + 1$ linearly independent vectors
\begin{equation*}
\left( \begin{array}{cc} 1 & 0 \\ 0 & \delta_{ij} \end{array}, \;
\begin{array}{c} \Pi_{0j} \\ \Pi_{ij} \end{array} \right).
\end{equation*}
Denoting by $\Lambda_{\mathfrak{m}}$ the lattice spanned by $2 \pi$
multiples of these vectors, the generalised Jacobian can therefore be
defined as the quotient
\begin{equation} \label{generalised jacobian}
J_{\mathfrak{m}}(C) \equiv \mathbb{C}^{g+1}/ \Lambda_{\mathfrak{m}}.
\end{equation}
Because the lattice $\Lambda_{\mathfrak{m}}$ is only spanned by $2g +
1$ vectors linearly independent over $\mathbb{R}$ and $\mathbb{C}^g$
has dimension $2 g + 2$ over $\mathbb{R}$, the lattice
$\Lambda_{\mathfrak{m}}$ is in some sense too small and as a result
the quotient \eqref{generalised jacobian} is a non-compact algebraic
group. Topologically it is simply $J_{\mathfrak{m}}(C) \simeq J(C)
\times \mathbb{C}^{\ast}$ with the $\mathbb{C}^{\ast}$ factor being
the origin of non-compactness.

\begin{definition} \label{def: generalised Abel map}
The \dub{generalised Abel map} $\vec{\mathcal{A}} : C \setminus S
\rightarrow J_{\mathfrak{m}}(C)$ is defined relative to some base
point $P_0 \in C$ by
\begin{equation} \label{Abel map}
P \mapsto \vec{\mathcal{A}}(P) = 2 \pi \int_{P_0}^P \vec{\omega} \;
\mod \Lambda_{\mathfrak{m}},
\end{equation}
where $\vec{\omega} = (\omega_0, \omega_1, \ldots, \omega_g)^{\sf T}$
is the vector of basis holomorphic forms together with the Abelian
differential of the third kind $\omega_0$.
\end{definition}
\begin{remark}
As for the usual Abel map, the integrals $\int_{P_0}^P
\vec{\omega}$ are not well defined since they depend on the path
$\gamma$ joining the base point $P_0$ to $P$. But if $\gamma'$ is
another such path then it is straightforward to see that $2 \pi
\int_{\gamma} \vec{\omega} - 2 \pi \int_{\gamma'} \vec{\omega} \in
\Lambda_{\mathfrak{m}}$.
\end{remark}

\begin{remark}
Note that the points in the support $S$ of the modulus $\mathfrak{m}$
are avoided in the definition of the generalised Abel map since
$\omega_0$ has poles there and so $\vec{\mathcal{A}}(P)$ would diverge
there.
\end{remark}

The generalised Abel map can also be extended to the group $\Div(C
\setminus S)$ of divisors prime to $S$ in the obvious way such that
$\vec{\mathcal{A}} : \Div(C \setminus S) \rightarrow
J_{\mathfrak{m}}(C)$ is a group homomorphism. When acting on divisors
of degree zero the Abel map $\vec{\mathcal{A}} : \Div^0(C \setminus S)
\rightarrow J_{\mathfrak{m}}(C)$ it does not depend on
$P_0$. Moreover, by theorem \ref{theorem: generalised Abel} and
theorem \ref{theorem: generalised Jacobi} below which are
generalisations of Abel and Jacobi's theorems, this map on divisors of
$C$ prime to $S$ descends to an isomorphism
\begin{equation} \label{Pic_m iso J_m}
\vec{\mathcal{A}} : \Pic^0_{\mathfrak{m}}(C) \longrightarrow
J_{\mathfrak{m}}(C)
\end{equation}
between the \dub{generalised Picard group} $\Pic^0_{\mathfrak{m}}(C) =
\Div(C \setminus S) / \sim_{\mathfrak{m}}$ of degree zero divisors
prime to $S$ modulo $\mathfrak{m}$-equivalence and the generalised
Jacobian $J_{\mathfrak{m}}(C)$.

\begin{theorem}[generalised Abel] \label{theorem: generalised Abel}
A divisor $D \in \Div(C \setminus S)$ is of the form $D = (f)$ for
some meromorphic function $f$ with $f(P_1) = f(P_2)$ if and only if
$\deg D = 0$ and $\vec{\mathcal{A}}(D) = 0$.
\begin{proof}
By Abel's theorem we have $D = (f)$ for some meromorphic function $f$
if and only if $\deg D = 0$ and $\bm{\mathcal{A}}(D) =
0$. Furthermore, it is immediate from lemma \ref{lemma: f(P)/f(Q)=r}
that $f(P_1) = f(P_2)$ if and only if $\mathcal{A}_0(D) = 0$.
\end{proof}
\end{theorem}

We also have the following generalisation of Jacobi's theorem
\cite{Fedorov}.

\begin{theorem}[generalised Jacobi] \label{theorem: generalised Jacobi}
Every point in $J_{\mathfrak{m}}(C)$ is the image of an integral
divisor of degree $g+1$.
\end{theorem}

\subsection*{$\theta$-functions}

\begin{definition}
The \dub{Riemann $\theta$-function} $\theta : \mathbb{C}^g \rightarrow
\mathbb{C}$ is given by
\begin{equation} \label{theta function def}
\bm{z} \mapsto \theta(\bm{z}; \Pi) = \sum_{\bm{m} \in
\mathbb{Z}^g} \exp \left\{ i \langle \bm{m}, \bm{z} \rangle + \pi i
\langle \Pi \bm{m}, \bm{m} \rangle \right\}.
\end{equation}
where $\langle \bm{x}, \bm{y} \rangle = \sum_{i = 1}^g x_i y_i$. When
it is clear which period matrix we are using we shall omit it form the
arguments and simply write $\theta(\bm{z}) = \theta(\bm{z}; \Pi)$.
\end{definition}

It can be shown \cite[pp.299--300]{Farkas} that the sum converges absolutely and
uniformly on any compact subset of $\mathbb{C}^g$ and thus the Riemann
$\theta$-function is homolorphic on the whole of
$\mathbb{C}^g$. Furthermore, it is obviously even $\theta(-\bm{z}) =
\theta(\bm{z})$ and has the following important \dub{automorphy
property} under translation by lattice vectors $2 \pi \bm{n} + 2 \pi
\Pi \bm{m} \in \Lambda$,
\begin{equation} \label{automorphy}
\theta(\bm{z} + 2 \pi \bm{n} + 2 \pi \Pi \bm{m}) = \exp \left\{ -i
\langle \bm{m}, \bm{z} \rangle - \pi i \langle \Pi \bm{m}, \bm{m}
\rangle \right\} \theta(\bm{z}).
\end{equation}
Note that although the Riemann $\theta$-function is defined on
$\mathbb{C}^g$, by the automorphy property its zeroes naturally live
on the Jacobian $J(M)$.

Combining the Riemann $\theta$-function with the Abel map
$\bm{\mathcal{A}}$ we can define an interesting multi-valued
function on $M$. Let $\bm{w} \in \mathbb{C}^g$ be an arbitrary
vector and consider the function $P \mapsto \theta \left(
\bm{\mathcal{A}}(P) - \bm{w} \right)$. Its zeroes are well defined
on $M$ and are characterised by the fundamental theorem of
Riemann,

\begin{theorem}[Riemann] \label{theorem: Riemann}
If $P \mapsto \theta(\bm{\mathcal{A}}(P) - \bm{w})$ does not vanish
identically then it has exactly $g$ zeroes $P_1, \ldots, P_g \in M$
satisfying
\begin{equation} \label{g zeroes of Riemann}
\bm{\mathcal{A}}(P_1) + \ldots + \bm{\mathcal{A}}(P_g) = \bm{w} -
\bm{\mathcal{K}},
\end{equation}
where $\bm{\mathcal{K}}$ is the \dub{vector of Riemann's constants}
which depends only on $M$ and the base point $P_0$ of the
Abel map, given explicitly in components by
\begin{equation} \label{vector of Riemann's constants}
\mathcal{K}_k = 2 \pi \left[ \frac{1 + \Pi_{kk}}{2} - \sum_{j = 1, j
\neq k}^g \int_{a_j} \left( \int_{P_0}^P \omega_k \right) \omega_j
\right].
\end{equation}
\end{theorem}

Now let $D > 0$ be an integral divisor of degree $\deg D = g$ and
in view of equation \eqref{g zeroes of Riemann} introduce the
notation $\bm{\zeta}_D \equiv \bm{\mathcal{A}}(D) +
\bm{\mathcal{K}}$. An important function that constitutes the
building block for constructing functions on $M$ with specified
poles and zeroes is the following multi-valued function
\begin{equation*}
\psi_D : P \mapsto \theta \left( \bm{\mathcal{A}}(P) - \bm{\zeta}_D
\right).
\end{equation*}

The following theorem \cite[p.313]{Farkas} asserts that a necessary
and sufficient condition for $\psi_D$ to vanish identically is that the
divisor $D$ be special.
\begin{theorem} \label{special equiv theta vanish}
$\psi_D \not \equiv 0$ if and only if $i(D) = 0$.
\end{theorem}

Since the Riemann $\theta$-function is holomorphic the function
$\psi_D$ has no poles, and by the automorphy property its zeroes
are well defined on $M$. Therefore although $\psi_D$ is
multi-valued its divisor $(\psi_D)$ is well defined on $M$ and we
have

\begin{corollary} \label{cor: theta zeroes}
If $D$ is non-special then $(\psi_D) = D$.
\begin{proof}
Since $D$ is non-special we have $i(D) = 0$ so that $\psi_D \not
\equiv 0$ by theorem \ref{special equiv theta vanish}. But then
Riemann's theorem tells us that $\psi_D$ has exactly $g$ zeros $P_1,
\ldots, P_g$ subject to the condition $\bm{\mathcal{A}}(P_1) + \ldots
+ \bm{\mathcal{A}}(P_g) = \bm{\zeta}_D - \bm{\mathcal{K}} =
\bm{\mathcal{A}}(D)$, namely
\begin{equation*}
\bm{\mathcal{A}}(P_1 + \ldots + P_g - D) = 0.
\end{equation*}
Now $\deg D = g$ implies $\deg \left( P_1 + \ldots + P_g - D
\right) = 0$ and so by Abel's theorem the divisor $P_1 + \ldots + P_g
- D = (f)$ is principal, for some meromorphic function $f$. But since
$i(D) = 0$ and $\deg D = g$, by the Riemann-Roch theorem $r(-D) = 1$
so that $f$ is constant and hence $P_1 + \ldots + P_g = D$.
\end{proof}
\end{corollary}

%% file: Semiclassical.tex
\newpage

\chapter{Semiclassical Approximations} \label{chapter: semiclassical}

In this chapter we review the necessary notions from semiclassical
quantisation of finite-dimensional systems, based on
\cite{Bates:1997kc, Sjostrand, Martinez, BerryTabor1, BerryTabor2,
Voros, Voros1, Voros2, VuNgoc1, VuNgoc2}, relevant for Part \ref{part:
Semicla}.

Consider a classical Hamiltonian system described by a $2n$
dimensional phase-space $(T^{\ast} X, \omega)$ with Hamiltonian $H :
T^{\ast} X \rightarrow \mathbb{R}$. Classically we are interested in
the trajectories of $H$, namely the integral curves of the vector
field $X_H$ on $T^{\ast} X$ which solves Hamilton's equation
\begin{equation} \label{Hamliton's equation}
\iota_{X_H} \omega = - dH.
\end{equation}
The Hamiltonian is conserved along any trajectory since $X_H H =
dH(X_H) = 0$. This constant value $E \in \mathbb{R}$ of $H$ defines
the `energy' of the trajectory which must therefore be constrained to
the codimension one level set $\Sigma_E \equiv H^{-1}(E) \subset
T^{\ast} X$.

Assume also that we have a desired quantisation of the system, that is
we are given a self-adjoint operator $\hat{H} = H(x, - i \hbar
\partial_x)$, for some choice of operator ordering, acting on
$L^2(X)$. Quantum mechanically we are interested in the spectrum of
this operator, namely the values of $E$ for which there exists a $\psi
\in L^2(X)$ which solves Schr\"odinger's equation
\begin{equation} \label{Schrodinger's equation}
(\hat{H} - E) \psi = 0.
\end{equation}

The subject of semiclassical analysis is to understand how the two
regimes are related in the limit $\hbar \rightarrow 0$.
Therefore the immediate goal of semiclassical quantisation is to
obtain the spectrum of $\hat{H}$ to leading order in $\hbar$ by
solving the Schr\"odinger equation to that order,
\begin{equation} \label{Schrodinger's equation h^2}
(\hat{H} - E) \psi = O(\hbar^2).
\end{equation}
The values $\{ E_j^{\hbar} \}_{j = 0}^{\infty}$ of $E$ for which this
equation admits a solution for $\psi$ approximate the spectrum of
$\hat{H}$ to order $O(\hbar)$.

One possible approach to obtain these values is to use what are known
as \dub{trace formulae}. The basic idea is to encode the spectrum in
terms of a single function $n(E) \equiv \sum_{j=0}^{\infty} \delta(E -
E_j^{\hbar}) = \text{tr} \; \delta(E - \hat{H})$ which one rewrites as
\begin{equation} \label{trace formula idea}
n(E) = \text{Re} \frac{1}{\pi \hbar} \int_0^{\infty} dt \;
\text{tr} \, e^{\frac{i}{\hbar} (E-\hat{H}) t} = \text{Re}
\frac{1}{\pi \hbar} \int_0^{\infty} dt \; e^{\frac{i E t}{\hbar}}
\int_{{\tiny \begin{array}{c}\text{p.o. } \gamma\\
\text{period }t \end{array}}} [d \gamma] e^{- \frac{i}{\hbar}
\int_\gamma \mathcal{L}},
\end{equation}
where the path integral is over \textit{closed} paths
$\gamma$ of period $t$ to account for the trace. In the semiclassical
limit $\hbar \rightarrow 0$ we can evaluate the integral in the
stationary phase approximation. If we assume that every periodic
trajectory of the flow $X_H$ is isolated on the level set $\Sigma_E$
then the dominant contributions to the path integral will come from each
isolated periodic orbit of the classical system. The result is known
as the Gutzwiller trace formula. It `associates' to each periodic orbit
of the classical system a tower of semiclassical energy eigenvalues
$\{ E_j^{\hbar} \}$ of $\hat{H}$.

The connection between a periodic orbit $\gamma \in \Sigma_E$ and its
associated spectrum $\{ E_j^{\hbar} \}$ determined by the Gutzwiller
trace formula is best understood in terms of the classical cylinder
theorem \cite[p576]{Marsden}.

\begin{figure}[ht]
\centering \psfrag{S}{\tiny $H^{-1}(E)$} \psfrag{g}{\tiny
$\gamma_E$} \psfrag{gp}{\tiny $\gamma_{E + \epsilon}$}
\psfrag{gm}{\tiny $\gamma_{E - \epsilon}$}
\includegraphics[height=40mm]{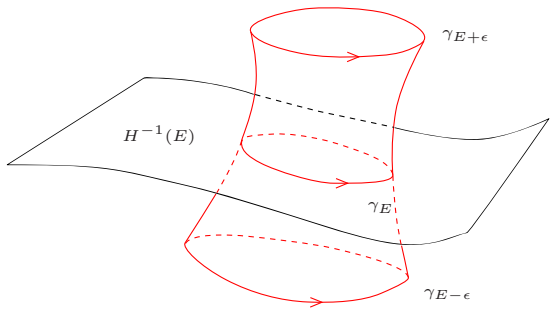}
\caption{Cylinder theorem: a periodic solution $\gamma_E$ on the
energy level $H^{-1}(E)$ is contained in a one parameter family of
periodic solutions of varying energy in the range $[E - \epsilon,
E + \epsilon]$.} \label{figure: cylinder}
\end{figure}

\begin{theorem}[Cylinder theorem] \label{thm: cylinder}
Let $\gamma \in H^{-1}(E)$ be a non-degenerate periodic orbit of
$X_H$. Then there exists $\epsilon > 0$ and $\Gamma : [E - \epsilon, E
+ \epsilon] \times S^1 \rightarrow T^{\ast} X$ such that for any $E_0
\in [E - \epsilon, E + \epsilon]$ the closed curve $\gamma_{E_0} =
\Gamma(E_0, \cdot)$ is a periodic orbit of $X_H$ in $H^{-1}(E_0)$, see
Figure \ref{figure: cylinder}.
\end{theorem}

\noindent Now the Gutzwiller trace formula can also be written as a
Bohr-Sommerfeld quantisation condition \eqref{BS5} which essentially
says that $\int_{\gamma_E} \lambda_{\text{BS}} \in \mathbb{Z}$ for a
specific 1-form $\lambda_{\text{BS}}$ to be identified later. The
Bohr-Sommerfeld condition therefore picks out a discrete set
$\gamma_{E_j^{\hbar}}$ of periodic orbits from the cylinder of theorem
\ref{thm: cylinder} whose energies $\{ E_j^{\hbar} \}$ are
semiclassical approximations to eigenvalues of $\hat{H}$. This
illustrates a very general feature of semiclassical analysis whereby
\textit{analytic} data of the quantum theory (here the spectrum of the
operator $\hat{H}$) is related to \textit{geometric} data of the
classical theory (here the periodic orbits of the classical
Hamiltonian $H$).

\begin{figure}[ht]
\begin{center}
\psfrag{S}{\tiny $\Sigma_E$} \psfrag{g}{\tiny $\gamma$}
\psfrag{X}{\tiny $X_F$}
\includegraphics[height=22mm]{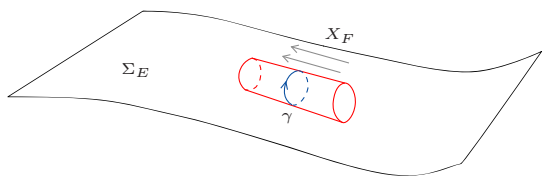}
\caption{Degeneracy of periodic orbit on $\Sigma_E$ in presence of
a symmetry.} \label{degeneracy}
\end{center}
\end{figure}

The assumption of non-degeneracy of the periodic orbits of the
Hamiltonian flow $X_H$ on the energy surface $\Sigma_E$ is crucial in
discussing the Gutzwiller trace formula: without it certain periodic
orbits might no longer be isolated local minima of the action which
complicates the stationary phase approximation. Yet this assumption
easily breaks down, for instance when the system possesses just a
single other first integral of motion\footnote{$F$ is an
\dub{integral} of $X_H$ if $dF \neq 0$ almost everywhere and $X_H F =
\{ H,F \} = 0$.}, say $F$, since its flow $X_F$ then generates from
$\gamma \subset \Sigma_E$ a continuous family of periodic orbits on
the hypersurface $\Sigma_E$ itself. Indeed, if $\phi^H_t$ denotes the
flow of $X_H$, so that $\gamma = \{ \phi^H_t(p_0) \}_{t = 0}^T$ is a
periodic orbit through $p_0$, and $\phi^F_s$ the flow of $X_F$, then
$\gamma_s = \{ \phi^H_t \circ \phi^F_s(p_0)\}_{t = 0}^T$ is a
continuous family of periodic orbits containing $\gamma = \gamma_0$
(see Figure \ref{degeneracy}).
Therefore the semiclassical approximation of the path integral in
\eqref{trace formula idea} will lead to a different semiclassical
trace formula in the degenerate case.

Suppose the Hamiltonian system locally possesses a total of $p$
independent integrals of motion $F_1, \ldots, F_p$ where $1 < p \leq
n$. Classically it is natural to consider all these integrals on the
same footing as the Hamiltonian $H = H(F_1, \ldots, F_p)$. So rather
than focusing on the Hamiltonian and its flow $X_H$ one should instead
use the moment map
\begin{equation*}
\bm{F} \equiv (F_1, \ldots, F_p) : T^{\ast} X \rightarrow \mathbb{R}^p
\end{equation*}
which generates a $p$-parameter flow $X_{F_i}$ through Hamilton's
equation
\begin{equation} \label{Hamliton's equation 2}
\iota_{X_{F_i}} \omega = - dF_i, \qquad i = 1, \ldots, p.
\end{equation}
Each integral $F_j$ is conserved along these flows since $X_{F_i} F_j
= dF_j (X_{F_i}) = 0$. In other words for any $f \in \mathbb{R}^p$ the
level set $\Sigma_f \equiv \bm{F}^{-1}(f) \subset T^{\ast} X$ is
invariant under the flows $X_{F_i}$. The objects of classical interest
here are the integral manifolds $\Lambda_f \subset \Sigma_f$ of the vector
fields $X_{F_i}$. From now on we assume $f$ to be a regular value of
$\bm{F}$ meaning that $d\bm{F}$ has maximal rank $p$. Then the integral
manifold $\Lambda_f$ is a $p$-dimensional manifold and assuming it is
compact it must be a $p$-torus $\mathbb{T}^p$. Furthermore, the level set
$\Sigma_f$ is of codimension $p$ in $T^{\ast} X$. The proper
generalisation of the cylinder theorem \ref{thm: cylinder} to
Hamiltonian systems with integrals is \cite[theorem 2.4 and lemma 2.6
pp.89-94]{Moser}

\begin{theorem}[Generalised cylinder theorem] \label{thm: cylinder gen}
With the above assumptions, let $\Lambda_f \in \Sigma_f$ be an
integral manifold of the $X_{F_i}$. Then there exists a small
neighbourhood $\mathcal{U}_f$ of $f \in \mathbb{R}^p$ and $\Gamma :
\mathcal{U}_f \times \mathbb{T}^p \rightarrow T^{\ast} X$ such that
for any $f_0 \in \mathcal{U}_f$ the $p$-torus $\Lambda_{f_0} =
\Gamma(f_0, \cdot)$ is an integral manifold of the $X_{F_i}$ in
$\Sigma_{f_0}$.
\end{theorem}

\noindent Once again the Bohr-Sommerfeld conditions for such a system
should pick out a discrete set of $p$-torii whose levels $f_i$ provide
semiclassical approximations to eigenvalues of the operators
$\hat{F}_i$. In other words these levels are such that the joint
system of Schr\"odinger equations admits a solution for $\psi \in
L^2(X)$,
\begin{equation} \label{Schrodinger's equation h^2 2}
(\hat{F}_i - f_i) \psi = O(\hbar^2), \qquad i = 1, \ldots, p.
\end{equation}
This is the analogue of \eqref{Schrodinger's equation h^2} for a
Hamiltonian system with symmetries. Note that for
\eqref{Schrodinger's equation h^2 2} to admit a solution at all
requires that $[\hat{F}_i , \hat{F_j}] = O(\hbar^3)$ which is the
semiclassical analogue of $\{ F_i, F_j \} = 0$.

The extreme case $p=n$ corresponds to an integrable system which
possesses the maximum number $n$ of independent Poisson commuting
first integrals. For such a system we must have $\Lambda_f =
\Sigma_f$ since $\Lambda_f \subset \Sigma_f$ and both manifolds
have the same dimension $n$. In the following we will focus on
this case since all intermediate cases can be obtained from it as
we will see in section \ref{section: Voros}. The path integral
treatment of integrable systems would lead to a semi-classical
trace formula known as the Berry-Tabor formula \cite{BerryTabor1,
BerryTabor2}.

Despite the geometrical appeal of the path integral approach to
semiclassical quantisation it is hard to discuss the issues of
operator ordering within this framework. Indeed, thinking in terms of
phase-space path integrals, since everything in the integrand itself
is classical, any information about quantum ordering is neatly tucked
away in the definition of the regularisation used in the phase-space
path integral measure $[d \gamma]$. The standard choice of
discretisation of the path integral measure involves the mid-point
prescription which corresponds to the Weyl-ordering prescription in
the operator formalism. In particular the quantum Hamiltonian is the
Weyl-ordered classical Hamiltonian, i.e. $\hat{H} =
\text{Op}^W_{\hbar}(H)$. In order to deal with operator ordering
issues, it is therefore more convenient to work directly with
operators and attempt to solve Schr\"odinger's equation
\eqref{Schrodinger's equation h^2} or \eqref{Schrodinger's equation
h^2 2} order by order in $\hbar$. This is also mathematically better
defined than path integral methods, although both lead to the same
Bohr-Sommerfeld conditions which are necessary and sufficient
conditions on the energy $E$ (respectively the levels $f$) for the
existence of a solution to \eqref{Schrodinger's equation h^2}
(respectively \eqref{Schrodinger's equation h^2 2}).

A convenient operator formalism for discussing semiclassical
quantisation involves pseudo-differential operators (referred to
as $\Psi$DOs for short). In section \ref{section: PsiDO} we give
a very brief introduction to $\Psi$DOs and their relevance for
treating semiclassical quantisation. We use it to discuss the issue of
operator ordering in an integrable system in section \ref{section:
operator ordering}. In section \ref{section: BS} we will show how the
Bohr-Sommerfeld quantisation conditions are modified by the presence
of a subprincipal symbol which reflects a choice of ordering.

To get an intuitive idea of how operator ordering ambiguities arise
even at the semiclassical level to affect the quantisation conditions,
it is instructive to consider the simple example of the harmonic
oscillator for which the leading order quantisation is exact. The
classical harmonic oscillator Hamiltonian is $H = \frac{p^2}{2 m} +
\frac{1}{2} m \omega^2 x^2$, and the action variable of the closed
path of energy $E$ is given by
\begin{equation*}
I = \frac{1}{2 \pi} \oint_{H = E} p dx = \frac{E}{\omega}.
\end{equation*}
By promoting the variables $x,p$ to operators $\hat{x},\hat{p}$
there is only one reasonable choice of ordering in the
Hamiltonian, namely the Weyl-ordered Hamiltonian $\hat{H} =
\frac{\hat{p}^2}{2 m} + \frac{1}{2} m \omega^2 \hat{x}^2$. The
spectrum of such an operator is well known to be $E_n = \left(n +
\frac{1}{2}\right) \hbar \omega, n \in \mathbb{N}$ so that the
spectrum of the Weyl-ordered action variable $\hat{I} =
\frac{1}{\omega} \hat{H}$ is simply,
\begin{equation*}
\text{Spec } (\hat{I}) \subset \left( \mathbb{Z} + \frac{1}{2}
\right) \hbar,
\end{equation*}
where the index of $\frac{1}{2}$ by which the spectrum is shifted
from $\hbar \mathbb{Z}$ is known as the Maslov index in the
context of Bohr-Sommerfeld quantisation. Now since we are given at
the outset only the classical Hamiltonian, we could always choose
to quantise it with a more perverse choice of ordering. For
instance, if we rewrite the classical Hamiltonian as $H = \omega a
a^{\ast}$ where $a \equiv \sqrt{\frac{m \omega}{2 \hbar}} \left( x
+ \frac{ip}{2 m} \right)$ and after promoting everything to
operators request that in the quantum Hamiltonian the $\hat{a}$
sits to the right of the $\hat{a}^{\dag}$ then we obtain the
normal-ordered Hamiltonian $:\!\hat{H}\!: \; = \omega \hbar
\hat{a}^{\dag} \hat{a}$, where $[\hat{a}, \hat{a}^{\dag}] = 1$.
The corresponding normal-ordered action operator is given by
$:\!\hat{I}\!: \; = \hbar \hat{a}^{\dag} \hat{a}$ whose spectrum
is easily seen to consist of integer multiples of $\hbar$,
\begin{equation*}
\text{Spec } (:\!\hat{I}\!:) \subset \mathbb{Z} \hbar.
\end{equation*}
We observe that the Maslov index is precisely cancelled by the
shift from Weyl-ordering to normal-ordering.


\section{Pseudo-differential operators} \label{section: PsiDO}


The passage from a classical system on phase-space $T^{\ast} X$ to
its quantum counterpart involves promoting the algebra of
classical observable $C(T^{\ast} X)$ to a noncommutative algebra
$\mathcal{A}$ of operators. Classically, the Poisson algebra of
observables is uniquely specified by the choice of a symplectic
structure $\omega = \sum_i d\xi_i \wedge dx_i$ and the Poisson
bracket of two observables $f,g \in C(T^{\ast} X)$ is then defined
by $\{ f, g\} = \omega(X_f, X_g)$, where $X_H$ denotes the
Hamiltonian vector field associated to any function $H \in
C(T^{\ast} X)$ satisfying $i_{X_H} \omega = - dH$. To pass to
quantum mechanics, the prescription of \dub{canonical
quantisation} is to promote the special functions $x_i, \xi_i \in
C(T^{\ast} X)$ to operators $\hat{x}_i,\hat{\xi}_i$ and the
symplectic structure $\omega = \sum_i d\xi_i \wedge dx_i$ to the
Weyl algebra $[ \hat{x}_i, \hat{\xi}_j] = i \hbar \delta_{ij}$
which admits the unique representation $\hat{x}_i = x_i,
\hat{\xi}_i = -i \hbar \partial/\partial x_i \equiv -i \hbar
\partial_i$ in terms of differential operators on $L^2(X)$. The
problem that remains after canonical quantisation is to associate
with any other given observable $f \in C(T^{\ast} X)$ (function of
$x_i, \xi_i$) a (pseudo-)differential operator $\hat{f}$ on $L^2(X)$,
and it is immediately obvious that this is by no means unique. Many
different operators correspond to the same classical function: for
instance, given any $t \in \mathbb{R}$, the differential operator $t
x_1 \partial_1 + (1-t) \partial_1 \cdot x_1$ is a possible candidate
for the quantisation of the function $x_1 \xi_1$. In other words, it
is not possible to specify the operator ordering in an operator
$\hat{f}$ starting from just single function $f \in C(T^{\ast}
X)$. However, with an infinite set of functions $f_k \in C(T^{\ast}
X)$ it turns out to be possible to associate a unique operator
$\hat{f}$ by canonical quantisation. Such a set defines a function of
$\hbar$ through the asymptotic expansion
\begin{equation} \label{classical Weyl symbol}
f(x,\xi;\hbar) \underset{\hbar \rightarrow 0}\sim \sum_{k \geq 0}
f_k(x,\xi) \hbar^k.
\end{equation}
We refer to such a $\hbar$-dependent function $f(\hbar) \in
C(T^{\ast} X)$ as a \dub{classical (Weyl) symbol}, which is
technically required to satisfy certain estimates, such as all its
partial derivatives being uniformly bounded by some order
function \cite{Martinez}.

Without going into details of the construction, we now state the
map from symbols to \dub{pseudo-differential
operators}\footnote{When the symbol $f(x, \xi; \hbar)$ is a
polynomial in $x, \xi$ the associated operator is an ordinary
partial differential operator. To include the more general case
when $f(x, \xi; \hbar)$ might not be a polynomial we talk about
\textit{pseudo}-differential operators.} ($\Psi$DO for short). Given a
symbol $f(\hbar)$, we define the corresponding $\Psi$DO by
specifying its action on $u \in L^2(X)$ using the \dub{Weyl
quantisation} formula (see \cite[chapter 2]{Martinez} for details)
\begin{equation*}
\left( \text{Op}_{\hbar}^W (f(\hbar)) u \right) (x) = \frac{1}{(2
\pi \hbar)^n} \int_{\mathbb{R}^{2n}} e^{\frac{i}{\hbar}(x-y)
\cdot \xi} f\left( \frac{x+y}{2}, \xi ; \hbar \right) u(y) dy
d\xi.
\end{equation*}
It is important to note here that the choice of Weyl quantisation
in the definition of the $\Psi$DO from its symbol does not limit
us to having only Weyl ordered $\Psi$DOs. Indeed, the operator
$\text{Op}_{\hbar}^W (f(\hbar))$ is Weyl ordered only when the
corresponding Weyl symbol is $\hbar$-independent. So it is
precisely the subleading terms in the asymptotic expansion
\eqref{classical Weyl symbol} of the symbol $f(x, \xi; \hbar)$
which account for the different possible choices of orderings in
the definition of the $\Psi$DO. For example, the Weyl ordered
operator of the classical observable $x_1 \xi_1$ is given simply
by the Weyl symbol $x_1 \xi_1$, namely
\begin{equation*}
\text{Op}_{\hbar}^W (x_1 \xi_1) = \frac{-i \hbar}{2}\left( x_1
\partial_1 + \partial_1 \cdot x_1 \right),
\end{equation*}
whereas the left ordered operator $-i \hbar x_1 \partial_1$ which
corresponds to the same classical observable $x_1 \xi_1$ as
$\text{Op}_{\hbar}^W (x_1 \xi_1)$ is given by a Weyl symbol with a
subleading term in $\hbar$ since
\begin{equation*}
\text{Op}_{\hbar}^W \left(x_1 \xi_1 + \frac{i \hbar}{2}\right) =
-i \hbar x_1 \partial_1.
\end{equation*}
Naturally the right ordered operator $-i \hbar \partial_1 \cdot
x_1$ has Weyl symbol $x_1 \xi_1 - \frac{i \hbar}{2}$. A general
$\Psi$DO $A$ always has a unique Weyl symbol, which is a
$\hbar$-dependent function $f(x,\xi;\hbar)$ denoted $\sigma^W(A)$.
The leading non-zero term in the asymptotic expansion
\eqref{classical Weyl symbol} of this Weyl symbol is called the
\dub{principal symbol}, denoted $\sigma_0^W(A)$, and the
subleading term is called the \dub{subprincipal symbol},
denoted $\sigma_{\text{sub}}^W(A)$. For instance, if
$f_0(x,\xi)\neq 0$ then $\sigma_0^W(A) = f_0(x,\xi)$ and
$\sigma_{\text{sub}}^W(A) = f_1(x,\xi) \hbar$.

An important object for the study of quantum integrability is the
commutator $[A,B]$ of two operators $A$ and $B$. In the present
context of $\Psi$DOs one can show that if $A,B$ are $\Psi$DOs then
their commutator $[A,B]$ is also a $\Psi$DO with principal symbol
\begin{equation} \label{comm p-symbol}
\sigma_0^W([A,B]) = -i \hbar \left\{ \sigma_0^W(A), \sigma_0^W(B)
\right\},
\end{equation}
(so that $-i \hbar \sigma_0^W$ is a Lie algebra homomorphism) and
subprincipal symbol
\begin{equation} \label{comm s-symbol}
\sigma_{\text{sub}}^W([A,B]) = -i \hbar \left\{ \sigma_0^W(A),
\sigma_{\text{sub}}^W(B) \right\} - i \hbar \left\{
\sigma_{\text{sub}}^W(A), \sigma_0^W(B) \right\}.
\end{equation}


\section{Integrable systems} \label{section: operator ordering}


As explained in section \ref{section: PsiDO}, one can keep track
of operator orderings in the language of pseudo-differential
operators by retaining subleading terms beyond the principal
symbol in the full Weyl symbol of an operator. In most
applications of the theory of $\Psi$DOs the quantities of interest
are specified as $\Psi$DOs at the outset so that their full Weyl
symbol is known. In the present case however we start from a
classical system specified by its phase-space $(T^{\ast} X,
\omega)$ and the set of classical observables of interest are
$F_1,\ldots,F_n, H$. Quantising this classical system requires an
operator ordering prescription for obtaining operators from the
corresponding classical observables. At the semiclassical level
this boils down to the specification of an extra function, the
subprincipal symbol, for each classical observable. Specifically,
given a classical observable $f_0 \in C(T^{\ast} X)$, we construct
\begin{equation*}
\hat{f} = \text{Op}_{\hbar}^W (f_0 + f_1 \hbar),
\end{equation*}
where the presence of the subprincipal symbol $f_1 \in C(T^{\ast}
X)$ reflects the operator ordering ambiguities already manifesting
themselves at the semiclassical level. Every possible choice of a
function $f_1 \in C(T^{\ast} X)$ corresponds to a different
prescription for the operator ordering in $\hat{f}$ at order
$O(\hbar)$. The principal symbol $f_0 = \sigma_0^W(\hat{f})$ is
the corresponding classical observable.

Recall the definition of an integrable system, which roughly
speaking is one which possesses the maximum possible number of
independent integrals of motion.

\begin{definition}
A Hamiltonian system $(T^{\ast} X,H)$ is said to be \dub{classically
integrable} if there exists $n = \dim X$ functions $F_1, \ldots, F_n
\in C(T^{\ast} X)$ such that
\begin{itemize}
  \item[$(1')$] $d F_1 \wedge \ldots \wedge d F_n \neq 0$ almost
everywhere,
  \item[$(2')$] $\{ F_i, F_j \} = 0, \; \forall i,j =
1,\ldots,n$,
  \item[$(3')$] $H = H(F_1,\ldots,F_n)$.
\end{itemize}
\end{definition}
Conditions $(2')$ and $(3')$ together imply that the $F_i$ are in
fact integrals of motion, $X_H F_i = 0$. In other words, $T^{\ast}
X$ admits a torus action with moment map
\begin{equation*}
\bm{F} \equiv (F_1, \ldots, F_n) : T^{\ast} X \rightarrow
\mathbb{R}^n.
\end{equation*}
When dealing with an integrable system it is convenient to treat all
the integrals of motion on the same footing as the Hamiltonian $H$
itself. At regular values $f$ of $\bm{F}$, the level sets
$\bm{F}^{-1}(f)$ define $n$-torii (in the compact case) and foliate
$T^{\ast} X$
\begin{equation*}
\mathbb{T}^n \hookrightarrow T^{\ast} X \overset{\bm{F}}\rightarrow
\mathbb{R}^n.
\end{equation*}
This foliation allows one to define canonical action-angle coordinates
with the action variables $\{ I_i \}_{i=1}^n$ parametrising the base
$\mathbb{R}^n$ and the conjugate angle variables $\{ \theta_i
\}_{i=1}^n$, each taking values in $[0,2 \pi]$, parametrising the
independent cycles of the torus $\mathbb{T}^n$. The condition $(2')$
can be phrased as $\omega(X_{F_i}, X_{F_j}) = 0$ which says that the
pullback of the symplectic form $\omega$ to a level set $\Lambda_f$
vanishes. In other words, the Liouville form $\alpha$ defined as a
primitive of $\omega = d \alpha$ is closed on $\Lambda_f$.

\begin{definition}
A $\Psi$DO $\hat{H}$ is \dub{semiclassically integrable} if there
exists $n$ $\Psi$DOs $\hat{F}_1, \ldots, \hat{F}_n$ with principal
symbols $F_i = \sigma_0^W(\hat{F}_i)$ such that
\begin{itemize}
  \item[$(1)$] $d F_1 \wedge \ldots \wedge d F_n \neq 0$ almost
everywhere,
  \item[$(2)$] $[ \hat{F}_i, \hat{F}_j ] = O(\hbar^3), \;
\forall i,j = 1,\ldots,n$,
  \item[$(3)$] $\hat{H} =
H(\hat{F}_1,\ldots, \hat{F}_n) + O(\hbar^2)$ for some function
$H$.
\end{itemize}
\end{definition}
Notice that we only require commutativity modulo $O(\hbar^3)$ in
property $(2)$; it guarantees in particular that the operator
$H(\hat{F}_1,\ldots, \hat{F}_n)$ in $(3)$ is free of operator
ordering ambiguities certainly up to $O(\hbar^3)$, so that
property $(3)$ makes sense. Property $(2)$ is to be contrasted
with the definition of full quantum integrability which requires
exact commutativity $[ \hat{F}_i, \hat{F}_j ] = 0$. Now since
$\sigma_0^W([\hat{F}_i,\hat{F}_j]) = -i \hbar \{ F_i, F_j \}$ by
\eqref{comm p-symbol} and $\sigma_0^W(\hat{H}) =
H(F_1,\ldots,F_n)$, it follows that the principal symbols $F_i =
\sigma_0^W(\hat{F}_i)$ satisfy all three properties $(1')$-$(3')$
above for a classically integrable system with Hamiltonian $H =
\sigma_0^W(\hat{H})$. This means that any semiclassically
integrable system exhibits at leading order the full geometric
structure of the underlying classically integrable system given by
its principal symbols. In particular, the level set $\Lambda_f
\equiv \bm{F}^{-1}(f)$ of the moment map $\bm{F} =
(F_1,\ldots,F_n) : T^{\ast} X \rightarrow \mathbb{R}^n$ is a
Lagrangian $n$-torus and foliates phase-space $T^{\ast} X$ as we
let $f$ vary.

But the notion of semiclassical integrability contains more
information than that of its underlying classical integrable
structure \cite{VuNgoc1, VuNgoc2}. Property $(1)$ only
contributes at leading order since it is a statement about the
principal symbols $F_i$ alone, whereas property $(2)$ at
$O(\hbar^2)$ yields an equation for the subprincipal symbols
$F^s_i = \sigma_{\text{sub}}^W(\hat{F}_i)$ of the $\hat{F}_i$ using
\eqref{comm s-symbol},
\begin{equation} \label{subprincipal symbol rel}
0 = \frac{i}{\hbar} \sigma_{\text{sub}}^W([\hat{F}_i,\hat{F}_j]) =
\left\{ F_i, F^s_j \right\} + \left\{ F^s_i, F_j \right\}.
\end{equation}
It is possible to interpret these equations geometrically so as to
supplement the geometrical structure of the underlying classical
integrable system defined by principal symbols. For this we
define the \dub{subprincipal form} $\kappa$ on $\Lambda_f$ by
specifying its action on the basis vectors $X_{F_i}$ at any point of
$\Lambda_f$ through \cite{VuNgoc1}
\begin{equation} \label{subprincipal form}
\kappa(X_{F_i}) = - F^s_i, \quad i = 1,\ldots,n.
\end{equation}
It then follows immediately from \eqref{subprincipal symbol rel}
that $\kappa$ is closed since
\begin{align*}
d\kappa(X_{F_i},X_{F_j}) &= X_{F_i} \kappa(X_{F_j}) - X_{F_j}
\kappa(X_{F_i}) - \kappa([X_{F_i}, X_{F_j}])\\
&= - X_{F_i} F^s_j + X_{F_j} F^s_i - \kappa(X_{\{ F_i,F_j\}}) = -
\{F_i,F^s_j\} + \{F_j,F^s_i\} - 0 = 0.
\end{align*}
Hence the operator ordering in the $\hat{F}_i$ can be accounted
for at the semiclassical level by specifying a closed 1-form
$\kappa$ on the Liouville $n$-torus $\Lambda_f$. And in fact it is
clear from \eqref{subprincipal form} that every choice of a closed
1-form $\kappa \in \Omega^1(\Lambda_f)$ corresponds to a different
choice of operator ordering in the definition of the $\hat{F}_i$.

To summarise, the classical and semiclassical integrability conditions
can both be expressed as the closure of the Liouville form $\alpha$
and subprincipal form $\kappa$ respectively on the level set
$\Lambda_f$,
\begin{subequations} \label{integrability conditions closure}
\begin{align}
\label{integrability conditions closure a} \textit{Classical} &:
\qquad d \alpha = 0  \quad \text{on } \Lambda_f,\\
\label{integrability conditions closure b} \textit{Semiclassical} &:
\qquad d \kappa = 0  \quad \text{on } \Lambda_f.
\end{align}
\end{subequations}


\section{Bohr-Sommerfeld conditions} \label{section: BS}


We are interested in the joint spectrum of the $\hat{F}_i$ up to
$O(\hbar)$ which requires solving the eigenvalue problem to that
order
\begin{equation} \label{eigenvalue problem}
(\hat{F}_i - f_i)\psi = O(\hbar^2).
\end{equation}
The Bohr-Sommerfeld conditions are conditions on the set $\{ f_i \}$
for the existence of a solution to these coupled pseudo-differential
equations. Their rigourous derivation is rather involved (see for
instance \cite{VuNgoc1,VuNgoc2}) so here we would just like to outline
how the subprincipal symbol comes about in these conditions. To solve
\eqref{eigenvalue problem} locally one considers a local patch $V
\subset \Lambda_f$ on which $\pi : T^{\ast} X \rightarrow X$ is a
diffeomorphism and uses the WKB ansatz
\begin{equation} \label{WKB ansatz}
\psi_{WKB} = e^{\frac{i}{\hbar} \phi_{-1} + i \phi_0} \rho +
O(\hbar)
\end{equation}
on $U = \pi(V) \subset X$ where the nature of $\rho$ will be
specified shortly. If we denote by $\iota_{d \phi_{-1}} : U
\hookrightarrow T^{\ast} X$ the 1-form $d\phi_{-1}$ viewed
as a map then equation \eqref{eigenvalue problem} can be shown
\cite{Bates:1997kc} at leading order in $\hbar$ to imply to the so
called \dub{eikonal equation}
\begin{equation} \label{eikonal equation}
\text{im } \iota_{d \phi_{-1}} = V \subset \Lambda_f.
\end{equation}
Therefore $\iota_{d \phi_{-1}} : U \rightarrow V$ with $\pi \circ
\iota_{d \phi_{-1}} = \text{id}_U$ so that $\iota_{d\phi_{-1}} =
\pi|_V^{-1}$. By a property of the tautological 1-form $\alpha$
\cite[lemma 3.23 p29]{Bates:1997kc}, namely $d \phi_{-1} =
\iota_{d\phi_{-1}}^{\ast} \alpha$, we then have
\begin{equation*}
d \pi|_V^{\ast} \phi_{-1} = \alpha.
\end{equation*}
In other words, $\pi|_V^{\ast} \phi_{-1}$ is a local solution to
the classical integrability condition \eqref{integrability conditions
closure a}. If $\rho$ is a half-density\footnote{Since the
product of two half-densities is a density of weight one there is
a natural inner-product on half densities $\langle \rho_1, \rho_2
\rangle = \int_M \rho_1 \rho_2$ which makes the completion into a
Hilbert space.} on $U \subset X$ then the subleading order of
\eqref{eigenvalue problem} implies the so called \dub{transport
equation} which can be written invariantly as \cite[theorem 11.11
p126]{Sjostrand}
\begin{equation} \label{transport equation}
\left(-i \mathcal{L}_{X_{F_i}} + F^s_i\right) \left(\pi|_V^{\ast}
e^{i \phi_0} \rho\right) = 0.
\end{equation}
Writing $a = \pi|_V^{\ast} \phi_0$, since $e^{i a}$ is a function we
have $\mathcal{L}_{X_{F_i}} e^{i a} = \iota_{X_{F_i}} de^{i a}$. Now
using \eqref{subprincipal form} we can rewrite \eqref{transport
equation} as
\begin{equation} \label{transport equation 2}
[d\pi|_V^{\ast} \phi_0 (X_{F_i}) - \kappa(X_{F_i})]
\left( \pi|_V^{\ast} \rho \right) = i \mathcal{L}_{X_{F_i}} \left(
\pi|_V^{\ast} \rho \right).
\end{equation}
Therefore provided the subprincipal symbols are real this equation
implies on the one hand that $\pi|_V^{\ast} \rho$ is an invariant
half-density on $\Lambda_f$, \textit{i.e.} $\mathcal{L}_{X_{F_i}}
\pi|_V^{\ast} \rho = 0$, and on the other hand that
\begin{equation*}
d \pi|_V^{\ast} \phi_0 = \kappa.
\end{equation*}
But this just says that $\pi|_V^{\ast} \phi_0$ is a local solution to
the subleading integrability condition \eqref{integrability conditions
closure b}. To summarise, in a neighbourhood $V \subset
\Lambda_f$ where $\pi|_V$ is a diffeomorphism the eigenvalue equation
\eqref{eigenvalue problem} is solved by \eqref{WKB ansatz} if
$\phi_{-1}$ and $\phi_0$ are primitives of the Liouville form
$\alpha$ and the subprincipal form $\kappa$ respectively.

\begin{figure}[h]
\begin{center}
\psfrag{pi}{\tiny $\pi$} \psfrag{L}{\tiny $\Lambda_f \subset
T^{\ast} X$} \psfrag{M}{\tiny $X$} \psfrag{sing}{\tiny
singularity} \psfrag{caus}{\tiny caustic}
\includegraphics[height=40mm]{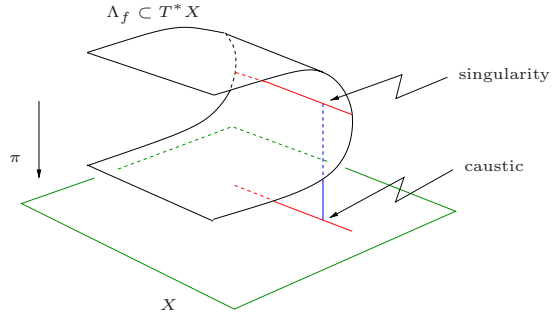}
\caption{Caustics of the Lagrangian submanifold $\Lambda_f$}
\label{Maslov}
\end{center}
\end{figure}

However, one runs into problems at caustic points where $\pi$ is
singular (see Figure \ref{Maslov}).
A way around this problem was proposed by Maslov, the idea being
to obtain a solution of \eqref{eigenvalue problem} which is
localised and defined patchwise on $\Lambda_f$ (near caustics one
uses the ``momentum'' projection $\pi_p$ of $T^{\ast} X$ onto a
typical fibre of $T^{\ast} X$ instead of $\pi$). Since this
wave-function is defined on the whole of $\Lambda_f$ and the level
set $\Lambda_f$ is compact, the single-valuedness of this global
solution requires its phase to be an integer multiple of $2\pi$.
The phase turns out to be that of the local WKB solutions
$\psi_{WKB}$ introduced above but with additional Maslov index
corrections (coming from the caustics). The single-valuedness of
this phase leads to the Bohr-Sommerfeld-Maslov quantisation
conditions (see \cite{VuNgoc1} for a nice review).

\begin{theorem}[Bohr-Sommerfeld-Maslov]
The eigenvalue problem \eqref{eigenvalue problem} has a solution if
and only if
\begin{equation} \label{BS3a}
\frac{1}{2 \pi \hbar} \int_{\gamma_i} \alpha + \frac{1}{2 \pi}
\int_{\gamma_i} \kappa = N_i + \frac{\mu_{\gamma_i}}{4} +
O(\hbar), \quad i=1,\ldots,n
\end{equation}
where $\gamma_i$ is a basis of $H_1(\Lambda_f, \mathbb{R})$ with
Maslov indices $\mu_{\gamma_i} \in \mathbb{Z}_4$ and integers $N_i
\in \mathbb{Z}$.
\end{theorem}
Note in particular the presence of the subprincipal form $\kappa$
which as we have argued is related to operator ordering ambiguities in
going from a classically integrable system to its quantum (or just
semiclassically) integrable counterpart. It has the effect of shifting
the spectrum of the action variables similar to what happened in the
case of the harmonic oscillator when we changed quantisation from Weyl
to normal ordering. In the cases where all the operators are chosen to
be Weyl ordered, in particular the $\hat{F}_i$, we have $\kappa = 0$
and \eqref{BS3a} reduces to the EBK quantisation conditions. From now
on we shall always assume that the cohomology class $[\kappa] \in
H^1(\Lambda_f)$ of the subprincipal form $\kappa$ vanishes. The reason
for this assumption is that the result is simpler to express in this
case and moreover it will give results that agree with those of
\cite{Gromov+Vieira1, Gromov+Vieira2, Gromov+Vieira3}. With this
assumption, the Bohr-Sommerfeld-Maslov conditions simplify
\begin{equation} \label{BS3}
\frac{1}{2 \pi \hbar} \int_{\gamma_i} \alpha = N_i +
\frac{\mu_{\gamma_i}}{4} + O(\hbar), \quad i=1,\ldots,n.
\end{equation}
We stress that this assumption does \textit{not} imply the choice
of Weyl ordering since it only corresponds to setting the
subprincipal symbol to zero, whereas Weyl ordering corresponds to
setting all the lower order Weyl symbols to zero as well.


\section{Bohr-Sommerfeld for degenerate torii} \label{section: Voros}

\begin{figure}[h]
\begin{center}
\psfrag{p1}{\tiny $p_1$} \psfrag{x1}{\tiny $x_1$}
\psfrag{p2}{\tiny $p_2$} \psfrag{x2}{\tiny $\omega_2 x_2$}
\psfrag{T1}{\tiny $T_1 = \frac{2 \pi}{\omega_1}$}
\psfrag{T2}{\tiny $T_2 = \frac{2 \pi}{\omega_2}$}
\includegraphics[height=30mm]{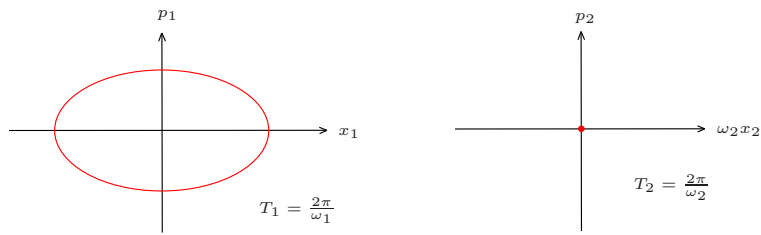}
\end{center}
\caption{Periodic orbit with $H_2 = 0$ of energy $H = H_1 = E$.}
\label{HOstab1}
\end{figure}
\begin{figure}[h]
\begin{center}
\psfrag{S1}{\tiny $E_1$} \psfrag{S2}{\tiny $E_2$}
\includegraphics[height=50mm]{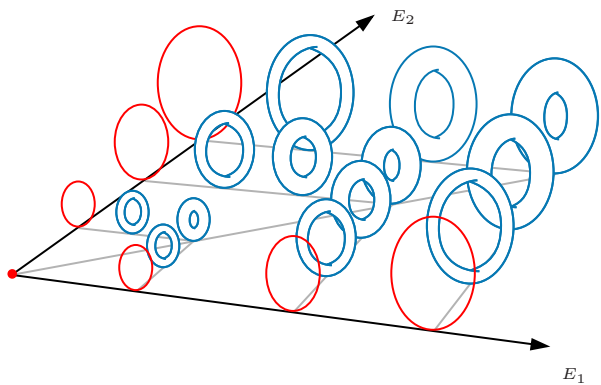}
\caption{The phase-space of the two-dimensional harmonic
oscillator.} \label{Phase_space}
\end{center}
\end{figure}

The derivation of the Bohr-Sommerfeld-Maslov conditions
\eqref{BS3a} or \eqref{BS3} essentially consisted in quantising a
Lagrangian $n$-torus $\Lambda_f$ by constructing a wave-function
localised around it. However, even though the level set $\Lambda_f
\equiv \bm{F}^{-1}(f)$ is indeed a Lagrangian $n$-torus for almost
every value of the integrals of motion $f_1, \ldots, f_n$ in an
integrable system, there exists interesting level sets
$\bm{F}^{-1}(f)$ in phase-space where this is not the case. This
happens at the (measure zero) set of critical values of the map
$\bm{F} = (F_1, \ldots, F_n)$. Consider for instance the
two-dimensional harmonic oscillator with different frequencies and
total Hamiltonian
\begin{equation} \label{2d HO}
H = \frac{p_1^2}{2} + \frac{1}{2} \omega_1^2 x_1^2 +
\frac{p_2^2}{2} + \frac{1}{2} \omega_2^2 x_2^2 = H_1 + H_2,
\end{equation}
whose integrals of motion are given by $H_1, H_2$. For non-zero
values $E_1, E_2 \neq 0$ of $H_1, H_2$ the level sets
$\bm{H}^{-1}(E_1,E_2)$ consists of two ellipses, in other words a
Lagrangian 2-torus. However, if say $E_2 = 0$ the level set
$\bm{H}^{-1}(E_1,0)$ consists of just a single ellipse (Figure
\ref{HOstab1}).
The same thing is true when $E_1 = 0$ and at the point where $E_1
= E_2 = 0$ the level set consists of just a single point. One can
draw a picture of the phase-space in the region where $\mathcal{E}
\equiv \{ (E_1,E_2) : E_i \geq 0, i=1,2\}$ which is foliated by
2-torii in the interior of $\mathcal{E}$ but with the fibres over
the boundary $\partial \mathcal{E} \setminus \{ (0,0) \}$ being
ellipses and the fibre over the point $(0,0)$ being just a single
point, see Figure \ref{Phase_space}.
Note that the set of critical values $\partial \mathcal{E}$ is of
measure zero. However, if we are interested in the semiclassical
spectrum of the two-dimensional harmonic oscillator in the region
near $\partial \mathcal{E}$ then a modification of the
Bohr-Sommerfeld-Maslov quantisation conditions \eqref{BS3} is
required so that it applies to isotropic $p$-torii which are the
level sets of a limited number $p < n$ of integrals of motion
$F_1, \ldots, F_p$.

It was pointed out by Voros \cite{Voros, Voros1} that the
Bohr-Sommerfeld conditions \eqref{BS3} for the apparently more
restrictive case of an integrable system may be used to obtain the
Bohr-Sommerfeld conditions in all other intermediate cases, namely
the partially integrable one (with $p < n$ integrals of motion)
and even the non-degenerate case $p=1$ (where $H$ is the only
integral). If the system has $p$ independent observables $\bm{F} =
(F_1, \ldots, F_p)$ in involution (with $H = H(\bm{F})$), then on
each codimension $p$ level set $\Sigma_f = \bm{F}^{-1}(f)$ the
system has a $p$-torus $\Lambda_f \subset \Sigma_f$ generated by
the vector fields $X_{F_i}$. Each of these $p$-torii is surrounded
by an $n$-torus of the linearised system to which the
Bohr-Sommerfeld-Maslov conditions \eqref{BS3} may be applied. This
results in a set of Bohr-Sommerfeld conditions for the cycles on
the $p$-torus which include stability angles for the small
fluctuations in the directions transverse to this $p$-torus. The
derivation of these Bohr-Sommerfeld conditions from those in the
integrable case \eqref{BS3} are a bit lengthy but the derivation
in the more general case $1 < p < n$ is conceptually the same as
the $p=1$ case. We will therefore outline the proof \cite{Voros,
Voros1} only in the latter case.

Let $\gamma \subset \Sigma_E$ be a periodic orbit of energy $E$. We
henceforth assume that $E$ is a regular value of $H$ so that
$\Sigma_E$ is a smooth codimension one submanifold of $T^{\ast}
X$. Given a point $p_0 \in \gamma$, we call a \dub{section} of
$\gamma$ at $p_0$ a smooth codimension one
\begin{figure}[h]
\begin{center}
\psfrag{g}{\tiny $\gamma$} \psfrag{p0}{\tiny $p_0$}
\psfrag{p}{\tiny $p'$} \psfrag{p2}{\tiny $p$} \psfrag{S}{\tiny
$S$}
\includegraphics[height=45mm]{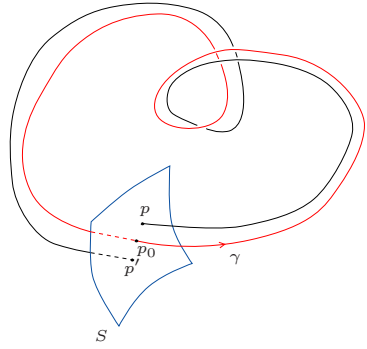}
\caption{Poincar\'e map \cite{Moser}: global perturbations of a
periodic orbit $\gamma$ can be studied locally in terms of a map $\psi
: S \rightarrow S$ defined by the flow of $X_H$.} \label{Poinc}
\end{center}
\end{figure}
surface $S \subset \Sigma_E$ transverse to $\gamma$ and
intersecting it at $p_0$. We then define the local map $\psi : S
\rightarrow S$ near $p_0$ by letting $p' = \psi(p)$ be the unique
point obtained by following $p \in S$ around the Hamiltonian flow
$X_H$ for a time close to the period $T_\gamma$ of $\gamma$ (see
Figure \ref{Poinc}). Note that fixed points $p = \psi(p)$
(respectively periodic points $p = \psi^k(p), k \geq 2$) of $\psi$
correspond to periodic orbits of the Hamiltonian flow $X_H$ of
period close to $T_\gamma$ (respectively close to $k T_\gamma$).
In particular, since $p_0 = \psi(p_0)$ we define the \dub{Poincar\'e
map} as the differential of $\psi$ at $p_0$ \cite{Moser}
\begin{equation*}
P = d\psi_{p_0} : T_{p_0} S \rightarrow T_{p_0} S.
\end{equation*}
We say that the periodic orbit $\gamma$ is \dub{non-degenerate} if and
only if $1$ is not an eigenvalue of the Poincar\'e map. This is a
way of saying that $\gamma$ is isolated on $\Sigma_E$ in the sense
that there are no periodic orbits on $\Sigma_E$ arbitrarily close
to it. The cylinder theorem then applies to $\gamma$ which therefore
belongs to a family $\gamma_E$. Furthermore, $\gamma$ is said to be
\dub{stable} when the eigenvalues of the Poincar\'e map come in
complex conjugate pairs of the form $(e^{i \nu_{\alpha}}, e^{- i
\nu_{\alpha}})$ with $\nu_{\alpha} \in \mathbb{R}$. The angles
$\nu_{\alpha}$ are then called the \dub{stability angles}. In
particular, for a non-degenerate curve all the stability angles are
non-zero.

\begin{theorem} \label{theorem: Voros}
Let $\gamma \in H^{-1}(E)$ be a stable non-degenerate periodic orbit
of $X_H$. Then
\begin{equation} \label{BS5}
\int_{\gamma} \alpha = \left[ 2 \pi \left( N +
\frac{\mu_{\gamma}}{4}\right) + \sum_{\alpha = 2}^n \left(
n_{\alpha} + \frac{1}{2} \right) \nu_{\alpha} \right] \hbar +
O(\hbar^2),
\end{equation}
with $N \in \mathbb{Z}$, $n_{\alpha} \in \mathbb{N}$ and $n_{\alpha}
\ll |N|$, is a sufficient condition on $E$ for the existence of a
solution to the Schr\"odinger equation in \eqref{Schrodinger's
equation h^2}.

\begin{proof}
\cite{Voros, Voros1}
Since $\gamma$ is stable the Poincar\'e map is merely a product of
rotations by angles $\nu_{\alpha}$ in $n-1$ disjoint planes
$\mathbb{R}^2_{\alpha} \subset T_{p_0} S$. In other words, every point
$p_0 \in \gamma$ of the stable isolated periodic orbit $\gamma$ is
surrounded by an infinitesimal torus $S^1_{F_2} \times \ldots \times
S^1_{F_n}$, where $S^1_{F_{\alpha}} = \{ x_{\alpha} \in
\mathbb{R}^2_{\alpha} \; | \; ||x_{\alpha}||^2 = F_{\alpha} \} \subset
\mathbb{R}^2_{\alpha}$, which is preserved by the Poincar\'e map
to first approximation in $F_{\alpha} \ll 1$. By the cylinder
theorem the periodic orbit $\gamma$ belongs to a continuous family
$\gamma_E$ parametrised by the energy $E$, and so one could now
apply the Bohr-Sommerfeld-Maslov quantisation conditions to the
family of torii $\Lambda \equiv \gamma_E \times S^1_{F_2} \times
\ldots \times S^1_{F_n}$ just constructed (see Figure
\ref{local_torus}).
\begin{figure}[h]
\centering
\begin{tabular}{ccc}
\psfrag{g}{\tiny $\gamma$} \psfrag{R}{\tiny $T_{p_0} S =
\mathbb{R}^2_{\alpha}$} \psfrag{p}{\tiny $p_0$} \psfrag{S}{\tiny
$S^1_{F_{\alpha}}$}
\includegraphics[height=40mm]{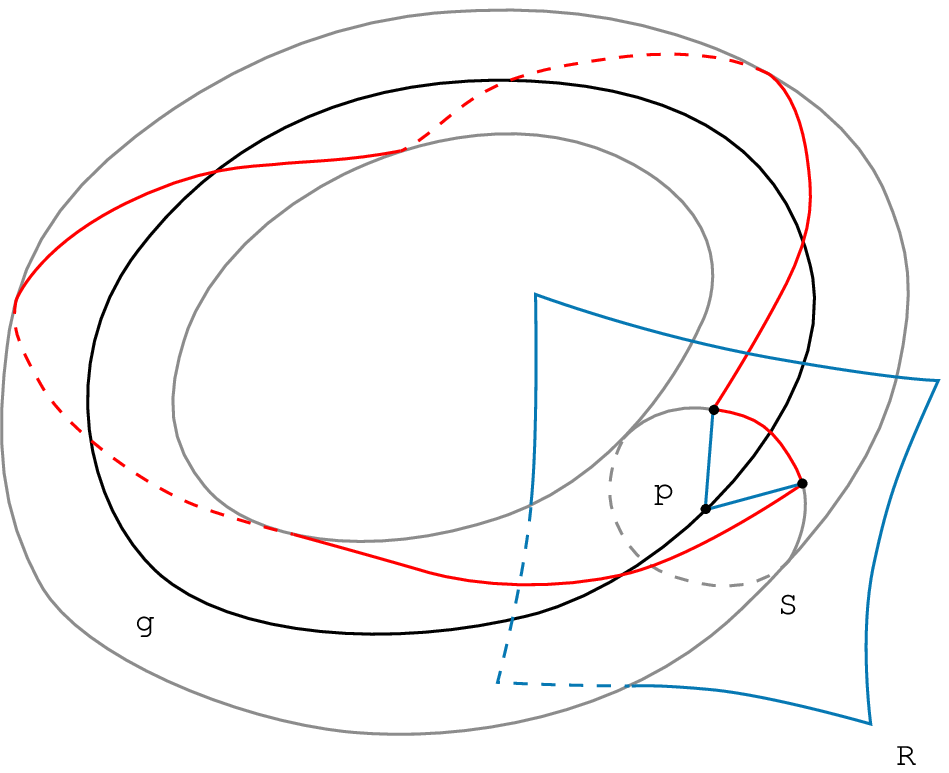}
& $\qquad\qquad$ & \psfrag{p}{\footnotesize $p_0$}
\psfrag{nu}{\footnotesize $\nu_\alpha$}
\raisebox{5mm}{\includegraphics[height=30mm]{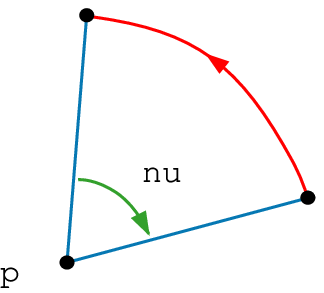}}
\end{tabular}
\caption{The infinitesimal torus around a stable isolated periodic
orbit $\gamma$ illustrated in the case $n=2$ where there
is only one stability angle $\nu_{\alpha}$ and $T_{p_0} S =
\mathbb{R}^2_{\alpha}$.} \label{local_torus}
\end{figure}
They read
\begin{align*}
\int_{S^1_{F_{\alpha}}} \alpha &= 2 \pi \left( n_{\alpha} +
\frac{1}{2} \right) \hbar + O(\hbar^2), \quad \alpha = 2,\ldots,n\\
\int_{\tilde{\gamma}} \alpha &= 2 \pi \left( N +
\frac{\mu_{\gamma}}{4} \right) \hbar + O(\hbar^2),
\end{align*}
where $\tilde{\gamma}$ is the closed path on $\Lambda$ consisting
of a classical path going from $T_{p_0} S$ once around $\Lambda$
back to $T_{p_0} S$ and the set of arcs of angles $- \nu_{\alpha}$
on $T_{p_0} S$ to close off this classical path (see red curve in
Figure \ref{local_torus}).

Consider the 2-dimensional surface $\Gamma$ bounded by the
periodic orbit $\gamma$ and the closed curve $\tilde{\gamma}$,
constructed in the obvious way: at any point $t \neq 0$ along the
curve $\gamma(t)$, $\Gamma$ looks locally like $\{ \gamma(t) +
\tau y(t) | 0 < t < T, 0\leq \tau \leq 1\}$ where $y(t)$ is the
transversal vector to $\gamma$ joining the points $\gamma(t)$ and
$\tilde{\gamma}(t)$. At $t=0$ we complete the surface by adding
the sections of the disc of angle $- \nu_{\alpha}$ on $T_{p_0} S$.
Then by Stokes's theorem we have
\begin{equation*}
\left( \int_{\tilde{\gamma}} - \int_{\gamma} \right) \alpha =
\int_{\partial \Gamma} \alpha = \int_{\Gamma} \omega.
\end{equation*}
On the part of $\Gamma$ corresponding to $t \neq 0$ we have
$\omega|_{\Gamma} = 0$ since the tangent space to $\Gamma$ is
spanned by $X_H$ and the transversal vector $y$ ($i_y i_{X_H}
\omega = i_y dH = y(H) = 0$ since $y$ lies in the energy surface
$\Sigma_E$). And since $\Gamma_{t=0}$ looks like sections of angle
$-\nu_{\alpha}$ of the disc of radius $\sqrt{F_{\alpha}}$ it
follows that
\begin{equation*}
\left( \int_{\tilde{\gamma}} - \int_{\gamma} \right) \alpha =
\int_{\Gamma_{t=0}} \omega = - \sum_{\alpha = 2}^n \nu_{\alpha}
F_{\alpha}.
\end{equation*}
On the other hand we have that
\begin{equation*}
\int_{S^1_{F_{\alpha}}} \alpha = \int_{D^1_{F_{\alpha}}} \omega =
2 \pi F_{\alpha},
\end{equation*}
where $D^1_{F_{\alpha}}$ is the disc in $\mathbb{R}^2_{\alpha}$
bounded by the circle $S^1_{F_{\alpha}}$. The last equality
follows by a direct computation. Combining everything we obtain
\eqref{BS5}.
\end{proof}
\end{theorem}

\begin{figure}[ht]
\begin{center}
\psfrag{S}{\tiny $\Sigma_E$}
\includegraphics[height=40mm]{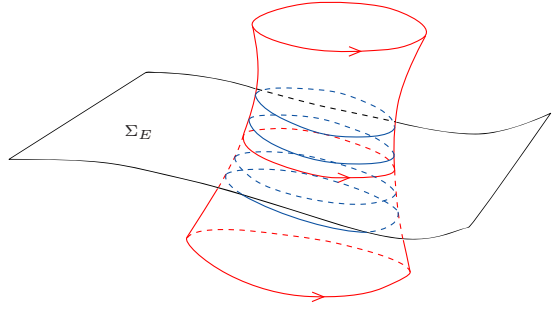}
\caption{Bohr-Sommerfeld semi-classical spectrum: the discrete set
of periodic orbits $\gamma_{E^{\hbar}_j}$ shown in blue have
energies $E^{\hbar}_j$ approximating eigenvalues of $\hat{H}$ to
$O(\hbar^2)$.} \label{BS spectrum}
\end{center}
\end{figure}

Since the periodic orbit $\gamma \subset \Sigma_E$ in fact belongs to
a continuous $1$-parameter family $\gamma_E$ of periodic orbits
parametrised by the energy $E$ according to the cylinder theorem
\ref{thm: cylinder}, what the condition \eqref{BS5} does is pick out a
discrete set of periodic orbits $\gamma_{E^{\hbar}_j}$, in a
neighbourhood of the level set $\Sigma_E$, whose energies
$E^{\hbar}_j$ approximate eigenvalues of $\hat{H}$ to leading order in
$\hbar$ (see Figure \ref{BS spectrum}).

The more general case of a system which has a total of $p$ independent
observables $F_1, \ldots, F_p$ in involution (with $H = H(F_1,
\ldots, F_p)$), where $p$ lies in the range $1 < p < n$ is a
straightforward generalisation (see \cite{Voros, Voros1} for
details). In this case there is a different Poincar\'e map for each
basis cycle $\gamma_k \in H_1(\Lambda_f), k=1,\ldots,p$ on the
$p$-torus $\Lambda_f$. Each has its own stability angles
$\nu^k_{\alpha}$, $\alpha = p+1, \ldots, n$ for oscillations in the
transverse directions to the $p$-torus and $\Lambda_f$ is \dub{stable}
if these are all real.

\begin{theorem}
Let $\Lambda_f \subset \Sigma_f$ be a stable integral manifold of the
$X_{F_i}$. Then
\begin{equation} \label{BS6}
\int_{\gamma_k} \alpha = \left[ 2 \pi \left( N_k +
\frac{\mu_{\gamma_k}}{4}\right) + \sum_{\alpha = p+1}^n \left(
n^k_{\alpha} + \frac{1}{2} \right) \nu^k_{\alpha} \right] \hbar +
O(\hbar^2),
\end{equation}
with $N_k \in \mathbb{Z}$, $n^k_{\alpha} \in \mathbb{N}$ and
$n^k_{\alpha} \ll |N_k|$, are sufficient conditions on
$f_1,\ldots,f_p$ for the existence of a solution to the Schr\"odinger
equations \eqref{Schrodinger's equation h^2 2}.
\end{theorem}

\begin{figure}[h]
\begin{center}
\psfrag{p1}{\tiny $p_1$} \psfrag{x1}{\tiny $x_1$}
\psfrag{p2}{\tiny $p_2$} \psfrag{x2}{\tiny $\omega_2 x_2$}
\psfrag{T1}{\tiny $T_1 = \frac{2 \pi}{\omega_1}$}
\psfrag{T2}{\tiny $T_2 = \frac{2 \pi}{\omega_2}$} \psfrag{n}{\tiny
$\nu$}
\includegraphics[height=30mm]{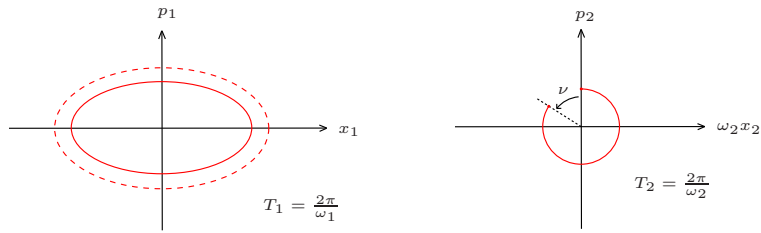}
\end{center}
\caption{Perturbed trajectory of energy $H = H_1 + H_2 = E$.}
\label{HOstab2}
\end{figure}

To illustrate the use of the modified Bohr-Sommerfeld conditions
\eqref{BS5} for an isolated orbit let use go back to the case of
the two-dimensional harmonic oscillators \eqref{2d HO}. This
system is obviously integrable and the exact spectrum of $H$ is
\begin{equation*}
E_{n_1,n_2} = \left( n_1 + \frac{1}{2} \right) \hbar \omega_1 +
\left( n_2 + \frac{1}{2} \right) \hbar \omega_2.
\end{equation*}
However, suppose for the sake of argument that we can only solve
classically for the Hamiltonian $H_1$ and wish to obtain the
spectrum of $H = H_1 + H_2$ by perturbation as describe above.
Then consider a particular motion of the Hamiltonian $H_1$ of
total energy $H_1 = E$, through the point $(p_1,x_1,p_2,x_2) =
(p_0,0,0,0)$ say, see Figure \ref{HOstab1}. This defines a
1-parameter family of periodic orbits parametrised by their energy
$H = H_1 = E$. It is clear that the $(p_2,x_2)$-plane gives a
Poincar\'e section of the orbit through the point $(p_0,0,0,0)$
since all orbits of $H_1$ have the same period $T_1 = \frac{2
\pi}{\omega_1}$. The prescription for determining the stability
angles of this orbit is to consider small perturbations around it
within the same energy level $H = E$. If the periods of the two
harmonic oscillators are different, $T_1 \neq T_2$, then after a
length of time $T_1$, the motion in the $(p_2,x_2)$-plane does not
close and there is a deficit angle of $\nu = \omega_2 \cdot T_1$,
see Figure \ref{HOstab2}.
The tower of energy levels corresponding to the periodic motion in
Figure \ref{HOstab1} is therefore given by the Bohr-Sommerfeld
condition \eqref{BS5} which reads
\begin{equation*}
I_1 = \left[ \left(n_1 + \frac{1}{2}\right) + \left( n_2 +
\frac{1}{2} \right) \frac{\nu}{2 \pi} \right] \hbar + O(\hbar^2)
\end{equation*}
and hence $E_{n_1,n_2} = \omega_1 \cdot I_1 = \left(n_1 +
\frac{1}{2}\right) \hbar \omega_1 + \left( n_2 + \frac{1}{2}
\right) \hbar \omega_2 + O(\hbar^2)$ so that the Bohr-Sommerfeld
condition is actually exact to first order in $\hbar$ on the
harmonic oscillator.

%% file: Strings_on_S3.tex
\newpage

\chapter{Strings on $\mathbb{R} \times S^3$} \label{chapter: strings on RxS3}

In this chapter we start by presenting two equivalent ways of
modelling bosonic strings moving on $\mathbb{R} \times S^3$. One
can either view $S^3$ as embedded in $\mathbb{R}^4$ and describe
the string by a $\sigma$-model action, or view $S^3$ as the group
manifold $SU(2)$ and describe the string in terms of a principal
chiral model action. We subsequently only study the latter in
great detail. It has a number of gauge symmetries which are
unphysical and thus it is desirable to fix these in order to be
left only with the physical degrees of freedom. At the end of the
day the gauge fixed string is described by a principal chiral
model action with flat metric subject to an added constraint.


\section{Action} \label{section: PCM}


\subsection*{$\sigma$-model on $\mathbb{R} \times S^3$}

Consider a bosonic string moving on $\mathbb{R} \times S^3$, where
the factor $\mathbb{R}$ corresponds to time. This string is
described by the embedding of a two dimensional worldsheet $W$
into the target manifold $\mathbb{R} \times S^3$. If we think of
$S^3$ as the unit sphere in $\mathbb{R}^4$ then the configuration
of such a string is specified by a field $X_0$ describing the
embedding into $\mathbb{R}$ and fields $X_i, i = 1, \ldots, 4$
subject to the constraint $\sum_i X_i^2 = 1$ describing the
embedding into $S^3$. To fix the metric conventions, we choose the
signatures $(- \, +)$ on the worldsheet and $(- \, + \, + \, + \,
+)$ on $\mathbb{R} \times S^3$.

The action for such a string is given by
\begin{equation} \label{sigma-model on RxS3}
S = - \frac{\sqrt{\lambda}}{4 \pi} \int d \sigma d \tau \left[
\sqrt{-\gamma} \gamma^{\alpha \beta} \left( \sum_{i=1}^4
\partial_{\alpha} X_i \partial_{\beta} X_i - \partial_{\alpha} X_0
\partial_{\beta} X_0 \right) + \Lambda \left(\sum_{j = 1}^4 X_j^2 - 1
\right) \right].
\end{equation}
Here $\gamma_{\alpha \beta}$ is the worldsheet metric, $\gamma =
\det (\gamma_{\alpha \beta})$ and $\Lambda$ is a Lagrange
multiplier constraining the string to the unit sphere $S^3 \subset
\mathbb{R}^4$. The equations of motion for the various fields are
\begin{subequations}
\begin{align}
\label{Xi eq} X_i : \qquad &\partial_{\alpha} \partial^{\alpha}
X_i -
\Lambda X_i = 0,\\
\label{X0 eq1} X_0 : \qquad &\partial_{\alpha} \partial^{\alpha} X_0 = 0,\\
\gamma_{\alpha \beta} : \qquad &T^{\alpha \beta} = 0,\\
\label{Lambda eq} \Lambda : \qquad &\sum_j X_j^2 = 1,
\end{align}
\end{subequations}
where
\begin{equation} \label{energy-momentum tensor}
T^{\alpha \beta} \equiv \frac{\partial \mathcal{L}}{\partial
\gamma_{\alpha \beta}} = -\frac{\sqrt{\lambda}}{4 \pi}
\sqrt{-\gamma} \left( G^{\alpha \beta} - \frac{1}{2}
\gamma^{\alpha \beta} \gamma_{\rho \sigma} G^{\rho \sigma} \right)
\end{equation}
is the energy-momentum tensor and $G_{\alpha \beta} \equiv
\sum_{i=1}^4 \partial_{\alpha} X_i \partial_{\beta} X_i -
\partial_{\alpha} X_0 \partial_{\beta} X_0$ is the pullback of the
target space metric to the worldsheet. Multiplying equation
\eqref{Xi eq} by $X_i$, summing over $i = 1, \ldots, 4$ and making
use of \eqref{Lambda eq} yields the value of the Lagrange
multiplier $\Lambda = - \sum_j \partial_{\alpha} X_j
\partial^{\alpha} X_j$. Substituting this value of the Lagrange
multiplier back into \eqref{Xi eq} gives rise to a set of
nonlinear differential equations for the fields $X_i$
\begin{equation} \label{nonlinear EOM}
\partial_{\alpha} \partial^{\alpha} X_i + \left( \sum_j
\partial_{\alpha} X_j \partial^{\alpha} X_j \right) X_i = 0.
\end{equation}
The nonlinearity is a consequence of the curvature of the
background $S^3$ on which the string is moving. Unlike the linear
equations for a string moving through flat space, the equations of
motion \eqref{nonlinear EOM} are a lot harder to solve in full
generality. Yet we will show in Chapter \ref{chapter:
integrability} that these equations are integrable which means
that they can in principle be solved.

Although the non-linear equations \eqref{nonlinear EOM} have been
solved explicitly using algebro-geometric methods
\cite{Krichever:1994} we shall work instead with a different model
for strings moving through $\mathbb{R} \times S^3$. We shall
exploit the group structure of $S^3$ and rewrite the string action
as a principal chiral model on $SU(2)$. This is mainly to follow
the literature on AdS/CFT \cite{Metsaev:1998it} in which
superstring theory on $AdS_5 \times S^5$ is described by a coset
superspace model with target space $\frac{SU(2,2 | 4)}{SO(4,1)
\times SO(5)}$. Moreover, in terms of this description the
algebro-geometric construction had already been initiated in
\cite{KMMZ, Kazakov:2004nh, Beisert:2005bm} for various subsectors
as well as for the full theory.

\subsection*{$SU(2)$ principal chiral model}

Since the sphere $S^3$ is isomorphic to the group $SU(2)$, the
motion of the bosonic string in the $S^3$ manifold can also be
formulated in terms of a field $g$ taking values in $SU(2)$ by
defining
\begin{equation} \label{matrix g}
g = \left( \begin{array}{cc} X_1 + i X_2 & X_3 + i X_4\\ -X_3 + i
X_4 & X_1 - i X_2 \end{array} \right).
\end{equation}
We immediately observe that $\det g = \sum_j X_j^2$ so that the
constraint for the string to lie on $S^3$ is solved when $g \in
SU(2)$. Furthermore, rewriting the $S^3$ part of the action
\eqref{sigma-model on RxS3} in terms of this new field one finds
\begin{equation*}
\sum_{i = 1}^4 \partial_{\alpha} X_i \partial_{\beta} X_i = -
\frac{1}{2} \tr \left( g^{-1} \partial_{\alpha} g g^{-1}
\partial_{\beta} g\right).
\end{equation*}
This is precisely the principal chiral model action for the
$SU(2)$-valued field $g$. Defining the $\mathfrak{su}(2)$-valued
worldsheet current $j_{\alpha} = - g^{-1} \partial_{\alpha} g$ we
can rewrite the $\sigma$-model action \eqref{sigma-model on RxS3}
as the following principal chiral model action
\begin{equation} \label{PCM on SU(2)}
S = \frac{\sqrt{\lambda}}{4 \pi} \int d \sigma d \tau
\sqrt{-\gamma} \gamma^{\alpha \beta} \left[ \frac{1}{2} \tr \left(
j_{\alpha} j_{\beta} \right) + \partial_{\alpha} X_0
\partial_{\beta} X_0 \right].
\end{equation}
Introducing the form notation $j = - g^{-1} dg = j_0 d\tau + j_1
d\sigma$ one can rewrite \eqref{PCM on SU(2)} more compactly as
\begin{equation} \label{PCM on SU(2) forms}
S = \frac{\sqrt{\lambda}}{4 \pi} \int \left[ \frac{1}{2} \tr
\left( j \wedge  \ast j \right) + dX_0 \wedge \ast dX_0 \right].
\end{equation}
The dependence on the worldsheet metric $\gamma_{\alpha \beta}$ is
now hidden in the Hodge $\ast$ operation. The current $j$ is
identically flat from its definition so the equations of motion
now read
\begin{subequations}
\begin{align}
\label{j eom} g : \qquad &d \ast j = 0, \quad dj - j \wedge j = 0,\\
\label{X0 eq2} X_0 : \qquad &d \ast d X_0 = 0,\\
\gamma_{\alpha \beta} : \qquad &T^{\alpha \beta} = 0,
\end{align}
\end{subequations}
where the induced metric is
\begin{equation} \label{induced metric}
G_{\alpha \beta} = - \frac{1}{2} \tr (j_{\alpha} j_{\beta}) -
\partial_{\alpha} X_0 \partial_{\beta} X_0
\end{equation}
when expressed in terms of the principal chiral model fields.

\begin{remark}
The second equation in \eqref{j eom} is the condition for the
existence of a matrix $g \in SU(2)$ such that $j = -g^{-1} dg$.
Indeed, $dg + g j = 0$ implies $dj - j \wedge j = 0$ and
conversely, if $dj - j \wedge j = 0$ then $j$ is a flat connection
so the path ordered exponential $g = P \overrightarrow{\exp}
\int^x -j$ is path independent and solves $dg + g j = 0$. Thus
\eqref{j eom} is equivalent to $d(\ast g^{-1} dg) = 0$ which in
turn is equivalent to \eqref{nonlinear EOM}.
\end{remark}

From now on we shall treat only this model of strings on
$\mathbb{R} \times S^3$.


\section{Symmetries} \label{section: symmetries}


\subsection*{Global}

The action \eqref{PCM on SU(2)} is invariant under constant shifts
in the time $X_0$. The Noether current is $\frac{\sqrt{\lambda}}{2
\pi} \sqrt{- \gamma} \partial_{\alpha} X_0$, and the corresponding
Noether charge is the space-time energy of the string
\begin{equation*}
\Delta = \frac{\sqrt{\lambda}}{2 \pi} \int_0^{2 \pi} d \sigma
\sqrt{- \gamma} \partial_0 X_0.
\end{equation*}

The action \eqref{PCM on SU(2)} also has a global $SU(2)_L \times
SU(2)_R$ symmetry
\begin{equation*}
g \mapsto U_L g U_R,
\end{equation*}
where $U_L$ and $U_R$ are constant matrices. The Noether current
corresponding to $SU(2)_R$ is the current $j = -g^{-1} dg$
introduced above whereas the Noether current for the $SU(2)_L$
symmetry is $l = - dg \, g^{-1} = g j g^{-1}$. The corresponding
Noether charges are
\begin{subequations} \label{global Noether charges}
\begin{align}
\label{Q_R def} &SU(2)_R : \quad Q_R = \frac{\sqrt{\lambda}}{4
\pi}
\int_{\gamma} \ast j, \\
\label{Q_L def} &SU(2)_L : \quad Q_L = \frac{\sqrt{\lambda}}{4
\pi} \int_{\gamma} \ast l,
\end{align}
\end{subequations}
where $\gamma$ is any curve winding once around the worldsheet,
expressing the conservation of these Noether charges,
\textit{e.g.}
\begin{equation*}
\raisebox{-14mm}{\psfrag{g1}{\footnotesize $\gamma_1$}
\psfrag{g2}{\footnotesize $\gamma_2$} \psfrag{D}{\footnotesize
$D$} \includegraphics[height=30mm]{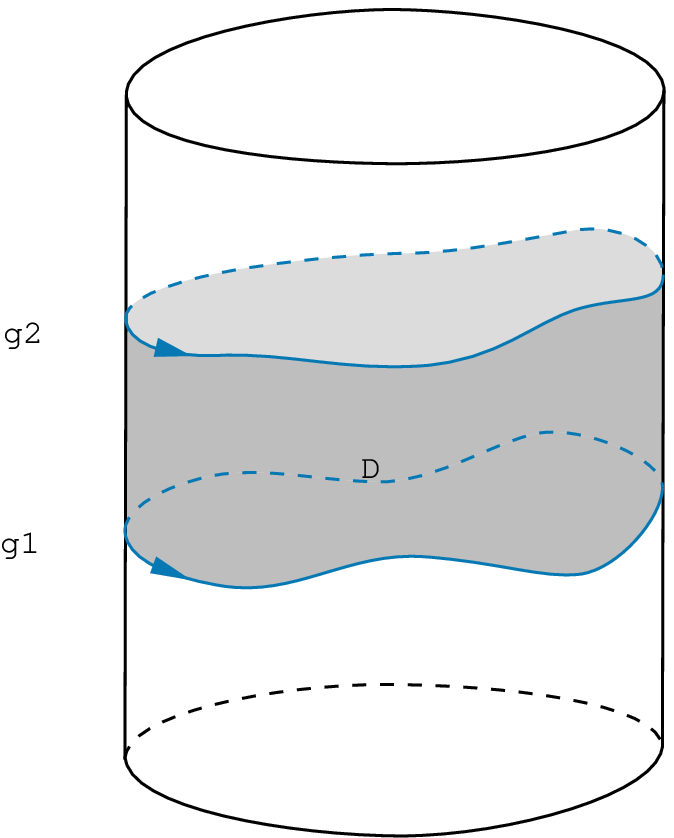}}
\qquad \qquad \int_{\gamma_2} \ast j - \int_{\gamma_1} \ast j =
\int_{\partial D} \ast j = \int_D d \ast j = 0.
\end{equation*}
Notice that the $SU(2)_R$ current $j$ which appears in the action
\eqref{PCM on SU(2) forms} is invariant under the action of
$SU(2)_L$. On the other hand the $SU(2)_R$ symmetry acts
non-trivially on the current
\begin{equation} \label{R symmetry on j}
j \mapsto U_R^{-1} j U_R.
\end{equation}

\subsection*{Local}

The string action in either of the above forms \eqref{sigma-model
on RxS3} or \eqref{PCM on SU(2)} is invariant under general
reparametrisations of the worldsheet
\begin{equation} \label{Diff inv.}
(\sigma, \tau) \mapsto (\sigma', \tau'),
\end{equation}
with the fields $X_0$ and $X_i$ (or equivalently $g$ in
\eqref{matrix g}) transforming as scalars and $\gamma_{\alpha
\beta}$ as the components of a {\tiny $\left( \!\!\!
\begin{array}{c} 0\\ 2
\end{array} \!\!\! \right)$} tensor. That is, if $\sigma^{\alpha}
\mapsto \sigma^{\alpha} + \epsilon^{\alpha}(\sigma,\tau)$ denotes
the infinitesimal version of \eqref{Diff inv.} then
\begin{align*}
\delta_{\epsilon} X_i &= \epsilon^{\alpha} \partial_{\alpha} X_i,
\quad \delta_{\epsilon} X_0 = \epsilon^{\alpha} \partial_{\alpha}
X_0,\\
\delta_{\epsilon} \gamma_{\alpha \beta} &= - \nabla_{\alpha}
\epsilon_{\beta} - \nabla_{\beta} \epsilon_{\alpha},
\end{align*}
where $\nabla_{\alpha}$ is the covariant derivative for the metric
$\gamma_{\alpha \beta}$. Also, since $\sqrt{-\gamma}$ is a scalar
density it behaves as $\delta_{\epsilon} \sqrt{-\gamma} =
\partial_{\alpha} (\epsilon^{\alpha} \sqrt{-\gamma})$ under
infinitesimal diffeomorphisms. Noting that the Lagrangian in
either \eqref{sigma-model on RxS3} or \eqref{PCM on SU(2) forms}
is a scalar density, so that $\delta_{\epsilon} \mathcal{L} =
\partial_{\alpha} (\epsilon^{\alpha} \mathcal{L})$, leads to the
following on-shell conserved current by Noether's theorem
\begin{equation} \label{gauge Noether theorem}
\nabla_{\alpha} j^{\alpha} \simeq 0 , \qquad j^{\alpha} = 2
\epsilon_{\beta} T^{\alpha \beta},
\end{equation}
where $\simeq$ indicates an on-shell equality. However, we are
dealing with a gauge transformation, since $\epsilon_{\beta}$ is
an arbitrary function of $(\sigma, \tau)$, and so expanding the
conservation equation \eqref{gauge Noether theorem} in derivatives
of $\epsilon_{\beta}$ leads to two equations, known as Noether
identities,
\begin{subequations}
\begin{align}
\label{T conserved} \epsilon_{\beta} &: \quad \nabla_{\alpha}
T^{\alpha \beta} \simeq 0,\\
\label{T vanishing} \nabla_{\alpha} \epsilon_{\beta} &: \quad
T^{\alpha \beta} \simeq 0.
\end{align}
\end{subequations}
Equation \eqref{T conserved} says that the energy-momentum tensor
$T^{\alpha \beta}$ is conserved on-shell as it should be since
$T^{\alpha \beta}$ is the Noether current for the global part of
the diffeomorphism group with $\epsilon^{\alpha} = \text{const}$.
The corresponding Noether charges are the components of the
worldsheet energy-momentum vector, generating $\sigma^{\alpha}
\mapsto \sigma^{\alpha} + \epsilon^{\alpha}$ and given by
\begin{equation} \label{energy-momentum}
P_{\alpha} = \int_0^{2 \pi} d \sigma 2 T^0_{\;\; \alpha}.
\end{equation}
However, equation \eqref{T vanishing} shows that in fact
$T^{\alpha \beta}$ itself vanishes on-shell. This we already knew
from the equation of motion for $\gamma_{\alpha \beta}$ but the
statement that the conserved charges vanish on-shell is
reminiscent of gauge theories: as we will see in the next chapter
\eqref{T vanishing} corresponds to a secondary constraint in the
Hamiltonian formalism.

The string action is also invariant under Weyl transformations of
the metric
\begin{equation} \label{Weyl}
\gamma_{\alpha \beta} \mapsto e^{\phi} \gamma_{\alpha \beta},
\end{equation}
where $\phi(\sigma, \tau)$ is an arbitrary function on the
worldsheet. This is a symmetry because the metric always enters
the action, \eqref{sigma-model on RxS3} or \eqref{PCM on SU(2)},
in the Weyl invariant form $\sqrt{- \gamma} \gamma_{\alpha
\beta}$.


\section{Gauge fixing} \label{section: gauge fixing}


The reason for wanting to fix the gauge, \textit{i.e.} the
coordinates on the worldsheet, is that all the remaining degrees
of freedom will be physical.

\subsection*{Conformal gauge}

Since the worldsheet is topologically a sphere every metric on it
is conformally equivalent. This is certainly true for Riemannian
metrics (Euclidean signature) but can also be shown \cite{Brink} in
the case at hand of a pseudo-Riemannian metric (Lorentzian
signature). In other words, it is possible to choose coordinates
$(\sigma, \tau)$ on the worldsheet with respect to which the metric
$\gamma_{\alpha \beta}$ assumes the conformal form
\begin{equation} \label{conformal gauge}
\gamma_{\alpha \beta} = e^{\phi} \eta_{\alpha \beta} = e^{\phi}
\left(
\begin{array}{cc} -1 & 0 \\ 0 & 1 \end{array} \right).
\end{equation}
We shall henceforth always assume such a choice of coordinates,
known as \dub{conformal gauge}. The presence of the prefactor
$e^{\phi}$ is irrelevant, at least classically, because of Weyl
invariance. In this gauge the worldsheet energy and momentum
defined in \eqref{energy-momentum} read
\begin{subequations} \label{energy and momentum}
\begin{align}
P_0 &= -\frac{\sqrt{\lambda}}{4 \pi} \int_0^{2 \pi} d \sigma
\left[ \frac{1}{2} \tr(j_0^2 + j_1^2) + (\partial_0 X_0)^2 +
(\partial_1 X_0)^2 \right],\\
P_1 &= -\frac{\sqrt{\lambda}}{2 \pi} \int_0^{2 \pi} d \sigma
\left[ \frac{1}{2} \tr(j_0 j_1) + \partial_0 X_0 \partial_1 X_0
\right].
\end{align}
\end{subequations}
Note also for later that in conformal gauge, the global Noether
charges $Q_R, Q_L$ defined in \eqref{global Noether charges} read
\begin{equation} \label{global Noether charges CG}
Q_R = - \frac{\sqrt{\lambda}}{4 \pi} \int_0^{2 \pi} d\sigma j_0,
\qquad Q_L = - \frac{\sqrt{\lambda}}{4 \pi} \int_0^{2 \pi} d\sigma
l_0.
\end{equation}

The gauge choice \eqref{conformal gauge} however does not
completely fix the gauge. Indeed, any coordinate transformation
$(\sigma, \tau) \mapsto (\sigma', \tau')$ that changes the metric
$\gamma_{\alpha \beta}$ only up to an overall factor
\begin{equation} \label{conformal transf}
\gamma_{\alpha \beta} \mapsto \gamma'_{\alpha \beta} = \Lambda
\gamma_{\alpha \beta},
\end{equation}
will not affect the gauge choice \eqref{conformal gauge} since the
factor $e^{\phi}$ is arbitrary and can absorb $\Lambda$. Such a
coordinate transformation is known as a \dub{conformal
transformation}. The group of conformal transformations is an
infinite dimensional subgroup of the diffeomorphism group
\eqref{Diff inv.} which possesses the following infinite number of
Noether currents
\begin{equation*}
j^{\alpha} = 2 \epsilon_{\beta} T^{\alpha \beta},
\end{equation*}
where $\epsilon_{\alpha}$ is a \dub{conformal Killing vector},
\textit{i.e.} it satisfies the infinitesimal form of
\eqref{conformal transf} which is $\nabla_{\alpha}
\epsilon_{\beta} + \nabla_{\beta} \epsilon_{\alpha} = \lambda
\gamma_{\alpha \beta}$. In particular, for $\epsilon_{\alpha} =
\text{const}$ we recover the global diffeomorphisms of equation
\eqref{energy-momentum} which are nothing but rigid translations
in $\sigma$ and $\tau$. Thus even after imposing the condition
\eqref{conformal gauge} there remains a residual gauge invariance
in the form of the conformal group, which we will have to fix by
imposing further gauge conditions. But in order to determine these
extra conditions we have to first understand what the general
conformal transformation looks like, which requires solving
\eqref{conformal transf}.

\subsection*{Static gauge}

A more suitable coordinate system for discussing conformal
transformations of the metric \eqref{conformal transf} are
light-cone coordinates: given a coordinate system $(\sigma, \tau)$
for which the metric takes the form \eqref{conformal gauge},
\textit{i.e.} $ds^2 = e^{\phi} (-d \tau^2 + d \sigma^2)$, we
define $\sigma^{\pm} = \frac{1}{2}(\tau \pm \sigma)$. In these
coordinate the metric becomes
\begin{equation} \label{flat metric LC}
ds^2 = - 4 e^{\phi} d\sigma^+ d\sigma^-.
\end{equation}
It follows easily that the only way for a transformation
$\sigma^{\pm} \mapsto \sigma'^{\pm}$ to be conformal is to have
$\frac{\partial \sigma'^+}{\partial \sigma^-} = \frac{\partial
\sigma'^-}{\partial \sigma^+} = 0$, up to the trivial permutation
$\sigma'^+ \leftrightarrow \sigma'^-$. In other words,
\begin{equation} \label{residual gauge symmetry 0}
\sigma^+ \mapsto \sigma'^+ = f^+(\sigma^+), \qquad \sigma^-
\mapsto \sigma'^- = f^-(\sigma^-),
\end{equation}
where $f^{\pm}$ are arbitrary invertible functions. We will now
fix this residual gauge symmetry by imposing a second condition on
top of \eqref{conformal gauge}.

Note that the equation of motion \eqref{X0 eq2} for $X_0$ is
decoupled from the other fields and hence can be solved
separately. Written in terms of light-cone coordinates it reads
$\partial_+ \partial_- X_0 = 0$ and has the general solution
\begin{equation*}
X^{\text{sol}}_0(\sigma, \tau) = X_0^+(\sigma^+) +
X_0^-(\sigma^-).
\end{equation*}
One would now like to apply a residual gauge transformation
$(\sigma,\tau) \mapsto (\sigma', \tau')$ given by \eqref{residual
gauge symmetry 0} with $\kappa f^{\pm} = X_0^{\pm}$ ($\kappa \neq
0$) in order to bring $X^{\text{sol}}_0$ to the simpler form
\begin{equation} \label{static gauge}
X^{\text{sol}}_0(\sigma', \tau') = \kappa \sigma'^+ + \kappa
\sigma'^- = \kappa \tau'.
\end{equation}
The coefficient $\kappa$ is not arbitrary but proportional the
space-time energy $\Delta$ of the string since $\Delta =
\frac{\sqrt{\lambda}}{2 \pi} \int_0^{2 \pi} d\sigma
\dot{X}^{\text{sol}}_0(\sigma,\tau) = \kappa \sqrt{\lambda}$ using
the coordinate system $(\sigma', \tau')$. The condition
\eqref{static gauge} is called the \dub{static gauge} condition.

\begin{remark}
Note that since the $f^{\pm}$ in \eqref{residual gauge symmetry 0}
must be invertible functions, the transformation to \eqref{static
gauge} just described is possible only if the $X_0^{\pm}$ are
themselves invertible \cite{Brink}. We will assume from now on
that this is the case for the solution $X^{\text{sol}}_0$.
\end{remark}

In static gauge the worldsheet energy and momentum \eqref{energy
and momentum} simplify further to
\begin{subequations} \label{energy and momentum static}
\begin{align}
P_0 &= -\frac{\sqrt{\lambda}}{4 \pi} \int_0^{2 \pi} d \sigma
\frac{1}{2} \tr(j_0^2 + j_1^2) - \frac{\sqrt{\lambda} \kappa^2}{2},\\
P_1 &= -\frac{\sqrt{\lambda}}{4 \pi} \int_0^{2 \pi} d \sigma
\tr(j_0 j_1).
\end{align}
\end{subequations}

\subsection*{Symplectic reduction}

By definition of the static gauge condition \eqref{static gauge}
the $\tau$ coordinate is now completely fixed. But this still
leaves the possibility of performing conformal transformations
\eqref{residual gauge symmetry} that fix $\tau$. It is easy to
show that the only such transformations are rigid translations in
$\sigma$
\begin{equation} \label{rigid residual symmetry}
\tau' \mapsto \tilde{\tau}, \qquad \sigma' \mapsto \tilde{\sigma}
+ b, \quad b \in \mathbb{R}.
\end{equation}
This is generated by the worldsheet momentum $P_1$. Thus, working
in conformal static gauge, \eqref{conformal gauge} together with
\eqref{static gauge}, the original gauge invariance of the full
string action is completely fixed except for the global
transformation \eqref{rigid residual symmetry}. We will therefore
have to make sure that physical states are invariant under this
global symmetry. This can be achieved by symplectic reduction onto
the level set $P_1 = 0$.


\section{Virasoro constraints} \label{section: Vir constraints}


It is important to note that even after fixing the metric using
conformal gauge \eqref{conformal gauge}, the equations of motion
for the metric $\gamma_{\alpha \beta}$ still carry nontrivial
information and must therefore be retained. They become
constraints on the other dynamical fields known as the
\dub{Virasoro constraints}.

When working in conformal gauge a lot of expressions simplify if
we use light-cone coordinates in which the metric \eqref{flat
metric LC} is off-diagonal $\gamma_{\pm \pm} = 0, \gamma_{+-} =
\gamma_{-+} = -2$. For instance, the tracelessness of the energy
momentum tensor $\gamma_{\alpha \beta} T^{\alpha \beta} = 0$
implies in light-cone coordinates that $T^{+-} = T^{-+} = 0$.
Moreover, from its definition \eqref{energy-momentum tensor} the
remaining components of $T^{\alpha \beta}$ take on the simple form
\begin{equation} \label{LC energy-momentum tensor}
T_{\pm \pm} = \frac{\sqrt{\lambda}}{4 \pi} \left( \frac{1}{2} \tr
j^2_{\pm} + (\partial_{\pm} X_0)^2 \right),
\end{equation}
where $j_{\pm} = j_0 \pm j_1$ are the components of the current
$j$ in light-cone coordinates.

In static gauge, since $X_0$ has been used to specify the
worldsheet $\tau$ coordinate, only the current $j$ remains and the
Virasoro constraints simplify to
\begin{equation} \label{Vir 3}
\frac{1}{2} \tr j^2_{\pm} = - \kappa^2.
\end{equation}
In fact, since the static gauge condition has fixed all the
residual gauge invariance except for the rigid
$\sigma$-translation of equation \eqref{rigid residual symmetry},
and since the latter is generated by the worldsheet momentum
$P_1$, it is convenient to postpone imposing the condition $P_1 =
0$, which is one of the Virasoro constraints \eqref{Vir 3}. Thus
we split the Virasoro constraints \eqref{Vir 3} into two parts.
The first set of constraints read,
\begin{equation} \label{Vir without P=0}
\frac{1}{2} \tr j^2_{\pm} = - \kappa_{\pm}^2.
\end{equation}
where $\kappa_{\pm}$ are two independent constants. After imposing
\eqref{Vir without P=0} the worldsheet energy and momentum
\eqref{energy and momentum static} become $P_0 = \mathcal{E} -
\sqrt{\lambda} \kappa^2 / 2$ and $P_1 = \mathcal{P}$ respectively,
where
\begin{equation} \label{Vir E and P}
\mathcal{E} = \frac{\sqrt{\lambda}}{4} (\kappa_+^2 + \kappa_-^2),
\qquad \mathcal{P} = \frac{\sqrt{\lambda}}{4} (\kappa_+^2 -
\kappa_-^2)
\end{equation}
are the energy and momentum of the principal chiral field $j$. The
remaining Virasoro constraint is the vanishing of the worldsheet
momentum $\mathcal{P} = 0$ which corresponds to setting $\kappa_+
= \kappa_- = \kappa$. When imposing this last Virasoro constraint
one must also identify string configurations related by rigid
$\sigma$-translations, which amounts to performing the symplectic
reduction of the previous subsection. We note finally that the
vanishing of the worldsheet energy $P_0 = 0$ gives the string
mass-shell condition, relating the energy of the principal chiral
model $\mathcal{E}$ to the space time energy
\begin{equation} \label{ws vs st energy}
\mathcal{E} = \frac{\sqrt{\lambda}}{2} \kappa^2 =
\frac{\Delta^2}{2 \sqrt{\lambda}}.
\end{equation}

%% file: Hamiltonian.tex
\newpage

\chapter{Hamiltonian formalism} \label{chapter: hamiltonian}

In the following chapter we will set up the Hamiltonian formalism
for the action \eqref{PCM on SU(2)}. The gauge invariance of the
string is generated by some primary and secondary first-class
constraints. The primary ones $p^{\alpha \beta} \approx 0$ are
completely fixed by imposing conformal gauge in section
\ref{section: fixing conformal gauge} whereas the secondary ones
$T^{\alpha \beta} \approx 0$ are almost entirely fixed using
static gauge in section \ref{section: fixing static gauge}. The
rigid $\sigma$-translation generated by the constraint $P_1
\approx 0$ remains and has to be fixed by a final symplectic
reduction. The complete procedure for gauge fixing the string is
summarised in the following diagram,
\begin{diagram}
p^{\alpha \beta} \approx 0, T^{\alpha \beta} \approx 0\\
\dTo^{\text{Conformal gauge}}_{\gamma_{\alpha \beta} \approx
  \eta_{\alpha \beta}}\\
T^{\alpha \beta} \approx 0\\
\dTo^{\text{Static gauge}}_{X_0 - \frac{p_0}{\sqrt{\lambda}}\tau
  \approx \pi^0 - \frac{p_0}{2 \pi} \approx 0}\\
P_1 \approx 0\\
\dTo^{\text{Symplectic reduction}}\\
\text{Physical d.o.f.}
\end{diagram}
As a result of fixing the gauge we must replace the Poisson
bracket by a Dirac bracket which we introduce in section
\ref{section: fixing static gauge}. We also explain how the
reduced dynamics for the physical degrees of freedom arises from
the time-dependence of the static gauge condition.


\section{$SU(2)$ principal chiral model} \label{section: hamiltonian}


To set up the Hamiltonian formalism for the action \eqref{PCM on
SU(2)} we start by identifying the canonical variables. For this
we need to choose the variables we shall take as our canonical
coordinates. Let us first choose a particular basis $t_a$ of the
Lie algebra $\mathfrak{su}(2)$ with structure constants $f_{abc}$
and normalised such that
\begin{equation*}
[t_a, t_b] = f_{ab}^{\;\;\;c} t_c, \quad \tr(t_a t_b) \equiv k_{ab} =
- \delta_{ab}, \qquad a,b,c = 1,2,3.
\end{equation*}
Note that $\tr (A B) = A_a B^a = - A_a B_a$ in terms of components $A
= A^a t_a, B = B^a t_a$ with respect to this basis $t_a$. For
concreteness we set $t_a = \frac{i}{\sqrt{2}} \sigma_a$ where
$\sigma_a$ are the Pauli matrices.

Following \cite{Faddeev}, we take the components of the spatial
part of the current $j$ as the first set of canonical variables
$q^a(\sigma) = j_1^a(\sigma)$. We choose the other canonical
coordinates to be the target-space time coordinate $q^0(\sigma) =
X_0(\sigma)$ and the components of the worldsheet metric
$\gamma_{\alpha \beta}$. The components $j_0^a(\sigma)$ are
related to the time derivative of $q^a$ and hence are not
independent coordinates. Indeed, by the flatness of the current
$j$ we have $\partial_0 j_1 - \partial_1 j_0 = [j_0, j_1]$ so that
\begin{equation} \label{q^a dot}
\dot{q}^a = \partial_{\sigma} j_0^a - [j_1, j_0]^a = \nabla_1
j_0^a,
\end{equation}
where $\nabla_1$ is the covariant derivative for the connection
$j_1 = j_1^a t_a$.

We now determine the conjugate momenta. The absence of derivatives
of the worldsheet metric in the action immediately implies that
the conjugate momentum to $\gamma_{\alpha \beta}$ vanishes,
\begin{equation} \label{p^ab = 0}
p^{\alpha \beta} = \frac{\delta S}{\delta \dot{\gamma}_{\alpha
\beta}} \equiv 0.
\end{equation}
This is a primary constraint on the Hamiltonian system which will
be partly responsible for the gauge invariance. The conjugate
momentum of the time coordinate $X_0$ is given by
\begin{equation} \label{pi0}
\pi_0(\sigma) = \frac{\delta S}{\delta \dot{X}_0(\sigma)} =
\frac{\sqrt{\lambda}}{2 \pi} \sqrt{- \gamma} \gamma^{0 \alpha}
\partial_{\alpha} X_0(\sigma).
\end{equation}
Finally, the computation of the conjugate momenta of the
coordinates $q^a$ is a little bit more involved. One has
\begin{align*}
\pi_a(\sigma) &= \frac{\delta S}{\delta \dot{q}^a(\sigma)} =
\frac{\sqrt{\lambda}}{8 \pi} \int d\sigma' d\tau' \sqrt{- \gamma}
\gamma^{\alpha \beta} \frac{\delta j_{\alpha}^b(\sigma') j_{\beta
b}(\sigma')}{\delta \dot{q}^a(\sigma)},\\
&= \frac{\sqrt{\lambda}}{4 \pi} \int d\sigma' d\tau' \sqrt{-
\gamma} \gamma^{0 \alpha} j_{\alpha b}(\sigma') \frac{\delta
j_0^b(\sigma')}{\delta \dot{q}^a(\sigma)},
\end{align*}
and using equation \eqref{q^a dot} one can write $\frac{\delta
j_0^b(\sigma')}{\delta \dot{q}^a(\sigma)} = \nabla_1^{-1}\left(
\delta_a^b \delta(\sigma - \sigma') \delta(\tau - \tau') \right)$.
Then using this relation and integrating by parts we end up with
\begin{equation} \label{D pi = j}
\pi_a(\sigma) = - \frac{\sqrt{\lambda}}{4 \pi} \nabla_1^{-1} \left(
\sqrt{- \gamma} \gamma^{0 \alpha} j_{\alpha a}(\sigma) \right).
\end{equation}
In other words, $\nabla_1 \pi^a(\sigma) = - \frac{\sqrt{\lambda}}{4
\pi} \sqrt{- \gamma} \gamma^{0 \alpha} j_{\alpha}^a(\sigma)$ for
$a = 1,2,3$.

\subsection*{Hamiltonian}

We now have enough information to define the Hamiltonian
corresponding to the action \eqref{PCM on SU(2)}. Introducing
capital letter indices $A = 0,1,2,3$, it is given by
\begin{equation*}
H_0 = \int d\sigma \left( \pi_A(\sigma) \partial_0 q^A(\sigma) -
\mathcal{L} \right),
\end{equation*}
where $\mathcal{L} = \frac{\sqrt{\lambda}}{4 \pi} \sqrt{- \gamma}
\gamma^{\alpha \beta} \left( \frac{1}{2} \tr(j_{\alpha} j_{\beta})
+
\partial_{\alpha} X_0 \partial_{\beta} X_0\right)$ is the Lagrangian.
After a little algebra it can be simplified to
\begin{multline*}
H_0 = \int d\sigma \Bigg[ -\frac{\sqrt{- \gamma}}{\gamma_{11}}
\left( \frac{2 \pi}{\sqrt{\lambda}} \nabla_1 \pi^a \nabla_1
\pi_a + \frac{\sqrt{\lambda}}{8 \pi} j_1^a j_{1 a} +
\frac{\pi}{\sqrt{\lambda}} (\pi^0)^2 + \frac{\sqrt{\lambda}}{4
\pi} (\partial_1 X_0)^2  \right)\\
+ \frac{\gamma_{01}}{\gamma_{11}} \left( - \nabla_1 \pi_a j_1^a +
\pi^0 \partial_1 X_0 \right) \Bigg].
\end{multline*}
It is convenient to define a different parametrisation of the
metric $\gamma_{\alpha \beta}$ as
\begin{equation} \label{new metric variables}
\lambda^{\pm} = \frac{- \sqrt{- \gamma} \pm
\gamma_{01}}{\gamma_{11}}, \qquad \xi = \ln \gamma_{11}.
\end{equation}
We see immediately that the first two parameters $\lambda^{\pm}$
are invariant under Weyl transformations \eqref{Weyl} whereas
$\xi$ transforms as $\xi \mapsto \xi + \phi$. Just as the action
\eqref{PCM on SU(2)} was Weyl invariant since it depended only on
the Weyl invariant combination $\sqrt{- \gamma} \gamma^{\alpha
\beta}$, the Weyl invariance of the Hamiltonian is explicit from
its sole dependence on the Weyl invariant variables
$\lambda^{\pm}$. Indeed, if we define the worldsheet energy and
momentum densities as
\begin{subequations} \label{H_0 and H_1}
\begin{align}
\mathcal{H}_0 &= - \frac{2 \pi}{\sqrt{\lambda}} \nabla_1 \pi^a
\nabla_1 \pi_a - \frac{\sqrt{\lambda}}{8 \pi} j_1^a j_{1 a} -
\frac{\pi}{\sqrt{\lambda}} (\pi^0)^2 - \frac{\sqrt{\lambda}}{4
\pi} (\partial_1 X_0)^2,\\
\mathcal{H}_1 &= - \nabla_1 \pi_a j_1^a + \pi^0 \partial_1 X_0,
\end{align}
\end{subequations}
and define the combinations $T_{\pm} = - \mathcal{H}_0 \pm
\mathcal{H}_1$ given explicitly by
\begin{equation} \label{Vir 4}
T_{\pm} = \frac{\sqrt{\lambda}}{4 \pi} \left[ \frac{1}{2} \tr
\left( \frac{4 \pi}{\sqrt{\lambda}} \nabla_1 \pi \mp j_1 \right)^2
+ \left( \frac{2 \pi}{\sqrt{\lambda}} \pi^0 \pm \partial_1 X_0
\right)^2 \right],
\end{equation}
then the Hamiltonian reads
\begin{equation} \label{Hamiltonian H_0}
H_0 = \int d\sigma \left( \frac{\lambda^+}{2} T_+ +
\frac{\lambda^-}{2} T_- \right).
\end{equation}

\subsection*{Poisson brackets}

The full set of canonical Poisson brackets between the generalised
coordinates and their conjugate momenta are,
\begin{equation} \label{PB def}
\begin{split}
&\left\{ q^A(\sigma), q^B(\sigma')\right\} = \left\{
\pi_A(\sigma),
\pi_B(\sigma')\right\} = 0 \\
&\left\{ \pi_B(\sigma), q^A(\sigma')\right\} = \delta^A_B
\delta(\sigma - \sigma').
\end{split}
\end{equation}
There are also Poisson brackets between the metric variables
$\lambda^{\pm}$ and their conjugate momenta (defined later in
\eqref{p^ab = 0 bis}) but we won't be needing those at any stage.

We can derive from \eqref{PB def} the Poisson brackets between the
variables $\nabla_1 \pi_a$ and $q^b$ that appear in the
Hamiltonian, for example
\begin{subequations} \label{PB pi q}
\begin{align}
\{ \nabla_1 \pi^a(\sigma), q^b(\sigma') \} &= \left\{
q^b(\sigma'),
\partial_{\sigma} \pi^a(\sigma) \right\} - f^{adc} \left\{
q^b(\sigma'), q_d(\sigma) \pi_c(\sigma)\right\} \notag\\
&= \partial_{\sigma}\left( k^{ab} \delta(\sigma' -
\sigma)\right) - f^{adc} q_d(\sigma) \delta^b_{\;\;c}
\delta(\sigma' - \sigma) \notag\\
\label{PB pi q 1} &= f^{abc} q_c(\sigma) \delta(\sigma - \sigma')
+ k^{ab} \delta'(\sigma - \sigma').
\end{align}
Similarly we have
\begin{equation}
\label{PB pi q 2} \{ \nabla_1 \pi^a(\sigma), \nabla_1
\pi^b(\sigma') \} = f^{abc} \nabla_1 \pi_c(\sigma) \delta(\sigma -
\sigma').
\end{equation}
\end{subequations}
As for the canonical variables $\pi^0, X_0$, since the coordinate
$X_0$ only appears differentiated with respect to $\sigma$, the
following Poisson bracket is more useful
\begin{equation} \label{PB pi q 3}
\{ \pi^0(\sigma), \partial_1 X_0(\sigma') \} = - \delta'(\sigma -
\sigma').
\end{equation}
Using \eqref{PB pi q} and \eqref{PB pi q 3} one can derive the
following algebra for the variables $T_{\pm}$,
\begin{equation} \label{Virasoro algebra}
\begin{split}
\{ T_{\pm}(\sigma), T_{\pm}(\sigma') \} &= \pm
\frac{\sqrt{\lambda}}{8 \pi} \left[ T_{\pm}(\sigma) +
T_{\pm}(\sigma') \right] \delta'(\sigma
- \sigma'),\\
\{ T_+(\sigma), T_-(\sigma') \} &= 0.
\end{split}
\end{equation}

\subsection*{Constraints}

The next step in the Hamiltonian analysis is to determine the
constraints. In terms of the new variables \eqref{new metric
variables}, the vanishing of the conjugate momentum of the metric
$\gamma_{\alpha \beta}$ in \eqref{p^ab = 0} reads
\begin{equation} \label{p^ab = 0 bis}
\pi_{\pm}^{\lambda} = \frac{\delta S}{\delta \dot{\lambda}_{\pm}}
\equiv 0, \qquad \pi_{\xi} = \frac{\delta S}{\delta \dot{\xi}}
\equiv 0.
\end{equation}
These are three primary constraints of the Hamiltonian system.
According to the general theory of constrained Hamiltonian systems
\cite{Dirac, Henneaux}, one must demand that these constraints be
preserved in time under the Hamiltonian \eqref{Hamiltonian H_0},
which can lead to a further set of constraints. Indeed here we
find
\begin{equation} \label{Vir 5}
\dot{\pi}_{\pm}^{\lambda} \approx 0 \quad \Rightarrow \quad
T_{\pm} \approx 0,
\end{equation}
whereas $\dot{\pi}_{\xi} \approx 0$ and $\dot{T}_{\pm} \approx 0$
(which follows from \eqref{Virasoro algebra}) do not lead to any
further constraints. One can do away with the canonical variables
$\xi, \pi_{\xi}$ very easily: together they form a pair of
second-class constraints since by definition $\{ \xi, \pi_{\xi} \}
= 1$ but since they do not appear in any of the physical variables
(everything is Weyl invariant and $\pi_{\xi} \equiv 0$) they can
simply be discarded (formally by defining an appropriate Dirac
bracket).

The new constraints in \eqref{Vir 5} are called secondary
constraints because they follow from the equations of motion as
opposed to primary constraints which follow from the definitions
of the conjugate momenta. However, equation \eqref{Virasoro
algebra} shows that these constraints are first-class constraints
since they form a closed algebra. In fact, the constraints
\eqref{Vir 5} are simply the Virasoro constraints again. A simple
way to see this is to go back to the primary constraints but in
the form \eqref{p^ab = 0} and again determine the condition for
their preservation in time,
\begin{equation*}
0 \approx \dot{p}^{\alpha \beta} = \frac{\partial}{\partial \tau}
\left( \frac{\partial \mathcal{L}}{\partial \dot{\gamma}_{\alpha
\beta}} \right) = \frac{\partial \mathcal{L}}{\partial
\gamma_{\alpha \beta}} \equiv T^{\alpha \beta}.
\end{equation*}
In the second last equality we have made use of the Euler-Lagrange
equations of motion and the fact that the Lagrangian $\mathcal{L}$
is independent of $\partial_{\sigma} \gamma_{\alpha \beta}$,
whereas the last equality is the definition of the energy-momentum
tensor.

In the theory of constrained Hamiltonians \cite{Dirac, Henneaux}
one should always include the constraints in the Hamiltonian
itself by the method of Lagrange multipliers. Thus one replaces
the original Hamiltonian \eqref{Hamiltonian H_0} with the total
Hamiltonian
\begin{align} \label{Hamiltonian H_T}
H_T &= H_0 + \int d\sigma \left[ \rho_+ T_+ + \rho_- T_- +
\rho_+^{\lambda} \pi_+^{\lambda} + \rho_-^{\lambda}
\pi_-^{\lambda}
\right],\\
&= \int d\sigma \left[ \left( \frac{\lambda^+}{2} + \rho_+ \right)
T_+ + \left( \frac{\lambda^-}{2} + \rho_- \right) T_- +
\rho_+^{\lambda} \pi_+^{\lambda} + \rho_-^{\lambda}
\pi_-^{\lambda} \right].
\end{align}
The effect of the constraints in Hamilton's equations corresponds
to the ability to perform arbitrary gauge transformations on top
of the true dynamical evolution of the system. Notice though that
the original Hamiltonian $H_0$ in \eqref{Hamiltonian H_0} is
itself a combination of the Virasoro constraints and hence
vanishes on the constraint surface. This situation is typical of
generally covariant theories.


\section{Conformal gauge} \label{section: fixing conformal gauge}


We are now in a position to discuss conformal gauge fixing. Using
the coordinate invariance generated by $T_{\alpha \beta}$ we wish
to fix $\gamma_{\alpha \beta}$ to the flat metric $\eta_{\alpha
\beta} = \diag(-1, 1)$. This can be done in the Hamiltonian
formalism by imposing the constraint $c_{\alpha \beta} =
\gamma_{\alpha \beta} - \eta_{\alpha \beta} \approx 0$ by hand,
which in terms of the metric variables \eqref{new metric
variables} reads
\begin{equation} \label{conformal gauge Hamilton}
c_{\pm} = \lambda^{\pm} + 1 \approx 0.
\end{equation}
This gauge fixing condition is second-class with respect to the
constraints \eqref{p^ab = 0 bis} since
\begin{equation*}
\left\{ \pi^{\lambda}_{\pm}, c_{\pm} \right\} = \left\{
\pi^{\lambda}_{\pm}, \lambda_{\pm} \right\} = 1.
\end{equation*}
However, both constraints $c_{\pm} \approx \pi^{\lambda}_{\pm}
\approx 0$ commute with the Virasoro constraints \eqref{Vir 4}
since the latter doesn't have any explicit dependence on the
metric variables $\lambda_{\pm}$ nor on their conjugate momenta
$\pi^{\lambda}_{\pm}$, as can be seen in \eqref{Vir 4}. It thus
follows that the matrix of Poisson brackets $\mathcal{C}_{ab} = \{
\phi_a, \phi_b \}$ between all the constraints $\phi_a = (T_{\pm},
c_{\pm}, \pi^{\lambda}_{\pm})$ takes the following schematic form
\begin{equation*}
\begin{array}{ccc}
& \begin{array}{ccc}\!\!T & \;c & \;\;\pi^{\lambda} \end{array} & \\
\mathcal{C}_{ab} = \{ \phi_a, \phi_b \} = & \left( \begin{array}{ccc}\ast & 0 & 0\\
0 & 0 & -1\\
0 & 1 & 0
\end{array} \right) & \begin{array}{l}\!\!\!\!T\\ \!\!\!\!c\\
\!\!\!\!\pi^{\lambda} \end{array}
\end{array}
\end{equation*}
the important point being that the second class constraints
$c_{\pm} \approx \pi^{\lambda}_{\pm} \approx 0$ form an
independent block of their own in the matrix of Poisson brackets
$\mathcal{C}_{ab}$. In fact it follows that the inverse matrix has
the same property
\begin{equation*}
\begin{array}{ccc}
& \begin{array}{ccc}\!\!T & c & \; \pi^{\lambda} \end{array} & \\
\mathcal{C}^{-1}_{ab} = & \left( \begin{array}{ccc}\ast & 0 & 0\\
0 & 0 & 1\\
0 & -1 & 0
\end{array} \right) & \begin{array}{l}\!\!\!\!T\\ \!\!\!\!c\\
\!\!\!\!\pi^{\lambda} \end{array}
\end{array}
\end{equation*}
so that the constraints $c_{\pm} \approx \pi^{\lambda}_{\pm}
\approx 0$ can be dealt with by defining a Dirac bracket
\begin{align*}
\{ F, G \}^{\ast} = \{ F, G \} &- \{ F, c_+ \} \{ \pi_+^{\lambda},
G \} + \{ F, \pi_+^{\lambda} \} \{ c_+, G \}\\ &- \{ F, c_- \} \{
\pi_-^{\lambda}, G \} + \{ F, \pi_-^{\lambda} \} \{ c_-, G \}.
\end{align*}
Clearly $\{ F, G\}^{\ast} = \{ F, G\}$ whenever $\{ F, c_{\pm}\} =
\{ G, c_{\pm}\} = 0$. But this is the case for arbitrary functions
$F,G$ of the canonical variables $\lambda^{\pm}, \pi_A, q^A$,
\textit{i.e.} that do not depend on $\pi^{\lambda}_{\pm}$. After
imposing conformal gauge we will retain the notation $\{ \cdot ,
\cdot \}$ for the Dirac bracket instead of $\{ \cdot , \cdot
\}^{\ast}$ since the Poisson bracket won't be needed any longer.

One can thus impose the constraints and thereafter forget about
the metric degrees of freedom $\lambda_{\pm}, \pi^{\lambda}_{\pm}$
(\textit{i.e.} $\gamma_{\alpha \beta}, p^{\alpha \beta}$)
altogether. Therefore even in the Hamiltonian framework it is
legitimate to work in the conformal gauge right from the outset,
and set the worldsheet metric to be flat in the Hamiltonian. This
corresponds in the variables \eqref{new metric variables} to
setting \eqref{conformal gauge Hamilton}, that is $\lambda^{\pm}
\approx -1$, and the preservation of this gauge condition in time
requires that $\dot{c}_{\pm} = \dot{\lambda}^{\pm} \approx 0$
which implies $\rho_{\pm}^{\lambda} = 0$. The total Hamiltonian
\eqref{Hamiltonian H_T} then becomes
\begin{equation*}
H_T = \int d\sigma \left( \rho_+ T_+ + \rho_- T_- \right),
\end{equation*}
where we have shifted the definitions of $\rho_{\pm}$ by
$-\frac{1}{2}$.


\section{Current algebra}


From now on we shall assume that the metric is flat. In this case
the covariant derivative of the momenta variables $\nabla_1
\pi_a(\sigma)$ are related to $j_0^a$ alone, as equation \eqref{D
pi = j} in conformal gauge shows
\begin{equation} \label{D pi = j conf}
\nabla_1 \pi^a(\sigma) = \frac{\sqrt{\lambda}}{4 \pi}
j_0^a(\sigma).
\end{equation}
In fact, owing to the fact that the momenta $\pi^a$ never appear
without a covariant derivative, it is possible to rewrite every
expression in terms of the current components $j_0^a, j_1^a$ alone
rather than the canonically conjugate variables $\pi^a, j_1^a$ and
it will be convenient to do so. The Hamiltonian, given by
\eqref{H_0 and H_1}, for example reads
\begin{subequations} \label{H_0 and H_1 bis}
\begin{align}
\mathcal{H}_0 &= - \frac{\sqrt{\lambda}}{8 \pi} \left( j_0^a j_{0 a}
- j_1^a j_{1 a} \right) - \frac{\pi}{\sqrt{\lambda}} (\pi^0)^2 -
\frac{\sqrt{\lambda}}{4 \pi} (\partial_1 X_0)^2,\\
\mathcal{H}_1 &= - \frac{\sqrt{\lambda}}{4 \pi} j_{0 a} j_1^a + \pi^0
\partial_1 X_0.
\end{align}
\end{subequations}
Equivalently, the Virasoro constraints \eqref{Vir 4} now read
\begin{equation} \label{Vir 4 bis}
T_{\pm} = \frac{\sqrt{\lambda}}{4 \pi} \left[ \frac{1}{2} \tr
j_{\mp}^2 + \left( \frac{2 \pi}{\sqrt{\lambda}} \pi^0 \pm
\partial_1 X_0 \right)^2 \right] \approx 0,
\end{equation}
where $j_{\pm} = j_0 \pm j_1$ are the components of the current
$j$ in light-cone coordinates. With the substitution $\pi^0 = -
\frac{\sqrt{\lambda}}{2 \pi} \dot{X}_0$ we notice that the
variables $T_{\pm}$ are nothing but the light-cone components
$T_{\mp \mp}$ of the energy-momentum tensor \eqref{LC
energy-momentum tensor}.

Rewriting also the Poisson brackets \eqref{PB pi q} by eliminating
the three conjugate momenta $\pi_a$ in favour of the current
components $j^a_0$ we obtain,
\begin{subequations} \label{jj Poisson brackets}
\begin{align}
\label{jj PB1} \left\{ j_1^a(\sigma), j_1^b(\sigma')\right\} &= 0,\\
\label{jj PB2} \frac{\sqrt{\lambda}}{4 \pi} \left\{ j_0^a(\sigma),
j_1^b(\sigma')\right\} &= f^{abc} j_{1 c}(\sigma) \delta(\sigma -
\sigma') + k^{ab} \delta'(\sigma - \sigma'),\\
\label{jj PB3} \frac{\sqrt{\lambda}}{4 \pi} \left\{ j_0^a(\sigma),
j_0^b(\sigma')\right\} &= f^{abc} j_{0 c}(\sigma) \delta(\sigma
- \sigma').
\end{align}
\end{subequations}
The key feature to note about these fundamental brackets is the
presence of the derivative of a delta function $\delta'(\sigma -
\sigma')$ on the right hand side of \eqref{jj PB2}. Because of
this term the brackets \eqref{jj Poisson brackets} are usually
described as \dub{non-ultralocal}. As we will see in the next
chapter, the non-ultra local term will be the main source of
problems in proving integrability of string theory on $\mathbb{R}
\times S^3$, giving rise to ambiguities which will have to be
dealt with properly.

At this stage however there is no apparent difficulty in dealing
with the brackets \eqref{jj Poisson brackets}. For example, one
can use them to show that the $SU(2)_R$ symmetry is generated by
the Noether charge $Q_R$ defined in \eqref{global Noether
charges}. Indeed, we find from the last two brackets \eqref{jj
PB2}, \eqref{jj PB3} that the Noether charge $Q_R$ acts on the
$SU(2)_R$ current $j$ as expected
\begin{equation} \label{SU(2)_R action}
\left\{ \epsilon \cdot Q_R, j \right\} = \left[ \epsilon, j
\right] = \delta_{\epsilon} j,
\end{equation}
where $\epsilon = \epsilon^a t_a \in \mathfrak{su}(2)$ is
infinitesimal, $\epsilon \cdot Q_R = \tr (\epsilon Q_R) = \epsilon_a
Q_R^a$ and $Q_R$ is given in conformal gauge by \eqref{global Noether
charges CG}. Moreover, the brackets \eqref{jj Poisson brackets}
correctly leads to the Hamiltonian version of the equations of motion
\eqref{j eom}, namely \begin{subequations} \label{j eom Hamilton}
\begin{align}
\label{P0,j0} \{ P_0, j_0\} &= \partial_1 j_1,\\
\label{P0,j1} \{ P_0, j_1\} &= \partial_1 j_0 + [j_0,j_1],\\
\label{P1,j} \{ P_1, j_{\alpha}\} &= \partial_1 j_{\alpha}, \quad
\alpha = 0,1.
\end{align}
\end{subequations}
where $P_{\alpha} = \int d\sigma \mathcal{H}_{\alpha}$ is the
worldsheet energy-momentum vector and $\mathcal{H}_{\alpha}$ are
given in \eqref{H_0 and H_1 bis}. If we interpret $P_0$ as
generating the $\tau$-flow on phase-space, \textit{i.e.} $\{ P_0,
j_{\alpha} \} =
\partial_0 j_{\alpha}$, then equations \eqref{P0,j0} and \eqref{P0,j1}
are equivalent to $\partial_0 j_0 = \partial_1 j_1$ and
$\partial_0 j_1 - \partial_1 j_0 = [j_0,j_1]$ respectively, which
are the equations of motion \eqref{j eom} for $j$ in components.


\section{Static gauge} \label{section: fixing static gauge}


As already discussed in section \ref{section: gauge fixing}, the
constraint \eqref{conformal gauge Hamilton} by itself isn't
sufficient to fix the gauge invariance since the group of
conformal transformations that leave the metric $\eta_{\alpha
\beta}$ invariant up to an overall factor remains as the residual
gauge group. We therefore have to impose further gauge fixing
conditions.

The static gauge condition was defined by the single equation
\eqref{static gauge} for the general solution
$X_0^{\text{sol}}(\sigma, \tau)$ of the field $X_0$. However, at
any given time $\tau$, a solution $X_0^{\text{sol}}$ not only
determines the configuration of the field $X_0(\sigma, \tau) =
X_0^{\text{sol}}(\sigma, \tau)$ but also its momentum through the
defining formula \eqref{pi0} which in conformal gauge reads
$\pi^0(\sigma, \tau) = - \frac{\sqrt{\lambda}}{2 \pi} \partial_0
X_0^{\text{sol}}(\sigma, \tau)$. Therefore in the Hamiltonian
formalism the static gauge condition really consists of two
constraints,
\begin{equation} \label{static gauge hamilton}
X_0 + \frac{p_0}{\sqrt{\lambda}} \tau \approx 0, \quad \pi^0 -
\frac{p_0}{2 \pi} \approx 0.
\end{equation}
As before, the constant of proportionality, which here we denote
$p_0$ since it is the zero-mode of the momentum $\pi^0$, is
constrained by the space time energy $\Delta$ of the string since
\begin{equation*}
p_0 = \int_0^{2 \pi} d \sigma \pi^0(\sigma,\tau) = -
\frac{\sqrt{\lambda}}{2 \pi} \int_0^{2 \pi} d \sigma
\dot{X}_0(\sigma,\tau) = - \Delta.
\end{equation*}

In section \ref{section: fixing conformal gauge} we imposed
conformal gauge $\gamma_{\alpha \beta} = \eta_{\alpha \beta}$
which had the effect of fixing the gauge invariance generated by
the primary constraints $p^{\alpha \beta}$. But there are also
secondary constraints, the Virasoro constraints \eqref{Vir 4 bis}
which remain unfixed and generate a residual gauge invariance.
This will be fixed by imposing static gauge. Even though the
Virasoro constraints $T_{\pm}$ by themselves are first class by
equation \eqref{Virasoro algebra}, the static gauge conditions
fail to commute with these and among themselves (since $\{
\pi^0(\sigma), X_0(\sigma') \} = \delta(\sigma - \sigma') \not
\approx 0$), so that the full set of constraints becomes
second-class.

However, as discussed in section \ref{section: gauge fixing}, the
static gauge still doesn't completely fix the residual gauge
invariance since it leaves the possibility of performing a rigid
$\sigma$-translation, which is generated by the worldsheet
momentum. Thus we start by isolating this generator among the
Virasoro constraints, which we do by decomposing both the Virasoro
constraints \eqref{Vir 4 bis} and static gauge conditions
\eqref{static gauge hamilton} into Fourier modes.

\subsection*{Fourier modes}

Introduce the modes $L_n, \tilde{L}_n$ of the current part of the
$T_{\pm}$ in \eqref{Vir 4 bis}, namely $\frac{1}{2} \text{tr}
j^2_{\pm}$, by
\begin{equation} \label{Virasoro modes}
L_n = \frac{\sqrt{\lambda}}{8 \pi} \int_0^{2 \pi} e^{i n \sigma}
\frac{1}{2} \text{tr}j_+^2(\sigma) d\sigma, \qquad \tilde{L}_n =
\frac{\sqrt{\lambda}}{8 \pi} \int_0^{2 \pi} e^{- i n \sigma}
\frac{1}{2} \text{tr}j_-^2(\sigma) d\sigma.
\end{equation}
These are easily seen to satisfy the following algebra
\begin{equation} \label{Virasoro algebra: modes}
\begin{split}
\{ L_m, L_n \} &= i (m - n) L_{m + n},\\
\{ L_m, \tilde{L}_n \} &= 0,\\
\{ \tilde{L}_m, \tilde{L}_n \} &= i (m - n) \tilde{L}_{m + n}.
\end{split}
\end{equation}
which follows from the Virasoro algebra \eqref{Virasoro algebra}
for the $T_{\pm}$. Define also the modes $\alpha_n,
\tilde{\alpha}_n$ of $X_0$ and $\pi^0$ as
\begin{equation} \label{extract data: flat}
\begin{split}
&\alpha_n = \frac{\lambda^{\frac{1}{4}}}{\sqrt{2} \pi} \int_0^{2
\pi} e^{-i n \sigma} \frac{1}{2} \left( - \frac{2
\pi}{\sqrt{\lambda}} \pi^0(\sigma) -
\partial_{\sigma} X_0(\sigma) \right) d\sigma, \quad n \neq 0 \\
&\tilde{\alpha}_n = \frac{\lambda^{\frac{1}{4}}}{\sqrt{2} \pi}
\int_0^{2 \pi} e^{i n \sigma} \frac{1}{2} \left( - \frac{2
\pi}{\sqrt{\lambda}} \pi^0(\sigma) +
\partial_{\sigma} X_0(\sigma) \right) d\sigma, \quad n \neq 0 \\
&x_0 = \frac{1}{2 \pi} \int_0^{2 \pi} X_0(\sigma) d\sigma, \qquad
p_0 = \int_0^{2 \pi} \pi^0(\sigma) d\sigma.
\end{split}
\end{equation}
Their algebra easily follows from the defining bracket $\{
\pi^0(\sigma), X_0(\sigma') \} = \delta(\sigma - \sigma')$, namely
\begin{equation*}
\begin{split}
\{ \alpha_m, \alpha_n \} &= i m \delta_{m + n}, \quad \{ \alpha_m,
\tilde{\alpha}_n \} = 0,\\ \{ \tilde{\alpha}_m, \tilde{\alpha}_n
\} &= i m \delta_{m + n}, \quad \{ p_0, x_0 \} = 1.
\end{split}
\end{equation*}

In terms of these modes, the Virasoro constraints \eqref{Vir 4
bis} and static gauge fixing conditions \eqref{static gauge
hamilton} read
\begin{align*}
\text{Virasoro} : \quad &L_n \approx \tilde{L}_n \approx 0 \; (n
\neq 0), \qquad L_0
\approx \tilde{L}_0 \approx -\frac{p^2_0}{4 \sqrt{\lambda}},\\
\text{Static gauge} : \quad &\alpha_n \approx \tilde{\alpha}_n
\approx 0 \; (n \neq 0), \qquad x_0 + \frac{p_0}{\sqrt{\lambda}}
\tau \approx 0,
\end{align*}
Yet these include the generator $L_0 - \tilde{L}_0$ of rigid
translations $\sigma \rightarrow \sigma + b$. Therefore setting
aside this rigid transformation to deal with it later by
symplectic reduction, the set of relevant constraints now read
\begin{subequations} \label{Vir + Static modes}
\begin{align}
\label{Vir modes} \text{Virasoro} : \quad &L_n \approx \tilde{L}_n
\approx 0 \; (n \neq 0), \qquad \gamma_0 \equiv (L_0 +
\tilde{L}_0) + \frac{p^2_0}{2 \sqrt{\lambda}} \approx 0,\\
\label{Static modes} \text{Static gauge} : \quad &\alpha_n \approx
\tilde{\alpha}_n \approx 0 \; (n \neq 0), \qquad c_0 \equiv x_0 +
\frac{p_0}{\sqrt{\lambda}} \tau \approx 0,
\end{align}
\end{subequations}
This separation of the constraint $P_1 \approx 0$ from the
Virasoro constraints is just a rephrasing in Hamiltonian terms of
equation \eqref{Vir without P=0} in section \ref{section: gauge
fixing}. Indeed, in the present language we have $L_0 = -
\frac{\sqrt{\lambda}}{4} \kappa_+^2, \tilde{L}_0 = -
\frac{\sqrt{\lambda}}{4} \kappa_-^2$ and $p_0 = - \sqrt{\lambda}
\kappa$ so that
\begin{equation*}
P_0 = - L_0 - \tilde{L}_0 - \frac{p_0^2}{2 \sqrt{\lambda}}, \qquad
P_1 = - L_0 + \tilde{L}_0,
\end{equation*}
is equivalent to equation \eqref{Vir E and P}. The energy and
momentum of the principal chiral field are $\mathcal{E} = - L_0 -
\tilde{L}_0$ and $\mathcal{P} = -L_0 + \tilde{L}_0$ respectively.
Now although we postpone imposing the Virasoro constraint
$\mathcal{P} \approx 0$ (because there is no corresponding gauge
fixing condition in static gauge \eqref{static gauge hamilton}),
the Virasoro constraint $P_0 \approx 0$ \textit{is} imposed
alongside the static gauge fixing conditions \eqref{static gauge
hamilton}. As we saw in section \ref{section: Vir constraints}
this condition has the effect of equating the principal chiral
model energy with the space-time energy of the string,
\begin{equation} \label{E vs p_0^2}
\mathcal{E} \approx \frac{p_0^2}{2 \sqrt{\lambda}} =
\frac{\Delta^2}{2 \sqrt{\lambda}}.
\end{equation}

\subsection*{Dirac brackets}

The static gauge condition \eqref{Static modes} fixes all the
modes of $X_0, \pi^0$ except for $p_0$ which leaves the degrees of
freedom of the principal chiral fields $j$ and $p_0$. But the last
Virasoro constraint in \eqref{Vir modes} determines $p_0$ as a
function of $j$ through the combination $L_0 + \tilde{L}_0$. We
shall refer to the degrees of freedom remaining after imposing
conformal static gauge and the Virasoro constraints \eqref{Vir +
Static modes} as the reduced phase-space.
\begin{definition}
The \dub{reduced phase-space} $\mathcal{P}^{\infty}$ is
parameterised by the current $j(\sigma)$ subject to the
constraints $L_n \approx \tilde{L}_n \approx 0, n \neq 0$.
\end{definition}
\noindent The physical degrees of freedom can now be described by
a simple symplectic reduction of the reduced phase-space
$\mathcal{P}^{\infty}$ onto the level set $P_1 \approx 0$:
\begin{diagram}
P_1 \approx 0 &\rInto^{\iota} \mathcal{P}^{\infty}\\
\dTo^{\pi}\\
\text{Physical phase-space}
\end{diagram}

Since the constraints \eqref{Vir + Static modes} defining
$\mathcal{P}^{\infty}$ are second-class, fixing them requires
introducing a Dirac bracket. The matrix of Poisson brackets
$\mathcal{C}'_{ab} = \{ \chi_a, \chi_b \}$ between all the
second-class constraints $\chi_a$ in \eqref{Vir + Static modes}
takes the following schematic form \textit{weakly} (\textit{i.e.} on the
constraint surface $\chi_a \approx 0$)
\begin{equation} \label{matrix of brackets}
\begin{array}{rl}
\mathcal{C}'_{ab} = \{ \chi_a, \chi_b \} \approx \left(
\begin{array}{cccccc}
\,0\, & \,\ast\, & \,0\, & \,0\, & \,0\, & \,0\,\\
\ast & 0 & 0 & 0 & 0 & 0\\
0 & 0 & \ast & 0 & 0 & 0\\
0 & 0 & 0 & \ast & 0 & 0\\
0 & 0 & 0 & 0 & \ast & 0\\
0 & 0 & 0 & 0 & 0 & \ast
\end{array}
\right) & \begin{array}{l} \gamma_0\\ c_0\\ L_n\\ \tilde{L}_n\\
\alpha_n\\ \tilde{\alpha}_n \end{array}
\end{array}
\end{equation}
with inverse $\mathcal{C}'^{-1}_{ab}$ of exactly the same form.
But when working in conformal static gauge it is enough to
consider functions $F,G$ of $j$ which are independent of $X_0,
\pi^0$ (and therefore commute with the constraints $\alpha_n,
\tilde{\alpha}_n, c_0 = x_0 + p_0 \tau / \sqrt{\lambda}$). It
follows from \eqref{matrix of brackets} that for such functions
the Dirac bracket takes the form
\begin{equation} \label{Dirac bracket SG pre}
\begin{split}
\{ F, G \}_{\text{D.B.}} = \{ F, G \} &- \sum_{n,m \neq 0} \{ F,
L_n \} \{ L_n, L_m \}^{-1} \{ L_m, G \}\\ &- \sum_{n,m \neq 0} \{ F,
\tilde{L}_n \} \{ \tilde{L}_n, \tilde{L}_m \}^{-1} \{ \tilde{L}_m,
G \}.
\end{split}
\end{equation}
Here $\{ L_n, L_m \}^{-1}$ denotes the matrix inverse of $\{ L_n, L_m
\}$, likewise for $\{ \tilde{L}_n, \tilde{L}_m \}^{-1}$. There
are no terms involving $L_0 + \tilde{L}_0$ because the corresponding
components in the inverse matrix $\mathcal{C}'^{-1}_{ab}$ all
vanish. If either of the two functions $F,G$ happen to be invariant
under residual gauge transformations generated by $L_n, \tilde{L}_n, n
\neq 0$ then their Dirac and Poisson brackets are equal
\begin{equation*}
\{ F, G \}_{\text{D.B.}} = \{ F, G \}.
\end{equation*}

The expression \eqref{Dirac bracket SG pre} for the Dirac bracket can be
simplified further. Using the Virasoro algebra \eqref{Virasoro
algebra: modes} one finds the weak equalities $\{ L_n, L_m \} \approx
2 i n L_0 \delta_{m+n}$ and $\{ \tilde{L}_n, \tilde{L}_m \} \approx
2 i n \tilde{L}_0 \delta_{m+n}$, the (matrix) inverses of which are
\begin{equation*}
\{ L_n, L_m \}^{-1} \approx \frac{i}{2 n L_0} \delta_{m+n}, \qquad \{
\tilde{L}_n, \tilde{L}_m \}^{-1} \approx \frac{i}{2 n \tilde{L}_0}
\delta_{m+n}.
\end{equation*}
The Dirac bracket \eqref{Dirac bracket SG pre} then takes the simpler form
\begin{equation} \label{Dirac bracket SG}
\begin{split}
\{ F, G \}_{\text{D.B.}} \approx \{ F, G \} &- \frac{i}{2 L_0} \sum_{n
\neq 0} \frac{1}{n} \{ F, L_n \} \{ L_{-n}, G \}\\ &- \frac{i}{2
\tilde{L}_0} \sum_{n \neq 0} \frac{1}{n} \{ F, \tilde{L}_n \} \{
\tilde{L}_{-n}, G \}.
\end{split}
\end{equation}
This bracket provides a non-degenerate symplectic structure on the
reduced phase-space $\mathcal{P}^{\infty}$. To close this chapter we
determine the reduced dynamics on $\mathcal{P}^{\infty}$ with respect
to this Dirac bracket \eqref{Dirac bracket SG}.

\subsection*{Reduced dynamics}

In a generally covariant theory such as string theory, `time'
cannot be an observable since arbitrary time-reparametrisations
are allowed. In other words time is pure-gauge and the only
quantities one can talk about are constants of the motion. But
instead of talking about gauge-invariant quantities we have chosen
to isolate the physical degrees of freedom by explicitly breaking
the time-reparametrisation invariance through the use of gauge
fixing conditions. And because such gauge conditions single out a
special time, it makes sense to talk about the reduced dynamics,
with respect to this time, of the degrees of freedom
parameterising the reduced phase-space $\mathcal{P^{\infty}}$.

Naively one would guess that the dynamics on
$\mathcal{P^{\infty}}$ is generated simply by the total
Hamiltonian $H_T$ if we use the Dirac brackets. From a physical
point of view this must obviously be wrong since otherwise the
Hamiltonian being weakly zero $H_T \approx 0$ would imply that
every function $F$ with no explicit time dependence is actually
time independent $\dot{F} \approx \{ H_T, F \}_{\text{D.B.}}
\approx 0$. The reason why $H_T$ gives the wrong dynamics on
$\mathcal{P^{\infty}}$ is because the static gauge fixing
conditions \eqref{static gauge hamilton} are
$\tau$-dependent\footnote{Any complete gauge fixing in a generally
  covariant theory always requires imposing time-dependent gauge
  fixing conditions.} and implementing such constraints in Dirac's
theory of constrained Hamiltonian systems turns out to be far from
obvious. Indeed, using the usual equations of motion the
$\tau$-dependent constraint $c_0 = x_0 +
\frac{p_0}{\sqrt{\lambda}} \tau \approx 0$ is not preserved under
time evolution because
\begin{equation*}
\frac{d c_0}{d \tau} = \frac{\partial c_0}{\partial \tau} + \{
H_T, c_0 \}_{\text{D.B.}} \approx \frac{\partial c_0}{\partial
\tau} = \frac{p_0}{\sqrt{\lambda}} \not \approx 0.
\end{equation*}
A correction term needs to be added to the equations of motion in
order to accommodate for the $\tau$-dependence of the constraint
$c_0 \approx 0$. For an arbitrary functions $F$ with explicit time
dependence the equations of motion now read \cite[p110, ex. 4.8]{Henneaux}
\begin{subequations} \label{corrected eom}
\begin{align}
\label{corrected eom a} \frac{d F}{d \tau} &= \frac{\partial
F}{\partial \tau} + \{ H_T, F \}_{\text{D.B.}} - \frac{\partial
c_0}{\partial
\tau} \{ \gamma_0, c_0 \}^{-1} \{ \gamma_0 , F\},\\
\label{corrected eom b} &= \frac{\partial F}{\partial \tau} + \{
H_T, F \}_{\text{D.B.}} - \{ \gamma_0 , F \},
\end{align}
\end{subequations}
where $ \gamma_0 = L_0 + \tilde{L}_0 + \frac{p_0^2}{2
\sqrt{\lambda}} \approx 0$ is the only Virasoro constraint that
has a non-zero Poisson bracket with $c_0 \approx 0$. It is
immediate from \eqref{corrected eom a} that now we have $\frac{d
c_0}{d \tau} \approx 0$ and all the other constraints are also
preserved, since their Poisson bracket with $\gamma_0$ is weakly
zero. Note that the correction term is just a gauge transformation
whose role is to maintain the dynamics on the constraint surface
$c_0 \approx 0$, much like the Dirac bracket ensures that
time-independent second-class constraints are preserved in time.

An undesirable feature of \eqref{corrected eom} is that it isn't
written in terms of the Dirac bracket. However, for functions $F$
which only depend on the principal chiral fields $j$ one can
show that $\{ p_0 , F \}_{\text{D.B.}} = - \frac{\sqrt{\lambda}}{p_0}
\{ L_0 + \tilde{L}_0 , F \}$. Indeed, going back to the matrix of
Poisson brackets \eqref{matrix of brackets} we have $\mathcal{C}'_{c_0
\gamma_0} = \{ c_0, \gamma_0 \}$ but $\mathcal{C}'^{-1}_{c_0 \gamma_0}
= 1/ \{ \gamma_0, c_0 \}$, thus
\begin{equation*}
\{ p_0, F \}_{\text{D.B.}} = \{ p_0 , F\} - \{ p_0, c_0 \} \frac{1}{\{
\gamma_0 , c_0 \}} \{ \gamma_0, F \}.
\end{equation*}
Now $\{ p_0, c_0 \} = 1$, $\{ \gamma_0, c_0 \} =
\frac{p_0}{\sqrt{\lambda}}$ and $\{ p_0, F \} = 0$ by assumption on
$F$ so the result follows. Using this result the equation of motion
\eqref{corrected eom b} can be rewritten for such functions of the
physical variables as
\begin{equation} \label{corrected eom 2}
\frac{d F}{d \tau} = \frac{\partial F}{\partial \tau} + \left\{
H_T + \frac{p_0^2}{2 \sqrt{\lambda}}, F \right\}_{\text{D.B.}}.
\end{equation}
Thus we observe that the equations of motion on the reduced
phase-space $\mathcal{P}^{\infty}$ are generated not by the total
Hamiltonian $H_T$ (which is weakly zero) but by a shifted
Hamiltonian
\begin{equation} \label{reduced Hamiltonian}
H^{\ast} \equiv H_T + \frac{p_0^2}{2 \sqrt{\lambda}} \approx
\frac{p_0^2}{2 \sqrt{\lambda}}.
\end{equation}
A careful generalisation of Dirac's analysis of constrained
Hamiltonian systems to include time-dependent constraints (hence
allowing the use of time-dependent gauge fixing conditions) was
given in \cite{Evans:1993dq,Evans:1994ks} and also leads to the
same conclusion. There the presence of time-dependent constraints
leads to a shift in the $1$-form $dH_T \mapsto dH_T + A$ so that
the reduced dynamics $\iota(v) \omega^{\ast} = dH_T + A$ can still
be described by Hamilton's equations in terms of the Dirac bracket
provided $A$ is locally exact. A simple computation in the
formalism of \cite{Evans:1993dq,Evans:1994ks} shows that $A = -
d(L_0 + \tilde{L}_0)$ and hence the total Hamiltonian gets shifted
by the same amount \eqref{reduced Hamiltonian} since $- L_0 -
\tilde{L}_0 \approx \frac{p_0^2}{2 \sqrt{\lambda}}$ by the
Virasoro constraints \eqref{Vir modes}.

The equation of motion for the reduced dynamics \eqref{corrected
eom 2} has an obvious interpretation. It says that the energy
$\mathcal{E} = - L_0 - \tilde{L}_0$ of the principal chiral model
generates worldsheet $\tau$-translations on $\mathcal{P}^{\infty}$
(from now on we assume $F$ has no explicit dependence of $\tau$)
\begin{subequations} \label{reduced eom E and P}
\begin{equation} \label{reduced eom E}
\frac{d F}{d \tau} = \{ \mathcal{E}, F \}_{\text{D.B.}}.
\end{equation}
But using the zero-mode parts of the Virasoro constraints \eqref{E
vs p_0^2} and static gauge fixing conditions \eqref{Static modes}
we see that the dynamics \eqref{reduced eom E} is equivalent to
the global translation symmetry in the target time $X_0$,
\begin{equation*}
\frac{d F}{d x_0} = \{ \Delta, F \}_{\text{D.B.}},
\end{equation*}
which is generated by the space-time energy $\Delta = - p_0$ of
the string. In conclusion, although the worldsheet coordinates
have been fixed, we have done so using the $\tau$-dependent static
gauge fixing conditions which relate the worldsheet time $\tau$ to
the target time $X_0$. As a result, the global $X_0$-translation
symmetry gives rise to non-trivial $\tau$-dynamics for the
remaining degrees of freedom of the string. Since at this stage
the vanishing of the worldsheet momentum $\mathcal{P} \approx 0$
hasn't yet been imposed, the momentum $\mathcal{P}$ of the
principal chiral model still generates worldsheet
$\sigma$-translations as in \eqref{P1,j},
\begin{equation} \label{reduced eom P}
\frac{d F}{d \sigma} = \{ \mathcal{P}, F \}_{\text{D.B.}}.
\end{equation}
\end{subequations}
Unlike \eqref{reduced eom E} however the $\sigma$-dynamics
\eqref{reduced eom P} are not physical and must be removed at the
end of the day by symplectic reduction to the level set
$\mathcal{P} \approx 0$.

%% file: Integrability.tex
\newpage

\chapter{Integrability} \label{chapter: integrability}

\begin{flushright}
{\small \textit{``Ce qui embellit le d\'esert, dit le petit
prince, c'est qu'il cache un puits quelque part...''}}
\footnote{\textit{``What makes the desert beautiful, said the
little prince, is that somewhere it hides a well...''}}\\
{\small Antoine de Saint-Exup\'ery, {\it Le Petit Prince}}
\end{flushright}
\vspace{1cm}


\section{Conserved charges} \label{section: conserved charges}


When working in conformal and static gauge, the only field that
remains unfixed is the principal chiral field $j$. The equations
of motion of the string reduce to $j$ being both conserved and
flat \eqref{j eom}
\begin{subequations} \label{j eom bis}
\begin{align}
\label{conserved ab} d \ast j &= 0,\\
\label{conserved non-ab} dj - j \wedge j &= 0.
\end{align}
\end{subequations}
These are two first-order differential equations for the current
$j$ which express abelian and non-abelian conservation laws for
$\ast j$ and $j$ respectively as we now explain.

First of all, as we discussed in section \ref{section:
symmetries}, the current $j$ is actually the Noether current for
the global $SU(2)_R$ symmetry whose conservation is equivalent to
equation \eqref{conserved ab}. The corresponding Noether charge
defined in \eqref{Q_R def} is the integral $\int_{\gamma} \ast j$
around a closed loop $\gamma$ of non-trivial homotopy on the
worldsheet. In geometrical terms its conservation is a consequence
of Stokes' theorem as already discussed in section \ref{section:
symmetries},
\begin{equation} \label{Abelian Stokes}
\raisebox{-14mm}{\psfrag{g1}{\footnotesize $\gamma_1$}
\psfrag{g2}{\footnotesize $\gamma_2$} \psfrag{D}{\footnotesize
$D$} \includegraphics[height=30mm]{Figures/Noether_charge.eps}}
\qquad \qquad \int_{\gamma_2} \ast j - \int_{\gamma_1} \ast j =
\int_{\partial D} \ast j = \int_D d \ast j = 0.
\end{equation}

Secondly, the current $j$ is flat by equation \eqref{conserved
non-ab}. This was a consequence of its definition $j = - g^{-1}
dg$. But this property also leads to a very nice conservation law.
Indeed, consider the parallel transporter $\widehat{\Psi}(\gamma)$
with $j$ as connection along a path $\gamma$ on the worldsheet,
\begin{equation} \label{parallel transporter j}
\raisebox{-14mm}{\psfrag{g}{\footnotesize $\gamma$}
\psfrag{x}{\footnotesize $x$} \psfrag{y}{\footnotesize $y$}
\includegraphics[height=30mm]{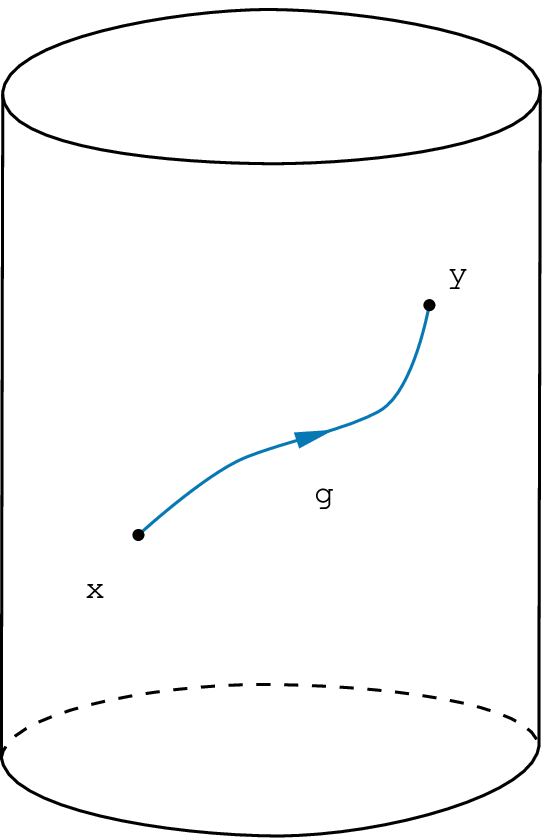}} \qquad \qquad
\widehat{\Psi}(\gamma) = P \overleftarrow{\exp} \int_{\gamma} j.
\end{equation}
\begin{theorem}[non-abelian Stokes']
If $j$ is a lie-algebra valued $1$-form and $D$ is a simply
connected region then
\begin{equation*}
P \, \overleftarrow{\exp} \int_{\partial D} j = A \, \exp \int_{D}
\widehat{\Psi}(\gamma)^{-1} (dj - j \wedge j)
\widehat{\Psi}(\gamma),
\end{equation*}
where $A$ is some ``surface ordering'' and $\gamma$ is a path
joining the base point of $\partial D$ to the integration point $x
\in D$.
\end{theorem}
\begin{corollary} \label{corollary: non-ab Stokes}
If $j$ is flat and $D$ is simply connected then
\begin{equation*}
P \, \overleftarrow{\exp} \int_{\partial D} j = {\bf 1}.
\end{equation*}
\end{corollary}
We deduce from corollary \ref{corollary: non-ab Stokes} that the
parallel transporter $\widehat{\Psi}(\gamma)$ defined by
\eqref{parallel transporter j} only depends on the homotopy class
of $\gamma$ with fixed endpoints $x,y$. Now consider the parallel
transporter around a closed loop $\gamma_x$ based at $x$ and
winding once around the worldsheet. Note that the base-point $x$
is important here since we are considering path-ordered
exponentials. Corollary \ref{corollary: non-ab Stokes} implies
that $\widehat{\Psi}(\gamma_x)$ is independent of the path,
provided it still starts and ends at $x$ after winding a single
time around the worldsheet. This is not quite a conservation law
in the sense of \eqref{Abelian Stokes} since it only gives
$\widehat{\Psi}(\gamma_2) = \widehat{\Psi}(\gamma_1)$ if the paths
$\gamma_1$ and $\gamma_2$ are both bound at the same point $x$. We
would like a relation between $\widehat{\Psi}(\gamma_1)$ and
$\widehat{\Psi}(\gamma_2)$ for two general loops $\gamma_1,
\gamma_2$ as in \eqref{Abelian Stokes}. But corollary
\ref{corollary: non-ab Stokes} also provides such a relation when
the base points $x$ and $y$ of $\gamma_1$ and $\gamma_2$ are
different, namely
\begin{equation*}
\raisebox{-14mm}{\psfrag{g1}{\footnotesize $\gamma_1$}
\psfrag{g2}{\footnotesize $\gamma_2$} \psfrag{g}{\footnotesize
$\gamma$} \psfrag{D}{\footnotesize $D$}
\includegraphics[height=30mm]{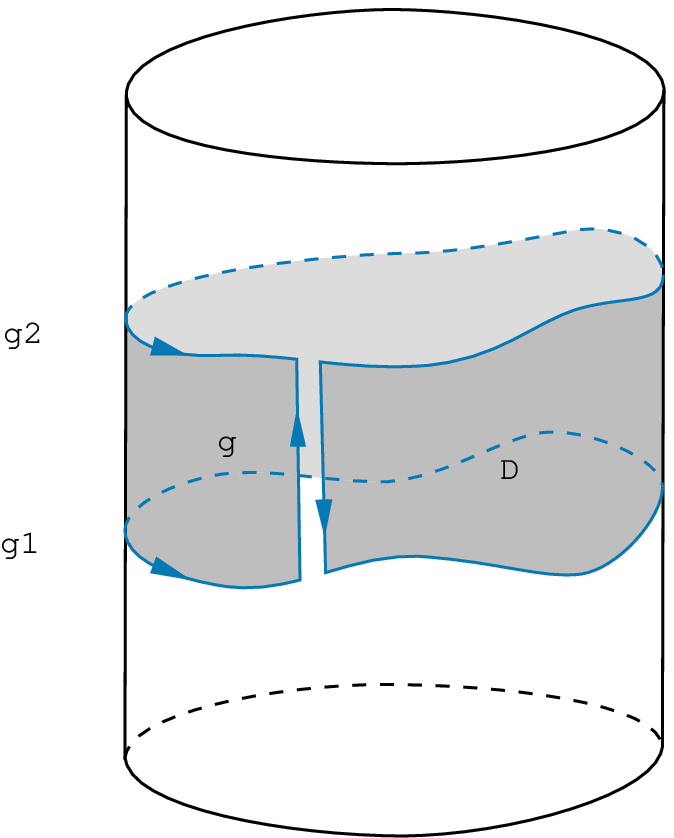}} \qquad \qquad
\widehat{\Psi}(\gamma_2) = \widehat{\Psi}(\gamma)
\widehat{\Psi}(\gamma_1) \widehat{\Psi}(\gamma)^{-1},
\end{equation*}
where $\gamma$ is a path connecting the base points $x$ and $y$.
This is a sort of ``non-abelian'' conservation law. In fact it
implies more than one ``abelian'' conservation law because each
eigenvalue of $\widehat{\Psi}(\gamma_i)$ is separately conserved,
which follows from
\begin{equation} \label{not a curve}
\det \left( \lambda {\bf 1} - \widehat{\Psi}(\gamma_1) \right) =
\det \left( \lambda {\bf 1} - \widehat{\Psi}(\gamma_2) \right).
\end{equation}

\subsection*{Lax connection}

Since flat currents lead to conserved quantities it makes sense to
consider the combination
\begin{equation} \label{pre Lax connection}
J = \alpha j + \beta \ast j,
\end{equation}
and try to adjust the constants $\alpha, \beta$ to render $J$
flat. It is straightforward to show from \eqref{j eom bis} using
the rules $\ast \ast = + 1$ and $a \wedge \ast b = - \ast a \wedge
b$ for any (lie-algebra valued) $1$-forms $a,b$ that
\begin{equation*}
dJ - J \wedge J = -(\alpha^2 - \alpha - \beta^2) j \wedge j.
\end{equation*}
Notice that the right hand side is proportional to the amount by
which $\ast j$ fails to be flat, namely $\ast j \wedge \ast j = -
j \wedge j$. We see that $J$ is flat provided $\alpha^2 - \alpha -
\beta^2 = 0$. This is a single constraint on the two parameters of
\eqref{pre Lax connection} admitting two solutions $\alpha  =
\frac{1}{1 - x^2}$, $\beta = \pm \frac{x}{1 - x^2}$, parameterised
by a single variable $x$. Both solutions are related by $x
\rightarrow -x$ so this construction provides a $1$-parameter
family of flat currents $J(x)$.
\begin{definition}
The \dub{Lax connection} is the $1$-parameter family of $1$-forms
on the worldsheet
\begin{equation} \label{Lax connection}
J(x) = \frac{1}{1 - x^2} (j - x \ast j), \qquad x \in \mathbb{C}.
\end{equation}
\end{definition}

By construction $dJ - J \wedge J = \alpha (dj - j \wedge j) +
\beta d \ast j$ and so we have the following
\begin{lemma}
The Lax connection $J(x)$ is flat if and only if $j$ is on-shell,
\textit{i.e.}
\begin{equation*}
dJ(x) - J(x) \wedge J(x) = 0 \qquad \Leftrightarrow \qquad \left\{
\begin{array}{l} d \ast j = 0,\\ dj - j \wedge j = 0.\end{array}
\right.
\end{equation*}
\end{lemma}

Note that the flatness condition (along with the whole formalism that
will follow from it) is invariant under gauge transformations
\begin{equation} \label{gauge transformation general}
J(x) \mapsto \tilde{g} J(x) \tilde{g}^{-1} + \left(
d \tilde{g} \right) \tilde{g}^{-1},
\end{equation}
where the matrix $\tilde{g}(x,\sigma, \tau)$ is an arbitrary function
of the spectral parameter $x$ and the worldsheet space and time
coordinates $\sigma, \tau$. In particular, for the purpose of
discussing the integrals of motion the Lax connection \eqref{Lax
connection} is by no means special. Indeed in section \ref{section:
local charges} we shall make use of the gauge freedom \eqref{gauge
transformation general} to move to a more appropriate gauge for
identifying the local conserved charges.

\subsection*{Monodromy}

Owing to the flatness of the current $J(x)$, it is now natural to
consider parallel transporters on the worldsheet using $J(x)$ as
the connection,
\begin{equation} \label{parallel transporter}
\raisebox{-14mm}{\psfrag{g}{\footnotesize $\gamma$}
\psfrag{x}{\footnotesize $x$} \psfrag{y}{\footnotesize $y$}
\includegraphics[height=30mm]{Figures/ParallelTransport.eps}} \qquad \qquad
\widehat{\Psi}(\gamma,x) = P \overleftarrow{\exp} \int_{\gamma}
J(x).
\end{equation}
As before, the object leading to ``non-abelian'' conservation laws
is the transporter around a path of non-trivial homotopy.
\begin{definition}
The \dub{Monodromy matrix} is the parallel transporter
\begin{equation*}
\raisebox{-14mm}{\psfrag{x}{\tiny $(\sigma, \tau)$}
\psfrag{g}{\footnotesize $\gamma(\sigma,\tau)$}
\includegraphics[height=30mm]{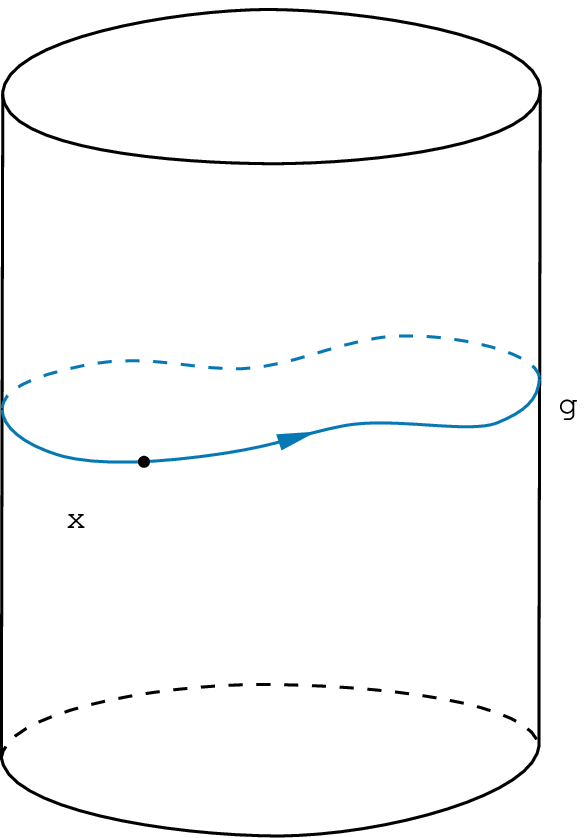}} \qquad \qquad
\Omega(x,\sigma,\tau) = P \overleftarrow{\exp}
\int_{[\gamma(\sigma,\tau)]} J(x),
\end{equation*}
where $\gamma(\sigma,\tau)$ is a loop starting and ending at
$(\sigma,\tau)$ that winds once around the worldsheet.
\end{definition}
By the non-abelian Stokes' theorem this definition only depends on
the homotopy class $[\gamma(\sigma,\tau)]$ of the curve
$\gamma(\sigma,\tau)$ with both end-points fixed at
$(\sigma,\tau)$. In particular, since the path $\sigma \in [0,2
\pi]$ at fixed time $\tau$ is in this homotopy class, if we write
the Lax connection in components as $J(x) = J_0(x) d\tau + J_1(x)
d\sigma$ then we can write the monodromy matrix as
\begin{equation} \label{monodromy with J1}
\Omega(x,\sigma,\tau) = P \overleftarrow{\text{exp}}\left[
\int_{\sigma}^{\sigma + 2 \pi} d\sigma' J_1(x,\sigma',\tau)
\right].
\end{equation}
Furthermore, by using the non-ablelian Stokes' theorem to change
the base point $(\sigma, \tau)$, the monodromy matrix has the
following immediate property
\begin{lemma} \label{lemma: isospectral}
The $(\sigma,\tau)$-evolution of $\Omega(x,\sigma,\tau)$ is
isospectral, \textit{i.e.}
\begin{equation} \label{isospectral}
\raisebox{-14mm}{\psfrag{g1}{\footnotesize $\tau$}
\psfrag{g2}{\footnotesize $\tau'$} \psfrag{g}{\footnotesize
$\gamma$} \psfrag{D}{\footnotesize $ $}
\includegraphics[height=30mm]{Figures/NonAbStokes.eps}} \qquad \qquad
\Omega(x,\sigma',\tau') = \widehat{\Psi}(\gamma,x)
\Omega(x,\sigma,\tau) \widehat{\Psi}(\gamma,x)^{-1},
\end{equation}
where $\gamma$ is a path from $(\sigma, \tau)$ to $(\sigma',
\tau')$.
\end{lemma}
From now on we may sometimes omit the explicit dependence on the
base point $(\sigma,\tau)$ when it is clear and abbreviate
$\Omega(x, \sigma, \tau)$ as $\Omega(x)$.

\subsection*{Integrals of motion}

Once more the isospectral evolution of the $\Omega(x)$ in lemma
\ref{lemma: isospectral} means that all its eigenvalues are
conserved since the characteristic polynomial
\begin{equation} \label{spectral curve 0}
\Gamma(x,\Lambda) = \det \left( \Lambda {\bf 1} - \Omega(x, \sigma,
\tau) \right),
\end{equation}
is independent of $(\sigma, \tau)$. However, the novelty when
considering $J(x)$ as connection instead of $j$ is that the
conserved eigenvalues are now functions of $x \in \mathbb{C}$.
Taylor expanding each eigenvalue in $x$ therefore provides an
infinite number of integrals of motion.

The infinitesimal version of \eqref{isospectral}, that is its
leading order in $\delta \sigma = \sigma' - \sigma$ and $\delta
\tau = \tau' - \tau$, gives a differential equation governing the
$(\sigma, \tau)$-dependence of the monodromy matrix,
\begin{equation} \label{monodromy evolution}
[d - J(x), \Omega(x)] = 0.
\end{equation}
It is evident that any power of $\Omega(x)$ also satisfies the
same equation, or equivalently in components $\partial_{\alpha}
\Omega(x)^n = [J_{\alpha}(x), \Omega(x)^n]$, $\alpha  = 0,1$.
Taking the trace yields another way of characterising the
conservation of the eigenvalues of $\Omega(x)$,
\begin{equation} \label{trace monodromy}
\partial_0 \tr \Omega(x)^n = \partial_1 \tr \Omega(x)^n = 0.
\end{equation}


\section{Local conserved charges} \label{section: local charges}


Conserved charges can be of two different types: local or
non-local. A conserved charge is \dub{local} if it is the integral
of a local density, otherwise it is \dub{non-local}, such as when
the density itself is an integral. Any charge arising from a
continuous symmetry through the use of Noether's theorem is always
local since it is the integral $\int_{\gamma} \ast j$ of a current
$\ast j$ which is a local expression of the fields and whose local
conservation is expressed as $d \ast j = 0$. It is clear also that
any closed $1$-form which is a local expression of the fields
would give rise to a local conserved charge. However, since the
conserved charges $\tr \Omega(x)^n$ arose from a non-abelian
conservation law $dJ(x) - J(x) \wedge J(x) = 0$, it is not obvious
that any of them are local. Although non-local charges are very
interesting we will not be concerned with them here and so we
would like a way of extracting only local charges from the
generator of charges $\tr \Omega(x)^n$. If it were possible to
diagonalise the Lax connection $J(x)$ then the non-abelian
conservation law for the current $J(x)$ would reduce to $dJ(x) =
0$ and immediately provide infinitely many local charges. The
following theorem \cite[p66]{Babelon} shows that this
``abelianisation'' is possible in a neighbourhood of the
singularities $x = \pm 1$ of the Lax connection.

\begin{theorem} \label{theorem: local charges}
Around the points $x = \pm 1$ there exists regular local periodic
gauge transformations
\begin{equation} \label{gauge redundancy}
J(x) \mapsto J'_{(\pm)}(x) = g_{(\pm)}(x) J(x) g_{(\pm)}(x)^{-1} +
dg_{(\pm)}(x) g_{(\pm)}(x)^{-1},
\end{equation}
such that $J'_{(\pm)}(x) = \sum_{n = -1}^{\infty} J^{(\pm)}_n (x
\mp 1)^n$ are diagonal. In particular, $d J'_{(\pm)}(x) = 0$.
\end{theorem}

In the gauge of theorem \ref{theorem: local charges} the
conservation laws become abelian $d J'_{(\pm)}(x) = 0$ and so one
can immediately assert the existence of an infinite number of
local charges
\begin{equation} \label{local charges}
\mathcal{Q}^{(\pm)}_n = \int_{\gamma} J^{(\pm)}_n, \quad n = - 1,
0, \ldots
\end{equation}
These are the coefficients of $\mathcal{Q}^{(\pm)}(x) = \int_{\gamma}
J'_{(\pm)}(x) = \sum_{n = -1}^{\infty} (x \mp 1)^n \int_{\gamma}
J^{(\pm)}_n$ which is conserved by Stokes' theorem,
\begin{equation*}
\raisebox{-14mm}{\psfrag{g1}{\footnotesize $\gamma_1$}
\psfrag{g2}{\footnotesize $\gamma_2$} \psfrag{D}{\footnotesize
$D$} \includegraphics[height=30mm]{Figures/Noether_charge.eps}}
\qquad \qquad \int_{\gamma_2} J'_{(\pm)}(x) - \int_{\gamma_1}
J'_{(\pm)}(x) = \int_{\partial D} J'_{(\pm)}(x) = \int_D d
J'_{(\pm)}(x) = 0.
\end{equation*}
Moreover, because the connection is diagonal, the path ordering in
the definition of the monodromy matrix is not necessary in this
gauge. Therefore around $x = \pm 1$ the monodromy matrix
transforms under the gauge transformation of theorem \ref{theorem:
local charges} to the following very simple diagonal form,
\begin{equation} \label{diagonal Omega pm 1}
\Omega(x) \mapsto g_{(\pm)}(x) \Omega(x) g_{(\pm)}(x)^{-1} = \exp
\left[ \sum_{n = -1}^{\infty} \mathcal{Q}^{(\pm)}_n (x \mp 1)^n
\right],
\end{equation}
where we have used the fact that $g_{(\pm)}(x)$ are periodic in
$\sigma$. In particular, the local charges \eqref{local charges}
can be extracted from $\tr \Omega(x)^n$ by expanding around $x =
\pm 1$, as claimed. Since $j_{\pm} \in \mathfrak{su}(2)$ implies
$\det \Omega(x) = 1$, it follows that all the diagonal matrices
$\mathcal{Q}^{(\pm)}_n$ are proportional to the third Pauli matrix
$\sigma_3 = \text{diag}(1,-1)$.
\begin{definition}
The \dub{local charges} are given by
\begin{equation*}
Q^{(\pm)}_n = \frac{1}{2 i} \tr \left( \mathcal{Q}^{(\pm)}_n \sigma_3
\right), \quad n = -1, 0, \ldots
\end{equation*}
\end{definition}

Recall that apart from satisfying the equations of motion \eqref{j
eom bis}, the current $j$ must also solve the Virasoro constraints
\eqref{Vir without P=0}
\begin{equation*}
\frac{1}{2} \tr j_{\pm}^2 = - \kappa_{\pm}^2.
\end{equation*}
Up to now we have not yet implemented these in the Lax formalism.
The next proposition makes first use of these constraints to
compute the first local charges $Q^{(\pm)}_{-1}$. Note however
that a complete treatment of the Virasoro constraints will have to
wait until we switch over to the Hamiltonian formalism in the next
section.

\begin{proposition} \label{proposition: monodromy asymptotics}
The first charges are equal to $Q^{(\pm)}_{-1} = - \pi
\kappa_{\pm}$. In particular,
\begin{equation} \label{monodromy asymptotics}
g_{(\pm)}(x) \Omega(x) g_{(\pm)}(x)^{-1} = \exp \left[ - \frac{i
\pi \kappa_{\pm}}{x \mp 1} \sigma_3 + O\left( (x \mp 1)^0 \right)
\right] \qquad \text{as} \; x \rightarrow \pm 1.
\end{equation}
\begin{proof}
The asymptotics of the first component $J_1(x)$ of the Lax
connection near $x = \pm 1$ are
\begin{equation*}
J_1(x) = - \frac{1}{2} \frac{j_{\pm}}{x \mp 1} + O\left( (x \mp 1)^0
\right), \qquad \text{as} \; x \rightarrow \pm 1.
\end{equation*}
But because the gauge parameters $g_{(\pm)}(x) = \sum_{n =
0}^{\infty} g^{(\pm)}_n (x \mp 1)^n$ are regular it follows from
\eqref{gauge redundancy} that $J'_{(\pm)}(x)$ has a simple pole at
$x = \pm 1$ and the coefficients $g^{(\pm)}_0$ are the matrices
diagonalising $j_{\pm}$. In other words,
\begin{equation*}
J'_{(\pm)}(x) = - \frac{1}{2} \frac{j^{\text{diag}}_{\pm}}{x \mp
1} + O\left( (x \mp 1)^0 \right), \qquad \text{as} \; x
\rightarrow \pm 1.
\end{equation*}
It remains to compute the eigenvalues of $j_{\pm}$. But since
$j_{\pm} \in \mathfrak{su}(2)$ one has $\det j_{\pm} =
-\frac{1}{2} \tr j_{\pm}^2$. So the Virasoro constraint \eqref{Vir
without P=0} may be rewritten as $\det j_{\pm} = \kappa_{\pm}^2$
and using $\tr j_{\pm} = 0$ the eigenvalues of $j_{\pm}$ are
therefore $i \kappa_{\pm}$ and $- i \kappa_{\pm}$.
\end{proof}
\end{proposition}

Recall from equation \eqref{Vir E and P} that the energy and
momentum of the principal chiral field $j$ are given by
$\mathcal{E} \pm \mathcal{P} = \frac{\sqrt{\lambda}}{2}
\kappa^2_{\pm}$ and therefore are directly related to the squares
of the first charges $Q^{(\pm)}_{-1}$, namely
\begin{equation} \label{E and P as local charges}
\mathcal{E} \pm \mathcal{P} = \frac{\sqrt{\lambda}}{2 \pi^2}
\left(Q^{(\pm)}_{-1}\right)^2.
\end{equation}
We note for later that the light-cone components $J_{\pm}(x) =
J_0(x) \pm J_1(x) = \frac{j_{\pm}}{1 \mp x}$ of the Lax connection
are diagonalised by $g^{(\pm)}_0 = g_{(\pm)}(\pm 1)$ and thus take
on the following simple form
\begin{equation} \label{Lax connection near pm 1}
J_{\pm}(x') = \frac{i \kappa_{\pm}}{1 \mp x'} g^{(\pm) \; -1}_0
\sigma_3 g^{(\pm)}_0.
\end{equation}

\subsection*{Noether charges}

We have just seen that an infinite number of local charges can be
extracted from the expansion of the monodromy matrix at the
special points $x = \pm 1$, in particular the principal chiral
field energy and momentum \eqref{E and P as local charges}. It
turns out that the Noether charges of the global $SU(2)_R \times
SU(2)_L$ symmetries can be easily extracted from asymptotics of
the monodromy matrix at other points. As we now show, the Noether
charge $Q_R$ (resp. $Q_L$) of the $SU(2)_R$ (resp. $SU(2)_L$)
symmetry is the first non-trivial coefficient in the expansion of
$\Omega(x)$ at $x = \infty$ (resp. $x = 0$). The higher
coefficients of the expansions at these points are all related to
non-local charges \cite{MacKay:2004tc} and will therefore not
interest us.

The asymptotic expansion of the connection \eqref{Lax connection}
at $x = \infty$
\begin{equation*}
J(x) = \frac{1}{x} \ast j + O\left( \frac{1}{x^2} \right),
\end{equation*}
leads to the following asymptotic expansion of the monodromy
matrix at $x = \infty$
\begin{subequations} \label{monodromy asymptotics inf & 0}
\begin{equation} \label{monodromy asymptotics at infty}
\begin{split}
\Omega(x) &= P \overleftarrow{\exp} \int_{[\gamma(\sigma,\tau)]}
\left( \frac{1}{x}
\ast j + O\left( \frac{1}{x^2} \right) \right) \\
&= {\bf 1} + \frac{1}{x} \frac{4 \pi Q_R}{\sqrt{\lambda}} +
O\left( \frac{1}{x^2} \right), \quad \text{as } x \rightarrow
\infty.
\end{split}
\end{equation}
The asymptotics of the connection at $x = 0$ is $J(x) = j - x \ast
j + O\left( x^2 \right)$, so that
\begin{align*}
d - J(x) &= d - j + x \ast j + O\left( x^2 \right),\\
&= g^{-1} \left( d + x \ast l + O\left( x^2 \right) \right) g,
\end{align*}
where $l = - dg \, g^{-1} = g j g^{-1}$. Now because the field
$g(\sigma,\tau)$ is periodic in $\sigma$ it follows that the
asymptotic expansion of the monodromy matrix near $x = 0$ is given
by
\begin{equation} \label{monodromy asymptotics at zero}
\begin{split}
g \Omega(x) g^{-1} &= P \overleftarrow{\exp} \left(
\int_{[\gamma(\sigma,\tau)]} - x
\ast l + O\left( x^2 \right) \right),\\
&= {\bf 1} - x \frac{4 \pi Q_L}{\sqrt{\lambda}} + O\left( x^2
\right), \quad \text{as } x \rightarrow 0.
\end{split}
\end{equation}
\end{subequations}

Since the Noether charges $Q_R$ and $Q_L$ are conserved
classically, we may fix them to lie in a particular direction of
$\mathfrak{su}(2)$ and take them for example to be proportional to
the third Pauli matrix $\sigma_3$
\begin{equation*}
Q_R = \frac{1}{2i}\, R\sigma_3, \; Q_L =\frac{1}{2i}\, L\sigma_3 ,
\quad R,L \in \mathbb{R}_+.
\end{equation*}
where $R$ and $L$ are constants of the motion. By restricting the
Noether charges in this way we focus on the subset of `highest
weight' solutions to the equations of motion. There is however no
loss of generality in doing so since all other solutions can be
obtained by applying a combination of $SU(2)_R$ and $SU(2)_L$ to
such a `highest weight' solution. With this restriction the
asymptotic expansions \eqref{monodromy asymptotics inf & 0} reduce to
\begin{subequations} \label{monodromy asymptotics inf & 0 2}
\begin{align}
\label{monodromy asymptotics at infty 2} \Omega(x) &= {\bf 1} -
\frac{1}{x} \frac{2 \pi i R}{\sqrt{\lambda}} \sigma_3 + O\left(
\frac{1}{x^2} \right), \quad \text{as } x \rightarrow \infty,
\\
\label{monodromy asymptotics at zero 2} g \Omega(x) g^{-1} &= {\bf
1} + x \frac{2 \pi i L}{\sqrt{\lambda}} \sigma_3 + O\left( x^2
\right), \quad \text{as } x \rightarrow 0.
\end{align}
\end{subequations}


\section{Involution of conserved charges} \label{section: involution}


In section \ref{section: conserved charges} we saw that given a
solution $j$ to the equations of motion \eqref{j eom bis} one
could construct a $1$-parameter family of flat $1$-forms on the
worldsheet, which in turn lead to the existence of an infinite
number of integrals. However, for the system in question to be
integrable requires also that these integrals of motion be in
pairwise involution. To study this question we must now turn to
the Hamiltonian framework, introduced in chapter \ref{chapter:
hamiltonian}.

After having studied solutions of the equations of motion it is
straightforward to pass to a Hamiltonian analysis once we realise
that the space of solutions of the equations of motion \eqref{j
eom bis} is in one to one correspondence with phase-space. Indeed,
any given solution $j^{\text{sol}}(\sigma, \tau)$ of \eqref{j eom
bis} determines a point in phase-space by restriction to a chosen
time slice, say $\tau = 0$, that is $j(\sigma) =
(j_0^{\text{sol}}(\sigma,0), j_1^{\text{sol}}(\sigma,0))$.
Conversely, any point $j(\sigma)$ in phase-space determines a
unique solution $j^{\text{sol}}(\sigma,\tau)$ whose initial
condition at $\tau = 0$ is $j^{\text{sol}}(\sigma,0) = j(\sigma)$.
However, as we saw in chapter \ref{chapter: strings on RxS3} the
current $j$ must also satisfy the Virasoro constraints \eqref{Vir
without P=0}
\begin{equation*}
\frac{1}{2} \tr j_{\pm}^2 = - \kappa_{\pm}^2.
\end{equation*}
If these constraints are satisfied by the initial conditions
$j^{\text{sol}}_{\pm}(\sigma,0)$ then using the equations of
motion we have $\partial_0 \frac{1}{2} \tr
(j^{\text{sol}}_{\pm})^2 = \pm
\partial_1 \frac{1}{2} \tr (j^{\text{sol}}_{\pm})^2$ so that the same
constraints are also satisfied by full solution
$j^{\text{sol}}_{\pm}(\sigma, \tau)$. Therefore the space of
solutions satisfying \eqref{Vir without P=0} is in one to one
correspondence with the reduced phase-space $\mathcal{P}^{\infty}$
introduced in chapter \ref{chapter: hamiltonian}.

Rephrased in the Hamiltonian formalism, the content of section
\ref{section: conserved charges} is as follows. One can define a
$1$-parameter family of $\mathfrak{su}(2)$-valued functions on
phase-space
\begin{equation} \label{Lax J1 Hamilton}
j(\sigma) \mapsto J_1(\sigma,x) = \frac{1}{1 - x^2} (j_1(\sigma) +
x j_0(\sigma)), \qquad x \in \mathbb{C},
\end{equation}
with the property, following from lemma \ref{lemma: isospectral},
that its path ordered exponential \eqref{monodromy with J1}
\begin{equation} \label{Monodromy Hamilton}
j(\sigma) \mapsto \Omega(x,\sigma) = P
\overleftarrow{\text{exp}}\left[ \int_{\sigma}^{\sigma + 2 \pi}
d\sigma' J_1(x,\sigma') \right],
\end{equation}
has a simple isospectral evolution under Hamilton's equations
\eqref{reduced eom E and P}. And in particular, equation \eqref{trace
monodromy} shows that the $1$-parameter family of functions
$j(\sigma) \mapsto \tr \Omega(x)$ is invariant under the $\tau$-
and $\sigma$-flows which in the Hamiltonian formalism are
generated by $\mathcal{E}$ and $\mathcal{P}$ respectively. In
other words, the upshot of section \ref{section: conserved
charges} rephrased in Hamiltonian terms should read
\begin{equation} \label{trace monodromy Hamiltonian}
\left\{ \mathcal{E}, \tr \Omega(x)^n \right\}_{\text{D.B.}} =
\left\{ \mathcal{P}, \tr \Omega(x)^n \right\}_{\text{D.B.}} = 0,
\end{equation}
We will rederive this result within the Hamiltonian formalism by
in fact proving a much stronger result.

What we are seeking to show using the Hamiltonian formalism is
that the conserved charges obtained in the previous section are in
pairwise involution. But this statement is equivalent to showing
that
\begin{align} \label{involution property}
\left\{ \tr \Omega(x)^n, \tr \Omega(x')^m \right\} = 0, \quad
\forall n,m \in \mathbb{N}.
\end{align}
However, since we are working on the reduced phase-space all
statements must be made with respect to the Dirac bracket instead
of the Poisson bracket. So the ultimate goal of this section is to
show that \eqref{involution property} also holds for Dirac
brackets,
\begin{theorem} \label{theorem: complete integrability}
The traces of powers of the monodromy matrix generate quantities
in involution with respect to the Dirac bracket \eqref{Dirac
bracket SG}, \textit{i.e.}
\begin{equation} \label{complete integrability}
\left\{ \tr \Omega(x)^n, \tr \Omega(x')^m \right\}_{\text{D.B.}} =
0, \quad \forall n,m \in \mathbb{N}.
\end{equation}
\end{theorem}
\noindent This is the full statement of Liouville integrability of
string theory on $\mathbb{R} \times S^3$ in conformal static
gauge. From section \ref{section: local charges} we know that
$\mathcal{E}$ and $\mathcal{P}$ can be obtained from $\tr
\Omega(x)^n$ in the limit $x \rightarrow \pm 1$ and thus
\eqref{trace monodromy Hamiltonian} is a trivial consequence of
\eqref{complete integrability}.

In the following we shall adopt tensor notation for all brackets.
We define the Poisson bracket between two $2 \times 2$ matrices
$A$ and $B$ as
\begin{equation} \label{PB definition}
\left\{ A \overset{\otimes}, B \right\} = \int d\sigma \left(
\frac{\delta A}{\delta \pi^a(\sigma)} \otimes \frac{\delta
B}{\delta q^a(\sigma)} - \frac{\delta A}{\delta q^a(\sigma)}
\otimes \frac{\delta B}{\delta \pi^a(\sigma)} \right),
\end{equation}
where the operation $\otimes$ on the right hand side denotes the
usual tensor product. This notation conveniently encodes all the
Poisson brackets between the various components of $A$ and $B$.
For example, if $A,B \in \mathfrak{su}(2)$ in components are $A =
A^a t_a$ and $B = B^a t_a$ then by definition \eqref{PB
definition} we have $\{ A \overset{\otimes}, B \} = \{ A^a , B^b
\} t_a \otimes t_b$.

\subsection*{$\bm{\{ J_1, J_1 \}}$ algebra}

The monodromy matrix \eqref{Monodromy Hamilton} being the path
ordered exponential of the space component \eqref{Lax J1 Hamilton}
of the Lax connection, we will need the Poisson bracket $\{
J_1,J_1 \}$ in order to construct the Poisson bracket of monodromy
matrices.

The set of Poisson brackets $\left\{
J_1^a(\sigma,x),J_1^b(\sigma',x')\right\}$ can be easily obtained
from the fundamental brackets of currents $\{
j^a_{\alpha}(\sigma), j^b_{\beta}(\sigma') \}$ in \eqref{jj
Poisson brackets}. Introducing $\eta := k^{ab} t_a \otimes t_b = t_a
\otimes t^a$, called
the Casimir tensor, they can be written as
\begin{multline} \label{Lax connection PB}
\frac{\sqrt{\lambda}}{4 \pi} \left\{ J_1(\sigma,x)
\mathop{,}^{\otimes} J_1(\sigma',x')\right\}\\ = \left[
-\frac{\eta}{x - x'}, \frac{{x'}^2}{1 - {x'}^2} J_1(\sigma,x)
\otimes {\bf 1} + \frac{x^2}{1 - x^2} {\bf 1}
\otimes J_1(\sigma,x') \right] \delta(\sigma - \sigma') \\
+ \frac{x + x'}{(1 - x^2)(1 - {x'}^2)} \eta \delta'(\sigma -
\sigma').
\end{multline}
This bracket has the form of the fundamental Poisson bracket $\{
J_1,J_1 \}$ for a non-ultralocal integrable system formulated by
Maillet \cite{Maillet, BFLS}
\begin{align} \label{Maillet bracket}
\left\{ J_1(\sigma,x) \mathop{,}^{\otimes} J_1(\sigma',x')\right\}
&= \left[ r(\sigma,x,x'), J_1(\sigma,x)\otimes \mathbf{1} +
\mathbf{1} \otimes J_1(\sigma',x') \right] \delta(\sigma -
\sigma') \notag \\ &- \left[ s(\sigma,x,x'), J_1(\sigma,x)\otimes
\mathbf{1} - \mathbf{1} \otimes J_1(\sigma',x') \right]
\delta(\sigma - \sigma')\\ &- \left( r(\sigma,x,x') +
s(\sigma,x,x') - r(\sigma',x,x') + s(\sigma',x,x')
\right)\delta'(\sigma - \sigma'), \notag
\end{align}
These brackets involve a pair of matrices $r$ and $s$. Notice that
the $r$ matrix can be removed from the $\delta'$-term using the
identity $\left( r(\sigma') - r(\sigma) \right)\delta'(\sigma -
\sigma') = r'(\sigma) \delta(\sigma - \sigma')$ valid for any
function $r$ (as can be seen by integrating the left hand side against
a test function $\psi(\sigma)$). Thus the non-ultralocality of the
bracket is accounted for by the matrix $s$ alone. Indeed, the bracket
\eqref{Maillet bracket} is a non-trivial generalisation of the
standard ultralocal bracket which corresponds to setting $s =
\partial_{\sigma} r = 0$. In the present case the matrices $r$ and
$s$ are constant (independent of $\sigma$ and $\tau$)
\begin{subequations} \label{r,s-matrices}
\begin{align}
s(x,x') &= - \frac{2 \pi}{\sqrt{\lambda}} \frac{x+x'}{(1 - x^2)(1
-
{x'}^2)} \eta,\\
r(x,x') &= - \frac{2 \pi}{\sqrt{\lambda}} \frac{x^2 + {x'}^2 - 2
x^2 {x'}^2}{(x-x')(1 - x^2)(1 - {x'}^2)} \eta.
\end{align}
\end{subequations}
The description of the principal chiral model in terms of
Maillet's $(r,s)$-matrix formalism and the corresponding formulae
\eqref{r,s-matrices} for the $(r,s)$-matrices were first obtained
in \cite{Maillet3}.

\subsection*{$\bm{\{ T, T \}}$ and $\bm{\{ T, J_1 \}}$ algebras}
     \label{section: transition matrices PB}

The next step towards the algebra of monodromy matrices is the
algebra of transition matrices. A \dub{transition matrix} is
defined relative to an interval $[\sigma_1, \sigma_2]$ as
\begin{equation} \label{def: transition matrix}
T(\sigma_1,\sigma_2,x) = P \overleftarrow{\exp}
\int_{\sigma_2}^{\sigma_1} d\sigma J_1(\sigma,x).
\end{equation}
The monodromy matrix is then simply a special transition matrix
whose interval wraps the circle fully once, that is
$\Omega(x,\sigma) = T(\sigma + 2 \pi,\sigma,x)$.

Now the transition matrix \eqref{def: transition matrix} is the
unique solution of either of the two following differential
equations with boundary condition $T(\sigma_2,\sigma_2,x) = {\bf
1}$,
\begin{equation} \label{eq for transition matrix}
\frac{\partial T}{\partial \sigma_1}(\sigma_1,\sigma_2,x) =
J_1(\sigma_1,x) T(\sigma_1,\sigma_2,x), \qquad \frac{\partial
T}{\partial \sigma_2}(\sigma_1,\sigma_2,x) = -
T(\sigma_1,\sigma_2,x) J_1(\sigma_2,x).
\end{equation}
Considering the first of these, its variation is
\begin{equation*}
\frac{\partial \delta T}{\partial \sigma_1}(\sigma_1,\sigma_2,x) =
\delta J_1(\sigma_1,x) T(\sigma_1,\sigma_2,x) + J_1(\sigma_1,x)
\delta T(\sigma_1,\sigma_2,x)
\end{equation*}
with initial condition $\delta T(\sigma_1,\sigma_1,x) = 0$, to which
the unique solution is easily seen to be \cite{Faddeev}
\begin{equation} \label{dT in relation to dJ}
\delta T(\sigma_1,\sigma_2,x) = \int_{\sigma_2}^{\sigma_1} d\sigma
T(\sigma_1,\sigma,x) \delta J_1(\sigma,x) T(\sigma,\sigma_2,x).
\end{equation}
But now using the definition of the Poisson bracket \eqref{PB
definition} along with equation \eqref{dT in relation to dJ} it is
easy to relate the bracket of transition matrices $\{ T, T\}$ or
the bracket $\{ T, J_1 \}$ to the bracket of currents $\{ J_1,
J_1\}$. Specifically we find
\begin{subequations}
\begin{multline} \label{TT from JJ}
\left\{ T(\sigma_1,\sigma_2,x) \mathop{,}^{\otimes}
T(\sigma'_1,\sigma'_2,x') \right\} = \int_{\sigma_2}^{\sigma_1}
d\sigma \int_{\sigma'_2}^{\sigma'_1} d\sigma' \left(
T(\sigma_1,\sigma,x) \otimes T(\sigma'_1,\sigma',x') \right)\\
\times \left\{ J_1(\sigma,x) \mathop{,}^{\otimes} J_1(\sigma',x')
\right\} \left( T(\sigma,\sigma_2,x) \otimes
T(\sigma',\sigma'_2,x') \right),
\end{multline}
\begin{multline} \label{TJ}
\{ T(\sigma_1, \sigma_2, x) \overset{\otimes}, J_1(\sigma_3, x')
\} \\ = \int_{\sigma_2}^{\sigma_1} d\sigma (T(\sigma_1,\sigma,x)
\otimes {\bf 1}) \{ J_1(\sigma, x) \overset{\otimes},
J_1(\sigma_3, x') \} (T(\sigma,\sigma_2,x) \otimes {\bf 1}).
\end{multline}
\end{subequations}
Plugging the bracket \eqref{Maillet bracket} into these equations
one finds after a bit of algebra \cite{BFLS}
\begin{subequations}
\begin{multline} \label{transition matrix PB}
\left\{ T(\sigma_1,\sigma_2,x) \mathop{,}^{\otimes}
T(\sigma'_1,\sigma'_2,x') \right\} \\
\begin{split}
= &+ \epsilon(\sigma'_1 - \sigma'_2) \chi(\sigma;
\sigma'_1,\sigma'_2)
\\ &\times \left. T(\sigma_1,\sigma,x) \otimes
T(\sigma'_1,\sigma,x') \left( r(\sigma,x,x') - s(\sigma,x,x')
\right) T(\sigma,\sigma_2,x) \otimes T(\sigma,\sigma'_2,x')
\right|_{\sigma = \sigma_2}^{\sigma =
\sigma_1} \\
&+ \epsilon(\sigma_1 - \sigma_2) \chi(\sigma; \sigma_1,\sigma_2)
\\ &\times \left. T(\sigma_1,\sigma,x) \otimes T(\sigma'_1,\sigma,x')
\left( r(\sigma,x,x') + s(\sigma,x,x') \right)
T(\sigma,\sigma_2,x) \otimes T(\sigma,\sigma'_2,x')
\right|_{\sigma = \sigma'_2}^{\sigma = \sigma'_1},
\end{split}
\end{multline}
where $\epsilon(\sigma) = \text{sign}(\sigma)$ is the usual sign
function and $\chi(\sigma; \sigma_1,\sigma_2)$ is the
characteristic function of the interval $(\sigma_1, \sigma_2)$,
and
\begin{multline} \label{PB T,J}
\{ T(\sigma_1, \sigma_2, x) \overset{\otimes}, J_1(\sigma_3, x')
\} \\
\begin{split} = &-2 (\delta(\sigma_3 - \sigma_1) -
\delta(\sigma_3 - \sigma_2)) (T(\sigma_1,\sigma_3,x) \otimes {\bf
1}) s(x,x') (T(\sigma_3,\sigma_2,x) \otimes {\bf 1})\\ &+
\epsilon(\sigma_1 - \sigma_2) \chi(\sigma_3;
\sigma_1,\sigma_2) (T(\sigma_1,\sigma_3,x) \otimes {\bf 1})\\
&\times [ (r + s)(x,x'), J_1(\sigma_3, x) \otimes {\bf 1} + {\bf
1} \otimes J_1(\sigma_3, x') ] (T(\sigma_3,\sigma_2,x) \otimes
{\bf 1}).
\end{split}
\end{multline}
\end{subequations}

\subsection*{Maillet regularisation} \label{section: Maillet}

It follows from the algebra \eqref{transition matrix PB} that the
function,
\begin{equation*}
\Delta^{(1)}(\sigma_1,\sigma_2,\sigma'_1,\sigma'_2; x,x') = \{
T(\sigma_1,\sigma_2,x) \, \overset{\otimes}, \,
T(\sigma'_1,\sigma'_2,x') \}
\end{equation*}
is well defined and continuous where $\sigma_1, \sigma_2,
\sigma'_1, \sigma'_2$ are all distinct, but it has discontinuities
proportional to $2 s$ precisely across the hyperplanes
corresponding to some of the $\sigma_1, \sigma_2, \sigma'_1,
\sigma'_2$ being equal. Defining the Poisson bracket $\{ T \,
\overset{\otimes}, \,  T\}$ for coinciding intervals ($\sigma_1 =
\sigma'_1, \sigma_2 = \sigma'_2$) or adjacent intervals
($\sigma'_1 = \sigma_2$ or $\sigma_1 = \sigma'_2$) requires
defining the value of the discontinuous matrix-valued function
$\Delta^{(1)}$ at its discontinuities.

\begin{remark}
The discontinuities encountered here are all proportional to the
matrix $s$ and hence are absent in the ultralocal case ($s =
\partial_{\sigma} r = 0$), as it should be. Heuristically, the reason
for this difference can be understood from equation \eqref{TT from
JJ} which expresses the $\{ T, T \}$ bracket as a double integral
of the $\{ J_1, J_1 \}$ bracket. In the ultralocal case where the
bracket $\{ J_1, J_1 \}$ contains only $\delta$-singularities, the
bracket $\{ T, T \}$ is thus a continuous function. However, in
the non-ultralocal case where the bracket $\{ J_1, J_1 \}$
contains also $\delta'$-singularities, its double integral $\{ T,
T \}$ will still be a distribution, and indeed it contains
characteristic functions $\chi$ which are discontinuous.
\end{remark}

It is shown in \cite{Maillet} that requiring antisymmetry of the
Poisson bracket and the derivation rule to hold imposes the
symmetric definition of $\Delta^{(1)}$ at its discontinuous
points; for example at $\sigma_1 = \sigma'_1$ we must define
\begin{multline*}
\Delta^{(1)}(\sigma_1,\sigma_2,\sigma_1,\sigma'_2; x,x') \\ =
\lim_{\epsilon \rightarrow 0^+} \frac{1}{2} \left(
\Delta^{(1)}(\sigma_1,\sigma_2,\sigma_1 + \epsilon,\sigma'_2;
x,x') + \Delta^{(1)}(\sigma_1,\sigma_2,\sigma_1 -
\epsilon,\sigma'_2; x,x') \right),
\end{multline*}
and likewise for all other possible coinciding endpoints. This
definition is equivalent to assigning the value of $\frac{1}{2}$
to the characteristic function $\chi$ at its discontinuities.
Having thus defined $\Delta^{(1)}$ at its discontinuities we now
have a definition of the Poisson bracket $\{ T \,
\overset{\otimes}, \, T\}$ for coinciding and adjacent intervals
consistent with the antisymmetry of the Poisson bracket and the
derivation rule. However this definition of the $\{ T \,
\overset{\otimes}, \, T\}$ Poisson bracket does not satisfy the
Jacobi identity as is shown in \cite{Maillet}, so that in fact no
strong definition of the bracket $\{ T \, \overset{\otimes}, \,
T\}$ with coinciding or adjacent intervals can be given without
violating the Jacobi identity \cite{Maillet}. It is nevertheless
possible \cite{Maillet, Maillet2} to give a \dub{weak}\footnote{The
bracket is \textit{weak} in the sense that any multiple Poisson
bracket of $T$'s can be given a meaning which cannot be reduced to its
similarly defined constituent Poisson brackets, i.e. the multiple
Poisson bracket $\{ T \, \overset{\otimes}, \, \{ \ldots \{ T \,
\overset{\otimes}, \, T \} \ldots \} \}$ with $n$ factors of $T$
must be separately defined for each $n$.} definition of this
bracket for coinciding or adjacent intervals in a way that is
consistent with the Jacobi identity as follows: consider the
multiple Poisson bracket of $(n+1)$ transition matrices
\begin{multline*}
\Delta^{(n)}\left(\sigma^{(1)}_1, \sigma^{(1)}_2, \ldots,
\sigma^{(n+1)}_1, \sigma^{(n+1)}_2; x^{(1)}, \ldots,x^{(n+1)}\right) \\
= \left\{ T\left(\sigma^{(1)}_1,\sigma^{(1)}_2,x^{(1)}\right) \,
\overset{\otimes}, \, \left\{ \ldots \, \overset{\otimes}, \,
\left\{ T\left(\sigma^{(n)}_1,\sigma^{(n)}_2,x^{(n)}\right) \,
\overset{\otimes}, \,
T\left(\sigma^{(n+1)}_1,\sigma^{(n+1)}_2,x^{(n+1)}\right) \right\}
\ldots \right\} \right\},
\end{multline*}
which is unambiguously defined and continuous where
$\sigma^{(1)}_1, \sigma^{(1)}_2, \ldots, \sigma^{(n+1)}_1,
\sigma^{(n+1)}_2$ are all distinct, but again is discontinuous
across the hyperplanes defined by some of the points
$\sigma^{(1)}_1, \sigma^{(1)}_2, \ldots, \sigma^{(n+1)}_1,
\sigma^{(n+1)}_2$ being equal. The values of $\Delta^{(n)}$ at its
discontinuities are defined by employing a point splitting
regularisation followed by a total symmetrisation limit
\cite{Maillet}. For example, we define its value at
$\sigma^{(i)}_1 = \sigma_1, i = 1,\ldots,n+1$ by
\begin{multline*}
\Delta^{(n)}\left(\sigma_1, \sigma^{(1)}_2, \ldots, \sigma_1,
\sigma^{(n+1)}_2; x^{(1)}, \ldots,x^{(n+1)}\right) \\ =
\lim_{\epsilon \rightarrow 0^+} \frac{1}{(n+1)!} \sum_{p \in
S_{n+1}} \Delta^{(n)}\left(\sigma_1 + p(1) \epsilon,
\sigma^{(1)}_2, \ldots, \sigma_1 + p(n+1) \epsilon,
\sigma^{(n+1)}_2; x^{(1)}, \ldots,x^{(n+1)}\right),
\end{multline*}
and similarly one defines the value of $\Delta^{(n)}$ at all other
discontinuities. With the function $\Delta^{(n)}$ being defined at
its discontinuities we now have the definition of a weak bracket
which reduces to the normal Poisson bracket on quantities for
which the latter is continuous. It is shown in \cite{Maillet} that
the Jacobi identity for transition matrices with coinciding or
adjacent interval is now satisfied in terms of this weak bracket
($\Delta^{(2)}$ being the relevant quantity in this case).

\subsection*{$\bm{\{ \Omega, \Omega \}}$ algebra}

\begin{figure}
\centering \psfrag{g}{\footnotesize $\gamma$}
\psfrag{gp}{\footnotesize $\gamma'$} \psfrag{S1}{\footnotesize
$S^1$} \psfrag{g2}{\footnotesize $\tilde{\gamma}$}
\psfrag{gp2}{\footnotesize $\tilde{\gamma}'$}
\psfrag{gp3}{\footnotesize $\tilde{\gamma}'$}
\psfrag{0}{\footnotesize $0$} \psfrag{2p}{\footnotesize$2 \pi$}
\psfrag{R}{\footnotesize $\mathbb{R}$}
\includegraphics[height=40mm,width=110mm]{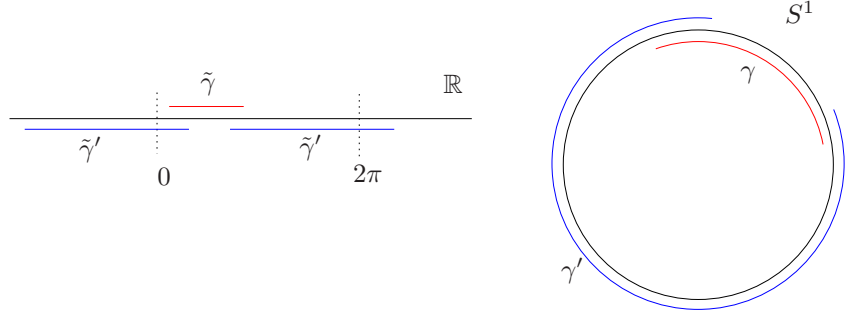}
\caption{Example of a path lifting required in computing Poisson
brackets of transition matrices on $S^1$ of the form $\{
T(\gamma,x) \, \overset{\otimes}, \, T(\gamma',x') \}$.}
\label{circle PB}
\end{figure}

Using this regularisation procedure we now derive an expression
for the Poisson bracket between two monodromy matrices in the
periodic case under consideration, a result which was first
obtained in \cite{Maillet2, Maillet}. To begin with consider the
Poisson bracket $\{ T(\gamma,x) \, \overset{\otimes}, \,
T(\gamma',x') \}$ between two generic transition matrices
$T(\gamma,x)$ and $T(\gamma',x')$ on the circle $S^1$, defined
relative to two different paths $\gamma$ and $\gamma'$ on $S^1$,
\textit{e.g.}
\begin{equation} \label{generic transition matrix}
T(\gamma,x) = P \overleftarrow{\exp} \int_{\gamma} d\sigma
J_1(\sigma,x).
\end{equation}
We would like to compute this bracket by working on the universal
cover $\mathbb{R}$ of $S^1$. So we choose a lift $\tilde{\gamma}$
of the path $\gamma$ to $\mathbb{R}$. Then because the only
contribution to the Poisson bracket comes from the region of
overlap between $\gamma$ and $\gamma'$ on $S^1$ (by \eqref{transition
matrix PB}) we have that \begin{equation} \label{PB on S^1}
\{ T(\gamma,x) \, \overset{\otimes}, \, T(\gamma',x') \} =
\sum_{\tilde{\gamma}' \text{ lift of } \gamma'} \{
T(\tilde{\gamma},x) \, \overset{\otimes}, \, T(\tilde{\gamma}',x')
\},
\end{equation}
where the sum is over lifts $\tilde{\gamma}'$ of $\gamma'$ to
$\mathbb{R}$. An example of these lifted paths is shown in Figure
\ref{circle PB}.
Let us now apply this formula to compute the Poisson bracket
between two transition matrices $\Omega(x,\sigma)$ and
$\Omega(x',\sigma)$ on $S^1$. The common interval $\gamma$ of both
matrices stretches once around the full circle and so it follows
that if we take $\tilde{\gamma} = [\sigma,\sigma + 2\pi]$ to be
the lift of the interval of $\Omega(x,\sigma)$ then there are only
three possibilities for the lift $\tilde{\gamma}'$ of the interval
of $\Omega(x',\sigma)$ which give a non-zero contribution to the
right hand side of \eqref{PB on S^1}, namely
\begin{equation} \label{contributing intervals}
[\sigma - 2 \pi,\sigma], \quad [\sigma,\sigma + 2 \pi], \quad
[\sigma + 2 \pi,\sigma + 4 \pi].
\end{equation}
Since the corresponding three brackets $\{ T(\tilde{\gamma},x) \,
\overset{\otimes}, \, T(\tilde{\gamma}',x') \}$ on $\mathbb{R}$
are over coinciding or adjacent intervals they need to be
regularised by the procedure described above. Let us start by
considering the coinciding interval bracket $\{ T(\sigma + 2
\pi,\sigma,x) \, \overset{\otimes}, \, T(\sigma + 2 \pi,\sigma,x')
\}$. There are 4 different possible point splittings of the
endpoints, each giving the same contribution (using
\eqref{transition matrix PB})
\begin{equation*}
r(x,x') \left( \Omega(x,\sigma) \otimes \Omega(x',\sigma) \right)
- \left( \Omega(x,\sigma) \otimes \Omega(x',\sigma) \right)
r(x,x')
\end{equation*}
in the limit of coinciding points. On the other hand, the adjacent
interval brackets (corresponding to the first and last choices for
$\tilde{\gamma}'$ in \eqref{contributing intervals}) each have two
possible point splittings and together they contribute, in the
coinciding end-point limit,
\begin{equation*}
\left( \Omega(x,\sigma) \otimes {\bf 1} \right) s(x,x') \left(
{\bf 1} \otimes \Omega(x',\sigma) \right) - \left( {\bf 1} \otimes
\Omega(x',\sigma) \right) s(x,x') \left( \Omega(x,\sigma) \otimes
{\bf 1} \right)
\end{equation*}
to the Poisson bracket of two monodromy matrices. The sum of the
last two expressions gives the right hand side of \eqref{PB on
S^1} which yields the sought-after (weak) Poisson bracket between
two monodromy matrices on $S^1$
\begin{equation} \label{fundamental Poisson bracket}
\begin{split}
\left\{ \Omega(x,\sigma) \mathop{,}^{\otimes} \Omega(x',\sigma)
\right\} = &[r(x,x'), \Omega(x,\sigma) \otimes \Omega(x',\sigma)] \\
+ &\left(\Omega(x,\sigma) \otimes {\bf 1}\right) s(x,x') \left( {\bf
1} \otimes \Omega(x',\sigma) \right) \\
- &\left( {\bf 1} \otimes \Omega(x',\sigma) \right) s(x,x') \left(
\Omega(x,\sigma) \otimes {\bf 1} \right).
\end{split}
\end{equation}

As a specific check of \eqref{fundamental Poisson bracket} we show
that the $SU(2)_R$ symmetry is canonically realised on $\Omega(x)$
via the weak Poisson bracket \cite{Maillet2}. Recall from equation
\eqref{monodromy asymptotics at infty} that the global Noether
charge $Q_R$ can be read off from the asymptotic expansion of the
monodromy matrix at $x = \infty$. Then starting with equation
\eqref{fundamental Poisson bracket} multiplied by $x \, (\epsilon
\otimes {\bf 1})$ and taking the trace over the first tensor
product space followed by the limit $x \rightarrow \infty$ one
deduces, using also the asymptotics $r(x,x') \sim_{x \rightarrow
\infty} \frac{2 \pi}{\sqrt{\lambda}} \frac{1-2 x'^2}{x(1-x'^2)} \eta$
and $s(x,x') \sim_{x \rightarrow \infty} \frac{2 \pi}{\sqrt{\lambda}}
\frac{1}{x(1-x'^2)} \eta$, that
\begin{equation*}
\left\{ \epsilon \cdot Q_R , \Omega(x') \right\} = \left[
\epsilon, \Omega(x') \right].
\end{equation*}
In other words, the right Noether charge $Q_R$ generates the
correct transformation on $\Omega(x)$, which we expect to be
\begin{equation*}
\Omega(x) \rightarrow U_R^{-1} \Omega(x) U_R,
\end{equation*}
provided we use the weak bracket \eqref{fundamental Poisson bracket}.

\subsection*{$\bm{\{ \text{tr} \, \Omega, \text{tr} \, \Omega \}}$ algebra}

Now consider the bracket $\left\{ \Omega(x,\sigma)^n
\overset{\otimes}, \Omega(x',\sigma)^m \right\}$ for any $n,m
\in \mathbb{N}$, which can easily be reduced to \eqref{fundamental
Poisson bracket} as follows (omitting the $\sigma$-dependence)
\begin{equation*}
\left\{ \Omega(x)^n \mathop{,}^{\otimes} \Omega(x')^m \right\} =
nm \left( \Omega(x)^{n-1} \otimes {\bf 1} \right) \left\{
\Omega(x) \mathop{,}^{\otimes} \Omega(x') \right\} \left( {\bf 1}
\otimes \Omega(x)^{m-1} \right).
\end{equation*}
Then using the standard notational shorthands $\overset{1}A = A
\otimes {\bf 1}$ and $\overset{2}A = {\bf 1} \otimes A$, and
taking the trace over both factors of the tensor product we find
\begin{align*}
\left\{ \tr \Omega(x)^n, \tr \Omega(x')^m \right\} &= nm \tr_{12}
\left( \overset{1}\Omega(x)^{n-1} \overset{2}\Omega(x')^{m-1}
\left\{ \overset{1} \Omega(x),
\overset{2}\Omega(x') \right\} \right) \\
&= nm \tr_{12} \left[r(x,x') + s(x,x'), \overset{1}\Omega(x)^n
\overset{2}\Omega(x')^m \right],
\end{align*}
where in the second line we have used \eqref{fundamental Poisson
bracket}. In conclusion we have arrived at the desired Poisson
bracket
\begin{equation} \label{involution of Omega}
\left\{ \tr \Omega(x)^n, \tr \Omega(x')^m \right\} = 0.
\end{equation}
Because this bracket is zero it can be understood as defining a
bracket in the \textit{strong} sense and without recourse to any
regularisation. We deduce from this last relation that the
invariants of the system encoded in the quantity $\tr \Omega(x)^n$
are in involution with respect to the Poisson bracket.

\subsection*{$\bm{\{ \text{tr} \, \Omega, \text{tr} \, \Omega \}_{\text{D.B.}}}$ algebra}

As explained in chapter \ref{chapter: strings on RxS3} we always
choose to work in conformal static gauge in order to isolate the
physical degrees of freedom of the string. This is done by
imposing static gauge conditions to fix the gauge invariance
generated by the Virasoro constraints. But within the Hamiltonian
description of chapter \ref{chapter: hamiltonian} these
constraints together form a set of second class constraints and so
to consistently impose them one must replace Poisson brackets by
the Dirac bracket \eqref{Dirac bracket SG}. However, as formula
\eqref{Dirac bracket SG} shows, this distinction between Poisson
and Dirac is unnecessary when one of the arguments is invariant
under conformal transformations generated by $L_n, \tilde{L}_n, n
\neq 0$. We now show that the generator of conserved charges $\tr
\Omega(x)^n$ is conformally invariant so that
\begin{equation*}
\{ \tr \Omega(x)^n, F \}_{\text{D.B.}} = \{ \tr \Omega(x)^n, F \}
\end{equation*}
for an arbitrary function $F$ of the principal chiral model fields
$j$. As a special case we deduce that the involution property
\eqref{involution of Omega} also holds with respect to the Dirac
bracket.

To show the conformal invariance of $\tr \Omega(x)^n$ let us start
with the Poisson bracket \eqref{PB T,J}. Once again, Poisson
brackets on $S^1$ are computed by working on the universal cover
$\mathbb{R}$. So let $\sigma_1 = \sigma + 2 \pi$, $\sigma_2 =
\sigma$ and $\sigma_3 = \sigma'$ in \eqref{PB T,J} to obtain the
Poisson bracket $\{ \Omega(\sigma,x) \overset{\otimes},
J_1(\sigma',x') \}$. This easily leads to the Poisson brackets $\{
\Omega(\sigma,x) \overset{\otimes}, j_{\pm}(\sigma') \}$ after
noting from the definition of $J_1(x)$ that $J_1(0) =
\frac{1}{2}(j_+ - j_-)$ and $ \lim_{x \rightarrow \infty} (-x)
J_1(x) = \frac{1}{2}(j_+ + j_-)$, in particular
\begin{multline*}
\{ \Omega(\sigma, x) \overset{\otimes}, j_{\pm}(\sigma') \} =
(T(\sigma + 2 \pi,\sigma',x) \otimes {\bf 1}) \times \\ \left(
(\delta(\sigma' - \sigma - 2 \pi) - \delta(\sigma' - \sigma))
\frac{4 \pi}{\sqrt{\lambda}} \frac{1 \pm x}{1 - x^2} \eta
+ \chi(\sigma'; \sigma + 2 \pi,\sigma) \times \quad \qquad \right.\\
\left. \quad \qquad \left[ -
\frac{2 \pi}{\sqrt{\lambda}} \frac{2 x}{1 - x^2} \eta, (x \pm 1)
J_1(\sigma', x) \otimes {\bf 1} \pm {\bf 1} \otimes \frac{1}{2}
(j_+(\sigma') - j_-(\sigma')) \right] \right) \\ \times
(T(\sigma',\sigma,x) \otimes {\bf 1}).
\end{multline*}
Using the identity $\text{tr}_2 (\eta {\bf 1} \otimes A) = A$ for
any matrix $A \in \mathfrak{su}(2)$ one can show that after
multiplying the above equation by ${\bf 1} \otimes
j_{\pm}(\sigma')$ and taking the trace $\text{tr}_2$ over the
second tensor factor the commutator disappears and we are left
with
\begin{multline*}
\left\{ \Omega(\sigma, x), \frac{1}{2} \text{tr}\,
j^2_{\pm}(\sigma') \right\}\\ = \frac{4 \pi}{\sqrt{\lambda}}
(\delta(\sigma' - \sigma - 2 \pi) - \delta(\sigma' - \sigma))
T(\sigma + 2 \pi,\sigma',x) J_{\pm}(\sigma',x)
T(\sigma',\sigma,x),
\end{multline*}
where $J_{\pm}(\sigma',x) = j_{\pm}(\sigma')/(1 \mp x)$. Next we
multiply both sides by $e^{\pm i n \sigma'}$ and integrate over
$\sigma'$ from $0$ to $2 \pi$. However, since we are on the
universal cover $\mathbb{R}$ of $S^1$ we get two non-zero
contributions, namely from the integrations over the two lifts
$[0, 2\pi]$ and $[2\pi, 4\pi]$ (assuming $\sigma \in (0, 2\pi)$).
From the definition \eqref{Virasoro modes} of the Virasoro
generators we can write the result as follows
\begin{equation} \label{monodromy conformal transf}
\begin{split}
\{ \Omega(\sigma, x), L_n \} &= \frac{1}{2} e^{i n
\sigma} [J_+(\sigma,x), \Omega(\sigma,x)], \\
\{ \Omega(\sigma, x), \tilde{L}_n \} &= \frac{1}{2} e^{-
i n \sigma} [J_-(\sigma,x), \Omega(\sigma,x)].
\end{split}
\end{equation}
Note that in the above calculation it is because of the presence
of the $s$-matrix, which arises from non-ultralocality of the
Poisson brackets of the model, that we end up with the correct
transformation property for $\Omega(x)$ under conformal
transformations. Finally, since the right hand sides are
commutators, taking the trace shows that $\tr \Omega(x)^m$ is
invariant under conformal transformations generated by $L_n,
\tilde{L}_n$, namely
\begin{equation*}
\{ \tr \Omega(x)^m, L_n \} = \{ \tr \Omega(x)^m, \tilde{L}_n \} =
0.
\end{equation*}
As we have already argued, this immediately implies the involution
of the conserved charges with respect to the Dirac bracket
\begin{equation*}
\left\{ \tr \Omega(x)^n, \tr \Omega(x')^m \right\}_{\text{D.B.}} =
0.
\end{equation*}
This completes the proof of theorem \ref{theorem: complete
integrability}.

\subsection*{$\bm{\{ \Omega, \Omega \}_{\text{D.B.}}}$ algebra}

In fact we can prove a much stronger result that will be useful later
in chapter \ref{chapter: symplectic}. Combining the relations
\eqref{monodromy conformal transf} for the conformal transformation of
the monodromy matrix with the definition \eqref{Dirac bracket
SG} of the Dirac bracket we can compute the Dirac algebra $\{
\Omega(x), \Omega(x') \}_{\text{D.B.}}$. One finds, using the fact that
the partial sums of $\sum_{n \neq 0} \frac{1}{n}$ vanish, that it is
identical to the Poisson algebra, namely
\begin{proposition}
The Dirac bracket between two monodromy matrices on $S^1$ is given on
the reduced phase-space $\mathcal{P}^{\infty}$ by
\begin{equation} \label{fundamental Dirac bracket}
\begin{split}
\{ \Omega(x) \overset{\otimes}, \Omega(x') \}_{\text{D.B.}} \approx
&[r(x,x'), \Omega(x) \otimes \Omega(x')] \\
+ &\left(\Omega(x) \otimes {\bf 1}\right) s(x,x') \left( {\bf 1}
\otimes \Omega(x') \right)\\
- &\left( {\bf 1} \otimes \Omega(x') \right) s(x,x') \left( \Omega(x)
  \otimes {\bf 1} \right).
\end{split}
\end{equation}
\end{proposition}


\section{The string hierarchy} \label{section: hierarchy}


An immediate consequence of theorem \ref{theorem: complete
integrability} is that the charges $\tr \Omega(x)^n$ are not only
conserved under the $\tau$- and $\sigma$-flows generated by
$\mathcal{E}$ and $\mathcal{P}$ but also under the flows generated
by the infinite number of charges $\tr \Omega(x')^m$ themselves.
In particular this is true for the flows generated by all the
local charges $Q_n^{(\pm)}$. It follows that if we treat any two
local charges $Q_m^{(\pm)}$ and $Q_n^{(\pm)}$ as Hamiltonians
instead of $\mathcal{E}$ and $\mathcal{P}$ then the corresponding
equations of motion will be integrable since they also admit the
infinite number of conserved charges $\tr \Omega(x)^n$. In the
light of section \ref{section: conserved charges} we therefore
expect these equations of motion to admit a Lax representation in
terms of some Lax connection with components $J_{m, \pm}(x)$ and
$J_{n, \pm}(x)$. In the following section we show that this is indeed
the case and derive the corresponding expressions for the Lax
matrix $J_{n, \pm}(x)$ associated with the local charge
$Q_n^{(\pm)}$.

Our starting point is the $\{ T, J_1 \}$ Poisson bracket \eqref{PB
T,J}. Let $\sigma_1 = \sigma + 2 \pi, \sigma_2 = \sigma, \sigma_3
= \sigma'$ in \eqref{PB T,J} and identify the monodromy matrix as
$\Omega(\sigma,x) = T(\sigma + 2 \pi, \sigma, x)$ then
\begin{multline*}
\{ \Omega(\sigma,x) \overset{\otimes}, J_1(\sigma', x') \}\\
\begin{split}
= &(T(\sigma  + 2 \pi,\sigma',x) \otimes {\bf 1}) [ (r + s)(x,x'),
J_1(\sigma', x) \otimes {\bf 1} + {\bf 1} \otimes J_1(\sigma', x')
] (T(\sigma',\sigma,x) \otimes {\bf 1})\\
&-2 (\delta(\sigma' - \sigma - 2 \pi) - \delta(\sigma' - \sigma))
(T(\sigma + 2 \pi,\sigma',x) \otimes {\bf 1}) s(x,x')
(T(\sigma',\sigma,x) \otimes {\bf 1}).
\end{split}
\end{multline*}
Taking the trace over the first factor of the tensor product we
observe that the left hand side of this equation becomes
independent of $\sigma$. Likewise, the first term on the right
hand side also becomes independent of $\sigma$ using the
translation invariance of the transition matrix $T$ by $2 \pi$
since we are working on the circle $S^1$, \textit{i.e.} $T(\sigma
+ 2 \pi, \sigma', x) = T(\sigma, \sigma' - 2 \pi, x)$. This shows
that the last term must also be independent of $\sigma$ after
taking the trace over the first tensor factor and hence one can
substitute its value at $\sigma \neq \sigma'$ which is zero. We
therefore end up with, after using \eqref{eq for transition
matrix}
\begin{equation} \label{pre zero-curvature}
\{ \tr \Omega(x) , J_1(\sigma', x') \} =
\partial_{\sigma'} \mathcal{J}(\sigma', x, x') +
[\mathcal{J}(\sigma', x, x'), J_1(\sigma', x')],
\end{equation}
where
\begin{equation} \label{Lax connection 0}
\mathcal{J}(\sigma', x, x') = \text{tr}_{1} \left[
(\Omega(\sigma',x) \otimes {\bf 1}) (r + s)(x,x') \right].
\end{equation}
If we interpret the Poisson bracket $\{ \tr \Omega(x) ,
J_1(\sigma', x') \}$ in \eqref{pre zero-curvature} as the ``time''
derivative of $J_1(\sigma', x')$ with respect to the ``time''
generated by the Hamiltonian $\text{tr}\, \Omega(x)$ then
\eqref{pre zero-curvature} takes exactly the form of a
zero-curvature equation. One can also obtain the equations of
motion for the monodromy matrix with respect to the Hamiltonian
$\text{tr}\, \Omega(x)$. Starting from the Poisson algebra of the
monodromies \eqref{fundamental Poisson bracket} and taking the
trace over the first factor of the tensor product as above yields
\begin{equation} \label{pre monodromy evolution}
\{ \tr \Omega(x) , \Omega(\sigma', x') \} = [\mathcal{J}(\sigma',
x, x'), \Omega(\sigma', x')].
\end{equation}
Once again, if we interpret the Poisson bracket $\{ \tr \Omega(x)
, \Omega(x') \}$ as a ``time'' derivative, this last equation take
the same form as the $(\sigma,\tau)$-evolution equation of the
monodromy matrix \eqref{monodromy evolution}. So equations
\eqref{pre zero-curvature} and \eqref{pre monodromy evolution}
both suggest that \eqref{Lax connection 0} is the Lax matrix
corresponding to all the higher order flows generated by the
Hamiltonians $\tr \Omega(x)$, just as $J_0$ and $J_1$ were the Lax
matrices generating $\tau$- and $\sigma$-flows respectively.
However, what we are really interested in are the Lax matrices
corresponding to the local charges of section \ref{section: local
charges}. And according to theorem \ref{theorem: local charges}
these are related to the coefficients of the Taylor expansion of
\eqref{Lax connection 0} around $x = \pm 1$.

\subsection*{Lax matrices}

Using expressions \eqref{r,s-matrices} for the $(r,s)$-matrices
their sum which enters in \eqref{Lax connection 0} is given by
$(r+s)(x,x') = - \frac{2 \pi}{\sqrt{\lambda}} \frac{2
x^2}{(x-x')(1 - x^2)} \eta$. Now by definition, $\eta = t_a
\otimes t^a$ where the $\mathfrak{su}(2)$ generator $t^a$ is
related to the Pauli matrices as $t^a = \frac{i}{\sqrt{2}}
\sigma_a$. Therefore the Lax matrix \eqref{Lax connection 0} can be
written more explicitly as
\begin{equation} \label{Lax connection 2}
\mathcal{J}(\sigma', x, x') = - \frac{\pi}{\sqrt{\lambda}} \frac{2
x^2}{(x-x')(1 - x^2)} \tr \left[ \Omega(\sigma',x) \, \sigma_a
\right] \sigma_a.
\end{equation}
Now it is straightforward to show that for any matrix $A \in
SL(2,\mathbb{C})$ the following is true
\begin{equation} \label{matrix identity}
V^{-1} \frac{\tr [A \sigma_a] \sigma_a}{\lambda_+ - \lambda_-} V =
\sigma_3, \qquad \text{where} \quad V^{-1} A V = \text{diag}\,
(\lambda_+,\lambda_-),
\end{equation}
\textit{i.e.} $V$ is the matrix of eigenvectors of $A$ and
$\lambda_{\pm}$ are the eigenvalues. Since $\Omega(\sigma', x)$
has unit determinant let us denote its eigenvalues by $e^{\pm i
p(x)}$. Let us also denote the corresponding matrix of
eigenvectors as $\Psi(\sigma', x)$ or simply $\Psi(x)$, omitting
the $\sigma'$ dependence for clarity. In particular we know from
equation \eqref{diagonal Omega pm 1} that in a neighbourhood of $x
= \pm 1$ we have
\begin{equation} \label{Psi near pm 1}
\Psi(x) = g_{(\pm)}(x)^{-1}, \qquad p(x) = \sum_{n = -1}^{\infty}
Q^{(\pm)}_n (x \mp 1)^n.
\end{equation}
The identity \eqref{matrix identity} applied to the Lax matrix
\eqref{Lax connection 2} corresponding to $\tr \Omega(x)$ yields
\begin{equation} \label{Lax connection 3}
\tr \Omega(x) \quad \longleftrightarrow \quad \mathcal{J}(x, x') =
\frac{4 \pi i}{\sqrt{\lambda}} \frac{\sin p(x)}{1 - 1/x^2}
\frac{\Psi(x) \sigma_3 \Psi(x)^{-1}}{x-x'}.
\end{equation}

\begin{remark}
From now on we indicate the correspondence between a charge $Q$
and its Lax matrix $L(x')$ by the shorthand notation $Q
\longleftrightarrow L(x')$. It is to be understood as meaning that
$Q$ and $L(x')$ are related by an equation of the form $\{ Q,
J_1(x') \} =
\partial_{\sigma'} L(x') + [L(x'), J_1(x')]$. For instance \eqref{Lax
connection 3} is to be read as \eqref{pre zero-curvature}.
\end{remark}

But since $\tr \Omega(x) = 2 \cos p(x)$, it follows that the Lax
matrix responsible for the flow of the Hamiltonian $p(x)$ is
\begin{equation} \label{Lax connection 4}
p(x) \quad \longleftrightarrow \quad J(x, x') = - \frac{2 \pi
i}{\sqrt{\lambda}} \frac{x^2}{x^2 - 1} \frac{\Psi(x) \sigma_3
\Psi(x)^{-1}}{x-x'}.
\end{equation}
Now by expanding this around $x = \pm 1$ and using \eqref{Psi near
pm 1} we can extract the Lax matrices associated with each local
charge $Q_{n-1}^{(\pm)}$, namely for $n \geq 0$
\begin{equation} \label{Lax connection 5}
Q_{n-1}^{(\pm)} \quad \longleftrightarrow \quad
\tilde{J}_{n,\pm}(x') = \res_{x = \pm 1} \, (x \mp 1)^{-n}
J(x,x').
\end{equation}
Using the straightforward identity for any rational matrix $M(x)$
with singularities at $x = \pm 1$
\begin{equation} \label{res singular part identity}
\res_{x = \pm 1} \, \frac{M(x)}{x - x'} dx = - \left( M(x')
\right)_{\pm 1},
\end{equation}
where $\left( M(x') \right)_{\pm 1}$ denotes the pole part of
$M(x')$ at $x' = \pm 1$, one can recast the Lax matrix \eqref{Lax
connection 5} in the much more useful form
\begin{equation} \label{Lax connection 6}
Q_{n-1}^{(\pm)} \quad \longleftrightarrow \quad
\tilde{J}_{n,\pm}(x') = \left( \frac{2 \pi i}{\sqrt{\lambda}}
\frac{x'^2}{x'^2 - 1} \frac{g_{(\pm)}(x')^{-1} \sigma_3
g_{(\pm)}(x')}{(x' \mp 1)^n} \right)_{\pm 1},
\end{equation}
where we have used the asymptotics \eqref{Psi near pm 1} of
$\Psi(x')$ near $x' = \pm 1$. At the zeroth level $n=0$ equation
\eqref{Lax connection 6} reads
\begin{equation*}
Q_{-1}^{(\pm)} \quad \longleftrightarrow \quad
\tilde{J}_{0,\pm}(x') = \pm \frac{\pi i}{\sqrt{\lambda}}
\frac{g_{(\pm)}(\pm 1)^{-1} \sigma_3 g_{(\pm)}(\pm 1)}{x' \mp 1}.
\end{equation*}
This Lax matrix is almost equal to $J_{\pm}(x')$ given in
\eqref{Lax connection near pm 1}. So let us introduce an
alternative basis $J_{n,\pm}$ of Lax matrices whose zeroth level
$n=0$ will correspond exactly to the components of the Lax
connection $J_{\pm}$. It follows from \eqref{Lax connection 6}
that we have the following correspondence between integral of
motion and Lax matrix
\begin{equation} \label{Lax connection 7}
\frac{\sqrt{\lambda}}{2 \pi^2} Q_{-1}^{(\pm)} Q_{n-1}^{(\pm)}
\quad \longleftrightarrow \quad J_{n,\pm} \equiv
\frac{\sqrt{\lambda}}{2 \pi^2} \left( Q_{-1}^{(\pm)}
\tilde{J}_{n,\pm} + Q_{n-1}^{(\pm)} \tilde{J}_{0,\pm} \right).
\end{equation}
We see from \eqref{E and P as local charges} and \eqref{Lax
connection near pm 1} that the zeroth level $n = 0$ of this
hierarchy is precisely the Lax connection $J_{\pm}$ associated
with $\mathcal{E} \pm \mathcal{P}$, hence $J_{0,\pm} = J_{\pm}$ as
desired. So we define,
\begin{definition}
The \dub{string hierarchy} is generated by the Hamiltonians
\begin{equation} \label{string hierarchy Hamiltonians}
H_{n,\pm} \equiv \frac{\sqrt{\lambda}}{2 \pi^2} Q_{-1}^{(\pm)}
Q_{n-1}^{(\pm)}.
\end{equation}
\end{definition}

\subsection*{Higher times}

At this point we can also define a hierarchy of times
$\tilde{t}_{n,\pm}$ parameterising the flows generated by the
Hamiltonians $Q_{n-1}^{(\pm)}$ of \eqref{Lax connection 5}, namely
we define
\begin{equation*}
\partial_{\tilde{t}_{n,\pm}} = \left\{ Q_{n-1}^{(\pm)}, \cdot
\right\}_{\text{D.B.}}.
\end{equation*}
However since it is preferable to work in terms of the alternative
basis of Lax matrices $J_{n, \pm}(x')$ which reduced to the Lax
connection $J_{\pm}(x')$ at the zeroth level, we define the
corresponding higher times,
\begin{definition} \label{hierarchy of times}
The \dub{hierarchy of times} $t_{n, \pm}$ of the hierarchy are
defined by
\begin{equation*}
\partial_{t_{n,\pm}} = \left\{ H_{n,\pm} , \cdot
\right\}_{\text{D.B.}}.
\end{equation*}
\end{definition}

When we will need to be explicit about the dependence of a function
$f$ on all the higher times we will write simply $f(t)$ using the
notation $\{t\}$ for the complete set of times $\{ t_{0,\pm},
t_{1,\pm}, \ldots \}$. Let us also denote the multi-indices labelling
the hierarchy, such as $(n,+)$, using capital letters, \textit{e.g.}
$N = (n,s)$ where $n \in \mathbb{N}$ and $s = \pm 1$.

\subsection*{Zero-curvature}

Going back to equation \eqref{pre monodromy evolution}, if we
follow the prescription that lead us from \eqref{Lax connection 3}
to \eqref{Lax connection 5}, namely of dividing through by $- 2
\sin p(x)$ and taking the residue at $x = \pm 1$ one easily shows,
\begin{equation} \label{monodromy evolution hierarchy 2}
[\partial_{\tilde{t}_N} - \tilde{J}_N(x'), \Omega(x')] = 0.
\end{equation}
By linearity of the definition \eqref{Lax connection 7} of the Lax
matrices $J_N(x')$ in terms of the $\tilde{J}_N(x')$ and using the
fact that the local charges $Q_{n-1}^{(\pm)}$ are constant with
respect to the higher times $\tilde{t}_M$ we deduce the following,
\begin{subequations} \label{evolution hierarchy}
\begin{proposition} \label{proposition: monodromy evolution hierarchy}
The evolution of the monodromy matrix under the hierarchy of times
\eqref{hierarchy of times} is governed by
\begin{equation} \label{monodromy evolution hierarchy}
[\partial_{t_N} - J_N(x'), \Omega(x')] = 0,
\end{equation}
which is exactly of the form \eqref{monodromy evolution}.
\begin{proof}
$\partial_{t_{n,\pm}} \Omega(x') = \frac{\sqrt{\lambda}}{2 \pi^2}
Q_{-1}^{(\pm)} \{ Q_{n-1}^{(\pm)} , \Omega(x') \}_{\text{D.B.}} +
\frac{\sqrt{\lambda}}{2 \pi^2} Q_{n-1}^{(\pm)} \{ Q_{-1}^{(\pm)} ,
\Omega(x') \}_{\text{D.B.}}$, which using \eqref{monodromy evolution
hierarchy 2} equals $\frac{\sqrt{\lambda}}{2 \pi^2}
[Q_{-1}^{(\pm)} \tilde{J}_{n,\pm}(x') + Q_{n-1}^{(\pm)}
\tilde{J}_{0,\pm}(x'), \Omega(x')] = [J_{n,\pm}(x'), \Omega(x')]$.
\end{proof}
\end{proposition}

Finally we derive the evolution equations for the Lax matrices
\eqref{Lax connection 7} under the hierarchy of times in
definition \ref{hierarchy of times} and show that they take the
zero-curvature form. We closely follow an argument given in
\cite[p51-52]{Babelon} for finite-dimensional systems which applies
readily here.
\begin{proposition} \label{proposition: zero-curvature hierarchy}
The Lax matrices \eqref{Lax connection 7} satisfy the
zero-curvature condition
\begin{equation} \label{zero-curvature hierarchy}
[\partial_{t_N} - J_N(x'), \partial_{t_M} - J_M(x')] = 0.
\end{equation}
\end{proposition}
\end{subequations}
\begin{proof}
As for proposition \ref{proposition: monodromy evolution
hierarchy} we first prove that the zero-curvature equation holds
for the matrices $\tilde{J}_M(x')$ and times $\tilde{t}_M$, namely
\begin{equation} \label{zero-curvature}
\partial_{\tilde{t}_M} \tilde{J}_N(x') - \partial_{\tilde{t}_N}
\tilde{J}_M(x') = [\tilde{J}_M(x'), \tilde{J}_N(x')].
\end{equation}
Equation \eqref{zero-curvature hierarchy} will then follow by
linearity and the constancy of the $Q_{n-1}^{(\pm)}$. Writing the
monodromy matrix as $\Omega(x') = \Psi(x')\, \text{diag}(e^{i
p(x)}, e^{-i p(x)}) \Psi(x')^{-1}$, equation \eqref{monodromy
evolution hierarchy 2} implies
\begin{equation} \label{Psi evolution}
\left[ \Psi(x')^{-1} \left(\partial_{\tilde{t}_N} \Psi(x') \right)
- \Psi(x')^{-1} \tilde{J}_N(x') \Psi(x'), \text{diag}(e^{i p(x)},
e^{-i p(x)}) \right] = 0.
\end{equation}
But any $2 \times 2$ matrix commuting with a diagonal matrix must
itself be diagonal, and therefore we may write
\begin{equation} \label{Psi evolution 2}
\partial_{\tilde{t}_N} \Psi(x') = \tilde{J}_N(x') \Psi(x') + \Psi(x')
D(x'),
\end{equation}
for some unknown diagonal $2 \times 2$ matrix $D(x')$. Let $N =
(n,s_n)$ and $M = (m,s_m)$, then
\begin{equation} \label{J^N evolution}
\partial_{\tilde{t}_M} \tilde{J}_N(x') = \left[ \tilde{J}_M(x'), \frac{2 \pi
i}{\sqrt{\lambda}} \frac{x'^2}{x'^2 - 1} \frac{\Psi(x') \sigma_3
\Psi(x')^{-1}}{(x' - s_n)^n} \right]_{s_n},
\end{equation}
where we have made use of \eqref{Psi evolution 2} and the
subscript on the commutator means we take the pole part of the
whole commutator at $x' = s_n$. Let us start by assuming that $s_n
\neq s_m$, then $\tilde{J}_M(x')$ is regular at $x' = s_n$ and
only the pole part at $x' = s_n$ of the second term in the
commutator contributes which is just $\tilde{J}_N(x')$, so
\begin{equation*}
\partial_{\tilde{t}_M} \tilde{J}_N(x') = [\tilde{J}_M(x'), \tilde{J}_N(x')]_{s_n},
\end{equation*}
and likewise we also have $\partial_{\tilde{t}_N} \tilde{J}_M(x')
= [\tilde{J}_N(x'), \tilde{J}_M(x')]_{s_m}$. Since
$[\tilde{J}_M(x'), \tilde{J}_N(x')]$ is rational with poles only
at $x' = \pm 1$ and vanishes at $x' = \infty$ it can be written as
a sum over its pole parts, namely
\begin{equation*}
[\tilde{J}_M(x'), \tilde{J}_N(x')] = [\tilde{J}_M(x'),
\tilde{J}_N(x')]_{+1} + [\tilde{J}_M(x'), \tilde{J}_N(x')]_{-1}.
\end{equation*}
But because $s_n \neq s_m$ we have $\{ s_m, s_n\} = \{ \pm 1\}$
and the zero-curvature condition \eqref{zero-curvature} follows.
If instead we assume that $s_n = s_m$, then we have
\begin{equation*}
\left[ \tilde{J}_N(x') -  \frac{2 \pi i}{\sqrt{\lambda}}
\frac{x'^2}{x'^2 - 1} \frac{\Psi(x') \sigma_3 \Psi(x')^{-1}}{(x' -
s_n)^n}, \tilde{J}_M(x') -  \frac{2 \pi i}{\sqrt{\lambda}}
\frac{x'^2}{x'^2 - 1} \frac{\Psi(x') \sigma_3 \Psi(x')^{-1}}{(x' -
s_n)^m} \right]_{s_n} = 0
\end{equation*}
since both arguments in the commutator are regular at $x' = s_n =
s_m$. The zero-curvature equation \eqref{zero-curvature} again
readily follows from the above equation and \eqref{J^N evolution}.
\end{proof}

\subsection*{Gauge redundancy}

The form of the zero-curvature equations \eqref{zero-curvature
hierarchy} is invariant under gauge transformations. If $\tilde{g}(t)$
is an arbitrary matrix depending on all the higher times $\{ t \}$
then the new Lax connections defined by the transformation
\begin{equation} \label{gauge transformation}
J_M(x') \mapsto \tilde{g} J_M(x') \tilde{g}^{-1} + \left(
\partial_{t_M} \tilde{g} \right) \tilde{g}^{-1}
\end{equation}
also satisfy the zero-curvature equations \eqref{zero-curvature
hierarchy}. The gauge transformation parameter $\tilde{g}$ will always
be taken to be independent of the spectral parameter $x$. This choice
obviously preserves the analytic properties of the Lax
matrices\footnote{It is also possible to choose $\tilde{g}$ to depend on
$x$ and still preserve the analytic properties of the Lax
matrices. The corresponding gauge transformations \eqref{gauge
transformation} give rise to \dub{B\"acklund transformations} which
allow one to construct new solutions from old ones. For a review of
such \dub{dressing methods} see \cite{Fordy, Manas} and
\cite[p74-79]{Babelon} as well as \cite{Spradlin:2006wk,
Kalousios:2006xy} for an application in the context of strings on
$\mathbb{R} \times S^5$.}. However, starting from the form \eqref{Lax
connection 7} of the Lax matrices which are all expressed as singular
parts, the transformation \eqref{gauge transformation} will
generically add a term constant in $x$. Therefore the Lax matrices
\eqref{Lax connection 7} correspond to the gauge choice
\begin{equation} \label{gauge choice}
J_M(\infty) = 0.
\end{equation}
When solving the zero-curvature equation we will use this gauge choice
to extract the Lax matrices in the form \eqref{Lax connection 7}. In
particular, extracting the current $j$ from the Lax connection $J(x)$
will require bringing the latter to the defining form \eqref{Lax
connection} and this is achieved by imposing $J(\infty) = 0$.

Even after imposing the gauge choice \eqref{gauge choice} there
remains a residual gauge transformation by \textit{constant}
matrices $\tilde{g}$. Requiring also that gauge transformations
preserve the reality conditions on the Lax matrices will lead to
the further restriction $\tilde{g} \in SU(2)$. But this residual
symmetry is nothing but the global $SU(2)_R$ symmetry \eqref{R
symmetry on j} of the original equations of motion.

%% file: Curves.tex
\newpage

\chapter{Some curves} \label{chapter: curves}

\begin{flushright}
{\small \textit{``Drama is life with the dull bits cut out.''}}\\
{\small Sir Alfred Joseph Hitchcock}
\end{flushright}
\vspace{1cm}

\noindent One of the key ingredients of chapter \ref{chapter:
integrability} that eventually lead to complete integrability was the
$1$-parameter family of flat currents $J(x)$ on the worldsheet
which crucially depended on an auxiliary complex parameter $x \in
\mathbb{C}$, called the \dub{spectral parameter}. Expanding the
eigenvalues of $\Omega(x)$ in this variable produced an infinite
number of integrals of motion. Now instead of expanding in the
spectral parameter to extract individual integrals of motion, consider
the characteristic polynomial \eqref{spectral curve 0} of the
monodromy matrix $\Omega(x)$ which neatly encodes all the integrals of
motion,
\begin{equation*}
\Gamma(x,\Lambda) \equiv \det(\Lambda {\bf 1} - \Omega(x)).
\end{equation*}
The presence of a spectral parameter makes the characteristic
polynomial depend on two complex variables which therefore defines
a curve $\Gamma \subset \mathbb{C}^2$ via the equation
\begin{equation*}
\Gamma(x,\Lambda) = 0.
\end{equation*}
It follows that to every solution $j$ of the equations of motion
one can assign an invariant curve $\Gamma$ which encodes all its
integrals of motion. The major problem with this curve however is
that it is infinitely singular and non-algebraic so the object of
this chapter is to desingularise it and obtain a Riemann surface
on which we can perform complex analysis in the subsequent
chapters.


\section{The spectral curve}  \label{section: spectral curve}


Since the evolution of the monodromy matrix $\Omega(x)$ with respect
to all the higher times $\{ t \}$ is isospectral by \eqref{monodromy
evolution hierarchy}, its characteristic equation defines a complex
curve in $\mathbb{C}^2$ independent of all the higher times,
\begin{definition}
The \dub{spectral curve} $\Gamma$ is a curve in $\mathbb{C}^2$
defined by
\begin{equation} \label{spectral curve 1}
\Gamma : \;\; \Gamma(x,\Lambda) \equiv \det(\Lambda {\bf 1} -
\Omega(x)) = 0.
\end{equation}
\end{definition}
\noindent It is a $2$-sheeted branched cover in the sense that
$\hat{\pi} : \Gamma \rightarrow \mathbb{C}, (x,\Lambda) \mapsto x$ is
surjective and two to one (almost everywhere). Indeed, since
$\Omega(x)$ is $2 \times 2$ it has at most two distinct
eigenvalues $\Lambda_{\pm}(x)$ with corresponding points
$\mathfrak{P}_{\pm} = (x,\Lambda_{\pm}) \in \Gamma$ in
$\hat{\pi}^{-1}(x)$. Note that $\Omega(x)$ having unit determinant
means
\begin{equation} \label{unimodular evalues}
\Lambda_+(x) \Lambda_-(x) = 1.
\end{equation}
But at values of $x$ for which these eigenvalues coincide,
$\hat{\pi}^{-1}(x)$ is a single point on $\Gamma$ which can be
either a branch point or a singular point. Note also that $\Gamma$
admits a natural holomorphic involution
\begin{equation} \label{holomorphic involution}
\hat{\sigma} : \;\; \Gamma \rightarrow \Gamma, \quad (x,
\Lambda) \mapsto (x, \Lambda^{-1})
\end{equation}
with the property that $\hat{\pi} \circ \hat{\sigma} = \hat{\pi}$ and
it is clear from \eqref{unimodular evalues} that $\hat{\sigma}$
interchanges the points $\mathfrak{P}_{\pm} \in \hat{\pi}^{-1}(x)$ for
any $x \in \mathbb{C}$. Moreover, the fixed points of $\hat{\sigma}$
precisely correspond to the branch points and singular points of
$\Gamma$.

Although the spectral curve is a very natural curve to consider it
is not algebraic. To see this recall that the evolution of the
monodromy matrix can be expressed in terms of the differential
equation \eqref{monodromy evolution hierarchy}.
We deduce from Poincar\'e's theorem on the analytic dependence of
solutions on the initial conditions and parameters\footnote{The
  solution $\bm{x}(t) \in \mathbb{C}^n$ to the differential equation
  $\frac{d \bm{x}}{dt} = \bm{F}(t,\bm{x})$ depends holomorphically
  on the initial value $\bm{x}_0 \in \mathbb{C}^n$ and on any other
  parameter provided the vector function $\bm{F}$ itself depends
  holomorphically on these parameters.} that $\Omega(x)$ is
holomorphic in $\mathbb{C} \setminus \{ \pm 1 \}$. From its
asymptotics at infinity \eqref{monodromy asymptotics at infty},
$\Omega(x)$ is also holomorphic at $x = \infty$. On the other hand,
proposition \ref{proposition: monodromy asymptotics} shows that in a
neighbourhood of the points $x = \pm 1$ the eigenvalues
$\Lambda_{\pm}(x)$ have essential singularities from which it follows
that \eqref{spectral curve 1} does not define an algebraic curve since
$\Gamma(x,\Lambda) = (\Lambda_+(x) - \Lambda)(\Lambda_-(x) - \Lambda)$
is not rational in $x \in \mathbb{C}$.

To determine the values of $x \in\mathbb{C}$ over which the cover
$\hat{\pi} : \Gamma \rightarrow \mathbb{C}$ branches we consider
the discriminant of the polynomial $\Gamma(x,\cdot)$,
\begin{equation} \label{discriminant Gamma}
\Delta_{\Gamma}(x) = \left( \Lambda_+(x) - \Lambda_-(x) \right)^2
\end{equation}
and let $\mathcal{Z}_{\Gamma} = \{ x \in \mathbb{C} \,|\,
\Delta_{\Gamma}(x) = 0 \}$ be its set of zeroes. This corresponds
to the set of $x \in \mathbb{C}$ where the two eigenvalues
coincide $\Lambda_+(x) = \Lambda_-(x)$. In particular, $\Omega(x)$ is
diagonalisable for all $x \in \mathbb{C} \setminus
\mathcal{Z}_{\Gamma}$ and at any $x_0 \in \mathcal{Z}_{\Gamma}$ we
have $\Lambda_+(x_0) = \pm 1$ by \eqref{unimodular evalues}. Now since
$\Delta_{\Gamma}(x)$ is meromorphic on $\mathbb{C} \setminus \{
\pm 1 \}$ its zeroes must be isolated so that the set
$\mathcal{Z}_{\Gamma}$ is discrete. However, since
$\Delta_{\Gamma}(x)$ has essential singularities at $x = \pm 1$ it
follows that the set $\mathcal{Z}_{\Gamma}$ accumulates at both
these points.

Consider a point $x_0 \in \mathcal{Z}_{\Gamma}$. The order of the
zero $x_0$ of $\Delta_{\Gamma}(x)$ determines the behaviour of the
eigenvalues $\Lambda_{\pm}(x)$ near $x_0$ because from
\eqref{unimodular evalues} and \eqref{discriminant Gamma} we find
\begin{equation} \label{general singularity}
\Lambda_{\pm}(x) = \Lambda_+(x_0) \pm \sqrt{\Delta_{\Gamma}(x)} + O(x
- x_0).
\end{equation}
In particular, if $\Delta_{\Gamma}(x) = O(x-x_0)$ then $x_0$
corresponds to a branch point since
\begin{equation} \label{square-root singularity}
\Lambda_{\pm}(x) = \Lambda_+(x_0) \pm \alpha \sqrt{x-x_0},
\end{equation}
and analytic continuation around $x_0$ locally interchanges the
two eigenvalues.

\begin{remark}
This means that the functions $\Lambda_{\pm}(x)$ are not globally well
defined in the $x$-plane, and so expressions involving them should
be handled with care. Nevertheless their sum $\left(\Lambda_+(x) +
\Lambda_-(x)\right)$ and product $\Lambda_+(x) \Lambda_-(x)$ are well
defined analytic functions in $x \in \mathbb{C} \setminus \{ \pm 1\}$
since $\Gamma(x,\Lambda) = (\Lambda_+(x) - \Lambda)(\Lambda_-(x) -
\Lambda)$ is. For instance $\Delta_{\Gamma}(x) = \left(\Lambda_+(x) +
\Lambda_-(x)\right)^2 - 4 \Lambda_+(x) \Lambda_-(x)$ is well defined
for $x \in \mathbb{C} \setminus \{ \pm 1\}$.
\end{remark}

On the other hand,

\begin{proposition} \label{proposition: singular points}
If $\Delta_{\Gamma}(x) = O\left( (x-x_0)^n \right), n > 1$ then
$\mathfrak{P}_0 = (x_0,\Lambda_+(x_0)) \in \Gamma$ is a singular
point.
\begin{proof}
From the definition \eqref{spectral curve 1} we have
$\Gamma(x,\Lambda) = (\Lambda_+(x) - \Lambda)(\Lambda_-(x) - \Lambda)$
so
\begin{align*}
\frac{\partial \Gamma}{\partial x}(x_0,\Lambda_+(x_0))
&= \left. \left[- \Lambda_+(x_0) \frac{d}{dx}\left(\Lambda_+(x) +
\Lambda_-(x)\right) + \frac{d}{dx}\left(\Lambda_+(x) \Lambda_-(x)\right) \right]
\right|_{\footnotesize x=x_0}\\
&= - \frac{1}{4} \left. \frac{d}{dx}\left[ \left(\Lambda_+(x) +
\Lambda_-(x)\right)^2 - 4 \Lambda_+(x) \Lambda_-(x) \right] \right|_{\footnotesize
x=x_0}\\
&= - \frac{1}{4} \frac{d \Delta_{\Gamma}}{dx}(x_0) = 0,
\end{align*}
where in the second line we have written $\Lambda_+(x_0)$ as
$\frac{1}{2}\left(\Lambda_+(x_0) + \Lambda_-(x_0)\right)$, and in the
last line we used the definition of $\Delta_{\Gamma}(x)$. The last
equality follows because by assumption $x_0$ is a multiple root of
$\Delta_{\Gamma}(x)$. But since $\Lambda = \Lambda_+(x_0)$ is a double
root of $\Gamma(x_0,\Lambda)$,
\begin{equation*}
\frac{\partial \Gamma}{\partial \Lambda}(x_0,\Lambda_+(x_0)) = 0,
\end{equation*}
and so we conclude that $(x_0, \Lambda_+(x_0)) \in \Gamma$ is indeed a
singular point.
\end{proof}
\end{proposition}

It follows that all points $\mathfrak{P} \in \Gamma$ with
$\hat{\pi}(\mathfrak{P}) \in \mathcal{Z}_{\Gamma}$ are either branch
points or singular points. By equation \eqref{general singularity} and
proposition \ref{proposition: singular points} the singular points are
locally of the form
\begin{equation*}
(\Lambda - \Lambda_0)^2 = (x - x_0)^n, \qquad n \geq 2
\end{equation*}
where $\Lambda_0 = \Lambda_+(x_0)$. These types of singularities were
discussed in chapter \ref{chapter: Riemann surfaces}. When $n = 2$
this is a node and for $n = 3$ it is a cusp. Higher order
singularities are either higher nodes or higher cusps depending on
whether $n$ is even or odd respectively.

Since every point $\mathfrak{P} \in \Gamma$ corresponds to an
eigenvalue of $\Omega(\hat{\pi}(\mathfrak{P}))$, let us denote by
$\mathcal{E}_{\Gamma}(\mathfrak{P})$ the corresponding eigenspace with
$\dim \mathcal{E}_{\Gamma}(\mathfrak{P}) \leq 2$. The following
proposition exhibits a fundamental difference between node-like and
cusp-like singularities with regard to their respective eigenspaces.

\begin{proposition} \label{proposition: unique eigenvector}
If $\mathfrak{P} \in \Gamma$ is not a node-like singularity then $\dim
\mathcal{E}_{\Gamma}(\mathfrak{P}) = 1$.
\begin{proof}
This is obvious for $\hat{\pi}(\mathfrak{P}) \in \mathbb{C} \setminus
\mathcal{Z}_{\Gamma}$. So let $x_0 \in \mathcal{Z}_{\Gamma}$ and
$\mathfrak{P}_0 = (x_0, \Lambda_+(x_0)) \in \Gamma$. Assume that
$\mathfrak{P}_0$ is either a branch point, a cusp or a higher
cusp. All these cases fall into the same category for which
$\Delta_{\Gamma}(x) = O(x - x_0)^{2 r + 1}, r \in \mathbb{N}$. Let us
also denote the components of the monodromy matrix as
\begin{equation*}
\Omega(x) = \left( \begin{array}{cc} \mathcal{A}(x) & \mathcal{B}(x) \\
\mathcal{C}(x) & \mathcal{D}(x) \end{array} \right).
\end{equation*}
Since zeroes of $\Delta_{\Gamma}(x)$ are isolated, in a small
enough neighbourhood of $x_0$ the monodromy matrix $\Omega(x)$ has
two distinct eigenvectors which are easily shown to be
\begin{equation*}
\psi_{\pm}(x) = \left( \; 1, \quad  \text{\footnotesize
$\frac{\mathcal{D}(x) - \mathcal{A}(x)}{2 \mathcal{B}(x)} \pm
\frac{\sqrt{\Delta_{\Gamma}(x)}}{\mathcal{B}(x)}$} \;
\right)^{\sf{T}}.
\end{equation*}
Now since we are assuming $\Delta_{\Gamma}(x) = O(x - x_0)^{2 r + 1}$
and because $\mathcal{B}(x) = O\left((x - x_0)^n \right)$,
$\mathcal{A}(x) - \mathcal{D}(x) = O\left((x - x_0)^m \right)$ for some
non-negative integer $n$ and $m$, it follows that $\psi_{\pm}
\rightarrow {\tiny \left( \!\! \begin{array}{c} 1 \\ 0 \end{array}
\!\! \right)}$ or $\psi_{\pm} \rightarrow {\tiny \left( \!\!
\begin{array}{c} 0 \\ 1 \end{array} \!\! \right)}$ and either way
$\dim \mathcal{E}_{\Gamma}(\mathfrak{P}_0) = 1$.
\end{proof}
\end{proposition}

Whereas proposition \ref{proposition: unique eigenvector}
establishes that most points of $\Gamma$ correspond to a single
eigenvector of the monodromy matrix, it does not forbid nodes and
higher nodes to have a two-dimensional eigenspace. But as we know
from chapter \ref{chapter: Riemann surfaces} these node-like
singularities are blown up into a pair of regular points upon
desingularisation. We therefore anticipate a crucial property of
the normalisation of the spectral curve, namely that to each of
its points corresponds a unique eigenvector.

However if we want the curve $\Gamma$ to have finite topological
genus (\textit{i.e.} finitely many branch points and cusp-like
singularities) then it must have an infinite number of node-like
singularities accumulating at $x = \pm 1$. With $\Gamma$ being so
singular it is not obvious how to normalise it. In the next
section we follow a standard approach for obtaining an algebraic
curve $\Sigma$ from $\Gamma$ which can then be normalised in the usual
way to obtain a Riemann surface $\hat{\Sigma}$.


\section{The algebraic curve}  \label{section: algebraic curve}


Thus far we have constructed a 1-dimensional complex curve $\Gamma$
from any given solution. The problem however is that this
curve is either of infinite genus or highly singular and in order
to make use of the powerful tools of complex analysis we need
instead a finite-genus Riemann surface. But for the normalisation of
$\Gamma$ to have finite genus, the curve $\Gamma$ itself must also be
finite-genus. The class of solutions giving rise to such
\textit{finite}-genus spectral curves will be called
\textit{finite}-gap solutions\footnote{The notion of a `gap'
  originates from the KdV equation for which these methods were first
  developed. There the branch cuts of the spectral curve all lie on the
  real axis and correspond to forbidden gaps in the spectrum of some
  operator. Unfortunately this terminology does not reflect the
  general situation for which a more suggestive term would be
  `finite-genus' or perhaps `finite-g'.}. The standard way to
introduce these solutions is as follows,

\begin{definition}
A \dub{finite-gap solution} is one that is independent of some
given combination of the higher-times of the hierarchy,
\textit{i.e.} $\sum_N c_N \partial_{t_N} j = 0, \, c_N \in
\mathbb{C}$.
\end{definition}

Consider the zero-curvature equations from the hierarchy
\begin{equation*}
\partial_{t_N} J_M(x) - \partial_{t_M} J_N(x) +
[J_M(x), J_N(x)] = 0.
\end{equation*}
Taking the sum over $N$ weighted by the coefficients $c_N$ and using the
finite-gap condition $\sum_N c_N \partial_{t_N} J_M(x) = 0$ we obtain an
equation of the form
\begin{equation} \label{Lax evolution}
\partial_{t_M} L(x) = [J_M(x), L(x)],
\end{equation}
where we have introduced the Lax matrix $L(x) \equiv \sum_N c_N
J_N(x)$. Equation \eqref{Lax evolution} takes exactly the same
form as the evolution equation \eqref{monodromy evolution hierarchy}
for $\Omega(x)$. It says that the evolution of the Lax matrix $L(x)$
with respect to all the higher times is also isospectral which once
more provides an invariant curve $\Sigma$ in $\mathbb{C}^2$. However,
because $L(x)$ is rational in $x$ with poles of finite order at $x =
\pm 1$ the resulting curve $\Sigma$ is now \textit{algebraic}, as
opposed to the spectral curve defined in terms of $\Omega(x)$ with
essential singularities at $x = \pm 1$.
\begin{definition}
The \dub{algebraic curve} $\Sigma \subset \mathbb{C}^2$ is defined
by
\begin{equation} \label{algebraic curve 1}
\Sigma : \;\; \Sigma(x,y) \equiv \det(y {\bf 1} - L(x)) = 0.
\end{equation}
\end{definition}
\noindent Since this new curve is algebraic it may be normalised
in the usual way to obtain a Riemann surface. We denote the
\dub{normalised algebraic curve} as $\hat{\Sigma}$, equipped with
the normalisation map
\begin{equation} \label{alg curve norm}
\pi_{\Sigma} : \hat{\Sigma} \rightarrow \Sigma.
\end{equation}
We now ask how this Riemann surface $\hat{\Sigma}$ is related to
the spectral curve $\Gamma$.

Just as for the spectral curve $\Gamma$ one can define the
discriminant
\begin{equation*}
\Delta_{\Sigma}(x) = (y_+(x) - y_-(x))^2
\end{equation*}
of the polynomial $\Sigma(x,\cdot)$ as well as its set of zeroes
$\mathcal{Z}_{\Sigma}$. Defining also the eigenspace
$\mathcal{E}_{\Sigma}(P)$ corresponding to a point $P = (x, y) \in
\Sigma$, propositions \ref{proposition: singular points} and
\ref{proposition: unique eigenvector} readily apply to the algebraic
curve $\Sigma$ without modification. In particular, since the
normalisation \eqref{alg curve norm} blows up each node-like
singularity of $\Sigma$ to a pair of regular points on $\hat{\Sigma}$,
if we define the eigenspaces $\mathcal{E}_{\hat{\Sigma}}(P)$
corresponding to points of $\hat{\Sigma}$ in the obvious way then we
have the following important result,
\begin{proposition}
$\forall P \in \hat{\Sigma}, \quad \dim \mathcal{E}_{\hat{\Sigma}}(P)
= 1$.
\end{proposition}

Going back to the evolution equation \eqref{monodromy evolution
hierarchy} of the monodromy matrix under the higher times we see
that in the case of a finite-gap solution, for which $\sum_N c_N
\partial_{t_N} \Omega(x) = 0$, one has
\begin{equation*}
[L(x), \Omega(x)] = 0.
\end{equation*}
It follows that if $\bm{\psi}(\mathfrak{P})$ where $\mathfrak{P} = (x,
\Lambda) \in \Gamma$ is an eigenvector of $\Omega(x)$ with eigenvalue
$\Lambda$ then
\begin{equation*}
(\Omega(x) - \Lambda {\bf 1}) (L(x) \bm{\psi}(\mathfrak{P})) = 0.
\end{equation*}
So if $\mathfrak{P} = (x,\Lambda) \in \Gamma$ is not a node-like
singularity then proposition \ref{proposition: unique eigenvector}
implies that $L(x) \bm{\psi}(\mathfrak{P})$ must be proportional to
$\bm{\psi}(\mathfrak{P})$ so that
\begin{subequations} \label{Lpsi=y'psi}
\begin{equation}
L(x) \bm{\psi}(\mathfrak{P}) = y \bm{\psi}(\mathfrak{P}), \qquad
\hat{\pi}(\mathfrak{P}) = x,
\end{equation}
where $y$ is one of the two eigenvalues of $L(x)$. If $P = (x, y) \in
\Sigma$ also isn't a node-like singularity of $\Sigma$ then the
analogue of proposition \ref{proposition: unique eigenvector} for
$\Sigma$ implies that there exists a unique eigenvector
$\bm{\psi}'(P)$ of $L(x)$ with eigenvalue $y$ such that
\begin{equation}
L(x) \bm{\psi}'(P) = y \bm{\psi}'(P), \qquad \hat{\pi}(P) = x,
\end{equation}
\end{subequations}
and hence from equations \eqref{Lpsi=y'psi} we have
$\bm{\psi}'(P) = \bm{\psi}(\mathfrak{P})$. But then by continuity at
the node-like singularities of $\Gamma$ and $\Sigma$ this equality
must also hold at these points. We conclude that $\Omega(x)$ and
$L(x)$ have the same eigenvectors for all $x \in \mathbb{C}$ even
though they do not have the same eigenvalues (because they define
different curves). We shall denote the eigenvector by the same
symbol $\bm{\psi}$ whether it lives on $\Gamma$, $\Sigma$ or
$\hat{\Sigma}$.

Now consider the eigenvector $\bm{\psi}(P)$ at any $P \in
\hat{\Sigma}$. By definition, the monodromy matrix is the parallel
transporter around the worldsheet of the string and so writing the
dependence on $\sigma$ explicitly we have
\begin{equation} \label{y function on Sigma}
\bm{\psi}(P, \sigma + 2 \pi) = \Omega(x, \sigma) \bm{\psi}(P, \sigma)
= \Lambda \bm{\psi}(P, \sigma).
\end{equation}
Hence $\Lambda$ can be written as a quotient of two functions of
$P \in \hat{\Sigma}$ and so is well defined on $\hat{\Sigma}$. Thus
$\hat{\Sigma}$ can be thought of as an (infinite) normalisation of
$\Gamma$ in the sense that there is a continuous surjection
\begin{equation} \label{normalisation of Gamma}
\pi_{\Gamma} : \hat{\Sigma} \rightarrow \Gamma, \quad P =
(x, y) \mapsto \mathfrak{P} = (x, \Lambda(P)),
\end{equation}
whose restriction $\pi_{\Gamma} : \hat{\Sigma} \setminus
\pi^{-1}(S) \rightarrow \Gamma \setminus S$, with $S$ denoting the
set of singular points of $\Gamma$, is a holomorphic bijection. In
particular the spectral curve $\Gamma$ has the same finite
topological genus as $\hat{\Sigma}$, \textit{i.e.} $g_{\Gamma} =
g_{\Sigma} = g_{\hat{\Sigma}} \equiv g$.

\subsection*{The Riemann surface}

\begin{figure}[ht]
\centering
\begin{tabular}{ccc}
\includegraphics[height=30mm]{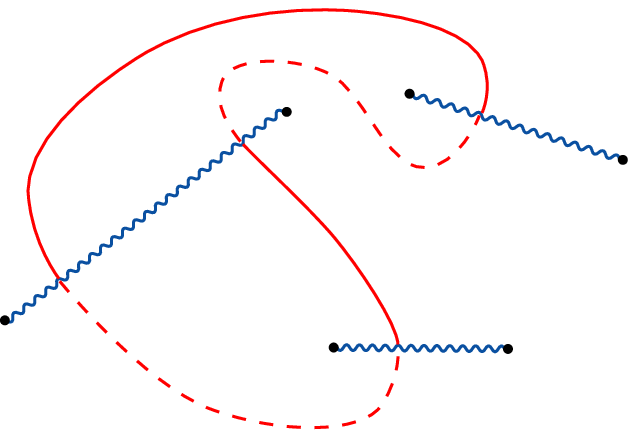} &
$\qquad \qquad$ &
\includegraphics[height=30mm]{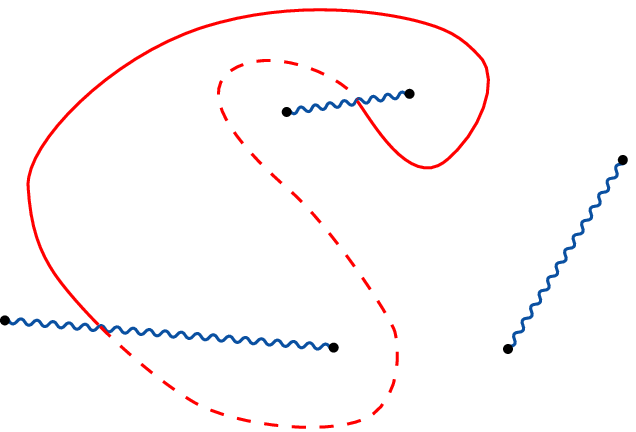}
\end{tabular}
\caption{Two equivalent ways of introducing branch cuts to
represent a genus two curve $\hat{\Sigma}$ of the form
\eqref{algebraic curve 4} with six given branch points $\{ u_i,
v_i \}_{i=1}^3$. The example cycle drawn in both cases represents
exactly the same curve on $\hat{\Sigma}$. Dashed lines are used to
represent parts of a curve lying on the lower sheet.}
\label{figure: branch cuts}
\end{figure}

We can be more explicit about the algebraic form of the normalisation
$\hat{\Sigma}$ of the algebraic curve $\Sigma$. Since the matrix $L(x)
= \sum_N c_N J_N(x)$ appearing in \eqref{algebraic curve 1} is
traceless (because all the $J_N$ are) it can be written out in
components as
\begin{equation*}
L(x) = \left( \begin{array}{cc} a(x) & b(x) \\ c(x)
& - a(x) \end{array} \right)
\end{equation*}
where each entry is a rational function of $x \in \mathbb{C}$. The
defining equation \eqref{algebraic curve 1} for $\Sigma$ then
simplifies to
\begin{equation} \label{algebraic curve 3}
\Sigma : \;\; y^2 = - \det L(x) = a(x)^2 + b(x) c(x).
\end{equation}
Multiplying this equation throughout by an appropriate perfect square
$(Q(x))^2$, where $Q(x)$ is a polynomial, and redefining $y
\mapsto Q(x) y$ it is possible to turn the right hand side of
\eqref{algebraic curve 3} into a polynomial, say $P(x)$. If this
polynomial contains any repeated factors one may further divide
throughout by another perfect square to eliminate them and so we may
assume without loss of generality that $P(x)$ contains no repeated
factors. The resulting non-singular curve $y^2 = P(x)$ is simply the
normalisation $\hat{\Sigma}$ in algebraic form and the various
redefinitions of $y$ to achieve this form are nothing but the
birational transformations required to normalise $\Sigma$. Since
$x = \infty$ is not a branch point of the spectral curve\footnote{It
follows from the asymptotics \eqref{monodromy asymptotics at infty 2}
of the monodromy matrix at $x = \infty$ that the spectral curve
generically takes the form $(\Lambda - 1)^2 = C^2/x^2$ near $x =
\infty$ where $C \neq 0$ is a constant, hence $\mathfrak{P}_{\infty} =
(\infty, 1) \in \Gamma$ corresponds to a node. By the same token
$\mathfrak{P}_0 = (0, 1) \in \Gamma$ is shown not to be a branch point
using the asymptotics \eqref{monodromy asymptotics at zero 2} of the
monodromy matrix at $x = 0$.} it can't be a branch point of
$\hat{\Sigma}$ which means that the polynomial $P$ must be of even
degree. And because the curve $\hat{\Sigma}$ has genus $g$ by
definition, the degree of $P$ must be precisely $2g+2$ (by the
Riemann-Hurwitz formula \eqref{Riemann-Hurwitz formula} since in the
hyperelliptic case the total branching number $b$ appearing in the
formula is equal to the number of branch points). Therefore the
normalised algebraic curve takes the following final form
\begin{equation} \label{algebraic curve 4}
\hat{\Sigma} : \;\; y^2 = \prod_{i = 1}^{g + 1} (x - u_i)(x - v_i),
\end{equation}
where at this stage the branch points $\{ u_i, v_i \}_{i = 1}^{g + 1}$
are arbitrary complex numbers. It is evident that one can always
represent the curve \eqref{algebraic curve 4} by introducing
\dub{branch cuts} in the complex plane, joining up the $2g + 2$ branch
points in pairs (Figure \ref{figure: branch cuts}).
Figure \ref{figure: branch cuts} shows that statements such as ``the
point $P \in \hat{\Sigma}$ lies on the top sheet'' are \textit{not}
invariant under changes of the representation of $\hat{\Sigma}$ in
terms of cuts.

The involution \eqref{holomorphic involution} of $\Gamma$ induces a
holomorphic involution of $\hat{\Sigma}$, called the
\dub{hyperelliptic involution},
\begin{equation*}
\hat{\sigma} : \;\; \hat{\Sigma} \rightarrow \hat{\Sigma}, \quad (x,y)
\mapsto (x,-y).
\end{equation*}
It has the effect of interchanging the two sheets of \eqref{algebraic
curve 4} with the branch points $\{ u_i, v_i \}_{i = 1}^{g + 1}$ as
fixed points.


\section{Quasi-momentum}  \label{section: quasimomentum}


The normalisation $\hat{\Sigma}$ of the spectral curve $\Gamma$ being
a Riemann surface, it is a much more desirable curve to work with than
the spectral curve itself. Therefore when discussing finite-gap
solutions we will always work with $\hat{\Sigma}$ and forget about the
spectral curve altogether. This is a legitimate step to take provided
we have a way of recovering the spectral curve from the Riemann
surface. For instance if we specify the function $\Lambda(P)$ on the
curve $\hat{\Sigma}$ then the spectral curve is simply the image of
$\hat{\Sigma}$ under the normalisation map \eqref{normalisation of
Gamma}. Therefore the pair $(\hat{\Sigma}, \Lambda)$ contains
sufficient information to characterise the spectral curve. However, as
we explain below, the function $\Lambda(P)$ is not meromorphic since
its `branches' $\Lambda_{\pm}(x)$ have essential singularities at $x =
\pm 1$. The goal of this section is to replace $\Lambda(P)$ by an
Abelian differential $dp$ on $\hat{\Sigma}$.

\begin{figure}[ht]
\centering
\begin{tabular}{ccc}
\psfrag{x1}{\tiny $x = -1$}
\includegraphics[height=30mm]{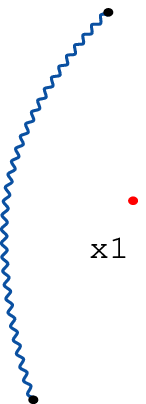} &
\raisebox{14mm}{$\quad \qquad \qquad \longrightarrow \qquad
\qquad$} & \psfrag{x1}{\tiny $x = -1$}
\includegraphics[height=30mm]{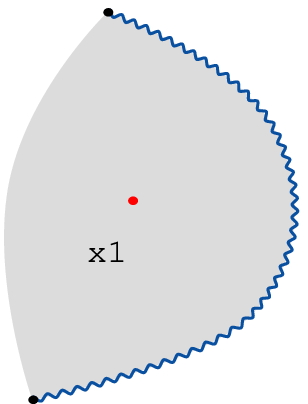}
\end{tabular}
\caption{Moving a single cut over the point $x = - 1$ as in the
figure interchanges the values of the branches $\Lambda_+(x)
\leftrightarrow \Lambda_-(x)$ in the shaded region.}
\label{figure: cuts equiv}
\end{figure}

Since $\Lambda(P)$ is well defined on $\hat{\Sigma}$ it can obviously
be represented by two functions $\Lambda_{\pm}(x)$ living on the top
and bottom sheets respectively which `match up' along the cuts. These
are the same `branches' of $\Lambda(P)$ as in section \ref{section:
spectral curve}, but the advantage of having introduced branch cuts is
that these functions $\Lambda_{\pm}(x)$ are now well defined on the cut
planes and moreover they are distinct from one another (see equation
\eqref{square-root singularity} and the remark following it). We are
now able to unambiguously specify the essential singularities of the
function $\Lambda$ by giving them for its branches
$\Lambda_{\pm}$. Because these essential singularities are located at
$x = \pm 1$, we need to be specific about the position of the
different cuts relative to the points $x = \pm 1$ since moving a cut
over either of these points will swap the relative definitions of
$\Lambda_+(x)$ and $\Lambda_-(x)$ at these points (see Figure
\ref{figure: cuts equiv}).
We therefore introduce an equivalence relation on representations
of $\hat{\Sigma}$ in terms of cuts, where two representations are
equivalent if the cuts of one can be deformed within the punctured
Riemann sphere $\mathbb{C}P^1 \setminus \{ \pm 1 \}$ to the cuts
of the other. It is straightforward to see that there are only two
such equivalence classes and that they obviously both specify the
same Riemann surface $\hat{\Sigma}$. Given a representative of one
equivalence class, one can obtain a representative of the other
class by crossing say $x = -1$ with just a single cut.

Now with respect to a given equivalence class of cuts, the essential
singularities of the function $\Lambda$ on $\hat{\Sigma}$ can be
uniquely specified in terms of those of its branches
$\Lambda_{\pm}(x)$, which can be read off from \eqref{monodromy
asymptotics}. Among the two equivalence classes of cuts to choose
from, we shall pick the one with respect to which the asymptotics of
$\Lambda_{\pm}(x)$ near $x = \pm 1$ take the following form
\begin{equation} \label{Lambda essential singularity}
\begin{split}
\Lambda_{\pm}(x) &= \exp \left[ \mp \frac{i \pi \kappa_+}{x - 1} + O\left(
(x - 1)^0 \right) \right], \quad \text{as} \;\; x \rightarrow +1,\\
\Lambda_{\pm}(x) &= \exp \left[ \mp \frac{i \pi \kappa_-}{x + 1} + O\left(
(x + 1)^0 \right) \right], \quad \text{as} \;\; x \rightarrow -1.
\end{split}
\end{equation}
To obtain the representation of $\Lambda(P)$ with respect to the
other equivalence class of cuts one simply flips the sign in the
exponent at $x = -1$.

\begin{remark}
Since the equations of motion are invariant under the interchange $\sigma
\leftrightarrow \tau$ of worldsheet space and time coordinates,
applying such a transformation to a given solution will generate
another solution. It will turn out that when applied to the present
solution this transformation will change the definition of
$\Lambda(P)$ so that the asymptotics \eqref{Lambda essential
singularity} will now be valid with respect to the other equivalence
class of cuts. Thus the two different equivalence classes of cuts give
two different ways of defining the function $\Lambda$ on
$\hat{\Sigma}$ by \eqref{Lambda essential singularity}. Both will turn
out to give solutions related by the discrete symmetry $\sigma
\leftrightarrow \tau$.
\end{remark}

\noindent Since the functions $\Lambda_{\pm}(x)$ are by definition the
eigenvalues of the monodromy matrix $\Omega(x)$ it follows using also
\eqref{unimodular evalues} that
\begin{equation*}
\Lambda(P) + \frac{1}{\Lambda(P)} = \tr \Omega(\hat{\pi}(P)).
\end{equation*}
And because $\Omega(x)$ is holomorphic in $\mathbb{C} \setminus \{
\pm 1 \}$ we conclude that the function $\Lambda$ can have no
poles or zeroes in $\hat{\Sigma} \setminus \hat{\pi}^{-1}(\{ \pm 1
\})$.

Although the function $\Lambda$ is enough to recover the spectral
curve $\Gamma$ from $\hat{\Sigma}$ as we have already discussed, its
essential singularities are not a very desirable feature. It is best
therefore to replace the function $\Lambda$ with a meromorphic
differential $dp$ defined by
\begin{equation} \label{quasi-momentum definition}
dp = -i \frac{d\Lambda}{\Lambda} = -i d \log \Lambda.
\end{equation}
Since $\Lambda$ has no poles or zeroes in $\hat{\Sigma} \setminus
\hat{\pi}^{-1}(\{ \pm 1 \})$ it follows that the poles of $dp$ can
only come from the points $P \in \hat{\Sigma}$ with $\hat{\pi}(P)
= \pm 1$. In fact these poles are easily derived from the
behaviour \eqref{Lambda essential singularity} of $\Lambda$ at
these points and one finds
\begin{equation} \label{quasi-momentum asymptotics pm1}
\begin{split}
dp(x^{\pm}) &= \mp d \left( \frac{\pi \kappa_+}{x - 1} \right) +
O\left( (x-1)^2 \right), \quad \text{as} \;\; x \rightarrow +1,\\
dp(x^{\pm}) &= \mp d \left( \frac{\pi \kappa_-}{x + 1} \right) +
O\left( (x+1)^2 \right), \quad \text{as} \;\; x \rightarrow -1.
\end{split}
\end{equation}
Here we have introduced the following notation: if $x$ is not a
branch point then $x^{\pm} \in \hat{\Sigma}$ denotes the pair of
points in $\hat{\pi}^{-1}(x)$ ($x^+$ living on one of the cut planes,
$x^-$ on the other), whereas if $x$ is a branch point then $x^+ = x^-
= \hat{\pi}^{-1}(x)$ is a single point. For instance we can rewrite
the branches of the function $\Lambda$ as $\Lambda_{\pm}(x) =
\Lambda(x^{\pm})$. Because $\Lambda$ is a well defined function on
$\hat{\Sigma}$, it follows from the definition \eqref{quasi-momentum
definition} of $dp$ that its integral around any closed loop is an
integer multiple of $2 \pi$, and in particular
\begin{equation*}
\int_{a_i} dp = 2 \pi m_i, \quad \int_{b_i} dp = 2 \pi n_i, \qquad
m_i, n_i \in \mathbb{Z}.
\end{equation*}
At this point we must be more specific about the choice of
homology basis on $\hat{\Sigma}$. We define the basis of
$\bm{a}$-cycles as loops encircling $g$ different cuts. The
$\bm{b}$-cycles are defined from the $g$ remaining independent
cycles such that they have canonical intersections with the
$\bm{a}$-cycles, \textit{i.e.} $a_i \cap b_j = \delta_{ij}$, $i,j
= 1, \ldots, g$. The resulting basis of $\bm{a}$- and $\bm{b}$-cycles
is called \dub{canonical}. An example is shown in Figure \ref{figure:
a b cyles}.

\begin{figure}[ht]
\centering \psfrag{a1}{\footnotesize \red $a_1$}
\psfrag{a2}{\footnotesize \red $a_2$} \psfrag{b1}{\footnotesize
\green $b_1$} \psfrag{b2}{\footnotesize \green $b_2$}
\includegraphics[height=40mm]{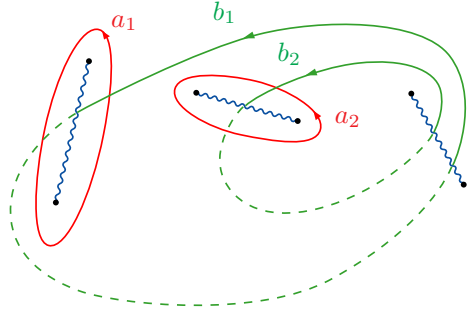}
\caption{Canonical $\bm{a}$- and $\bm{b}$-cycles for a genus two
curve.} \label{figure: a b cyles}
\end{figure}

\begin{remark}
Such a basis always exists (indeed the homology basis constructed in
section \ref{section: topology} is canonical) but is by no means
unique: if $\kappa = (\bm{a}, \bm{b})^{\sf T}$ and $\kappa' = (\bm{a}',
\bm{b}')^{\sf T}$ are homology bases then $\kappa' = X \kappa$ with $X
\in SL(2, \mathbb{Z})$. The condition for $\kappa$ being canonical is
$\kappa \circ \kappa^{\sf T} = J = {\tiny \left( \!\!
\begin{array}{cc} 0 & {\bf 1}\\ - {\bf 1} & 0 \end{array}
\!\!\right)}$ so that the new basis $\kappa'$ is also canonical
provided $X J X^{\sf T} = J$, that is $X \in Sp(2g,\mathbb{Z})$.
\end{remark}

As usual we may choose to normalise this differential (and we
denote the normalised differential by the same symbol) so that
\begin{equation} \label{periods of dp}
\int_{a_i} dp = 0, \quad \int_{b_i} dp = 2 \pi n_i, \qquad n_i \in
\mathbb{Z}.
\end{equation}
According to lemma \ref{lemma: unique normalised} it is then uniquely
defined by its asymptotics at $x = \pm 1$ given in
\eqref{quasi-momentum asymptotics pm1}. As an immediate consequence it
is easy to see that it has the following behaviour under pullback by
the hyperelliptic involution,
\begin{equation*}
\hat{\sigma}^{\ast} dp = - dp.
\end{equation*}

\begin{remark}
For the same reason as with $dp$, this property under pullback by the
hyperellitpic involution also holds for any \textit{normalised}
Abelian differential of the \textit{second} or \textit{third} kind
whose poles are invariant under $\hat{\sigma}$ and whose singular
parts at the poles $x^+$ and $x^- = \hat{\sigma} x^+$ are
opposite. This is the case for the third kind Abelian differential
$\omega_{x^+ x^-}$ and we have $\hat{\sigma}^{\ast} \omega_{x^+ x^-} =
- \omega_{x^+ x^-}$. In fact, it even holds for \textit{normalised}
Abelian differentials of the \textit{first} kind. Indeed,
the holomorphic basis differentials can be locally written as
$\omega_i = df_i$ for some holomorphic functions $f_i$ so that
$\hat{\sigma}^{\ast} \omega_i = d (f_i \circ \hat{\sigma})$. Since
$f_i \circ \hat{\sigma}$ are still holomorphic so are
$\hat{\sigma}^{\ast} \omega_i$. But by the choice of $\bm{a}$-periods
$\hat{\sigma} a_i = - a_i$ and hence $\int_{a_i} \hat{\sigma}^{\ast}
\omega_j = \int_{\hat{\sigma} a_i} \omega_j = - \int_{a_i} \omega_j =
- \delta_{ij}$. Lemma \ref{lemma: unique holo basis} implies
$\hat{\sigma}^{\ast} \omega_j = - \omega_j$.
\end{remark}

The normalised differential $dp$ and its Abelian integral will play a
fundamental role in the sequel.
\begin{definition}
The \dub{quasi-momentum} is the Abelian integral $p(P) = \int^P dp$.
\end{definition}
\noindent A consequence of normalising $dp$ with respect to the chosen
set of $\bm{a}$-cycles is that the branches $p_{\pm}(x) = p(x^{\pm})$
of the quasi-momentum now define single-valued functions on the
complex plane with cuts, even though the Abelian integral $p(P)$
itself is multi-valued on the whole of $\hat{\Sigma}$.

\subsection*{Asymptotics}

The asymptotics of the differential $dp$ near the points
$0^{\pm}$, $\infty^{\pm}$ can be deduced from the asymptotics of
the monodromy matrix $\Omega(x)$ near $x = 0, \infty$ for `highest
weight' solutions, namely equations \eqref{monodromy asymptotics
inf & 0 2} of chapter \ref{chapter: integrability}. They are
directly expressed in terms of the Casimirs $R^2$, $L^2$ of
$SU(2)_R \times SU(2)_L$ as follows
\begin{equation} \label{dp asymp at 0,infty}
\begin{split}
dp(x^{\pm}) &= \mp d \left[\frac{1}{x} \frac{2 \pi R}{\sqrt{\lambda}}
+ O\left( \frac{1}{x^2} \right)\right], \quad \text{as } x \rightarrow
\infty, \\
dp(x^{\pm}) &= \pm d \left[x \frac{2 \pi L}{\sqrt{\lambda}} + O\left(
x^2 \right)\right], \quad \text{as } x \rightarrow 0.
\end{split}
\end{equation}

\begin{remark}
Just as when specifying the asymptotics \eqref{Lambda essential
singularity} near $x = \pm 1$, here the positions of the different cuts
relative to the points $x = \infty$ and $x = 0$ are important. Both
asymptotics in \eqref{dp asymp at 0,infty} are valid with respect to
one of two equivalence classes of cuts, two sets of cuts being
equivalent if they can be deformed within $\mathbb{C}P^1 \setminus \{
\infty^{\pm} \}$ or $\mathbb{C}P^1 \setminus \{ 0^{\pm} \}$
respectively.
\end{remark}

We will always choose the base point for the quasi-momentum to be
$\infty^+$, which fixes the additive constant such that $p(\infty^+) =
0$, namely
\begin{equation*}
p(P) = \int_{\infty^+}^P dp.
\end{equation*}
Considering only points $P = x^+$ on the upper sheet and restricting
also the integration path to lie on the upper sheet we find the
following asymptotics for the quasi-momentum $p(x) \equiv
\int_{\infty^+}^{x^+} dp$ itself,
\begin{equation} \label{p asymp at 0,infty}
\begin{split}
p(x) &= - \frac{1}{x} \frac{2 \pi R}{\sqrt{\lambda}} + O\left(
\frac{1}{x^2} \right), \quad \text{as } x \rightarrow \infty, \\
p(x) &= 2 \pi m + x \frac{2 \pi L}{\sqrt{\lambda}} + O\left( x^2
\right), \quad \text{as } x \rightarrow 0,
\end{split}
\end{equation}
where $m \in \mathbb{Z}$.

\subsection*{The logarithmic derivative curve}

Another way to obtain an algebraic curve from the spectral curve
is to define a new matrix $L'(x)$ by \cite{Beisert:2004ag, Beisert:2005bm}
\begin{equation*}
\Psi(x)^{-1}L'(x)\Psi(x) = -i \frac{\partial}{\partial x} \log
\left( \Psi(x)^{-1}\Omega(x)\Psi(x) \right),
\end{equation*}
where $\Psi(x)$ is the matrix of eigenvectors of the monodromy
matrix $\Omega(x)$. This way the eigenvalues of $L'(x)$ are the
logarithmic derivatives\footnote{the logarithm $\text{log}\,f$ of
a function is not well defined (it requires branch cuts) but its
derivative $(\text{log}\,f)'$ is well defined since the values of
$\text{log}\,f$ on different branches differ by constants.} of
those of $\Omega(x)$, but the corresponding eigenvectors are
unchanged. By the above discussion it is clear that the eigenvalues
$\lambda_{\pm}(x) \equiv -i (\log \Lambda_{\pm}(x))'$ of $L'(x)$ are
rational because they can be written as the quotient of two meromorphic
differentials, namely $\lambda_{\pm} = \frac{dp}{dx}(x^{\pm})$. The
characteristic equation for $L'(x)$ thus defines another algebraic
curve in $\mathbb{C}^2$,
\begin{definition}
The \dub{logarithmic derivative curve} $\Sigma' \subset
\mathbb{C}^2$ is defined by
\begin{equation} \label{algebraic curve 2}
\Sigma' : \;\; \Sigma'(x,\lambda) \equiv \det(\lambda {\bf 1} -
L'(x)) = 0.
\end{equation}
\end{definition}
\noindent This curve has the same normalisation $\hat{\Sigma}$ as the
curves $\Gamma$ and $\Sigma$ with the obvious normalisation map
\begin{equation*}
\pi_{\Sigma'} : \hat{\Sigma} \rightarrow \Sigma', \quad P = (x,y)
\mapsto (x, \lambda(P)), \qquad \text{where} \quad \lambda(P) =
\frac{dp}{dx}(P).
\end{equation*}

To understand how the new curve $\Sigma'$ relates to the spectral
curve $\Gamma$ we first relate the set of zeroes
$\mathcal{Z}_{\Gamma}$ of $\Delta_{\Gamma}(x)$ to the set of zeroes
$\mathcal{Z}_{\Sigma'}$ of the discriminant
\begin{equation*}
\Delta_{\Sigma'}(x) = \left( \lambda_+(x) - \lambda_-(x) \right)^2
\end{equation*}
of the curve $\Sigma'$. So consider a point $x^{\ast} \in
\mathcal{Z}_{\Gamma}$, then $\left( \Lambda_+(x) - \Lambda_-(x)
\right)^2 = O \left( (x - x^{\ast})^n \right)$ with $n \geq 1$ so that
$\Lambda_+(x)/\Lambda_-(x) = 1 + O \left( (x - x^{\ast})^{\frac{n}{2}}
\right)$. After taking the logarithmic derivative this leads to
$\left( \lambda_+(x) - \lambda_-(x) \right)^2 = O \left( (x -
x^{\ast})^{n-2} \right)$, from which we read:
\begin{itemize}
  \item[$\bullet$] \underline{$n = 1$}: branch points of $\Gamma$
become square-root singularities of $\Sigma'$,
  \item[$\bullet$] \underline{$n = 2$}: nodes of $\Gamma$ all disappear on $\Sigma'$,
  \item[$\bullet$] \underline{$n = 3$}: cusps of $\Gamma$ become ordinary branch points of $\Sigma'$,
  \item[$\bullet$] \underline{$n \geq 4$}: higher order singularities of $\Gamma$
persist on $\Sigma'$ with order $n-2$.
\end{itemize}
Now because the curve $\Sigma'$ is algebraic, the discriminant
$\Delta_{\Sigma'}(x)$ of the polynomial $\Sigma'(x,\cdot)$ is
meromorphic on $\mathbb{CP}^1$ and so its set of zeros
$\mathcal{Z}_{\Sigma'} \subset \mathbb{CP}^1$ is finite. This shows
that the spectral curve $\Gamma$ has only a finite number of singular
points of order $n>2$, so that the singular points accumulating at $x
= \pm 1$ must be nodes.


\section{Moduli} \label{section: moduli}


At this point we have now replaced the spectral curve $\Gamma$ by a
Riemann surface $\hat{\Sigma}$ equipped with an Abelian integral $p$
called the quasi-momentum. The purpose of this section is to count the
number of independent moduli of the spectral curve and introduce a
`good' set of coordinates on the moduli space. This problem was solved
in great generality by Krichever and Phong in \cite{Krichever+Phong,
Krichever+Phong2} where they devised a `universal' and more systematic
description of the moduli spaces of the spectral data for a large
class of integrable systems. Specifically, the spectral data of those
systems covered by \cite{Krichever+Phong} all consist of a Riemann
surface $\hat{\Sigma}$ with $N$ punctures $(P_{\alpha})_{\alpha =
1}^N$ and two Abelian integrals $E$ and $Q$ with poles of orders at
most $n = (n_{\alpha})_{\alpha = 1}^N$ and $m = (m_{\alpha})_{\alpha =
1}^N$ at the punctures. So the strategy of \cite{Krichever+Phong} is
to consider the moduli space of \textit{all} such Riemann surfaces
(with the discrete parameters $g = \text{genus}(\hat{\Sigma}),N,n,m$
held fixed) called the \dub{universal configuration space}
$\mathcal{M}_g(n,m)$ and introduce an explicit set of local
coordinates on it. The moduli space for the spectral data of a
specific integrable system then consists of a leaf in a foliation of
$\mathcal{M}_g(n,m)$ for some $g,n,m$. Remarkably, or perhaps not so
surprisingly, we will find that the moduli space for the spectral data
at hand also admits such a description.
We start by reviewing the construction of the universal configuration
space $\mathcal{M}_g(n,m)$ and the definition of a set of local
coordinates \cite{Krichever+Phong}.

\subsection*{The universal configuration space}

In the present subsection we closely follow the discussion in
\cite{Krichever+Phong2}. The first immediate goal is to determine
the dimension of the universal configuration space
$\mathcal{M}_g(n,m)$. This is an easy consequence of the
Riemann-Roch theorem.

\begin{lemma}
$\dim_{\mathbb{C}} \, \mathcal{M}_g(n,m) = 5 g - 3 + 3N + \sum_{\alpha = 1}^N
(n_{\alpha} + m_{\alpha})$.
\begin{proof}
By corollary \ref{corollary: i(-D)} of the Riemann-Roch theorem, the
number of degrees of freedom of the Abelian differential $dE$ with
poles of order at most $n_{\alpha} + 1$ at $P_{\alpha}, \alpha = 1,
\ldots, N$ is $\sum_{\alpha = 1}^N (n_{\alpha} + 1) - 1 + g = N - 1 +
g + \sum_{\alpha = 1}^N n_{\alpha}$. The Abelian integral $E(P) =
\int^P_{P_0} dE$ has one extra degree of freedom corresponding to the
choice of $P_0$ so $E$ has a total of $N + g + \sum_{\alpha = 1}^N
n_{\alpha}$ free parameters. Likewise the Abelian integral $Q$ has $N
+ g + \sum_{\alpha = 1}^N m_{\alpha}$ degrees of freedom. Finally, by
corollary \ref{corollary: moduli punctured RS} the dimension of the
moduli space of Riemann surfaces of genus $g$ with $N$ punctures is
$3g - 3 + N$ for all $g \geq 0$.
\end{proof}
\end{lemma}

The next goal is to determine a set of $5 g - 3 + 3N +
\sum_{\alpha = 1}^N (n_{\alpha} + m_{\alpha})$ functions on
$\mathcal{M}_g(n,m)$ with linearly independent differentials which
would thus define a set of homolorphic coordinates on
$\mathcal{M}_g(n,m)$. Krichever and Phong introduced in
\cite{Krichever+Phong, Krichever+Phong2} a convenient set of such
functions with respect to which the moduli spaces of the spectral
data for many integrable systems locally correspond to level sets
of some of these coordinates (\textit{i.e.} to leaves in
$\mathcal{M}_g(n,m)$). We now review the construction of this
coordinate system.

A fundamental ingredient for defining these coordinates is a certain
meromorphic differential $d \lambda$ which is central to the study of
many integrable systems. It will also turn up naturally in chapter
\ref{chapter: symplectic} and play a crucial role there when we come
to study the symplectic structure of the string in the
algebro-geometric context. Although the Abelian integrals $E,Q$ are
potentially multi-valued on $\hat{\Sigma}$, they define single-valued
branches on the normal form $\hat{\Sigma}_{\text{cut}}$ (see
definition \ref{def: normal form of RS}) with extra cuts between the
various punctures (for instance by joining $P_1$ to $P_{\alpha}$ for
each $\alpha = 2, \ldots, N$). We make a choice of branch for the
Abelian integral $Q$ and define the $1$-form
\begin{equation} \label{1-form dlambda}
d \lambda = Q dE
\end{equation}
on $\hat{\Sigma}_{\text{cut}}$, which has a pole at each puncture
$P_{\alpha}$ of order $n_{\alpha} + m_{\alpha} + 1$. This construction
for defining $d \lambda$ should be carried out in a continuous way
locally on the universal configuration space $\mathcal{M}_g(n,m)$.
Next, in order to discuss the local behaviours of the various
differentials $d \lambda$, $dE$, $dQ$ and Abelian integrals $Q$, $E$
at the punctures we also need to introduce a local set of charts
$w_{\alpha}$ near each puncture $P_{\alpha}$. Such local charts are
naturally provided by one of the Abelian integrals, say $E$.

We are now in a position to define the set of local coordinates on
$\mathcal{M}_g(n,m)$ of \cite{Krichever+Phong, Krichever+Phong2}. The
first set of $\sum_{\alpha = 1}^N (n_{\alpha} + m_{\alpha})$
coordinates are given by
\begin{subequations} \label{K+P coords}
\begin{equation} \label{K+P coords 1}
T_{\alpha,k} = \frac{1}{k} \res_{P_{\alpha}} (w_{\alpha}^k
d\lambda), \qquad \alpha = 1, \ldots, N, \quad k = 1, \ldots,
n_{\alpha} + m_{\alpha}.
\end{equation}
The next set of $3 N - 3$ coordinates are given by the residues of the
differentials $d \lambda$, $dE$ and $dQ$ at the punctures\footnote{For the
differentials $dE$ and $dQ$ which are well defined on $\hat{\Sigma}$
only $N-1$ residues can be specified since the total sum of their
residues must add up to zero by proposition \ref{proposition: sum of
residues}.}
\begin{equation} \label{K+P coords 2}
R_{\alpha}^{\lambda} = \res_{P_{\alpha}} d\lambda, \quad R_{\alpha}^E
= \res_{P_{\alpha}} dE, \quad R_{\alpha}^Q = \res_{P_{\alpha}} dQ,
\qquad \alpha = 2, \ldots, N.
\end{equation}
Finally the remaining $5g$ coordinates are given by periods of the
differentials $d \lambda$, $dE$ and $dQ$, namely
\begin{gather}
\tau_{a_i, E} = \int_{a_i} dE, \quad \tau_{b_i, E} = \int_{b_i}
dE, \label{K+P coords 3}\\
\tau_{a_i, Q} = \int_{a_i} dQ, \quad \tau_{b_i, Q} = \int_{b_i}
dQ, \label{K+P coords 4}\\
s_i = \int_{a_i} d\lambda, \qquad i = 1, \ldots, g. \label{K+P coords 5}
\end{gather}
\end{subequations}
It is proved in \cite{Krichever+Phong} that these $5g -3 + 3N +
\sum_{\alpha} (n_{\alpha} + m_{\alpha})$ functions \eqref{K+P
coords} have linearly independent differentials and thus define a
local holomorphic coordinate system for $\mathcal{M}_g(n,m)$.
Given such a coordinate system, one can consider the joint level
set of all but the last $g$ coordinates \eqref{K+P coords 5} and
excluding also a certain number $l \leq N - 1$ of residues
$R_{\alpha}^{\lambda}$. This defines a smooth foliation of
$\mathcal{M}_g(n,m)$ with the remaining $g+l$ coordinates defining
a coordinate system $\{(s_i)_{i = 1}^g,
(R_{\alpha}^{\lambda})_{\alpha = 2}^{l+1} \}$ on each
$(g+l)$-dimensional leaf.

\subsection*{The leaf}

We now want to make use of the general framework reviewed in the
previous subsection to count the independent moduli of the spectral
data $\{ \hat{\Sigma}, p \}$. Let us identify the quasi-momentum with
the first Abelian integral, namely $E \equiv p$. The general setup
requires choosing another Abelian integral $Q$. Our choice at this
stage might seem rather \textit{ad'hoc} but it is guided by the
results on the symplectic structure to be derived in chapter
\ref{chapter: symplectic}. Indeed it will turn out that the moduli
defined in this section are precisely the action variables of the
string.

Since $\hat{\Sigma}$ is hyperelliptic it also comes equipped with a
holomorphic function $x : \hat{\Sigma} \rightarrow \mathbb{C}P^1$ of
degree two which provides a coordinate chart in the neighbourhood of
any point $P \in \hat{\Sigma}$ that isn't a branch point of the cover
given by $x$. The appropriate choice for the Abelian integral $Q$ is
the following meromorphic \textit{function} on $\hat{\Sigma}$
\begin{equation} \label{definition of z}
z = x + \frac{1}{x}.
\end{equation}
This function clearly defines a double cover of the $x$-plane and
thus has degree four on $\hat{\Sigma}$. To make contact with the
general construction we make the following identifications
\begin{equation*}
E \equiv p, \quad Q \equiv z.
\end{equation*}
By definition of the quasi-momentum \eqref{quasi-momentum asymptotics
pm1} it has simple poles at the four points $\{ (+1)^{\pm}, (-1)^{\pm}
\} \in \hat{\pi}^{-1}(\{\pm 1\})$ above $x = \pm 1$. And by
\eqref{definition of z} we see that the function $z$ has simple poles
at the four points $\{ 0^{\pm}, \infty^{\pm} \} \in
\hat{\pi}^{-1}(\{0, \infty\})$ above $x = 0, \infty$. Therefore we
have a total of $N = 8$ punctures. Because the Abelian integral $Q =
z$ is actually a function on $\hat{\Sigma}$, here the $1$-form
\eqref{1-form dlambda} is a well defined and single-valued meromorphic
differential on $\hat{\Sigma}$,
\begin{equation*}
d \lambda \equiv z dp.
\end{equation*}
From the asymptotics of the quasi-momentum at $x = \pm 1$ we can
define local coordinates $w_{\pm}$ near these points by  setting
$E = 1/w_{\pm}$. Local coordinates around $0^{\pm}$ and
$\infty^{\pm}$ are provided by $w_0 = x$ and $w_{\infty} = 1/x$
respectively.

The residues of the differentials $dz$, $dp$, $d \lambda$ and
$w_{\alpha} d \lambda$ can be easily computed at all these punctures,
for instance
\begin{equation*}
T_{\pm, 1} = \res_{(\pm 1)} w_{\pm} d\lambda = \res_{(\pm 1)} w_{\pm}
z \; d \left( \frac{1}{w_{\pm}} \right) = - \res_{(\pm 1)} z \frac{d
w_{\pm}}{w_{\pm}} = - z(\pm 1) = \mp 2.
\end{equation*}
For the residues at $0^{\pm}$, $\infty^{\pm}$ one must use the
asymptotics \eqref{dp asymp at 0,infty} of the quasi-momentum at $0$
and $\infty$ respectively. All the residues are summarised in table
\ref{table of residues} using the notation of the general construction.
\begin{table}[ht]
\centering
\begin{tabular}{c|cccc}
$P_{\alpha}$ & $(+1)^{\pm}$ & $(-1)^{\pm}$ & $0^{\pm}$ & $\infty^{\pm}$\\
\hline
$m$ & $1$ & $1$ & $0$ & $0$\\
$n$ & $0$ & $0$ & $1$ & $1$\\
$R_{\alpha}^E$ & $0$ & $0$ & $0$ & $0$\\
$R_{\alpha}^Q$ & $0$ & $0$ & $0$ & $0$\\
$R_{\alpha}^{\lambda}$ & $0$ & $0$ & $\pm \frac{2 \pi L}{\sqrt{\lambda}}$
& $\mp \frac{2 \pi R}{\sqrt{\lambda}}$\\
$T_{\alpha, 1}$ & $-2$ & $2$ & $0$ & $0$
\end{tabular}
\caption{Residues at the eight punctures.}
\label{table of residues}
\end{table}
Furthermore, since the function $z$ is single-valued on $\hat{\Sigma}$
all the periods of $dz$ are zero whereas those of the normalised
differential of the quasi-momentum $dp$ are determined by
\eqref{periods of dp} so we have
\begin{equation} \label{periods defining leaf}
\tau_{a_i, E} = \tau_{b_i, E} = 0, \quad \tau_{a_i, Q} = 0, \tau_{b_i,
Q} = 2 \pi n_i.
\end{equation}
The remaining $g$ coordinates were defined in \eqref{K+P coords
5}. However, for conventional reasons we will scale these coordinates
differently and set
\begin{equation} \label{pre filling fractions}
S_i = \frac{\sqrt{\lambda}}{8 \pi^2 i} \int_{a_i} z dp, \quad i = 1,
\ldots, g.
\end{equation}

\begin{figure}[ht]
\centering \psfrag{a}{$\mathcal{A}_I$} \psfrag{b}{$\mathcal{B}_I$}
\psfrag{c}{$\mathcal{C}_I$} \psfrag{pinf}{$\infty^+$}
\psfrag{minf}{$\infty^-$}
\includegraphics[height=35mm]{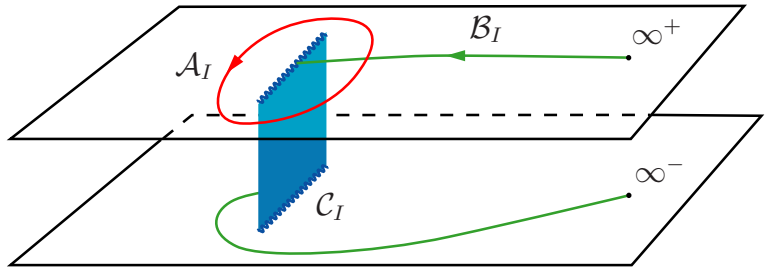}
\caption{The cycle $\mathcal{A}_I$ and path $\mathcal{B}_I$ for
the cut $\mathcal{C}_I$.} \label{AnBcycles}
\end{figure}

We see from table \ref{table of residues} that besides these $g$
coordinates there are only two other tunable parameters in the general
solution, namely the Casimirs of the global $SU(2)_R$ and $SU(2)_L$
symmetries which are expressible here in terms of residues on the top
sheet at infinity and zero respectively,
\begin{equation*}
R = - \frac{\sqrt{\lambda}}{2 \pi} \res_{\infty^+} z dp, \qquad L =
\frac{\sqrt{\lambda}}{2 \pi} \res_{0^+} z dp.
\end{equation*}
Recall however from section \ref{section: symmetries} that the current
$j$ is invariant under $SU(2)_L$ but still transforms under $SU(2)_R$
by \eqref{R symmetry on j}. Since its components parametrise
phase-space it follows that the action of the $SU(2)_L$ symmetry on
phase-space is trivial and does not play any part in the Hamiltonian
formalism. We therefore fix the parameter $L$ to define a leaf
$\mathcal{L}$ as the joint level set of all but the $g+1$ remaining
parameters $\{ S_i \}_{i = 1}^g$ and $R$. Defining the following
differential on $\hat{\Sigma}$
\begin{equation} \label{symplectic 1-form}
\alpha = \frac{\sqrt{\lambda}}{4 \pi} z dp,
\end{equation}
the remaining $g+1$ coordinates parametrising the leaf are
\begin{equation} \label{independent moduli}
S_i = \frac{1}{2 \pi i} \int_{a_i} \alpha, \; i = 1, \ldots, g,
\qquad \frac{R}{2} = - \res_{\infty^{+}} \alpha.
\end{equation}
Equivalently, since the number of moduli precisely coincides with the
number of cuts in the algebraic curve \eqref{algebraic curve 4} one
can parametrise $\mathcal{L}$ by assigning a modulus to each
cut. Specifically, for $I = 1, \ldots, g+1$ we define a cycle
$\mathcal{A}_I$ to encircle the $I^{\text{th}}$ cut $\mathcal{C}_I$
once counterclockwise on the top sheet. We can also define the dual
cycles $\mathcal{B}_I$ as the contour going from $\infty^+$ to
$\infty^-$ through the $I^{\text{th}}$ cut, see Figure
\ref{AnBcycles}.

\begin{definition}
The \dub{filling fraction} for the $I^{\text{th}}$ cut is given by,
\begin{equation} \label{filling fractions}
\mathcal{S}_I = \frac{1}{2 \pi i} \int_{\mathcal{A}_I} \alpha =
\frac{\sqrt{\lambda}}{8 \pi^2 i} \int_{\mathcal{A}_I} z dp.
\end{equation}
\end{definition}

\noindent The filling fractions are related to the variable $R$ and
the parameter $L$ by
\begin{equation} \label{sum of fillings}
\sum_{I=1}^{g+1} \mathcal{S}_I = - \res_{\infty^+} \alpha - \res_{0^+}
\alpha = \frac{1}{2} (R - L).
\end{equation}
The moduli space $\mathcal{L}$ is therefore a complex manifold
with only orbifold singularities of dimension
\begin{equation*}
\dim_{\mathbb{C}} \mathcal{L} = g+1,
\end{equation*}
every point of which corresponds to an admissible curve $\hat{\Sigma}$
of genus $g$.

%% file: FiniteGap.tex
\newpage

\chapter{Algebro-geometric solutions} \label{chapter: finite-gap}

\begin{figure}[ht]
\begin{flushright}
\includegraphics[width=100mm]{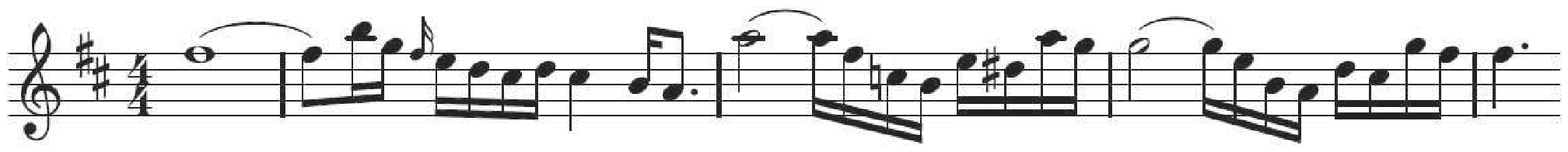}\\
J.~S.~Bach - {\it Air on a G string}
\end{flushright}
\end{figure}
\vspace{1cm}

\noindent Given any (finite-gap) solution to the hierarchy of
zero-curvature equations \eqref{zero-curvature hierarchy} we have
shown how to construct a Riemann surface $\hat{\Sigma}$ equipped
with an Abelian differential $dp$, both of which are independent
of the hierarchy of times. The goal of \dub{finite-gap
integration} (or \dub{algebro-geometric methods})
\cite{Krichever2, Krichever3, Krichever4, Belokolos, Babelon, Gesztesy} is
to reconstruct the (finite-gap) solution itself after specifying
further analytic data on $\hat{\Sigma}$. A key part of the theory
of Riemann surfaces which underlies this method of finite-gap
integration is the construction of functions and differentials on
a Riemann surface with prescribed singularities. The idea of
finite-gap integration therefore is to identify a finite set of
points on $\hat{\Sigma}$ that will be the zeroes and poles of
certain functions in terms of which the solution can be expressed.
If this data is sufficient to uniquely determine these functions
then it will also be enough to recover the solution.

In chapter \ref{chapter: curves} we have focused mostly on the
integrals of motion of the solution, namely the eigenvalues of the
monodromy matrix, which we showed were encoded in the data $\{
\hat{\Sigma}, dp \}$. To completely encode the monodromy matrix we are
missing its dynamical part, which corresponds to its
eigenvectors. However we have already argued in chapter \ref{chapter:
curves} that these eigenvectors define a single-valued vector function
$\bm{\psi}$ on $\hat{\Sigma}$. To remove the arbitrary normalisation
of $\bm{\psi}$ we introduce the normalised eigenvector denoted $\bm{h}$,
with the suitable choices of normalisation conditions to be discussed
later. As it turns out $\bm{h}(P,t)$ is in fact meromorphic
in $P \in \hat{\Sigma}$ with precisely $g+1$ poles
$\hat{\gamma}_1(t), \ldots, \hat{\gamma}_{g+1}(t)$ (and hence also
$g+1$ zeroes) which explicitly depend on the hierarchy of times $\{ t
\}$. We can conveniently gather these points by defining the
\dub{dynamical divisor}
\begin{equation} \label{dyn div intro}
\hat{\gamma}(t) \equiv \hat{\gamma}_1(t) + \cdots +
\hat{\gamma}_{g+1}(t),
\end{equation}
which, as its name suggests, encodes the dynamics of the monodromy
matrix. After making use of the gauge symmetry to set $h_i(P_j) =
\delta_{ij}$ where $P_{1,2} = \infty^{\pm}$, it follows from the
Riemann-Roch theorem that this data is enough to uniquely specify the
components $h_1$ and $h_2$ of the normalised eigenvector $\bm{h}$ and
hence also $\Omega(x)$.

One would like to construct a similar set of functions that can be
uniquely specified by some analytic data but in terms of which we can
also write the solution. For this we exploit the hierarchy equations
\eqref{evolution hierarchy} which express the fact that the operators
$\partial_{t_M} - J_M(x)$ and $\Omega(x)$ all commute among themselves
and can thus be simultaneously diagonalised. Thus there exists an
alternative normalisation of the eigenvector $\bm{\psi}(P,t) =
\varphi(P,t) \bm{h}(P,t)$ such that it solves the following linear
system
\begin{equation*}
\left( \partial_{t_M} - J_M(x) \right) \bm{\psi}(P,t) = 0, \quad
\forall M.
\end{equation*}
Unlike the normalised eigenvector $\bm{h}(P,t)$ above, the eigenvector
$\bm{\psi}(P,t)$ is not meromorphic. Instead its components have
essential singularities at the poles $x = \pm 1$ of the Lax matrices
and define what are called \dub{Baker-Akhiezer functions}. If we are
able to identify a set of analytic data which uniquely characterises
this vector $\bm{\psi}$ then the Lax connection could be recovered
from it by the formula
\begin{equation} \label{pre reconstruction formula}
J(x) = d \Psi(x) \Psi(x)^{-1},
\end{equation}
where $\Psi(x) = (\bm{\psi}(x^+), \bm{\psi}(x^-))$ is the matrix
constructed out of the pair of column eigenvectors at the points
$x^{\pm}$ above $x \in \mathbb{C}$. Remarkably it turns out the
only extra data needed to uniquely characterise the vector
$\bm{\psi}$ is the initial condition $\hat{\gamma}(0)$ of the
dynamical divisor \eqref{dyn div intro}. All the dynamics can be
recovered uniquely once the constant data $\{ (\hat{\Sigma}, dp),
\hat{\gamma}(0) \}$ has been specified. In particular, the
time-dependence of the dynamical divisor \eqref{dyn div intro} can
be inferred from that of the vector $\bm{\psi}$. The idea of
finite-gap integration is illustrated in Figure \ref{FG
integration}:
\begin{figure}[h]
\begin{gather*}
\begin{tabular}{c}
\psfrag{S}{\small $\hat{\Sigma}$} \psfrag{g}{\small \red \small
$\hat{\gamma}$} \includegraphics[height=2cm]{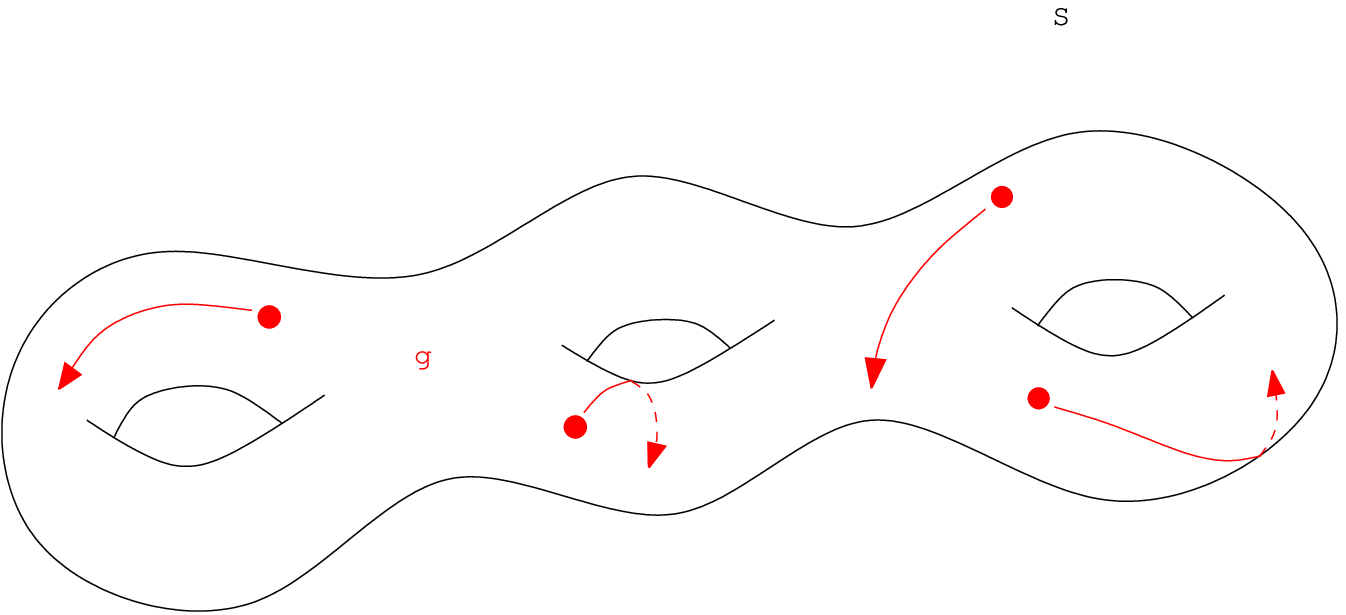}
\end{tabular} \quad \Leftrightarrow \quad \text{finite-gap solution}\\
\hspace{-70mm} \qquad \qquad \qquad \psfrag{A}{\small
$\vec{\mathcal{A}}$}
\includegraphics[height=1cm]{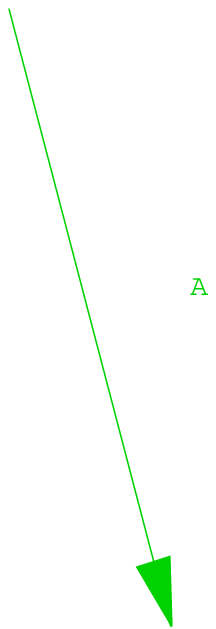}\\ \psfrag{J}{\small
$J(\hat{\Sigma})$} \psfrag{C}{\small $\!\!\!\!\!
\times \mathbb{C}^{\ast}$}
\includegraphics[height=1.5cm]{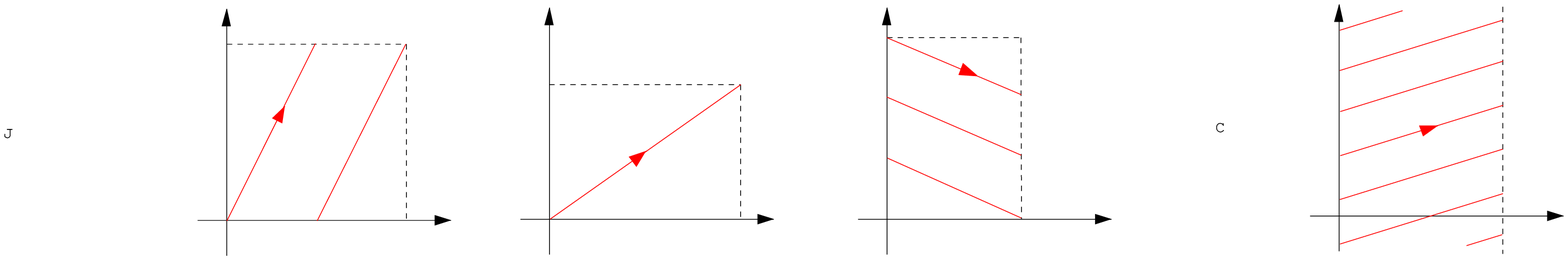}
\end{gather*}
\caption{Idea of finite-gap integration.} \label{FG integration}
\end{figure}
Every finite-gap solution is in one-to-one correspondence with a
smooth Riemann surface $\hat{\Sigma}$ (of genus three in Figure
\ref{FG integration}) equipped with a set of marked points
$\hat{\gamma}(0)$ (four of them in Figure \ref{FG integration}). The
pair $(\hat{\Sigma}, dp)$ encodes the integrals of motion of the
solution whereas the dynamics is encoded in the marked points
$\hat{\gamma}(t)$. Their exact motion on $\hat{\Sigma}$ is very
complex, but what we find is that if we map $\hat{\Sigma}$ to its
generalised Jacobian (which is topologically a $g$-torus times a
$\mathbb{C}^{\ast}$ factor) via the generalised Abel map then the
motion becomes extremely simple, namely it linearises. The $(\sigma,
\tau)$-motion of the string on the generalised Jacobian is like that
of an infinitely rigid string wrapping one cycle of the torus and
moving linearly in time along another direction.


\section{The normalised eigenvector} \label{section: normalised eigenvector}


In order to discuss the analytic properties of the eigenvector
$\bm{\psi}(P)$ at every $P \in \hat{\Sigma}$ we must first fix its
normalisation. There are many ways one could normalise $\bm{\psi}$
but to keep things simple we choose a linear normalisation
condition and define the \dub{normalised eigenvector} $\bm{h}$ to
satisfy
\begin{equation} \label{normalisation}
\bm{\alpha} \cdot \bm{h}(P) = 1,
\end{equation}
where $\bm{\alpha} \in \mathbb{C}^2$ is an arbitrary two component
constant row vector. A common choice is $\bm{\alpha} = (1,0)$ which
has the effect of setting the first component $h_1$ of $\bm{h}$ equal
to one. Although this might be the simplest possible normalisation
condition it is not the most symmetric one. A more symmetric condition
is $\bm{\alpha} = (1,1)$ which sets both components on an equal
footing,
\begin{equation} \label{normalisation explicit}
h_1(P) + h_2(P) = 1.
\end{equation}
From now on we shall always stick to this normalisation for $\bm{h}$.

\begin{lemma} \label{lem: normalized ev is mero}
The components of $\bm{h}$ are meromorphic functions on
$\hat{\Sigma}$.
\begin{proof}
Let $\widehat{\Delta}(x,y)$ be the matrix of cofactors of $(L(x) -
y \textbf{1})$, which satisfy $(L(x) - y \textbf{1})
\widehat{\Delta}(x,y) = \Sigma(x,y) \textbf{1}$. Then for $(x,y)
\in \Sigma$ we have $(L(x) - y \textbf{1}) \widehat{\Delta}(x,y) =
0$, so that every column of $\widehat{\Delta}(P)$ is proportional
to the eigenvectors at $P$. Since $\widehat{\Delta}(P)$ is
meromorphic (\textit{i.e.} rational in $x$ and $y$) the result
follows.
\end{proof}
\end{lemma}

\begin{definition}
A vector $\bm{v}(P)$ is said to have a \dub{pole} at $Q \in
\hat{\Sigma}$ if at least one of its components $v_i(P)$ has a pole at
$Q$.
\end{definition}

\begin{proposition} \label{prop: g+1 poles}
The normalised eigenvector $\bm{h}$ has $g + 1$ poles on
$\hat{\Sigma}$.

\begin{proof}
Consider the function $W(x) = \left( \det H(x) \right)^2$ where
$H(x) = \left( \bm{h}(x^+), \bm{h}(x^-) \right)$ is the matrix of
the normalised eigenvectors at $x$.
The function $W(x)$ is a well defined meromorphic function on the
Riemann sphere since it depends symmetrically on the points $x^{\pm}$
above $x$. Counting multiplicities, it is obvious that
\begin{equation*}
(\# \; \text{poles of} \; W(x)) = 2 \times (\# \;
\text{poles of} \; \bm{h}(P)).
\end{equation*}

Since the eigenvectors $\bm{h}(x^{\pm})$ corresponding to distinct
eigenvalues $y_{\pm}$ of $L(x)$ are linearly independent it
follows that $W(x)$ vanishes if and only if $x$ corresponds to a
branch point, where two columns of $H(x)$ coincide ($\hat{\Sigma}$
is non-singular). Now suppose $x_0$ corresponds to a branch point
$P \in \hat{\Sigma}$, and let $z$ be a local parameter on
$\hat{\Sigma}$ around $P$. In this coordinate, the covering map $P
\mapsto x$ takes the form $x - x_0 = O(z^2)$ near $z = 0$. Also,
$\det H(x) = O(z)$ near $z=0$ and so $W(x) = O(z^2) = O(x-x_0)$,
and hence the multiplicity of the zero $x_0$ of $W(x)$ is equal to
one, which is also the branching number of the corresponding
branch point $P$.
So summing over all branch points we have
\begin{equation*}
(\# \; \text{zeros of} \; W(x)) = (\text{total branching number of} \;
P \mapsto x) = 2(N + g - 1),
\end{equation*}
where the last equality follows from the Riemann-Hurwitz relation
\eqref{Riemann-Hurwitz formula}. But since $W(x)$ is a function
meromorphic on the Riemann sphere, it has as many poles as zeroes
(counting multiplicities) and the result follows.
\end{proof}
\end{proposition}

Recall that the present formalism derives from the hierarchy of
zero-curvature equations \eqref{zero-curvature hierarchy} which are
invariant under gauge transformations \eqref{gauge transformation}. We
now make use of this freedom to fix the normalised eigenvectors at
infinity. Specifically, we apply the gauge transformation with
parameter\footnote{$\Omega(\infty)$ is clearly invertible because
$\Omega(\infty) - {\bf 1} = 0$. This in turn means that the
eigenvectors $\bm{h}(\infty^+)$ and $\bm{h}(\infty^-)$ are linearly
independent which implies $H(\infty)$ is invertible.} $\tilde{g} =
H(\infty)^{-1}$ where $H(x) = \left( \bm{h}(x^+),
\bm{h}(x^-)\right)$. Because eigenvectors of the monodromy matrix
transform as $\bm{h} \mapsto \tilde{g} \bm{h}$, we have in the new
gauge
\begin{equation} \label{h at infty}
\bm{h}(\infty^+) = {\tiny \left(\!\!\begin{array}{c} 1 \\ 0
\end{array}\!\!\right)}, \qquad \bm{h}(\infty^-) = {\tiny
\left(\!\!\begin{array}{c} 0 \\ 1 \end{array}\!\!\right)}.
\end{equation}
Notice that this gauge transformation preserves the normalisation of
$\bm{h}$ because by the special choice $\bm{\alpha} = (1,1)$ of
normalisation in \eqref{normalisation} we have $\bm{\alpha}H(x) =
\bm{\alpha}$.

\begin{remark}
The gauge fixing condition \eqref{h at infty} imposed so far also
fixes part of the global $SU(2)_R$ symmetry of the equations of
motion (since the latter also acts on the eigenvectors as $\bm{h} \mapsto
\tilde{g} \bm{h}$). Specifically, since in this gauge the eigenvectors
of the monodromy matrix $\Omega(x)$ are of the canonical form \eqref{h
at infty} it follows from the general form \eqref{monodromy
asymptotics at infty} of $\Omega(x)$ at $x = \infty$ that the right
Noether charge $Q_R$ must be diagonal in this gauge. Thus the gauge
fixing condition \eqref{h at infty} restricts the $SU(2)_R$ current
$j$ to the level set $Q_R = \frac{1}{2 i} R \sigma_3$, thereby
breaking $SU(2)_R$ to a diagonal $U(1)_R$.
\end{remark}

The residual gauge symmetry which preserves \eqref{h at infty}
consists of diagonal matrices $\tilde{g}(t) = \diag(d_1,
d_2)$ whose action on the normalised eigenvector $\bm{h}$ is
\begin{equation} \label{residual gauge symmetry}
\bm{h} \mapsto f(P)^{-1} \tilde{g} \bm{h},
\end{equation}
where $f(P) = \bm{\alpha} \cdot (\tilde{g} \bm{h}(P)) = d_1 h_1(P) +
d_2 h_2(P)$. The role of the function $f(P)$ is to keep $\bm{h}$
normalised by \eqref{normalisation}. Since its poles are the same as
those of $\bm{h}$ it has the effect of changing the divisor
$\hat{\gamma}(t)$ of poles of $\bm{h}$ to the equivalent
divisor $\hat{\gamma}'(t)$ ($\sim \hat{\gamma}(t)$) of zeroes of
$f$. Let $[\hat{\gamma}(t)]$ denote the equivalence class of
such divisors $\hat{\gamma}(t)$.

\begin{proposition} \label{prop: residual gauge}
There is a $1-1$ correspondence between residual gauges and
representatives of the equivalence class $[\hat{\gamma}(t)]$.
\begin{proof}
A specific representative $\hat{\gamma}'(t) = \hat{\gamma}'_1(t) +
\ldots + \hat{\gamma}'_{g+1}(t)$ of the equivalence class
$[\hat{\gamma}(t)]$ is uniquely specified by a single one of its
points. Thus it suffices to show that for an arbitrary point
$\hat{\gamma}'_1 \in \hat{\Sigma}$ there exists a unique $\tilde{g} =
\diag(d_1, d_2)$ such that $0 = f(\hat{\gamma}'_1) = d_1
h_1(\hat{\gamma}'_1) + d_2 h_2(\hat{\gamma}'_1)$. But since we can
scale away $d_1$ in \eqref{residual gauge symmetry} this has the
unique solution $d_2 = - h_1(\hat{\gamma}'_1) /
h_2(\hat{\gamma}'_1)$.
\end{proof}
\end{proposition}

From now on we fix the residual gauge by choosing a representative
$\hat{\gamma}(t)$ from the equivalence class $[\hat{\gamma}(t)]$. It
follows from proposition \ref{prop: g+1 poles} and equation \eqref{h
at infty} that the components of the eigenvector $\bm{h}$ satisfy the
following properties,
\begin{equation} \label{normalised ev properties}
\begin{split}
(h_1) &\geq - \hat{\gamma}(t) + \infty^-, \quad h_1(\infty^+) = 1, \\
(h_2) &\geq - \hat{\gamma}(t) + \infty^+, \quad h_2(\infty^-) = 1.
\end{split}
\end{equation}
From now on we assume the divisor $\hat{\gamma}(t)$ to be in general
position. Let $\gamma_-(t) + \infty^-$ and $\gamma_+(t) + \infty^+$
($\deg \gamma_{\pm}(t) = g$) be the divisors of zeroes of $h_1$
and $h_2$ respectively. We also assume $\gamma_{\pm}(t)$ to be
non-special, \textit{i.e.} $r(- \hat{\gamma}_+(t)) = 1$, which implies
that $\hat{\gamma}(t)$ is non-special (by the remark following
definition \ref{def: non-special divisors}). The divisors
$\gamma_{\pm}(t)$ are defined uniquely by the following equivalence of
divisors,
\begin{equation} \label{divisor of zeroes}
\hat{\gamma}(t) \sim \gamma_{\pm}(t) + \infty^{\pm}.
\end{equation}

\begin{proposition} \label{prop: uniqueness of h}
Conditions \eqref{normalised ev properties} uniquely specify a
meromorphic vector $\bm{h}$.
\begin{proof}
Suppose not and let $\bm{h}$ and $\bm{h}'$ be two vectors satisfying
conditions \eqref{normalised ev properties}. Consider the meromorphic
function $f_i(P) \equiv h_i(P)/h'_i(P)$, $i = 1, 2$ of degree $g$. Its
divisor or poles is contained in the divisor $\gamma'_-(t)$ or
$\gamma'_+(t)$ of zeroes of $h'_1$ or $h'_2$ which by assumption is in
general position. Thus $r(-\gamma'_{\pm}) = 1$ and $f_i$ must be
constant by Riemann-Roch. But $f_1(\infty^+) = f_2(\infty^-) = 1$ so
$f_i = 1$ and hence $\bm{h} = \bm{h}'$.
\end{proof}
\end{proposition}

\begin{remark}
Suppose we chose to normalise $\bm{h}$ by the condition $h_1(P) = 1$
instead of \eqref{normalisation explicit}. This corresponds to
multiplying the above eigenvector $\bm{h}$ by $\frac{1}{h_1}$. The
second component would then have divisor $\geq - \gamma_-(t) -
\infty^- + \infty^+$ now with a pole forced at $\infty^-$.
\end{remark}

We now show that there exists a pair of functions $h_1, h_2$ which
satisfy the conditions \eqref{normalised ev properties}. To do this we
construct explicit functions on $\hat{\Sigma}$ with the properties
\eqref{normalised ev properties} in terms of Riemann
$\theta$-functions. By proposition \ref{prop: uniqueness of h} these
constructed functions must therefore be equal to the components of the
normalised eigenvector. Proposition \ref{prop: existence of
h} below provides a \dub{reconstruction formula} for
\textit{reconstructing} the normalised eigenvector $\bm{h}$ from its
analytic data, namely a divisor $\hat{\gamma}(t)$ on the Riemann
surface $\hat{\Sigma}$. Let $w_1, w_{g+1}, w^{\pm}_{\infty}, w^{\pm}
\in \mathbb{C}^g$ be defined as follows,
\begin{subequations} \label{four vectors w}
\begin{align}
w_1 &= \sum_{i = 1}^g \bm{\mathcal{A}}(\hat{\gamma}_i(t)) +
\bm{\mathcal{K}},\\
w_{g+1} &= \sum_{i = 2}^{g+1} \bm{\mathcal{A}}(\hat{\gamma}_i(t)) +
\bm{\mathcal{K}},\\
w^{\pm}_{\infty} &= \bm{\mathcal{A}}(\infty^{\pm}) + \sum_{i = 2}^g
\bm{\mathcal{A}}(\hat{\gamma}_i(t)) + \bm{\mathcal{K}},\\
w^{\pm} &= w_1 + w_{g+1} - w^{\pm}_{\infty} =
\bm{\mathcal{A}}(\hat{\gamma}) - \bm{\mathcal{A}}(\infty^{\pm}) +
\bm{\mathcal{K}}. \label{four vectors w 4}
\end{align}
\end{subequations}

\begin{proposition} \label{prop: existence of h}
The components $h_1, h_2$ of the normalised eigenvector $\bm{h}$ are
given by $h_1(P) = h_-(P)$ and $h_2(P) = h_+(P)$ where
\begin{equation*}
h_{\pm}(P) = \frac{\theta\left( \bm{\mathcal{A}}(\infty^{\mp}) - w_1
\right) \theta\left( \bm{\mathcal{A}}(\infty^{\mp}) - w_{g+1}
\right)}{\theta\left( \bm{\mathcal{A}}(\infty^{\mp}) -
w^{\pm}_{\infty} \right) \theta\left( \bm{\mathcal{A}}(\infty^{\mp}) -
w^{\pm} \right)} \cdot \frac{\theta\left( \bm{\mathcal{A}}(P) -
w^{\pm}_{\infty} \right) \theta\left( \bm{\mathcal{A}}(P) - w^{\pm}
\right)}{\theta\left( \bm{\mathcal{A}}(P) - w_1 \right) \theta\left(
\bm{\mathcal{A}}(P) - w_{g+1} \right)}.
\end{equation*}
\begin{proof}
The first factor in this formula is merely a constant ensuring
$h_{\pm}(\infty^{\mp}) = 1$. So we need to show that the second factor
is a well defined function of $P$ and has the right divisor. But as
$P$ is taken around a $\bm{b}$-cycle the $\theta$-functions change by
various factors \eqref{automorphy} which cancel however by \eqref{four
vectors w 4}. As a result $h_{\pm}(P)$ are well defined.

The two $\theta$-functions in the denominator
vanish at the points $\hat{\gamma}_1(t), \ldots, \hat{\gamma}_g(t)$
and $\hat{\gamma}_2(t), \ldots, \hat{\gamma}_{g+1}(t)$
respectively by corollary \ref{cor: theta zeroes}. Likewise the first
$\theta$-function in the numerator vanishes at $\infty^{\pm}$ and
$\hat{\gamma}_2(t), \ldots, \hat{\gamma}_g(t)$ but the latter $g-1$
zeroes cancel with the same zeroes in the denominator so that
$(h_{\pm}) \geq - \hat{\gamma}(t) + \infty^{\pm}$.
\end{proof}
\end{proposition}


\section{Baker-Akhiezer vector and linearisation} \label{section: B-A vector}


Equations \eqref{evolution hierarchy} express the fact that the
operators $\partial_{t_M} - J_M(x)$ all commute among themselves as
well as individually with the monodromy matrix $\Omega(x)$. This means
they can all be simultaneously diagonalised and there exists an
eigenvector $\bm{\psi}(P,t)$ at every $P \in \hat{\Sigma}$ with
$\hat{\pi}(P) = x$ which solves the following linear system
\begin{equation} \label{simultaneous}
\big(\partial_{t_M} - J_M(x,t)\big) \bm{\psi}(P,t) = 0, \quad \forall
M.
\end{equation}

\begin{remark}
Note that the vector equations \eqref{simultaneous} might not have
global solution $\bm{\psi}$ for topological reasons: if the base
space is not simply connected (which is the case here since the
$\sigma$ coordinate on the worldsheet is periodic), then even
though the Lax connection $J(x)$ is flat there are still closed
paths with non-trivial holonomy, and hence a covariantly constant
vector cannot exist globally on the base space. Thus if
$\bm{\psi}(P)$ is a local solution on a neighbourhood
$\mathcal{U}$ of the base space, then $\Omega(x)\bm{\psi}(P) =
\Lambda(P) \bm{\psi}(P)$ describes the same solution on
$\mathcal{U}$.
\end{remark}

Just as in the case of the normalised eigenvector $\bm{h}$ the aim is
to identify the analytic properties of the vector $\bm{\psi}(P, t)$
which specify it uniquely.
Since $\bm{\psi}(P,t)$ is an eigenvector it can be written as a
multiple of the normalised eigenvector,
\begin{subequations} \label{psi vs h}
\begin{equation} \label{psi vs h 1}
\bm{\psi}(P,t) = \varphi(P,t) \bm{h}(P,t).
\end{equation}
Alternatively we can also write the solution to \eqref{simultaneous}
in the form
\begin{equation} \label{psi vs h 2}
\bm{\psi}(P, t) = \widehat{\Psi}(x, t) \bm{h}(P, 0),
\end{equation}
\end{subequations}
where $\widehat{\Psi}(x, t)$ is a formal solution to the matrix
analogue of equation \eqref{simultaneous}, namely
\begin{equation} \label{simultaneous matrix}
\big(\partial_{t_M} - J_M(x,t)\big) \widehat{\Psi}(x,t) = 0, \quad
\forall M.
\end{equation}
Indeed the vector $\bm{\psi}$ defined by \eqref{psi vs h 2} then trivially
satisfies \eqref{simultaneous}. Furthermore, if we fix the initial
condition to be $\bm{\psi}(P, 0) = \bm{h}(P, 0)$ so that
$\widehat{\Psi}(x, 0) = {\bf 1}$,
then by uniqueness of the solution to \eqref{simultaneous matrix} with
initial condition $\Omega(x, 0)$ it follows that $\widehat{\Psi}(x, t)
\Omega(x, 0) = \Omega(x, t) \widehat{\Psi}(x, t)$ and therefore
\eqref{psi vs h 2} is indeed also an eigenvector of the monodromy
matrix $\Omega(x, t)$ with eigenvalue $\Lambda(P)$. Having already
identified the defining analytic properties of $\bm{h}$ we now
use \eqref{psi vs h} to determine those of $\bm{\psi}$.

The hierarchy of Lax matrices can be rewritten in the more transparent
form
\begin{equation} \label{Lax matrix singular parts}
J_{n,\pm}(x) = \left( \Psi(x) s_{n,\pm}(x) \sigma_3 \Psi(x)^{-1}
\right)_{\pm 1}.
\end{equation}
\begin{definition}
The \dub{singular parts} $s_{n,\pm}(x)$ are given by
\begin{equation} \label{singular parts}
s_{n,\pm}(x) = \left( \frac{i}{\pi} \frac{x^2}{x^2 - 1}
\left(Q^{(\pm)}_{n-1} + \frac{Q^{(\pm)}_{-1}}{(x \mp 1)^n}\right)
\right)_{\pm 1}.
\end{equation}
\end{definition}
In the particular case of the zeroth level $n = 0$ where the Lax
matrix becomes the Lax connection $J_{0,\pm}(x) = J_{\pm}(x)$, the
singular parts read
\begin{equation} \label{singular parts J}
s_{0,\pm}(x) = s_{\pm}(x) = \frac{i \kappa_{\pm}}{1 \mp x}.
\end{equation}
\begin{lemma} \label{lemma: BA vector properties}
Let $\bm{\psi}(P,t)$ be the eigenvector which solves
\eqref{simultaneous} with initial condition $\bm{\psi}(P,0) =
\bm{h}(P,0)$ then it is meromorphic on $\hat{\Sigma} \setminus \{ (\pm
1)^{\pm} \}$ with
\begin{subequations} \label{Baker-Akhiezer vec def}
\begin{equation} \label{Baker-Akhiezer vec def 1}
\begin{split}
&(\psi_1) \geq - \hat{\gamma}(0) + \infty^-, \quad \psi_1(\infty^+) =
1, \\
&(\psi_2) \geq - \hat{\gamma}(0) + \infty^+, \quad \psi_2(\infty^-) =
1,
\end{split}
\end{equation}
and has the following asymptotic behaviour in a neighbourhood of $(\pm
1)^{\pm} \in \hat{\Sigma}$,
\begin{equation} \label{Baker-Akhiezer vec def 2}
\left\{
\begin{split}
&\psi_i(x^{\pm},t) e^{\mp \sum_n s_{n,+}(x) t_{n,+}} = O(1), \quad
\text{as }\; x \rightarrow +1,\\
&\psi_i(x^{\pm},t) e^{\mp \sum_n s_{n,-}(x) t_{n,-}} = O(1), \quad
\text{as }\; x \rightarrow -1.
\end{split}
\right.
\end{equation}
\end{subequations}
\begin{proof}
Because $J_M(x)$ only has poles at $x = \pm 1$ it follows by
Poincar\'e's theorem on holomorphic differential equations that
$\widehat{\Psi}(x,t)$ is holomorphic outside $x = \pm 1$ since the
initial condition $\widehat{\Psi}(x,0) = {\bf 1}$ is\footnote{The same
conclusion does not hold for the vector $\bm{\psi}(P,t)$ even though
it satisfies the system \eqref{simultaneous}. Indeed, we chose its initial
condition to be $\bm{\psi}(P,0) = \bm{h}(P,0)$ which has poles at
$\hat{\gamma}(0)$. Therefore we conclude that the components of
$\bm{\psi}(P,t)$ are holomorphic in $P$ away from both
$\hat{\pi}^{-1}(\pm 1)$ and the points of $\hat{\gamma}(0)$.}. It
directly follows from \eqref{psi vs h 2} that $\bm{\psi}(P,t)$ is
meromorphic outside $\hat{\pi}^{-1}(\pm 1)$ with poles at
$\hat{\gamma}(0)$. Moreover, using the gauge fixing condition
$J_M(\infty) = 0$ we observe that $\partial_{t_M}
\widehat{\Psi}(\infty, t) = 0$ and hence $\widehat{\Psi}(\infty, t) =
{\bf 1}$ by the choice of initial conditions. Equations
\eqref{Baker-Akhiezer vec def 1} now follow from \eqref{normalised ev
properties} at $t = 0$.

Consider now the representation \eqref{psi vs h 1} of
$\bm{\psi}(P,t)$ which we can write as $\Psi(x) = H(x) \Phi(x)$
where $\Psi(x)$ and $H(x)$ are the matrix of column eigenvectors
$\bm{\psi}$ and $\bm{h}$ at $x$ respectively and $\Phi(x) =
\diag(\varphi(x^+), \varphi(x^-))$. Since $\bm{h}$ is holomorphic
in a neighbourhood of $\hat{\pi}^{-1}(\pm 1)$ this means that
$H(x)$ is holomorphic near $x = \pm 1$. Rewriting
\eqref{simultaneous} as a matrix equation $\Psi^{-1}(x)
\partial_{t_{m,\pm}} \Psi(x) = \Psi(x)^{-1} J_{m,\pm}(x) \Psi(x)$
we study it in a neighbourhood of $x = \pm 1$. It can be written as
\begin{equation} \label{singular parts proof}
\left(\partial_{t_{m,\pm}} \Phi(x) \right) \Phi(x)^{-1} + H(x)^{-1}
\partial_{t_{m,\pm}} H(x) = s_{m,\pm}(x) \sigma_3 + H(x)^{-1} V(x) H(x),
\end{equation}
where we have set $J_{m,\pm}(x) = H(x) s_{m,\pm}(x) \sigma_3
H(x)^{-1} + V(x)$ with $V(x)$ being the negative of the holomorphic
part of $H(x) s_{m,\pm}(x) \sigma_3 H(x)^{-1}$ at $x = \pm
1$. The second term on the right hand side is clearly holomorphic at
$x = \pm 1$. One can show that the second term on the left hand
side also is. For this we need the evolution equation
\eqref{normalised evolution} of the normalised eigenvector that we
will derive later in the proof of theorem \ref{thm: linear motion}. It
reads in matrix form
\begin{equation*}
H(x)^{-1} \partial_{t_{m,\pm}} H(x) = H(x)^{-1} J_{m,\pm}(x) H(x) -
\diag(C(x^+), C(x^-)),
\end{equation*}
where $C(P) = \bm{\alpha} \cdot J_{m,\pm}(x) \bm{h}(P)$. Therefore
\begin{equation*}
\diag(C(x^+),C(x^-)) = s_{m,\pm}(x) \sigma_3 + \diag(\bm{\alpha}
\cdot V(x) \bm{h}(x^+), \bm{\alpha} \cdot V(x) \bm{h}(x^-) ),
\end{equation*}
where the second term is holomorphic at $x = \pm 1$. The first term is
singular but cancels with the corresponding term in $H(x)^{-1}
J_{m,\pm}(x) H(x) = s_{m,\pm}(x) \sigma_3 + H(x)^{-1} V(x)
H(x)$. Hence the second terms in both the left and right hand sides of
\eqref{singular parts proof} are holomorphic at $x = \pm 1$ so that
$\varphi(x^{\pm})^{-1} \partial_{t_{m,\pm}} \varphi(x^{\pm}) = \pm
s_{m,\pm}(x) + O(1)$ from which \eqref{Baker-Akhiezer vec def 2}
follows.
\end{proof}
\end{lemma}

Functions on a Riemann surface $\hat{\Sigma}$ satisfying
properties like those in \eqref{Baker-Akhiezer vec def} are known
as \dub{Baker-Akhiezer functions}. They have essential
singularities at certain punctures \eqref{Baker-Akhiezer vec def
2} generalising the exponential map $z \mapsto \text{exp}\, z$
which is holomorphic in $\mathbb{C}$ but has an essential
singularity at $z = \infty$. Despite the fact that these functions
are not meromorphic on $\hat{\Sigma}$ they still admit the notion
of a degree since,

\begin{lemma}
The Baker-Akhiezer functions $\psi_i$ have an equal number of zeroes
and poles (counting multiplicities).
\begin{proof}
Consider the differential $d\log \psi_i = d\psi_i/\psi_i$ on
$\hat{\Sigma}$. It is straightforward to show using property
\eqref{Baker-Akhiezer vec def 2} that $d \log \psi_i$ is meromorphic
in a neighbourhood of the punctures $\hat{\pi}^{-1}(\pm 1)$. But since
it is also meromorphic away from the punctures on $\hat{\Sigma}
\setminus \hat{\pi}^{-1}(\pm 1)$, $d\log \psi_i$ defines a meromorphic
differential on $\hat{\Sigma}$. The lemma follows using
$\int_{\partial \hat{\Sigma}_{\text{cut}}} d\log \psi_i = 0$ and the
fact that $d\log \psi_i$ has no residues at $\hat{\pi}^{-1}(\pm 1)$.
\end{proof}
\end{lemma}

Since we are assuming $\hat{\gamma}(t)$ to be non-special the divisor
$\hat{\gamma}(0)$ of poles of $\psi_i$ is also in general position
which allows us to use the Riemann-Roch theorem to prove,

\begin{proposition} \label{uniqueness of psi}
Conditions \eqref{Baker-Akhiezer vec def} uniquely specify a
Baker-Akhiezer vector $\bm{\psi}$.
\begin{proof}
Suppose there are two vectors $\bm{\psi}$ and $\bm{\psi}'$ satisfying
conditions \eqref{Baker-Akhiezer vec def} and consider the function
$f_i(P) \equiv \psi_i(P)/\psi'_i(P)$, $i = 1, 2$. Since $\psi_i$ and
$\psi'_i$ have the same essential singularities \eqref{Baker-Akhiezer
vec def 2} at $\hat{\pi}^{-1}(\pm 1)$ they cancel in the definition of
$f_i$ which is therefore meromorphic. Its divisor of poles is
contained in the divisor of zeroes of $\psi'_i$ which is of degree $g$
and by assumption is in general position. Thus $f_i$ must be constants
which are fixed to one by the conditions $f_1(\infty^+) = f_2(\infty^-)
= 1$.
\end{proof}
\end{proposition}

It remains to show that there exists a pair of function $\psi_1,
\psi_2$ which satisfy all the conditions of \eqref{Baker-Akhiezer
vec def}. Once again existence is shown by explicit construction
of such functions using the Riemann $\theta$-function as a
building block. It follows from proposition \ref{uniqueness of
psi} that the functions constructed below must be equal to the
components of the Baker-Akhiezer vector $\bm{\psi}$ thus providing
\dub{reconstruction formulae}.

The main ingredient of these formulae is a certain normalised Abelian
differential of the second kind $d \mathcal{Q}$. We let $d
\mathcal{Q}$ have poles at the points $\hat{\pi}^{-1}(\pm 1) \in
\hat{\Sigma}$ with singular parts defined in terms of \eqref{singular
parts} by
\begin{equation*}
d \mathcal{Q} = - i dS_{\pm}, \quad \text{as} \; x \rightarrow \pm 1,
\qquad \text{where} \;
\left\{
\begin{array}{l}
S_+(x^{\pm}, t) = \pm \sum_n s_{n,+}(x) t_{n,+}, \\
S_-(x^{\pm}, t) = \pm \sum_n s_{n,-}(x) t_{n,-}.
\end{array}
\right.
\end{equation*}
Its regular part is fixed uniquely by the normalisation condition
$\int_{a_i} d \mathcal{Q} = 0$. The $\bm{b}$-periods define a vector
in $\mathbb{C}^g$. As in chapter \ref{chapter: Riemann surfaces} we
denote $\bm{\zeta}_D = \bm{\mathcal{A}}(D) + \bm{\mathcal{K}}$.

\begin{proposition} \label{prop: existence of psi}
The components $\psi_1, \psi_2$ of the Baker-Akhiezer vector
$\bm{\psi}$ are given by $\psi_1(P) = \psi_+(P)$ and $\psi_2(P) =
\psi_-(P)$ where
\begin{equation*}
\psi_{\pm}(P) = h_{\mp}(P,0) \frac{\theta \left( \bm{\mathcal{A}}(P) +
\int_{\bm{b}} d \mathcal{Q} - \bm{\zeta}_{\gamma_{\mp}(0)} \right) \theta
\left( \bm{\mathcal{A}}(\infty^{\pm}) - \bm{\zeta}_{\gamma_{\mp}(0)}
\right)}{\theta \left( \bm{\mathcal{A}}(P) - \bm{\zeta}_{\gamma_{\mp}(0)}
\right) \theta \left( \bm{\mathcal{A}}(\infty^{\pm}) + \int_{\bm{b}} d
\mathcal{Q} - \bm{\zeta}_{\gamma_{\mp}(0)} \right)} \; \exp \left( i
\int_{\infty^{\pm}}^P d\mathcal{Q} \right).
\end{equation*}
\begin{proof}
Since the $\theta$-functions are all holomorphic in a
neighbourhood of $x = \pm 1$, it follows by definition of $d
\mathcal{Q}$ that $\psi_{\pm}$ have the right asymptotics
\eqref{Baker-Akhiezer vec def 2}.

Among the four $\theta$-functions present only two of them depend on
$P$. The other two merely define overall constants ensuring
$\psi_{\pm}(\infty^{\pm}) = 1$. So focusing on the $P$ dependence we
need to show that $\psi_{\pm}(P)$ is a well defined function of $P$,
has the right divisor and the right asymptotics at $x = \pm 1$.

When $P$ is taken around an $\bm{a}$-cycle nothing changes because $d
\mathcal{Q}$ is normalised and the $\theta$-functions are
$\bm{a}$-periodic. As $P$ goes around the $b_k$-cycle the ratio of
$\theta$-functions gets multiplied by $\exp \left(- i \int_{b_k} d
\mathcal{Q} \right)$ which exactly cancels with the shift in the
exponential of $\psi_{\pm}(P)$, which is therefore well defined.

The $\theta$-function in the denominator vanishes at the $g$ points
of $\gamma_{\mp}(0)$ which all cancel with the corresponding zeroes of
$h_{\mp}(P,0)$ to leave $(\psi_{\pm}) \geq - \hat{\gamma}(0) +
\infty^{\pm}$.
\end{proof}
\end{proposition}

Recall from proposition \ref{prop: residual gauge} that the choice of
a dynamical divisor $\hat{\gamma}(t)$ for the normalised eigenvector
$\bm{h}$ corresponded to a choice of residual gauge. However the
Baker-Akhiezer only depends on the initial value $\hat{\gamma}(0)$ of
the divisor. Thus the choice of an initial divisor $\hat{\gamma}(0)$
in the construction of the Baker-Akhiezer vector should correspond to
fixing only the constant part of the residual gauge. But the constant
part of the residual gauge symmetry \eqref{residual gauge symmetry}
corresponds precisely to the unfixed $U(1)_R$ subgroup of the global
$SU(2)_R$ (in fact, before imposing reality conditions we are really
dealing with a $\mathbb{C}^{\ast}$ subgroup of $SL(2,\mathbb{C})_R$),
therefore

\begin{proposition} \label{prop: U(1)_R angle}
The choice of an initial divisor in $[\hat{\gamma}(0)]$ corresponds to
a choice of initial value for the $U(1)_R$ angle.
\end{proposition}

We can be a bit more specific about this connection between the
divisor $\hat{\gamma}(0)$ and the $U(1)_R$ angle. Since the
Baker-Akhiezer vector is defined as the solution to the linear
system \eqref{simultaneous} with initial condition $\bm{\psi}(P,0)
= \bm{h}(P,0)$ it is easy to determine how it transforms under
$U(1)_R$. Indeed, the Lax matrices all transform by conjugation
$J_M(x) \mapsto \tilde{g} J_M(x) \tilde{g}^{-1}$ where $\tilde{g}
= \diag(W, W^{-1}) \in SL(2, \mathbb{C})$. The initial condition
being the normalised eigenvector it transforms as in
\eqref{residual gauge symmetry}, namely $\bm{h}(P,0) \mapsto
f(P,0)^{-1} \tilde{g} \bm{h}(P,0)$ where $f(P,0) = W h_1(P,0) +
W^{-1} h_2(P,0)$. It follows then that the Baker-Akhiezer vector
transforms as $\bm{\psi}(P,t) \mapsto f(P,0)^{-1} \tilde{g}
\bm{\psi}(P,t)$ or equivalently in terms of the reconstructed
components $\psi_{\pm}$ of proposition \ref{prop: existence of
psi},
\begin{equation} \label{U1R of reconstructed psi}
\psi_{\pm}(P,t) \mapsto f(P,0)^{-1} W^{\pm 1} \psi_{\pm}(P,t),
\end{equation}
where we can write $f(P,0) = W \psi_1(P,0) + W^{-1} \psi_2(P,0)$.
Proposition \ref{U1R divisors} below expresses exactly how the
parameter $W$ of a $U(1)_R$ transformation depends on the two
divisors $\hat{\gamma}(0)$ and $\hat{\gamma}'(0)$ related through
this $U(1)_R$ transform. We first need to define a normalised
Abelian differential of the third kind $\omega_{\infty}$ that will
be essential in the description of the $U(1)_R$ degree of freedom.
It is defined by the residues $\pm \frac{1}{2 \pi i}$ at its
simple poles $\infty^{\pm} \in \hat{\Sigma}$. Using the notation
of chapter \ref{chapter: Riemann surfaces} for the basis of
normalised Abelian differentials of the third kind $\omega_{PQ}$
it can also be written more explicitly as
\begin{equation} \label{third kind diff at infty}
\omega_{\infty} = \frac{1}{2 \pi i} \omega_{\infty^+ \infty^-}.
\end{equation}

\begin{proposition} \label{U1R divisors}
The $U(1)_R$ transformation $\tilde{g} = \diag(W, W^{-1})$ which
takes the initial divisor from $\hat{\gamma}(0)$ to
$\hat{\gamma}'(0)$ is given explicitly by
\begin{equation*}
W = \exp \frac{i}{2} \left( 2 \pi \sum_{j = 1}^{g+1}
\int_{\hat{\gamma}_i(0)}^{\hat{\gamma}'_i(0)} \omega_{\infty}
\right).
\end{equation*}
\begin{proof}
Recall that the function $f(P,0)$ has poles at the initial divisor
$\hat{\gamma}(0)$ and its zeroes define the `new' initial divisor
$\hat{\gamma}'(0)$. Furthermore it takes the values $f(\infty^{\pm}) =
W^{\pm 1}$ at the points $\infty^{\pm}$. The result is now immediate
by lemma \ref{lemma: f(P)/f(Q)=r}.
\end{proof}
\end{proposition}

\subsection*{Linearisation}

Notice that the hierarchy of times enters linearly in the
definition of the Baker-Akhiezer vector $\bm{\psi}(P, t)$
through the essential singularity, which is a usual trait of
finite-gap integration. All the time dependence of the Baker-Akhiezer
vector, and hence of the solution, is encoded in the meromorphic
differential $d \mathcal{Q}$ which is linear in the hierarchy of
times. In fact, we can define a differential associated to each time
of the hierarchy by writing
\begin{equation} \label{time-differential coupling}
d \mathcal{Q} = \sum_n t_{n,+} d \Omega_{n,+} + \sum_n t_{n,-} d
\Omega_{n,-} = \sum_N t_N d \Omega_N,
\end{equation}
using the multi-index notation, where the normalised Abelian
differentials of the second kind $d \Omega_{n,\pm}$ are defined
uniquely by their respective behaviours at the points $x = \pm 1$,
namely
\begin{equation} \label{dOmega asymptotics pm 1}
\begin{split}
d \Omega_{n,+}(x^{\pm}) = \mp i d s_{n,+}(x) \quad \text{as} \;
x \rightarrow +1,\\
d \Omega_{n,-}(x^{\pm}) = \mp i d s_{n,-}(x) \quad \text{as} \; x
\rightarrow -1.
\end{split}
\end{equation}
This correspondence between times of the hierarchy and Abelian
differentials on $\hat{\Sigma}$
\begin{equation*}
t_{n,\pm} \mapsto d \Omega_{n,\pm}
\end{equation*}
is a very general feature of finite-gap integration. In standard
terminology one says that the differential \dub{couples} to the
time for obvious reasons from \eqref{time-differential coupling}.
As we saw in section \ref{section: hierarchy} of chapter
\ref{chapter: integrability} every Hamiltonian corresponds to a
Lax matrix which is responsible for generating the corresponding
time in the Lax formalism. Here we see that every Hamiltonian also
corresponds to a meromorphic differential on $\hat{\Sigma}$
responsible for generating the corresponding time in the
finite-gap language. Notice the splitting between differentials
singular at $x = +1$ and those singular at $x = -1$. These are
related to \dub{left and right movers} of the string. For
instance, at the zeroth level $n=0$ we have $\sigma^{\pm} \equiv
\frac{1}{2} (\tau \pm \sigma) = t_{0,\pm}$ and $dq_{\pm} \equiv dq
\pm dp = 2 \pi d \Omega_{0,\pm}$, so
\begin{equation*}
t_{0,+} d\Omega_{0,+} + t_{0,-} d\Omega_{0,-} = \frac{1}{2 \pi} (
\sigma dp + \tau dq).
\end{equation*}
The normalised Abelian differential $dp = \pi d \Omega_{0,+} - \pi
d \Omega_{0,-}$ is nothing but the differential of the
quasi-momentum defined by its asymptotics in \eqref{quasi-momentum
asymptotics pm1}. We see here that it couples to the worldsheet
spatial coordinate $\sigma$ which justifies the nomenclature
`quasi-momentum' for its Abelian integral. The differential $dq =
\pi d \Omega_{0,+} + \pi d \Omega_{0,-}$ on the other hand couples
to the worldsheet time coordinate $\tau$ suggesting that,
\begin{definition}
The \dub{quasi-energy} is the Abelian integral $q(P) = \int^P dq$.
\end{definition}

\noindent Its differential $dq$ is the unique normalised Abelian
differential of the second kind defined by the following asymptotics,
\begin{equation} \label{quasi-energy asymptotics pm1}
\begin{split}
dq(x^{\pm}) &= \mp d \left( \frac{\pi \kappa_+}{x - 1} \right) +
O\left( (x-1)^2 \right), \quad \text{as} \;\; x \rightarrow +1,\\
dq(x^{\pm}) &= \pm d \left( \frac{\pi \kappa_-}{x + 1} \right) +
O\left( (x+1)^2 \right), \quad \text{as} \;\; x \rightarrow -1.
\end{split}
\end{equation}

The linear time-dependence of the singular parts (\textit{i.e.} of
the exponents of the Baker-Akhiezer vector) has the profound
consequence that the motion of the system can be mapped to a
linear motion in an appropriate space, which is characteristic of
all integrable systems. This is the statement of theorem \ref{thm:
linear motion} below. Before we can state the theorem we need to
introduce some notation. It is evident from proposition \ref{U1R
divisors} that the points $\infty^{\pm}$ will play a particular
role in characterising the $U(1)_R$ degree of freedom. In
particular the differential \eqref{third kind diff at infty} plays
an essential part. As in chapter \ref{chapter: Riemann surfaces}
we therefore introduce a modulus
\begin{equation*}
\mathfrak{m} = \infty^+ + \infty^-
\end{equation*}
(which is an integral divisor) to encapsulate these
special points at infinity. The generalised Jacobian
$J_{\mathfrak{m}}(\hat{\Sigma})$ (sometimes also denoted
$J(\hat{\Sigma}, \infty^{\pm})$) relative to this modulus was
defined in chapter \ref{chapter: Riemann surfaces} as well. It can
be understood as the Jacobian associated to the singular algebraic
curve obtained by identifying the points $\infty^{\pm}$ on
$\hat{\Sigma}$. Besides the $g$ canonical $\bm{b}$-cycles we
introduce a degenerate $b_{\infty}$-cycle starting at $\infty^-$
and ending at $\infty^+$ and combine these into a
$(g+1)$-dimensional vector $\vec{b} = (b_1, \ldots, b_g,
b_{\infty})^{\sf T}$. Following definition \ref{def: generalised
Abel map} of chapter \ref{chapter: Riemann surfaces} we also
introduce the generalised Abel map $\vec{\mathcal{A}}(P) = 2 \pi
\int_{P_0}^P \vec{\omega}$ where here $\vec{\omega} = (\omega_1,
\ldots, \omega_g, \omega_{\infty})^{\sf T}$. Note that here we let
the extra $b$-period $b_{\infty}$ and the third kind Abelian
differential $\omega_{\infty}$ be the $(g+1)^{\text{st}}$
component and not the $0^{\text{th}}$, just for notational
convenience. Recall from chapter \ref{chapter: Riemann surfaces}
that the generalised Jacobian is isomorphic via the generalised
Abel map to the generalised Picard group of degree zero divisors
on $\hat{\Sigma} \setminus \{ \infty^{\pm} \}$ modulo
$\mathfrak{m}$-equivalence. Thus the divisor $\hat{\gamma}(t) -
\hat{\gamma}(0)$ represents a point in
$J_{\mathfrak{m}}(\hat{\Sigma})$ which by the following theorem
has the amazing property that its motion is linear on
$J_{\mathfrak{m}}(\hat{\Sigma})$. The quite lengthy proof is an
adaptation of that in \cite[pp.142--145]{Babelon} to include the
$U(1)_R$ degree of freedom which as we have already know
corresponds to a choice of divisor in the class
$[\hat{\gamma}(0)]$.

\begin{theorem} \label{thm: linear motion}
The motion of the dynamical divisor $\hat{\gamma}(t)$ on
$\hat{\Sigma}$ is mapped by the generalised Abel map
$\vec{\mathcal{A}}$ to a linear motion on the generalised Jacobian
$J_{\mathfrak{m}}(\hat{\Sigma})$,
\begin{equation} \label{linear dynamical divisor}
\vec{\mathcal{A}}(\hat{\gamma}(t)) =
\vec{\mathcal{A}}(\hat{\gamma}(0)) - \int_{\vec{b}} d\mathcal{Q}.
\end{equation}
\begin{proof}
Consider the equation $\Omega(x) \bm{h}(P) = \Lambda(P) \bm{h}(P)$
for the normalised eigenvector. Differentiating this equation with
respect to the higher time $t_N$ (with $N = (n, s)$ where $n \in
\mathbb{N}$ and $s = \pm 1$) and using the evolution equation
\eqref{monodromy evolution hierarchy} for the monodromy matrix we
find
\begin{equation*}
\left( \Omega(x) - \Lambda(P) \right) \left(
\partial_{t_N} \bm{h}(P) - J_N(x) \bm{h}(P) \right) = 0.
\end{equation*}
It follows then by uniqueness of the eigenvector at each point $P
\in \hat{\Sigma}$ that
\begin{equation} \label{normalised evolution}
\partial_{t_N} \bm{h}(P) = \left[ J_N(x) - C(P) \right] \bm{h}(P),
\end{equation}
for some scalar function $C(P,t) \in \mathbb{C}$. Using the fact
that the eigenvector $\bm{h}(P)$ is normalised by the condition
\eqref{normalisation} we obtain an expression for this scalar,
namely $C(P) = \bm{\alpha} \cdot J_N(x) \bm{h}(P)$. Next we
introduce the following function depending on a small time
difference $\delta t$,
\begin{equation} \label{N definition}
\mathcal{N}(t,\delta t,P) = 1 + \delta t C(P,t) = 1 + \delta t
\bm{\alpha} \cdot J_N(x) \bm{h}(P).
\end{equation}
Working to first order in $\delta t$ one can then rewrite equation
\eqref{normalised evolution} in terms of this function as follows,
\begin{equation} \label{N relation}
\mathcal{N}(t, \delta t, P) \bm{h}(P,t + \delta t) = \left( {\bf
1} + \delta t J_N(x) \right) \bm{h}(P,t) + O(\delta t^2).
\end{equation}
This relation allows us to read off the pole structure of
$\mathcal{N}$. Indeed, the right hand side of \eqref{N relation}
has simple poles at $\hat{\gamma}(t)$ from $\bm{h}(P,t)$ and poles
of order $n+1$ at $s^+, s^-$ from $J_N(x) = J_{n,s}(x)$. Since the
left hand side must have the same poles this implies that the
function $\mathcal{N}$ must have simple poles at $\hat{\gamma}(t)$
as well as poles of order $n+1$ at $s^+, s^-$. Furthermore, to
cancel off the undesired poles at $\hat{\gamma}(t + \delta t)$
coming from $\bm{h}(P,t + \delta t)$ the function $\mathcal{N}$
must also have simple zeroes at $\hat{\gamma}(t + \delta t)$. We
denote its remaining $2n+2$ zeroes as $s_{\alpha}^{\pm}(t), \alpha
= 1, \ldots, n+1$. By continuity, as $\delta t \rightarrow 0$ the
zeroes $s_{\alpha}^{\pm}(t)$ must converge to the poles $s^{\pm}$
respectively since $\mathcal{N}(t,0,P) = 1$. Moreover we also note
from \eqref{N definition} that at both points $\infty^{\pm}$ the
function $\mathcal{N}$ takes the value one (since $J_N(\infty) =
0$ and the components of $\bm{h}$ are regular at $\infty^{\pm}$).
Thus
\begin{gather*}
(\mathcal{N}) = \hat{\gamma}(t + \delta t) - \hat{\gamma}(t) +
\sum_{\alpha = 1}^{n+1} s_{\alpha}^+(t) + \sum_{\alpha = 1}^{n+1}
s_{\alpha}^-(t) - (n+1) s^+ - (n+1) s^-,\\
\mathcal{N}(t, \delta t, \infty^+) = \mathcal{N}(t, \delta t,
\infty^-) = 1.
\end{gather*}
It now follows by the generalised Abel theorem \ref{theorem:
generalised Abel} that
\begin{equation*}
\sum_{j = 1}^{g+1} \int_{\hat{\gamma}_j(t)}^{\hat{\gamma}_j(t +
\delta t)} \vec{\omega} = - \sum_{\alpha = 1}^{n+1}
\int_{s^+}^{s_{\alpha}^+(t)} \vec{\omega} - \sum_{\alpha = 1}^{n+1}
\int_{s^-}^{s_{\alpha}^-(t)} \vec{\omega}.
\end{equation*}
The left hand sides of the above equations multiplied by $(\delta
t)^{-1}$ tend to the time-derivative of the generalised Abel map of
the divisor $\hat{\gamma}(t)$ in the limit $\delta t \rightarrow
0$. To show \eqref{linear dynamical divisor} we therefore compute the
right hand sides in this limit. We take $\delta t$ sufficiently small
so that all the zeroes $s_{\alpha}^{\pm}$ are within a small
neighbourhood $U^{\pm}$ of the corresponding poles $s^{\pm}$. Note
that the differentials $\vec{\omega}$ are all holomorphic in the
neighbourhoods $U^{\pm}$. Now letting $\vec{\sigma}^{\pm}(P) =
\int_{s^{\pm}}^P \vec{\omega}$ be the local integral of $\vec{\omega}$ in
$U^{\pm}$,
\begin{equation*}
\frac{1}{\delta t} \sum_{\alpha = 1}^{n+1}
\int_{s^{\pm}}^{s_{\alpha}^{\pm}(t)} \vec{\omega} = \frac{1}{\delta t}
\sum_{\alpha = 1}^{n+1} \vec{\sigma}^{\pm}(s_{\alpha}^{\pm}(t))
= \frac{1}{2 \pi i} \int_{\partial U^{\pm}} \frac{\vec{\sigma}^{\pm}
dC}{1 + (\delta t) C},
\end{equation*}
using the fact that the zeroes $s_{\alpha}^{\pm}(t)$ satisfy
$C(s_{\alpha}^{\pm},t) + (\delta t)^{-1} = 0$ and
$\vec{\sigma}^{\pm}(s^{\pm}) = 0$ so that within the neighbourhood
$U^{\pm}$ only the zeroes $s_{\alpha}^{\pm}(t)$ contribute. In the
limit $\delta t \rightarrow 0$ all the zeroes tend to the single point
$s^{\pm}$ so that
\begin{equation*}
\lim_{\delta t \rightarrow 0} \frac{1}{\delta t} \sum_{\alpha =
1}^{n+1} \int_{s^{\pm}}^{s_{\alpha}^{\pm}(t)} \vec{\omega} =
\res_{s^{\pm}} \left( \vec{\sigma}^{\pm} dC \right) = - \res_{s^{\pm}}
\left( C \vec{\omega} \right).
\end{equation*}
These last residues can be computed explicitly using the definition of
the function $C(P) = \bm{\alpha} \cdot J_{n,s}(x) \bm{h}(P)$ and the
Lax matrices $J_{n,s} = \left( H(x) s_{n,s}(x) \sigma_3 H(x)^{-1}
\right)_s$. When computing the residue at $x = s$ one need not take
the pole part in the expression for the Lax matrix. Thus we can write
$J_{n,s} = H(x) s_{n,s}(x) \sigma_3 H(x)^{-1}$ and
\begin{equation*}
\res_{s^{\pm}} \left( C \vec{\omega} \right) = \res_{s^{\pm}} \left(
\bm{\alpha} \cdot H(x) s_{n,s}(x) \sigma_3 H(x)^{-1} \bm{h}(P)
\vec{\omega} \right) = \pm \res_{s^{\pm}} \left( s_{n,s}(x)
\vec{\omega} \right).
\end{equation*}
Here we have also made use of the definition $H(x) = \left(
\bm{h}(x^+), \bm{h}(x^-) \right)$. This last expression can be
rewritten in terms of the Abelian integral of the differentials $d
\Omega_{n,s}$, namely $\pm \res_{s^{\pm}} \left( s_{n,s}(x)
\vec{\omega} \right) = i \res_{s^{\pm}} \left( \Omega_{n,s}(P)
\vec{\omega} \right)$. Finally we arrive at the following simple
expressions for the time-derivatives of the generalised Abel maps
$\vec{\mathcal{A}}(\hat{\gamma}(t))$,
\begin{equation} \label{linear motion proof}
\frac{\partial}{\partial t_{n,s}} \vec{\mathcal{A}}(\hat{\gamma}(t)) =
2 \pi i (\res_{s^+} + \res_{s^-}) \Omega_{n,s}(P) \vec{\omega}.
\end{equation}
Notice first of all that the left hand side is independent of the
higher times $\{ t \}$ and hence the dynamics of $\hat{\gamma}(t)$ is
mapped to a linear flow under the generalised Abel map. Considering
the first $g$ components of \eqref{linear motion proof} and using the
Riemann bilinear identity \eqref{Riemann bilinear 3} with $d\Omega_1 =
d\Omega_{n,s}$ and $d\Omega_2 = \omega_i$ we find that
\begin{equation*}
2 \pi i (\res_{s^+} + \res_{s^-}) \Omega_{n,s}(P) \omega_i =
- \int_{b_i} d \Omega_{n,s}.
\end{equation*}
Finally for the $(g+1)^{\text{st}}$ component we use again the Riemann
bilinear identity \eqref{Riemann bilinear 3} but with $d\Omega_1 =
d\Omega_{n,s}$ and $d\Omega_2 = \omega_{\infty}$ which reads
\begin{equation*}
2 \pi i (\res_{s^+} + \res_{s^-}) \Omega_{n,s}(P) \omega_{\infty} = -
2 \pi i (\res_{\infty^+} + \res_{\infty^-}) \Omega_{n,s}(P)
\omega_{\infty}.
\end{equation*}
The left hand side is easily evaluated using the definition
$\omega_{\infty} = \frac{1}{2 \pi i} \omega_{\infty^+ \infty^-}$ to give
\begin{equation*}
- 2 \pi i (\res_{\infty^+} + \res_{\infty^-}) \Omega_{n,s}(P)
\omega_{\infty} = - \left( \Omega_{n,s}(\infty^+) -
\Omega_{n,s}(\infty^-) \right) = - \int_{b_{\infty}} d\Omega_{n,s}
\end{equation*}
and the theorem is proved.
\end{proof}
\end{theorem}


\section{The dual linear system} \label{section: dual system}


We now introduce the concept of the dual normalised eigenvector
$\bm{h}^+$ and the dual Baker-Akhiezer vector $\bm{\psi}^+$. The
purpose of these vectors is two-fold. First of all they will provide
useful formulae for the inverses $H(x)^{-1}$ and $\Psi(x)^{-1}$ of the
matrices $H(x) = (\bm{h}(x^+), \bm{h}(x^-))$ and $\Psi(x) =
(\bm{\psi}(x^+),  \bm{\psi}(x^-))$ which appear in most of the
reconstruction formulae such as \eqref{pre reconstruction formula}.
Secondly the dual Baker-Akhiezer vector will be very useful in
discussing reality conditions in chapter \ref{chapter: reality}. Since
the matrix $g$ defined in \eqref{matrix g} is $SU(2)$-valued $g^{\dag}
= g^{-1}$ its inverse will be expressible in terms of the dual
Baker-Akhiezer vector.

The dual vectors $\bm{h}^+$ and $\bm{\psi}^+$ are defined in
essentially the same way as their usual counterparts $\bm{h}$ and
$\bm{\psi}$ except that they are taken to be \textit{left}
eigenvectors of the monodromy matrix $\Omega(x)$ as opposed to
\textit{right} eigenvectors. Specifically we have,
\begin{equation*}
\bm{h}^+(P,t) \left( \Omega(x,t) - \Lambda(P) {\bf 1} \right) = 0.
\end{equation*}
and $\bm{\psi}^+(P,t) = \varphi^+(P,t) \bm{h}^+(P,t)$. They are
both row vectors and $\bm{h}^+$ can be normalised by the condition
$\bm{h}^+ \cdot \bm{\alpha}^{\sf T} = 1$. The reason these dual
eigenvectors provide formulae for $H(x)^{-1}$ and $\Psi(x)^{-1}$
respectively essentially boils down to,
\begin{lemma} \label{lemma: orthogonal}
$\forall P \in \hat{\Sigma}$, $\quad \bm{h}^+(P) \cdot
\bm{h}(\hat{\sigma} P) = 0$.
\begin{proof}
Recall that $\hat{\sigma}$ denotes the hyperelliptic involution. If
$P$ corresponds to a point of the spectral curve for which $\Lambda(P)
\neq \Lambda(\hat{\sigma} P)$ then
\begin{equation*}
\Lambda(\hat{\sigma} P) \bm{h}^+(P) \cdot \bm{h}(\hat{\sigma} P) =
\bm{h}^+(P) \Omega(x) \bm{h}(\hat{\sigma} P) = \Lambda(P)
\bm{h}^+(P) \cdot \bm{h}(\hat{\sigma} P).
\end{equation*}
Thus the result holds at such points and remains true for all
$P \in \hat{\Sigma}$ by continuity.
\end{proof}
\end{lemma}

\begin{remark}
Note in particular that since a branch point $Q$ of $\hat{\Sigma}$
is a fixed point of the hyperellitpic involution $\hat{\sigma}$,
namely $\hat{\sigma} Q = Q$, it follows that $\bm{h}^+(Q) \cdot
\bm{h}(Q) = 0$.
\end{remark}

If we define the meromorphic function $\eta(P) = \bm{h}^+(P) \cdot
\bm{h}(P)$ then the row vector $\bm{H}^+(P) = \eta(P)^{-1}
\bm{h}^+(P)$ satisfies the following orthogonality conditions with
the normalised eigenvector $\bm{h}$,
\begin{equation} \label{orthogonality conditions}
\bm{H}^+(P) \cdot \bm{h}(P) = 1, \qquad \bm{H}^+(P) \cdot
\bm{h}(\hat{\sigma} P) = 0.
\end{equation}
Since by definition the matrix $H(x)$ is built out of the column
vector $\bm{h}(P)$ it follows from \eqref{orthogonality
conditions} that its inverse $H(x)^{-1}$ can be constructed using
the row vector $\bm{H}^+(P)$. Specifically we have proved the
following
\begin{proposition} \label{left inverse of H}
$H(x)^{-1} = \left( \bm{H}^+(x^+)^{\sf T}, \bm{H}^+(x^-)^{\sf T}
\right)^{\sf T}$.
\end{proposition}

We would like to obtain the analytic properties of the dual vectors
$\bm{h}^+(P)$ and $\bm{\psi}^+(P)$ on the Riemann surface
$\hat{\Sigma}$ so that they may also be reconstructed from a set of
algebro-geometric data. Let us begin with the dual normalised
eigenvector $\bm{h}^+$.

\subsection*{The dual normalised eigenvector}

We can extract the algebro-geometric data of the dual normalised
eigenvector $\bm{h}^+$ in a similar way to section \ref{section:
normalised eigenvector} for the normalised eigenvector $\bm{h}$.
It is straightforward to see that lemma \ref{lem: normalized ev is
mero} and proposition \ref{prop: g+1 poles} both remain true for
$\bm{h}^+$. We therefore define the \dub{dual dynamical divisor}
$\hat{\gamma}^+(t)$ to be the divisor of poles of $\bm{h}^+(P,t)$
whose degree is again $\deg \hat{\gamma}^+(t) = g + 1$. Its
equivalence class $[\hat{\gamma}^+(t)]$ is conveniently
characterised by the following,

\begin{lemma} \label{lemma: dual dyn characterisation}
Let $\Omega$ be a meromorphic differential with double poles at
$\infty^{\pm}$ and zeroes at $\hat{\gamma}(t)$. Its $g+1$
remaining zeroes are equivalent to the divisor
$\hat{\gamma}^+(t)$ whose image under the generalised Abel map
satisfies (where $B$ denotes the divisor of branch points of
$\hat{\Sigma}$)
\begin{equation} \label{Abel dyn + dyn^+}
\vec{\mathcal{A}}(\hat{\gamma}(t)) +
\vec{\mathcal{A}}(\hat{\gamma}^+(t)) = \vec{\mathcal{A}}(B).
\end{equation}
\begin{proof}
Equation \eqref{Abel dyn + dyn^+} easily follows from consideration of
the function $\eta(P)$ which has poles at $\hat{\gamma}(t) +
\hat{\gamma}^+(t)$, zeroes at the branch points of $\hat{\Sigma}$ and
satisfies $\eta(\infty^{\pm}) = 1$.

Now consider the differential $\tilde{\Omega} = \eta(P)^{-1} dx$. It
is easy to show that $dx$ has zeroes at the branch points and double
poles at $\infty^{\pm}$. Thus $\tilde{\Omega}$ is of the form
prescribed by the lemma with double poles at $\infty^{\pm}$ and zeroes
at $\hat{\gamma}(t) + \hat{\gamma}^+(t)$.

Now let $\Omega$ be any other differential with double poles at
$\infty^{\pm}$, zeroes at $\hat{\gamma}(t)$ and some other $g+1$
zeroes at $\hat{\gamma}'^+(t)$. Then $\Omega/\tilde{\Omega}$ is a
meromorphic function with divisor $\hat{\gamma}'^+(t) -
\hat{\gamma}^+(t)$ which gives the required equivalence
$\hat{\gamma}'^+(t) \sim \hat{\gamma}^+(t)$.
\end{proof}
\end{lemma}

There is however one notable difference with the procedure of
section \ref{section: normalised eigenvector} for extracting the
analytic data of the normalised eigenvector $\bm{h}$. In that
section we already exploited the gauge freedom of the
zero-curvature equations which by now is completely fixed. Indeed
in \eqref{h at infty} we had used a gauge transformation to set
the normalised eigenvectors at $x = \infty$ equal to the canonical
basis, and then we used the residual gauge symmetry to pick a
particular divisor $\hat{\gamma}(t)$ from the equivalence class
$[\hat{\gamma}(t)]$. Thus when determining the analytic properties
of the dual normalised eigenvector $\bm{h}^+$ there is no longer
any gauge freedom to exploit and we must stick to the gauge
conditions used up to this point.

However, we know from lemma \ref{lemma: orthogonal} that for
instance $\bm{h}^+(\infty^+)$ should be orthogonal to
$\bm{h}(\infty^-) = {\tiny \left(\!\!\begin{array}{c} 0 \\ 1
\end{array}\!\!\right)}$ and should be normalised by the condition
$\bm{h}^+ \cdot \bm{\alpha}^{\sf T} = 1$. From these conditions
and the corresponding conditions on $\bm{h}^+(\infty^-)$ we
conclude
\begin{equation} \label{h^+ at infty}
\bm{h}^+(\infty^+) = {\tiny \left( 1, 0 \right)}, \qquad
\bm{h}^+(\infty^-) = {\tiny \left( 0, 1 \right)}.
\end{equation}
Therefore the gauge transformation that brought the normalised
eigenvectors $\bm{h}(\infty^{\pm})$ to the canonical form \eqref{h
at infty} at the same time puts the dual normalised eigenvectors
$\bm{h}^+(\infty^{\pm})$ in the desired form \eqref{h^+ at infty}.

As in the case of the normalised eigenvector, equation \eqref{h^+
at infty} is invariant under residual gauge transformations which
applied to dual vectors looks like,
\begin{equation} \label{residual gauge symmetry^+}
\bm{h}^+ \mapsto f(P)^{-1} \bm{h}^+ \tilde{g},
\end{equation}
where $\tilde{g} = \diag(d_1, d_2)$ is diagonal and $f(P) =
(\bm{h}^+(P) \tilde{g}) \cdot \bm{\alpha}^{\sf T} = d_1 h^+_1(P) +
d_2 h^+_2(P)$. It has the effect of swapping the pole divisor
$\hat{\gamma}^+(t)$ of $\bm{h}^+$ for an equivalent divisor
$\hat{\gamma}'^+(t) \sim \hat{\gamma}^+(t)$. The difference now is
that this residual gauge invariance \eqref{residual gauge
symmetry} has already been used on the normalised eigenvector to
pick its divisor of poles $\hat{\gamma}(t)$ from the equivalence
class $[\hat{\gamma}(t)]$. Hence there is no freedom left to move
around the dual dynamical divisor $\hat{\gamma}^+(t)$ in the
equivalence class $[\hat{\gamma}^+(t)]$. Indeed changing
$\hat{\gamma}^+(t)$ is equivalent to multiplying the dual
normalised eigenvector $\bm{h}^+$ by a diagonal $\tilde{g}$ which
will affect the orthogonality condition of lemma \ref{lemma:
orthogonal}.

Nevertheless, let $\hat{\gamma}^+(t)$ be \textit{any} divisor in
the equivalence class $[\hat{\gamma}^+(t)]$. The corresponding
dual normalised eigenvector $\tilde{\bm{h}}^+$ is likely to be
expressed in the `wrong' residual gauge and needs to be
transformed by \eqref{residual gauge symmetry^+} so as to satisfy
lemma \ref{lemma: orthogonal}.

\begin{proposition} \label{prop: rows of H^-1}
The rows $\bm{H}^+(P)$ of the inverse matrix $H(x)^{-1}$ are given
by
\begin{equation} \label{rows H^-1}
H^+_1(P) = \chi(P) \tilde{h}^+_1(P), \qquad H^+_2(P) =
\frac{\chi(P)}{\chi(\infty^-)} \tilde{h}^+_2(P),
\end{equation}
where $\chi$ is the meromorphic function with zeroes at
$\hat{\gamma}(t) + \hat{\gamma}^+(t)$, poles at the branch points
and normalised by $\chi(\infty^+) = 1$.
\begin{proof}
Applying a residual gauge transformation \eqref{residual gauge
symmetry^+} to $\tilde{\bm{h}}^+$, the new normalised eigenvector
$\bm{h}^+ = f(P)^{-1} \tilde{\bm{h}}^+ \tilde{g}$ should satisfy
the orthogonality condition \eqref{orthogonality conditions}. But this
condition is equivalent to the statement of proposition \ref{left
inverse of H} that $\eta(P)^{-1} \bm{h}^+(P)$ constitutes the rows of
the \textit{left} inverse of the matrix $H(x)$. Since the left
inverse is equal to the right inverse for finite dimensional matrices
we also have
\begin{equation*}
\sum_{P \in \hat{\pi}^{-1}(x)} \eta(P)^{-1} h_i^+(P) h_j(P) =
\delta_{ij}.
\end{equation*}
Written in terms of the components of $\tilde{\bm{h}}^+$ this
condition reads
\begin{equation} \label{right orthogonality}
\sum_{P \in \hat{\pi}^{-1}(x)} \chi(P) d_i \tilde{h}_i^+(P) h_j(P) =
\delta_{ij},
\end{equation}
where $\chi(P) = \left( \eta(P) f(P) \right)^{-1}$. The parameters
$d_i$ of the residual gauge transformation $\tilde{g} = \diag(d_1,
d_2)$ can now be deduced from \eqref{right orthogonality} by taking
the $x \rightarrow \infty$ limit. In particular since
$\tilde{\bm{h}}^+$ also satisfies \eqref{h^+ at infty} one finds $d_1 =
1/\chi(\infty^+)$ and $d_2 = 1/\chi(\infty^-)$.

Now since $(\eta) = B - \hat{\gamma}(t) - \hat{\gamma}'^+(t)$ and $(f)
= \hat{\gamma}'^+(t) - \hat{\gamma}^+(t)$ where $B$ is the divisor of
branch points and $\hat{\gamma}'^+(t)$ the pole divisor of $\bm{h}^+$,
we deduce that $(\chi) = \hat{\gamma}(t) + \hat{\gamma}^+(t) -
B$. Normalising $\chi$ such that $\chi(\infty^+) = 1$ we find
\eqref{rows H^-1}.
\end{proof}
\end{proposition}

\begin{remark}
The factor of $1 / \chi(\infty^-)$ in the second component of
\eqref{rows H^-1} corresponds to the residual gauge transformation
required to turn $\tilde{\bm{h}}^+$ into the correct eigenvector
$\bm{h}^+$ satisfying \eqref{orthogonality conditions}. Thus the
upshot of proposition \ref{prop: rows of H^-1} is that we may pick any
divisor $\hat{\gamma}^+(t)$ from the equivalence class
$[\hat{\gamma}^+(t)]$ to be the dual dynamical divisor. The
corresponding dual normalised eigenvector $\bm{h}^+$ then needs to be
adjusted by a residual gauge transformation, determined by proposition
\ref{prop: rows of H^-1}, before it can provide the rows of the inverse
matrix $H(x)^{-1}$.
\end{remark}

After choosing a divisor $\hat{\gamma}^+(t)$ from the equivalence class
$[\hat{\gamma}^+(t)]$ it follows from the analogue of proposition
\ref{prop: g+1 poles} for $\tilde{\bm{h}}^+$ and equation \eqref{h^+
at infty} that the components of $\tilde{\bm{h}}^+$ satisfy the
following properties,
\begin{equation} \label{normalised ev^+ properties}
\begin{split}
(\tilde{h}^+_1) &\geq - \hat{\gamma}^+(t) + \infty^-, \quad
\tilde{h}^+_1(\infty^+) = 1, \\
(\tilde{h}^+_2) &\geq - \hat{\gamma}^+(t) + \infty^+, \quad
\tilde{h}^+_2(\infty^-) = 1.
\end{split}
\end{equation}
The remainder of the analysis of the dual eigenvector
$\tilde{\bm{h}}^+$ is now identical to that of the eigenvector
$\bm{h}$ but with $\hat{\gamma}(t)$ replaced everywhere by
$\hat{\gamma}^+(t)$. In particular, proposition \ref{prop:
uniqueness of h} says that the conditions \eqref{normalised ev^+
properties} uniquely specify $\tilde{\bm{h}}^+$ and an analogous
reconstruction formula as in proposition \ref{prop: existence of
h} can be obtained for this vector. Specifically, defining the
vectors $v_1, v_{g+1}, v^{\pm}_{\infty}, v^{\pm} \in \mathbb{C}^g$
as follows,
\begin{subequations}
\begin{align*}
v_1 &= \sum_{i = 1}^g \bm{\mathcal{A}}(\hat{\gamma}^+_i(t)) +
\bm{\mathcal{K}}, \quad
v_{g+1} = \sum_{i = 2}^{g+1} \bm{\mathcal{A}}(\hat{\gamma}^+_i(t)) +
\bm{\mathcal{K}},\\
v^{\pm}_{\infty} &= \bm{\mathcal{A}}(\infty^{\pm}) + \sum_{i = 2}^g
\bm{\mathcal{A}}(\hat{\gamma}^+_i(t)) + \bm{\mathcal{K}}, \quad
v^{\pm} = v_1 + v_{g+1} - v^{\pm}_{\infty},
\end{align*}
\end{subequations}
\begin{proposition} \label{prop: existence of h^+}
The components $\tilde{h}^+_1, \tilde{h}^+_2$ of the dual normalised
eigenvector $\tilde{\bm{h}}^+$ are given by $\tilde{h}^+_1(P) =
k_-(P)$ and $\tilde{h}^+_2(P) = k_+(P)$ where
\begin{equation*}
k_{\pm}(P) = \frac{\theta\left( \bm{\mathcal{A}}(\infty^{\mp}) - v_1
\right) \theta\left( \bm{\mathcal{A}}(\infty^{\mp}) - v_{g+1}
\right)}{\theta\left( \bm{\mathcal{A}}(\infty^{\mp}) -
v^{\pm}_{\infty} \right) \theta\left( \bm{\mathcal{A}}(\infty^{\mp}) -
v^{\pm} \right)} \cdot \frac{\theta\left( \bm{\mathcal{A}}(P) -
v^{\pm}_{\infty} \right) \theta\left( \bm{\mathcal{A}}(P) - v^{\pm}
\right)}{\theta\left( \bm{\mathcal{A}}(P) - v_1 \right) \theta\left(
\bm{\mathcal{A}}(P) - v_{g+1} \right)}.
\end{equation*}
\end{proposition}

\subsection*{The dual Baker-Akhiezer vector}

We now wish to obtain a formula for the inverse of the matrix
$\Psi(x)$ constructed from the Baker-Akhiezer vector
$\bm{\psi}(P,t)$. Since $\bm{\psi}$ satisfies the linear system
\eqref{simultaneous} it follows that $\Psi(x)$ satisfies the matrix
analogue $\big(\partial_{t_M} - J_M(x)\big) \Psi(x) = 0, \forall M$.
The inverse matrix then solves the \dub{dual linear system}
\begin{equation} \label{dual simultaneous matrix}
\partial_{t_M} \Psi(x)^{-1} + \Psi(x)^{-1} J_M(x) = 0, \quad \forall
M.
\end{equation}
Thus if $\Psi(x)^{-1}$ is to be built out of row vectors
$\bm{\Psi}^+(P)$ these should satisfy the analogue of this equation
for row vectors, namely
\begin{equation} \label{dual simultaneous}
\partial_{t_M} \bm{\Psi}^+(P) + \bm{\Psi}^+(P) J_M(x) = 0, \quad
\forall M.
\end{equation}

\begin{proposition} \label{prop: inverse of Psi}
Let $\bm{\Psi}^+(P,t)$ be the row vector solution to \eqref{dual
simultaneous} with initial condition $\bm{\Psi}^+(P,0) =
\bm{H}^+(P,0)$ then its components can be written as
\begin{equation} \label{rows Psi^-1}
\Psi_1^+(P) = \chi_0(P) \widetilde{\psi}_1^+(P), \qquad \Psi_2^+(P) =
\frac{\chi_0(P)}{\chi_0(\infty^-)} \widetilde{\psi}_2^+(P),
\end{equation}
where $\chi_0$ is the function $\chi$ taken at $t = 0$. Moreover, the
functions $\widetilde{\psi}_i^+$ are meromorphic on $\hat{\Sigma}
\setminus \{ (\pm 1)^{\pm} \}$ with
\begin{subequations} \label{dual Baker-Akhiezer vec def}
\begin{equation} \label{dual Baker-Akhiezer vec def 1}
\begin{split}
&(\widetilde{\psi}^+_1) \geq - \hat{\gamma}^+(0) + \infty^-, \quad
\widetilde{\psi}^+_1(\infty^+) = 1, \\
&(\widetilde{\psi}^+_2) \geq - \hat{\gamma}^+(0) + \infty^+, \quad
\widetilde{\psi}^+_2(\infty^-) = 1,
\end{split}
\end{equation}
and have the following asymptotic behaviour in a neighbourhood of
$(\pm 1)^{\pm} \in \hat{\Sigma}$,
\begin{equation} \label{dual Baker-Akhiezer vec def 2}
\left\{
\begin{split}
&\widetilde{\psi}^+_i(x^{\pm},t) e^{\pm \sum_n s_{n,+}(x) t_{n,+}} =
O(1), \quad \text{as }\; x \rightarrow +1,\\
&\widetilde{\psi}^+_i(x^{\pm},t) e^{\pm \sum_n s_{n,-}(x) t_{n,-}} =
O(1), \quad \text{as }\; x \rightarrow -1.
\end{split}
\right.
\end{equation}
\end{subequations}
\begin{proof}
Let $\widehat{\Psi}(x)$ be the formal matrix solution of the linear
system \eqref{simultaneous matrix} with initial condition
$\widehat{\Psi}(x,0) = {\bf 1}$. It follows that
$\widehat{\Psi}(x)^{-1}$ is a formal matrix solution to \eqref{dual
simultaneous matrix} with the same initial condition. We may then
write the solution to \eqref{dual simultaneous} with the initial
condition $\bm{\Psi}^+(P,0) = \bm{H}^+(P,0)$ as $\bm{\Psi}^+(P,t) =
\bm{H}^+(P,0) \widehat{\Psi}(x,t)^{-1}$. Taking \eqref{rows Psi^-1} as
defining the functions $\widetilde{\psi}^+_i$ and using \eqref{rows
H^-1} this can be rewritten
\begin{equation} \label{psi^+ vs h^+}
\left( \widetilde{\psi}_1^+(P,t), \;
\frac{1}{\chi_0(\infty^-)} \widetilde{\psi}_2^+(P,t) \right) = \left(
\tilde{h}_1^+(P,0), \; \frac{1}{\chi_0(\infty^-)} \tilde{h}_2^+(P,0)
\right) \widehat{\Psi}(x,t)^{-1}.
\end{equation}

Now since $\widehat{\Psi}(x,t)$ is holomorphic outside $x = \pm 1$
it follows from \eqref{psi^+ vs h^+} that
$\widetilde{\bm{\psi}}^+(P,t)$ is meromorphic outside
$\hat{\pi}^{-1}(\pm 1)$ with the same pole divisor as
$\tilde{\bm{h}}^+(P,0)$, namely $\hat{\gamma}^+(0)$. Moreover, since
$\widehat{\Psi}(\infty, t) = {\bf 1}$ we have
$\widetilde{\psi}_i^+(\infty^{\pm},t) = \tilde{h}_i^+(\infty^{\pm},0)$
and equations \eqref{dual Baker-Akhiezer vec def 1} readily follow
from \eqref{normalised ev^+ properties} by setting $t = 0$.

Recall the matrix equation $\Psi(x) = H(x) \Phi(x)$ which was used in
the proof of lemma \ref{lemma: BA vector properties} where $\Phi(x) =
\diag(\varphi(x^+), \varphi(x^-))$. We are now interested in the
inverse matrices, namely $\Psi(x)^{-1} = \Phi(x)^{-1} H(x)^{-1}$. But this
immediately shows that the singular parts of
$\widetilde{\bm{\psi}}^+$, encoded in $\Phi(x)^{-1}$, are opposite to
those of $\bm{\psi}$, which were encoded in $\Phi(x)$, and \eqref{dual
Baker-Akhiezer vec def 2} follows.
\end{proof}
\end{proposition}

The conditions \eqref{dual Baker-Akhiezer vec def} are those of a
Baker-Akhiezer vector (with respect to different data) and just as in
proposition \ref{uniqueness of psi} they uniquely specify the vector
$\widetilde{\bm{\psi}}^+$. This vector will be called the \dub{dual
Baker-Akhiezer vector}. One can also write down explicit formulae
in terms of Riemann $\theta$-functions which satisfy \eqref{dual
Baker-Akhiezer vec def}, giving rise to reconstruction formulae for
the components of $\widetilde{\bm{\psi}}^+$. Specifically, defining
the divisors $\delta_{\pm}(t)$ by the following equivalence
\begin{equation} \label{divisor of zeroes dual}
\hat{\gamma}^+(t) \sim \delta_{\pm}(t) + \infty^{\pm},
\end{equation}
then the analogue of proposition \ref{prop: existence of psi} is
obtained simply by making the replacements $h_{\pm} \rightarrow
k_{\pm}$, $\gamma_{\pm} \rightarrow \delta_{\pm}$ and $d \mathcal{Q}
\rightarrow - d \mathcal{Q}$. The result is the following,
\begin{proposition} \label{prop: existence of psi^+}
The components $\widetilde{\psi}^+_1, \widetilde{\psi}^+_2$ of the dual
Baker-Akhiezer vector $\widetilde{\bm{\psi}}^+$ are given by
$\widetilde{\psi}^+_1(P) = \phi_+(P)$ and $\widetilde{\psi}^+_2(P) =
\phi_-(P)$ where
\begin{equation*}
\phi_{\pm}(P) = k_{\mp}(P,0) \frac{\theta \left( \bm{\mathcal{A}}(P) -
\int_{\bm{b}} d \mathcal{Q} - \bm{\zeta}_{\delta_{\mp}(0)} \right) \theta
\left( \bm{\mathcal{A}}(\infty^{\pm}) - \bm{\zeta}_{\delta_{\mp}(0)}
\right)}{\theta \left( \bm{\mathcal{A}}(P) - \bm{\zeta}_{\delta_{\mp}(0)}
\right) \theta \left( \bm{\mathcal{A}}(\infty^{\pm}) - \int_{\bm{b}} d
\mathcal{Q} - \bm{\zeta}_{\delta_{\mp}(0)} \right)} \; \exp \left( - i
\int_{\infty^{\pm}}^P d\mathcal{Q} \right).
\end{equation*}
\end{proposition}


\section{Reconstruction formulae} \label{section: reconstruction}


\subsection*{The $SL(2, \mathbb{C})_R$ current $j$}

The Lax connection $J(x)$ can be reconstructed from the formula
\begin{equation} \label{reconstructed Lax connection}
J(x) = d \Psi(x) \Psi(x)^{-1},
\end{equation}
where $\Psi(x) = \left( \bm{\psi}(x^+), \bm{\psi}(x^-) \right)$ is the
matrix of Baker-Akhiezer column vectors $\bm{\psi}$ above
$x$. However, in order to obtain expressions for the components $j_0,
j_1$ of the current $j$ we must first show that the reconstructed
Lax connection \eqref{reconstructed Lax connection} takes the original
form \eqref{Lax connection} for some current $j$. This is the content
of theorem \ref{thm: reconstructing j} below. The crux of the proof is a
standard argument based on the uniqueness of the Baker-Akhiezer vector
(see for instance \cite[pp.93--94]{Belokolos}).

\begin{remark}
Note that even though the definition of $\Psi(x)$ depends on the
order of the rows (so $\Psi(x)$ isn't a properly defined function
of $x$), the definitions \eqref{reconstructed Lax connection} of
$J(x)$ in terms of this matrix do not depend of the ordering of
its columns and therefore the connection $J(x)$ obtained this way
is a well defined function of the spectral parameter $x$.
\end{remark}

\begin{remark}
It was noted that a solution $\bm{\psi}(P)$ to the auxiliary linear
system \eqref{simultaneous} is determined locally only up to a power
of the eigenvalue $\Lambda(P)$ of $\Omega(x)$ so that $\Psi(x)$ is
also determined locally only up to right multiplication by a diagonal
matrix $\diag\,(\Lambda(x^+)^{n_+}, \Lambda(x^-)^{n_-})$. But this
constant right diagonal matrix cancels out in the definitions
\eqref{reconstructed Lax connection} of the Lax connection in terms of
$\Psi(x)$ so that $J(x)$ is well defined globally on the base space.
\end{remark}

\begin{theorem} \label{thm: reconstructing j}
Given the Baker-Akhiezer vector $\bm{\psi}$, the light-cone components
$j_{\pm}$ of the $SL(2,\mathbb{C})_R$ current can be recovered by the
formula
\begin{equation} \label{reconstruction formula for j components}
j_{\pm} = i \kappa_{\pm} \lim_{x \rightarrow \pm 1} \left(\Psi(x)
\sigma_3 \Psi(x)^{-1} \right).
\end{equation}
\begin{proof}
Equation \eqref{Baker-Akhiezer vec def 2} together with \eqref{singular
parts J} gave the behaviour of the eigenvector $\bm{\psi}$ near the
essential singularities at $x = \pm 1$. Focusing on the zeroth level
$n = 0$ of the hierarchy, namely the $(\sigma, \tau)$-dependence, we
have
\begin{equation*}
\bm{\psi}(x^{\pm}) \underset{x \rightarrow 1}\sim O(1) e^{\pm
\frac{i \kappa_+ \sigma^+}{1 - x}}, \quad \bm{\psi}(x^{\pm})
\underset{x \rightarrow -1}\sim O(1) e^{\pm \frac{i \kappa_-
\sigma^-}{1 + x}}.
\end{equation*}
We may rewrite this behaviour in terms of the matrix $\Psi(x) =
(\bm{\psi}(x^+), \bm{\psi}(x^-))$ near $x = \pm 1$ as follows
\begin{equation} \label{Psi essential singularity}
\begin{split}
\Psi(x,\sigma,\tau) &= \left( \Psi_0(\sigma,\tau) + \sum_{s =
1}^{\infty} \Psi_s(\sigma,\tau) (x - 1)^s \right) e^{\frac{i
\kappa_+ \sigma^+}{1 - x} \sigma_3} \quad \text{as } x
\rightarrow 1, \\
\Psi(x,\sigma,\tau) &= \left( \Phi_0(\sigma,\tau) + \sum_{s =
1}^{\infty} \Phi_s(\sigma,\tau) (x + 1)^s \right) e^{\frac{i
\kappa_- \sigma^-}{1 + x} \sigma_3}  \quad \text{as } x \rightarrow
-1.
\end{split}
\end{equation}
It is straightforward to derive from these expansions the asymptotics
near $x = +1$
\begin{equation*}
\left\{
\begin{split}
\left( \partial_+ \Psi \right) \Psi^{-1} &= \frac{i \kappa_+}{1-x}
\left( \Psi_0 \sigma_3 \Psi_0^{-1} \right) + O(1) \\
\left( \partial_- \Psi \right) \Psi^{-1} &= O(1)
\end{split}
\right. \quad \text{as } x \rightarrow 1
\end{equation*}
and likewise near $x = -1$,
\begin{equation*}
\left\{
\begin{split}
\left( \partial_+ \Psi \right) \Psi^{-1} &= O(1) \\
\left( \partial_- \Psi \right) \Psi^{-1} &= \frac{i \kappa_-}{1 + x}
\left( \Phi_0 \sigma_3 \Phi_0^{-1} \right) + O(1)
\end{split}
\right. \quad \text{as } x \rightarrow -1.
\end{equation*}
However we also find from \eqref{Baker-Akhiezer vec def 1}
that $\Psi(x) = {\bf 1} + O\left( \frac{1}{x} \right)$ as $x
\rightarrow \infty$ so that
\begin{equation*}
\left( \partial_{\pm} \Psi \right) \Psi^{-1} = O\left( \frac{1}{x}
\right) \quad \text{as } x \rightarrow \infty.
\end{equation*}
Thus the above asymptotics at $x = \pm 1, \infty$ take the following
form
\begin{subequations} \label{f_pm essential singularities}
\begin{gather}
\left( \partial_+ \Psi \right) \Psi^{-1} = J_+(x) + O(1), \quad
\left( \partial_- \Psi \right) \Psi^{-1} = J_-(1)
+ O(1) \quad \text{as } x \rightarrow 1 \label{f_pm at p1} \\
\left( \partial_+ \Psi \right) \Psi^{-1} = J_+(-1) + O(1), \quad
\left( \partial_- \Psi \right) \Psi^{-1} = J_-(x) + O(1) \quad
\text{as } x
\rightarrow -1 \label{f_pm at m1} \\
\left( \partial_{\pm} \Psi \right) \Psi^{-1} = J_{\pm}(\infty) +
O\left( \frac{1}{x} \right) \quad \text{as } x \rightarrow \infty,
\label{f_pm at inf}
\end{gather}
\end{subequations}
where the matrices $J_{\pm}(x)$ here have been defined as
\begin{equation} \label{reconstruction formula for j}
J_+(x) = \frac{i \kappa_+}{1 - x} \left( \Psi_0 \sigma_3 \Psi_0^{-1}
\right), \qquad J_-(x) = \frac{i \kappa_-}{1 + x} \left( \Phi_0
\sigma_3 \Phi_0^{-1} \right).
\end{equation}
To show that these are in fact the light-cone components of the Lax
connection consider the following vector-valued functions
\begin{subequations} \label{vector f}
\begin{align}
\label{vector f 1} \bm{f}_{\pm}(P) &= \left( \partial_{\pm} -
J_{\pm}(x) \right) \bm{\psi}(P)\\
\label{vector f 2} &= \left[ \left(\partial_{\pm} \Psi(x) \right)
\Psi(x)^{-1} - J_{\pm}(x) \right] \bm{\psi}(P),
\end{align}
\end{subequations}
where $\hat{\pi}(P) = x$. From \eqref{vector f 1} we see that on
$\hat{\Sigma} \setminus \hat{\pi}^{-1}(\pm 1)$ the components of the
vectors $\bm{f}_{\pm}(P)$ have exactly the same \textit{constant}
poles as $\bm{\psi}(P)$ at $\hat{\gamma}(0)$ as well as the same
\textit{constant} zeroes as the components of $\bm{\psi}(P)$ at
$\infty^{\pm}$ (see \eqref{Baker-Akhiezer vec def 1}) using the same
gauge fixing condition $J_{\pm}(\infty) = 0$ as usual. Also from
\eqref{vector f 2} and using the asymptotics at $x = \pm 1$ in
\eqref{f_pm at p1} and \eqref{f_pm at m1}, these vectors have
essential singularities at $x = \pm 1$ of exactly the same form as
those of the vector $\bm{\psi}$. Unlike the Baker-Akhiezer vector
$\bm{\psi}(P)$ however, the vector $\bm{f}_{\pm}(P)$ may take on
arbitrary $\{ t \}$-dependent values at $\infty^{\pm}$. Thus by
the uniqueness of the Baker-Akhiezer vector we must have
$\bm{f}_{\pm}(P) = D(t) \bm{\psi}(P)$ where $D(t) =
\diag(f_{\pm}(\infty^+), f_{\pm}(\infty^-))$ is an undetermined
diagonal matrix independent of $P \in \hat{\Sigma}$. But the
asymptotics at $x = \infty$ in \eqref{f_pm at inf} together with
\eqref{vector f 2} now show that in fact $D(t)$ must be zero, so we
conclude
\begin{equation*}
\bm{f}_{\pm}(P) \equiv 0.
\end{equation*}
Going back to the definition \eqref{vector f} of these vectors this
implies that $J_{\pm}(x)$ defined in \eqref{reconstruction formula for
j} is exactly the reconstructed Lax connection \eqref{reconstructed
Lax connection}, and hence the latter is indeed of the form \eqref{Lax
connection}.
\end{proof}
\end{theorem}

One can easily check that the reconstructed currents
\eqref{reconstruction formula for j components} satisfy the first set
of Virasoro constraints \eqref{Vir without P=0} since $j_{\pm}^2 = -
\kappa_{\pm}^2 {\bf 1}$ so that
\begin{equation*}
\tr j_{\pm}^2 = - \kappa_{\pm}^2 \tr {\bf 1} = - 2 \kappa_{\pm}^2.
\end{equation*}
Also, before having imposed any reality conditions on the
algebro-geometric data the reconstructed current
\eqref{reconstruction formula for j components} takes values in
$\mathfrak{sl}(2, \mathbb{C})$ since it is obviously invertible
and traceless,
\begin{equation*}
\tr j_{\pm} = i \kappa_{\pm} \tr \sigma_3 = 0.
\end{equation*}

\subsection*{The $SL(2, \mathbb{C})$ embedding $g$}

Having shown that the Lax connection \eqref{reconstructed Lax
connection} reconstructed out of Baker-Akhiezer vectors takes
precisely the form of a Lax connection constructed from a current $j$
we were able to express the current $j$ itself in terms of
Baker-Akhiezer functions. Now since the current $j$ is really of the
form $j = -g^{-1} dg$ for some $g$, we would like to extract now a
formula for the matrix $g$ in terms of Baker-Akhiezer functions. For
this we can go back to equation \eqref{reconstructed Lax connection}
for $J(x)$ and rewrite it as
\begin{equation*}
d \Psi(x)^{-1} + \Psi(x)^{-1} J(x) = 0.
\end{equation*}
And since we know $J(x)$ is of the form $J(x) = \frac{1}{1 - x^2}(j -
x \ast j)$ we have $j = J(0)$ and setting $x = 0$ in the above
equation we find
\begin{equation*}
d \Psi(0)^{-1} + \Psi(0)^{-1} j = 0.
\end{equation*}
This is to be compared with the defining equation $dg + g j = 0$ for
the matrix $g$. We see immediately from this comparison that the
matrix $g$ can be reconstructed in terms of $\Psi(0)^{-1}$ whose rows
we showed were dual Baker-Akhiezer vectors. Because $\det g = 1$ we
would need to divide $\Psi(0)^{-1}$ by the square root of its
determinant, but $d \det \Psi(0)^{-1} = \det \Psi(0)^{-1} \tr (\Psi(0)
d \Psi(0)^{-1}) = - \det \Psi(0)^{-1} \tr j = 0$ so this is
possible. However since we haven't yet imposed reality conditions, at
this stage we can only require that $g \in SL(2, \mathbb{C})$. In
particular $g$ could be of the general form $g^{-1}_L \Psi(0)^{-1}
g^{-1}_R$ where $g_R, g_L \in SL(2,\mathbb{C})$ are constant diagonal
matrices. Such issues will only be resolved later in chapter
\ref{chapter: reality} when we come to discuss reality conditions. We
postpone the complete reconstruction of $g$ until then. At this point
we have,
\begin{proposition} \label{thm: reconstructing g}
Given the dual Baker-Akhiezer vector $\widetilde{\bm{\psi}}^+$, the
matrix $g \in SL(2,\mathbb{C})_R$ can be recovered by the formula
\begin{equation*}
g(t) = \sqrt{\det \Psi(0,t)} \cdot g^{-1}_L
\Psi(0,t)^{-1} g^{-1}_R,
\end{equation*}
where $g_R, g_L \in SL(2,\mathbb{C})$ are constant diagonal matrices.
\end{proposition}

\begin{remark}
Recall that $\Psi(x)$ is determined locally only up to right
multiplication by a diagonal matrix $\diag\,(\Lambda(x^+)^{n_+},
\Lambda(x^-)^{n_-})$. However since $\Lambda(0^{\pm}) = 1$ and the
reconstruction formula for $g(t)$ only depends on $\Psi(0)$ and it
follows that this ambiguity is absent in $g(t)$.
\end{remark}

%% file: Symplectic.tex
\newpage

\chapter{Symplectic structure} \label{chapter: symplectic}

The subject of the previous chapter was the reconstruction of the
general finite-gap solution from the following piece of
\dub{algebro-geometric data}:
\begin{itemize}
  \item[$\bullet$] A smooth algebraic curve $\hat{\Sigma}$ of genus
$g$ equipped with a differential $dp$.
  \item[$\bullet$] A generic set of $g+1$ points $\hat{\gamma}(0)$ on
this curve.
\end{itemize}
At fixed genus $g$, different finite-gap solutions are obtained by
varying the moduli of the pair $(\hat{\Sigma}, dp)$ and choosing
different initial divisors $\hat{\gamma}(0)$ on this curve. As we
saw in chapter \ref{chapter: curves} the correct interpretation of
the moduli space of curves is as a $g+1$ dimensional leaf
$\mathcal{L}$ in the universal configuration space. Furthermore,
since a non-special divisor of degree $g+1$ uniquely determines a
point in the generalised Jacobian $J_{\mathfrak{m}}(\hat{\Sigma})$
via the generalised Abel map, a more natural description for the
initial divisor $\hat{\gamma}(0)$ is as the point
$\vec{\mathcal{A}}(\hat{\gamma}(0))$ in the generalised Jacobian
$J_{\mathfrak{m}}(\hat{\Sigma})$. Then by theorem \ref{thm: linear
motion} the locus of the dynamical divisor $\hat{\gamma}(t)$ in
$J_{\mathfrak{m}}(\hat{\Sigma})$ is a straight line through this
point. The above algebro-geometric data at genus $g$ therefore
corresponds to a point in the \dub{Jacobian bundle} over
$\mathcal{L}$ whose fibre over any point $\hat{\Sigma}$ of the
base $\mathcal{L}$ is the generalised Jacobian
$J_{\mathfrak{m}}(\hat{\Sigma})$.

\begin{figure}[ht]
\centering
\begin{tabular}{cc}
\psfrag{J}{\small $\qquad \mathbb{T}^{g+1}$} \psfrag{L}{\small
$\qquad \mathcal{L}_{\mathbb{R}}$}
\includegraphics[height=2cm]{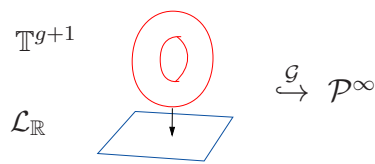} &
\raisebox{.8cm}{$\; \overset{\mathcal{G}}\hookrightarrow \;
\mathcal{P}^{\infty}$}
\end{tabular}
\caption{The algebro-geometric data as a $(2g + 2)$-dimensional
phase-space.} \label{finite-dim phase-space}
\end{figure}

This suggests an alternative way of picturing finite-gap solutions
that will be useful later. Once we will have imposed reality
conditions in chapter \ref{chapter: reality} the real slice of the
generalised Jacobian will turn out to be a $(g+1)$-dimensional real
torus and the base $\mathcal{L}_{\mathbb{R}}$ will become
$(g+1)$-dimensional over the reals. Therefore the dynamics of a
finite-gap solution will correspond to linear motion on a
$(g+1)$-torus which is very reminiscent of a finite-dimensional
integrable system. In fact one can view the Jacobian bundle as the
phase-space of a $(g+1)$-dimensional dynamical system. But if the
algebro-geometric data is to be thought of as a finite-dimensional
phase-space it must be equipped with a natural symplectic
structure. Now the finite-gap construction provides a (reconstruction)
map $\mathcal{G}$ from the Jacobian bundle to the reduced phase-space
$\mathcal{P}^{\infty}$ which was introduced in chapter \ref{chapter:
hamiltonian} as the space of solutions to the equations of motion
satisfying the Virasoro and static gauge constraints \eqref{Vir
without P=0}, see Figure \ref{finite-dim phase-space}.
But the space $\mathcal{P}^{\infty}$ is equipped with a Dirac bracket
from chapter \ref{chapter: hamiltonian} which can thus be pulled back
to the algebro-geometric data using the map $\mathcal{G}$. This way we
obtain a `natural' symplectic structure on the Jacobian bundle.

In practise the pullback will be achieved as follows. Recall from
chapter \ref{chapter: integrability} that the integrals of motion
which parameterise the base $\mathcal{L}$ can be obtained from the
trace of the monodromy matrix $\Omega(x)$. On the other hand, using a
trick due to Sklyanin (see \cite{Sklyanin:1995bm} for a review) we
will show how to also extract the initial divisor $\hat{\gamma}(0)$ from
$\Omega(x)$. Therefore the Dirac bracket $\{ \Omega(x), \Omega(x')
\}_{\text{D.B.}}$ between monodromies appropriately regularised \`a la
Maillet (see chapter \ref{chapter: integrability}) can be used to read
off the Dirac brackets of the algebro-geometric data, much like we
obtained the Dirac brackets between integrals of motion already in
chapter \ref{chapter: integrability}. As we will see below, our
analysis for strings moving on ${\mathbb{R}}\times S^{3}$ can be
thought of as a non-linear generalisation of the more familiar
Hamiltonian analysis of strings in flat space. We will therefore begin
by reviewing the standard discussion of the flat space case following
\cite{GSWI, Henneaux2}.


\section{Strings in flat space} \label{section: flat space}


Consider a closed bosonic string moving on $(D+1)$-dimensional
Minkowski space with worldsheet fields $X^{\mu}(\sigma,\tau)$,
$\mu=0,1,\ldots,D$. In conformal gauge, the worldsheet equation of
motion is the two-dimensional Laplace equation $\partial_+ \partial_-
X^{\mu} = 0$. As the equation is linear, the general
solution for closed string boundary conditions is given by the
Fourier series,
\begin{equation} \label{General solution: flat}
X^{\mu}(\sigma,\tau) = x^{\mu} + p^{\mu} \tau + i \sum_{n \neq 0}
\frac{1}{n} \alpha_n^{\mu} e^{-i n(\tau - \sigma)} + i \sum_{n
\neq 0} \frac{1}{n} \tilde{\alpha}_n^{\mu} e^{-i n(\tau +
\sigma)}.
\end{equation}
where the Fourier coefficients $\alpha^{\mu}_n$ and
$\tilde{\alpha}^{\mu}_n$  correspond to classical oscillator
coordinates for left- and right-moving modes respectively.
For the purpose of drawing the analogy with finite-gap solutions it
will be convenient to restrict attention to classical solutions with a
finite number of oscillators turned on. Indeed we will see that these
`finite-oscillator' solutions are close analogs of the finite-gap
solutions to string theory on $\mathbb{R}\times S^3$ and more
generally on classically integrable backgrounds. Generic solutions can
then be obtained as a limiting case.

Since \eqref{General solution: flat} is the general solution to
the field equations, the fields $X^{\mu}(\sigma) = X^{\mu}(\sigma,0)$
and $P^{\mu}(\sigma) = \dot{X}^{\mu}(\sigma,0)$ restricted to a
$\tau$-slice (taken at $\tau = 0$ without loss of generality) give
a convenient parametrisation of the phase-space of the
string (since we have not yet fixed the gauge this is not the physical
phase-space). Written in terms of the oscillator coordinates we find,
\begin{equation} \label{generic PS configuration}
\begin{split}
X^{\mu}(\sigma) &= x^{\mu} + i \sum_{n \neq 0} \frac{1}{n}
\alpha_n^{\mu} e^{i n\sigma} + i \sum_{n \neq 0}
\frac{1}{n} \tilde{\alpha}_n^{\mu} e^{-i n\sigma},\\
P^{\mu}(\sigma) &= p^{\mu} + \sum_{n \neq 0} \alpha_n^{\mu} e^{i
n\sigma} + \sum_{n \neq 0} \tilde{\alpha}_n^{\mu} e^{-i n\sigma}.
\end{split}
\end{equation}
Conversely the oscillator coefficients $\alpha_n^{\mu},
\tilde{\alpha}_n^{\mu}$ as well as the centre of mass position and
momenta $x^{\mu}, p^{\mu}$ can be extracted from a generic phase-space
configuration $X^{\mu}(\sigma), P^{\mu}(\sigma)$ by the following
relations
\begin{equation} \label{extract data: flat}
\left\{
\begin{split}
&\alpha_m^{\mu} = \frac{1}{2 \pi} \int_0^{2 \pi} e^{- i m \sigma}
\frac{1}{2} \left( P^{\mu}(\sigma) - \partial_{\sigma}
X^{\mu}(\sigma)
\right) d\sigma, \quad m \neq 0 \\
&\tilde{\alpha}_m^{\mu} = \frac{1}{2 \pi} \int_0^{2 \pi} e^{i m
\sigma} \frac{1}{2} \left( P^{\mu}(\sigma) + \partial_{\sigma}
X^{\mu}(\sigma) \right) d\sigma, \quad m \neq 0 \\
&x^{\mu} = \frac{1}{2 \pi} \int_0^{2 \pi} X^{\mu}(\sigma) d\sigma,
\qquad p^{\mu} = \frac{1}{2 \pi} \int_0^{2 \pi} P^{\mu}(\sigma)
d\sigma.
\end{split}
\right.
\end{equation}
Equations \eqref{extract data: flat} are the inverse of the equations
\eqref{generic PS configuration} and the transformation
\begin{equation} \label{flat space data}
\left\{ X^{\mu}(\sigma), P^{\mu}(\sigma) \right\} \mapsto
\left\{ x^{\mu}, p^{\mu}, \alpha_n^{\mu}, \tilde{\alpha}_n^{\mu}
\right\}
\end{equation}
is simply a change of variable on phase-space. The Poisson brackets
which follow from the string action take the form,
\begin{equation} \label{canonical PB: flat}
\{ X^{\mu}(\sigma), X^{\nu}(\sigma') \} = \{
P^{\mu}(\sigma), P^{\nu}(\sigma') \} = 0, \quad
\{ P^{\mu}(\sigma), X^{\nu}(\sigma') \} = \eta^{\mu
\nu} \delta(\sigma - \sigma'),
\end{equation}
and it is straightforward to rewrite these brackets in the new
coordinate system as,
\begin{equation} \label{data PB: flat}
\begin{split}
\{ \alpha^{\mu}_m, \alpha^{\nu}_n \} &=
i m \delta_{m+n} \eta^{\mu \nu}, \{ \alpha^{\mu}_m,
\tilde{\alpha}^{\nu}_n \} = 0,\\
\{ \tilde{\alpha}^{\mu}_m, \tilde{\alpha}^{\nu}_n
\} &= i m \delta_{m+n} \eta^{\mu \nu}, \{ p^{\mu}, x^{\nu} \} =
\eta^{\mu \nu}.
\end{split}
\end{equation}

So far we have discussed the full solution space of the equations
of motion. The next step is to restrict to physical configurations
of the string by fixing the residual gauge symmetry and imposing
the Virasoro constraints. The former is achieved by defining
light-cone coordinate $X^{\pm}=X^0 \pm X^D$ and imposing the
light-cone gauge condition $X^+ + \tau p^+ = 0$, $P^+ = \frac{p^+}{2
\pi}$ which fixes all the oscillator modes $\alpha_n^+,
\tilde{\alpha}_n^+$ of $X^+$ to zero and $x^+ = - \tau p^+$. With this
choice, it is possible to solve the Virasoro constraint $(P^- \pm
X'^-) = (P^i \pm X'^i)^2 / 2 p^+$ explicitly to eliminate $p^-$ and
all the oscillator modes of $X^-$ as well. The remaining independent
degrees of freedom are
\begin{equation} \label{reduced phase space: flat}
\{x^i, p^i, x^-, p^+, \alpha^i_n, \tilde{\alpha}^i_n \}
\end{equation}
where the index $i = 1, 2, \ldots, D-1$ runs over the transverse
spacetime dimensions. To find the Poisson brackets of the physical
degrees of freedom one must follow the standard Dirac procedure
for constrained systems. In the present case this is described in
detail in \cite{Henneaux2}. The Virasoro constraint and the light
cone gauge fixing condition together correspond to a system of
second class constraints on phase-space. Fortunately, the
resulting Dirac bracket for the transverse degrees of freedom is
the same as their naive Poisson bracket, namely
\begin{equation} \label{Dirac brackets: flat case}
\begin{split}
\{ \alpha^i_m, \alpha^j_n \}_{\text{D.B.}} &= i m \delta_{m+n}
\delta^{ij}, \{ \alpha^i_m,
\tilde{\alpha}^j_n \}_{\text{D.B.}} = 0,\\
\{ \tilde{\alpha}^i_m, \tilde{\alpha}^j_n \}_{\text{D.B.}} &=
i m \delta_{m+n} \delta^{i j}, \{ p^i, x^j \}_{\text{D.B.}} =
\delta^{i j}.
\end{split}
\end{equation}
These brackets are the starting point for canonical quantisation
of the string which proceeds by the usual recipe of promoting
brackets to commutators.

Classical string theory in flat space is trivially integrable as
the corresponding equations of motion are linear. For comparison
with the non-linear case, it will be convenient to exhibit
integrability explicitly by constructing the corresponding
action-angle variables. While keeping the centre of mass variables $\{
x^j, p^j \}$ we introduce a new set of variables $\{\theta ^j_n,
S^j_n, \tilde{\theta}^j_n, \tilde{S}^j_n \}_{j = 1}^{D-1}$ for the
oscillator degrees of freedom by setting
\begin{equation} \label{action-angle def: flat}
\alpha_n^j = \sqrt{n S_n^j} e^{i \theta_n^j}, \quad
\tilde{\alpha}_n^j = \sqrt{n \tilde{S}_n^j} e^{i
\tilde{\theta}_n^j}.
\end{equation}
The variables $S^j_n$ and $\tilde{S}^j_n$ correspond to the
classical analogs of the occupation numbers for the transverse
oscillators. They can be shown to commute with the light-cone
Hamiltonian governing the dynamics on reduced phase-space and
therefore correspond to conserved charges. One may also check the
involution condition
\begin{equation} \label{involution: flat}
\{S^i_n, S^j_m\}_{\text{D.B.}} = \{S^i_n,
\tilde{S}^j_m\}_{\text{D.B.}} = \{\tilde{S}^i_n,
\tilde{S}^j_m\}_{\text{D.B.}} = 0.
\end{equation}
Together with the momentum variables $p^j$ these are the action
variables of the flat space string. Furthermore, the angular variables
$\theta^j_n$ and $\tilde{\theta}^j_n$ each have period $2\pi$ and are
canonically conjugate to the corresponding action variables
$S^j_n$ since their non-vanishing Dirac brackets are
\begin{equation} \label{action-angle bracket: flat}
\{S^i_n, \theta^j_m\}_{\text{D.B.}} = \{\tilde{S}^i_n,
\tilde{\theta}^j_m\}_{\text{D.B.}} = \delta_{nm}\delta^{ij}.
\end{equation}
Likewise from \eqref{Dirac brackets: flat case} the variables
$x^j$ are canonically conjugate to the $p^j$. It follows
immediately from Hamilton's equations that the angle variables $\{
x^j, \theta_n^j, \tilde{\theta}_n^j \}$ evolve linearly in time while,
as above, the conjugate action variables $\{ p^j, S_n^j, \tilde{S}_n^j
\}$ remain constant, thus
\begin{equation} \label{linear motion: flat}
\begin{split}
x^j(\tau) &= x^j(0) + p^j \tau, \quad p^j = \text{const.}\\
\theta_n^j(\tau) &= \theta_n^j(0) - n \tau, \quad S_n^j =
\text{const.}\\
\tilde{\theta}_n^j(\tau) &= \tilde{\theta}_n^j(0) - n \tau,
\quad \tilde{S}_n^j = \text{const.}
\end{split}
\end{equation}

We can summarise these results in a language more suitable for drawing
the analogy with the non-linear case of strings on $\mathbb{R} \times
S^3$. Using the change of variables \eqref{action-angle def: flat} we
can rewrite the general phase-space configuration \eqref{generic PS
configuration} as
\begin{equation} \label{reconstruction: flat}
\begin{split}
X^j (\sigma) &= x^j + i \sum_{n \neq 0} \frac{1}{n} \sqrt{n S_n^j}
e^{i \theta^j_n + i n \sigma} + i \sum_{n \neq 0} \frac{1}{n} \sqrt{n
\tilde{S}_n^j} e^{i \tilde{\theta}^j_n - i n \sigma},\\
P^j (\sigma) &= p^j + \sum_{n \neq 0} \sqrt{n S_n^j} e^{i \theta^j_n +
i n \sigma} + \sum_{n \neq 0} \sqrt{n \tilde{S}_n^j} e^{i
\tilde{\theta}^j_n - i n \sigma}.
\end{split}
\end{equation}
Recall that we are only considering `finite-oscillator' solutions with
a finite number of oscillator modes turned on. Therefore the sums in
\eqref{reconstruction: flat} are finite and the functions
$X^j$, $P^j$ depend only on a finite number of parameters. These are
the angle variables $\{ x^j, \theta^j_n, \tilde{\theta}^j_n \}$ on the
one hand and the action variables $\{ p^j, S^j_n, \tilde{S}^j_n \}$ on
the other. Thus the pair $X^j$, $P^j$ in \eqref{reconstruction: flat}
can be thought of as a map $\mathcal{G} : \mathcal{P}^{(k)}
\rightarrow \mathcal{P}^{\infty}$ from a finite-dimensional
phase-space $\mathcal{P}^{(k)}$ comprised of these non-vanishing modes
of the string to the actual physical phase-space
$\mathcal{P}^{\infty}$ of the string. Moreover, the linear
$\tau$-evolution \eqref{linear motion: flat} through the
finite-dimensional phase-space gets mapped by \eqref{reconstruction:
flat} to the physical motion in $\mathcal{P}^{\infty}$. In particular
the function $X^j$ alone maps the linear motion \eqref{linear motion:
flat} to the general solution \eqref{General solution: flat} in
configuration space. We can say that a $k$-oscillator phase-space
solution $(X^j,P^j) : \mathbb{R}_{\tau} \rightarrow
\mathcal{P}^{\infty}$ factors through the finite-dimensional
phase-space $\mathcal{P}^{(k)}$ parameterised by $\{ x^j, \theta^j_n,
\tilde{\theta}^j_n, p^j, S^j_n, \tilde{S}^j_n \}$ since it decomposes
as
\begin{equation} \label{solution factors: flat}
(X^j,P^j) : \mathbb{R}_{\tau}
\overset{\vec{\theta}}\longrightarrow \mathcal{P}^{(k)}
\overset{\mathcal{G}}\longrightarrow \mathcal{P}^{\infty},
\end{equation}
where the first map $\vec{\theta}$ is linear and the second is given
by \eqref{reconstruction: flat}. This picture for strings in flat
space is therefore very similar to the one presented at the start of
this chapter for strings on $\mathbb{R} \times S^3$ where we had a
finite-gap solution mapping the Jacobian bundle to the reduced
phase-space $\mathcal{P}^{\infty}$, as illustrated in Figure
\ref{finite-dim phase-space}. Equation \eqref{reconstruction: flat}
can be thought of as the flat space analogue of the reconstruction
formula for the current $j(\sigma)$ (theorem \ref{thm: reconstructing
j}) in that it expresses the general phase-space configuration
$(X^i(\sigma), P^i(\sigma))$ in terms of the finite data $\{ x^j,
\theta^j_n, \tilde{\theta}^j_n, p^j, S^j_n, \tilde{S}^j_n \}$.


\section{The geometric map} \label{section: geometric map}


As we argued at the start of this chapter, the complete set of
algebro-geometric data $\{ (\hat{\Sigma}, dp), \hat{\gamma}(0) \}$ for an
arbitrary finite-gap solution $j$ can be succinctly described as a
point in the Jacobian bundle $\mathcal{M}^{(2g + 2)}_{\mathbb{C}}$
over $\mathcal{L}$,
\begin{equation} \label{bundle M}
J_{\mathfrak{m}}(\hat{\Sigma}) \rightarrow \mathcal{M}^{(2g +
2)}_{\mathbb{C}} \rightarrow \mathcal{L},
\end{equation}
whose fibre over every point of the base, specified by a pair
$(\hat{\Sigma}, dp)$, is the generalised Jacobian
$J_{\mathfrak{m}}(\hat{\Sigma})$ relative the usual modulus
$\mathfrak{m} = \infty^+ + \infty^-$. The finite-gap construction
of chapter \ref{chapter: finite-gap} (in particular theorem \ref{thm:
reconstructing j}) thus defines an injective map, called the
\dub{geometric map} in the terminology of \cite{Krichever+Phong2},
from the algebro-geometric data $\mathcal{M}^{(2g + 2)}_{\mathbb{C}}$
into the space $\mathcal{S}^{\infty}_{\mathbb{C}}$ of complexified
solutions $j \in \mathfrak{sl}(2, \mathbb{C})$ to the equations of
motion of a string moving on $\mathbb{R} \times S^3$ which also
satisfy the Virasoro and static gauge conditions \eqref{Vir without
P=0},
\begin{equation} \label{geometric map}
\mathcal{G}' : \mathcal{M}^{(2g+2)}_{\mathbb{C}} \hookrightarrow
\mathcal{S}^{\infty}_{\mathbb{C}}.
\end{equation}
At the start of section \ref{section: involution} we described the
reduced phase-space $\mathcal{P}^{\infty}$ as the restriction of the
space of solutions $\mathcal{S}^{\infty}$ satisfying \eqref{Vir
without P=0} to a chosen time slice. That is to say, setting all the
higher times in the reconstruction map \eqref{geometric map} to zero
except for the worldsheet $\sigma$-coordinate provides an embedding
of the Jacobian bundle into (complexified) reduced phase-space
$\mathcal{P}^{\infty}_{\mathbb{C}}$, namely
\begin{equation} \label{geometric map 2}
\mathcal{G} : \mathcal{M}^{(2g+2)}_{\mathbb{C}} \hookrightarrow
\mathcal{P}^{\infty}_{\mathbb{C}}.
\end{equation}
However, by virtue of theorem \ref{thm: linear motion} the dependence
on all the higher times can be recovered simply by composing the
phase-space map \eqref{geometric map 2} with a linear map determined
by \eqref{linear dynamical divisor},
\begin{equation*}
\vec{\theta} : \mathbb{R}^N \rightarrow
\mathcal{M}^{(2g+2)}_{\mathbb{C}},
\end{equation*}
which given a set of $N$ higher times $\{ t_i \}_{i = 1}^N$ translates
the Abel map $\vec{\mathcal{A}}(\hat{\gamma})$ of a divisor
$\hat{\gamma}$ to the point $\vec{\mathcal{A}}(\hat{\gamma}) +
\int_{\vec{b}} d\mathcal{Q}$ while staying on the same fibre of
$\mathcal{M}^{(2g+2)}_{\mathbb{C}}$. So much like equation
\eqref{solution factors: flat} in the flat space case, here the
solution factors through the Jacobian bundle,
\begin{equation} \label{solution factors}
j : \mathbb{R}^N \overset{\vec{\theta}}\longrightarrow
\mathcal{M}^{(2g+2)}_{\mathbb{C}} \overset{\mathcal{G}}\longrightarrow
\mathcal{P}^{\infty}.
\end{equation}
The domain $\mathbb{R}^N$ could be restricted to just
$\mathbb{R}_{\tau}$ if we are only interested in $\tau$-evolution.

Let $\hat{\omega}_{\infty}$ denote the symplectic form on the reduced
phase-space $\mathcal{P}^{\infty}$ corresponding to the Dirac
bracket \eqref{Dirac bracket SG} introduced in chapter
\ref{chapter: hamiltonian}. The goal of the remainder of this
chapter will be to compute the pullback of $\hat{\omega}_{\infty}$ to
the Jacobian bundle $J_{\mathfrak{m}}(\hat{\Sigma})$ by the
geometric map \eqref{geometric map 2}. The result is the
following,
\begin{theorem} \label{thm: pullback}
The pullback of the Dirac bracket on the reduced phase-space
$\mathcal{P}^{\infty}$ by the geometric map $\mathcal{G}$ takes
the simple form,
\begin{equation} \label{symplectic form pullback}
\hat{\omega}_{2g+2} \equiv \mathcal{G}^{\ast} \hat{\omega}_{\infty} =
\sum_{I = 1}^{g+1} \delta \mathcal{S}_I \wedge \delta \varphi_I,
\end{equation}
where $\mathcal{S}_I$ are the filling fractions \eqref{filling
fractions}. In particular we see that they precisely correspond to
the \dub{action variables} of the string. The conjugate \dub{angle
variables} $\varphi_I$ are defined in terms of the divisor
$\hat{\gamma}(t)$ by
\begin{equation} \label{angle variables def}
\begin{split}
\varphi_i &= \mathcal{A}_i(\hat{\gamma}(t)) -
\mathcal{A}_{g+1}(\hat{\gamma}(t)), \quad  i = 1, \ldots, g\\
\varphi_{g+1} &= - \mathcal{A}_{g+1}(\hat{\gamma}(t)).
\end{split}
\end{equation}
\end{theorem}

To prove this theorem we will show how to express the
algebro-geometric data in terms of the monodromy matrix $\Omega(x)$,
analogously to \eqref{extract data: flat}, and use this to read off
their Dirac brackets from $\{ \Omega(x), \Omega(x')
\}_{\text{D.B.}}$. We already know from chapter \ref{chapter:
integrability} how to read off the integrals of motion from
$\Omega(x)$ and we have obtained their Dirac bracket \eqref{complete
integrability}, which is the analogue of \eqref{involution: flat} in
flat space.

\subsection*{Extracting data}

The divisor $\hat{\gamma}(t)$ of poles of $\bm{h}(P,t)$ can be
extracted from $\Omega(x)$ using a method due to Sklyanin
\cite{Sklyanin:1995bm} as follows. We perform a similarity
transformation on the monodromy matrix $\Omega(x)$ by $L = {\tiny
\left( \begin{array}{cc} 1&1\\ 0&1 \end{array} \right)} \in SL(2,
\mathbb{C})$ and define $\widetilde{\Omega}(x) = L \Omega(x) L^{-1}$
with components
\begin{equation} \label{mono tilde comp}
\widetilde{\Omega}(x) = \left( \begin{array}{cc}
\widetilde{\mathcal{A}}(x) & \widetilde{\mathcal{B}}(x)\\
\widetilde{\mathcal{C}}(x) & \widetilde{\mathcal{D}}(x) \end{array}
\right).
\end{equation}

\begin{lemma} \label{lemma: extracting oscillators: S^3}
Let $\hat{\gamma}_i \in \hat{\Sigma}$, $i = 1, \ldots, g+1$ be the
points of the divisor $\hat{\gamma} \equiv \hat{\gamma}(t)$ of poles
of the normalised eigenvector $\bm{h}$. Then the coordinates of the
corresponding points on the spectral curve $\mathfrak{P}_i =
(x(\hat{\gamma}_i), \Lambda(\hat{\gamma}_i)) \in \Gamma$ satisfy
\begin{equation} \label{extracting oscillators: S^3}
\widetilde{\mathcal{B}}(x(\hat{\gamma}_i)) = 0, \qquad
\Lambda(\hat{\gamma}_i) =
\widetilde{\mathcal{D}}(x(\hat{\gamma}_i)) =
\widetilde{\mathcal{A}}(x(\hat{\gamma}_i))^{-1}.
\end{equation}
\begin{proof}
The normalised eigenvector $\bm{h}(P)$ satisfies the system of equations
\begin{equation*}
\Omega(x) \bm{h}(P) = \Lambda(P) \bm{h}(P), \qquad \bm{\alpha}
\cdot \bm{h}(P) = 1,
\end{equation*}
where as before $\bm{\alpha} = (1,1)$. Since the components of
$\bm{h}(P)$ have poles at $\hat{\gamma}_i$ we introduce the residue
vectors $\bm{h}_i = \res_{\hat{\gamma}_i} \bm{h}(P)$ which satisfy
the Sklyanin system
\begin{equation} \label{Sklyanin system}
\Omega(x(\hat{\gamma}_i)) \bm{h}_i = \Lambda(\hat{\gamma}_i)
\bm{h}_i, \qquad
\bm{\alpha} \cdot \bm{h}_i = 0.
\end{equation}
After the similarity transformation $\tilde{\bm{h}}_i = L \bm{h}_i$,
$\widetilde{\Omega}(x(\hat{\gamma}_i)) = L \Omega(x(\hat{\gamma}_i))
L^{-1}$ the system of equations \eqref{Sklyanin system} can be
rewritten as
\begin{equation*}
\widetilde{\Omega}(x(\hat{\gamma}_i)) \tilde{\bm{h}}_i =
\Lambda(\hat{\gamma}_i) \tilde{\bm{h}}_i, \qquad
\left(\tilde{\bm{h}}_i\right)_1 = 0.
\end{equation*}
When expressed in terms of components \eqref{mono tilde comp} this
immediately implies \eqref{extracting oscillators: S^3}.
\end{proof}
\end{lemma}

Notice that the variables $\{ \Lambda(\hat{\gamma}_i) \}_{i=1}^{g+1}$
form a set of $g+1$ integrals of motion of the genus $g$ finite-gap
solution. For generic divisors $\hat{\gamma} = \hat{\gamma}_1 +
\ldots + \hat{\gamma}_{g+1}$ these variables are independent and
therefore parameterise the base $\mathcal{L}$ of the bundle
$\mathcal{M}^{(2g+2)}_{\mathbb{C}}$. Since the knowledge of the
remaining coordinates $\{ x(\hat{\gamma}_i) \}_{i=1}^{g+1}$ completely
specifies the divisor $\hat{\gamma}$ they naturally provide
coordinates along the fibres $J_{\mathfrak{m}}(\hat{\Sigma})$. The
full set of coordinates $\{ x(\hat{\gamma}_i), \Lambda(\hat{\gamma}_i)
\}_{i=1}^{g+1}$ of the points $\mathfrak{P}_i \in \Gamma$ can be thus
be thought of as the complete algebro-geometric data for genus $g$
finite-gap solutions. Equations \eqref{extracting oscillators: S^3}
will be our way of extracting the algebro-geometric data of a generic
$g$-gap string. This is the non-linear analogue of extracting the
Fourier coefficients of a finite-oscillator solution in the flat space
case \eqref{extract data: flat}.

Because the matrix from which one reads off the divisor
$\hat{\gamma}$ isn't exactly the monodromy matrix $\Omega(x)$
but the similar matrix $\tilde{\Omega}(x) = L \Omega(x) L^{-1}$, we
will need the Dirac bracket $\{ \widetilde{\Omega}(x)
\overset{\otimes}, \widetilde{\Omega}(x') \}_{\text{D.B.}}$ between
these new matrices.

\begin{lemma}
The Dirac algebra between monodromy matrices \eqref{fundamental Dirac
bracket} is invariant under $SL(2, \mathbb{C})$ similarity
transformations. In particular,
\begin{equation} \label{fundamental Dirac bracket SL2C}
\begin{split}
\{ \widetilde{\Omega}(x) \mathop{,}^{\otimes}
\widetilde{\Omega}(x') \}_{\text{D.B.}} \approx &[r(x,x'),
\widetilde{\Omega}(x) \otimes \widetilde{\Omega}(x')] \\
+ &( \widetilde{\Omega}(x) \otimes {\bf 1} ) s(x,x')
( {\bf 1} \otimes \widetilde{\Omega}(x') ) \\
- & ( {\bf 1} \otimes \widetilde{\Omega}(x') ) s(x,x')
( \widetilde{\Omega}(x) \otimes {\bf 1} ).
\end{split}
\end{equation}
\begin{proof}
Let $L \in SL(2,\mathbb{C})$. Conjugating the Dirac bracket
\eqref{fundamental Dirac bracket} by $L \otimes L$ has the effect
of replacing $\Omega(x)$ by $\widetilde{\Omega}(x) = L \Omega(x)
L^{-1}$ but also $r(x,x')$ and $s(x,x')$ by $\widetilde{r}(x,x') =
(L \otimes L) r(x,x') (L^{-1} \otimes L^{-1})$ and
$\widetilde{s}(x,x') = (L \otimes L) s(x,x') (L^{-1} \otimes
L^{-1})$ respectively. However $r(x,x')$ and $s(x,x')$ are both
multiples of $\eta = t_a \otimes t^a$ which is invariant under
$SL(2, \mathbb{C})$, \textit{i.e.} $(L \otimes L) \eta (L^{-1}
\otimes L^{-1}) = \eta$, since infinitesimally for any $\alpha \in
\mathfrak{sl}(2, \mathbb{C})$ we have $\left[ {\bf 1} \otimes
\alpha + \alpha \otimes {\bf 1}, \eta \right] = 0$. Therefore
$\widetilde{r}(x,x') = r(x,x')$ and $\widetilde{s}(x,x') =
s(x,x')$ and \eqref{fundamental Dirac bracket SL2C} follows.
\end{proof}
\end{lemma}


\section{Dirac brackets of algebro-geometric data} \label{section: PB data 2}


By lemma \ref{lemma: extracting oscillators: S^3} the relevant
components of $\widetilde{\Omega}(x)$ for extracting the
algebro-geometric data are $\widetilde{\mathcal{A}}(x)$ and
$\widetilde{\mathcal{B}}(x)$. Their Dirac brackets can then be deduced
from the Dirac algebra between monodromies \eqref{fundamental Dirac
bracket SL2C}.

\begin{lemma}
Let $\hat{r}(x,x')$ and $\hat{s}(x,x')$ be defined as $r(x,x')$
and $s(x,x')$ respectively but without the factors of $\eta$,
\textit{i.e.} $r(x,x') = \hat{r}(x,x') \eta$ and $s(x,x') =
\hat{s}(x,x') \eta$. Then
\begin{subequations} \label{A and B PB}
\begin{align}
\label{PB1}
\left\{ \widetilde{\mathcal{A}}(x),\widetilde{\mathcal{A}}(x')
\right\}_{\text{D.B.}} &= \left( \widetilde{\mathcal{B}}(x)
\widetilde{\mathcal{C}}(x') - \widetilde{\mathcal{B}}(x')
\widetilde{\mathcal{C}}(x) \right) \hat{s}(x,x'),\\
\begin{split} \label{PB2}
\left\{ \widetilde{\mathcal{A}}(x),\widetilde{\mathcal{B}}(x')
\right\}_{\text{D.B.}} &= \left( \widetilde{\mathcal{A}}(x)
\widetilde{\mathcal{B}}(x') + \widetilde{\mathcal{A}}(x')
\widetilde{\mathcal{B}}(x) \right) \hat{r}(x,x') \\
&\qquad\qquad + \left(\widetilde{\mathcal{A}}(x)
\widetilde{\mathcal{B}}(x') + \widetilde{\mathcal{D}}(x')
\widetilde{\mathcal{B}}(x) \right) \hat{s}(x,x'),
\end{split}
\\
\label{PB3}
\left\{ \widetilde{\mathcal{B}}(x),\widetilde{\mathcal{B}}(x')
\right\}_{\text{D.B.}} &= 0.
\end{align}
\end{subequations}
\begin{proof}
Let us express the right hand side of \eqref{fundamental Dirac bracket
SL2C} in terms of the components \eqref{mono tilde comp} of
$\widetilde{\Omega}(x)$. This requires the following ingredients
\begin{gather*}
\eta = {\footnotesize \frac{1}{2} \left(
\begin{array}{cc}
\sigma_3 & \sigma_1 - i \sigma_2\\
\sigma_1 + i \sigma_2 & -\sigma_3
\end{array}
\right)}, \qquad \widetilde{\Omega}(x) \otimes \widetilde{\Omega}(x') = {\footnotesize \left(
\begin{array}{cc}
\widetilde{\mathcal{A}}(x)\widetilde{\Omega}(x') &
\widetilde{\mathcal{B}}(x)\widetilde{\Omega}(x')\\
\widetilde{\mathcal{C}}(x)\widetilde{\Omega}(x') &
\widetilde{\mathcal{D}}(x)\widetilde{\Omega}(x')
\end{array}
\right)}, \\ {\bf 1} \otimes \widetilde{\Omega}(x') = {\footnotesize \left(
\begin{array}{cc}
\widetilde{\Omega}(x') & 0\\
0 & \widetilde{\Omega}(x')
\end{array}
\right)}, \qquad \widetilde{\Omega}(x) \otimes {\bf 1} =
{\footnotesize \left(
\begin{array}{cc}
\widetilde{\mathcal{A}}(x) {\bf 1} & \widetilde{\mathcal{B}}(x) {\bf 1}\\
\widetilde{\mathcal{C}}(x) {\bf 1} & \widetilde{\mathcal{D}}(x)
{\bf 1}
\end{array}
\right)}.
\end{gather*}
Using these one can easily compute the following quantities
\begin{equation*}
{\fontsize{8}{12} \eta \widetilde{\Omega}(x) \otimes
\widetilde{\Omega}(x') = \frac{1}{2} \left(
\begin{array}{cc}
\widetilde{\mathcal{A}}(x)\sigma_3\widetilde{\Omega}(x') +
\widetilde{\mathcal{C}}(x)(\sigma_1 - i
\sigma_2)\widetilde{\Omega}(x') &
\widetilde{\mathcal{B}}(x)\sigma_3\widetilde{\Omega}(x')
+ \widetilde{\mathcal{D}}(x)(\sigma_1 - i \sigma_2)\widetilde{\Omega}(x')\\
\widetilde{\mathcal{A}}(x)(\sigma_1 + i
\sigma_2)\widetilde{\Omega}(x') -
\widetilde{\mathcal{C}}(x)\sigma_3\widetilde{\Omega}(x') &
\widetilde{\mathcal{B}}(x)(\sigma_1 + i
\sigma_2)\widetilde{\Omega}(x') -
\widetilde{\mathcal{D}}(x)\sigma_3\widetilde{\Omega}(x')
\end{array}\right)},
\end{equation*}
\begin{equation*}
{\fontsize{8}{12} \widetilde{\Omega}(x) \otimes \widetilde{\Omega}(x') \eta =
\frac{1}{2} \left(
\begin{array}{cc} \widetilde{\mathcal{A}}(x)\widetilde{\Omega}(x')\sigma_3 +
\widetilde{\mathcal{B}}(x)\widetilde{\Omega}(x')(\sigma_1 + i
\sigma_2) &
\widetilde{\mathcal{A}}(x)\widetilde{\Omega}(x')(\sigma_1 - i
\sigma_2)
- \widetilde{\mathcal{B}}(x)\widetilde{\Omega}(x')\sigma_3\\
\widetilde{\mathcal{C}}(x)\widetilde{\Omega}(x')\sigma_3 +
\widetilde{\mathcal{D}}(x)\widetilde{\Omega}(x')(\sigma_1 + i
\sigma_2) &
\widetilde{\mathcal{C}}(x)\widetilde{\Omega}(x')(\sigma_1 - i
\sigma_2) -
\widetilde{\mathcal{D}}(x)\widetilde{\Omega}(x')\sigma_3
\end{array}
\right)},
\end{equation*}
\begin{equation*}
{\fontsize{7}{12} \widetilde{\Omega}(x) \otimes {\bf 1} \eta {\bf 1} \otimes
\widetilde{\Omega}(x') = \frac{1}{2} \left(
\begin{array}{cc} \widetilde{\mathcal{A}}(x)\sigma_3 \widetilde{\Omega}(x') +
\widetilde{\mathcal{B}}(x)(\sigma_1 + i
\sigma_2)\widetilde{\Omega}(x') &
\widetilde{\mathcal{A}}(x)(\sigma_1 - i
\sigma_2)\widetilde{\Omega}(x')
- \widetilde{\mathcal{B}}(x)\sigma_3\widetilde{\Omega}(x')\\
\widetilde{\mathcal{C}}(x)\sigma_3 \widetilde{\Omega}(x') +
\widetilde{\mathcal{D}}(x)(\sigma_1 + i
\sigma_2)\widetilde{\Omega}(x') &
\widetilde{\mathcal{C}}(x)(\sigma_1 - i
\sigma_2)\widetilde{\Omega}(x') -
\widetilde{\mathcal{D}}(x)\sigma_3\widetilde{\Omega}(x')
\end{array}
\right)},
\end{equation*}
\begin{equation*}
{\fontsize{7}{12} {\bf 1} \otimes \widetilde{\Omega}(x')
\eta \widetilde{\Omega}(x) \otimes {\bf 1} = \frac{1}{2} \left(
\begin{array}{cc} \widetilde{\mathcal{A}}(x)\widetilde{\Omega}(x')\sigma_3 +
\widetilde{\mathcal{C}}(x)\widetilde{\Omega}(x')(\sigma_1 - i
\sigma_2) &
\widetilde{\mathcal{B}}(x)\widetilde{\Omega}(x')\sigma_3 +
\widetilde{\mathcal{D}}(x)\widetilde{\Omega}(x')(\sigma_1 - i \sigma_2)\\
\widetilde{\mathcal{A}}(x)\widetilde{\Omega}(x')(\sigma_1 + i
\sigma_2) -
\widetilde{\mathcal{C}}(x)\widetilde{\Omega}(x')\sigma_3 &
\widetilde{\mathcal{B}}(x)\widetilde{\Omega}(x')(\sigma_1 + i
\sigma_2) -
\widetilde{\mathcal{D}}(x)\widetilde{\Omega}(x')\sigma_3
\end{array}
\right)}.
\end{equation*}
The Dirac brackets \eqref{A and B PB} of various components of
$\widetilde{\Omega}(x)$ can now be read off from \eqref{fundamental
Dirac bracket SL2C} using the above. In particular for the Dirac
brackets
\begin{equation*}
\left\{ \widetilde{\mathcal{A}}(x),\widetilde{\mathcal{A}}(x')
\right\}_{\text{D.B.}}, \quad \left\{
\widetilde{\mathcal{A}}(x),\widetilde{\mathcal{B}}(x')
\right\}_{\text{D.B.}} \; \text{and} \; \left\{
\widetilde{\mathcal{B}}(x),\widetilde{\mathcal{B}}(x')
\right\}_{\text{D.B.}}
\end{equation*}
we take respectively the components $(11,11)$,
$(11,12)$ and $(12,12)$ of the tensor product relation
\eqref{fundamental Dirac bracket SL2C}.
\end{proof}
\end{lemma}

Next we show that the relations \eqref{A and B PB} imply
non-trivial Dirac brackets between the complex variables $\{
x(\hat{\gamma}_i), \Lambda(\hat{\gamma}_i) \}_{i = 1}^{g+1}$
comprising the algebro-geometric data.

\begin{proposition} \label{DB alg-geom data}
The Dirac brackets of the algebro-geometric data are
\begin{subequations} \label{DB alg data}
\begin{align}
\label{DB alg data 1} \{ \Lambda(\hat{\gamma}_j),
\Lambda(\hat{\gamma}_k) \}_{\text{D.B.}} &= 0,\\
\label{DB alg data 2} \frac{\sqrt{\lambda}}{4 \pi} \left\{
\Lambda(\hat{\gamma}_j), x(\hat{\gamma}_k) \right\}_{\text{D.B.}} &=
\Lambda(\hat{\gamma}_j) \frac{x(\hat{\gamma}_j)^2}{1 -
x(\hat{\gamma}_j)^2} \delta_{jk},\\
\label{DB alg data 3} \left\{ x(\hat{\gamma}_j), x(\hat{\gamma}_k)
\right\}_{\text{D.B.}} &= 0.
\end{align}
\end{subequations}
\begin{proof}
We will consider the implications of the three relations \eqref{A and
B PB} in turn. First we take the limit $x' \rightarrow
x_{\hat{\gamma}_k} \equiv x(\hat{\gamma}_k)$ of \eqref{PB1}. Using
\eqref{extracting oscillators: S^3} this gives
\begin{equation*}
\{ \widetilde{\mathcal{A}}(x), \Lambda(\hat{\gamma}_k)^{-1}
\}_{\text{D.B.}} = \widetilde{\mathcal{B}}(x)
\widetilde{\mathcal{C}}(x_{\hat{\gamma}_k})
\hat{s}\left(x, x_{\hat{\gamma}_k} \right).
\end{equation*}
Taking the limit $x \rightarrow x_{\hat{\gamma}_j}$ yields $\{
\Lambda(\hat{\gamma}_j)^{-1}, \Lambda(\hat{\gamma}_k)^{-1}
\}_{\text{D.B.}} = 0$, or equivalently \eqref{DB alg data 1}.

We now turn to the Poisson bracket \eqref{PB2}. Taking the limit
$x \rightarrow x_{\hat{\gamma}_j}$ first gets rid of the terms
proportional to $\widetilde{\mathcal{B}}(x)$ (using
$\widetilde{\mathcal{B}}(x_{\hat{\gamma}_j}) = 0$) and leaves
\begin{equation*}
\left\{
\widetilde{\mathcal{A}}(x_{\hat{\gamma}_j}),
\widetilde{\mathcal{B}}(x') \right\}_{\text{D.B.}} =
\widetilde{\mathcal{A}}(x_{\hat{\gamma}_j})
\widetilde{\mathcal{B}}(x') \left( \hat{r}(x_{\hat{\gamma}_j},x') +
\hat{s}(x_{\hat{\gamma}_j},x') \right).
\end{equation*}
Now using \eqref{extracting oscillators: S^3} we can write
$\widetilde{\mathcal{B}}(x') = (x' - x_{\hat{\gamma}_k})
\widetilde{\mathcal{B}}_k(x')$ with
$\widetilde{\mathcal{B}}_k(x_{\hat{\gamma}_k}) \neq 0$, so that
\begin{multline*}
(x' - x_{\hat{\gamma}_k}) \left\{
\widetilde{\mathcal{A}}(x_{\hat{\gamma}_j}),
\widetilde{\mathcal{B}}_k(x') \right\}_{\text{D.B.}} - \left\{
\widetilde{\mathcal{A}}(x_{\hat{\gamma}_j}),x_{\hat{\gamma}_k}
\right\}_{\text{D.B.}} \widetilde{\mathcal{B}}_k(x') \\ =
\widetilde{\mathcal{A}}(x_{\hat{\gamma}_j}) (x' -
x_{\hat{\gamma}_k}) \widetilde{\mathcal{B}}_k(x') \left(
\hat{r}(x_{\hat{\gamma}_j},x') +
\hat{s}(x_{\hat{\gamma}_j},x')\right),
\end{multline*}
where
\begin{equation*}
\hat{r}(x_{\hat{\gamma}_j},x') + \hat{s}(x_{\hat{\gamma}_j},x') =
- \frac{2 \pi}{\sqrt{\lambda}} \frac{x_{\hat{\gamma}_j}^2 + x'^2 -
2 x_{\hat{\gamma}_j}^2 x'^2}{(x_{\hat{\gamma}_j} - x')(1 -
x_{\hat{\gamma}_j}^2)(1 - x'^2)} - \frac{2 \pi}{\sqrt{\lambda}}
\frac{x_{\hat{\gamma}_j} + x'}{(1 - x_{\hat{\gamma}_j}^2)(1 -
x'^2)}.
\end{equation*}
Taking the limit $x' \rightarrow x_{\hat{\gamma}_k}$ with $k \neq j$
kills everything but the second term on the left hand side, leaving
$\{ \widetilde{\mathcal{A}}(x_{\hat{\gamma}_j}), x_{\hat{\gamma}_k}
\}_{\text{D.B.}} = 0, \, k \neq j$. Now setting $k=j$ and taking the
limit $x' \rightarrow x_{\hat{\gamma}_j}$ kills the $\hat{s}$ term
leaving $-\{ \widetilde{\mathcal{A}}(x_{\hat{\gamma}_j}),
x_{\hat{\gamma}_j} \}_{\text{D.B.}} = \frac{4 \pi}{\sqrt{\lambda}}
\widetilde{\mathcal{A}}(x_{\hat{\gamma}_j})
\frac{x_{\hat{\gamma}_j}^2}{1 - x_{\hat{\gamma}_j}^2}$ which is
equivalent to \eqref{DB alg data 2} by \eqref{extracting oscillators:
S^3}.

Finally, writing again $\widetilde{\mathcal{B}}(x) = (x -
x_{\hat{\gamma}_j}) \widetilde{\mathcal{B}}_j(x)$, equation
\eqref{PB3} immediately leads to $\{ x_{\hat{\gamma}_j},
\widetilde{\mathcal{B}}(x') \}_{\text{D.B.}} = 0$ which in turn
implies \eqref{DB alg data 3}.
\end{proof}
\end{proposition}

The algebro-geometric data needed to reconstruct a $g$-gap
solution is a point on the Jacobian bundle
$\mathcal{M}^{(2g+2)}_{\mathbb{C}}$ specified by $2g+2$ complex
coordinates $\left\{ x_{\hat{\gamma}_i}, \Lambda(\hat{\gamma}_i)
\right\}_{i=1}^{g+1}$. Proposition \ref{DB alg-geom data} gives
the complete set of Dirac brackets for these variables. To write
these brackets in canonical form we perform the change of spectral
parameter,
\begin{equation*}
z = x + \frac{1}{x}.
\end{equation*}
We have already introduced this function in equation \eqref{definition
of z} of chapter \ref{chapter: curves} to discuss the moduli of the
spectral curve. However, there the change of variable $x \mapsto z$
was unjustified. Here we see from \eqref{DB alg data bis 2} below that
the new spectral parameter $z$ is much better suited for discussions
of the symplectic structure. Recalling also the definition
\eqref{quasi-momentum definition} of the quasi-momentum we can rewrite
the brackets \eqref{DB alg data} as
\begin{subequations} \label{DB alg data bis}
\begin{align}
\label{DB alg data bis 1} \{ p(\hat{\gamma}_j),
p(\hat{\gamma}_k) \}_{\text{D.B.}} &= 0,\\
\label{DB alg data bis 2} \frac{\sqrt{\lambda}}{4 \pi i} \left\{
p(\hat{\gamma}_j), z(\hat{\gamma}_k) \right\}_{\text{D.B.}} &=
\delta_{jk},\\
\label{DB alg data bis 3} \left\{ z(\hat{\gamma}_j), z(\hat{\gamma}_k)
\right\}_{\text{D.B.}} &= 0.
\end{align}
\end{subequations}

\begin{corollary}
The pullback of the Dirac bracket on the reduced phase-space
$\mathcal{P}^{\infty}$ by the geometric map $\mathcal{G}$ is
\begin{equation} \label{alg-geom sympl form}
\hat{\omega}_{2g + 2} \equiv \mathcal{G}^{\ast} \hat{\omega}_{\infty} = -
\frac{\sqrt{\lambda}}{4 \pi i} \sum_{i=1}^{g+1} \delta
p(\hat{\gamma}_i) \wedge \delta z(\hat{\gamma}_i).
\end{equation}
\end{corollary}

\begin{remark}
Recall that the Abel map defines a local isomorphism from the group of
divisors of degree $g+1$ in the neighbourhood of a non-special divisor
to the generalised Jacobian $J_{\mathfrak{m}}(\hat{\Sigma})$. But a
divisor $\hat{\gamma} = \hat{\gamma}_1 + \ldots +
\hat{\gamma}_{g+1}$ of degree $g+1$ is nothing but an unordered set of
$g+1$ points $\{ \hat{\gamma}_i \}_{i = 1}^{g+1}$ on
$\hat{\Sigma}$. Therefore the $(g+1)$-symmetric product of the curve
$\hat{\Sigma}$ can be locally identified via the Abel map with the
generalised Jacobian $J_{\mathfrak{m}}(\hat{\Sigma})$. In particular,
any symmetric expression in $\{ \hat{\gamma}_i \}_{i = 1}^{g+1}$, such
as \eqref{alg-geom sympl form}, naturally lives on
$J_{\mathfrak{m}}(\hat{\Sigma})$.
\end{remark}

\subsection*{Action-angle variables}

The change of coordinates to action-angle variables is fairly
standard (see for instance \cite{Krichever+Phong, Krichever+Phong2}
and \cite[p.16]{Krichever-1998}). We shall construct the
complete set of action-angle variables starting from the
algebro-geometric symplectic form \eqref{alg-geom sympl form} on
$\mathcal{M}_{\mathbb{C}}^{(2g+2)}$.

It is useful at first to consider the universal curve bundle
$\mathcal{N}$ over the leaf $\mathcal{L}$
\begin{equation*}
\hat{\Sigma} \rightarrow \mathcal{N} \rightarrow \mathcal{L},
\end{equation*}
whose fibre over every point of the base $\mathcal{L}$ is the
corresponding curve $\hat{\Sigma}$. Now recall from chapter
\ref{chapter: curves} that the $\{ S_i \}_{i=1}^g$ and $R$ defined in
\eqref{independent moduli} form a set of coordinates on the base
$\mathcal{L}$. They can be expressed in terms of the $a_i$-periods and
residue at $\infty^+$ or $\infty^-$ of the differential\footnote{Note
that $\tilde{\alpha} + i \alpha \equiv d \beta$ is locally exact where
$\beta = i \frac{\sqrt{\lambda}}{4 \pi} p z$ and $\alpha$ was defined
in \eqref{symplectic 1-form}.} $\tilde{\alpha} \equiv -
\frac{\sqrt{\lambda}}{4 \pi i} p dz$ respectively,
\begin{equation} \label{action variables}
S_i = \frac{1}{2 \pi} \int_{a_i} \tilde{\alpha}, \; i = 1, \ldots,
g, \quad \frac{R}{2} = \mp \frac{1}{2 \pi} \oint_{\infty^{\pm}}
\tilde{\alpha},
\end{equation}
where the contour integrals around the points $\infty^{\pm} \in
\hat{\Sigma}$ are taken counterclockwise. Note also that $z$ can be
taken as a local coordinate along the fibres of
$\mathcal{N}$. Denoting then by $\delta$ the exterior derivative on
the total space $\mathcal{N}$, the differentials $\delta z$, $\delta
R$ and $\{ \delta S_i \}_{i = 1}^g$ form a basis of differentials at
every point of $\mathcal{N}$. In this basis, the total exterior
derivative of any function (or 1-form) $f$ on $\mathcal{N}$ can be
separated as
\begin{equation*}
\delta f = \delta z \wedge \partial_z f + \frac{1}{2} \delta R \wedge
\partial_{\frac{R}{2}} f + \sum_{i = 1}^g \delta S_i \wedge
\partial_{S_i} f \equiv df + \delta^{\mathcal{L}} f,
\end{equation*}
where $\delta^{\mathcal{L}}$ denotes the exterior derivative along the
leaf $\mathcal{L}$. In particular $\delta z = dz$, $\delta R =
\delta^{\mathcal{L}} R$ and $\delta S_i = \delta^{\mathcal{L}}
S_i$. The differential $\tilde{\alpha}$ on $\hat{\Sigma}$, as in fact
any differential on $\hat{\Sigma}$, can be extended to a differential
on $\mathcal{N}$ by setting it to zero along $\delta R$ and $\{ \delta
S_i \}_{i=1}^g$.

Consider now its exterior derivative $\delta \tilde{\alpha}$ on
$\mathcal{N}$
\begin{equation} \label{dpdz}
\delta \tilde{\alpha} = - \frac{\sqrt{\lambda}}{4 \pi i} \delta p
\wedge dz = \sum_{i=1}^g \delta S_i \wedge \partial_{S_i}
\tilde{\alpha} + \frac{1}{2} \delta R \wedge
\partial_{\frac{R}{2}} \tilde{\alpha}.
\end{equation}
Although $\tilde{\alpha}$ is not single valued on $\hat{\Sigma}$,
the ambiguities in its definition are constant along the leaf
$\mathcal{L}$. Indeed by equation \eqref{periods defining leaf}
one can add to the differential $\tilde{\alpha}$ any integer
multiple of $- \frac{\sqrt{\lambda}}{2 i} dz$, but this latter
differential depends neither on $R$ nor on $\{ S_i \}_{i=1}^g$. It
follows that $\partial_{S_i} \tilde{\alpha}$ and
$\partial_{\frac{R}{2}} \tilde{\alpha}$ are well defined.
Furthermore, since none of the residues in table \ref{table of
residues} depend on $S_i$, all the residues of $\tilde{\alpha} =
d\beta - i \alpha$ are independent of $S_i$ and it follows that
$\partial_{S_i} \tilde{\alpha}$ is holomorphic, \textit{i.e.}
$\partial_{S_i} \tilde{\alpha} = \sum_{j=1}^g c_{ij} \omega_j$.
Using \eqref{action variables} we obtain $c_{ij} = 2 \pi
\delta_{ij}$. In contrast, we notice from table \ref{table of
residues} that the residues of $\tilde{\alpha}$ at $\infty^{\pm}$
are proportional to $R$ so that $\partial_{\frac{R}{2}}
\tilde{\alpha}$ must have simple poles at these points.
Specifically, using \eqref{action variables} its residues are
found to be $\res_{\infty^{\pm}} \partial_{\frac{R}{2}}
\tilde{\alpha} = \pm i$. In conclusion
\begin{equation*}
\partial_{S_i} \tilde{\alpha} = 2 \pi \omega_i, \; i = 1, \ldots, g,
\qquad \partial_{\frac{R}{2}} \tilde{\alpha} = - 2 \pi
\omega_{\infty},
\end{equation*}
where $\omega_{\infty}$ was defined in \eqref{third kind diff at
infty}. Therefore \eqref{dpdz} simplifies to
\begin{equation*}
\delta \tilde{\alpha} = \sum_{i=1}^g \delta S_i \wedge 2 \pi
\omega_i - \frac{1}{2} \delta R \wedge 2 \pi \omega_{\infty}.
\end{equation*}

The differential $\delta \tilde{\alpha}$ on $\mathcal{N}$ can be used
to write down the symplectic form \eqref{alg-geom sympl form} as the
following expression symmetric in the points $\hat{\gamma}_j \in
\hat{\Sigma}$,
\begin{equation*}
\hat{\omega}_{2g+2} = \sum_{j = 1}^{g+1} \delta
\tilde{\alpha}(\hat{\gamma}_j) = \sum_{i=1}^g \delta S_i \wedge \left(
2 \pi \sum_{j = 1}^{g+1} \omega_i(\hat{\gamma}_j) \right) -
\frac{1}{2} \delta R \wedge \left( 2 \pi \sum_{j = 1}^{g+1}
\omega_{\infty}(\hat{\gamma}_j) \right).
\end{equation*}
Recall that such a symmetric expression in the $\hat{\gamma}_j \in
\hat{\Sigma}$ naturally lives on the generalised Jacobian
$J_{\mathfrak{m}}(\hat{\Sigma})$ which can be locally (in the
neighbourhood of a non-special divisor) identified with the
$(g+1)^{\text{st}}$ symmetric power of the curve $\hat{\Sigma}$
via the generalised Abel map $\mathcal{A} :
\hat{\Sigma}^{g+1}/S_{g+1} \rightarrow
J_{\mathfrak{m}}(\hat{\Sigma})$ given explicitly by
\begin{equation*}
\phi_i = \mathcal{A}_i( \hat{\gamma} ) = 2 \pi \sum_{j=1}^{g+1}
\int^{\hat{\gamma}_j} \omega_i, \qquad i = 1, \ldots, g+1,
\end{equation*}
where $\hat{\gamma} = \hat{\gamma}_1 + \ldots +
\hat{\gamma}_{g+1}$, $\omega_{g+1} = \omega_{\infty}$ and $\phi_i$ are
coordinates on $J_{\mathfrak{m}}(\hat{\Sigma})$. Now we can introduce
an exterior derivative $\delta$ on the Jacobian bundle
$\mathcal{M}^{(2g+2)}_{\mathbb{C}}$ by defining for any function (or
1-form) $f$,
\begin{equation*}
\delta f = \sum_{i=1}^{g+1} \delta z_i \wedge \partial_{z_i} f +
\frac{1}{2} \delta R \wedge \partial_{\frac{R}{2}} f + \sum_{i=1}^{g}
\delta S_i \wedge \partial_{S_i} f \equiv \sum_{i=1}^{g+1} d_i f +
\delta^{\mathcal{L}} f,
\end{equation*}
where $z_i$ is the local coordinate on the $i^{\text{th}}$ factor
$\hat{\Sigma}$ of $\hat{\Sigma}^{g+1}/S_{g+1}$. It follows that
\begin{equation*}
\delta \phi_i = \sum_{k=1}^{g+1} d_k \left( 2 \pi \sum_{j=1}^{g+1}
\int^{\hat{\gamma}_j} \omega_i \right) = 2 \pi \sum_{k=1}^{g+1}
\omega_i(\hat{\gamma}_k).
\end{equation*}
We can now write $\hat{\omega}_{2g+2}$ explicitly as a symplectic form
on $\mathcal{M}^{(2g+2)}_{\mathbb{C}}$, namely
\begin{equation*}
\hat{\omega}_{2g+2} = \sum_{i=1}^g \delta S_i \wedge \delta
\phi_i - \frac{1}{2} \delta R \wedge \delta \phi_{g+1}.
\end{equation*}
This can be further rewritten as
\begin{equation} \label{action-angle bracket 2}
\hat{\omega}_{2g+2} = \sum_{i=1}^g \delta S_i \wedge \delta (\phi_i -
\phi_{g+1}) + \delta \left( \frac{R - L}{2} - \sum_{i=1}^g S_i \right)
\wedge \delta (-\phi_{g+1}),
\end{equation}
where we have used the fact that $\delta L = 0$ since $L$ is fixed
along the leaf $\mathcal{L}$ by definition. Recalling the
definition of the $g+1$ filling fractions $\{ \mathcal{S}_I \}_{I =
1}^{g+1}$ in \eqref{filling fractions} and introducing the \dub{angle
variables} $\{ \varphi_I \}_{I = 1}^{g+1}$ as in \eqref{angle
variables def}, equation \eqref{action-angle bracket 2} becomes
equivalent to \eqref{symplectic form pullback} which completes the
proof of theorem \ref{thm: pullback}.


\section{Quasi-actions} \label{section: quasi-actions}


Remember that the Lax matrix in \eqref{Lax connection 4} is
responsible for the flow of the Hamiltonian $p(x)$. Thus going back to
the corresponding Hamilton equation written in Lax form we have
\begin{equation} \label{pre action Lax equation}
2 \pi i \left\{ - \frac{\sqrt{\lambda}}{8 \pi^2 i} \left( 1 -
\frac{1}{x^2} \right) p(x) , J_1(x') \right\}_{\text{D.B.}} = \left[
\partial_{\sigma} - J_1(x'), \frac{\Psi(x)
\frac{i}{2} \sigma_3 \Psi(x)^{-1}}{x-x'} \right].
\end{equation}
Integrating this equation in $x$ over the different $\bm{a}$-cycles,
and recalling the definition \eqref{independent moduli} of the first
$g$ action variables $S_i = - \frac{\sqrt{\lambda}}{8 \pi^2 i}
\int_{a_i} \left( 1 - \frac{1}{x^2} \right) p(x) dx$ we find
\begin{subequations} \label{action, R Lax eqs}
\begin{equation} \label{action Lax equation}
\{ S_i , J_1(x') \}_{\text{D.B.}} = \left[
\partial_{\sigma} - J_1(x'), \frac{1}{4 \pi}
\int_{a_i} \frac{\Psi(x) \sigma_3 \Psi(x)^{-1}}{x-x'} dx \right].
\end{equation}
Similarly, integrating around the point $x = \infty$ and
recalling the definition \eqref{independent moduli} of the global
$SU(2)_R$ charge $\frac{R}{2} = \frac{\sqrt{\lambda}}{8 \pi^2 i}
\oint_{\infty} \left( 1 - \frac{1}{x^2} \right) p(x) dx$ we find
\begin{equation} \label{R Lax equation}
\frac{1}{2} \{ R , J_1(x') \}_{\text{D.B.}} = \left[
\partial_{\sigma} - J_1(x'), - \frac{1}{4 \pi}
\oint_{\infty} \frac{\Psi(x) \sigma_3 \Psi(x)^{-1}}{x-x'} dx
\right].
\end{equation}
\end{subequations}
Equations \eqref{action, R Lax eqs} simply say that the
Hamiltonian flow of the action variables $S_i$ and $R$ are
generated by the following respective Lax matrices
\begin{equation} \label{action Lax}
\begin{split}
S_i \quad &\longleftrightarrow \quad A_i(x') = \frac{1}{4 \pi}
\int_{a_i} \frac{\Psi(x) \sigma_3 \Psi(x)^{-1}}{x-x'}
dx,\\
\frac{R}{2} \quad &\longleftrightarrow \quad - \frac{1}{4 \pi}
\oint_{\infty} \frac{\Psi(x) \sigma_3 \Psi(x)^{-1}}{x-x'} dx.
\end{split}
\end{equation}

Because any integral of motion can be expressed in terms of the
action variables, one ought to be able to use equation
\eqref{action Lax} to derive the Lax matrix for any other integral
of motion. We start with the following lemma, for which we
introduce the cohomology group $H^1(\hat{\Sigma}, \infty^{\pm})$
of the singular curve obtained from $\hat{\Sigma}$ by identifying
the points $\infty^{\pm}$.

\begin{lemma} \label{lemma: dP and dE}
The variation of the string hierarchy Hamiltonian $H_{n, \pm}$ along
the leaf $\mathcal{L}$ depends only on the cohomology class of
$d\Omega_{n, \pm}$ in $H^1(\hat{\Sigma}, \infty^{\pm})$,
\begin{subequations} \label{dP and dE}
\begin{equation} \label{dP and dE a}
\delta^{\mathcal{L}} H_{n, \pm} = \sum_{i=1}^g \delta^{\mathcal{L}}
S_i \int_{b_i} d\Omega_{n, \pm} - \frac{1}{2} \delta^{\mathcal{L}} R
\int_{\infty^-}^{\infty^+} d\Omega_{n, \pm}.
\end{equation}
Using the filling fractions to parametrise $\mathcal{L}$ this can
equivalently be written as
\begin{equation} \label{dP and dE b}
\delta^{\mathcal{L}} H_{n, \pm} = \sum_{I=1}^{g+1}
\delta^{\mathcal{L}} \mathcal{S}_I \int_{\mathcal{B}_I} d\Omega_{n,
\pm},
\end{equation}
where $\mathcal{B}_I$ is the contour going from $\infty^+$ to
$\infty^-$ through the $I^{\text{th}}$ cut.
\end{subequations}
\begin{proof}
Using the Riemann bilinear identity \eqref{Riemann bilinear 3} with $d
\Omega_1 = d\Omega_{n, \pm}$ and $d \Omega_2 = \delta^{\mathcal{L}}
\tilde{\alpha} = -\frac{\sqrt{\lambda}}{4 \pi i} \delta^{\mathcal{L}}
(p dz)$ we find
\begin{equation} \label{dP and dE proof 1}
- \sum_{i=1}^g \int_{b_i} d\Omega_{n, \pm} \int_{a_i} \delta^{\mathcal{L}}
\tilde{\alpha} = 2 \pi i \sum_{x = -1, +1, \infty} \left( \res_{x^+} +
\res_{x^-} \right) \Omega_{\pm} \delta^{\mathcal{L}} \tilde{\alpha}.
\end{equation}
Since $\Omega_{\pm}$ is regular at infinity but $\tilde{\alpha}$ has a
simple pole at $\infty^{\pm}$ with opposite residues, the contribution
from $x = \infty$ to \eqref{dP and dE proof 1} is easily evaluated to
be
\begin{multline*}
2 \pi i \left( \res_{\infty^+} + \res_{\infty^-} \right) \Omega_{\pm}
\delta^{\mathcal{L}} \tilde{\alpha} = 2 \pi i ( \Omega_{\pm}(\infty^+) -
\Omega_{\pm}(\infty^-) ) \res_{\infty^+} \delta^{\mathcal{L}}
\tilde{\alpha}\\
= \int_{\infty^-}^{\infty^+} d\Omega_{\pm} \oint_{\infty^+}
\delta^{\mathcal{L}} \tilde{\alpha}.
\end{multline*}
Equation \eqref{dP and dE proof 1} now simplifies using the
definitions \eqref{action variables} of the action variables $S_i$ and
$R$ to
\begin{equation} \label{dP and dE proof 2}
\sum_{i=1}^g \delta^{\mathcal{L}} S_i \int_{b_i} d\Omega_{\pm} -
\frac{1}{2} \delta^{\mathcal{L}} R \int_{\infty^-}^{\infty^+}
d\Omega_{\pm} = -i \sum_{x = -1, +1} \left( \res_{x^+} + \res_{x^-}
\right) \Omega_{\pm} \delta^{\mathcal{L}} \tilde{\alpha}.
\end{equation}
The left hand side can be evaluated using the asymptotics
\eqref{dOmega asymptotics pm 1} of the differentials $d \Omega_{n,
\pm}$ at $x = \pm 1$, the explicit form \eqref{singular parts} of
the singular parts $s_{n, \pm}(x)$, the expansion \eqref{Psi near
pm 1} of the quasi-momentum as well as the definition
\eqref{string hierarchy Hamiltonians} of the string hierarchy
Hamiltonians $H_{n, \pm}$. The final result is equation \eqref{dP
and dE a}. Rewriting this equation as
\begin{equation*}
\delta^{\mathcal{L}} H_{n,\pm} = \sum_{i=1}^g \bigg( \int_{b_i}
d\Omega_{n,\pm} - \int_{\infty^-}^{\infty^+} d\Omega_{n,\pm} \bigg)
\delta^{\mathcal{L}} S_i - \bigg( \int_{\infty^-}^{\infty^+}
d\Omega_{n,\pm} \bigg) \; \delta^{\mathcal{L}} \left( \frac{1}{2} R -
\sum_{i=1}^g S_i \right),
\end{equation*}
and using \eqref{sum of fillings} along with the fact that $L$ is
constant along the leaf $\mathcal{L}$ by definition gives equation
\eqref{dP and dE b}.
\end{proof}
\end{lemma}

As a special case of lemma \ref{lemma: dP and dE} consider the
$0^{\text{th}}$ level of the string hierarchy. We have $H_{0, \pm} =
\mathcal{E} \pm \mathcal{P}$ and $d \Omega_{0, \pm} = dq_{\pm}/ 2 \pi$
so that

\begin{corollary} \label{cor: dP and dE}
The variations of the worldsheet energy $\mathcal{E}$ and momentum
$\mathcal{P}$ along the leaf $\mathcal{L}$ depend only on the
cohomology class in $H^1(\hat{\Sigma}, \infty^{\pm})$ of the
differentials $dq$ and $dp$ of the quasi-energy and quasi-momentum
respectively, namely
\begin{subequations}  \label{dP and dE 2}
\begin{align}
\label{dP and dE 2a} \delta^{\mathcal{L}} \left(\mathcal{E} \pm \mathcal{P}\right) &=
\sum_{i=1}^g \left( \int_{b_i} \frac{dq_{\pm}}{2 \pi} \right)
\delta^{\mathcal{L}} S_i - \bigg( \int_{\infty^-}^{\infty^+}
\frac{dq_{\pm}}{2 \pi} \bigg) \; \frac{1}{2} \delta^{\mathcal{L}} R,\\
\label{dP and dE 2b} &= \sum_{I=1}^{g+1} \delta^{\mathcal{L}} \mathcal{S}_I
\int_{\mathcal{B}_I} \frac{dq_{\pm}}{2 \pi}.
\end{align}
\end{subequations}
\end{corollary}

\noindent It follows immediately from lemma \ref{lemma: dP and dE} that
\begin{equation*}
\{ H_{n, \pm}, \cdot \}_{\text{D.B.}} = \sum_{i=1}^g \left(
\int_{b_i} d\Omega_{n,\pm} \right) \{ S_i, \cdot \}_{\text{D.B.}} -
\bigg( \int_{\infty^-}^{\infty^+} d\Omega_{n,\pm} \bigg)
\frac{1}{2} \left\{ R, \cdot \right\}_{\text{D.B.}}.
\end{equation*}
Making use of the Lax matrix for the action variables \eqref{action
Lax} and the fact that the differentials $d\Omega_{n,\pm}$ are
normalised we can write the Lax matrix for $H_{n,\pm}$ as
\begin{multline*}
H_{n,\pm} \quad \longleftrightarrow\\
\frac{1}{4 \pi} \sum_{i=1}^g \left[ \int_{a_i} \frac{\Psi(x)
\sigma_3 \Psi(x)^{-1}}{x-x'} dx \int_{b_i} d\Omega_{n,\pm} -
\int_{b_i} \frac{\Psi(x) \sigma_3 \Psi(x)^{-1}}{x-x'} dx
\int_{a_i} d\Omega_{n,\pm} \right]\\ + \frac{1}{4 \pi}
\oint_{\infty} \frac{\Psi(x) \sigma_3 \Psi(x)^{-1}}{x-x'} dx
\int_{\infty^-}^{\infty^+} d\Omega_{n,\pm}.
\end{multline*}
Written in this form we can apply a Riemann bilinear
identity. Specifically we note that the Riemann bilinear identity
\eqref{dP and dE proof 1} and the equation following it in the
proof of lemma \ref{lemma: dP and dE} are valid for any differential
$\delta^{\mathcal{L}} \tilde{\alpha}$ which has simple poles at
$\infty^{\pm}$ of opposite residues there. But this is true of
$\frac{\Psi(x) \sigma_3 \Psi(x)^{-1}}{x-x'} dx$ which has simple poles
only at $x'^{\pm}, \infty^{\pm}$ since the poles of $\Psi(x) \sigma_3
\Psi(x)^{-1}$ at the branch points cancel with the zeroes of
$dx$. Furthermore, the residues at $\infty^{\pm}$ are opposite because
viewing $\Psi(x) = \left( \bm{\psi}(P), \bm{\psi}(\hat{\sigma} P)
\right)$ as a function $\Psi(P)$ on $\hat{\Sigma}$ we have
$\Psi(\hat{\sigma} P) = \Psi(P) \sigma_1$ and hence $\Psi(\hat{\sigma}
P) \sigma_3 \Psi(\hat{\sigma} P)^{-1} = - \Psi(P) \sigma_3
\Psi(P)^{-1}$. Therefore
\begin{equation} \label{energy momentum Lax}
H_{n,\pm} \quad \longleftrightarrow \quad - i
\Big( \text{res}_{x = 1} + \text{res}_{x = - 1} \Big)
\frac{\Psi(x) \sigma_3 \Psi(x)^{-1}}{x-x'} \Omega_{n,\pm}(x^+) dx,
\end{equation}
where an overall factor of two came from the fact that we get
equivalent contributions from both sheets, namely at $x^{\pm} =
(+1)^{\pm}$ and $x^{\pm} = (-1)^{\pm}$. Note also importantly that
there is no contribution from the apparent pole at $x = x'$ because
this is not actually a pole of the Lax equation itself. This can be
seen from \eqref{pre action Lax equation} which is perfectly regular
as $x$ approaches $x'$ since $[\partial_{t_{n,\pm}} - J_{n,\pm}(x'),
\Psi(x') \sigma_3 \Psi(x')^{-1}] = 0$ from \eqref{Psi evolution 2} and
the trivial fact that diagonal matrices commute. As already remarked
in section \ref{section: hierarchy}, an equation such as \eqref{energy
momentum Lax} relating an integral of motion to a Lax matrix should
really always be understood as a relation between two ingredients of a
Lax equation. To evaluate the residues in \eqref{energy momentum Lax}
we use the identity \eqref{res singular part identity} and the
asymptotics \eqref{dOmega asymptotics pm 1} of the differentials $d
\Omega_{n, \pm}$ at $x = \pm 1$. One finds
\begin{multline*}
H_{n, \pm} \quad \longleftrightarrow \quad \left(
\Psi(x') i \sigma_3 \Psi(x')^{-1} \Omega_{n,\pm}(x'^+) \right)_{\pm 1}\\
= \left( \Psi(x') s_{n,\pm}(x') \sigma_3 \Psi(x')^{-1} \right)_{\pm 1}
= J_{n,\pm}(x'),
\end{multline*}
which is exactly the expression \eqref{Lax matrix singular parts} for
the hierarchy of Lax matrices, \textit{c.f.} \eqref{Lax connection 7}.

It is important to note that it was the multi-valuedness of the
Abelian integral $\Omega_{n,\pm}(P) = \int^P d\Omega_{n,\pm}$ (or
equivalently the fact that $d\Omega_{n,\pm}$ had some non-trivial
periods) which resulted in a non-zero answer for the corresponding
Lax matrix. Indeed, the Lax matrix obtained by this argument
clearly depends only on the cohomology class $[d\Omega_{n,\pm}]
\in H^1(\hat{\Sigma}, \infty^{\pm})$ of the Abelian differential
$d\Omega_{n,\pm}$ one starts off with on the singular algebraic
curve $\hat{\Sigma} / \{ \infty^{\pm} \}$. One can see this explicitly
from the equation preceding \eqref{energy momentum Lax} or
otherwise from \eqref{energy momentum Lax} itself: suppose
$d\Omega_{n,\pm}, d\Omega'_{n,\pm}$ are two representatives of the
same cohomology class, then $d\Omega_{n,\pm} - d\Omega'_{n,\pm} =
df$ is exact with $f(\infty^+) = f(\infty^-)$ and the
corresponding difference of the expressions in \eqref{energy
momentum Lax} is
\begin{equation*}
- \frac{i}{2} \sum_{P \in \{ (\pm 1)^{\pm} \} } \text{res}_P
\frac{\Psi(P) \sigma_3 \Psi(P)^{-1}}{x(P) - x'} f(P) dx,
\end{equation*}
where $\Psi(P) = (\bm{\psi}(P),\bm{\psi}(\hat{\sigma} P))$. But this
is the sum over the residues of a well defined meromorphic
differential on $\hat{\Sigma} / \{ \infty^{\pm} \}$ (since $f(P)$ is
single-valued and the residues at $\infty^{\pm}$ cancel against each
other since $f(\infty^+) = f(\infty^-)$) and so is zero.

One could use the same trick as above to compute more explicitly
the Lax matrices for the action variables \eqref{action Lax}. To apply
the previous reasoning we write
\begin{equation*}
\delta \mathcal{S}_I = \sum_{J=1}^{g+1} \delta_{IJ} \delta
\mathcal{S}_J.
\end{equation*}
For the same argument to follow through we must introduce second kind
Abelian differentials $dq^{(J)}$ with specific periods
\begin{equation} \label{quasi-actions}
\int_{\mathcal{A}_I} dq^{(J)} = 0, \qquad \int_{\mathcal{B}_I}
dq^{(J)} = \delta_{IJ}.
\end{equation}
Such differentials exist: consider $g + 1$ independent differentials
$d \Omega_J$ from the hierarchy. Then $A_{IJ} = \int_{\mathcal{B}_I} d
\Omega_J$ is invertible and $dq^{(J)} = A^{-1}_{KJ} d\Omega_K$ have
the desired property. Yet since the conditions \eqref{quasi-actions}
on the differentials $dq^{(J)}$ uniquely specify their cohomology
class in $H^1(\hat{\Sigma}, \infty^{\pm})$, by the preceding remark
they are also sufficient to uniquely fix the resulting Lax matrix
\begin{equation*}
\mathcal{S}_I \quad \longleftrightarrow \quad \left( \Psi(x') i
\sigma_3 \Psi(x')^{-1} q^{(I)}(x'^+) \right)_{+1} + \left( \Psi(x') i
\sigma_3 \Psi(x')^{-1} q^{(I)}(x'^+) \right)_{-1}.
\end{equation*}
By the procedure of section \ref{section: B-A vector} (see in
particular the proof of lemma \ref{lemma: BA vector properties}) these
Lax matrices yield unique normalised Abelian differentials which
satisfy \eqref{quasi-actions}, which we still denote $dq^{(J)}$ by
abuse of notation. Since the operations of constructing a Lax matrix
from a given integral of motion and that of constructing an Abelian
differential from a given Lax matrix are both linear, it follows that
the equation for $H_{n.\pm}$ in \eqref{dP and dE b} translates into an
equation in terms of differential forms on $\hat{\Sigma} / \{
\infty^{\pm} \}$, namely
\begin{equation} \label{important formula dOmega}
d\Omega_{n,\pm} = \sum_{I=1}^{g+1} \left( \int_{\mathcal{B}_I}
d\Omega_{n,\pm} \right) dq^{(I)}.
\end{equation}
In particular at the $0^{\text{th}}$ level $n=0$ this equation
provides an important formula for the differential of the quasi-energy
that we will need in chapter \ref{chapter: semi},
\begin{equation} \label{important formula}
dq = \sum_{I=1}^{g+1} \left( \int_{\mathcal{B}_I} dq \right)
dq^{(I)}.
\end{equation}

%% file: Reality.tex
\newpage

\chapter{Real closed strings} \label{chapter: reality}

\begin{flushright}
{\small \it ``Reality continues to ruin my life.''}\\
{\small Calvin, {\it Calvin and Hobbes}}
\end{flushright}
\vspace{1cm}

\noindent The method of finite-gap integration described in
chapters \ref{chapter: curves} and \ref{chapter: finite-gap}
heavily relied on complex analysis and the theory of Riemann
surfaces. This was to make use of powerful theorems such as the
Riemann-Roch theorem to reconstruct solutions. All solutions
obtained by this method are build out of a combination of
meromorphic and Baker-Akhiezer functions $\hat{\Sigma} \rightarrow
\mathbb{C}$ from a Riemann surface $\hat{\Sigma}$ into
$\mathbb{C}$. In particular the phase-space coordinate of the
string $j$ reconstructed in theorem \ref{thm: reconstructing j} is
$\mathfrak{sl}(2, \mathbb{C})$-valued and the corresponding
embedding $g$ of proposition \ref{thm: reconstructing g} is
$SL(2,\mathbb{C})$-valued. However the differential equations we
set out to solve were all equations for physical strings whose
embedding into the target space $\mathbb{R} \times S^3$ is
described by an $SU(2)$-valued map $g(\sigma, \tau) \in SU(2)$.
Furthermore, the closed string boundary conditions require these
embeddings to be $2 \pi$-periodic in $\sigma$. It is therefore
important to identify the subset of solutions among all those
constructed by the finite-gap method which are both real
(\textit{i.e.} $SU(2)$-valued) and periodic in $\sigma$.

The way to obtain real periodic solutions will simply be to
restrict the allowed algebro-geometric data. In the language of
chapter \ref{chapter: symplectic} a genus $g$ finite-gap solution
is a geometric map \eqref{geometric map} from the $2g + 2$ complex
dimensional Jacobian bundle $\mathcal{M}^{(2g+2)}_{\mathbb{C}}$
into the space of complexified solutions
$\mathcal{S}^{\infty}_{\mathbb{C}}$. The restriction to \dub{real}
algebro-geometric data giving rise to \dub{real} solutions through
the geometric map \eqref{geometric map} can be identified with a
sub-bundle $\mathcal{M}^{(2g+2)}_{\mathbb{R}}$ of
$\mathcal{M}^{(2g+2)}_{\mathbb{C}}$. As we will see, the real
slice of the generalised Jacobian $J_{\mathfrak{m}}(\hat{\Sigma})$
is simply a real $(g+1)$-torus $\mathbb{T}^{g+1} = S^1 \times
\ldots \times S^1$ (with $g+1$ factors of $S^1$) and the real part
$\mathcal{L}_{\mathbb{R}}$ of the leaf $\mathcal{L}$ is
parametrised by real values of the filling fractions,
\begin{equation*}
\mathbb{T}^{g+1} \rightarrow \mathcal{M}^{(2g+2)}_{\mathbb{R}}
\rightarrow \mathcal{L}_{\mathbb{R}}.
\end{equation*}
The restriction $\mathcal{G}'_{\mathbb{R}} =
\mathcal{G}'|_{\mathcal{M}^{(2g+2)}_{\mathbb{R}}}$ of the geometric
map to the real bundle $\mathcal{M}^{(2g+2)}_{\mathbb{R}}$ is an
injective map from \textit{real} algebro-geometric data to the space
of \textit{real} solutions $\mathcal{S}^{\infty}_{\mathbb{R}}$,
\begin{equation} \label{geometric map real}
\begin{split}
\xymatrix{ \mathcal{M}^{(2g+2)}_{\mathbb{C}} \ar[rr]^{\mathcal{G}'} & &
\mathcal{S}^{\infty}_{\mathbb{C}}\\
\mathcal{M}^{(2g+2)}_{\mathbb{R}} \ar@{^{(}->}[u]
\ar[rr]^{\mathcal{G}'_{\mathbb{R}}} & &
\mathcal{S}^{\infty}_{\mathbb{R}} \ar@{^{(}->}[u] }
\end{split}
\end{equation}
By further restricting the the real geometric map
$\mathcal{G}'_{\mathbb{R}}$ to a sub-bundle of the real
algebro-geometric data $\mathcal{M}^{(2g+2)}_{\mathbb{R}}$
corresponding to data satisfying certain periodicity conditions,
its image will consist of real periodic solutions.


\section{Real curves} \label{section: real curves}


To identify the restrictions imposed by the reality conditions on the
various curves we go back to their respective definitions in chapter
\ref{chapter: curves}.

\subsection*{The spectral curve}

The spectral curve is defined by equation \eqref{spectral curve 1} in
terms of the monodromy matrix which was defined as the path-ordered
exponential of the Lax connection,
\begin{equation*}
\Omega(x,\sigma,\tau) = P \overleftarrow{\exp} \int_{\sigma}^{\sigma +
2 \pi} J(x), \qquad J(x) = \frac{1}{1 - x^2} (j - x \ast j).
\end{equation*}
Now the requirement that the current $j \in \mathfrak{su}(2)$ is
equivalent to $j^{\dag} = -j$ which implies reality conditions on
$J(x)$ and $\Omega(x)$ in turn. Specifically, for $J(x)$ we have
\begin{equation*}
J(x)^{\dag} = \frac{1}{1 - \bar{x}^2} (j^{\dag} - \bar{x} \ast
j^{\dag}) = - \frac{1}{1 - \bar{x}^2} (j - \bar{x} \ast j) = - J(\bar{x}).
\end{equation*}
This implies the following reality conditions on $\Omega(x)$
\begin{multline} \label{reality monodromy}
\Omega(x)^{\dag} = \left( P \overleftarrow{\exp} \int_{\sigma}^{\sigma
+ 2 \pi} J(x) \right)^{\dag} = P \overrightarrow{\exp}
\int_{\sigma}^{\sigma + 2 \pi} J(x)^{\dag}\\ = P \overrightarrow{\exp}
\int_{\sigma}^{\sigma + 2 \pi} -J(\bar{x}) = \left( P
\overleftarrow{\exp} \int_{\sigma}^{\sigma + 2 \pi} J(\bar{x})
\right)^{-1} = \Omega(\bar{x})^{-1}.
\end{multline}
In particular, for \textit{real} values of $x \in \mathbb{R}$ we have
$J(x) \in \mathfrak{su}(2)$ and $\Omega(x) \in SU(2)$.

\begin{definition}
A curve $C$ in $\mathbb{C}^2$ is \dub{real} if it admits an
anti-holomorphic involution $\hat{\tau} : C \rightarrow C$. That is,
$\hat{\tau}^2 = 1$ and for any function $f$ on $C$ holomorphic in a
neighbourhood $U \subset C$ the function $f \circ \hat{\tau}$ is
anti-holomorphic in $\hat{\tau}(U)$.
\end{definition}

A simple example of a real curve is the complex plane $\mathbb{C}$
itself, which obviously admits complex conjugation $\hat{\tau} : x
\rightarrow \bar{x}$ as an anti-holomorphic involution. In particular
a real curve is still `complex' in the sense that it locally looks
like $\mathbb{C}$.

\begin{lemma} \label{lemma: real spectral curve}
The spectral curve $\Gamma$ is real with anti-holomorphic involution
\begin{equation} \label{anti-holomorphic involution}
\hat{\tau} : \;\; \Gamma \rightarrow \Gamma, \quad (x,
\Lambda) \mapsto (\bar{x}, \bar{\Lambda}^{-1}).
\end{equation}
\begin{proof}
Let $(x, \Lambda) \in \Gamma$ then by definition $\det (\Lambda {\bf
1} - \Omega(x)) = 0$. Taking the complex conjugate and using
\eqref{reality monodromy} yields $\det (\bar{\Lambda} {\bf 1} -
\Omega(\bar{x})^{-1}) = 0$. Then provided $\Lambda \neq 0$ and since
$\det \Omega(x) = 1 \neq 0$ we have $\det (\Omega(\bar{x}) -
\bar{\Lambda}^{-1} {\bf 1}) = 0$ which means that $(\bar{x},
\bar{\Lambda}^{-1}) \in \Gamma$. Therefore the map $(x, \Lambda)
\mapsto (\bar{x}, \bar{\Lambda}^{-1})$ sends the curve $\Gamma$ to
itself. Moreover it clearly squares to one and is anti-holomorphic.
\end{proof}
\end{lemma}

This anti-holomorphic involution can be combined with the holomorphic
involution $\hat{\sigma}$ defined in \eqref{holomorphic
involution}. Together they generate a $\mathbb{Z}_2 \times
\mathbb{Z}_2$ group of involutions on $\Gamma$ such that $\hat{\sigma}
\hat{\tau} = \hat{\tau} \hat{\sigma}$. Recall that each point $x_0 \in
\mathcal{Z}_{\Gamma}$ corresponding to degenerate eigenvalues of the
monodromy matrix was a fixed point of $\hat{\sigma}$. A slightly
weaker statement is true for $\hat{\tau}$,

\begin{lemma} \label{lemma: disciminant tau}
The discriminant $\Delta_{\Gamma}(x)$ satisfies
$\overline{\Delta_{\Gamma}(x)} = \Delta_{\Gamma}(\bar{x})$. In
particular, its set of zeroes $\mathcal{Z}_{\Gamma}$ is invariant
under $\hat{\tau}$.
\begin{proof}
Consider $\Delta_{\Gamma}(\bar{x}) = (\Lambda_+(\bar{x}) -
\Lambda_-(\bar{x}))^2$. By lemma \ref{lemma: real spectral curve} the
eigenvalues $\Lambda_{\pm}(\bar{x})$ above $\bar{x}$ can equally be
written $\overline{\Lambda_{\pm}(x)}^{-1}$. Thus
$\Delta_{\Gamma}(\bar{x}) = \left( \overline{\Lambda_+(x)}^{-1}
- \overline{\Lambda_-(x)}^{-1} \right)^2$ which can be rewritten as
$\overline{\Delta_{\Gamma}(x)}$ using \eqref{unimodular evalues}.
Hence $x_0 \in \mathcal{Z}_{\Gamma} \Leftrightarrow \bar{x}_0
\in \mathcal{Z}_{\Gamma}$.
\end{proof}
\end{lemma}

Recall from chapter \ref{chapter: curves} that the points $x_0 \in
\mathcal{Z}_{\Gamma}$ fall into one of two categories:
\begin{itemize}
  \item[$(1)$] $\Delta_{\Gamma}(x) = O(x - x_0)^{2r+1}$, \textit{i.e.}
$x_0$ is a branch point or cusp-like singularity.
  \item[$(2)$] $\Delta_{\Gamma}(x) = O(x - x_0)^{2r}$, \textit{i.e.}
$x_0$ is a node-like singularity.
\end{itemize}
It follows from lemma \ref{lemma: disciminant tau} that the order of
the zero is also preserved under the action of $\hat{\tau}$. Thus
branch points are mapped to branch points, cusps to cusps, and so on.
The next lemma shows that branch points and cusp-like singularities
must all lie off the real axis.

\begin{lemma} \label{lemma: vertical cuts}
If $x_0 \in \mathcal{Z}_{\Gamma} \cap \mathbb{R}$ then $x_0$
corresponds to a node-like singularity.
\begin{proof}
Let $\mathfrak{P}_0 = (x_0, \Lambda_0) \in \Gamma$ with $x_0 \in
\mathcal{Z}_{\Gamma}$. We assume $\hat{\tau} \mathfrak{P}_0 =
\mathfrak{P}_0$ ($\Leftrightarrow x_0 \in \mathbb{R}$) and show
$\mathfrak{P}_0$ is node-like. Since $x_0 \in \mathbb{R}$, we can
write $\Omega(x_0) \in SU(2)$ as
\begin{equation*}
\Omega(x_0) = \left( \begin{array}{cc} \mathcal{A}(x_0) &
\mathcal{B}(x_0) \\ - \overline{\mathcal{B}(x_0)} &
\overline{\mathcal{A}(x_0)} \end{array} \right).
\end{equation*}
But then $|\mathcal{A}(x_0) - \Lambda_0|^2 + |\mathcal{B}(x_0)|^2 = 0$
since $\Lambda_0 = \pm 1 \in \mathbb{R}$. Hence $\mathcal{A}(x_0) = \Lambda_0$
and $\mathcal{B}(x_0) = 0$ which implies that $\Omega(x_0)$ is
diagonal. Therefore $\dim \mathcal{E}_{\Gamma}(\mathfrak{P}_0) = 2$
and $\mathfrak{P}_0$ must be node-like by proposition
\ref{proposition: unique eigenvector}.
\end{proof}
\end{lemma}

Recall that the involution $\hat{\sigma}$ had the effect of
interchanging the two sheets of the spectral curve $\Gamma$. We wish
to similarly describe the effect of $\hat{\tau}$ on the individual
sheets. Since the branch points and cusp-like singularities all come
in complex conjugate pairs by lemma \ref{lemma: vertical cuts} we
choose the cuts in the complex $x$-plane to be vertical, connecting a
branch point with its reflection through the real axis (see figure
\ref{fig: real canonical cycles}). With this choice the set of cuts is
invariant under $\hat{\tau}$ which allows us to describe its effect as
follows,

\begin{lemma} \label{lemma: involution sheets}
The involution $\hat{\tau}$ maps both sheets to themselves by $x
\mapsto \bar{x}$.
\begin{proof}
Consider the points $(x,\Lambda_{\pm}(x)) \in \Gamma$ above $x \in
\mathbb{C}$ on the upper and lower sheet of the spectral curve. When
$x \in \mathbb{R}$ we have $\Omega(x) \in SU(2)$ so that
$|\Lambda_{\pm}(x)| = 1$ and hence $\overline{\Lambda_{\pm}(x)}^{-1} =
\Lambda_{\pm}(x)$. Now let $x \in \mathbb{C}$. Then by equation
\eqref{reality monodromy} the eigenvalues $\{ \Lambda_{\pm}(\bar{x})
\}$ of $\Omega(\bar{x})$ can equally be written $\left\{
\overline{\Lambda_{\pm}(x)}^{-1} \right\}$. Therefore by continuity,
the equality $\overline{\Lambda_{\pm}(x)}^{-1} =
\Lambda_{\pm}(\bar{x})$ which holds for $x \in \mathbb{R}$ must also
hold for all $x$ in the cut plane. It follows that
\begin{equation*}
\hat{\tau}(x, \Lambda_{\pm}(x)) = \left(\bar{x},
\overline{\Lambda_{\pm}(x)}^{-1} \right) = (\bar{x},
\Lambda_{\pm}(\bar{x})).
\end{equation*}
In other words, the point above $x$ on the upper (respectively lower)
sheet is mapped by $\hat{\tau}$ to the point above $\bar{x}$ on the
upper (respectively lower) sheet.
\end{proof}
\end{lemma}

\subsection*{The algebraic curve}

The algebraic curve is defined by equation \eqref{algebraic curve 1} in
terms of a chosen combination of Lax matrices $L(x) = \sum_N c_N
J_N(x)$ where $J_N(x) = \left( \Psi(x) s_N(x) \sigma_3 \Psi(x)^{-1}
\right)_{\pm 1}$ and $s_N(x)$ are the singular parts defined in
\eqref{singular parts}. Since $\Psi(x)$ is the matrix of eigenvectors
of $\Omega(x)$ it satisfies $\Omega(x) \Psi(x) = \Psi(x)
\diag(\Lambda_+(x), \Lambda_-(x))$. Taking the hermitian conjugate
followed by the inverse yields $\Omega(\bar{x}) (\Psi(x)^{\dag})^{-1}
= (\Psi(x)^{\dag})^{-1} \diag(\Lambda_+(\bar{x}), \Lambda_-(\bar{x}))$
from which it follows that $\Psi(\bar{x}) = (\Psi(x)^{\dag})^{-1} D$
for some diagonal matrix $D$. It follows that $(\Psi(x)^{-1})^{\dag} =
\Psi(\bar{x}) D^{-1}$ and $\Psi(x)^{\dag} = D \Psi(\bar{x})^{-1}$
which combined with the fact that $\overline{s_N(x)} = - s_N(\bar{x})$
and the assumption that $c_N \in \mathbb{R}$ gives
\begin{equation} \label{reality Lax matrix}
L(x)^{\dag} = - L(\bar{x}).
\end{equation}
The anti-holomorphic involution \eqref{anti-holomorphic involution} of
the spectral curve $\Gamma$ induces an anti-holomorphic involution
(also denoted $\hat{\tau}$) on the algebraic curve $\Sigma$ which is
easily obtained using \eqref{reality Lax matrix}.

\begin{lemma} \label{lemma: real algebraic curve}
The algebraic curve $\Sigma$ is real with anti-holomorphic involution
\begin{equation} \label{anti-holomorphic involution 2}
\hat{\tau} : \;\; \Sigma \rightarrow \Sigma, \quad (x, y) \mapsto
(\bar{x}, -\bar{y}).
\end{equation}
\begin{proof}
Let $(x, y) \in \Sigma$ then by definition $\det (y {\bf
1} - L(x)) = 0$. Taking the complex conjugate and using
\eqref{reality Lax matrix} yields $\det (\bar{y} {\bf 1} +
L(\bar{x})) = 0$ so that $(\bar{x}, - \bar{y}) \in \Sigma$.
\end{proof}
\end{lemma}

\begin{remark}
It is straightforward to check that the statements of the lemmas
\ref{lemma: disciminant tau}, \ref{lemma: vertical cuts} and
\ref{lemma: involution sheets} equally apply to the algebraic
curve $\Sigma$ with the involution \eqref{anti-holomorphic
involution 2}. This is to be expected since $\Sigma$ is a
(partial) normalisation of $\Gamma$. The proofs of the lemmas for
$\Sigma$ are essentially the same as those for $\Gamma$ so we do
not repeat them. We simply note that when $x \in \mathbb{R}$ the
reality condition \eqref{reality Lax matrix} says that $L(x) \in
\mathfrak{su}(2)$ which can be used to prove the analogues of
lemmas \ref{lemma: vertical cuts} and \ref{lemma: involution
sheets}.
\end{remark}

\subsection*{The Riemann surface}

The Riemann surface $\hat{\Sigma}$ was defined by equation
\eqref{algebraic curve 4}. Because it is merely the normalisation of
$\Sigma$, the involution $\hat{\tau} : \Sigma \rightarrow \Sigma$
naturally induces an anti-holomorphic involution on the Riemann
surface $\hat{\Sigma}$ defined by exactly the same formula,
\begin{equation*}
\hat{\tau} : \;\; \hat{\Sigma} \rightarrow \hat{\Sigma}, \quad (x, y)
\mapsto (\bar{x}, -\bar{y}).
\end{equation*}
Since the full set of branch points $\{ u_I, v_I \}_{I=1}^{g+1}$
of the Riemann surface $\hat{\Sigma}$ must be invariant under
$\hat{\tau}$ by lemma \ref{lemma: disciminant tau} and none of
them can be real by lemma \ref{lemma: vertical cuts}, the only
possibility is that they form complex conjugate pairs. We can
therefore set $v_I = \bar{u}_I$ in \eqref{algebraic curve 4} so
that the general \dub{real} Riemann surface takes the form,
\begin{equation} \label{real RS}
\hat{\Sigma} : \;\; y^2 = \prod_{i = 1}^{g + 1} (x - u_i)(x -
\bar{u}_i).
\end{equation}

\begin{figure}[ht]
\centering \psfrag{a1}{\footnotesize \red $a_1$}
\psfrag{a2}{\footnotesize \red $a_2$} \psfrag{ag}{\footnotesize
\red $a_g$} \psfrag{b1}{\footnotesize \green $b_1$}
\psfrag{b2}{\footnotesize \green $b_2$} \psfrag{bg}{\footnotesize
\green $b_g$} \psfrag{d}{\footnotesize $\cdots$}
\includegraphics[height=50mm]{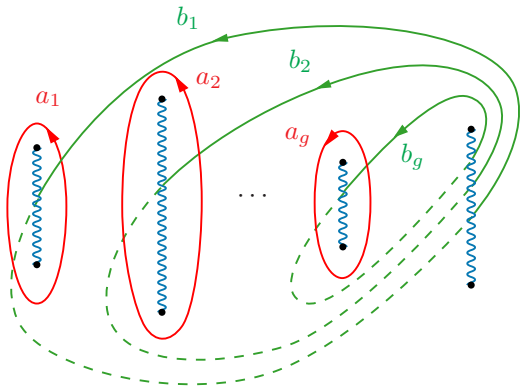}
\caption{Canonical choice of $\bm{a}$- and $\bm{b}$-cycles for a
real curve.} \label{fig: real canonical cycles}
\end{figure}

In order to specify the quasi-momentum which is normalised with
respect to the $\bm{a}$-cycles we must choose a canonical basis of $\bm{a}$- and
$\bm{b}$-cycles. As in section \ref{section:
quasimomentum} we will choose the $\bm{a}$-cycles to encircle $g$ of
the cuts. As for the canonically conjugate $\bm{b}$-cycles, in the
case of a real curve \eqref{real RS} it is convenient to choose them
as shown in figure \ref{fig: real canonical cycles}. The homology
classes of these basis cycles are easily shown to have the following
properties under the action of the anti-holomorphic involution
$\hat{\tau}$
\begin{equation} \label{reality: a b cycles}
\hat{\tau} a_i \sim - a_i, \qquad \hat{\tau} b_i \sim b_i + a_i +
\sum_{j=1}^{g+1} a_j,
\end{equation}
where $\sim$ denotes homology equivalence so that these expressions
are to be understood modulo cycles homologous to zero. In particular
the $\bm{a}$-cycles are pure imaginary.
The reality conditions \eqref{reality: a b cycles} on the
basis homology cycles then induce reality conditions on meromorphic
differentials. For instance the vector $\vec{\omega} = (\omega_1,
\ldots, \omega_g, \omega_{\infty})^{\sf T}$, made up of the
holomorphic differentials $\{ \omega_i \}_{i=1}^g$ and the third kind
Abelian differential $\omega_{\infty}$ defined in \eqref{third kind
diff at infty}, has the following transformation under pullback by
$\hat{\tau}$. These are to be compared with the transformation
property under pullback by $\hat{\sigma}$ which reads
$\hat{\sigma}^{\ast} \vec{\omega} = - \vec{\omega}$.

\begin{lemma} \label{lemma: omega imaginary}
$\overline{\hat{\tau}^{\ast} \vec{\omega}} = - \vec{\omega}$.
\begin{proof}
The differentials $\overline{\hat{\tau}^{\ast} \omega_i}$ are
holomorphic. Indeed, the holomorphic differentials $\omega_i$ can
locally be written as $df_i$ for some holomorphic $f_i$. Then
locally we have $\overline{\hat{\tau}^{\ast} \omega_i} = d \left(
\overline{f_i \circ \hat{\tau}} \right)$, where $\overline{f_i
\circ \hat{\tau}}$ are holomorphic. Furthermore,
\begin{equation*}
\int_{a_i} \overline{\hat{\tau}^{\ast} \omega_j} =
\overline{\int_{a_i} \hat{\tau}^{\ast} \omega_j} =
\overline{\int_{\hat{\tau} a_i} \omega_j} = - \overline{\int_{a_i}
\omega_j} = - \delta_{ij}.
\end{equation*}
Therefore by lemma \ref{lemma: unique holo basis} we have
$-\overline{\hat{\tau}^{\ast} \omega_i} = \omega_i$. As for the
third kind differential $\omega_{\infty}$, since its residues are
pure imaginary it is clear that $-\overline{\hat{\tau}^{\ast}
\omega_{\infty}}$ (which is also an Abelian differential of the
third kind) has the same residues at the poles $\infty^{\pm}$ and
is still normalised because
\begin{equation*}
\int_{a_i} \overline{\hat{\tau}^{\ast} \omega_{\infty}} =
\overline{\int_{a_i} \hat{\tau}^{\ast} \omega_{\infty}} =
\overline{\int_{\hat{\tau} a_i} \omega_{\infty}} = -
\overline{\int_{a_i} \omega_{\infty}} = 0.
\end{equation*}
It then follows by lemma \ref{lemma: unique normalised} that
$-\overline{\hat{\tau}^{\ast} \omega_{\infty}} = \omega_{\infty}$.
\end{proof}
\end{lemma}

\begin{corollary} \label{cor: real divisor}
Let $D \geq 0$ be an integral divisor. Then
$\vec{\mathcal{A}}(\hat{\tau} D) = -\overline{\vec{\mathcal{A}}(D)}$.
\begin{proof}
Let $D = P_1 + \ldots + P_n$ be an integral divisor of degree $n$.
Then
\begin{equation*}
\vec{\mathcal{A}}(\hat{\tau} D) = 2 \pi \sum_{i=1}^n
\int_{\infty^+}^{\hat{\tau} P_i} \vec{\omega} = 2 \pi
\sum_{i=1}^n \int_{\infty^+}^{P_i} \hat{\tau}^{\ast}
\vec{\omega} = - 2 \pi \sum_{i=1}^n
\int_{\infty^+}^{P_i} \overline{\vec{\omega}} =
-\overline{\vec{\mathcal{A}}(D)},
\end{equation*}
where the second equality holds on $J_{\mathfrak{m}}(\hat{\Sigma})$
since the path $[\hat{\tau} P_i, \infty^+]$ is only equal to the path
$\hat{\tau}[P_i, \infty^+]$ modulo $\bm{a}$- and $\bm{b}$-cycles.
\end{proof}
\end{corollary}

The reality condition on the homology basis also induces reality
conditions on the differential $d\mathcal{Q}$. Since this
differential generates the linear flow on the generalised Jacobian
$J_{\mathfrak{m}}(\hat{\Sigma})$ by theorem \ref{thm: linear
motion}, the reality conditions on $d\mathcal{Q}$ immediately
allow us to conclude that the motion of the dynamical divisor
$\hat{\gamma}(t)$ on $J_{\mathfrak{m}}(\hat{\Sigma})$ is
restricted to a real slice in the sense of the following lemma. In
the next section we will describe the real slice of the Jacobian
in more detail by discussing the reality conditions on the
dynamical divisor itself.
\begin{lemma} \label{lemma: real quasi-mom}
$\overline{\hat{\tau}^{\ast} d \mathcal{Q}} = d \mathcal{Q}$. In
particular $\int_{\vec{b}} d \mathcal{Q} \in \mathbb{R}^{g+1}$.
\begin{proof}
$\overline{\hat{\tau}^{\ast} d \mathcal{Q}}$ is a second kind Abelian
differential with the same singular parts as $d \mathcal{Q}$, so the
result follows by lemma \ref{lemma: unique normalised}. Moreover,
$\overline{\int_{\vec{b}} d \mathcal{Q}} = \int_{\vec{b}}
\hat{\tau}^{\ast} d \mathcal{Q} = \int_{\hat{\tau} \vec{b}} d
\mathcal{Q} = \int_{\vec{b}} d \mathcal{Q}$ using the fact that $d
\mathcal{Q}$ is normalised together with the reality conditions
\eqref{reality: a b cycles} on the $\bm{b}$-cycles and the fact that
$\hat{\tau} b_{\infty}$ is $b_{\infty}$ plus a linear combination of
$\bm{a}$-cycles.
\end{proof}
\end{lemma}

We can also show that the reality conditions restrict the base
$\mathcal{L}$ of the Jacobian bundle
$\mathcal{M}^{(2g+2)}_{\mathbb{C}}$ to a sub-leaf
$\mathcal{L}_{\mathbb{R}}$ as advertised at the start of this
chapter. The next lemma shows that $\mathcal{L}_{\mathbb{R}}
\subset \mathcal{L}$ is parameterised by \textit{real} filling
fractions.

\begin{lemma}
The filling fractions are real, namely $\mathcal{S}_I \in \mathbb{R}$,
$I = 1, \ldots, g+1$.
\begin{proof}
The differential $\alpha = \frac{\sqrt{\lambda}}{4 \pi} z dp$ is real
by lemma \ref{lemma: real quasi-mom}, that is one can show
$\overline{\hat{\tau}^{\ast} \alpha} = \alpha$. Hence
\begin{equation*}
\overline{\mathcal{S}_I} = - \frac{1}{2 \pi i} \int_{\mathcal{A}_I}
\bar{\alpha} = -\frac{1}{2 \pi i} \int_{\mathcal{A}_I}
\hat{\tau}^{\ast} \alpha =  -\frac{1}{2 \pi i} \int_{\hat{\tau}
\mathcal{A}_I} \alpha = \frac{1}{2 \pi i} \int_{\mathcal{A}_I} \alpha
= \mathcal{S}_I,
\end{equation*}
using the fact that $\hat{\tau} \mathcal{A}_I = - \mathcal{A}_I$, $I =
1, \ldots, g+1$.
\end{proof}
\end{lemma}


\section{Real divisor} \label{section: real divisor}


Let the dual dynamical divisor $\hat{\gamma}^+(t)$ be the poles of the
dual normalised eigenvector $\bm{h}^+$ satisfying,
\begin{equation} \label{reality: dual ev}
\bm{h}^+(P) \left( \Omega(x) - \Lambda(P) {\bf 1} \right) = 0, \qquad
\bm{h}^+(P) \cdot \bm{\alpha}^{\sf T} = 1.
\end{equation}
It is related in a very simple way to the dynamical divisor
$\hat{\gamma}(t)$ which was defined as the poles of the ordinary
normalised eigenvector $\bm{h}$ satisfying,
\begin{equation} \label{reality: ev}
\left( \Omega(x) - \Lambda(P) {\bf 1} \right) \bm{h}(P) = 0, \qquad
\bm{\alpha} \cdot \bm{h}(P) = 1.
\end{equation}

\begin{lemma} \label{lemma: real divisor}
The reality condition on the dynamical divisor is $\hat{\gamma}^+(t)
= \hat{\tau} \hat{\gamma}(t)$.
\begin{proof}
Taking the dual normalised eigenvector equation \eqref{reality: dual
ev} at the point $\hat{\tau} P$ it can be rewritten as
\begin{equation} \label{reality: dual ev 2}
\bm{h}^+(\hat{\tau} P) \Lambda(\hat{\tau} P)^{-1} -
\bm{h}^+(\hat{\tau} P) \Omega(\bar{x})^{-1} = 0, \qquad
\bm{h}^+(\hat{\tau} P) \cdot \bm{\alpha}^{\sf T} = 1.
\end{equation}
Using the reality conditions on the monodromy matrix we have
$\Omega(\bar{x})^{\dag} \cdot \Omega(x) = 1$. Then
$\bm{h}(\hat{\tau}(P))^{\dag} \cdot \bm{h}(P) =
\bm{h}(\hat{\tau}(P))^{\dag} \Omega(\bar{x})^{\dag} \cdot
\Omega(x) \bm{h}(P) = \left( \Omega(\bar{x}) \bm{h}(\hat{\tau}(P))
\right)^{\dag} \cdot \Omega(x) \bm{h}(P) =
\overline{\Lambda(\hat{\tau} P)} \Lambda(P)
\bm{h}(\hat{\tau}(P))^{\dag} \cdot \bm{h}(P)$ so that
$\Lambda(\hat{\tau} P)^{-1} = \overline{\Lambda(P)}$. Taking the
hermitian conjugate of \eqref{reality: dual ev 2} it can now be
rewritten as
\begin{equation} \label{reality: dual ev 3}
\left( \Omega(x) - \Lambda(P) {\bf 1} \right) \left(
\bm{h}^+(\hat{\tau} P) \right)^{\dag} = 0, \qquad \bm{\alpha} \cdot
\left( \bm{h}^+(\hat{\tau} P) \right)^{\dag} = 1.
\end{equation}
Now by proposition \ref{proposition: unique eigenvector} there is a
unique solution to \eqref{reality: ev} and hence \eqref{reality:
dual ev 3} implies that $\bm{h}^+(P) =
\bm{h}(\hat{\tau}P)^{\dag}$. These vectors have poles at
$\hat{\gamma}^+(t)$ and $\hat{\tau} \hat{\gamma}(t)$ respectively, so
the result follows.
\end{proof}
\end{lemma}

\begin{remark}
Recall that in chapter \ref{chapter: finite-gap} we could
characterise the equivalence class $[\hat{\gamma}^+(t)]$ of the
dual dynamical divisor using lemma \ref{lemma: dual dyn
characterisation}, although quite implicitly. The divisor
$\hat{\gamma}^+(t)$ then had to be chosen arbitrarily from the
class $[\hat{\gamma}^+(t)]$ and the reconstruction of the inverse
matrix in proposition \ref{prop: rows of H^-1} required a residual
gauge transformation because the corresponding normalised
eigenvector $\tilde{\bm{h}}^+(P)$ was expressed in the `wrong'
residual gauge. Here lemma \ref{lemma: real divisor} says that the
dual dynamical divisor $\hat{\gamma}^+(t)$ corresponding to the
poles of the dual normalised eigenvector $\bm{h}^+(P)$ in the
`correct' residual gauge can be immediately obtained from the
dynamical divisor $\hat{\gamma}(t)$ by complex conjugation. This
avoids the worry of having to correct the residual gauge in
reconstructing the inverse matrix as was done in proposition
\ref{prop: rows of H^-1} since one can simply use proposition
\ref{left inverse of H} if the correct dual dynamical divisor is
known.
\end{remark}

\begin{corollary}
Real divisors satisfy $2 \text{Im } \vec{\mathcal{A}}(\hat{\gamma}(t))
= \vec{\mathcal{A}}(B)$.
\begin{proof}
Combining equation \eqref{Abel dyn + dyn^+} of lemma \ref{lemma: dual
dyn characterisation} with lemma \ref{lemma: real divisor} and corollary
\ref{cor: real divisor} we have $\vec{\mathcal{A}}(B) =
\vec{\mathcal{A}}(\hat{\gamma}(t)) + \vec{\mathcal{A}}(\hat{\tau}
\hat{\gamma}(t)) = \vec{\mathcal{A}}(\hat{\gamma}(t)) -
\overline{\vec{\mathcal{A}}(\hat{\gamma}(t))} = 2 \text{Im }
\vec{\mathcal{A}}(\hat{\gamma}(t))$.
\end{proof}
\end{corollary}

Although the generalised Jacobian $J_{\mathfrak{m}}(\hat{\Sigma})$ is
a non-compact Abelian group it turns out that its real slice,
\textit{i.e.} the generalised Abel map of real divisors in lemma
\ref{lemma: real divisor}, is a real $(g+1)$-dimensional torus.

\begin{corollary} \label{cor: real Jacobian}
The real slice of $J_{\mathfrak{m}}(\hat{\Sigma})$ is a $(g+1)$-torus
$\mathbb{T}^{g+1}$ given explicitly by
\begin{equation} \label{Jacobian real slice}
\{ \vec{X} \in J_{\mathfrak{m}}(\hat{\Sigma}) \; |
\; 2 \text{Im } \vec{X} = \vec{\mathcal{A}}(B) \} \subset
J_{\mathfrak{m}}(\hat{\Sigma}).
\end{equation}
It is a translation of the real torus $\mathbb{R}^{g+1}/2 \pi
\mathbb{Z}^{g+1} \subset J_{\mathfrak{m}}(\hat{\Sigma})$ by the
vector $\vec{X}_0 \equiv \frac{1}{2} \vec{\mathcal{A}}(B)$.
\begin{proof}
Recall that the generalised Jacobian
$J_{\mathfrak{m}}(\hat{\Sigma})$ is the quotient of
$\mathbb{C}^{g+1}$ by the lattice $\Lambda_{\mathfrak{m}}$ spanned
by $2 \pi$ multiples of the $2 g + 1$ linearly independent vectors
(over $\mathbb{R}$)
\begin{equation} \label{gen Jac lattice}
\left( \begin{array}{cc} \delta_{ij} & 0 \\ 0 & 1 \end{array}, \;
\begin{array}{c} \Pi_{ij} \\ \Pi_j \end{array} \right),
\end{equation}
where $\Pi_j = \int_{b_j} \omega_{\infty}$ and $\Pi_{ij} =
\int_{b_j} \omega_i$. Since $\text{Im } \Pi_{ij}$ is positive
definite by lemma \ref{lemma: imaginary Pi} the last $g$ column
vectors in \eqref{gen Jac lattice} have a non-zero imaginary part.
Now any vector $\vec{X}$ belonging to \eqref{Jacobian real slice}
can be written as $\vec{X} = \vec{X_0} + \vec{V}$ where $\text{Im
} \vec{V} = 0$. Hence $\vec{V}$ must be a linear combination of
the first $g+1$ real columns in \eqref{gen Jac lattice} which span
the real torus $\mathbb{R}^{g+1}/2 \pi \mathbb{Z}^{g+1} \subset
J_{\mathfrak{m}}(\hat{\Sigma})$. The real slice \eqref{Jacobian
real slice} is a translation by $\vec{X}_0$.
\end{proof}
\end{corollary}


\section{Periodicity} \label{section: periodicity}


For closed strings the embedding field $g(\sigma, \tau)$ is periodic
under $\sigma \rightarrow \sigma + 2 \pi$. And just as for the reality
conditions on $g(\sigma, \tau)$, this periodicity condition imposes
restrictions on the allowed algebro-geometric data.

Because the configuration of a finite-gap string is specified by the
position of the point $\vec{\mathcal{A}}(\hat{\gamma}(t)) \in
J_{\mathfrak{m}}(\hat{\Sigma})$ on the generalised Jacobian, a
necessary condition for the string to be closed is that the motion
of this point be $\sigma$-periodic on
$J_{\mathfrak{m}}(\hat{\Sigma})$. Yet we know from theorem \ref{thm:
linear motion} that the motion of this point is linear on
$J_{\mathfrak{m}}(\hat{\Sigma})$ in all the higher times $\{ t \}$,
and in particular in $\sigma$ and $\tau$. For a generic complex
solution, since the generalised Jacobian is non-compact, the linear
motion could very well never come back to itself. However, by
corollary \ref{cor: real Jacobian} the real slice of the Jacobian is a
real $(g+1)$-torus. Since this is compact, in the real case the linear
motion of theorem \ref{thm: linear motion} must wrap densely on the
real slice of $J_{\mathfrak{m}}(\hat{\Sigma})$. Therefore all real
finite-gap strings are quasi-periodic in all higher times $\{ t
\}$. This is not too surprising since they describe a string moving on
$S^3$ which is itself compact. Exact periodicity in any of the higher
times $t_N$ with period $T_N$ is guaranteed if the vector $T_N
\int_{\vec{b}} d \Omega_N$, which is real by lemma \ref{lemma: real
quasi-mom}, happens to coincide with a lattice vector, namely
\begin{equation*}
T_N \int_{\vec{b}} d \Omega_N \in 2 \pi \mathbb{Z}^{g+1}.
\end{equation*}
In particular, a finite-gap string is \dub{closed} (\textit{i.e.}
invariant under $\sigma \rightarrow \sigma + 2 \pi$) if
\begin{equation} \label{closed string condition}
\int_{\vec{b}} dp \in 2 \pi \mathbb{Z}^{g+1}.
\end{equation}
Most of these conditions are automatically satisfied. Indeed all the
$\bm{b}$-periods of $dp$ are integer multiples of $2 \pi$ by equation
\eqref{periods of dp} which was a consequence of the single-valuedness
of $\Lambda(P)$ as a function on $\hat{\Sigma}$. The only non-trivial
condition in \eqref{closed string condition} is
\begin{equation} \label{closed string condition 2}
\frac{1}{2 \pi} \int_{\infty^-}^{\infty^+} dp \in \mathbb{Z}.
\end{equation}

In fact the linear motion on the generalised Jacobian only completely
encodes the dependence of the current $j = -g^{-1} dg$ on the higher
times, but it is not sufficient to encode the time dependence of the
embedding $g$. To obtain the complete set of periodicity conditions on
the algebro-geometric data one should instead consider the embedding
itself
\begin{equation*}
g(\sigma) = U_L \cdot P \overleftarrow{\exp} \int_{\sigma} j,
\end{equation*}
where we have only explicitly written the dependence on
$\sigma$. Comparing this expression to the same expression translated
by $\sigma \rightarrow \sigma + 2 \pi$ whose inverse is given by
$g^{-1}(\sigma + 2 \pi) = \left( P \overleftarrow{\exp} \int^{\sigma +
2 \pi} j \right) \cdot U^{-1}_L$ we find
\begin{equation*}
g^{-1}(\sigma + 2 \pi) g(\sigma) = P \overleftarrow{\exp}
\int^{\sigma + 2 \pi}_{\sigma} j = \Omega(0, \sigma).
\end{equation*}
Periodicity in $\sigma$ of the embedding field $g$ is therefore
guaranteed provided $\Omega(0) = {\bf 1}$. Since the eigenvalues of
the monodromy matrix at the origin are $\Lambda(0^{\pm})$ this
condition can equally be written as conditions on the periods of $dp$,
namely
\begin{equation} \label{closed string condition 3}
\frac{1}{2 \pi} \int_{\infty^+}^{0^{\pm}} dp \in \mathbb{Z}.
\end{equation}
These conditions imply the earlier conditions \eqref{closed string
condition 2} as it should be. Indeed, using the property
$\hat{\sigma}^{\ast} dp = - dp$ of the differential of the
quasi-momentum we find
\begin{equation} \label{dp 0 to infty relation}
\int_{\infty^{\pm}}^{0^{\mp}} dp = - \int_{\infty^{\pm}}^{0^{\mp}}
\hat{\sigma}^{\ast} dp = - \int_{\infty^{\mp}}^{0^{\pm}} dp,
\end{equation}
where the contour $[0^{\pm}, \infty^{\mp}]$ is simply the image of the
contour $[0^{\mp}, \infty^{\pm}]$ under the holomorphic involution
$\hat{\sigma}$. But now breaking up the integral
$\int_{\infty^+}^{0^+} dp$ as follows
\begin{equation*}
\int_{\infty^+}^{0^+} dp = \int_{\infty^+}^{\infty^-} dp +
\int_{\infty^-}^{0^-} dp + \int_{0^-}^{0^+} dp,
\end{equation*}
and doing the same for the integral $\int_{\infty^+}^{0^-} dp$ it
follows using \eqref{dp 0 to infty relation} that
\begin{equation*}
\int_{\infty^+}^{0^{\pm}} dp = - \frac{1}{2} \left(
\int_{\infty^-}^{\infty^+} dp \mp \int_{0^-}^{0^+} dp \right).
\end{equation*}
Thus we can write $\int_{\infty^-}^{\infty^+} dp = -
\int_{\infty^+}^{0^+} dp - \int_{\infty^+}^{0^-} dp$ and \eqref{closed
string condition 3} implies \eqref{closed string condition 2} as
claimed. Finally let us use \eqref{dp 0 to infty relation} to rewrite
the full set of periodicity conditions \eqref{closed string condition
3} slightly differently as follows,
\begin{equation} \label{closed string condition 4}
\frac{1}{2 \pi} \int_{\infty^{\pm}}^{0^+} dp \in \mathbb{Z}.
\end{equation}


\section{Real closed finite-g strings} \label{section: real strings}


In the previous sections we have obtained necessary conditions on the
algebro-geometric data for the finite-gap strings to be both real and
closed. In this section we show that these conditions are also
sufficient. That is, with algebro-geometric data satisfying the
reality conditions and the periodicity conditions, the reconstructed
current of theorem \ref{thm: reconstructing j} is both
$\mathfrak{su}(2)$-valued and $2 \pi$ periodic in $\sigma$. Moreover
the reconstructed embedding of theorem \ref{thm: reconstructing g} is
$SU(2)$-valued and $2 \pi$ periodic in $\sigma$.

\subsection*{The $SU(2)_R$ current $j$}

Let $\bm{\psi}^+(P)$ be the dual Baker-Akhiezer vector defined by
\eqref{dual Baker-Akhiezer vec def} with respect to the `correct' dual
dynamical divisor given in lemma \ref{lemma: real divisor} by
$\hat{\gamma}^+(0) = \hat{\tau} \hat{\gamma}(0)$.

\begin{lemma} \label{lemma: psi^+ = psi^dag}
$\bm{\psi}^+(P) = \bm{\psi}(\hat{\tau} P)^{\dag}$.
\begin{proof}
Consider the functions $f_i(P) = \psi^+_i(P) /
\overline{\psi_i(\hat{\tau} P)}$. These are meromorphic functions with
at most $g$ poles (in general position) and hence are constant by the
Riemann-Roch theorem. But by the normalisation conditions in
\eqref{dual Baker-Akhiezer vec def 1} and \eqref{Baker-Akhiezer vec
def 1} we have $f_1(\infty^+) = f_2(\infty^-) = 1$ so that $f_i(P)
\equiv 1$.
\end{proof}

\noindent \textup{It is instructive to give a second proof of lemma
\ref{lemma: psi^+ = psi^dag} but using the explicit reconstruction
formulae of the two vectors $\bm{\psi}$ and $\bm{\psi}^+$ in
propositions \ref{prop: existence of psi} and \ref{prop: existence of
psi^+}.}
\begin{proof}[Proof of lemma \ref{lemma: psi^+ = psi^dag} (using
reconstruction formulae)]
It follows from lemma \ref{lemma: omega imaginary} and the
reality condition \eqref{reality: a b cycles} on the $\bm{b}$-cycles
that the period matrix satisfies the following reality condition
\begin{equation} \label{imaginary period matrix}
\bar{\Pi} = - \Pi - \Pi_0,
\end{equation}
where $(\Pi_0)_{ij} = \sum_{k \neq i} \delta_{kj} + 2 \delta_{ij}$ has
$1$'s in all off-diagonal entries and $2$'s along the diagonal. Using
this, it follows from its definition \eqref{vector of Riemann's
constants} that the vector of Riemann's constants is pure imaginary
$\bar{\bm{\mathcal{K}}} = - \bm{\mathcal{K}}$. In particular we have
\begin{equation*}
\overline{\bm{\zeta}_{\gamma_{\mp}(0)}} =
\overline{\bm{\mathcal{A}}(\gamma_{\mp}(0))} +
\overline{\bm{\mathcal{K}}} = - \bm{\mathcal{A}}(\hat{\tau}
\gamma_{\mp}(0)) - \bm{\mathcal{K}} = -
\bm{\mathcal{A}}(\delta_{\mp}(0)) - \bm{\mathcal{K}} =
- \bm{\zeta}_{\delta_{\mp}(0)},
\end{equation*}
where the divisors $\gamma_{\pm}(t)$ and $\delta_{\pm}(t)$ of degree
$g$ were defined in \eqref{divisor of zeroes} and \eqref{divisor of
zeroes dual} respectively. It also follows from \eqref{imaginary
period matrix} that the $\theta$-function defined in \eqref{theta
function def} satisfies the reality condition $\overline{\theta(z)} =
\theta(- \bar{z})$. This comes down to the following identity for the
matrix $\Pi_0$,
\begin{multline*}
\exp \{ \pi i \langle \Pi_0 \bm{m}, \bm{m} \rangle \} = \exp \left\{
\pi i \sum_{i,j = 1}^g (\Pi_0)_{ij} m_j m_i \right\}\\ = \exp \left\{
\pi i \sum_{i = 1}^g \left( \sum_{k \neq i} m_k + 2 m_i \right) m_i
\right\} = \exp \left\{ 2 \pi i \sum_{i=1}^g \left( m_i^2 + \sum_{k >
i} m_k m_i \right) \right\} = 1.
\end{multline*}
Finally, using the above and corollary \ref{cor: real divisor} it is
easily shown directly from the reconstruction formulae for the
normalised eigenvector $\bm{h}$ and the dual normalised eigenvector
$\bm{h}^+$ in propositions \ref{prop: existence of h} and \ref{prop:
existence of h^+} that
\begin{equation*}
\overline{h_{\pm}(\hat{\tau} P)} = k_{\pm}(P).
\end{equation*}
Moreover, starting from the formulae in proposition
\ref{prop: existence of psi} for the components of the Baker-Akhiezer
vector, we can compute their conjugates evaluated at $\hat{\tau} P$
and obtain the formulae in proposition \ref{prop: existence of psi^+}
for the dual Baker-Akhiezer vector. Specifically,
\begin{align*}
\overline{\psi_{\pm}(\hat{\tau} P)} &= \overline{h_{\mp}(\hat{\tau}
P,0)} \frac{\theta \left( - \overline{\bm{\mathcal{A}}(\hat{\tau} P)}
- \int_{\bm{b}} \overline{d \mathcal{Q}} +
\overline{\bm{\zeta}_{\gamma_{\mp}(0)}} \right) \theta \left( -
\overline{\bm{\mathcal{A}}(\infty^{\pm})} +
\overline{\bm{\zeta}_{\gamma_{\mp}(0)}} \right)}{\theta \left( -
\overline{\bm{\mathcal{A}}(\hat{\tau} P)} +
\overline{\bm{\zeta}_{\gamma_{\mp}(0)}} \right) \theta \left( -
\overline{\bm{\mathcal{A}}(\infty^{\pm})} - \int_{\bm{b}} \overline{d
\mathcal{Q}} + \overline{\bm{\zeta}_{\gamma_{\mp}(0)}} \right)}\\
&\qquad\qquad\qquad\qquad\qquad\qquad\qquad\qquad\qquad\qquad\qquad
\times \exp \left( - i \int_{\infty^{\pm}}^{\hat{\tau} P}
\overline{d\mathcal{Q}} \right),\\
&= k_{\mp}(P,0) \frac{\theta \left( \bm{\mathcal{A}}(P) -
\int_{\bm{b}} d \mathcal{Q} - \bm{\zeta}_{\delta_{\mp}(0)} \right)
\theta \left( \bm{\mathcal{A}}(\infty^{\pm}) -
\bm{\zeta}_{\delta_{\mp}(0)} \right)}{\theta \left(
\bm{\mathcal{A}}(P) - \bm{\zeta}_{\delta_{\mp}(0)} \right) \theta
\left( \bm{\mathcal{A}}(\infty^{\pm}) - \int_{\bm{b}} d \mathcal{Q} -
\bm{\zeta}_{\delta_{\mp}(0)} \right)}\\
&\qquad\qquad\qquad\qquad\qquad\qquad\qquad\qquad\qquad\qquad\qquad
\times \exp \left( - i \int_{\infty^{\pm}}^P
\overline{\hat{\tau}^{\ast} d\mathcal{Q}} \right),
\end{align*}
which is the expression for $\phi_{\pm}(P)$ in proposition \ref{prop:
existence of psi^+} after using lemma \ref{lemma: real quasi-mom}.
\end{proof}
\end{lemma}

\begin{corollary} \label{cor: inverse Psi^-1}
The inverse matrix of $\Psi(x) = (\bm{\psi}(x^+), \bm{\psi}(x^-))$ can
be written as
\begin{equation} \label{inverse matrix Psi^-1}
\Psi(x)^{-1} = \diag( \chi_0(x^+), \chi_0(x^-)) \Psi(\bar{x})^{\dag},
\end{equation}
where $(\chi_0) = \hat{\gamma}(0) + \hat{\tau} \hat{\gamma}(0) - B$
and $\chi_0(\infty^{\pm}) = 1$.
\begin{proof}
By proposition \ref{prop: inverse of Psi} the inverse matrix
of $\Psi(x)$ can be written as
\begin{equation*}
\Psi(x)^{-1} = \diag( \chi_0(x^+), \chi_0(x^-)) \left(
\bm{\psi}^+(x^+)^{\sf T}, \bm{\psi}^+(x^-)^{\sf T} \right)^{\sf T},
\end{equation*}
where $\chi_0(P) = \eta(P,0)^{-1} = (\bm{h}^+(P,0) \cdot
\bm{h}(P,0))^{-1}$ has zeroes at $\hat{\gamma}(0)$ and $\hat{\tau}
\hat{\gamma}(0)$, poles at the divisor of branch points $B$ and is
normalised at infinity, \textit{i.e.}
\begin{equation}
(\chi_0) = \hat{\gamma}(0) + \hat{\tau} \hat{\gamma}(0) - B, \qquad
\chi_0(\infty^{\pm}) = 1.
\end{equation}
Using lemma \ref{lemma: psi^+ = psi^dag} we can now rewrite the matrix
of dual Baker-Akhiezer vectors in terms of Baker-Akhiezer vectors.
\end{proof}
\end{corollary}

\begin{theorem}
When using real and periodic algebro-geometric data, the reconstructed
current $j$ of theorem \ref{thm: reconstructing j} is
$\mathfrak{su}(2)$-valued and $\sigma$-periodic, \textit{i.e.}
\begin{equation*}
j_{\pm} \in \mathfrak{su}(2), \qquad j_{\pm}(\sigma + 2 \pi) =
j_{\pm}(\sigma).
\end{equation*}
\begin{proof}
Using corollary \ref{cor: inverse Psi^-1} the reconstructed current
\eqref{reconstruction formula for j components} can be written as
\begin{equation} \label{formula for j: real}
\begin{split}
j_+ = i \kappa_+ \Psi_0 \sigma_3 \diag(\chi_0((+1)^+), \chi_0((+1)^-))
\Psi_0^{\dag},\\
j_- = i \kappa_- \Phi_0 \sigma_3 \diag(\chi_0((-1)^+), \chi_0((-1)^-))
\Phi_0^{\dag},
\end{split}
\end{equation}
where $\Psi_0$, $\Phi_0$ were defined in \eqref{Psi essential
singularity} as the leading terms in the expansion of $\Psi(x)$ at $x
= \pm 1$. The defining properties of $\chi_0$ stated in corollary
\ref{cor: inverse Psi^-1} also imply that $\chi_0(\hat{\tau} P) =
\overline{\chi_0(P)}$ and hence $\chi_0((\pm 1)^+), \chi_0((\pm 1)^-)
\in \mathbb{R}$. It is then immediate from \eqref{formula for j: real}
that $j_{\pm}^{\dag} = - j_{\pm}$.

The $\sigma$-periodicity can be shown using the explicit
reconstruction formulae for the Baker-Akhiezer and dual Baker-Akhiezer
vectors in propositions \ref{prop: existence of psi} and \ref{prop:
existence of psi^+}. The arguments of the $\theta$-functions of both
these vectors depends on $\sigma$ only through the combination
$\frac{\sigma}{2 \pi} \int_{\bm{b}} dp \in \sigma \mathbb{Z}^g$. Then
by the automorphy property \eqref{automorphy} it follows that the
$\theta$-function parts of the expressions in propositions \ref{prop:
existence of psi} and \ref{prop: existence of psi^+} are invariant
under $\sigma \rightarrow \sigma + 2 \pi$. Now focusing on the
exponential parts we can write
\begin{equation*}
\begin{split}
\Psi(x) &= \diag\left( 1, e^{i \int_{\infty^-}^{\infty^+}
d\mathcal{Q}} \right) \Theta_+(x) \diag\left( e^{i
\int_{\infty^+}^{x^+} d\mathcal{Q}}, e^{i \int_{\infty^+}^{x^-}
d\mathcal{Q}} \right),\\
\Psi(x)^{-1} &= \diag\left( e^{-i \int_{\infty^+}^{x^+} d\mathcal{Q}},
e^{-i \int_{\infty^+}^{x^-} d\mathcal{Q}} \right) \Theta_-(x)
\diag\left( 1, e^{-i \int_{\infty^-}^{\infty^+} d\mathcal{Q}} \right),
\end{split}
\end{equation*}
where $\Theta_{\pm}(x)$ contains the $\theta$-function part of these
formulae. It now follows from the reconstruction formula
\eqref{reconstruction formula for j components} for $j_{\pm}$ that the
current also depends on $\sigma$ through $\exp \left( \frac{i
\sigma}{2 \pi} \int_{\infty^-}^{\infty^+} dp \right) = \exp \left( i
\sigma n \right)$ for $n \in \mathbb{Z}$, which is also invariant
under $\sigma \rightarrow \sigma + 2 \pi$. Therefore the full
reconstruction formula for $j_{\pm}$ is periodic in $\sigma$ of period
$2 \pi$.
\end{proof}
\end{theorem}

\subsection*{The $SU(2)$ embedding $g$}

\begin{lemma} \label{lemma: second column of Psi}
Let $P = \hat{\tau} P \in \hat{\Sigma}$ be a fixed point of
$\hat{\tau}$ then
\begin{equation*}
\psi_1(\hat{\sigma} P) = - A(P) \overline{\psi_2(P)}, \qquad
\psi_2(\hat{\sigma} P) = A(P) \overline{\psi_1(P)}
\end{equation*}
where $A(P) = \chi_0(P) \det(\bm{\psi}(P), \bm{\psi}(\hat{\sigma} P)
)$.
\begin{proof}
By lemma \ref{lemma: psi^+ = psi^dag} and equation \eqref{inverse
matrix Psi^-1} the rows of $\Psi(x)^{-1}$ take the form $\chi_0(P)
\bm{\psi}(P)^{\dag}$ and $\chi_0(\hat{\sigma} P)
\bm{\psi}(\hat{\sigma} P)^{\dag}$. It follows that
\begin{equation} \label{second column proof 1}
\left( \begin{array}{cc} \overline{\psi_1(P)} & \overline{\psi_2(P)}\\
\overline{\psi_1(\hat{\sigma} P)} & \overline{\psi_2(\hat{\sigma} P)}
\end{array} \right) \left( \begin{array}{c} \psi_1(P)\\ \psi_2(P)
\end{array} \right) = \left( \begin{array}{c} \frac{1}{\chi_0(P)} \\ 0
\end{array} \right).
\end{equation}
Multiplying by the inverse of the matrix on the left hand side we
obtain
\begin{equation*}
\left( \begin{array}{c} \psi_1(P)\\ \psi_2(P) \end{array} \right) =
-\frac{1}{\overline{D(P)} \chi_0(P)}
\left( \begin{array}{c} \overline{\psi_2(\hat{\sigma} P)} \\ -
\overline{\psi_1(\hat{\sigma} P)} \end{array} \right),
\end{equation*}
where $D(P) = \det(\bm{\psi}(P), \bm{\psi}(\hat{\sigma} P))$. Defining
$A(P) = \chi_0(P) D(P)$ its conjugate is $\overline{A(P)} =
\chi_0(P) \overline{D(P)}$ since $\chi_0(P) \in \mathbb{R}$ for
$\hat{\pi}(P) \in \mathbb{R}$ and the result follows.
\end{proof}
\end{lemma}

\begin{theorem} \label{thm: embedding real}
After a residual diagonal $SL(2,\mathbb{C})_L$ transformation $g_L$
the $SU(2)$ embedding can be recovered by the formula,
\begin{equation*}
g = \chi_0(0^+)^{\frac{1}{2}} \left( \begin{array}{cc}
\overline{\psi_1(0^+)} & \overline{\psi_2(0^+)}\\ - \psi_2(0^+) &
\psi_1(0^+) \end{array} \right) \in SU(2).
\end{equation*}
\begin{proof}
By lemma \ref{lemma: second column of Psi} we can write $\Psi(0)$ as
\begin{equation*}
\Psi(0) = \left( \begin{array}{cc} \psi_1(0^+) & -
\overline{\psi_2(0^+)}\\ \psi_2(0^+) & \overline{\psi_1(0^+)}
\end{array} \right) \; \diag(1, \chi_0(0^+) \det \Psi(0)).
\end{equation*}
This can equivalently be written as
\begin{equation} \label{pre embedding real}
\frac{1}{\sqrt{\det \Psi(0)}} \cdot \Psi(0) = \chi_0(0^+)^{\frac{1}{2}}
\left( \begin{array}{cc} \psi_1(0^+) & - \overline{\psi_2(0^+)}\\
\psi_2(0^+) & \overline{\psi_1(0^+)} \end{array} \right) \;
\diag(S, S^{-1}),
\end{equation}
where $S = \sqrt{\chi_0(0^+)^{-1} \det \Psi(0)}$.
The diagonal matrix $g_L = \diag(S, S^{-1})$ on the right hand side is
nothing but an $SL(2, \mathbb{C})_L$ residual transformation. Now the
first component of \eqref{second column proof 1} reads
$|\psi_1(0^+)|^2 + |\psi_2(0^+)|^2 = \frac{1}{\chi_0(0^+)}$ from which
it follows that $\chi_0(0^+) > 0$ and
\begin{equation*}
\chi_0(0^+)^{\frac{1}{2}} \left( \begin{array}{cc} \psi_1(0^+) & -
\overline{\psi_2(0^+)}\\ \psi_2(0^+) & \overline{\psi_1(0^+)}
\end{array} \right) \in SU(2).
\end{equation*}
Removing the residual gauge transformation in \eqref{pre embedding
real} we have by proposition \ref{thm: reconstructing g},
\begin{equation*}
g^{-1} = \frac{1}{\sqrt{\det \Psi(0)}} \cdot \Psi(0) g^{-1}_L =
\chi_0(0^+)^{\frac{1}{2}} \left( \begin{array}{cc} \psi_1(0^+) & -
\overline{\psi_2(0^+)}\\ \psi_2(0^+) & \overline{\psi_1(0^+)}
\end{array} \right).
\end{equation*}
Inverting this proves the theorem.
\end{proof}
\end{theorem}

Recall that the embedding matrix $g$ encoded the fields $X_i$, $i = 1,
\ldots, 4$ describing the embedding into $S^3 \subset \mathbb{R}^4$
through equation \eqref{matrix g}. Defining the complex fields $Z_1 =
X_1 + i X_2$ and $Z_2 = X_3 + i X_4$ we have
\begin{corollary}
The embedding fields $X_i$, $i = 1, \ldots, 4$ are recovered in terms
of the dual Baker-Akhiezer vector evaluated at $0^+$ by
\begin{equation} \label{reconstruction of Z_i}
Z_i = C \psi^+_i(0^+), \quad i = 1,2
\end{equation}
where $C = \chi_0(0^+) \in \mathbb{R}_+$ is a normalisation ensuring
that $|Z_1|^2 + |Z_2|^2 = 1$.
\end{corollary}

\begin{proposition}
The reconstruction formulae \eqref{reconstruction of Z_i} are
$2 \pi$ periodic in $\sigma$.
\begin{proof}
The arguments of the $\theta$-functions in the reconstruction formulae
for the dual Baker-Akhiezer vector in propositions \ref{prop:
existence of psi^+} depends on $\sigma$ only through the combination
$\frac{\sigma}{2 \pi} \int_{\bm{b}} dp \in \sigma \mathbb{Z}^g$. Then
by the automorphy property \eqref{automorphy} the $\theta$-function
part is invariant under $\sigma \rightarrow \sigma + 2 \pi$. As for
the exponentials, the $\sigma$-dependent parts are $\exp \left(
-\frac{i \sigma}{2 \pi} \int_{\infty^{\pm}}^{0^+} dp \right)$ which
are clearly invariant under $\sigma \rightarrow \sigma + 2 \pi$ by
\eqref{closed string condition 4}.
\end{proof}
\end{proposition}

%% file: PerturbFG.tex
\newpage

\chapter{Semiclassical strings on $\mathbb{R} \times S^3$} \label{chapter: semi}

The method of semiclassical quantisation in field theory has been
extensively developed by many authors in the 70's using different
approaches \cite{DHN1, DHN2, DHN3, Korepin1, Korepin2, Korepin3,
BerryTabor1, BerryTabor2} (see also the books
\cite{Coleman:1975qj, Rajaraman:1982is} for a more or less
complete survey and list of references). The aim of all these
methods is to give a quantum mechanical meaning to extended
classical solutions of the field equations which already
classically exhibit particle like properties. The role played by
such non-trivial classical solutions in the leading order
quantisation of any field theory is evident from the path integral
which is dominated by classical solutions in the $\hbar
\rightarrow 0$ limit. It follows then that the applicability of
semiclassical methods crucially relies on an explicit knowledge of
classical solutions. Having studied the general finite-gap string in
Part \ref{part: Algebro} we can now proceed with semiclassically
quantising the string on $\mathbb{R} \times S^3$.

An important part in any approach to semiclassical quantisation is the
treatment of the zero-modes (see \cite{Korepin3} for a clear
exposition of the problem and \cite{DHN1, DHN2, Coleman:1975qj,
Rajaraman:1982is} for various resolutions). Roughly speaking, if
$\phi_{\text{cl}}$ is a solution to the field equations derived from
an action $S[\phi]$ then a \dub{zero-mode} of $\phi_{\text{cl}}$ is a
(possibly hidden) symmetry of the equations of motion $S'[\phi] = 0$
which isn't a symmetry of $\phi_{\text{cl}}$ itself. If a classical
solution has zero-modes then a naive semiclassical quantisation of the
solution will fail. Indeed, suppose that $\phi_{\text{cl}}$ is not
invariant under an infinitesimal symmetry $v$ of the equations of
motion, then it follows immediately that $(v \phi_{\text{cl}}) \neq 0$
is in the kernel of the operator $S''[\phi_{\text{cl}}]$ which is
therefore not invertible and so the propagator of the theory in the
background $\phi_{\text{cl}}$ cannot be defined. The standard way
around this difficulty is to treat the zero-mode directions separately
using the method of \dub{collective coordinates}. In short, collective
coordinates parametrise the zero-mode directions, namely the flat
directions in field space, along which the wave function will tend to
spread out in the form of a plane wave. As a result the quantum
counterpart of the solution $\phi_{\text{cl}}$ will acquire dynamics
along these collective coordinates. Generally one has to perform a
change of variables in field space to include the collective
coordinates among the set of field variables and this can often only
be done implicitly. A nice feature of the finite-gap construction is
that it naturally lends itself to the separation of zero-modes.

To see why that is, recall from theorem \ref{thm: pullback} that
the action variables $\{ \mathcal{S}_I \}_{I=1}^{g+1}$ act
non-trivially on the angle variables $\{ \varphi_I
\}_{I=1}^{g+1}$, which parametrise the divisor $\hat{\gamma}(t)$
according to \eqref{angle variables def}. Thus although each
action variable generates an infinitesimal symmetry $v=
\partial/\partial \varphi_I$ of the string equations of motion,
the finite-gap string itself is not invariant under this symmetry.
Therefore any $g$-gap string always has $g+1$ zero-modes for which
the divisor $\hat{\gamma}(t)$ fills the role of collective
coordinates. Alternatively, as we saw in chapter \ref{chapter:
symplectic} the non-special divisor $\hat{\gamma}(t)$ can equally
be described as a point $\vec{\mathcal{A}}(\hat{\gamma}(t))$ on
the generalised Jacobian. Therefore any set of coordinates on the
generalised Jacobian can be used as collective coordinates.

This leads to a very nice picture of finite-gap strings which ties
in with the discussion of semiclassical quantisation of
finite-dimensional systems in chapter \ref{chapter:
semiclassical}. Indeed, the upshot of chapters \ref{chapter:
symplectic} and \ref{chapter: reality} was that a finite-gap string
could be thought of as an embedding $\mathcal{G}_{\mathbb{R}} :
\mathcal{M}^{(2g+2)}_{\mathbb{R}} \hookrightarrow
\mathcal{P}^{\infty}$ of a finite-dimensional integrable system
\begin{equation*}
\mathbb{T}^{g+1} \rightarrow \mathcal{M}^{(2g+2)}_{\mathbb{R}}
\rightarrow \mathcal{L}_{\mathbb{R}}
\end{equation*}
into the infinite dimensional reduced phase-space
$\mathcal{P}^{\infty}$ of the string. Or put another way, a finite-gap
string describes a $(g+1)$-parameter family of $(g+1)$-torii in
$\mathcal{P}^{\infty}$ parameterised by the filling fractions $\{
\mathcal{S}_I \}_{I=1}^{g+1}$. These torii are \textit{isotropic}
since the pullback \eqref{symplectic form pullback} of the symplectic
form $\hat{\omega}_{\infty}$ to them is identically zero. Moreover,
being finite-dimensional they are necessarily \textit{degenerate}
isotropic torii of $\mathcal{P}^{\infty}$. This is the necessary
set-up to apply the Bohr-Sommerfeld conditions \eqref{BS6} for the
quantisation of a $p$-torus in an $n$-dimensional phase-space,
where here the total phase-space is infinite dimensional so that
$n = \infty$ and $p = g+1$.

In section \ref{section: Breather} we start by recalling the
method of semiclassical quantisation \`a la Dashen, Hasslacher and
Neveu \cite{DHN1, DHN2, DHN3} when applied to the specific example
of the breather solution in Sine-Gordon theory. We reformulate
everything in a language that we hope will facilitate the
conceptual understanding of the method in the finite-gap setting.
In section \ref{section: perturbations} we will explicitly compute the
stability angles of perturbations around a given finite-gap solution
which appear in the Bohr-Sommerfeld conditions.


\section{Analogy with Sine-Gordon breathers} \label{section:
Breather}


Consider the example of the boosted Sine-Gordon breather solution
\cite{DHN2, Coleman:1975qj, Rajaraman:1982is}
\begin{equation} \label{breather}
\phi_{\tau,v}(x,t) = \frac{4 m}{\sqrt{\lambda}} \tan^{-1}\left\{
\frac{ \sqrt{\left(\frac{\tau m}{2 \pi}\right)^2 - 1} \cdot \sin \left[
\left(\frac{2 \pi}{\tau}\right) \cdot \frac{t - v x}{\sqrt{1 - v^2}}
\right] }{\cosh \left[ \sqrt{\left(\frac{\tau m}{2 \pi} \right)^2 - 1} \cdot
\left(\frac{2 \pi}{\tau}\right) \cdot \frac{x - v t}{\sqrt{1 - v^2}} \right]
} \right\}.
\end{equation}
This is really a two parameter family of solutions parametrised by
their proper period $\tau$ and their velocity $v$, or equivalently
by their energy $E$ and momentum $p$. To compute the (possibly
continuous) spectrum of the corresponding quantum states it is
always simpler at first to put the system in a very large but
finite box of length $L$ by identifying $x \sim x + L$ so as to
make the spectrum discrete, and then take the infinite volume
limit $L \rightarrow \infty$ at the end. In this closed-loop world
the breather solution \eqref{breather} is periodic in $t$ of
period $T$ provided $\tau$ and $v$ satisfy
\begin{equation*}
T = l \frac{\tau}{\sqrt{1-v^2}} = m \frac{L}{v}, \qquad l,m \in
\mathbb{N}.
\end{equation*}

If we were quantising the kink, we could move to its rest frame in
which it is static and study small fluctuations in terms of
eigenfrequencies. However, the breather is a little more
complicated since it is time dependent in its rest frame, and
because time dependent solutions are not point-like in field
space, we need a way to characterise perturbations of the orbit as
a whole. This was described in chapter \ref{chapter: semiclassical}
where we defined the Poincar\'e map. The idea was to consider the
perturbation of a specific point on the orbit, evolve that
perturbation under the equations of motion for roughly the period of
the underlying solution, and compare the final perturbation with the
original one. If the perturbation is stable then it will have merely
rotated in which case the angle of rotation is called the stability
angle. If instead the perturbation is unstable it will have grown
exponentially in magnitude, which corresponds to the case of a
complex stability angle. Finally, if the perturbation comes back
exactly to itself, this means it describes a nearby periodic
solution, and in general zero stability angles correspond to
symmetries. In the case of the Sine-Gordon breather we therefore
need to look for generic nearby solutions $\phi(x,t) =
\phi_{\tau,v}(x,t) + \delta \phi$. This perturbed solution won't
be periodic in general, yet because the linearised equation
\begin{equation} \label{linearised SG eq}
\Box \delta \phi = \left( \cos \phi_{\tau,v} \right) \delta \phi
\end{equation}
is invariant under time translation by $T$ we can always write its
solution as a superposition of eigenfunctions of time translation
$\delta \phi(x,t+T) = e^{-i \nu} \delta \phi(x,t)$, where $\nu$
are their stability angles (another way to say this is that the
time translation operator $\hat{T} : t \mapsto t + T$ commutes with
the linearised operator $\hat{L} = \Box - \cos \phi_{\tau,v}$ and
hence both operators can be simultaneously diagonalised. In particular
the kernel of $\hat{L}$ is spanned by eigenfunctions of
$\hat{T}$). Notice that the Sine-Gordon equation is invariant under
arbitrary space and time translations, but the breather solution
$\phi_{\tau,v}$ is not. As a result, $\partial \phi_{\tau,v}/\partial
x$ and $\partial \phi_{\tau,v}/\partial t$ are both \dub{zero-modes},
\textit{i.e.} perturbations with zero stability angles. This is a
special case of a much more general result,
\begin{lemma}
If a classical solution is not invariant under a symmetry of the
action then it has a zero-mode.
\begin{proof}
Consider a periodic solution $\phi_{\text{cl}}$ of a field equation
derived from an action $S[\phi]$, \textit{i.e.} $S'[\phi_{\text{cl}}]
= 0$, where $'$ denotes $\delta/\delta \phi$. If $v$ is an
infinitesimal symmetry of the equations of motion, \textit{i.e.}
$v(S'[\phi]) = S''[\phi](v\phi)$, and suppose that $\phi_{\text{cl}}$
is not invariant under the symmetry then it follows immediately that
$(v \phi_{\text{cl}}) \neq 0$ is in the kernel of the operator
$S''[\phi_{\text{cl}}]$. Clearly it is a zero-mode since
$v \phi_{\text{cl}}(t+T) = v \phi_{\text{cl}}(t)$.
\end{proof}
\end{lemma}

\begin{figure}[h]
\centering
\begin{tabular}{ccc}
\includegraphics[height=15mm]{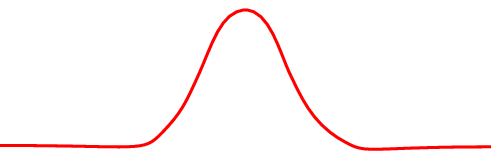} & \raisebox{5mm}{$\longrightarrow$} &
\includegraphics[height=15mm]{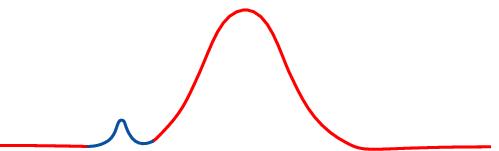}
\end{tabular}
\caption{Perturbing the breather by another small breather using
the B\"acklund transform} \label{figure: Backlund}
\end{figure}

The task of finding nearby solutions to the breather is greatly
facilitated by the fact that the Sine-Gordon equation is
integrable, since we can use the B\"acklund transform to get new
solutions from known solutions. In particular we can perturb our
breather by adding a little breather of small amplitude on top of
it (Figure \ref{figure: Backlund}).
Studying double breather solutions in the limit where the small
breather has vanishingly small amplitude corresponds to a
linearised study of the Sine-Gordon equation around the breather
solution. So integrability gives us a convenient way of writing
down explicit solutions to the linearised equation
\eqref{linearised SG eq} from which the stability angles of the
breather may be read off.

Identifying the space of classical solutions with phase-space, for
each $\tau, v$ (or equivalently $E,p$) the breather solution
\eqref{breather} is just a specific point in phase space. However,
the existence of two zero-modes $\partial \phi_{\tau,v}/\partial
x$ and $\partial \phi_{\tau,v}/\partial t$ for the breather
solution indicates that it really belongs to a two parameter
family of solutions with the same integrals of motion $E,p$. These
are the space and time translated breather solutions
\begin{equation} \label{breather family}
\phi_{\tau,v}(x + x_0,t + t_0).
\end{equation}
Since all the other stability angles of the breather are real,
when we include first order quantum corrections the wavefunction
will want to localise around not one breather, but around the
whole two parameter family \eqref{breather family} of breathers by
spreading along the flat directions, namely the $x_0$ and $t_0$
directions. Along these directions the wavefunction will therefore
be a plane wave, but since the $t_0$-direction is closed by
periodicity of the breather solution the plane wave along it must
have an integer number of peaks and troughs. In other words the
change of phase of the wavefunction around this closed direction
will have to be an integer multiple $n$ of $2 \pi$. Along all the
other non-zero stability angle directions the wavefunction will
decay rapidly and, intuitively, for states with higher excitation
number $n_i$ it will extend further in these directions. The
correct quantisation conditions encoding the semiclassical energy
spectrum of the wavefunction localised around the family of
breather solutions was first derived by Dashen, Hasslacher and
Neveu \cite{DHN1} and can be expressed as follows. If we define
the `action' of the breather solution as
\begin{subequations} \label{DHN}
\begin{equation} \label{DHN1}
W(E) = \int_0^T dt \int dx \pi_{\tau,v}(x,t) \partial_0
\phi_{\tau,v}(x,t),
\end{equation}
then the DHN quantisation conditions read
\begin{equation} \label{DHN2}
\frac{W(E)}{\hbar} = 2 \pi n + \sum_{\nu_i > 0} \left(n_i +
\frac{1}{2}\right) \nu_i + O(\hbar).
\end{equation}
\end{subequations}
Although the derivation of this formula is very complicated, it
intuitively makes a lot of sense. In general the phase of the
wavefunction in the semiclassial approximation is an action of the
form \eqref{DHN1} so the first term on the right hand side of
\eqref{DHN2} can be seen to come from the single-valuedness of
the wavefunction along the compact $t_0$-direction whereas the
correction from the sum over stability angles is related to the
small fluctuations transverse to the $t_0$ and $x_0$ directions.

For the purpose of drawing the analogy between Sine-Gordon breathers
and finite-gap strings it will be convenient to think of the
conditions \eqref{DHN} in more geometric terms in phase-space as
follows. Since the breather in \eqref{breather family} with $x_0 = 0$
is periodic, it can be thought of as a closed orbit on the level set
$\Sigma_{E,p}$ of fixed $E,p$. The direction along the orbit,
parametrised by $t_0$, corresponds to the zero-mode $\partial
\phi_{\tau,v}/\partial t$ of the breather. But since it has another
zero-mode, namely $\partial \phi_{\tau,v}/\partial x$, this orbit
really belongs to a continuous family of periodic orbits, parametrised
by $x_0$, all contained in $\Sigma_{E,p}$. However, because we are
working in a periodically identified finite box, this two parameter
($x_0, t_0$) family of breathers is in fact a torus
$\mathbb{T}^2_{E,p}$ lying within $\Sigma_{E,p}$. And since all the
other stability angles of the breather are non-zero, this means that
$\mathbb{T}^2_{E,p}$ is isolated on the level set $\Sigma_{E,p}$
in the sense that it does not belong to a larger continuous family
of periodic orbits within $\Sigma_{E,p}$. Yet if we leave the
level set $\Sigma_{E,p}$, one can show that in a neighbourhood of
$\Sigma_{E,p}$ the torus $\mathbb{T}^2_{E,p}$ persists, namely it
belongs to a two parameter family of torii parametrised by $E,p$.
This was the content of the generalised cylinder theorem \ref{thm:
cylinder gen} in chapter \ref{chapter: semiclassical}. Looking back at
the most general breather solution \eqref{breather family} it contains
four independent parameters: the two parameters $x_0, t_0$ are
parameters along the torus $\mathbb{T}^2_{E,p}$ whereas $E,p$
parameterise the family of torii of the generalised cylinder theorem
\ref{thm: cylinder gen}. Now the effect of the quantisation condition
\eqref{DHN} is to pick out a discrete set of breathers from this
generalised cylinder of breathers \eqref{breather}, the energy and
momentum of which approximate to order $O(\hbar)$ the semiclassical
energy spectrum of the quantum states localised around the breather
solution. For instance, when applied to the Sine-Gordon breather the
quantisation conditions \eqref{DHN} yield the following semiclassical
spectrum \cite{DHN2}
\begin{equation*}
E_{k,n} = (p_k^2 + M_n^2)^{\frac{1}{2}}, \quad p_k = \frac{2 \pi
k}{L},
\end{equation*}
where $M_n = \frac{16 m}{\gamma'} \sin \frac{n \gamma'}{16}$ and
$\gamma' = \frac{\lambda}{m^2} \left( 1 - \frac{\lambda}{8 \pi
m^2} \right)^{-1}$, and in the infinite volume limit $L
\rightarrow \infty$ the momentum becomes continuous as expected.

The analogy with the finite-gap construction is as follows. Just
as the generic breather \eqref{breather family} defined a
four-parameter family of solutions, a finite-gap string defines a
whole $(2g+2)$-parameter family of solutions parametrised by the
algebro-geometric data. It can be written schematically as
\begin{equation*}
g = g \Big( {\sum}_N t_N \vec{U}_N(\vec{\mathcal{S}}) + \vec{D} \Big|
\vec{\mathcal{S}} \Big),
\end{equation*}
where $t_N$ are a set of $g+1$ independent times (defined in
section \ref{section: hierarchy}), $\vec{U}_N(\vec{\mathcal{S}})$ is
some function of the filling fractions which play the role of the
parameters $(\tau, v)$ or $(E,p)$ here. The vector $\vec{D} \in
\mathbb{C}^{g+1}$ is related to the initial divisor $\hat{\gamma}(0)$
and is the exact analogue of the initial coordinates of the breather
$(x_0,t_0)$. As already explained at the start of this chapter the
$g+1$ components of this vector correspond to $g+1$ zero-modes of the
$g$-gap string, analogously to the breather case.


\section{Perturbations of finite-gap strings} \label{section:
perturbations}


In view of applying a semiclassical quantisation formula like the
one in \eqref{DHN} we must first determine all the stability
angles of a given finite-gap string. So just as in the case of
the Sine-Gordon breather, we would like to study perturbations of
finite-gap strings obtained in chapter \ref{chapter: finite-gap}. Once
again integrability will play a prominent role in solving the
linearised equations. In fact, finding solutions to the linearised
problem is very simple now that we have already fully exploited
integrability to construct the most general finite-gap string. A
perturbation of a given finite-gap string will simply be another
`nearby' finite-gap string. Recall from chapter \ref{chapter: curves}
that the algebraic curve is hyperelliptic and can be represented by a
set of $g+1$ vertical cuts in the complex plane. How can one describe
perturbations of the $g$-gap string corresponding to this curve?
Playing the same game as for the Sine-Gordon breather where we used
integrability to add another little breather on it, here we can just
take a solution corresponding to a curve of genus one higher, but make
the extra filling fraction very small, which corresponds to making the
cut very small, see Figure \ref{figure: perturb}.
\begin{figure}[h]
\centering
\includegraphics[height=30mm]{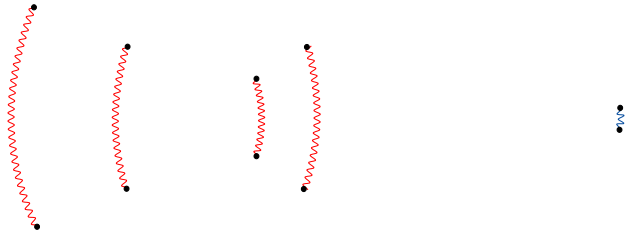}
\caption{Perturbation of a finite-gap solution.} \label{figure:
perturb}
\end{figure}
There is an obvious analogy here between breathers in Sine-Gordon
and cuts in bosonic strings on $\mathbb{R} \times S^3$ as one can
think of a finite-gap string as a multi-breather solution
consisting of finitely many breathers. Cuts with small filling
fractions are analogous to breathers of small amplitude as both
describe perturbations. If we define the $a_i$-cycle ($i = 1,
\ldots, g$) as in chapter \ref{chapter: curves} to encircle the
$i^{\text{th}}$ cut counterclockwise (on the upper sheet) then a
perturbation of this kind clearly corresponds to pinching an $a$-cycle
of the algebraic curve. So we want to take the difference between the
solution before pinching an $a$-cycle and the solution after pinching
the $a$-cycle; this will give us a perturbation of the latter and we
can then analyse its periodicity properties to extract the
corresponding stability angles. Notice however that any given
perturbation of a finite-gap string will have one stability angle
defined for each cycle on the generalised Jacobian, or equivalently
for each macroscopic cut.

So given a $g$-gap solution $Z_i$ with underlying algebraic curve
$\hat{\Sigma}$ of genus $g$ and filling fractions $\{ \mathcal{S}_I
\}_{I=1}^{g+1}$, we will obtain its stability angles by considering
nearby $(g+1)$-gap solutions $Z_i + \delta Z_i$ with algebraic curves
$\hat{\Sigma}^{\epsilon}$ of genus $g + 1$ with the \textit{same}
macroscopic filling fractions $\{ \mathcal{S}_I \}_{I=1}^{g+1}$ and an
extra small filling fraction $\mathcal{S}_0 = O(\epsilon)$. The limit
$\epsilon \rightarrow 0$ then corresponds to pinching the extra handle
to zero size, so that the limit curve $\hat{\Sigma}^0$ desingularises
to the original curve $\hat{\Sigma}$, see Figure \ref{Figure:
degeneration}.
\begin{figure}[ht]
\centering
\begin{tabular}{ccccc}
\psfrag{S}{\footnotesize $\Sigma^{\epsilon}$} \psfrag{a}{\tiny
\red $a_0$} \psfrag{b}{\tiny \lightblue $b_0$}
\includegraphics[height=25mm]{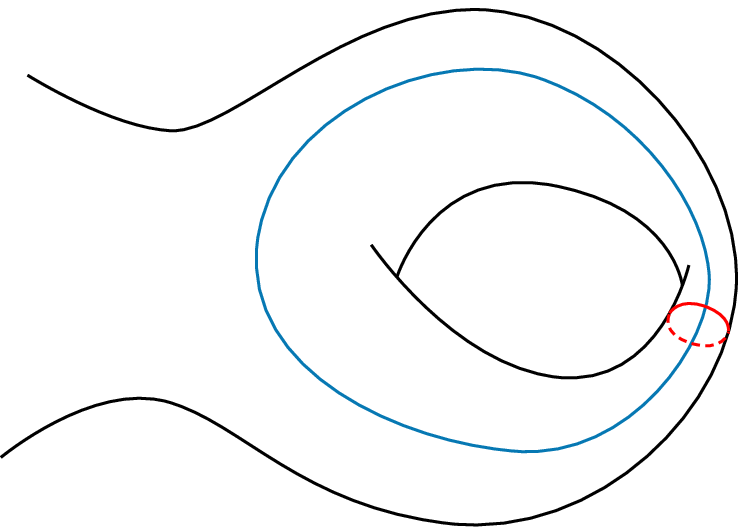} & $\quad$ &
\raisebox{10mm}{$\red \underset{\black \epsilon \rightarrow
0}\longrightarrow$} & $\quad$ & \psfrag{S}{\footnotesize $\Sigma$}
\psfrag{b}{\tiny \lightblue $b_0$}
\includegraphics[height=25mm]{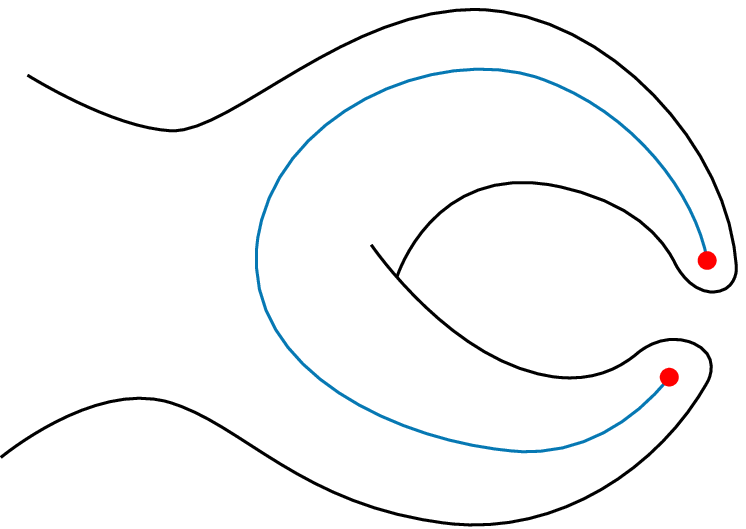}\\
\end{tabular}
\caption{Pinching an $a$-cycle.} \label{Figure: degeneration}
\end{figure}
The reason for wanting the macroscopic filling fractions $\{
\mathcal{S}_I \}_{I=1}^{g+1}$ to be fixed is that we need to
compute the perturbation of a finite-gap string \textit{within}
the level set determined by these filling fractions (see chapter
\ref{chapter: semiclassical} as well as section \ref{section:
Breather}).

Now since we are concerned with real finite-gap solutions,
constructed from real algebraic curves $\hat{\Sigma}$ (see chapter
\ref{chapter: reality}), the degeneration process in Figure
\ref{Figure: degeneration} describing the perturbation should respect
this reality condition. This forces us to consider degenerations
through the pinching of imaginary cycles, namely the
$a$-cycles. The process of pinching $a$-cycles on Riemann surfaces was
discussed in relation to generalised Jacobians in section
\ref{section: jacobian} of chapter \ref{chapter: Riemann surfaces}.

As we showed in chapter \ref{chapter: finite-gap} the dependence
of the general finite-gap solution on the hierarchy of times $\{ t
\}$ is entirely encoded in the normalised Abelian differential of
the second kind $d\mathcal{Q} = \sum_N t_N d\Omega_N$ defined in
\eqref{time-differential coupling} which enters the reconstruction
formula as follows
\begin{equation} \label{reconstruction formula for Z_i}
Z_i = C_i \; \frac{\theta \big( 2 \pi \int^{0^+}_{P_i} \bm{\omega}
- \int_{\bm{b}} d\mathcal{Q} - \bm{D} ; \Pi \big)}{\theta
\big(\int_{\bm{b}} d\mathcal{Q} + \bm{D} ; \Pi \big)} \; \exp
\left( - i \int^{0^+}_{P_i} d\mathcal{Q} \right),
\end{equation}
where $P_1 = \infty^+$ and $P_2 = \infty^-$. In this expression we
have hidden all the time independent part into the overall
constants $C_i$ for clarity. A nearby solution $Z_i + \delta Z_i$ is constructed
with the same formulae but using data on $\hat{\Sigma}^{\epsilon}$ to
be specified below,
\begin{equation} \label{reconstruction formula for Z_i + dZ_i}
Z_i + \delta Z_i = C^{\epsilon}_i \; \frac{\theta \big( 2 \pi
\int^{0^+}_{P_i} \vec{\omega}^{\epsilon} -
\int_{\vec{b}^{\epsilon}} d\mathcal{Q}^{\epsilon} - \vec{D} ;
\tilde{\Pi}^{\epsilon} \big)}{\theta
\big(\int_{\vec{b}^{\epsilon}} d\mathcal{Q}^{\epsilon} + \vec{D} ;
\tilde{\Pi}^{\epsilon} \big)} \; \exp \left( - i \int^{0^+}_{P_i}
d\mathcal{Q}^{\epsilon} \right).
\end{equation}
The ingredients of this deformed solution are as follows. First of
all, since the underlying curve $\hat{\Sigma}^{\epsilon}$ has genus
$g+1$, the arguments of the $\theta$-functions for this curve are
$(g+1)$-component vectors, namely $\vec{D} = ( D_0, \bm{D} )^{\sf T}
\in \mathbb{C}^{g+1}$, $\vec{b}^{\epsilon} = ( b_0^{\epsilon},
\bm{b}^{\epsilon} )^{\sf T} \in H^1(\hat{\Sigma}^{\epsilon})$ are the
$b$-periods of $\hat{\Sigma}^{\epsilon}$ and $\vec{\omega}^{\epsilon}
= ( \omega_0^{\epsilon}, \bm{\omega}^{\epsilon})^{\sf T}$ its
holomorphic differentials. In the singular limit $\epsilon \rightarrow
0$ one has $\bm{b}^{\epsilon} \rightarrow \bm{b}$ and
$\bm{\omega}^{\epsilon} \rightarrow \bm{\omega}$ which are the
$\bm{b}$-cycles and the $g$ holomorphic differentials on $\hat{\Sigma}$
respectively. The extra $b$-cycle $b_0^{\epsilon}$ becomes a
degenerate cycle on the curve $\hat{\Sigma}$, see Figure \ref{Figure:
degeneration}. As we showed in section \ref{section: jacobian} of
chapter \ref{chapter: Riemann surfaces}, in the limit $\epsilon
\rightarrow 0$ the extra holomorphic differential
$\omega_0^{\epsilon}$ on $\hat{\Sigma}^{\epsilon}$ acquires a simple
pole at the singular point and so becomes a normalised Abelian
differential of the third kind. The Abelian differential
$d\mathcal{Q}^{\epsilon}$ on $\hat{\Sigma}^{\epsilon}$ is defined by
the same singular parts \eqref{singular parts} as $d\mathcal{Q}$ at $x
= \pm 1$ but could potentially acquire an extra simple pole at the
singular point. However, because $d\mathcal{Q}^{\epsilon}$ is
normalised on $\hat{\Sigma}^{\epsilon}$, its residue there would
vanish in the $\epsilon \rightarrow 0$ limit, so that in fact
$d\mathcal{Q}^{\epsilon} \rightarrow d\mathcal{Q}$. One can also show
that $C^{\epsilon}_i \rightarrow C_i$.

The important object in \eqref{reconstruction formula for Z_i +
dZ_i} when considering the singular limit $\epsilon \rightarrow 0$
is the period matrix $\tilde{\Pi}^{\epsilon}$ which admits the natural
block form
\begin{equation} \label{period matrix g+1 curve}
\tilde{\Pi}^{\epsilon} = \int_{\vec{b}^{\epsilon}}
\vec{\omega}^{\epsilon} = \left(\begin{array}{cc}\Pi_{00}^{\epsilon} &
{\bm{\Pi}_0^{\epsilon}}^{\sf T}\\ \bm{\Pi}_0^{\epsilon} &
\Pi^{\epsilon}\end{array}\right).
\end{equation}
The singular limits of each block follow from the above considerations
of $\vec{b}^{\epsilon}, \vec{\omega}^{\epsilon}$ in the limit (see
section \ref{section: jacobian} of chapter \ref{chapter: Riemann
surfaces} for details). In particular, $\Pi^{\epsilon} \rightarrow
\Pi$ as $\epsilon \rightarrow 0$ which is simply the period matrix of
$\hat{\Sigma}$. The vectors $\bm{\Pi}_0^{\epsilon}$ also stay finite
in the limit. The top left component $\Pi_{00}^{\epsilon}$ on the
other hand diverges in this limit, leading to a simplification of the
Riemann $\theta$-function $\theta( \cdot ; \tilde{\Pi}^{\epsilon} )$
as $\epsilon \rightarrow 0$ which becomes expressible in terms of the
Riemann $\theta$-function $\theta( \cdot ; \Pi )$ of
$\hat{\Sigma}$. The result is expressed in the following lemma
\cite{Fay, McKean},

\begin{lemma} \label{lemma: theta reduction}
The behaviour of the Riemann $\theta$-function
$\theta(\vec{z};\tilde{\Pi}^{\epsilon})$ associated with
$\hat{\Sigma}^{\epsilon}$, where $\vec{z} = ( z_0, \bm{z} )^{\sf T}
\in \mathbb{C}^{g+1}$, has the following expansion in the limit
$\epsilon \rightarrow 0$
\begin{equation*}
\theta(\vec{z} ; \tilde{\Pi}^{\epsilon}) =
\theta(\bm{z} ; \Pi^{\epsilon})\\ + \left[ \theta(\bm{z} +
\bm{\Pi}_0^{\epsilon};\Pi^{\epsilon}) e^{i z_0} + \theta(\bm{z} -
\bm{\Pi}_0^{\epsilon};\Pi^{\epsilon}) e^{-i z_0} \right] e^{\pi i
\Pi_{00}^{\epsilon}}\\ + O\left(e^{2 \pi i
\Pi_{00}^{\epsilon}}\right).
\end{equation*}
\begin{proof}
Using the fact that the imaginary part $\text{Im}\,
\tilde{\Pi}^{\epsilon}$ of the period matrix $\tilde{\Pi}^{\epsilon}$
is positive definite we have $\text{Im}\, \Pi_{00}^{\epsilon} =
\text{Im}\,\langle \tilde{\Pi}^{\epsilon} e^{(0)},e^{(0)} \rangle >
0$, where $e^{(0)} = (1, 0, \ldots, 0)^{\sf T}$. It follows that the
quantity $e^{\pi i \Pi_{00}^{\epsilon}}$ tends to zero in the limit
$\epsilon \rightarrow 0$. The result then follows from a
straightforward expansion of $\theta(\vec{z};\tilde{\Pi}^{\epsilon})$
in terms of $e^{\pi i \Pi_{00}^{\epsilon}}$.
\end{proof}
\end{lemma}

Now taking into account all the above limits and dropping all terms of
order $O(\epsilon^2)$, a direct but tedious computation using lemma
\ref{lemma: theta reduction} shows that the difference $\delta Z_i$
between expressions \eqref{reconstruction formula for Z_i + dZ_i} and
\eqref{reconstruction formula for Z_i} contains three types of
contribution
\begin{equation} \label{variation}
\begin{split}
\delta Z_i &= \big( \{\text{periodic}\} + \{\text{periodic}\}
\times e^{i \int_{b_0} d\mathcal{Q}}\\ &\qquad\qquad\qquad\qquad\qquad
+ \{\text{periodic}\} \times e^{- i \int_{b_0} d\mathcal{Q}} \big)
\times e^{\pi i \Pi_{00}^{\epsilon}},\\
&= \delta Z^0_i + \delta Z^+_i + \delta Z^-_i,
\end{split}
\end{equation}
where ``$\left\{ \text{periodic} \right\}$'' denotes functions
periodic in all the angle variables $\varphi_I$ of the underlying
finite-gap solution \eqref{reconstruction formula for Z_i}. The
behaviour of each of the three perturbations in \eqref{variation}
under a shift $\varphi_I \rightarrow \varphi_I + 2 \pi$ of the
$I^{\text{th}}$ angle variable is then
\begin{equation} \label{variation2}
\begin{split}
\delta Z^0_i(\varphi_I + 2 \pi) &= \delta Z^0_i(\varphi_I),\\
\delta Z^{\pm}_i(\varphi_I + 2 \pi) &= e^{\pm 2 \pi i \int_{b_0}
dq^{(I)}} \delta Z^{\pm}_i(\varphi_I).
\end{split}
\end{equation}
The original perturbation $\delta Z_i$ defined by opening up a small
handle is therefore composed of three separate perturbations $\delta
Z^0_i$, $\delta Z^+_i$ and $\delta Z^-_i$, each corresponding to
different stability angles of the underlying solution
\eqref{reconstruction formula for Z_i}. These stability angles can be
read off directly from \eqref{variation2},
\begin{equation} \label{stability angles 1}
\nu^{(I)}_0 = 0, \qquad \nu^{(I)}_{\pm} = \pm 2 \pi \int_{b_0}
dq^{(I)}, \quad I = 1, \ldots, g+1.
\end{equation}
The zero stability angles $\nu^{(I)}_0$ are related to the
$\varphi_I$-translation invariance of the equations of motion
which is explicitly broken by the finite-gap string
\eqref{reconstruction formula for Z_i}. These zero stability angles
can be obtained much more directly by considering two neighbouring
finite-gap strings with the \textit{same} underlying curve
$\hat{\Sigma}$, but slightly different initial divisors
$\hat{\gamma}(0)$ and $\hat{\gamma}^{\epsilon}(0)$ near each
other on $\hat{\Sigma}$. Since there are $g+1$ degrees of freedom in
choosing the perturbed divisor $\hat{\gamma}^{\epsilon}(0)$, for each
angle $\varphi_I$, $I = 1, \ldots, g$ this gives $g+1$ zero-modes, as
one expects from the $\varphi_J$-translation invariance of the
equations of motion which the finite-gap string explicitly breaks,
\begin{equation} \label{stability angles 1}
\nu^{(I)}_{0, J} = 0, \quad J = 1, \ldots, g+1.
\end{equation}

\begin{figure}[h]
\begin{tabular}{ccc}
\psfrag{a0}{\tiny \red $a_0$} \psfrag{a1}{\tiny \red $a_1$}
\psfrag{ag}{\tiny \red $a_g$} \psfrag{b0}{\tiny \green $b_0$}
\psfrag{b1}{\tiny \green $b_1$} \psfrag{bg}{\tiny \green $b_g$}
\psfrag{d}{\tiny $\cdots$} \psfrag{ip}{\tiny $\infty^+$}
\psfrag{im}{\tiny $\infty^-$}
\includegraphics[height=40mm]{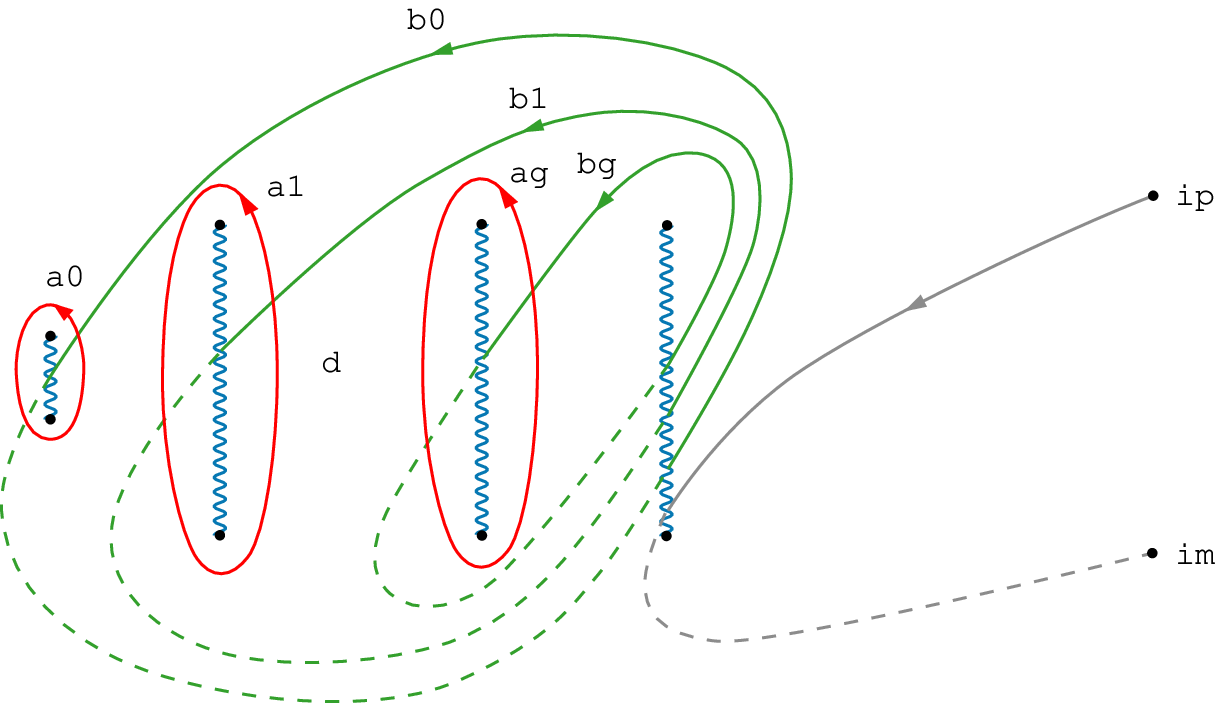} & \qquad &
\psfrag{P0}{\tiny $P_0$} \psfrag{a1}{\tiny \red $a_1$}
\psfrag{ag}{\tiny \red $a_g$} \psfrag{b0}{\tiny \green
$\mathcal{B}_0$} \psfrag{b1}{\tiny \green $b_1$} \psfrag{bg}{\tiny
\green $b_g$} \psfrag{d}{\tiny $\cdots$} \psfrag{ip}{\tiny
$\infty^+$} \psfrag{im}{\tiny $\infty^-$}
\includegraphics[height=40mm]{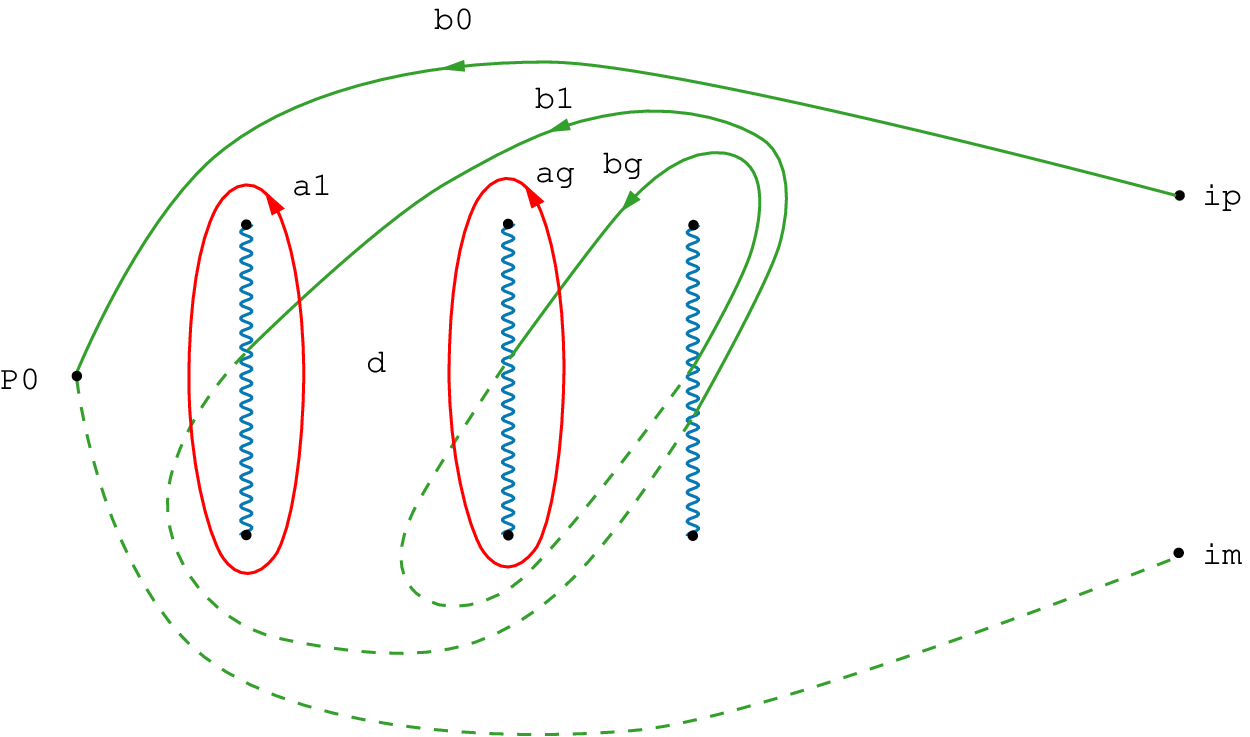}\\
& & \\
$(a)$ & & $(b)$
\end{tabular}
\caption{The canonical cycles before $(a)$ and after $(b)$
shrinking of the $0^{\text{th}}$ cut. Note that it doesn't matter
where this cut lies with respect to the other cuts, but
for the sake of clarity of the figure we chose it to be the
furthest to the left.} \label{Figure: cycles deg}
\end{figure}

Now stability angles are only defined modulo $2 \pi$. But recall from
section \ref{section: periodicity} of chapter \ref{chapter: reality}
that for the solution to be periodic under $\sigma \rightarrow \sigma
+ 2 \pi$ required that the quasi-momentum $dp$ satisfied the condition
\eqref{closed string condition 2}. Here we are interested in using the
$2 \pi$ periodicity of the underlying solution \eqref{reconstruction
formula for Z_i} in the angle variables. This statement is equivalent
to the quasi-actions satisfying
\begin{equation*}
2 \pi \int_{\infty^-}^{\infty^+} dq^{(I)} \in 2 \pi \mathbb{Z},
\qquad I = 1, \ldots, g+1.
\end{equation*}
Therefore we can redefine the stability angles $\nu^{(I)}_{\pm}$ as
\begin{equation} \label{stability angles 2}
\nu^{(I)}_{\pm} = \pm 2 \pi \left( \int_{b_0} dq^{(I)} +
\int_{\infty^+}^{\infty^-} dq^{(I)} \right) = \pm 2 \pi
\int_{\mathcal{B}_0} dq^{(I)},
\end{equation}
where the contour $\mathcal{B}_0$ runs from $\infty^+$ on the top
sheet to $\infty^-$ on the bottom sheet, by going through the
$0^{\text{th}}$ cut, see Figure \ref{Figure: cycles deg}. In the
singular limit $\epsilon \rightarrow 0$ the $0^{\text{th}}$ cut
shrinks to a point, say $P_0$ and so \eqref{stability angles 2}
yields
\begin{equation} \label{stability angles 3}
\begin{split}
\nu^{(I)}_{\pm} &= \pm 2 \pi \left( \int_{\infty^+}^{P_0} dq^{(I)}
+ \int_{\hat{\sigma} P_0}^{\infty^-} dq^{(I)} \right) = \pm 2 \pi
\left( \int_{\infty^+}^{P_0} dq^{(I)} -
\int_{\hat{\sigma} P_0}^{\infty^-} \hat{\sigma}^{\ast} dq^{(I)}
\right)\\
&= \pm 2 \pi \left( \int_{\infty^+}^{P_0} dq^{(I)} -
\int_{P_0}^{\infty^+} dq^{(I)} \right) = \pm 2 \pi
\left( \int_{\infty^+}^{P_0} dq^{(I)} + \int^{P_0}_{\infty^+} dq^{(I)}
\right)\\ &= \pm 4 \pi q^{(I)}(P_0),
\end{split}
\end{equation}
where $q^{(I)}(P) \equiv \int_{\infty^+}^P dq^{(I)}$ with the
integral running along the top sheet (the precise choice of
contour then doesn't matter since $dq^{(I)}$ is normalised).
By performing a similar calculation to the one in \eqref{stability
angles 3} but on $\int_{\mathcal{B}_0} dp = 2 \pi n_0$, $n_0 \in
\mathbb{Z}$ which comes from $2 \pi$ periodicity in $\sigma$, one
derives also an equation for the location of the singular point $P_0$,
namely
\begin{equation} \label{position of singular points}
p(P_0) = n_0 \pi.
\end{equation}
The above analysis shows that to this singular point $P_0$ there
corresponds two stability angles for each of the $g+1$ cuts
determined by the $\mathcal{B}_0$-period of corresponding
quasi-action $dq^{(I)}$ or
\begin{equation} \label{stability angles 4}
\nu^{(I)}_{\pm} = \pm 4 \pi q^{(I)}(P_0).
\end{equation}

\section{Semi-classical energy spectrum} \label{section: semiclassical 2}


Applying the Bohr-Sommerfeld conditions \eqref{BS6} of chapter
\ref{chapter: semiclassical} to the filling fractions, which were
shown in chapter \ref{chapter: symplectic} to be the action variables
of the string, we find
\begin{equation} \label{BS7}
\frac{\mathcal{S}_I}{\hbar} = N_I + \frac{1}{2} + \sum_{\alpha =
g+2}^{\infty} \left( n_{\alpha} + \frac{1}{2} \right)
\frac{\nu_{\alpha}^{(I)}}{2 \pi} + O(\hbar),
\end{equation}
where the sum is over positive stability angles and in the string
theory context we have $\hbar = \frac{1}{\sqrt{\lambda}}$. In
\eqref{BS7} we used the fact that the Maslov index for the
$\mathcal{A}_I$-cycle ($I = 1, \ldots, g+1$) in the generalised
Jacobian $J(\hat{\Sigma},\infty^{\pm})$ is simply $\mu_I = 2$.

Obtaining the energy spectrum from \eqref{BS7} is relatively
straightforward assuming the system is semiclassically integrable,
which guarantees that the action operators satisfy
$[\hat{\mathcal{S}}_i, \hat{\mathcal{S}}_j] = O(\hbar^3)$. In the
semiclassical regime the Hamiltonian is then defined by the same
classical function of the actions
$\mathcal{E}_{\text{cl}}[\mathcal{S}_1,\ldots,\mathcal{S}_{g+1}]$
but evaluated on the action operators, that is
\begin{equation*}
\hat{\mathcal{H}}_{\text{string}} =
\mathcal{E}_{\text{cl}}[\hat{\mathcal{S}}_1,\ldots,\hat{\mathcal{S}}_{g+1}]
+ O(\hbar^2).
\end{equation*}
It follows that the energy spectrum is simply the classical energy
$\mathcal{E}_{\text{cl}}$ evaluated on the eigenvalues of the action
variables \eqref{BS7} namely
\begin{multline*}
\mathcal{E} = \mathcal{E}_{\text{cl}} \left[ N_1 \hbar +
\frac{\hbar}{2} + \sum_{\alpha = g+2}^{\infty} \left( n_{\alpha} +
\frac{1}{2} \right) \frac{\nu_{\alpha}^{(1)}}{2 \pi} \hbar,\ldots,
\right. \\
\left. N_{g+1} \hbar + \frac{\hbar}{2} + \sum_{\alpha =
g+2}^{\infty} \left( n_{\alpha} + \frac{1}{2} \right)
\frac{\nu_{\alpha}^{(g+1)}}{2 \pi} \hbar \right] + O(\hbar^2).
\end{multline*}
We now Taylor expand this using the fact that $N_I \gg n_{\alpha}$ and
$\hbar \ll 1$ to obtain
\begin{equation*}
\mathcal{E} = \mathcal{E}_{\text{cl}} \left[\left(N_1 +
\frac{1}{2}\right) \hbar,\ldots, \left(N_{g+1} +
\frac{1}{2}\right)\hbar \right] + \sum_{I= 1}^{g+1} \sum_{\alpha =
g+2}^{\infty} \left( n_{\alpha} + \frac{1}{2} \right) \frac{\partial
\mathcal{E}_{\text{cl}}}{\partial \mathcal{S}_I}
\frac{\nu^{(I)}_{\alpha}}{2 \pi} \hbar.
\end{equation*}
Using \eqref{dP and dE 2b} and \eqref{stability angles 2} to
express $\partial \mathcal{E}_{\text{cl}}/\partial \mathcal{S}_I$ and
$\nu^{(I)}_{\alpha}$ respectively as $\mathcal{B}$-periods,
\begin{equation*}
\mathcal{E} = \mathcal{E}_{\text{cl}} \left[\left(N_1 +
\frac{1}{2}\right) \hbar,\ldots, \left(N_{g+1} +
\frac{1}{2}\right)\hbar \right] + \sum_{I= 1}^{g+1} \sum_{\alpha =
g+2}^{\infty} \left( n_{\alpha} + \frac{1}{2} \right)
\int_{\mathcal{B}_I} \frac{dq}{2 \pi} \int_{\mathcal{B}_{\alpha}}
dq^{(I)} \hbar,
\end{equation*}
where $\mathcal{B}_{\alpha}$ is the contour running from
$\infty^+$ to the singular point labelled $\alpha$ on the top
sheet, and back on the bottom sheet to $\infty^-$. The sum over
$I$ can now be performed using equation \eqref{important formula}
which yields
\begin{equation} \label{main result}
\mathcal{E} = \mathcal{E}_{\text{cl}} \left[\left(N_1 +
\frac{1}{2}\right) \hbar,\ldots, \left(N_{g+1} +
\frac{1}{2}\right)\hbar \right] + \sum_{\alpha = g+2}^{\infty} \left(
n_{\alpha} + \frac{1}{2} \right) \int_{\mathcal{B}_{\alpha}}
\frac{dq}{2 \pi} \hbar.
\end{equation}
This is the main result of this section. It expresses the
semiclassical energy spectrum corresponding to a finite-gap
solution as the sum of two terms. The order $O(1)$ term is the
classical energy of a finite-gap string evaluated on half-integer
quantised filling fractions and the order
$O(\frac{1}{\sqrt{\lambda}})$ term is an infinite sum over
fluctuation energies $\delta \mathcal{E}_{\alpha}$ for each
singular point $P_{\alpha}$ of the spectral curve.

Equation \eqref{main result} provides a closed form
expression for the fluctuation energy $\delta \mathcal{E}_{\alpha}$ of
any singular point $P_{\alpha}$. The required ingredient is the
differential of the quasi-energy $dq$ which is the Abelian
differential on $\hat{\Sigma}$ uniquely defined by its asymptotics
\eqref{quasi-energy asymptotics pm1} at $x = \pm 1$. The (pinched)
contour $\mathcal{B}_{\alpha}$ runs from $\infty^+$ to $P_{\alpha}$ on
the top sheet, then back from $\hat{\sigma} P_{\alpha}$ to $\infty^-$
on the bottom sheet. Thus the integral can be evaluated more
explicitly using the same argument as in \eqref{stability angles
3}. Combining this with the result of equation \eqref{position of
singular points} we have proved,
\begin{theorem} \label{thm: fluctuation energies}
Let $P_{\alpha} \in \Gamma$ be any singular point of the spectral
curve $\Gamma$. Then the value of the quasi-momentum at $P_{\alpha}$
is an integer multiple $n_{\alpha} \in \mathbb{Z}$ of $\pi$. Moreover,
the value of the quasi-energy at $P_{\alpha}$ gives the fluctuation
energy $\delta \mathcal{E}_{\alpha}$ of $P_{\alpha}$, namely
\begin{equation} \label{fluctuation energies}
p(P_{\alpha}) = n_{\alpha} \pi, \qquad \delta \mathcal{E}_{\alpha} =
\frac{q(P_{\alpha})}{\pi \sqrt{\lambda}}.
\end{equation}
\end{theorem}

\begin{remark}
These fluctuation energies can also be obtained more directly by
computing the stability angles of a finite-gap string \textit{periodic
in the worldsheet $\tau$-coordinate}. For this one repeats the
calculation of section \ref{section: perturbations} on a finite-gap
solution $Z_i$ with $Z_i(\tau + T) = Z_i(\tau)$. Its perturbations
$Z_i + \delta Z_i$ defined by opening up a cut on the underlying curve
$\hat{\Sigma}$ of $Z_i$ are in general not periodic in $\tau$. We
obtain the stability angles $\nu_{\alpha} = \frac{T
q(P_{\alpha})}{\pi}$ so that the fluctuation energies are given by
$\delta \mathcal{E}_{\alpha} = \frac{\nu_{\alpha}}{T}
\frac{1}{\sqrt{\lambda}}$ which reproduces \eqref{fluctuation
energies}.
\end{remark}

Note that the infinite sum $\sum_{\alpha =
g+2}^{\infty} \left( n_{\alpha} + \frac{1}{2} \right) \delta
\mathcal{E}_{\alpha}$ in \eqref{main result} is only formal and
requires regularisation. Still, we can formally rewrite the main
result \eqref{main result} in a way that makes the quantisation of all
the fillings apparent, including the fillings of the singular points.
If we formally think of the function $\mathcal{E}_{\text{cl}}$ as
depending on the infinite set of filling fractions $\{ \mathcal{S}_I
\}_{I=1}^{g+1}$, $\{ \mathcal{S}_{\alpha} \}_{\alpha = g+2}^{\infty}$
(all but finitely many of which are turned off for the classical
finite-gap solutions) then we can interpret the
$\mathcal{B}_{\alpha}$-period of $dq/2 \pi$ as $\partial
\mathcal{E}_{\text{cl}}/\partial \mathcal{S}_{\alpha}$ using a formal
analogue of \eqref{dP and dE 2b} for an \textit{infinite}-gap
solution. One can then resum the resulting Taylor expansion to obtain
the following formal expression for the semiclassical energy spectrum
\begin{equation} \label{formal result}
\mathcal{E} = \mathcal{E}_{\text{cl}} \left[\left(N_1 +
\frac{1}{2}\right) \hbar,\ldots, \left(N_{g+1} +
\frac{1}{2}\right)\hbar, \left(n_{g+2} + \frac{1}{2}\right) \hbar,
\ldots \right].
\end{equation}
We stress that this is only a formal derivation as rigorously one
would have to regularise the divergent infinite sum over stability
angles at the intermediate steps as well as subtract off the energy of
the vacuum (\textit{i.e.} the zero cut finite-gap solution). But
formally at least the result of the above derivation is the following:
\begin{itemize}
  \item The semiclassical energy spectrum is obtained by evaluating
the classical energy function of an infinite-gap solution on filling
fractions quantised to half-integer multiples of $\hbar$.
  \item The infinite number of singular points of the spectral curve
$\text{det}\, (\Omega(x) - y {\bf 1}) = 0$ which accumulate at $x
= \pm 1$ must be filled with half a unit of $\hbar$ in their
ground state with an additional integer multiple of $\hbar$ for
excitations.
\end{itemize}

\begin{remark}
The energy $\mathcal{E}_{\text{cl}}$ we have been using is not the
space-time energy $\Delta$ of the classical solution but rather the
worldsheet energy. They are related by the simple formula \eqref{ws
vs st energy}.
\end{remark}

\subsection*{Comparison with alternative approach}

In \cite{Gromov+Vieira1} an alternative method was proposed for
extracting the semiclassical energy spacing around any given
classical solution from the algebraic curve $\hat{\Sigma}$ itself,
without making use of the divisor $\hat{\gamma}(t)$ on $\hat{\Sigma}$
as we have done. The heart of the method resides in the assumption
that the filling fractions $\mathcal{S}_I$ become quantised in integer
units at least in a semiclassical approximation. This was interpreted
in the language of the gauge theory side by attributing to a single
Bethe root one unit of filling fraction. In the semiclassical
quantisation of a solution each cut of its algebraic curve thus turns
into a large clump of Bethe roots with the filling fraction counting
the number of such roots. The idea of \cite{Gromov+Vieira1} for
obtaining the semiclassical energy spacings is then to compare the
energies of two neighbouring classical solutions differing only by a
single Bethe root. If the underlying solution is characterised by the
quasi-momentum $p(x)$ and has $K = g+1$ cuts $\mathcal{C}_j$ with mode
numbers $n_j \in \mathbb{Z}, j = 1, \ldots, K$,
\begin{equation} \label{cuts}
p(x + i0) + p(x - i0) = 2 \pi n_j, \quad x \in \mathcal{C}_j, j =
1, \ldots, K,
\end{equation}
then its perturbation is characterised by a perturbed
quasi-momentum $p(x) + \delta p(x)$ with still the same $K$ cuts
but also with an extra isolated Bethe root at $x_{K+1}$ with mode
number $n_{K+1} \in \mathbb{Z}$,
\begin{subequations} \label{perturbed cuts}
\begin{equation} \label{perturbed cuts 1}
p(x + i0) + \delta p(x + i0) + p(x - i0) + \delta p(x - i0) = 2
\pi n_j, \quad x \in \mathcal{C}_j, j = 1, \ldots, K,
\end{equation}
\begin{equation} \label{perturbed cuts 2}
p(x_{K+1}) + \delta p(x_{K+1}) + p(x_{K+1}) + \delta p(x_{K+1}) =
2 \pi n_{K+1}.
\end{equation}
\end{subequations}
By using \eqref{cuts} we may simplify \eqref{perturbed cuts 1} to
\begin{subequations} \label{perturbed cuts bis}
\begin{equation} \label{perturbed cuts bis 1}
\delta p(x + i0) + \delta p(x - i0) = 0, \quad x \in
\mathcal{C}_j, j = 1, \ldots, K.
\end{equation}
and since $\delta p(x)$ is small, to lowest order equation
\eqref{perturbed cuts 2} yields
\begin{equation} \label{perturbed cuts bis 2}
p(x_{K+1}) = \pi n_{K+1},
\end{equation}
\end{subequations}
Equations \eqref{perturbed cuts bis} are the starting point in
\cite{Gromov+Vieira1} for obtaining the semiclassical energy
spacings by reading them off from $\delta p(x)$.

Let us now show that the semiclassical energy spacings obtained by
this method agrees with the fluctuation energies of theorem
\ref{thm: fluctuation energies}. We know from \eqref{dP and dE 2b}
that the variation of the energy $\mathcal{E}$ of a classical solution
as we vary the moduli $\mathcal{S}_I$ is
\begin{equation*}
\delta \mathcal{E} = \sum_{I = 1}^{g+1} \left( \int_{\mathcal{B}_I}
\frac{dq}{2\pi} \right) \delta \mathcal{S}_I.
\end{equation*}
It follows that adding a single Bethe root (which would correspond to
setting $\delta \mathcal{S}_J = \hbar$ for some $J$) should increase
the energy of the solution by
\begin{equation} \label{Energy spacing}
\delta \mathcal{E} = \int_{\mathcal{B}_J} \frac{dq}{2\pi} \hbar.
\end{equation}
This is exactly the formula \eqref{fluctuation energies} for the
fluctuation energies derived in this chapter. Moreover, equation
\eqref{perturbed cuts bis 2} is exactly the same formula as in
\eqref{fluctuation energies} for the value of the quasi-momentum
at a singular point. Thus theorem \ref{thm: fluctuation energies}
predicts the same energy spacing \eqref{Energy spacing} as we
would expect if Bethe roots carried $\hbar =
\frac{1}{\sqrt{\lambda}}$ units of filling fraction. Theorem
\ref{thm: fluctuation energies} however was proved without any
input from the gauge theory side and was derived by a purely
string theoretic calculation.

%% file: Conclusion.tex
\newpage

\noindent {\bf \large $\blacktriangleright$ Integrability of
string theory on $AdS_5 \times S^5$}

\noindent It is now a very well established fact that the
Metsaev-Tseytlin action \cite{Metsaev:1998it} for type IIB
superstrings on $AdS_5 \times S^5$ admits a Lax connection
\cite{Bena:2003wd}. This connection gives rise through the usual
construction of the monodromy matrix to a wealth of integrals of
motion. However the existence of a Lax connection is only half the
conditions required for Liouville integrability. Indeed, as we
have stressed in chapter \ref{chapter: integrability}, it is also
necessary that the integrals of motion be in pairwise involution
with respect to the Poisson structure.

{\bf Non-ultralocality.} The main obstacle in proving the
involution property was the non-ultralocal nature of the Poisson
brackets of the current \eqref{jj Poisson brackets}. The
problematic $\delta'$-term gives rise in the algebra of monodromy
matrices to ambiguous $\chi$-terms containing the value of the
characteristic functions $\chi(\sigma; \sigma_1, \sigma_2)$ at the
endpoints $\sigma = \sigma_1, \sigma_2$. Yet no value can be given
such that the anti-symmetry property and the derivation rule are
satisfied without violating the Jacobi identity for the Poisson
bracket of monodromies.

{\bf Maillet regularisation.} A way around this problem proposed
by Maillet \cite{Maillet, Maillet2, Maillet3} is to define a
\textit{weak} bracket by `temporarily' giving independent
definitions for each multiply nested Poisson bracket of
monodromies. Using this weak bracket consistent with all the
fundamental properties of the Poisson bracket one then follows the
usual arguments to show that $\{ \tr \Omega(x), \tr \Omega(x') \}
= 0$. But since this final bracket is equal to zero, the Jacobi
identity involving it obviously hold. This final bracket thus
holds in the usual \textit{strong} sense.

{\bf String theory.} In chapter \ref{chapter: integrability} we
applied Maillet's procedure to string theory on $\mathbb{R} \times
S^3$. In particular we showed that the integrals of motion are in
pairwise involution with respect to the Dirac bracket associated
with Virasoro constraints and static gauge fixing conditions, thus
proving the complete statement of integrability for strings on
$\mathbb{R} \times S^3$. These arguments were later generalised to
the case of bosonic strings on $AdS_5 \times S^5$ in a series of
papers by Kluso\v{n} \cite{Kluson:2007gp, Kluson:2007vw,
Kluson:2007md, Kluson:2007ua} (see also
\cite{Das:2004hy,Das:2005hp}).
\newline
\newline

\noindent {\bf \large $\blacktriangleright$ Finite-gap strings on
$\mathbb{R} \times S^3$}

\noindent The fact that superstring theory on $AdS_5 \times S^5$
possesses an infinite number of integrals of motion has been
thoroughly exploited in the literature (initiated by \cite{KMMZ}
in the $SU(2)$ sector and eventually in the general case by
\cite{Beisert:2005bm}) to completely classify the full set of
classical solutions on $AdS_5 \times S^5$. More precisely, every
\textit{finite}-gap solution was assigned a \textit{finite}-genus
algebraic curve whose moduli encodes the integrals of motion.
However, the algebraic curve is not enough to uniquely
characterise a specific solution. The identification of the extra
data and the reconstruction of the corresponding solution was the
subject of Part \ref{part: Algebro}.

{\bf Finite-gap integration.} The existence of a flat Lax
connection $J(x)$ is the starting point in the theory of finite-gap
integration \cite{Belokolos, Babelon, Krichever2, Krichever3,
Krichever4}. The key idea behind this method is that analytic
functions are uniquely specified by only a finite amount of data,
such as their poles and zeroes. In chapter \ref{chapter: curves}
we constructed the KMMZ curve $\hat{\Sigma}$, equipped with a
meromorphic differential $dp$, which provides an arena for doing
complex analysis. We also showed that the eigenvectors of the
monodromy matrix define a vector function $\bm{\psi}(P)$ on
$\hat{\Sigma}$. After normalising it we can determine its analytic
properties.

{\bf The divisor.} Choosing $\bm{\psi}(P)$ to solve the equation
$(d - J(x)) \bm{\psi}(P) = 0$ we find it is uniquely specified by
$g+1$ poles, its value ${\tiny \left( \!\! \begin{array}{c} 1 \\
0 \end{array} \!\! \right)}$ and ${\tiny \left( \!\!
\begin{array}{c} 0 \\ 1 \end{array} \!\! \right)}$ at
$\infty^{\pm} \in \hat{\Sigma}$ and essential singularities at $x
= \pm 1$. The remarkable fact is that its divisor of poles
$\hat{\gamma}(0)$ is \textit{static}. Since the Lax connection
$J(x)$ can be recovered from $\bm{\psi}(P)$ which in turn can be
reconstructed from its analytic data, we were able to reconstruct
the current $j$. As a quick check the general solution was shown
in \cite{Paper3} to reduce in the elliptic case ($g = 1$) to the
so called helical solutions of \cite{Okamura:2006zv} obtained by
the method of Pohlmeyer reduction. It would be very nice to extend
this construction to larger sectors and in particular to the full
case of bosonic strings on $AdS_5 \times S^5$.

{\bf Induced symplectic structure.} Since a finite-gap solution is
parametrised by the algebro-geometric data consisting of the KMMZ
curve and the divisor $\hat{\gamma}(0)$, it can be thought of as a
map $\{ (\hat{\Sigma}, dp) , \hat{\gamma}(0) \} \mapsto j$. In
chapter \ref{chapter: symplectic} we obtained the pullback of the
bracket \eqref{jj Poisson brackets} of currents $j$ to the
algebro-geometric data by making use of the Maillet regularised
bracket of monodromy matrices obtained in chapter \ref{chapter:
integrability}. The remarkable result is that the induced bracket
assumes the canonical Darboux form \eqref{DB alg data bis} when
expressed in terms of two special Abelian integrals on
$\hat{\Sigma}$: the quasi-momentum $p$ and the Zhukovsky transform
of the spectra parameter $x$,
\begin{equation*}
z = x + \frac{1}{x}.
\end{equation*}
It would be very interesting to check whether this is still true
for finite-gap strings on $AdS_5 \times S^5$. In view of ultimately
quantising the string directly, the fact that the symplectic structure
is canonical with respect to the spectral parameter $z$ strongly
suggest the right variables for an exact quantisation.

{\bf Reality conditions.} Since the method of finite-gap
integration is so firmly grounded in complex analysis, the general
solution it produces satisfies the complexification of the
equations we set out to solve. In chapter \ref{chapter: reality}
we obtained the necessary restrictions on the algebro-geometric
data $\{ (\hat{\Sigma}, dp) , \hat{\gamma}(0) \}$ for the
reconstructed solution to describe a closed string on $\mathbb{R}
\times S^3$. In particular the condition on the KMMZ curve is that
its branch points come in complex conjugate pairs. It would be
interesting to derive the analogous fact in the non-compact $AdS$
sectors where the dual gauge theory predicts that the branch
points should all be real \cite{Kazakov:2004nh}.
\newline
\newline

\noindent {\bf \large $\blacktriangleright$ Semiclassical strings
on $\mathbb{R} \times S^3$}

In chapter \ref{chapter: semi} we performed a first principle
semiclassical quantisation on the general finite-gap solution
constructed in Part \ref{part: Algebro}. The main result of this
analysis is the formula \eqref{fluctuation energies} for the
fluctuation energies around a generic finite-gap solution. It was
shown to agree with the implicit method of Gromov and Vieira
\cite{Gromov+Vieira1} for extracting fluctuation energies from the
spectral curve and on which the subsequent papers
\cite{Gromov+Vieira2, Gromov+Vieira3} relied. Our result
\eqref{main result} for the semiclassical spectrum is only formal
since one would need to regularise the infinite sum over
fluctuation energies as well as subtract from it the vacuum energy
given by a zero-gap solution (\textit{i.e.} the BMN string). In
any case, such a regularisation would only be interesting in the
full case of strings on $AdS_5 \times S^5$ where the fluctuations
transverse to the subsector $\mathbb{R} \times S^3$ are included along
with the fermions.
More formally still, we showed that the energy spectrum can be
obtained by evaluating the classical energy of an
\textit{infinite}-gap string \eqref{formal result} with all its
infinite filling fractions quantised to half-integer multiples of
$\hbar$, namely
\begin{equation*}
\mathcal{E} = \mathcal{E}_{\text{cl}}\left[ \left( N_1 +
\frac{1}{2} \right) \frac{1}{\sqrt{\lambda}}, \ldots \right].
\end{equation*}
This result is to be interpreted as a limit of expressions where a
finite but \textit{arbitrary} number of first entries are of order
$O(1)$ corresponding to the tree level order and the remaining
infinite number of entries encode the 1-loop corrections of order
$O(\frac{1}{\sqrt{\lambda}})$.

Finally, in view of ultimately obtaining an exact quantisation of
string theory on $AdS_5 \times S^5$ we have argued that operator
ordering issues will be of crucial importance since they already
appear in the semiclassical analysis. By assuming for simplicity
that the cohomology class of the subprincipal form vanished, our
results for the fluctuation energies for the $SU(2)$ sector agreed
with \cite{Gromov+Vieira1, Gromov+Vieira2, Gromov+Vieira3}. This
rules out many operator orderings for an exact quantisation and
provides further hints as to how one might go about quantising
string theory on $AdS_5 \times S^5$.